%% file: thesis.tex
\documentclass[twoside,12pt,a4paper]{report}

\usepackage{suthesis-2e,amsfonts,amscd,graphicx,rotating,afterpage,overpic}
\usepackage{times}
\begin{document}

%\setlength{\baselineskip}{1ex}

% TAKE OUT FOR LATEX, INCLUDE FOR PDFLATEX
%\DeclareGraphicsExtensions{.pdf,.png,.gif,.jpg,.tif,.pcx}

\def\be{\begin{equation}}
\def\ee{\end{equation}}

\def\bfig{\begin{figure}[h]}
\def\efig{\end{figure}}

\def\btab{\begin{table}[h]}
\def\btab{\end{figure}}

\def\bcen{\begin{center}}
\def\ecen{\end{center}}

\newcommand{\ket}[1]{\mbox{$|#1\rangle$}}
\newcommand{\bra}[1]{\mbox{$\langle#1|$}}

    \title{Experimental Quantum Computation with Nuclear Spins 
	in Liquid Solution}
    \author{Lieven M. K. Vandersypen}
    \dept{Electrical Engineering}
    \principaladviser{James S. Harris}
    \coprincipaladviser{Isaac L. Chuang}
%    \firstreader{Isaac L. Chuang}
    \secondreader{Yoshihisa Yamamoto}

    \submitdate{July 2001} 

    \figurespagefalse
    \tablespagefalse

\beforepreface

%\newpage~\newpage
    \prefacesection{Abstract}

\begin{doublespace}
%\vspace*{-1.8ex} 
Quantum computation offers the extraordinary promise
of solving mathematical and physical problems which are simply beyond
the reach of classical computers. However, the experimental
realization of quantum computers is extremely challenging, because of
the need to initialize, manipulate and measure the state of a set of
coupled quantum systems while maintaining fragile quantum coherence.

In this thesis work, we have taken significant steps towards the
realization of a practical quantum computer: using nuclear spins and
magnetic resonance techniques at room temperature, we provided proof
of principle of quantum computing in a series of experiments which
culminated in the implementation of the simplest instance of Shor's
quantum algorithm for prime factorization ($15=3\times5$), using a
seven-spin molecule. This algorithm achieves an exponential advantage
over the best known classical factoring algorithms and its
implementation represents a milestone in the experimental exploration
of quantum computation.

These remarkable successes have been made possible by the synthesis of
suitable molecules and the development of many novel techniques for
initialization, coherent control and readout of the state of multiple
coupled nuclear spins.  Furthermore, we devised and implemented a
model to simulate both unitary and decoherence processes in these
systems, in order to study and quantify the impact of various
technological as well as fundamental sources of errors.

In summary, this work has given us a much needed practical
appreciation of what it takes to build a quantum computer.
Furthermore, while liquid NMR quantum computing has well-understood
scaling limitations, many of the techniques that originated from this
research may find use in other, perhaps more scalable quantum computer
implementations.
\end{doublespace}

\newpage~\newpage
    \prefacesection{Preface}

A long, long time ago, in a land far away, a man was sentenced to
death. The man requested to speak to the King, and the King agreed to
hear him. ``If you let me live for one more year,'' offered the man,
``I promise to make your horse fly high above the land.'' The King
realized that a flying horse would be quite unique, and took immense
pleasure in the prospect of possesing the only flying horse in the
land. He agreed to set the man free and let him live one more year.

When the man came home and told his wife what he had promised the
King, she exclaimed in anguish: ``But you'll never be able to make the
King's horse fly!''. ``Well,'' said the man, ``I know that, and you
know that, but in the meantime many things can happen. The country may
go to war, the King may die, or the King's horse may die, but I will
still be alive for another year.''~\footnote{Free after James Harris.
Thanks to Nico for the drawing of the flying horse.}

\vspace*{2ex}
\begin{center}
\includegraphics*[width=10cm]{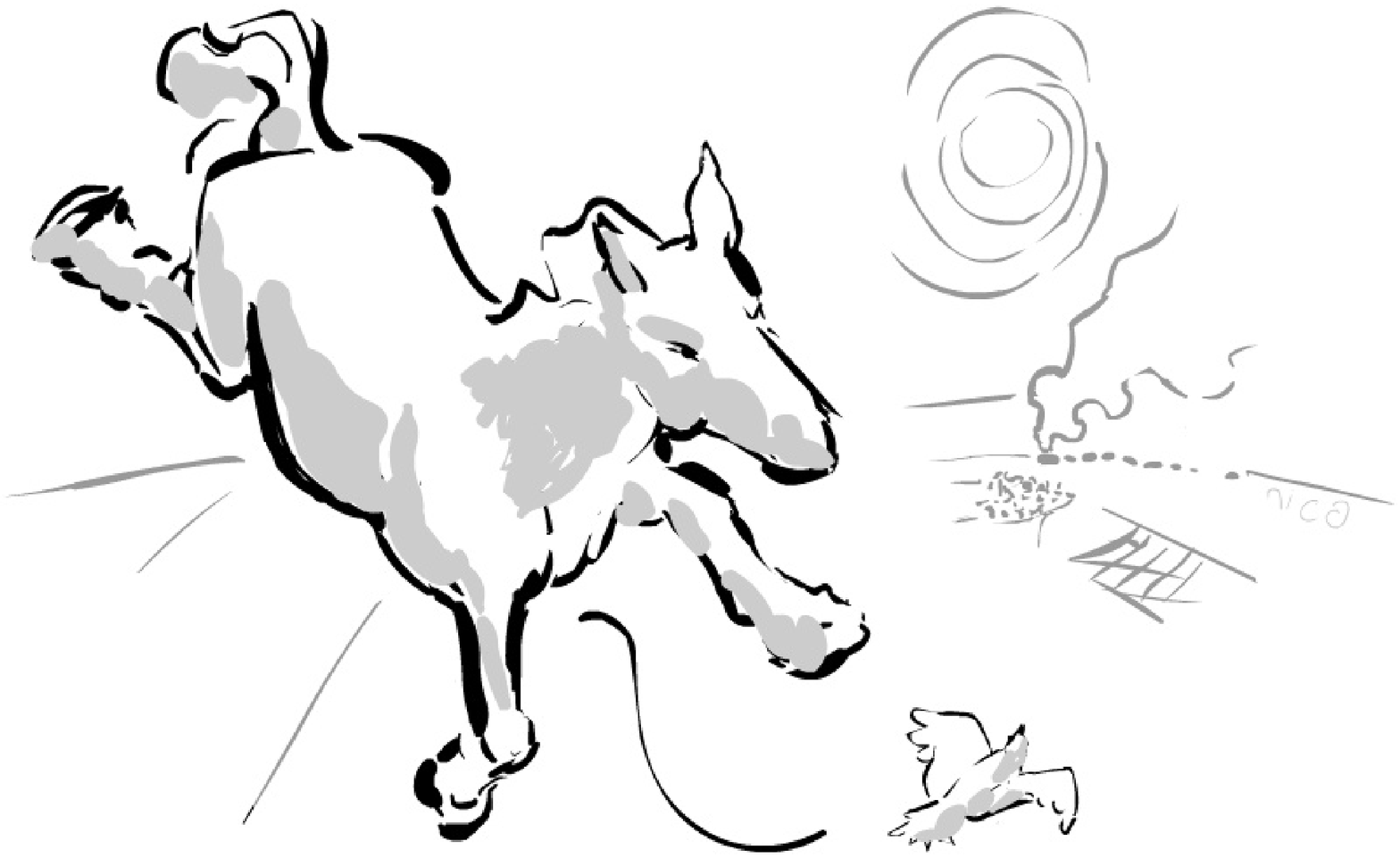} 
\end{center}

Now, can we build a quantum computer ? And should we promise to build
one ? These are the broad and ambitious questions underlying this
thesis work. The final verdict is not in yet, and fortunately we are
given more than one year. However, it is for certain that only by
studying quantum computation experimentally, can we begin to
understand and appreciate at a practical level what it would take to
turn the dream of quantum computation into reality. This is the
purpose of my work.

%\end{document}

%%%%%%%%%%%%%%%%%%%%%%%%%%%%%%%%%%%%%%%%%%%%%%%%%%%%%%%%%%%%%%%%%%%%%
\newpage~\newpage
    \prefacesection{Acknowledgements}
%\vspace*{-2ex}
As a mechanical engineering sophomore at the Katholieke Universiteit
Leuven, I took a great and inspiring class in quantum mechanics with
Guido Langouche. It was the beginning of a profound interest in
quantum mechanics, and a fascination which continues to this
day. Later, a second interest developed: to build mechanical systems
on a very small scale. Two other great courses, with Hendrik Van
Brussel and Willy Sansen, gave me the opportunity to further explore
this area.

When I came to Stanford University for graduate school, I was looking
for a project to combine these two interests, for example by studying
if and how quantum mechanical effects could be observed in
micromachined structures.  For one quarter, I worked on
microcantilevers in the most stimulating group of Yoshihisa Yamamoto.

Then I talked to Jim Harris, who suggested I try to microfabricate
components for an NMR quantum computer. Soon after I talked Ike
Chuang, the initiator of this project (then at Los Alamos, later at
IBM Almaden), and started reading about quantum computing, I became
increasingly fascinated by quantum computing itself. I knew this is
what I wanted to work on!

It was the beginning of an extraordinary four years, four years I am
very thankful for. The fact that this has been such a wonderful time
is due to many great people.

Jim Harris, my advisor and ``coach'', has generously introduced me to
the right people to make my work a success, and strongly supported me
in many ways, in and outside research. His view on life and on what is
really important is a true inspiration to me.
Ike Chuang, my co-advisor, gave me both guidance and independence, at
the right times. He taught me how to present my work and put it in
perspective, and also to think positive and creatively (to say ``what
would it take to {\em make} this work'' instead of ``this won't
work'').
Also, Coach's and Ike's strong support and belief in my abilities, as
well as their encouragement for my non-research activities, have been
invaluable to me.

Matthias Steffen worked very closely with me for about two years and a
half, first as an apprentice, but increasingly as a great
co-worker. He brought in many useful ideas and did a lot of the work
in the later experiments.  Xinlan Zhou kindly helped me out with
many theoretical questions throughout the past years.

Nino Yannoni discovered most of the molecules we have used, and also
provided me with many wise words. Mark Sherwood has greatly
contributed for concepts and techniques in NMR and for molecule
selection. Greg Breyta gave us a lot of the time he didn't have, to
synthesize the five and seven spin molecules.

I am very grateful to my other co-authors, in particular Richard
Cleve, with whom I worked so pleasantly. I am also indepted to the
many colleagues from whom I've learnt and with whom I had such nice
interactions (especially Dorit Aharonov, Michael Nielsen, David
DiVincenzo and Ray Freeman).

My close colleagues Anne Verhulst and Oskar Liivak have contributed to
a great working environment and provided many useful discussions. So
have the many summer students in the group.

Lois Durham made her NMR lab available for my early experiments, and
we have had a fruitful collaboration with the people at Varian NMR.

At Stanford, I have particularly enjoyed Tom Cover's lectures on
information theory. He gave me a deep apprecation for this beautiful
field, which is now being revisited in the context of quantum
information. Similarly, Rajeev Motwani's course on classical
complexity theory and automata helped me put quantum computation in
perspective.

Patricia Ryan, coach of the Stanford Improvisors, has affected my life
in a very positive way.  I say ``yes'' more often now, I have more
adventures, and I learnt it's ok if things don't always work out.
Philip Zimbardo's course on psychology has been an inspiration for
teaching and a lesson for life.

The financial support of a Francqui Fellowship of the Belgian-American
Educational Foundation, a Yansouni Family Stanford Graduate
Fellowship, DARPA, and IBM Research, have given me the opportunity to
freely pursue my interests.

The warm support of my parents, family, friends (especially PS, EV and PC)
and roommates has done me much good, especially in the days when things
didn't go so well. I cherish the many good moments I shared with all of
them.

%\end{document}

\afterpreface

%\end{document}
\include{intro}

\newpage~\newpage
\include{qct}
\newpage~\newpage
\include{impl}

%\newpage~\newpage
\include{nmrqc}
%\newpage~\newpage
\include{expt}
%\newpage~\newpage
\include{concl}

\newpage~\newpage
\appendix
\include{app}

%\newpage~\newpage
\small
\begin{singlespace}
\bibliographystyle{thesis}
\bibliography{thesis}
\end{singlespace}

\end{document}

%% file: intro.tex
    \chapter{Introduction}

\section{Historical background}

``There is plenty of room at the bottom.'' This was the title of a now
classic 1959 talk given by Richard Feynman at the annual meeting of
the American Physical Society~\cite{Feynman60a}. In this talk, Feynman
gave physicists and engineers a wonderful challenge: to manipulate and
control things on a {\em small} scale.  In particular, he challenged
his audience to think about building very small computers, with wires
just 10 or 100 atoms in diameter, and circuits just a few thousand
angstroms across.  Forty years later, semiconductor technology is
rapidly approaching these dimensions, driven by Moore's law.  But
Feynman didn't mean just small, he meant {\em really} small:

\begin{quote}
{\em ``When we get to the very, very small world --- say circuits of
seven atoms --- we have a lot of new things that would happen that
represent completely new opportunities for design. Atoms on a small
scale behave like {\em nothing} on a large scale, for they satisfy the
laws of quantum mechanics. So, as we go down and fiddle around with
the atoms down there, we are working with different laws, and we can
expect to do different things. We can manufacture in different
ways. We can use, not just circuits, but some system involving the
quantized energy levels, or the interactions of quantized spins,
etc.''}
\end{quote}

This is the earliest reference I am aware of that hints at the subject
matter of my thesis work. With reference to his daring ideas, Feynman
also made the following crucially important point:

\begin{quote}
{\em ``It is not an attempt to violate any laws; it is something, in
principle, that can be done; but in practice, it has not been done
because we are too big.''}
\end{quote}

So what {\em are} the laws which limit computation ? How much energy
does it take to compute, and how much time and space does a
computation require ?

The relationship between energy consumption and computation has been
studied in detail by Rolf Landauer. In a 1961
paper~\cite{Landauer61a}, he showed that the amount of energy
dissipated into the environment when a single bit of information is
{\em erased}, is at least $k_B T \ln 2$, where $k_B$ is Boltzman's
constant and $T$ is the temperature of the environment. As a result,
{\em irreversible} logic gates, such as the {\sc nand} gates in
today's computers, must dissipate a finite amount of energy, as
information is lost when executing the gate (it is not possible to run
the gate backwards and reconstruct the input from the output).
Remarkably, Lecerf~\cite{Lecerf63a} and Bennett~\cite{Bennett73a}
later showed that it is possible to perform {\em universal}
computation {\em reversibly}, without ever erasing information, and
furthermore that universal computation is possible without net
dissipation of energy.

The time and space resources needed for computation, and in particular
how the resources scale with the problem size, are the subject of
complexity theory. Arguably the most significant result of this field,
which started with Alan Turing's introduction of the Turing
machine~\cite{Turing36a}, is the strong Church-Turing
thesis~\cite{Church36a,Davis65a}. It states that ``Any model of
computation can be simulated on a probabilistic Turing machine with at
most a polynomial increase in the number of elementary operations
required.'' As a consequence, a mechanical machine\footnote{Provided
it has a large enough memory, similar to the tape of a Turing
machine.} such as Babbage's difference engine of the 1800's is
polynomially equivalent to the fastest supercomputer.

Polynomial differences in speed can of course still be significant,
and over the past decades, enormous increases in speed have been
realized by making devices that are smaller, consume less power and
are more highly integrated. However, no matter how impressive the
progress, the laws of physics underlying the operation of today's
computers are still the same as in computers fifty years ago, namely
the classical laws of physics.\\

In the early eighties, the quest for really small computers took on a
completely new face.  First, Paul Benioff showed that a {\em quantum
mechanical} Hamiltonian can represent a universal (classical) Turing
machine~\cite{Benioff80a}.  Then Richard Feynman conjectured that a
{\em quantum computer} might be able to do {\em more} than classical
Turing machines; it might for example efficiently simulate the
dynamics of another quantum system~\cite{Feynman82a,Feynman85a}, a
feat which is impossible on classical computers. David Deutsch then
fully developed the concept of a quantum Turing machine and
highlighted the potential of quantum computers to speed up
computations via {\em quantum parallellism}~\cite{Deutsch85a}.

Ten years later, the field of quantum computation really took off when
Peter Shor announced his quantum factoring
algorithm~\cite{Shor94a}. This was the first {\em quantum algorithm}
which exploited quantum parallellism to offer an exponential speed-up
over classical machines for solving an important mathematical problem
(prime factorization). Another two years later, Lov Grover invented a
quantum algorithm for unstructured search problems~\cite{Grover97a}
and Seth Lloyd~\cite{Lloyd96a} proved Feynman's conjecture on quantum
simulations.

Despite these spectacular results, the field of quantum computation
was regarded with much scepticism because of the difficulty of
maintaining coherent superposition states. However, much of the
scepticism was silenced when Peter Shor~\cite{Shor95a} and Andrew
Steane~\cite{Steane96a} discovered {\em quantum error correction} and
showed that random errors due to decoherence can in fact be
corrected. Furthermore, provided the probability of error per
computational step is low enough, the coding and decoding operations
associated with quantum error correction introduce fewer errors than
can be corrected, even with imperfect
operations~\cite{Aharonov97a,Kitaev97b,Knill98c}.\\

At this point, the {\em physical realization} of quantum computers
became another grand challenge, much like Feynman's challenge of
building a very, very small classical computer: to build a computer
capable of solving problems beyond the reach of classical computers,
by virtue of using quantum mechanical superpositions and entanglement.

Many physical systems have been proposed as potential quantum
computers, including trapped ions~\cite{Cirac95a}, cavity quantum
electrodynamics~\cite{Turchette95a}, electron spins in quantum
dots~\cite{Loss98a}, superconducting loops~\cite{Mooij99a} and nuclear
spins \cite{DiVincenzo95a}.  However, due to the
limited state of the art in any of these experimental techniques, a
demonstration of even the most modest quantum algorithm appeared to be
out of reach for a number of years.

This situation changed completely when Neil Gershenfeld and Isaac
Chuang~\cite{Gershenfeld97a} and independently David Cory, Timothy
Havel and Amr Fahmy~\cite{Cory97a} developed an explicit proposal to
build a simple quantum computer using nuclear spins in liquid
solution, requiring only standard nuclear magnetic resonance
technology. Fifty five years after nuclear spin states and spin echoes
were proposed for (classical) data storage~\cite{Anderson55a}, nuclear
magnetic resonance thus became the workhorse for the early exploration
of experimental quantum computation.

%%%%%%%%%%%%%%%%%%%%%%%%%%%%%%%%%%%%%%%%%%%%%%%%%%%%%%%%%%%%%%%%%

\subsubsection{Related fields}

In parallel with quantum computation, the related field of quantum
information theory developed, which forms the quantum analogue of
classical information theory~\cite{Cover91a}.  Quantum information
theory describes the notions of a quantum source and a quantum
channel, and studies techniques for quantum source and channel
coding. In particular, quantum information theory sets out to
understand how entanglement, which has no classical equivalent,
can be used as a resource in communication.

This field has produced spectacular results such as quantum
teleportation \cite{Bennett93a}, superdense coding \cite{Bennett92c}
and quantum cryptography \cite{Bennett84a,Bennett92b}. Several groups
have already teleported photon states \cite{Bouwmeester97a,Boschi98a},
and secure key distribution using quantum cryptography has been
demonstrated experimentally through optical fibers over tens of
kilometers \cite{Muller96a} and through space by daylight over a
distance of 1.6 km \cite{Buttler00a} (see \cite{Gisin01a} for a
review).  Certainly, quantum cryptography is at a more mature stage
than quantum computing.

%%%%%%%%%%%%%%%%%%%%%%%%%%%%%%%%%%%%%%%%%%%%%%%%%%%%%%%%%%%%%%%%%%%%%%
\section{Purpose of my work}

The main purpose of my work is to study quantum computation
experimentally, and to increase our understanding of what it would
take to build a practical quantum computer. To this purpose, I have
used nuclear spins in liquid solution as quantum bits, and
initialized, manipulated and read out the spin states using
adaptations of standard nuclear magnetic resonance techniques
\cite{Gershenfeld97a,Cory97a}.  Specifically, my
objectives have been \\

\noindent (1) To experimentally provide proof of principle of quantum computation.\\
\vspace*{-2ex}

Until 1997, quantum computers existed only on paper, and in people's
imagination. I wanted to test quantum computation in the lab, and see
various quantum algorithms at work for the first time. \\

\noindent (2) To stimulate theoretical questions by doing quantum computing
    experiments.\\ \vspace*{-2ex}

Interplay between theory and experiment is crucial for the healthy
development of any research field.  I hoped to stimulate theoretical
thinking about the fundamentals of quantum computing by doing actual
experiments. Furthermore, I hoped to interest theorists in helping
with quantum control and in explaining unexpected experimental
observations.\\

\noindent (3) To develop techniques for state initialization, 
coherent quantum control and read out of quantum states, useful in
many implementations of quantum computers.\\ \vspace*{-2ex}

It is clear that many of the challenges in building quantum computers
are similar across different proposed implementations. Therefore,
techniques and solutions invented in the context of NMR (nuclear
magnetic resonance) quantum computing have the potential to advance
other, perhaps more scalable, approaches to the realization of quantum
computers.\\

The general direction of my work has been to push the state of the art
towards more qubits, more gates and more complex algorithms. At each
stage, I have conciously paid attention to all three objectives. The
goal was not just to demonstrate ``the next algorithm'', but rather to
learn scientifically from the experiment, and in particular to
increase our understanding of how we can meet this wonderful challenge
of building a quantum computer, a computer capable of solving problems
beyond the reach of any classical machine.

%%%%%%%%%%%%%%%%%%%%%%%%%%%%%%%%%%%%%%%%%%%%%%%%%%%%%%%%%%%%%%%%%%%%%%

\section{Organization of the dissertation}

Chapter~\ref{ch:qct} lays out the principles of quantum computation,
introduces quantum circuits and quantum gates, and explains the
operation of quantum algorithms and quantum error correction. From
this theoretical discussion, five requirements for the implementation
of quantum computers naturally emerge. Those are discussed in
chapter~\ref{ch:impl}, along with a brief overview of the state of the
art. In chapter~\ref{ch:nmrqc}, we study in detail how those five
requirements can at least in principle be met in liquid solution NMR
experiments. Finally, we explore NMR quantum computing in practice in
a series of experiments, presented in chapter~\ref{ch:expt}. This
structure is illustrated in Fig.~\ref{fig:outline}.  

\bfig
\bcen
\vspace*{1ex}
\includegraphics*[width=6cm]{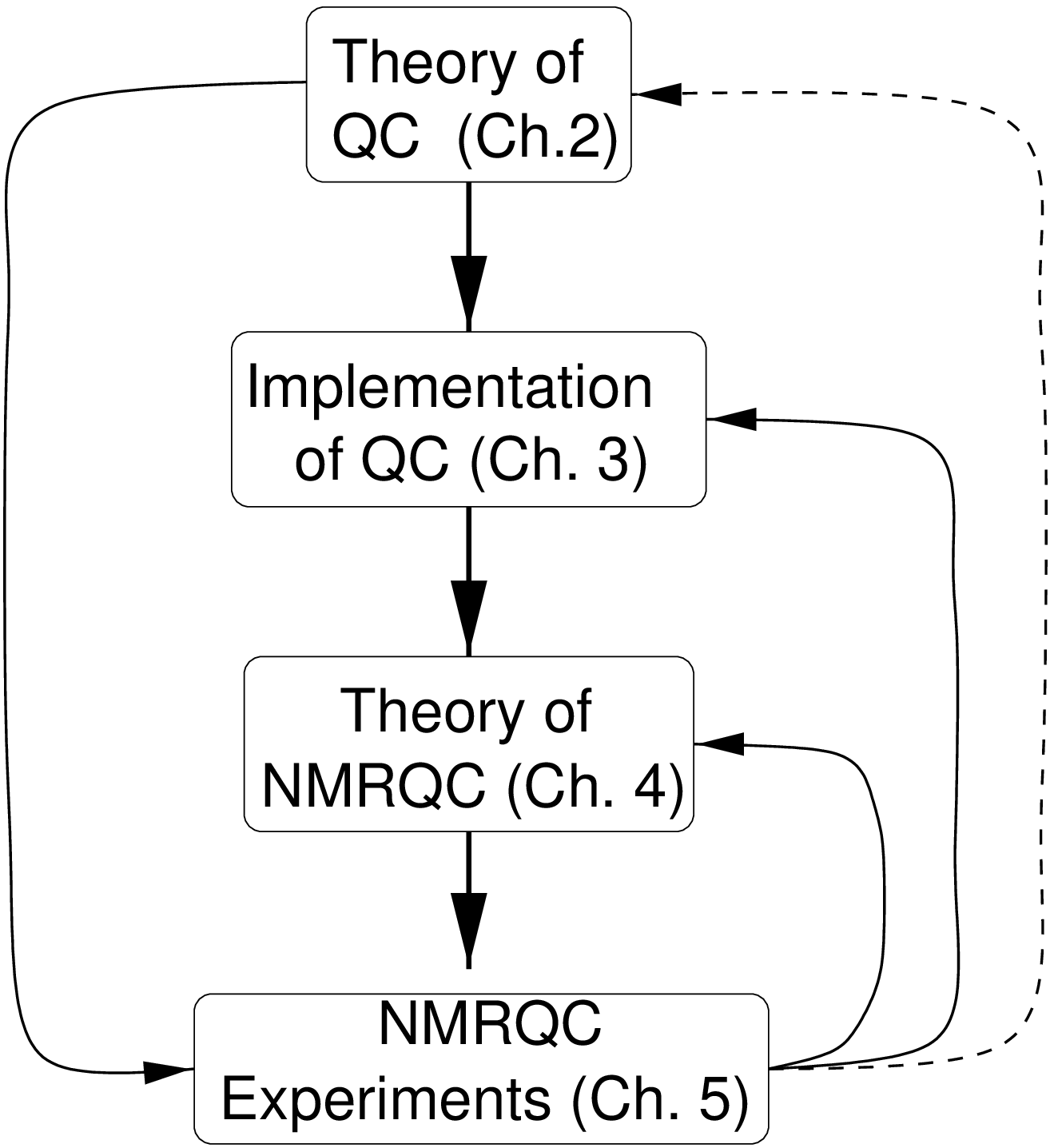} 
\vspace*{-2ex}
\ecen
\caption{Connections between the four main chapters of this thesis work.}
\label{fig:outline}
\efig

Additional connections between the chapters are as follows.  The
selection of topics and the choice of examples in chapter~\ref{ch:qct}
are in function of the experiments of chapter~\ref{ch:expt}.
Furthermore, several of the techniques and concepts for
initialization, control and readout of the spin states presented in
chapters~\ref{ch:impl} and~\ref{ch:nmrqc} were inspired by or invented
in the context of the experiments of chapter~\ref{ch:expt}.  Finally,
the NMR quantum computing experiments have raised theoretical
questions about where the power of quantum computation comes from, and
what the role of entanglement
is~\cite{Braunstein99a,Schack99a,Knill98d}\footnote{We have not gone
into these questions in this thesis work, hence the dashed line.}.
The bulk of my own contributions are in parts of
chapter~\ref{ch:nmrqc} and in chapter~\ref{ch:expt}, as indicated
there.

%%%%%%%%%%%%%%%%%%%%%%%%%%%%%%%%%%%%%%%%%%%%%%%%%%%%%%%%%%%%%%%%%%%%%%

\section{Literature}

Many references to original papers are included throughout. In
addition, the following review articles and books are particularly
relevant.

Michael Nielsen and Isaac Chuang's monumental text ``Quantum Computation
and Quantum Information''~\cite{Nielsen00b} has been an invaluable
resource as I was writing chapters~\ref{ch:qct} and~\ref{ch:impl}. In
addition to this book, the interested reader will find the following
review articles helpful. Bennett and DiVincenzo wrote a recent
authoritative review of quantum computation and
information~\cite{Bennett00a}.  An excellent extended pedagogic review
geared towards non-specialist physicists is given by
Steane~\cite{Steane98a} (although the section on implementations is
outdated). A great introduction to quantum computing and specifically
Shor's algorithm, also for physicists, is by Ekert and
Jozsa~\cite{Ekert96a}, and Lloyd wrote a good introduction for a
general audience~\cite{Lloyd95b}.  Introductions to a wide array of
quantum computer implementations are compiled in a recent special
issue of Fortschritte der Physik~\cite{Braunstein00a}.

Of the many excellent textbooks on quantum mechanics, very few cover
the topics most applicable to quantum computing.  Perhaps the most
helpful text for understanding the relevant concepts of quantum
mechanics is the great book by Peres~\cite{Peres93a}. Reference works
which cover some of the relevant ideas and notation of quantum
mechanics include Cohen-Tannoudji, Diu and
Lalo{\"e}~\cite{Cohen-Tannoudji77c}, Feynman, Leigthon and
Sands~\cite{Feynman65c} and Sakurai~\cite{Sakurai95a}.

Ray Freeman's ``Spin Choreography'' gives a marvelous and intuitive
overview of high-resolution solution {NMR} techniques and spin
dynamics~\cite{Freeman97a}. I found it the most helpful textbook for
the {NMR} techniques underlying chapter~\ref{ch:nmrqc}. A
classic and comprehensive treatise of {NMR} is Ernst, Bodenhausen and
Wokaun~\cite{Ernst87a}. Two other classic texts on {NMR} are
Slichter~\cite{Slichter96a} and Abragam \cite{Abragam61a}; both focus
on spin physics more than on spin dynamics. 

No textbooks exist specifically on NMR quantum computing, but there
are several good introductory review papers. A good introduction for a
general audience is~\cite{Gershenfeld98a}.  Jonathan Jones wrote an
introductory review for an NMR audience~\cite{Jones01a}, and so did
we. We also wrote an accessible introduction for electrical
engineers:

\begin{itemize}
\item L.M.K. Vandersypen, C.S. Yannoni, and I.L. Chuang, {\em to appear in The encyclopedia of NMR (supplement)}, 2001~\cite{Vandersypen01b}.
\item M.~Steffen, L.M.K. Vandersypen, and I.L. Chuang, {\em IEEE 
Micro}, 2001~\cite{Steffen01a}.
\end{itemize}

Each of the experiments presented in sections~\ref{expt:dj}
through~\ref{expt:order} has been published in refereed
journals. These papers also include many of the techniques presented
in chapter~\ref{ch:nmrqc}; only the technique of
section~\ref{nmrqc:simpulse_artefacts} was published separately.

\begin{itemize}
\item \ref{expt:dj}: I.~L. Chuang, L.~M.~K. Vandersypen, X.~L. Zhou, 
D.~W.  Leung, and S.~Lloyd, {\em Nature}, 1998. Reprinted by permission 
from~\cite{Chuang98c} \copyright $\,$ (1998) by Macmillan Magazines, Ltd.
\item \ref{expt:2bitcode}: D.~Leung, L.~Vandersypen, X.~Zhou, 
M.~Sherwood, C.~Yannoni, and I.~Chuang, {\em Phys. Rev. A},
1999. Reprinted by permission from~\cite{Leung99a} \copyright $\,$ (1999)
by The American Physical Society.
\item \ref{expt:labeling}: L.~M.~K. Vandersypen, C.~S. Yannoni, M.~H. 
Sherwood, and I.~L. Chuang, {\em Phys. Rev. Lett.}, 1999. Reprinted by
permission from~\cite{Vandersypen99a} \copyright $\,$ (1999) by The
American Physical Society.
\item \ref{expt:lc}: C.S. Yannoni, M.H. Sherwood, L.M.K. Vandersypen, 
M.G.  Kubinec, D.C. Miller, and I.L. Chuang, {\em Appl. Phys. Lett.},
1999. Reprinted by permission from~\cite{Yannoni99a} \copyright $\,$ (1999)
by The American Institute of Physics.
\item \ref{expt:grover3}: L.M.K. Vandersypen, M.~Steffen, M.~H. 
Sherwood, C.S. Yannoni, G.~Breyta, and I.~L. Chuang, {\em
Appl. Phys. Lett.}, 2000. Reprinted by permission
from~\cite{Vandersypen00a} \copyright $\,$ (2000) by The American Institute
of Physics.
\item \ref{expt:cooling}: D.E. Chang, L.M.K. Vandersypen, and 
M.~Steffen, {\em Chem. Phys. Lett.}, 2001. Reprinted by permission
from~\cite{Chang01a} \copyright $\,$ (2001) by Elsevier Science.
\item \ref{expt:order}: L.M.K. Vandersypen, M.~Steffen, G.~Breyta, 
C.S. Yannoni, R.~Cleve, and I.~L.  Chuang, {\em Phys. Rev. Lett.},
2000. Reprinted by permission from~\cite{Vandersypen00b} \copyright
(2000) by The American Physical Society.
\item \ref{nmrqc:simpulse_artefacts}:
M.~Steffen, L.M.K. Vandersypen, and I.L. Chuang.  {\em
J. Magn. Reson.}, 2000. Reprinted by permission from~\cite{Steffen00a}
\copyright $\,$ (2000) by Academic Press.
\item \ref{expt:shor}: L.M.K. Vandersypen, M.~Steffen, G.~Breyta, C.S. 
Yannoni, M.~Sherwood, and I.~L.  Chuang, {\em in preparation},
2001~\cite{Vandersypen01a}.
\end{itemize}

%% file: qct.tex
  \chapter{Theory of quantum computation}
    \label{ch:qct}

In this chapter, we review the principles of the theory of quantum
computation. From the outset, the presentation is directed towards a
practical appreciation and understanding of the subject. Our starting
point is the notion of quantum bits. We next present the language of
quantum gates and circuits, and use this language to outline the
operation of quantum algorithms and quantum error correction.

%%%%%%%%%%%%%%%%%%%%%%%%%%%%%%%%%%%%%%%%%%%%%%%%%%%%%%%%%%%%%%%%%%%%%%%%%
\section{Fundamental concepts}
\label{sec:principles}
%%%%%%%%%%%%%%%%%%%%%%%%%%%%%%%%%%%%%%%%%%%%%%%%%%%%%%%%%%%%%%%%%%%%%%%%%

\subsection{Quantum bits}
\label{qct:qubits}
\subsubsection{One quantum bit}

In today's digital computers, information is stored and processed in
the form of bits, entities which can take on only two values: logical
zero, $0$, or logical one, $1$.  These are typically represented by
the voltage at a node, or the alignment of a piece of magnetic
material, but any physical system with at least two distinct states
can serve to represent a bit, including two-level quantum systems such
as spins-1/2 and polarized photons. The quantum state $\ket{0}$
corresponds to $0$ and the state $\ket{1}$ corresponds to $1$. For a
spin-1/2 particle, the two computational basis states are represented
by the spin up and spin down state ($\ket{\!\uparrow}$ or
$\ket{\!\downarrow}$), and for photons by the vertical or horizontal
polarization state ($\ket{\!\updownarrow\,}$ or $\ket{\!\!\leftrightarrow}$).

In contrast to classical bits which can only exist as $0$ or $1$,
two-level quantum systems, called quantum bits or {\em qubits}, can
also exist in a superposition state of $\ket{0}$ and $\ket{1}$,
mathematically written as
\be
\ket{\psi} = a \ket{0} + b \ket{1} \,,
\label{eq:1qubit}
\ee
where $a$ and $b$ are complex numbers satisfying the normalization
condition $|a|^2 + |b|^2 = 1$.  The {\em overall} phase of
$\ket{\psi}$ is physically irrelevant as it cannot be revealed by any
measurement. Therefore, we can also write $\ket{\psi}$ as
\be
\ket{\psi} = \cos\frac{\theta}{2} \ket{0} 
+ e^{i\phi} \sin \frac{\theta}{2} \ket{1} \,,
\label{eq:bloch_qubit}
\ee
and visualize the state of a qubit as a point on a sphere, called the
Bloch sphere, as in Fig.~\ref{fig:bloch_sphere}.  This representation
may convey the impression that a qubit is very much like an analog
classical variable, with two degrees of freedom $\theta$ and
$\phi$. However, as we will see, qubits are in many ways very
different from such analog classical variables.  Rather than pointing
along a certain direction, a qubit in a superposition state $a \ket{0}
+ b \ket{1}$ is in some sense in both $\ket{0}$ and $\ket{1}$ {\em at
the same time}. Furthermore, as we shall see next, the number of
degrees of freedom in an $n$-qubit state increases exponentially with
$n$.

\bfig
\begin{center}
$ \begin{array}{cc}
\hspace*{0.2cm} \includegraphics*[width=4cm]{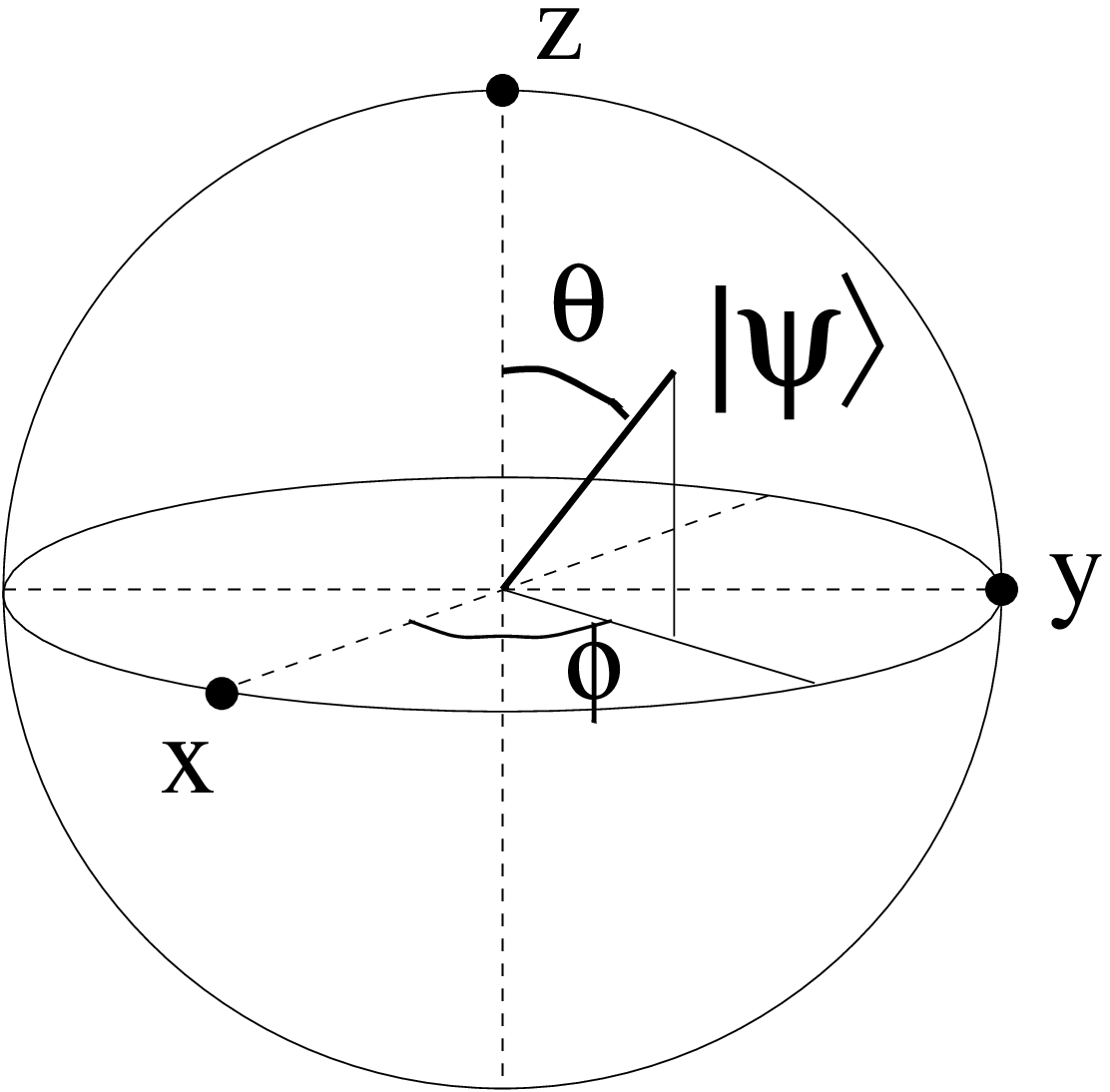} &
\hspace*{1cm} \includegraphics*[width=5cm]{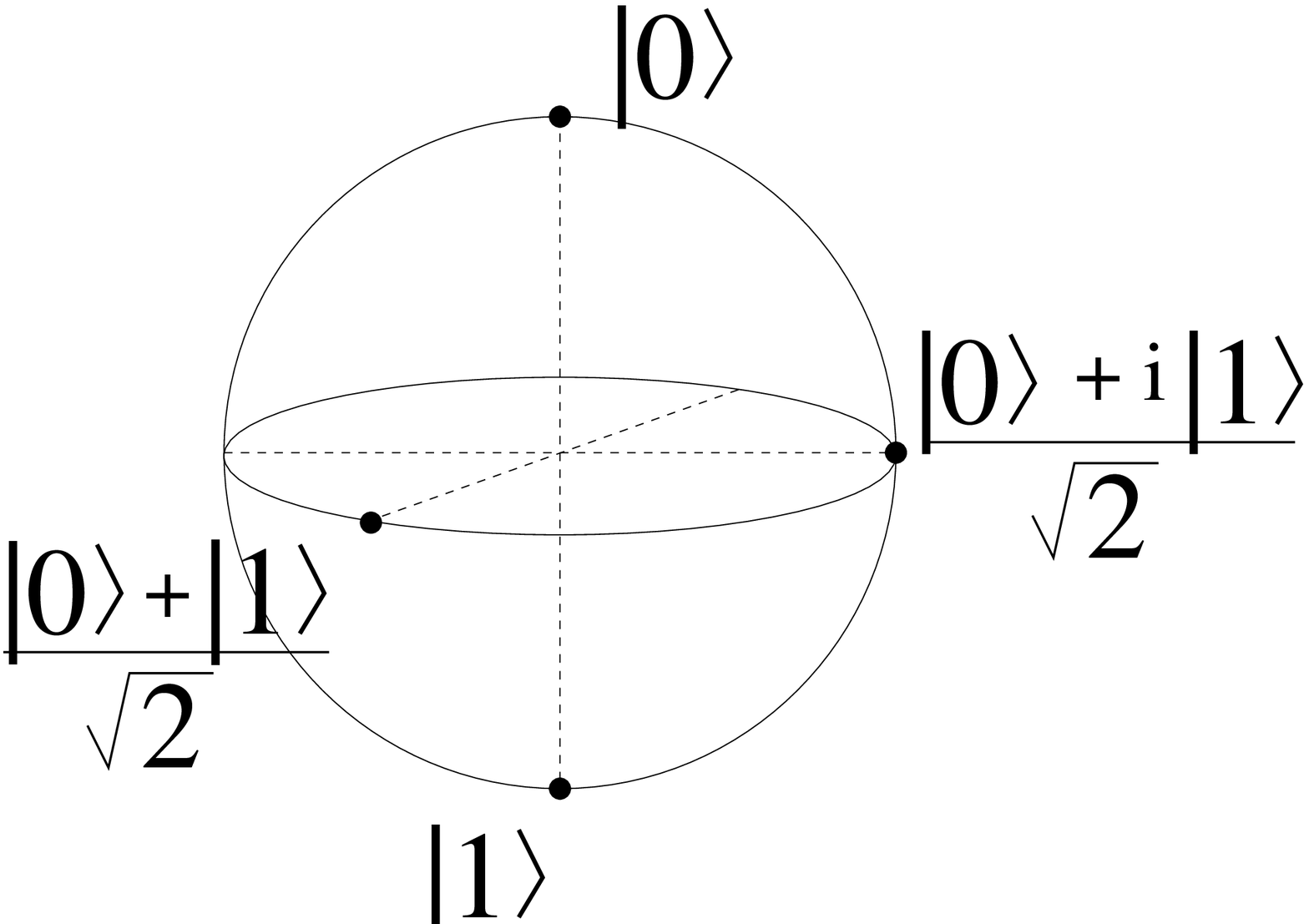} 
\vspace{1.5ex}  \\
\mbox{(a)} & \mbox{(b)}
\end{array} $
\end{center}
\caption{(a) Bloch sphere representation of the state $\ket{\psi}$ 
of a single qubit. (b) The position in the Bloch sphere of four
important states. By convention, we will always let $\ket{0}$ be along
$+\hat{z}$.}
\label{fig:bloch_sphere}
\efig

\subsubsection{Multiple qubits}

The state of two qubits, each in an arbitrary superposition state
$\ket{\psi}_1 = a_1 \ket{0} + b_1 \ket{1}$ and $\ket{\psi}_2 = a_2
\ket{0} + b_2 \ket{1}$, is written as
\be 
\ket{\psi} = \ket{\psi}_1 \otimes \ket{\psi}_2
= (a_1 \ket{0} + b_1 \ket{1}) \otimes (a_2 \ket{0} + b_2 \ket{1}) \,,
\label{eq:product_qubits}
\ee
where $\otimes$ is the {\em tensor product} or {\em Kronecker product}
symbol. We can rearrange this expression as
\be
\ket{\psi} = a_1 a_2 \ket{0} \otimes \ket{0}
           + a_1 b_2 \ket{0} \otimes \ket{1}
           + b_1 a_2 \ket{1} \otimes \ket{0}
           + b_1 b_2 \ket{1} \otimes \ket{1} \,.
\ee
From now on, we will leave the $\otimes$ symbol implicit, and
furthermore abbreviate $\ket{0} \ket{0}$ as $\ket{00}$, $ \ket{0}
\ket{1}$ as $\ket{01}$ and so forth. Thus,
\be
\ket{\psi} = 
a_1 a_2 \ket{00} + a_1 b_2 \ket{01} + 
b_1 a_2 \ket{10} + b_1 b_2 \ket{11} \,.
\ee
Remarkably and surprisingly, the coefficients of the terms in the
joint superposition state of the two qubits can in fact be chosen {\em
independently}. That is, they don't need to be the product of the
coefficients of two single-qubit states. We can express this by
writing the joint state of two qubits in the more general form
\be
\ket{\psi} = 
c_{00} \ket{00} + c_{01} \ket{01} + 
c_{10} \ket{10} + c_{11} \ket{11} \,.
\ee
\noindent or equivalently, if we represent the states in decimal instead of binary represenation,
\be
\ket{\psi} = 
c_{0} \ket{0} + c_{1} \ket{1} + c_{2} \ket{2} + c_{3} \ket{3} \,.
\label{eq:psi_c_i}
\ee
Similarly, a register of $n$ qubits can be in an arbitrary
superposition of $2^n$ states,
\be
\ket{\psi} = \sum_{k=0}^{2^n-1} c_k \ket{k} \,,
\label{eq:general_superposition}
\ee
where the only constraint on the complex amplitudes $c_k$ is that they
must satisfy the normalization condition
\be
\sum_k |c_k|^2 = 1 \,.
\label{eq:normalization}
\ee
As for single-qubit states, the overall phase is
irrelevant. Therefore,\\

{\em the description of a pure state of $n$ qubits requires $2^n-1$
complex numbers.\footnote{A mixed state of $n$ qubits has $4^n-1$
degrees of freedom. The distinction between pure and mixed states will
be discussed shortly.}}\\

\noindent This is manifestly different from classical systems. For 
example, the position of $n$ points on the sphere of
Fig.~\ref{fig:bloch_sphere} is described by only $n$ rather than $2^n$
complex numbers. In fact, the position of any $n$ classical particles
can always be described by a number of real or complex numbers that is
linear in $n$.

Since we cannot visualize the state of $n$ qubits via $n$ Bloch
spheres, or $n$ points on a single Bloch sphere, how {\em can} we
visualize their state? This is difficult --- our intuition fails at
the quantum level, because we didn't grow up with an intuition for
quantum mechanics, and because our observations of the every-day world
around us are observations of a classical world. Mathematically, the
extension of the Bloch sphere is called {\em Hilbert space}, a $2^n$
dimensional complex vector space with an inner product. The state of a
quantum system thus corresponds to a point in Hilbert space.

\subsubsection{Entanglement}

Since the number of degrees of freedom of $n$ quantum systems grows
exponentially more quickly than that of $n$ classical systems, surely
there must exist quantum states which have no classical
equivalent. The state of Eq.~\ref{eq:product_qubits} is a classical
state: this joint state of two qubits can be fully described via a
description of the individual qubits (which requires one complex
number, or two real numbers, for each qubit). We say that the state of
Eq.~\ref{eq:product_qubits} is {\em separable}.

In contrast, it is impossible to find two one-qubit states
$\ket{\psi}_1 = a_1 \ket{0} + b_1 \ket{1}$ and $\ket{\psi}_2 = a_2
\ket{0} + b_2 \ket{1}$, such that their tensor product gives the state
\be
\frac{\ket{00}+\ket{11}}{\sqrt{2}} \,.
\label{eq:epr}
\ee
In other words, this state cannot be written as a product of two
one-qubit states. We call such a state {\em non-separable} or {\em
entangled}. Let us give two more examples: the state $\frac{1}{2}
(\ket{00} - \ket{01} + \ket{10} - \ket{11}$ can be written as
$\frac{1}{2} (\ket{0} + \ket{1})(\ket{0}-\ket{1})$ and is thus
separable; the state $\frac{1}{2} (\ket{00} + \ket{01} +
\ket{10} - \ket{11})$ cannot be factored into two one-qubit states and
is thus entangled.

\subsubsection{Mixed states versus pure states}
\label{qct:mixed_vs_pure}

A quantum system in a well-defined and well-known state $\ket{\psi}$
is said to be in a {\em pure} state. If all we know about a quantum
system is that it is in one of several pure states $\ket{\psi_i}$, each
with certain probabilities $p_i$, we say the quantum system is in a
{\em statistical mixture} of these pure states, or for short that it
is in a {\em mixed} state.  The state of a quantum system in a
statistical mixture is conveniently described by its {\em density
operator} 
\be
\rho = \sum_i p_i \ket{\psi_i}\bra{\psi_i} \,,
\label{eq:mixed_rho}
\ee
where $\bra{\psi}$ represents the Hermitian conjugate of
$\ket{\psi}$, and $\ket{\psi}\bra{\phi}$ denotes the {\em outer
product} (a linear operator). Obviously, the probabilities $p_i$ must
satisfy $p_i \ge 0$ and $\sum_i p_i = 1$. For a pure state
$\ket{\psi}$, the density operator is simply $\rho =
\ket{\psi}\bra{\psi}$.

Every density operator satisfies
\be
\mbox{Tr}(\rho) = 1 \,,
\ee
since $\mbox{Tr}(\rho) = \sum_i p_i
\mbox{Tr}(\ket{\psi_i}\bra{\psi_i}) = \sum_i p_i$. Furthermore, the 
eigenvalues $\lambda_j$ of $\rho$ satisfy
\be
\lambda_j \ge 0 \,,
\ee
so $\rho$ is a {\em positive} operator, and one can decompose it as
\be
\rho = \sum_j \lambda_j \ket{j}\bra{j} \,,
\label{eq:spect_decomp_rho}
\ee
where the $\ket{j}$ are orthogonal eigenvectors of $\rho$ (the
$\ket{\psi_i}$ of Eq.~\ref{eq:mixed_rho} need not be orthogonal).  We
can thus also interpret a quantum system in $\rho$ to be in the state
$\ket{j}$ with probability $\lambda_j$, and make the important
observation that an arbitrary density matrix does thus {\em not} have
a {\em unique} decomposition into any specific mixture of states.

The mathematical distinction between pure and mixed states is that a
pure state density operator has only one non-zero eigenvalue
(necessarily equal to 1), whereas a mixed state density operator has
more than one non-zero eigenvalue. It follows that a convenient
criterion to distinguish pure and mixed states is
\begin{eqnarray}
\mbox{Tr}(\rho^2) = 1 & \Leftrightarrow & \rho \mbox{ is pure} \,\\
\mbox{Tr}(\rho^2) < 1 & \Leftrightarrow & \rho \mbox{ is mixed} \,.
\end{eqnarray}

Of course, any quantum system is really in just one state. We
emphasize therefore, that to say that a quantum system is in a mixed
state is merely a statement about our {\em knowledge} of the state of
the quantum system. As we shall see, the distinction between pure and
mixed states has important implications --- pure states are in
many respects more ``useful'' than mixed states.

\subsubsection{Promise of qubits}

It may appear at first sight that a bit which is simultaneously $0$
and $1$ is not very useful for computation, and is, in fact, rather
confusing. However, the exponential complexity of quantum systems also
suggests that perhaps quantum bits could be immensely useful for
computation. This observation led Richard Feynman to speculate that
``quantum computers'' might be able to solve certain problems
exponentially faster than any classical
machine~\cite{Feynman85a,Feynman96a}. In the next section, we first
verify that quantum systems can indeed be used for computation. In the
following section we explore the potential of quantum superpositions
and entanglement for performing massively parallel computations.

%%%%%%%%%%%%%%%%%%%%%%%%%%%%%%%%%%%%%%%%%%%%%%%%%%%%%%%%%%%%%%%%%%%%%%%%%

\subsection{Computation using quantum systems}
\label{qct:dyn&rev}

So far, we have merely given a {\em static} description of quantum
bits as two-level quantum systems which can hold
information in binary form. We will now look at the {\em dynamics} of
quantum bits, and examine whether we can perform computations by
evolving the state of a set of quantum systems in a controlled way.

\subsubsection{Unitary evolution}

One of the postulates of quantum mechanics dictates that the time
evolution of the state $\ket{\psi(t)}$ of a {\em closed} quantum
system (i.e. a system which does not interact with the environment,
the rest of the universe) is governed by Schr\"odinger's equation:
\be
i\hbar \frac{d \ket{\psi(t)}}{d t} \; = \; {\cal H} \ket{\psi(t)} \,,
\ee
where $\hbar$ is Plank's constant and ${\cal H}$ is the {\em
Hamiltonian}, an operator for the total energy of the system. For
time-{\em in}dependent Hamiltonians, the Schr\"odinger equation has a
straightforward solution:
\be
\ket{\psi(t)} \; = \; \mbox{exp} \left(\frac{-i {\cal H} t}{\hbar}\right) \; \ket{\psi(t=0)} \,.
\label{eq:sol_schroed}
\ee
If the Hamiltonian is time-dependent, the Schr\"odinger equation has no
easy solution, although the evolution can be approximated as a
sequence of evolutions under time-independent Hamiltonians.  We
usually denote the time-evolution operator as $U$, where
\be
U = \mbox{exp} (\frac{-i {\cal H} t}{\hbar}) \,,
\label{eq:U}
\ee
so
\be
\ket{\psi(t)} = U \ket{\psi(0)} \,.
\ee

Similarly, the time evolution of the density operator $\rho$ of a
quantum system is
\begin{eqnarray}
\rho(t) = \sum_i p_i \ket{\psi_i(t)}\bra{\psi_i(t)} = 
\sum_i p_i \; U \ket{\psi_i(0)}  \bra{\psi_i(0)} \; U^\dagger
= U \rho(0) \; U^\dagger \,,
\end{eqnarray}
where the $^\dagger$ symbol indicates the Hermitian conjugate.

From Eq.~\ref{eq:sol_schroed}, we can appreciate the important role of
the Hamiltonian of a quantum system: it controls the time evolution of
the quantum system. Since ${\cal H}$ is a {\em Hermitian} operator,
i.e. it is its own Hermitian conjugate ${\cal H} = {\cal H}^\dagger$,
the time evolution operator $U=\exp(-i{\cal H}t/\hbar)$ is {\em
unitary}, that is $U U^\dagger = e^{-i{\cal H}t/\hbar} e^{i{\cal
H^\dagger}t/\hbar} = I = e^{i{\cal H^\dagger}t/\hbar} e^{-i{\cal
H}t/\hbar} = U^\dagger U$. This implies that the evolution of a closed
quantum system is completely {\em reversible} . Indeed, we can {\em
unwind} any time evolution $U$ by a subsequent time evolution
$U^\dagger$.

Geometrically, we can visualize the unitary evolution of a single
qubit as a rotation in the Bloch sphere (Fig.~\ref{fig:bloch_sphere}),
a picture we will often use in chapter~\ref{ch:nmrqc}.  By extension,
the unitary evolution of multiple qubits corresponds to a rotation in
Hilbert space.

\subsubsection{Irreversible and reversible computation}

Today's classical computers do not at all operate in a reversible
manner. Note for example that your computer generates heat. Also note
that it is not possible to reconstruct the input of a traditional {\sc
and} gate from its output (Fig.~\ref{fig:and_gate} a). This is
obvious since the {\sc and} gate has two input bits and only one
output bit; it is never possible to reconstruct the value of two bits
starting from only one bit. But even if we introduce a second output
bit (Fig.~\ref{fig:and_gate} b), it is not possible to make the {\sc
and} gate reversible. This is because the {\sc and} gate is not
single-valued. If the output is 00 in the example of
Fig.~\ref{fig:and_gate} b, we cannot know whether the input was 00 or
01.

It is thus natural to ask whether universal computation can be done
reversibly. Rolf Landauer and Charles Bennett showed that indeed any
computation can be performed in a completely reversible manner, that
is without (or with infinitesimal) energy
dissipation~\cite{Landauer61a}\cite{Bennett73a}. The only time heat must
be dissipated is in the process of resetting a bit, which irreversibly
erases the information contained in the bit and thus necessarily
increases the entropy.

\bfig 
\begin{center}
\begin{tabular}{c|c}
$\rm In$  & $\rm Out$ \\ \hline
$\mathbf{0}$  $\mathbf{0}$  & $\mathbf{0}$  \\
$\mathbf{0}$  $\mathbf{1}$  & $\mathbf{0}$  \\ 
$\mathbf{1}$  $\mathbf{0}$  & $\mathbf{0}$  \\ 
$\mathbf{1}$  $\mathbf{1}$  & $\mathbf{1}$  \\ 
\end{tabular} 
\hspace*{1cm}
\begin{tabular}{c|c}
$\rm In$  & $\rm Out$ \\ \hline
$\mathbf{0}$  $\mathbf{0}$  & $0$ $\mathbf{0}$  \\
$\mathbf{0}$  $\mathbf{1}$  & $0$ $\mathbf{0}$  \\ 
$\mathbf{1}$  $\mathbf{0}$  & $1$ $\mathbf{0}$  \\ 
$\mathbf{1}$  $\mathbf{1}$  & $1$ $\mathbf{1}$  \\ 
\end{tabular}
\hspace*{1cm}
\begin{tabular}{c|c}
$\rm In$  & $\rm Out$ \\ \hline
$\mathbf{0}$  $\mathbf{0}$ $0$  & $0$ $0$ $\mathbf{0}$  \\
$\mathbf{0}$  $\mathbf{1}$ $0$  & $0$ $1$ $\mathbf{0}$  \\ 
$\mathbf{1}$  $\mathbf{0}$ $0$  & $1$ $0$ $\mathbf{0}$  \\ 
$\mathbf{1}$  $\mathbf{1}$ $0$  & $1$ $1$ $\mathbf{1}$  \\ 
$0$  $0$ $1$  & $0$ $0$ $1$  \\
$0$  $1$ $1$  & $0$ $1$ $1$  \\ 
$1$  $0$ $1$  & $1$ $0$ $1$  \\ 
$1$  $1$ $1$  & $1$ $1$ $0$  \\ 
\end{tabular} \\ \vspace*{-1cm}
\end{center}
\hspace*{3.5cm} (a) \hspace*{2.6cm} (b)\vspace*{0.5cm} \\\hspace*{10.8cm}(c)
\caption{Truth table for 
(a) The traditional {\sc and} gate; the output is $1$ when both inputs
are $1$, and the output is $0$ otherwise. (b) The extended {\sc and}
gate, with two output bits. (c) A reversible {\sc and} gate, also
known as the {\sc Toffoli} gate.}
\label{fig:and_gate}
\efig

Any multi-valued function $f$,
\be
x \mapsto f(x) 
\label{eq:map_f}
\ee
can be made reversible by introducing a second input bit string $y$ of
the same length as $f(x)$ and extending Eq.~\ref{eq:map_f} to
\be
(x,y) \mapsto (x, y \oplus f(x))  \,,
\label{eq:map_f_reversibly}
\ee
where $\oplus$ is the bitwise addition modulo two (equivalent to the
bitwise {\sc xor}).  If we set $y$ to 0, we simply obtain $f(x)$ in
the second register:
\be
(x,0) \mapsto (x,f(x))  \,.
\label{eq:make_f_reversible}
\ee
We can thus construct a reversible version of the {\sc and} gate for
example, by using an additional input bit (Fig.~\ref{fig:and_gate}
c). The third bit is always initialized to $0$, so in practice only
the top half of the truth table is ever used. The input can now always
be reconstructed from the output. Similarly, if we initialize the
third bit to $1$, and thus use only the bottom half of the truth table
of Fig.~\ref{fig:and_gate} c, we obtain a {\sc nand} gate. Since the
{\sc nand} gate is universal for classical logic, the {\sc Toffoli} is
universal for reversible classical computation~\cite{Toffoli80a}.

We will return to the implementation of quantum computers in
chapter~\ref{ch:impl}. For now, we will just state that it may be
possible to control the Hamiltonian in such a way that the time
evolution results in a transformation of the qubit states which
corresponds to the transformation of bit values in a classical truth
table~\cite{Benioff80a}.  That is, the computational basis states of
the qubits ($\ket{0}$ or $\ket{1}$ for each qubit) can be transformed
as
\be
\ket{x} \mapsto \ket{f(x)} 
\ee
for reversible $f$, or, by extension, as
\be
\ket{x}\ket{y} \mapsto \ket{x}\ket{y \oplus f(x)}  \,,
\ee
for irreversible $f$, corresponding to Eqs.~\ref{eq:map_f}
and~\ref{eq:map_f_reversibly} respectively. Therefore, we can say that

\begin{quote}
{\em quantum computation subsumes classical computation.}
\end{quote}

\noindent Now, what would happen if the quantum bits 
were initially in a superposition state of the computational basis
states, $\sum_{k=0}^{2^n-1} c_k
\ket{k}$ ? This is the subject of
the next section.

%%%%%%%%%%%%%%%%%%%%%%%%%%%%%%%%%%%%%%%%%%%%%%%%%%%%%%%%%%%%%%%%%%%%%%%%%

\subsection{Quantum parallellism}
\label{sec:parallel}

\subsubsection{Quantum parallellism}

Every computation can be seen as the concatenation of many logic
gates. Each logic gate produces an output which is a function of its
input. Now consider a classical and reversible\footnote{If $f$ is not
reversible, we know how to make it reversible from
section~\ref{qct:dyn&rev}.} logic gate which implements a function $f$
with one input bit $x$ and one output bit $y=f(x)$. If $x=0$, the gate
will output $f(0)$, and if $x=1$, the gate will output $f(1)$. Now
imagine we can implement an analogous quantum logic gate, which
transforms a qubit as $\ket{x} \mapsto
\ket{f(x)}$. By virtue of the linearity of quantum
mechanics, the same quantum gate transforms a qubit in a superposition
state as
\be
a \ket{0} + b \ket{1} \quad \stackrel{f}{\mapsto} \quad
 a \ket{f(0)} + b \ket{f(1)} \,.
\label{eq:1qubit_out}
\ee
In some sense, the function $f$ has been evaluated for both its input
values ($0$ and $1$) in one step!  Next consider a different logic
gate which implements a function $f$ with two input and two output
bits. If we prepare the two qubit system in the state
\be
\ket{\psi} = a \ket{00} + b \ket{01} + c \ket{10} + d \ket{11} \,,
\label{eq:2qubits_in}
\ee
evaluation of the function $f$ tranforms the state to
\be
a \ket{f(00)} + b \ket{f(01)} + c \ket{f(10)} + d \ket{f(11)} \,,
\label{eq:2qubits_out}
\ee
so $f$ has been evaluated for four input values in parallel. For every
additional input bit, the potential number of parallel function
evaluations doubles!  In general, a function of $n$ bits implemented
on a quantum computer can be evaluated for all $2^n$ possible input
values at the same time:
\be
\sum_{x=0}^{2^n-1} c_x \ket{x} \quad \stackrel{f}{\mapsto} \quad
\sum_{x=0}^{2^n-1} c_x \ket{f(x)} \,,
\label{eq:nqubits_out}
\ee
where $x$ is an integer encoded by a string of $n$ bits.  Thus,
whereas for classical computers the number of parallel function
evaluations increases at best linearly with their size, 

\begin{quote}
{\em the number of parallel function evaluations grows exponentially
with the size of the quantum computer (the number of qubits).}
\end{quote}

\noindent This truly spectacular notion was first introduced by 
David Deutsch in 1985, and termed {\em quantum
parallellism}~\cite{Deutsch85a}.

Machines based on quantum bits thus appear to be exponentially more
powerful than any machine using just classical bits. Of course, a
computation is only meaningful if the output result can be read out,
but how do we measure the state of quantum bits, and just what does
the measurement give when the qubit is in a superposition state?

%%%%%%%%%%%%%%%%%%%%%%%%%%%%%%%%%%%%%%%%%%%%%%%%%%%%%%%%%%%%%%%%%%%%%%%%%

\subsubsection{Measurement of quantum states}
\label{qct:measurement}

The postulates of quantum mechanics dictate that any measurement of a
quantum system can be described in terms of a set of {\em measurement
operators} $P_m$. Measurement of a quantum system in the state
$\ket{\psi}$ immediately before the measurement gives outcome $m$
with probability
\be
p(m) = \bra{\psi}P_m\ket{\psi} \,.
\label{eq:p_meas}
\ee
The state of the quantum system after the measurement is
\be
\frac{P_m \ket{\psi}}{\bra{\psi}P_m\ket{\psi}} \,.
\label{eq:post_meas}
\ee
The measurement operators must satisfy the completeness relation
\be
\sum_m P_m = I \,,
\ee
such that the probabilities $p_m$ sum to 1. Furthermore, for a
projective measurement, we also require that the operators $P_m$ be
Hermitian and that
\be
P_m P_{m'} = \delta_{mm'} P_m \,.
\label{eq:orthog_meas}
\ee
We can therefore associate an orthonormal basis of states $\ket{m}$
with any set of projective measurement operators $P_m$, such that $P_m
= \ket{m}\bra{m}$. Then, the probability of obtaining $m$ in a
measurement of a quantum system in $\ket{\psi}$ is $p(m) =
|\bra{m}\psi\rangle|^2$ (note that $0 \le p_m \le 1$, with equality only
if $\ket{\psi}=\ket{m}$), and the post-measurement state is $\ket{m}$.

For example, if we measure a single qubit in the state $\ket{0}$ in
the $\{\ket{0},\ket{1}\}$ basis (the computational basis), the
measurement always gives $0$. If we measure a qubit in $a\ket{0} +
b\ket{1}$ in the same basis, we obtain $0$ with probability $|a|^2$
and $1$ with probability $|b|^2$.

What happens if we measure a qubit in $\ket{\psi} = a\ket{0} +
b\ket{1}$ twice in a row, in the same basis?  From
Eqs.~\ref{eq:post_meas} and~\ref{eq:orthog_meas}, we see that the
result of the second measurement will always be identical to the
result of the first measurement. The first measurement will give $m$
and in the process {\em collapses}~\footnote{Collapse is only one of
several interpretations of the measurement process. Since they all
make the same predictions for the measurement statistics and outcomes,
we will not concern ourselves with interpretation issues.}
$\ket{\psi}$ onto the corresponding measurement basis state
$\ket{m}$. Since the $\ket{m}$ are orthogonal, the second measurement
will with certainty return $m$ as well.

Collapse of the quantum state implies that the information contained
in the coefficients $a$ and $b$ is instantaneously and irreversibly
destroyed.  As a result, an unknown quantum state cannot be fully
characterized even by repeated measurements, whether they take place
in the same basis or in different bases.  A second measurement of a
qubit in a different basis than the first measurement will {\em
project} the state (which now is $\ket{m}$, that is $\ket{0}$
or $\ket{1}$ in our example) onto the new measurement basis (for
example the $\{\frac{\ket{0}+\ket{1}}{\sqrt{2}}, \frac{\ket{0} -
\ket{1}}{\sqrt{2}}\}$ basis). This measurement does not yield any extra
information about $\ket{\psi}$, however, because $a$ and $b$ have
already been irretrievably lost; the second measurement is a
measurement on the state $\ket{m}$, not on $\ket{\psi}$.

It would be possible to determine $a$ and $b$ with good accuracy by
performing a properly designed series of measurements on a large
number of copies of the qubit in the same unknown state.  However, the
{\em no-cloning theorem}~\cite{Dieks82a,Wootters82a}
\label{page:no-cloning} forbids the creation of copies of a qubit in
an unknown state (it is of course possible to create many copies of a
qubit in a known state). In summary, \\

{\em no measurement can fully reveal the state of a qubit in an
unknown state.}\\

Furthermore, Eq.~\ref{eq:p_meas} implies that it is not possible
to reliably distinguish two non-orthogonal quantum states
$\ket{\psi_1}$ and $\ket{\psi_2}$: regardless of the basis we choose,
there must be a $\ket{m}$ for which $\langle\psi_1\ket{m}^2 \neq 0$
and $\langle\psi_2\ket{m}^2\neq 0$. Orthogonal states in contrast, can
be perfectly distinguished by a measurement in the appropriate basis.

We note that while the evolution of a closed quantum system is unitary
(see section~\ref{qct:dyn&rev}), the measurement process inherently
invokes an interaction with an external measuring device during which
the quantum system cannot remain closed. As a result, the measurement
process is non-unitary; it constitutes a projection onto a finite set
of basis states in Hilbert space, rather than a rotation in Hilbert
space.

Finally, we point out that we have restricted ourselves to projective
measurements, also known as {\em hard}\label{text:hard_meas}
measurements.  We postpone a discussion of {\em weak} measurements
until section~\ref{impl:readout} and apply it to the case of NMR in
section~\ref{nmrqc:meas}.

\subsubsection{Hidden variables and measurements on entangled particles} 

In light of the measurement process, it is natural to ask what the
meaning of superpositions is. Indeed, if upon measurement we obtain
only one of the terms in a superposition, wasn't the qubit perhaps
already in the corresponding state all along, instead of in several
states ``at the same time'' ? Isn't there some {\em hidden variable}
which predetermines what the measurement outcome will be ?  This
question has been the subject of much debate throughout the 20th
century (for a good introduction, see~\cite{Mermin85a}).

In a 1935 paper, Einstein, Podolsky and Rosen (EPR) considered what
would happen if a measurement is performed on one of two entangled
particles~\cite{Einstein35a}. Suppose we prepare two qubits in the
entangled state $\frac{1}{\sqrt{2}}\ket{01} -
\frac{1}{\sqrt{2}}\ket{10}$ (the singlet state, one of the four
so-called EPR states). If we measure qubit 1 in the computational
basis, the outcome will be $\ket{0}$ or $\ket{1}$. Now, what would be
the result of a subsequent measurement of qubit 2? Because of the
entanglement, the wavefunction of {\em both} particles collapses to
either $\ket{01}$ or $\ket{10}$ upon measuring the first qubit, and
therefore the outcome for the second qubit will always be opposite to
the outcome of the first qubit. In fact, for the singlet state, the
outcomes for the two particles will be opposite for a measurement in
{\em any} basis. Furthermore, this is true even if the two entangled
qubits are lightyears away from each other!  Actions in one location
would thus appear to have instanteneous consequences for observations
in a different location. Einstein rejected this ``spooky action at a
distance'' calling it absurd. He believed that quantum theory was
incomplete and had to be supplemented with a theory of local hidden
variables.

In 1964, John Bell proposed an actual experiment which would either
confirm or disprove local hidden variably
theories~\cite{Bell64a}~\footnote{Or to be precise, that hidden
variables can only exist if information can travel faster than light,
which we strongly believe is not the case.}.  This experiment was
carried out in 1980 by Alain Aspect~\cite{Aspect81a}, and has been
repeated many times since then, in attempts to close more and more
possible loopholes in the experiment. The experimental observations
have consistently refuted the existence of local hidden variables,
thereby confirming the validity of quantum theory.

\subsubsection{Implications of quantum measurements for quantum parallellism}

Returning to quantum parallellism, we see that measurement of two
qubits in a superposition state such as in Eq.~\ref{eq:2qubits_out},
collapses the state of the qubits and probabilistically returns
$f(00)$, $f(01)$, $f(10)$ or $f(11)$. In general, after performing a
phenomenal number of parallel function evalutions ($2^n$ for $n$
qubits), as in Eq.~\ref{eq:nqubits_out}, a measurement of the final
state will probabilistically give one of the $2^n$ terms in the output
superposition state.  It thus appears that the exponential
computational power of quantum computers is not accessible!

Remarkably, special quantum algorithms exist which allow one to take
advantage of the exponential complexity of quantum systems and
circumvent the limitations of quantum measurements and readout, in
order to signficantly speed up certain computational tasks.

%%%%%%%%%%%%%%%%%%%%%%%%%%%%%%%%%%%%%%%%%%%%%%%%%%%%%%%%%%%%%%%%%%%%%%%%%

\subsection{Quantum algorithms}

Quantum algorithms allow one to take advantage of quantum parallellism
and thereby solve certain problems in {\em far fewer steps} than is
possible classically. 

\subsubsection{Notions from complexity theory}

The basis for comparison of the power of quantum and classical
computers is provided by complexity theory, which analyzes how the
minimal physical resources (time, space, energy) required for an
algorithm to solve a given problem vary with the problem size
$n$~\cite{Hopcroft79a}. The key distinction is whether the resources
required are {\em polynomial} or {\em exponential} in $n$.

Adding two $n$ digit numbers, for example, can be done in ${\cal
O}(n)$ (a linear function of $n$) elementary operations such as {\sc
nand} gates. In contrast, factoring an $n$ digit integer number into
prime numbers is a task for which the best known classical algorithms
require exponentially many operations, about ${\cal
O}(e^{n^{1/3}})$~\cite{Knuth98a}.

Computer scientists call an algorithm {\em efficient} if the required
resources grow only polynomially with the problem size, and call it
{\em inefficient} if the resources increase exponentially.  Problems
for which there exist efficient algorithms are called {\em
tractable}. Problems for which no efficient algorithm exists are
called {\em intractable} or {\em hard}.

In order to drive home the significance of intractibility, consider
the travelling salesman problem, which is provably intractable
classically. Suppose that on a fast computer it takes one second to
find the shortest path connecting 100 cities. For 101 cities, it would
then take two seconds, for 102 cities four seconds and so forth. It
would then take over an hour to find the shortest path through 112
cities, and over a year for only 125 cities. The same fast computer
would need over 35 billion years to solve the travelling salesman
problem for only 160 cities, longer than the estimated age of
the universe $\ldots$

\subsubsection{Quantum algorithms}

The extraordinary promise of quantum  computing is that \\

{\em certain problems which appear intractable on any classical
computer are tractable on a quantum computer.}\\

Factorization of integers into products of prime numbers is an example
of such a problem.  This problem is believed to be intractable on any
classical machine (although this remains to be proven): a $400$-digit
integer cannot be factored with high probability of success in a
reasonable time, not with a handheld calculator, not with a personal
computer, not with a supercomputer, and not using all the fastest
supercomputers combined.

However, in 1994, almost 10 years after Deutsch introduced quantum
parallellism, Peter Shor stunned the world with an efficient quantum
algorithm for prime factorization and computation of discrete
logarithms~\cite{Shor94a,Shor97a}.  The practical importance of this
algorithm is that it could be used to break widely used cryptographic
codes, such as the RSA public key cryptographic
system~\cite{Rivest78a}. These codes are based precisely on the fact
that no efficient (classical) algorithm is known for factoring. At a
more fundamental level, Shor's algorithm is the most powerful
example of how quantum mechanics offers a new way of thinking about
information and computation. As a result, the announcement of Shor's
algorithm gave a tremendous boost to the interest in quantum computing
of both funding agencies and scientists.

Historically, the first quantum algorithm was invented by David
Deutsch and Richard Jozsa (1992)~\cite{Deutsch92a}.  This algorithm
allows a quantum computer to solve with certainty an artificial
mathematical problem known as Deutsch's problem. It provided the first
steps towards Simon's algorithm~\cite{Simon94a,Simon97a}, and later to
Shor's algorithm.  Furthermore, it is important as a simple quantum
algorithm that can be experimentally tested.

Another class of quantum algorithms was discovered in 1996 by Lov
Grover~\cite{Grover96a,Grover97a}.  These algorithms allow a quadratic
speed-up of unstructured search problems, for which there is no better
approach classically than to try all $L$ candidate solutions one at a
time. A quantum computer using Grover's algorithm needs to make only
$\sqrt{L}$ such trials.  Even though this speed-up is only quadratic
rather than exponential, it is still significant.

The last currently known algorithmic application of quantum computers
lies in the simulation of other quantum systems \cite{Lloyd96a}, as
Feynman conjectured. Even a computer consisting of no more than a few
dozen quantum bits could outperform the fastest classical computers in
solving relevant physics problems, such as calculating the energy
levels of an atom.

\subsubsection{Scope of quantum computing}

We close with two final remarks on quantum algorithms. 
\begin{enumerate}
\item Quantum
computing cannot offer any speed-up at all for many common tasks, such
as adding up two numbers or word processing, which can already be
done efficiently on a classical computer. 
\item There are many problems which are classically intractable, but 
for which no efficient quantum algorithm is possible
either~\cite{Bennett97b}.
\end{enumerate}

An efficient quantum algorithm for the travelling salesman problem or
a similar problem would have an enormous impact in the computer
science community and the computer industry.  Whether or not such a
breakthrough will be made, it would be somewhat disappointing from a
practical viewpoint if no other applications of quantum computers were
found than the ones currently known. Either way, it is for certain
that the developments in quantum computation have dramatically changed
our understanding of the connection between physics, information and
computation.\\

We gave here only a brief summary of the known quantum algorithms.
Section~\ref{qct:alg} explains the operation and steps of the
Deutsch-Jozsa algorithm, Grover's algorithm and Shor's algorithm in
detail. We will present experimental implementations of simple
instances of each of these algorithms in chapter~\ref{ch:expt}.

%%%%%%%%%%%%%%%%%%%%%%%%%%%%%%%%%%%%%%%%%%%%%%%%%%%%%%%%%%%%%%%%%%%%%%%%%

\subsection{Correcting quantum errors}
\label{sec:corr_q_errors}
\subsubsection{Quantum errors or decoherence}

Quantum parallellism and quantum algorithms inherently rely on quantum
mechanical superpositions.  However, in real quantum systems,
superposition states are preserved only for a limited time: quantum
bits gradually loose coherence due to unavoidable interactions with
the environment, so the information stored in the coefficients of the
terms in a superposition is lost. The environment in a sense acts as a
measuring device which alters the state of the quantum system.  This
non-unitary process is called {\em
decoherence}~\cite{Zurek82a,Zurek91a}.  The time for which
superposition states are preserved is called the {\em coherence time}.

From a fundamental point of view, decoherence is extremely
interesting, although little understood. It is our explanation for the
fact that we never see macroscopic objects in two states at the same
time. Based on quantum theory, we would in fact expect such
macroscopic superpositions, as Schr\"odinger pointed out in his famous
gedanken experiment~\cite{Schrodinger35a}. He imagined a cat in a
perfectly closed box, and in the same box an atom in a superposition
of its ground state and first excited state. If the atom decays, it
emits a photon which sets off a trigger which in turn releases a
poisonous gas in the box, that would kill the cat. If the atom is in a
superposition of having decayed and not decayed, the cat should be in
a superposition of its dead and alive states!

This prediction of quantum mechanics is contrary to our intuition
based on observations of the world around us --- if we see a cat, we
expect it to be either alive or dead, but not dead and alive at the
same time. The explanation is that a cat interacts so strongly with
the environment that it decoheres almost instantaneously into either
the dead or alive state, much too fast for superpositions of dead and
alive cats to be observed.

From a practical point of view, decoherence can be detrimental for
quantum computers, because it causes random errors in the state of the
qubits~\cite{Unruh95a}. Therefore, quantum computations must either be
completed within the coherence time or the errors resulting from
decoherence must be corrected before their effect is too severe.

\subsubsection{Quantum error correction}

The correction of quantum errors arising from decoherence is much more
complicated than the correction of classical errors. First, in
contrast to the classical case, quantum errors occur not only as bit
flips but can be arbitrary rotations in the Bloch sphere, which will
influence the outcome of a subsequent measurement.  Furthermore, a
measurement which obtains information about a quantum state inevitably
disturbs it.  Finally, the no-cloning theorem
(page~\pageref{page:no-cloning}) forbids making copies of unknown
quantum states.

For several years, the correction of truly random errors due to
decoherence looked hopeless. Therefore, the implementation of
practical quantum computers appeared virtually impossible since some
degree of decoherence is unavoidable.

The invention of quantum error correction in 1995, independently by
Peter Shor \cite{Shor95a} and by Andrew Steane \cite{Steane96a},
represented a crucial breakthrough which gave hope that practical
quantum computers may be feasible. The main steps in quantum error
correction are the same as in classical error correction: encoding, a
noisy process, decoding, correction
(Fig.~\ref{fig:error_corr}). Despite this similar general outline, the
difficulties in correcting quantum errors require fundamentally
different solutions.  These principles will be made explicit in
section~\ref{qct:qec}, where we explore quantum error correction in
more detail.

\bfig 
\begin{center}
\vspace*{2ex}
\includegraphics*[width=12cm]{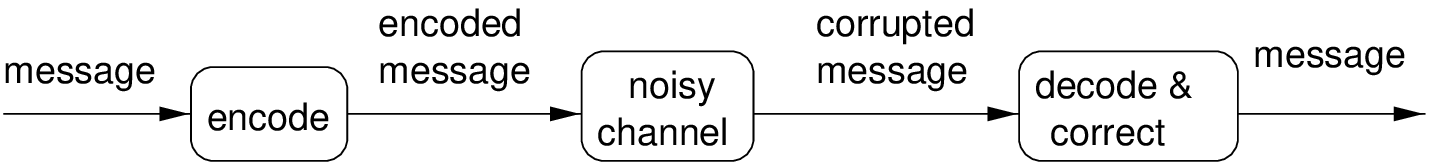} 
\end{center} 
\vspace*{-2ex}
\caption{A message (one or more bits or qubits) is first redundantly 
encoded, the encoded message then goes through the process of interest
(transmission over a noisy channel, a computation subject to errors,
etcetera), and finally the corrupted encoded message is decoded and
corrections are made if needed, based on the error syndrome
(information about which errors occurred, contained in the redudancy
bits.}
\label{fig:error_corr}
\efig

Compared to the classical case, quantum error correction involves an
even greater overhead: encoding, decoding and correction require many
additional qubits and operations. An important question then is
whether quantum error correction actually corrects for more errors
than it introduces, when the operations are carried out with faulty
components. The answer~\cite{Aharonov97a,Kitaev97b,Knill98c}, a second
key result for quantum computation, was that

\begin{quote}
{\em provided the error rate (probability of error per elementary
operation) is below a certain threshold, and given a fresh supply of
qubits in $\ket{0}$, it is possible to perform arbitrarily long
quantum computations.}
\end{quote}

The critical threshold is called the {\em accuracy threshold}
\label{page:threshold}. It is currently estimated to be between 
$10^{-4}$ and $10^{-6}$, depending on the assumptions made about the
nature of the errors and the process of interest.

%%%%%%%%%%%%%%%%%%%%%%%%%%%%%%%%%%%%%%%%%%%%%%%%%%%%%%%%%%%%%%%%%%%%%%%%%
%%%%%%%%%%%%%%%%%%%%%%%%%%%%%%%%%%%%%%%%%%%%%%%%%%%%%%%%%%%%%%%%%%%%%%%%%

\section{Quantum gates and circuits}
\label{qct:gates}

Any algorithm consists of a sequence of steps; in quantum algorithms,
each step is a unitary transformation, $U_k$. In theory, we could
implement the unitary transformation for each step, $U_k$, by letting
the qubits evolve under the Hamiltonian $i \ln U_k /\Delta t$ for a
duration $\Delta t$ (we recall Eq.~\ref{eq:U}). In actual experiments,
it is often not practical to turn on arbitrary Hamiltonians.
Fortunately, as we will see, a small set of Hamiltonians is sufficient
to generate arbitrary unitary transformations.

A convenient description of the steps in quantum algorithms at an
intermediate level of abstraction is based on quantum gates, analogous
to classical logic gates such as the {\sc not}, {\sc and} and {\sc
xor} gates. In this section, we will

\begin{enumerate}
\item describe quantum gates which can be directly implemented on
    realistic quantum computers,
\item provide a universal set of implementable quantum gates,
\item present methods to decompose complex quantum gates into sequences 
of directly-implementable gates, and
\item introduce the {\em quantum circuit} notation for the relevant
    quantum gates (quantum circuits are diagrams which represent a
    sequence of quantum gates applied to one or more
    qubits)~\cite{Deutsch89a,Yao93a}.
\end{enumerate}

The first three points are the subject of the following three
subsections. The fourth point will be covered throughout this section.

%%%%%%%%%%%%%%%%%%%%%%%%%%%%%%%%%%%%%%%%%%%%%%%%%%%%%%%%%%%%%%%%%%%%%%%%%

\subsection{Directly implementable quantum gates}
\label{sec:qct_direct_gates}

The Hamiltonians present in simple quantum systems contain
single-particle terms and interaction terms between two particles;
three-particle terms are normally not observed.  Therefore, the only
quantum gates which we can easily implement directly are gates which
act on one or two particles. If each particle represents a qubit, as
we shall assume for now, we can thus realize one- and two-qubit gates
directly.

\subsubsection{One-qubit gates}

Let us first introduce a convenient matrix representation in which
to describe quantum states and unitary transformations.  The quantum
state $\ket{\psi} = a \ket{0} + b
\ket{1}$ is written in matrix notation as
\be
\ket{\psi} = \left[\matrix{a \cr
 			   b }\right] \,,
\label{eq:psi_matrix}
\ee
a column vector containing the complex amplitudes of the $\ket{0}$ and
$\ket{1}$ terms.  The matrix representation of $\bra{\psi}$ is the
complex conjugate of the transpose of the vector $\ket{\psi}$. The
inner product $\bra{\psi_1}\psi_2\rangle$ and the outer product
$\ket{\psi_1}\bra{\psi_2}$ can be computed as the respective products
of the corresponding column and row matrices.

Now let us consider the simplest building block of quantum
computation, a one-qubit quantum gate, called the {\sc not} gate. It
maps $\ket{0}$ onto $\ket{1}$ and vice versa, similar to classical
inversion.  The unitary matrix which effects this transformation for
arbitrary input states is
\be
U_{\mbox{\sc not}}= \left[\matrix{0 & 1 \cr
 				  1 & 0 }\right] \,.
\ee
The action of a unitary operator $U$ on a quantum state $\ket{\psi}$,
\be
\ket{\psi}_{\mbox{final}} = U \,\ket{\psi}_{\mbox{initial}} \,,
\label{eq:action_U}
\ee
can be calculated by standard matrix multiplication. For example, the
output state obtained after applying $U_{\mbox{\sc not}}$ to
$\ket{\psi}$ of Eq.~\ref{eq:psi_matrix} is
\be
U_{\mbox{\sc not}}\, \ket{\psi} 
= \left[\matrix{0 & 1 \cr 1 & 0 }\right] \left[\matrix{a \cr b}\right]
= \left[\matrix{b \cr a}\right]
\,,
\ee
which is the state vector corresponding to the state $a\ket{1} +
b\ket{0}$, as expected given Eq.~\ref{eq:1qubit_out}.  \\

The rest of the discussion of one-qubit gates expands on the following
notion: any one-qubit unitary operator can be written in the form
\be
U = e^{i \alpha} R_{\hat{n}}(\theta) \,,
\label{eq:U_rot}
\ee
where $R_{\hat{n}}(\theta)$ corresponds to a rotation in the Bloch
sphere (Fig.~\ref{fig:bloch_sphere}) about the $\hat{n} =
(n_x,n_y,n_z)$ axis and over an angle $\theta$. If there is ambiguity
about which qubit $R$ acts on, we use a superscript to indicate the
label of the qubit, $R_{\hat{n}}^i(\theta)$.  In order to give an
explicit definition of $R_{\hat{n}}(\theta)$, let us define the usual
Pauli matrices
\be
\sigma_x \equiv \left[\matrix{0 & 1 \cr 1 & 0}\right] \,, \quad
\sigma_y \equiv \left[\matrix{0 &-i \cr i & 0}\right] \,, \quad
\sigma_z \equiv \left[\matrix{1 & 0 \cr 0 &-1}\right] \,,
\ee
which obey the relations
\be
\sigma_x \sigma_y = i \sigma_z \,, \quad
\sigma_x \sigma_z = - i \sigma_y \,, \quad
\sigma_y \sigma_z = i \sigma_x \,, \quad
\label{eq:pauli_rel1}
\ee
\be
\sigma_x^2 = \sigma_y^2 =  \sigma_z^2 = \sigma_I \,,
\label{eq:pauli_rel2}
\ee
where 
\be
\sigma_I \equiv \left[\matrix{1 & 0 \cr 0 & 1}\right] \,.
\ee
With ${\vec{\sigma}} = (\sigma_x, \sigma_y, \sigma_z)$, we can
construct $R_{\hat{n}}(\theta)$ by exponentiating the Pauli operators
as follows:
\be
R_{\hat{n}}(\theta) \equiv 
\mathrm{exp} \left(-i \frac{\theta \hat{n} {\vec{\sigma}}}{2}\;\right) =
\cos(\theta/2) \, \sigma_I \, - \, i\, \sin\,
(\theta/2) [n_x \sigma_x + n_y \sigma_y + n_z \sigma_z] \,.
\ee
Rotations about the $\hat{x}, \hat{y}$ and $\hat{z}$ axis
respectively, are thus given by
\be
R_x( \theta) = 
\mathrm{exp} \left(\frac{-i \theta \sigma_x}{2}\right) =
\cos(\theta/2) \, \sigma_I \, 
- \, i\, \sin\,(\theta/2) \sigma_x 
= \left[\matrix{\cos \frac{\theta}{2} &
          -i \, \sin \frac{\theta}{2} \cr 
          -i \, \sin \frac{\theta}{2} &
                \cos \frac{\theta}{2} }\right] \,,
\label{eq:R_x}
\ee
\be
R_y( \theta) = 
\mathrm{exp} \left(\frac{-i \theta \sigma_y}{2}\right) =
\cos(\theta/2) \, \sigma_I \, 
- \, i\, \sin\,(\theta/2) \sigma_y 
= \left[\matrix{\cos \frac{\theta}{2} &
           - \, \sin \frac{\theta}{2} \cr 
             \, \sin \frac{\theta}{2} &
                \cos \frac{\theta}{2} }\right] \,,
\ee
\be
R_z( \theta) = 
\mathrm{exp} \left(\frac{-i \theta \sigma_z}{2}\right) =
\cos(\theta/2) \, \sigma_I \, 
- \, i\, \sin\,(\theta/2) \sigma_z 
= \left[\matrix{ e^{-i \theta/2} & 0 \cr 
                 0 & e^{ i\theta/2} }\right] \,.
\label{eq:R_z}
\ee

A one-qubit gate which deserves special mention is the {\sc Hadamard}
gate, defined as
\be
H = \frac{1}{\sqrt{2}} \left[\matrix{1 & 1 \cr
 			             1 &-1 }\right] \,.
\ee
This gate transforms the computational basis states into the equal
superposition states, and back:
\be
\ket{0} \stackrel{\textstyle H}{\longleftrightarrow} \frac{\ket{0}+\ket{1}}{\sqrt{2}} 
\quad \mbox{and} \quad
\ket{1} \stackrel{\textstyle H}{\longleftrightarrow} \frac{\ket{0}-\ket{1}}{\sqrt{2}} \,.
\label{eq:Had_effect}
\ee
The {\sc hadamard} gate corresponds to a rotation over $180^\circ$
about an axis halfway between the $\hat{x}$ and the $\hat{z}$ axes.
The {\sc not} gate corresponds to a $180^\circ$ rotation about the
$\hat{x}$ axis, up to an overall phase factor, which is irrelevant.\\

The quantum circuit element for a qubit is a horizontal wire. An
arbitrary single-qubit gate $U$ is represented as shown in
Fig.~\ref{fig:U1_circuit} (a). The {\sc not} gate is
often represented by the $\oplus$ symbol, as in
Fig.~\ref{fig:U1_circuit} (b).

\bfig
\bcen
\vspace*{1ex}
\includegraphics*[width=9cm]{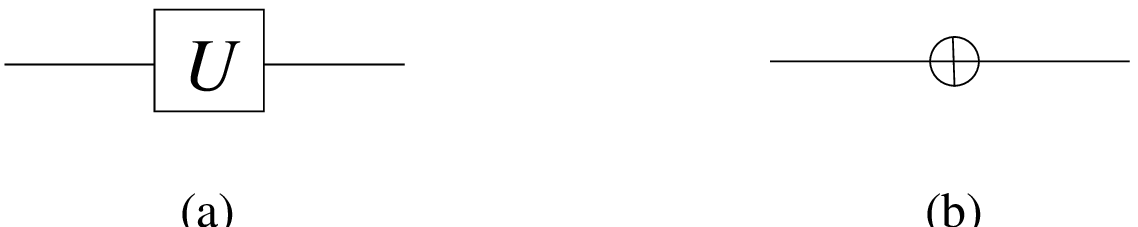} 
\vspace*{-2ex}
\ecen
\caption{The quantum circuit representation of (a) an arbitrary 
one-qubit gate $U$ and of (b) the {\sc not} gate.}
\label{fig:U1_circuit}
\efig

\subsubsection{Two-qubit gates}

The prototypical two-qubit gate, for historical reasons, is the
controlled-{\sc not} or {\sc cnot} gate; {\sc cnot}$_{ij}$ flips
(performs a {\sc not} operation on) qubit $j$, called the target, if
and only if qubit $i$, called the control qubit, is in the state
$\ket{1}$. The truth table is shown in Fig.~\ref{fig:cnot_table}.

\bfig
\bcen
\begin{tabular}{c|c}
$\rm In$  & $\rm Out$ \\ \hline
$0$  $0$  & $0$ $0$  \\
$0$  $1$  & $0$ $1$  \\ 
$1$  $0$  & $1$ $1$  \\ 
$1$  $1$  & $1$ $0$  \\ 
\end{tabular}
\hspace*{2cm}
\begin{tabular}{c|c}
$\rm In$  & $\rm Out$ \\ \hline
$0$  $0$  & $0$ $0$  \\
$0$  $1$  & $1$ $1$  \\ 
$1$  $0$  & $1$ $0$  \\ 
$1$  $1$  & $0$ $1$  
\end{tabular} \vspace*{2ex} \\
{\sc cnot}$_{12}$ \hspace*{2.8cm} {\sc cnot}$_{21}$
\ecen
\caption{Truth table of the {\sc cnot} gate with (Left) the first qubit 
in the role of the control qubit and (Right) the second qubit in
the role of the control qubit.}
\label{fig:cnot_table}
\efig

The matrix representation for an arbitary two-qubit state $\ket{\psi}
= c_0\ket{00} + c_1\ket{01} + c_2\ket{10} + c_3\ket{11}$ is
\be
\ket{\psi} = \left[\matrix{c_0 \cr c_1 \cr c_2 \cr c_3}\right] \,,
\ee
and accordingly, unitary matrices representing two-qubit gates are of
dimension $4 \times 4$. For example, the unitary matrices
corresponding to {\sc cnot}$_{12}$ and {\sc cnot}$_{21}$ are 
\be
U_{\mbox{\sc cnot}_{12}} = \left[\matrix{1 & 0 & 0 & 0 \cr 
			  	         0 & 1 & 0 & 0 \cr
				         0 & 0 & 0 & 1 \cr
				         0 & 0 & 1 & 0 }\right] \quad
\mbox{and} \quad
U_{\mbox{\sc cnot}_{21}} = \left[\matrix{1 & 0 & 0 & 0 \cr 
				         0 & 0 & 0 & 1 \cr
				         0 & 0 & 1 & 0 \cr
				         0 & 1 & 0 & 0 }\right] \,.
\label{eq:U_cnot}
\ee
An obvious extension of the {\sc cnot} gate is the controlled-$U$
gate, where a single-qubit operation $U$ is performed on the target
qubit if and only if the control qubit is in $\ket{1}$.  Analogous to
the controlled-$U$ gate, we also define the zero-controlled-$U$ gate,
in which $U$ is executed if and only if the control is $\ket{0}$. The
last two-qubit gate we wish to introduce here is the {\sc swap} gate,
\be
U_{\mbox{\sc swap}} = \left[\matrix{1 & 0 & 0 & 0 \cr 
				    0 & 0 & 1 & 0 \cr
				    0 & 1 & 0 & 0 \cr
				    0 & 0 & 0 & 1 }\right] \,, 
\label{eq:U_swap}
\ee
which, as the name suggests, swaps the state of the two qubits.  The
quantum circuit representations of all these two qubit operations are
collected in Fig.~\ref{fig:U2_circuit}.

\bfig
\bcen
\vspace*{1ex}
\includegraphics*[width=12cm]{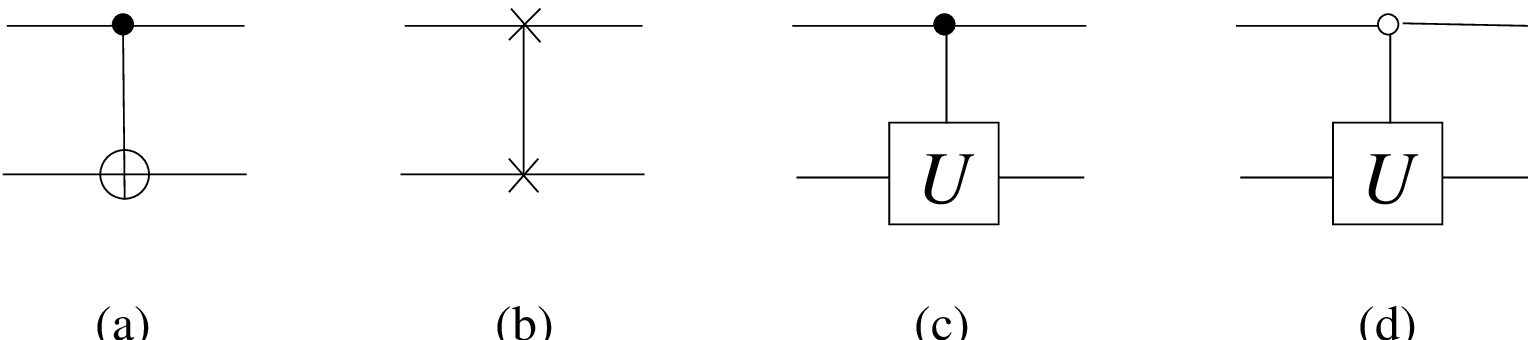} 
\vspace*{-2ex}
\ecen
\caption{Quantum circuits for (a) the {\sc cnot}$_{12}$ gate, (b) the 
{\sc swap} gate, (c) a controlled-$U$ gate and (d) a
zero-controlled-$U$ gate. The $\bullet$ symbol indicates the control
qubit; it controls the operation it is connected to via a vertical
line. The $\circ$ symbol indicates a zero-control qubit.  A vertical
line connecting two $\times$ symbols denotes a {\sc swap} operation of
the two qubits.}
\label{fig:U2_circuit}
\efig

Now that we have introduced a set of widely used and useful one- and
two-qubit gates, we will examine if and how we can construct a
universal set of quantum logic gates using only one- and two-qubit
operations.

%%%%%%%%%%%%%%%%%%%%%%%%%%%%%%%%%%%%%%%%%%%%%%%%%%%%%%%%%%%%%%%%%%%%%%%%%

\subsection{Universality}
\label{qct:universality}

A {\em universal set of classical logic gates} has the property that
any implementable Boolean function can be implemented by an
arrangement of gates belonging to this set. For example, the {\sc
nand} gate is in itself universal, and so is the {\sc nor}
gate. Obviously, the {\sc and} gate and the {\sc not} gate together
thus also constitute a universal set of logic gates. By extension, the
{\sc Toffoli} gate is in itself universal for reversible classical
logic.

A {\em universal set of quantum logic gates} has the property that any
unitary transformation can be implemented (or approximated to
arbitrary accuracy) by an arrangement of gates belonging to this
set. Clearly, the {\sc Toffoli} is not universal for quantum logic, as
it is for example not possible to create a superposition state
starting from the ground state using just {\sc Toffoli} gates. Deutsch
presented a three-qubit gate (a rotation of one qubit conditioned upon
two other qubits being $\ket{1}$) which is
universal~\cite{Deutsch89a}; DiVincenzo presented a universal set of
four two-qubit gates~\cite{DiVincenzo95b}. Lloyd~\cite{Lloyd95a} and
independently Deutsch, Barenco and Ekert~\cite{Deutsch95a}, extended
this result to show that almost any two-qubit gate is universal.  

Of course, certain sets of universal quantum gates are more practical
than others to work with, and we will come back to this in
section~\ref{impl:gates}. The most widely used result in the
theory of quantum gates is that 

\begin{quote}
{\em the combination of the {\sc cnot} gate with arbitrary single-qubit
rotations constitutes a set of universal quantum gates.}
\end{quote}

In fact, the {\sc cnot} along with arbitrary rotations about $\hat{x}$
and $\hat{y}$ is sufficient as well, as it can be shown that for any
single-qubit rotation $U$ there exist real numbers $\alpha, \beta,
\gamma$ and $\delta$ such that
\be
U = e^{i\alpha} R_x(\beta) R_y(\gamma) R_x(\delta) \,.
\label{eq:universal_1bitgate}
\ee
We will give examples of decompositions of multi-qubit gates into just
single-qubit gates and {\sc cnot}'s in
section~\ref{sec:multi-q-gates}.

We close by remarking that universality does not say anything about
{\em efficiency}. In fact, it has been proven that the required number
of elementary operations, such as {\sc cnot}s and single-qubit
rotations, increases exponentially with $n$ for almost all $n$-qubit
unitary operations. Therefore, a crucial part in the design of quantum
algorithms is to prove that each of the steps can be implemented
efficiently, i.e. in only polynomially many elementary operations.

%%%%%%%%%%%%%%%%%%%%%%%%%%%%%%%%%%%%%%%%%%%%%%%%%%%%%%%%%%%%%%%%%%%%%%%%%

\subsection{Remarks on unitary operators}
\label{sec:remarks_unitaries}

This section with technical remarks answers two questions: (1) how do
we compute the unitary operator corresponding to several consecutive
gates, and (2) given an operation which acts on a subset of $n$ qubits, how do we write the $2^n \times 2^n$ unitary matrix which describes the evolution of the $n$ qubits ?

\subsubsection{Multiplication and commutation of unitary operators}

The concatenation of several quantum logic gates is described by the
{\em product} of the corresponding unitary matrices, ordered such that
the operator of the {\em first gate is placed on the right}. Thus, the
unitary operator of $k$ consecutive operations $U_1, U_2, \ldots, U_k$
is written as
\be
U = U_k U_{k-1} \ldots U_2 U_1 \,.
\label{eq:multiply_Us}
\ee
This may seem awkward but makes sense if we recall
Eq.~\ref{eq:action_U}, because this way $U_1$ is applied to
$\ket{\psi}$ first, then $U_2$ and so forth.
The order is important since in general
\be
U_2 U_1 \neq U_1 U_2 \,,
\ee
that is, in general, two {\em unitary operators may not commute under
multiplication}. Hermitian matrices also may or may not commute with
each other. Furthermore, for two non-commuting Hermitian operators
$H_1$ and $H_2$,
\begin{eqnarray}
e^{-i H_1} e^{-i H_2} &\neq& e^{-i H_2} e^{-i H_1} \,, \\
e^{-i H_1} e^{-i H_2} &\neq& e^{-i (H_1 +H_2)} \,, \\
e^{-i H_2} e^{-i H_1} &\neq& e^{-i (H_1 +H_2)} \,.
\end{eqnarray}
These inequalities demonstrate the importance of commutation
properties for quantum computing. Turning on two terms in the
Hamiltonian at the same time does not have the same effect as turning
them on one after the other; and turning on one first and then the
other is not the same as the other first and then the one. In closing,
let us introduce a few simple practical commutation rules:
\label{page:commutation}

\begin{enumerate}
\item Any unitary operator commutes with itself.
\item Unitary operators acting on different qubits commute.
\item All diagonal operators commute with each other.
\end{enumerate}

\subsubsection{Tensor products and (non)-local operations}

If a one--qubit gate $U_1$ is applied to one qubit $l$ of an $n$ qubit
system, we can write the $2^n \times 2^n$ unitary matrix $U$ acting on
the $n$ qubits as
\be
U = 
\underbrace{\sigma_I \otimes \ldots \otimes \sigma_I}_{l-1\; \mathrm{factors}} 
\otimes \; U_1 \otimes
\underbrace{\sigma_I \otimes \ldots \otimes \sigma_I}_{n-l\; \mathrm{factors}} 
\,,
\ee
Similar extensions apply for any gate applied to any subsystem of a
larger system.  As an example of the effect of tensor products on
matrices, the $4 \times 4$ unitary matrices representing a {\sc not}
operation on the first respectively the second of two qubits are
\be
U_{\mbox{\sc not}_1} = \left[\matrix{0 & 0 & 1 & 0 \cr 
		  0 & 0 & 0 & 1 \cr
		  1 & 0 & 0 & 0 \cr
		  0 & 1 & 0 & 0 }\right] 
\quad \mbox{and} \quad
U_{\mbox{\sc not}_2} = \left[\matrix{0 & 1 & 0 & 0 \cr 
		  1 & 0 & 0 & 0 \cr
		  0 & 0 & 0 & 1 \cr
		  0 & 0 & 1 & 0 }\right] \,.
\ee
Finally, we point out that any concatenation of one-qubit gates $U_k$
on different qubits $k$ can be written in the form
\be
U = U_1 \otimes U_2 \otimes \ldots \otimes U_n
\,.
\ee
In contrast, a two-qubit or multi-qubit gate cannot in general be
factored into such a product of single-qubit operators. This
distinction is directly related to the distinction between separable
and non-separable states, mentioned in section~\ref{qct:qubits}. {\em
Local operations} (single-qubit operations) cannot create or undo
entanglement, whereas {\em non-local} (multi-qubit) operations can.
The {\sc cnot}$_{12}$ gate (Eq.~\ref{eq:U_cnot}) for example
transforms the non-entangled state
$\frac{1}{\sqrt{2}}(\ket{0}+\ket{1}) \ket{0} =
\frac{1}{\sqrt{2}}(\ket{00}+\ket{10})$ to the entangled state of
Eq.~\ref{eq:epr}, $\frac{1}{\sqrt{2}}(\ket{00}+\ket{11})$.

%%%%%%%%%%%%%%%%%%%%%%%%%%%%%%%%%%%%%%%%%%%%%%%%%%%%%%%%%%%%%%%%%%%%%%%%%

\subsection{Multi-qubit gates}
\label{sec:multi-q-gates}

We will refer to all gates which act on more than two-qubits as
multi-qubit gates. We already saw one example of such a gate in
section~\ref{qct:dyn&rev}, namely the {\sc Toffoli} gate. From its
truth table (Fig.~\ref{fig:and_gate}), we can see that this gate
flips the state of the target qubit (the third qubit in this case) if
and only if two control qubits are in $\ket{1}$. It is therefore also
called the doubly-controlled {\sc not} or {\sc ccnot} gate, as
conveyed in Fig.~\ref{fig:toffoli_circuit}.

\vspace*{1ex}
\bfig
\bcen
\includegraphics*[width=7cm]{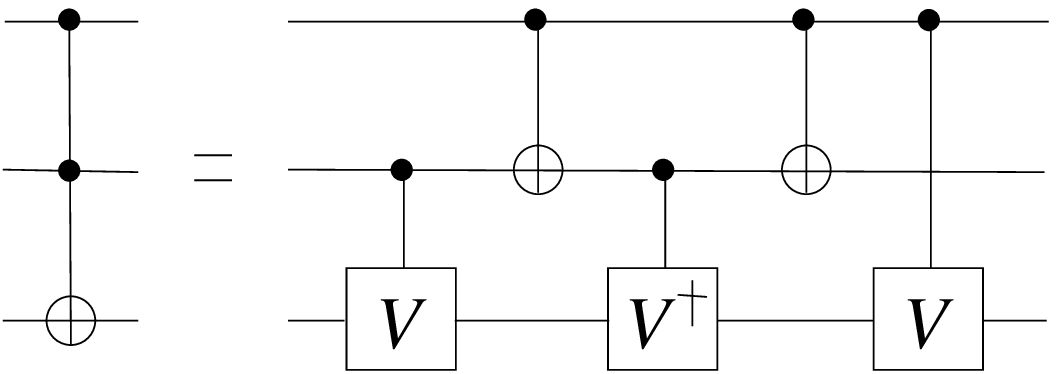} 
\hspace*{1cm}
\raisebox{1.3cm}{$V = 
\left[\matrix{\sqrt{\frac{i}{2}} & \sqrt{\frac{-i}{2}} \cr 
	      \sqrt{\frac{-i}{2}} & \sqrt{\frac{i}{2}}}\right] $}
\ecen
\vspace*{-2ex}
\caption{Quantum circuit representation of the {\sc Toffoli} or 
{\sc ccnot} gate, and its decomposition into two-qubit gates. We note
that $V^2 = U_{\mbox{\sc not}}$.}
\label{fig:toffoli_circuit}
\efig

The doubly-controlled {\sc not} gate has obvious extensions to
multiply-controlled {\sc not}'s. The number of elementary (one- and
two-qubit) operations needed to implement a gate with $n-1$ control
qubits is ${\cal O}(n^2)$. If we allow a scratch pad qubit (an
ancilla), the number of elementary operations is only ${\cal
O}(n)$~\cite{Barenco95a}.

Another useful and historically important gate is the {\sc Fredkin}
gate~\cite{Fredkin82a}, or controlled-{\sc swap} gate, which swaps the
state of two qubits if and only if a third qubit is in $\ket{1}$. The
two-qubit {\sc swap} gate (see page~\pageref{eq:U_swap}) between
qubits $2$ and $3$ can be implemented as {\sc cnot}$_{23}$ {\sc
cnot}$_{32}$ {\sc cnot}$_{23}$, or by symmetry also as {\sc
cnot}$_{32}$ {\sc cnot}$_{23}$ {\sc cnot}$_{32}$. Therefore, a
possible implementation of a {\sc swap} of qubits $2$ and $3$
controlled by qubit $1$ is as given in Fig.~\ref{fig:fredkin_circuit}.

\vspace*{1ex}
\bfig
\bcen
\includegraphics*[width=9cm]{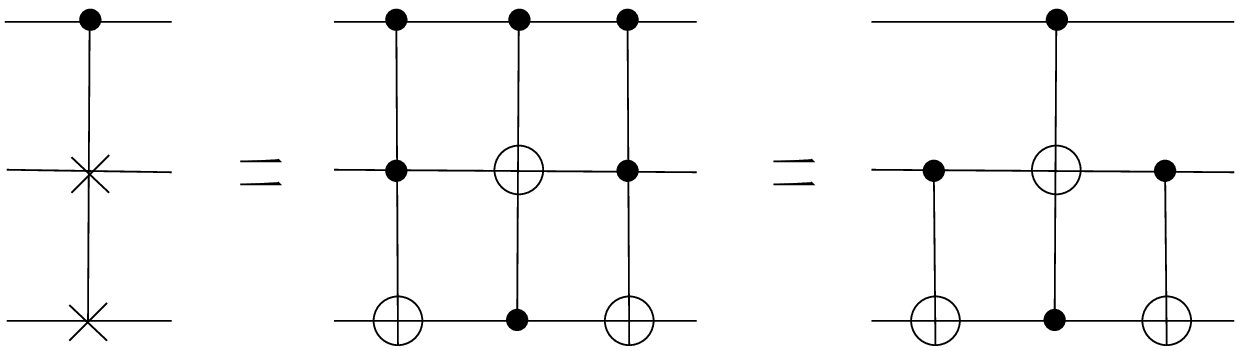} 
\ecen
\vspace*{-2ex}
\caption{Quantum circuit representation of the {\sc Fredkin} or 
{\sc cswap} gate, and two quantum circuits equivalent to the {\sc
Fredkin} gate.}
\label{fig:fredkin_circuit}
\efig

Efficient constructions for gates controlled by more than two qubits
are extensively described in the
literature~\cite{Barenco95a,DiVincenzo98c}.  In
sections~\ref{expt:order} and~\ref{expt:shor} respectively, we will
present quantum circuits for specific multi-qubit gates that were
implemented in our experiments.

%%%%%%%%%%%%%%%%%%%%%%%%%%%%%%%%%%%%%%%%%%%%%%%%%%%%%%%%%%%%%%%%%%%%%%%%%
%%%%%%%%%%%%%%%%%%%%%%%%%%%%%%%%%%%%%%%%%%%%%%%%%%%%%%%%%%%%%%%%%%%%%%%%%

\section{Quantum algorithms}
\label{qct:alg}

\subsection{The Deutsch-Jozsa algorithm}
\label{qct:dj}

In 1992, David Deutsch and Richard Jozsa invented the first ever
quantum algorithm~\cite{Deutsch92a}. The Deutsch-Jozsa algorithm
achieves an exponential advantage over classical algorithms in solving
Deutsch's problem~\cite{Deutsch95a} with certainty. Deutsch's problem
may be described as follows. You are given a black box or {\em oracle}
$f$ which takes $n$ input bits and returns one output bit. Furthermore
you are told that the black box either outputs the same value ($0$ or
$1$) for all possible input strings $x$, or outputs $0$ for exactly
half the possible input values and $1$ for the other input
values. Deutsch's problem is a thus a {\em promise} problem, and the
promise is that $f$ is either {\em constant} or {\em balanced}.

How many oracle queries do you need classically to solve Deutsch's
problem {\em with certainty}? As soon as you find that the oracle
returns $0$ for some inputs and $1$ for other inputs, you know for
certain that $f$ is balanced. However, if it the output is still the
same after trying $2^n/2$ different input values, the function $f$
might still be balanced, even though most likely it is constant. Only
when $2^n/2 + 1$ input values produce the same output, you can be sure
the function is really constant. Thus, in the worst case, you need
$2^n/2 + 1$ queries.

Using a quantum computer, the input of the oracle can be put in a
superposition of all possible input values, and a single oracle query
suffices to determine with certainty whether $f$ is constant or
balanced.  We note that rather than to compute individual $f(x)$,
which we know a quantum computer cannot do in fewer steps than a
classical computer, the task is to determine a {\em global property}
of the function, namely whether $f$ is constant or balanced. This is
a type of problem for which quantum computers may offer an
advantage.

\subsubsection{Procedure}

The steps of the Deutsch-Jozsa algorithm, as improved by Cleve {\em et
al.} \cite{Cleve98a} and Tapp \cite{Tapp98a}, are outlined in
Fig.~\ref{fig:DJ_circuit}.  The initial state is
\be
\ket{\psi_0} = \ket{0}^{\otimes n} \ket{1} \,,
\ee
where $^{\otimes n}$ indicates that the first register, the input
register, is of size $n$ (we will often leave this impicit). The
second register, the output register, contains only one qubit. First
we apply a {\sc hadamard} gate on each of the
$n+1$ qubits, resulting in the state
\be
\ket{\psi_1} = \sum_{x=0}^{2^n-1} \frac{\ket{x}}{\sqrt{2^n}} 
\left[ \frac{\ket{0} - \ket{1}}{\sqrt{2}} \right] \,.
\label{eq:dj_psi1}
\ee

\bfig
\bcen
\vspace*{1ex}
\includegraphics*[width=6.5cm]{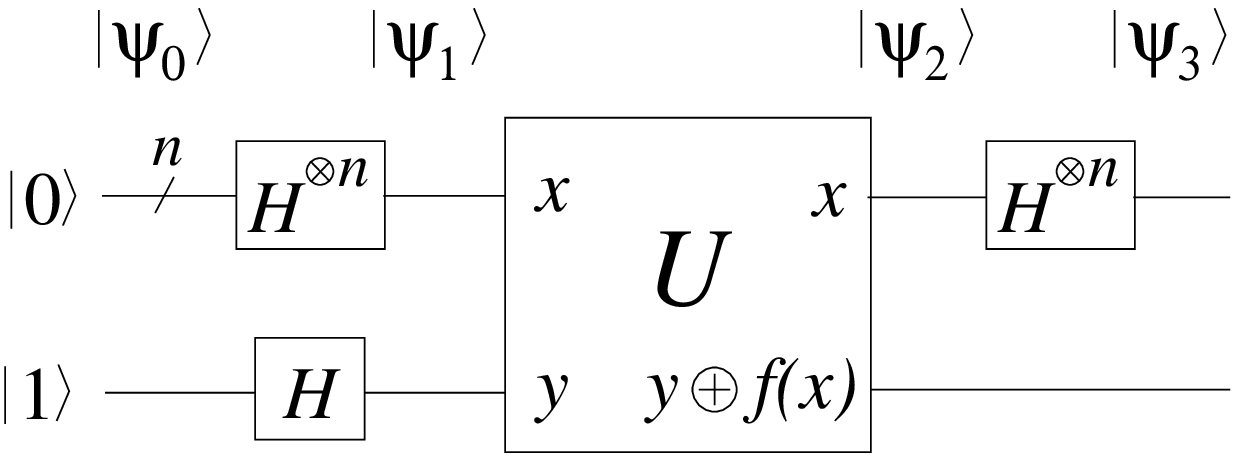} 
\vspace*{-2ex}
\ecen
\caption{Quantum circuit for the Deutsch-Jozsa algorithm.}
\label{fig:DJ_circuit}
\efig

The input register is now in an equal superposition of all possible
$x$. The reason why the output register is placed in $\frac{\ket{0} -
\ket{1}}{\sqrt{2}}$ will become clear shortly. Next we query the 
oracle $f$ (we come back to what it means in practice to query an
oracle on page~\pageref{page:oracle_call}), which effects the unitary
transformation
\be
U_f =
\ket{x}\ket{y} \stackrel{f}{\mapsto} \ket{x}\ket{y \oplus f(x)} \,,
\label{eq:dj_U_f}
\ee 
where $\oplus$ stands for addition modulo 2. 
The oracle thus transforms $\ket{\psi_1}$ to
\be
\ket{\psi_2} = \sum_x \frac{\ket{x}}{\sqrt{2^n}} 
\left[ \frac{\ket{0 \oplus f(x)} - \ket{1 \oplus f(x)}}{\sqrt{2}} \right]
\,.
\ee
This is an instance of quantum parallellism.  Now, we see that whenever
$f(x) = 0$, the output register does not change, and whenever $f(x) =
1$, the output register is changed to $\frac{\ket{1} -
\ket{0}}{\sqrt{2}} = - \frac{\ket{0} - \ket{1}}{\sqrt{2}}$. Thus the
oracle query has no net effect other than a sign flip whenever
$f(x)=1$ and we can rewrite $\ket{\psi_2}$ as
\be
\ket{\psi_2} = \sum_x \frac{(-1)^{f(x)} \ket{x}}{\sqrt{2^n}} 
\left[ \frac{\ket{0} - \ket{1}}{\sqrt{2}} \right] \,.
\ee
The value of $f(x)$ is thus encoded in the coefficient of
$\ket{x}$, by virtue of initializing the output qubit to
$\frac{\ket{0} - \ket{1}}{\sqrt{2}}$. Since the state of the output
qubit never changes, we could in fact leave this qubit out altogether
and implement $f$ via the unitary transformation $\ket{\psi}
\stackrel{f}{\mapsto} (-1)^{f(x)} \ket{x}$ \cite{Cleve98a}.

We already see that if $f$ is constant, the phase factor $(-1)^{f(x)}$
is constant as well, so it becomes a physically irrelevant overall
phase. In this case, the subsequent $H^{\otimes n}$ operation restores
the first register to the state $\ket{0}$. For the case of balanced
$f$, let us first calculate $H \ket{x_i}$ and then $H^{\otimes
n}\ket{x}$. From Eq.~\ref{eq:Had_effect}, we see that
\be
H \ket{x_i} = \frac{\ket{0} + (-1)^{x_i} \ket{1}}{\sqrt{2}} 
	= \sum_{z=0,1} \frac{(-1)^{x_i z} \ket{z}}{\sqrt{2}} \,.
\ee 
Therefore, 
\be
H^{\otimes n}\ket{x_1, \ldots, x_n} =
\frac{\sum_{z_1, \ldots, z_n} (-1)^{x_1 z_1 + \ldots + x_n z_n} 
\ket{z_1 \ldots z_n}}{\sqrt{2^n}} \;
= \; \frac{\sum_{z} (-1)^{x \cdot z} \ket{z}}{\sqrt{2^n}} \,,
\ee
where $x \cdot z$ is the bitwise inner product of $x$ and $z$, modulo
$2$. Using this result, we find that 
\be
\ket{\psi_3} = H^{\otimes n} \ket{\psi_2} = 
\sum_z \sum_x \frac{(-1)^{x \cdot z + f(x)} 
\ket{z}}{\sqrt{2^n}} 
\left[ \frac{\ket{0} - \ket{1}}{\sqrt{2}} \right] \,.
\ee
We now measure the first register. For constant $f$, the amplitude of
the $\ket{0}^{\otimes n}$ term, $\sum_x (-1)^{f(x)}$, is either $+1$
or $-1$, depending on the constant value $f$ takes. Given the
normalization condition of Eq.~\ref{eq:normalization}, the amplitude
of the remaining terms must thus be zero, like we anticipated. For
balanced $f$, we always have that $\sum_x (-1)^{f(x)} = 0$ as there
are as many positive as negative $f(x)$. The amplitude of the
$\ket{0}^{\otimes n}$ term is thus zero in this case.  In summary, if
the measurement of the first register gives all $0$'s we know $f$ is
constant, and otherwise $f$ is balanced.

\subsubsection{Significance}

We have thus shown that the Deutsch-Jozsa algorithm solves Deutsch's
problem exponentially faster than any classical machine. While this is
truly remarkable in itself, the practical importance of this algorithm
is limited. First, Deutsch's problem is an artificial mathematical
problem which has no known applications. Second, classical computers
can solve this problem quickly and with high probability of success by
asking the oracle what $f(x)$ is for a few random $x$: the probability
for obtaining $k$ times the same answer (either 0 or 1) if $f$ is
balanced decreases as $(1/2)^{k-1}$. Only if absolute certainty is
required, exponentially many oracle queries may be required
classically.

The significance of this algorithm therefore lies mostly in that it
inspired later, more useful algorithms, is relatively easily
understood, and can be used as a simple test for implementations of
quantum computers. In section~\ref{expt:dj}, we will present such
an experiment on a two-qubit NMR quantum computer.

%%%%%%%%%%%%%%%%%%%%%%%%%%%%%%%%%%%%%%%%%%%%%%%%%%%%%%%%%%%%%%%%%%%%%%%%%

\subsection{Grover's algorithm}
\label{qct:grover}

In 1996, Lov Grover invented a quantum algorithm for {\em unstructured
searches}~\cite{Grover96a,Grover97a}. An example of a structured
search is finding the phone number matching with a certain name using
a phone book with $N$ alphabetically listed names. An example of an
unstructured search is to find the name matching with a certain phone
number using the same phone book. The time this takes goes up linearly
with $N$: on average you will have to try $[N(N+1)/2-1]/N \approx N/2$
different names before you find the one with the desired number. In
contrast, using Grover's algorithm, such a search can be accomplished
in $\sqrt{N}$ attempts.

Mathematically, we can describe this as the following promise
problem. Given an oracle which returns $f(x)=0$ for all values of $x$
except for a unique entry $x=x_0$ for which $f(x)=1$ (there is a
unique name $x_0$ in the phone book which has the desired phone
number), find the special element $x_0$ in the least number of oracle
queries.

As in the Deutsch-Jozsa algorithm, the oracle query takes the form of
the transformation
\be
U_f = 
\ket{x}\ket{y} \stackrel{f}{\mapsto} \ket{x}\ket{x \oplus y} \,,
\label{eq:grover_U_f}
\ee
where we will initialize the state of the output qubit $\ket{y}$ to 
$\frac{\ket{0} - \ket{1}}{\sqrt{2}}$. As we have seen in
section~\ref{qct:dj}, the content of the output register in fact
doesn't change, and $f(x)$ is encoded in the sign of $\ket{x}$. We
will therefore leave out the second register and from now on only
consider the effect of the oracle call on $\ket{x}$.

\subsubsection{Procedure and performance}

The steps in Grover's algorithm for a search space of size
$N = 2^n$ are:\\

\noindent {\bf (a)} Initialize to $\ket{0}^{\otimes n}$.\\

\noindent {\bf (b)} Apply $H^{\otimes n}$ to obtain 
$\frac{1}{\sqrt{N}} \sum_{x=0}^{N-1} \ket{x}$. \\

\noindent {\bf (c)} Repeat the following subroutine, known as the 
{\em Grover iteration}, $\lceil \pi \sqrt{N}/4 \rceil$ times:

\begin{enumerate}
\item Query the oracle 
$U_f: \ket{x} \stackrel{f}{\mapsto} (-1)^{f(x)} \ket{x}$.
This flips the phase of the $\ket{x_0}$ term.
\item Apply $H^{\otimes n}$.
\item Flip the phase of all terms except the $\ket{0}$ term. 
Thus, $\forall x \neq 0: \ket{x} \mapsto - \ket{x}$; $\ket{0} \mapsto
    \ket{0}$.
\item Apply $H^{\otimes n}$.
\end{enumerate}

Steps 2, 3 and 4 together are often referred to as {\em inversion
about the average}, because their combined effect is to invert the
amplitude of each term $\ket{x}$ about the average amplitude of all
$2^n$ terms. 

\begin{figure}
\begin{center}
\includegraphics*[width=10cm]{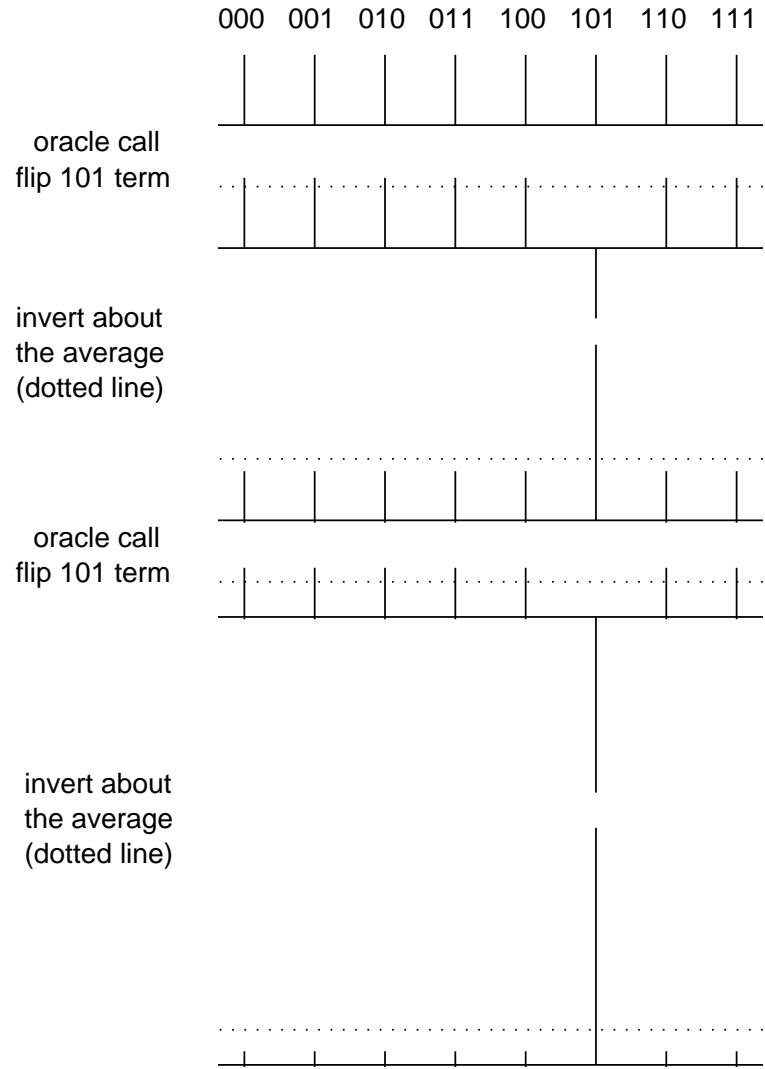} 
\end{center} 
\vspace*{-2ex}
\caption{Illustration of amplitude amplification in Grover's algorithm 
for $N=8$ ($n=3$) and $\ket{x_0} = \ket{101}$. The diagrams shows the
(real) amplitude of the eight terms $\ket{000}$ through
$\ket{111}$. The starting point is an equal superposition of all
terms. After each Grover iteration (an oracle call followed by
inversion about the average), the amplitude of the special element is
amplified. For the case $N=8$, the amplitude of the $\ket{x_0}$
reaches almost $1$ after two Grover iterations.}
\label{fig:grover_pic}
\end{figure}

Figure~\ref{fig:grover_pic} graphically illustrates the
operation of Grover's algorithm.  The amplitude of all terms
$\ket{x}$ are equal after step (b) in the algorithm. The amplitude of
$\ket{x_0}$ builds up after each Grover iteration, at the expense of
the amplitude of the remaining terms, until it reaches a maximum and
decreases again. For increasing numbers of Grover iterations, the
amplitude of the special element $\ket{x_0}$ oscillates
sinusoidally. The first maximum occurs after $\lceil \pi \sqrt{N} / 4
\rceil$ iterations. If we measure the $n$ qubits at this point, the
measurement result will be $x_0$ with high probability and the search
has succeeded.

How does the number of elementary operations required for a Grover
search scale with the size $N$ of the search space?  Steps 2 and 4
take $n = \log_2 N$ {\sc hadamard} gates each. Step 3, the conditional
phase flip, can be done in ${\cal O}(n) = {\cal O}(\log_2 N)$
operations, as noted in section~\ref{sec:multi-q-gates}. The cost of
the oracle depends on $f$ and we will come back to it shortly, but in
any case the oracle is called only once per iteration. The Grover
iteration must be repeated ${\cal O}(\sqrt{N})$ times, so the entire
algorithm requires ${\cal O}(\sqrt{N} \log_2 N)$ operations and ${\cal
O}(\sqrt{N})$ oracle calls, as opposed to ${\cal O}(N)$ calls
classically. We therefore say that Grover's algorithm achieves a
quadratic speed-up over classical search algorithms.

\subsubsection{Application and implementation}
\label{page:oracle_call}

What does it mean to call an oracle? In real life, we don't have
actual oracles available (much less oracles which interface with
quantum computers), so we need to implement $U_f$ ourselves. In the
phone book example, $U_f$ would have to reflect a phone book of $N$
entries and would therefore take at least ${\cal O}(N)$ operations to
implement.  This is not a useful application of Grover's algorithm: we
may have to make only ${\cal O}(\sqrt{N})$ oracle calls, but each
oracle call takes ${\cal O}(N)$ operations in itself.

Then, what are useful applications of Grover's algorithm?  The general
answer is: problems where we want to find $x_0=f^{-1}(y_0)$ where $f$
is easily computable (unlike the case of phonebooks) but $f^{-1}$ is
hard to compute, such that there is no better approach than to
evaluate $f(x)$ for random values of $x$ until we hit $x=x_0$ such
that $f(x_0)=y_0$.

As an example, consider the following instance of the satisfiability
problem: find the values of $x_1$, $x_2$ and $x_3$ which satisfy the
boolean expression $(x_1 \bar x_2 + x_3) \bar x_3 $. In this case, it
is easy to see that there is one unique solution, $x_1 x_2 x_3 =
100$. However, the effort needed to find a solution for an arbitrary
Boolean expression, or even to ascertain whether there is a solution,
increases exponentially with the problem size for any known classical
algorithm. The general satisfiability problem is thus
hard~\cite{Papadimitriou94a}~\footnote{A restricted case of the
satisfiability problem, 2-{\sc sat}, is not hard.}. A quantum computer
running Grover's algorithm could solve this problem in quadratically
fewer operations than is possible classically.

For many realistic applications, such as the satisfiability problem,
there may be more than one solution. However, if there are $M$
solutions, the amplitude of the solutions is highest after ${\cal
O}(\sqrt{N/M})$ Grover iterations~\cite{Boyer98a}. We thus need to
know $M$ in order to know the optimal number of iterations.
Fortunately, $M$ can be found in ${\cal O}(\sqrt{N})$ oracle calls as
well, via a procedure called quantum counting~\cite{Brassard98b}. In
summary, by combining quantum counting and quantum search,
unstructured searches with an unknown number of solutions can be sped
up quadratically compared to any classical algorithm.

In section~\ref{expt:grover3}, we will present an experimental
realization of Grover's algorithm on a search space of eight
elements. This experiment also nicely illustrates the oscillatory
behavior of the amplitude of $\ket{x_0}$ as a function of the number
of Grover iterations.

%%%%%%%%%%%%%%%%%%%%%%%%%%%%%%%%%%%%%%%%%%%%%%%%%%%%%%%%%%%%%%%%%%%%%%%%%

\subsection{Order-finding and Shor's algorithm}
\label{qct:shor}

In 1994, Peter Shor discovered an efficient quantum algorithm for
prime factorization and for computing discrete
logarithms~\cite{Shor94a,Shor97a}. This algorithm represented a
tremendous breakthrough, because it offered an exponential speed-up
over both deterministic and probabilistic classical algorithms for an
important mathematical problem.

Shor's algorithm was later generalized to an algorithm for
order-finding and the Abelian hidden-subgroup
problem~\cite{Kitaev95a}. The key step common to all algorithms in
this class is the quantum Fourier transform.  We will first introduce
the quantum Fourier transform (QFT), then describe an algorithm which
uses the QFT for order-finding, and finally present the factoring
algorithm as a specific instance of order-finding. For a good
introductory article on Shor's algorithm, see~\cite{Ekert96a}.

\subsubsection{The quantum Fourier transform}

The quantum Fourier transform (QFT) performs the same transformation
as the (classical) fast Fourier transform (FFT), but it can be
computed efficiently, which is classically not possible. Or more
precisely, the QFT allows us to efficiently {\em sample} the FFT. Even
this is impossible classically and, as we will see, being able to
sample the FFT is sufficient for order-finding.

The FFT$_N$ takes as input a string of $N$ complex numbers $x_j$ and
produces as output another string of $N$ complex numbers $y_k$, such
that
\be y_k =
\frac{1}{\sqrt{N}} \sum_{j=0}^{N-1} x_j e^{2\pi i jk/N} \,.
\ee 
For an input string with numbers which repeat themselves with period
$r$, the FFT$_N$ {\em inverts the periodicity}, i.e. it produces an
output string with period $N/r$, as illustrated in the following four
examples for $N=8$ (the output strings have been rescaled for clarity)
\begin{eqnarray}
\!\!r \quad \quad \mbox{input string} \;\; &&\; \mbox{output string} \quad
\;\;\;\;N/r \nonumber \\ 
8\quad \; \;1\;0\;0\;0\;0\;0\;0\;0\;&\mapsto&\;1\;1\;1\;1\;1\;1\;1\;1\; \quad 1 \quad \label{eq:qft_example_1} \\
4\quad \; \;1\;0\;0\;0\;1\;0\;0\;0\;&\mapsto&\;1\;0\;1\;0\;1\;0\;1\;0\; \quad 2 \quad  \\
2\quad \; \;1\;0\;1\;0\;1\;0\;1\;0\;&\mapsto&\;1\;0\;0\;0\;1\;0\;0\;0\; \quad 4 \quad  \\
1\quad \; \;1\;1\;1\;1\;1\;1\;1\;1\;&\mapsto&\;1\;0\;0\;0\;0\;0\;0\;0\; \quad 8 \quad 
\label{eq:qft_example_8}
\end{eqnarray}

If $r$ does not divide $N$, the inversion of the period is
approximate.  In addition to inverting the period, the FFT {\em
converts off-sets} in the locations of the numbers in the input string
{\em into phase factors} in front of the numbers in the output string:

\begin{eqnarray}
1\; 0\; 0\; 0\; 1\; 0\; 0\; 0\; \;&\mapsto& \; 1\;\;\; 0\;\;\; 1\;\;\; 0\;\;\; 1\;\;\; 0\;\;\; 1\;\;\; 0 \\
0\; 1\; 0\; 0\; 0\; 1\; 0\; 0\; \;&\mapsto& \; 1\;\;\; 0\;  -\!i\;\:   0\;-\!\!1\;\; 0\;\;\; i\;\;\:\, 0 \label{eq:fft_example} \\
0\; 0\; 1\; 0\; 0\; 0\; 1\; 0\; \;&\mapsto& \; 1\;\;\; 0\;-\!\!1\,\,\, 0\;\;\; 1\;\;\; 0\;-\!1\!\,\,\: 0 \\
0\; 0\; 0\; 1\; 0\; 0\; 0\; 1\; \;&\mapsto& \; 1\;\;\; 0\;\;\; i\;\;\;\, 0\;-\!\!1\;\; 0\;-\!i\;\: 0
\end{eqnarray}

The QFT performs exactly the same transformation, but differs from the
FFT in that the complex numbers are stored in the amplitude and phase
of the terms in a superposition state of $n=\log_2 N$ qubits. For $N=8$,
the qubit string is of length $\log_2 8 = 3$, and the amplitude of
the $\ket{000}$ term represents the first complex number, the
amplitude of the $\ket{001}$ term represents the second number and so
forth. As before, we will label the states $\ket{000},
\ket{001},\ldots \ket{111}$ as $\ket{0}, \ket{1}, \ldots \ket{7}$. As 
an example, we see from Eq.~\ref{eq:fft_example} that the QFT thus
transforms the state $(\ket{1} + \ket{5})\sqrt{2}$ to the state $(\ket{0} - i
\ket{2} - \ket{4} + i \ket{6})/2$.

The action of the QFT on a computational basis state of length $n$ is
\be
\ket{j} \mapsto \frac{1}{\sqrt{N}} \sum_{k=0}^{N-1} 
e^{2\pi i j k/N} \ket{k} \,,
\label{eq:QFT_1}
\ee
which we can rewrite (after some algebra) as
\be
\ket{j_1,\ldots,j_n} \mapsto
\frac{(\ket{0}+e^{2\pi i 0.j_n}\ket{1}) 
      (\ket{0}+e^{2\pi i 0.j_{n-1}j_n}\ket{1}) 
      \cdots
      (\ket{0}+e^{2\pi i 0.j_1 j_2 \cdots j_n}\ket{1}) }
      {\sqrt{2^n}} \,,
\label{eq:QFT_2}
\ee
where $0.j_1j_2\ldots j_n$ stands for $2^{-j_1} + 2^{-2 j_2} + \ldots
+ 2^{-n j_n}$. Suppose we now reverse the order of the qubits in the
output of the QFT, so output qubit $j_n$ becomes output qubit $1$,
qubit $j_{n-1}$ becomes qubit $j_2$ and so forth. The transformation
of Eq.~\ref{eq:QFT_2} then changes into
\be
\ket{j_1,\ldots,j_n} \mapsto
\frac{(\ket{0}+e^{2\pi i 0.j_1}\ket{1}) 
      (\ket{0}+e^{2\pi i 0.j_2j_1}\ket{1}) 
      \cdots
      (\ket{0}+e^{2\pi i 0.j_n j_{n-1} \cdots j_1}\ket{1}) }
      {\sqrt{2^n}} \,.
\label{eq:QFT_3}
\ee
%\be
%\ket{j_1,\ldots,j_n} \mapsto
%\frac{(\ket{0}+e^{2\pi i 0.j_1 j_2 \cdots j_n}\ket{1}) 
%      \cdots
%      (\ket{0}+e^{2\pi i 0.j_{n-1}j_n}\ket{1})       
%      (\ket{0}+e^{2\pi i 0.j_n}\ket{1}) }
%      {\sqrt{2^n}} \,,
%\ee
which we can easily implement as follows.  The first factor of this
expression represents a $180^\circ$ phase shift of qubit $1$
controlled by qubit $1$ itself; this is accomplished by a {\sc
hadamard} gate on qubit $1$. The next factor shifts the phase of qubit
$2$ over $180^\circ$ controlled by qubit $2$ (a {\sc hadamard} gate on
qubit $2$), and over another $90^\circ$ controlled by qubit $1$, which
corresponds to a controlled-$Z$ rotation. The next factor requires a
{\sc hadamard} gate on qubit $3$, a $Z$ rotation of qubit $3$
controlled by qubit $2$ and a $Z^{1/2}$ ($45^\circ$) rotation of $3$
controlled by $1$. The quantum circuit for the QFT acting on three
qubits is shown in Fig.~\ref{fig:qft_circuit}, and it can be
easily extended for $n>3$, using a total of $n$ {\sc hadamard} gates
and $n(n-1)/2$ controlled rotations.
We next see how the QFT is incorporated in a quantum algorithm for
order-finding. 

\bfig
\begin{center}
\includegraphics*[width=7cm]{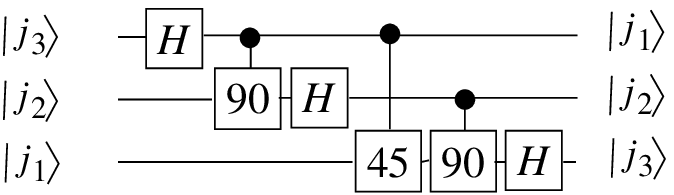} 
\end{center} 
\vspace*{-2ex}
\caption{Quantum circuit for the quantum Fourier transform (QFT) 
acting on three qubits. In this implementation of the QFT, due to
Coppersmith \protect\cite{Coppersmith94a}, the order of the qubits is
reversed at the output with respect to the input.}
\label{fig:qft_circuit}
\efig

\subsubsection{Order-finding}

The order of a permutation $\pi$ on $M$ elements can be understood
via the following analogy: imagine $M$ rooms and $M$ {\it one-way}
passages connecting the rooms, with {\it exactly} one entrance and one
exit in each room (for some rooms, the passage going out may loop back
to the room itself). The rooms are thus connected in subcycles, as
shown in Fig.~\ref{fig:cycles}. This configuration ensures that if you
start in some room $y$ and keep going from one room to the next using
the one-way passages, you must eventually come back to the room you
started from. We then define the order $r$ of the permutation $\pi$ as
the {\em minimum number of transitions needed to return to the
starting room $y$}, where $r$ may depend on $y$ but is never greater
than $M$.

\bfig
\begin{center}
\includegraphics*[width=12cm]{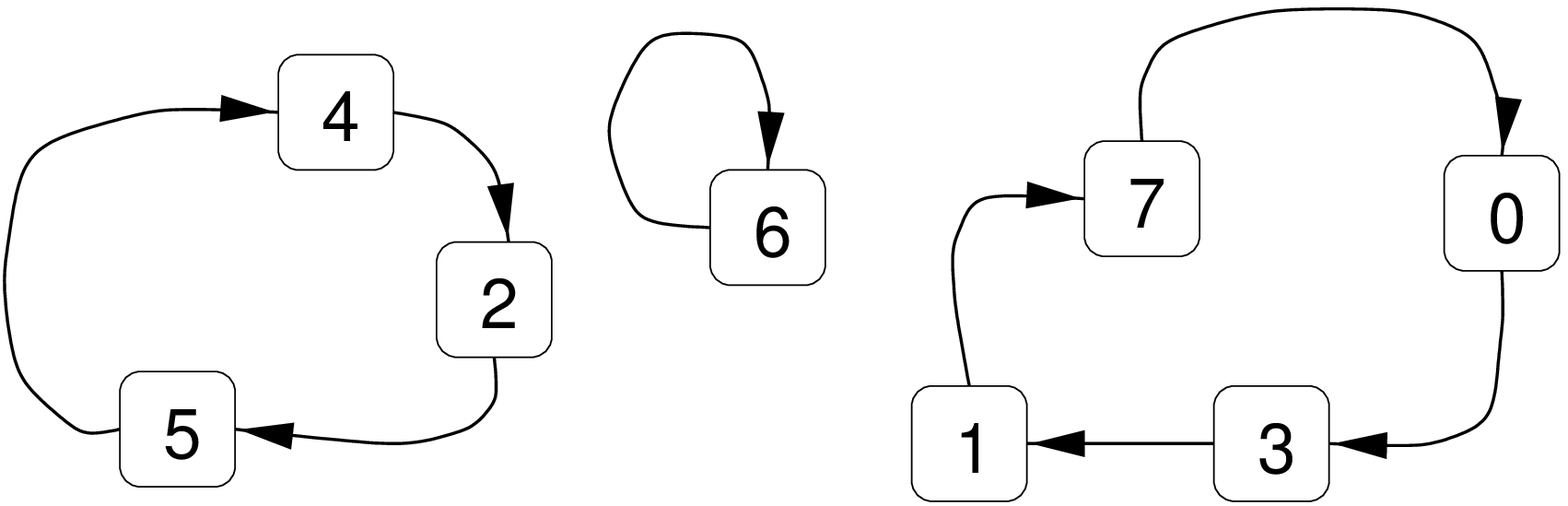} 
\end{center} 
\vspace*{-2ex}
\caption{Pictorial representation of a permutation $\pi$ on eight 
elements. The order $r$ is 3 if $y \in \{2,4,5\}$, $r=1$ if $y = 6$,
and $r=4$ if $y \in \{0,1,3,7\}$.}
\label{fig:cycles}
\efig

Suppose you are in a room $y$ and must determine the order $r$ solely
by making trials of the type ``make $x$ transitions starting from room
$y$ and check which room you are in''. Mathematically, we will
describe such trials as queries of an oracle or black box which
outputs $\pi^x(y)$, that is the element obtained after permuting $x$
times starting from $y$ using $\pi$ (so for the permutation of
Fig.~\ref{fig:cycles}, we have $\pi^1(5)=4,\; \pi^2(5)=2$ and so
forth).  How many such queries are needed in order to find $r$ with a
given probability of success ?

Richard Cleve~\cite{Cleve99b} showed that the minimum number
of classical oracle queries needed for a given probability of success
increases exponentially with the problem size $m=\lceil\log_2 M\rceil$ (the
number of bits needed to represent $M$ numbers).  In contrast, on a
quantum computer using a generalization of Shor's quantum algorithm,
the number of oracle queries needed in order to achieve a given
probability of success does not increase with $m$.  Thus, there is an
exponential gap in the number of oracle queries required between the
quantum and classical cases.\\

We now will explain the steps in the order-finding quantum algorithm via
an example where the permutation $\pi$ acts on $M=4$ elements
(Fig.~\ref{fig:order_circuit}).

First, initialize a register of $n=\log_2N=3$ qubits in the ground
state, and a second register of two qubits in the state $\ket{y_1
y_0}$ or for short $\ket{y}$, where $y_1 y_0$ is the binary
representation of $y$. Apply a {\sc hadamard} transformation to all
qubits in the first register.  The state of the quantum computer is
now
\be 
\ket{\psi_1} = 
(\ket{0} + \ket{1} + \ket{2} + \ket{3} + \ket{4} 
+ \ket{5} + \ket{6} + \ket{7}) \; \ket{y} \,,
\label{eq:input_order}
\ee
where we have left out the normalization factor.

Then query the oracle, i.e. evaluate the function $\pi^x(y)$, and
store the result in the second register, via the transformation
\be
\ket{x}\ket{y} \mapsto \ket{x}\ket{\pi^x(y)} \,,
\label{eq:oracle_orderfinding}
\ee
which is conveniently implemented in $n=3$ {\em exponentiated
permutations} (Fig.~\ref{fig:order_circuit}), using $x = 4 x_2 + 2 x_1
+ x_0$ and thus $\pi^x = \pi^{4 x_2} \pi^{2 x_1} \pi^{x_0}$.  Since
the first register is in an equal superposition of all values of $x$
between $0$ and $2^n$, the function is evaluated for all those values
of $x$ in parallel. In the analogy of the rooms and one-way passages,
the quantum computer thus makes transitions to many rooms at once.
For the sake of the argument, let us say for example that $y=3$, and
that $\pi^1(3)=1$ and $\pi^2(3)=3$.\footnote{This already tells us that the
order is $2$ (because after permuting twice we get back to the
starting element $y$), and we can also deduce that $\pi^3(3)=1$,
$\pi^4(3)=3$ and so forth. Of course, in a realistic application we
don't know the order in advance; we only have a description of the
permutation to help us determine $r$.} 
For this example, evaluation of
$\pi^x(y)$ transforms the state of Eq.~\ref{eq:input_order} into
\be
\ket{\psi_2} = 
\ket{0}\ket{3} + \ket{1}\ket{1} + \ket{2}\ket{3} + \ket{3}\ket{1} 
+ \ket{4}\ket{3} + \ket{5}\ket{1} + \ket{6}\ket{3} + \ket{7}\ket{1}
\ee 
\be =
(\ket{0} + \ket{2} + \ket{4} + \ket{6}) \ket{3}
+ (\ket{1} + \ket{3} + \ket{5} + \ket{7}) \ket{1} \,.
\ee 

\bfig
\bcen
\vspace*{1ex}
\includegraphics*[width=8.5cm]{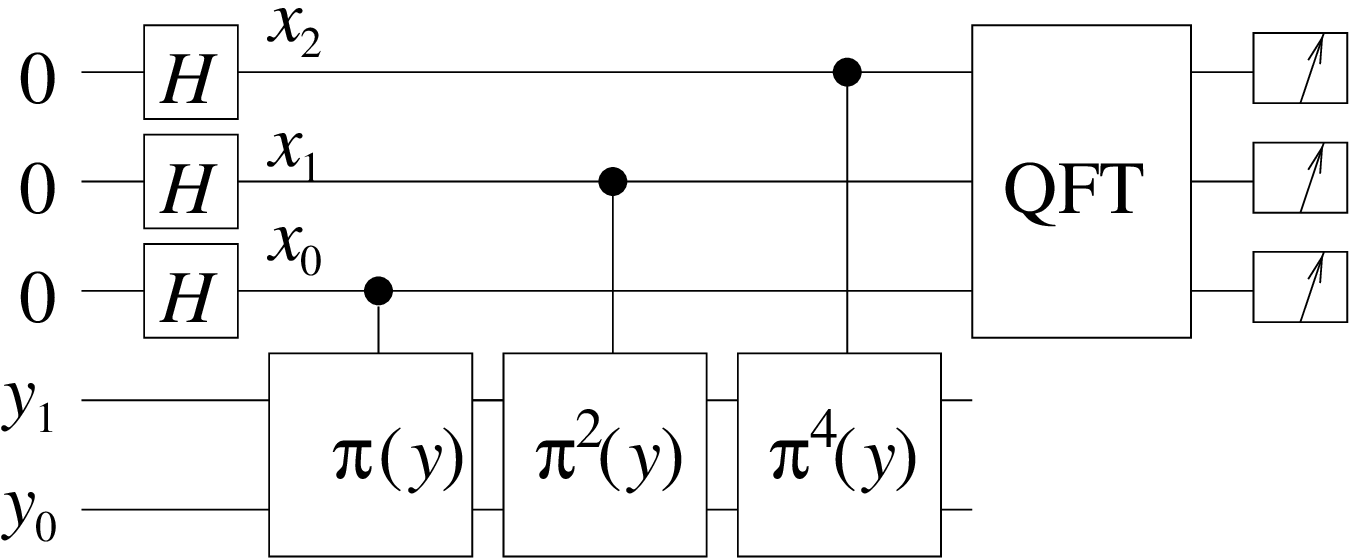} 
\vspace*{-2ex}
\ecen
\caption{Outline of the order-finding quantum algorithm.}
\label{fig:order_circuit}
\efig

Next, apply the QFT to the first register. In order to appreciate
the need for and role of the QFT, suppose we now measured
$\ket{\psi_2}$ directly. In $\ket{\psi_2}$, the first register is
still in a superposition of eight states, regardless of $\pi$ so this
measurement will not tell us much useful. Suppose we instead measure
the second register. The result will be either $1$ or $3$ but that in
itself isn't of much help either --- it is merely a sample of
$\pi^x(y)$ for a random $x$, which we could do equally well
classically. Now suppose we measure both registers. In the measurement
process of the second register, the state of the first register will
collapse to either
\be
\ket{0} + \ket{2} + \ket{4} + \ket{6} \quad \mbox{OR} \quad
\ket{1} + \ket{3} + \ket{5} + \ket{7} \,.
\ee
depending on whether the measurement of register $2$ gave 3 or
1. However at this point, a measurement of the first register still
does not yield any information at all, because all eight possible
outcomes are still equally likely, as we don't know from which subset
of values of $x$ the measured $x$ will come. But if we apply the QFT,
the first register will be transformed to~\footnote{Note that while
introducing the measurement of the second register makes it easier to
explain things, this measurement can actually be left out. The QFT
will still transform the first register as shown.}
\be 
\ket{0} + \ket{4} \quad \mbox{OR} \quad \ket{0} - \ket{4} \,.
\label{eq:output_qft}
\ee
Now a measurement of the first register does give useful information,
because only multiples of $N/r$ are possible outcomes, in this example
$0$ and $4$. In general, we can derive $r$ from the measurement
outcome with high probability of success, using the continued fraction
expansion \cite{Hardy60a,Koblitz94a}, provided $n \ge 2 m$, that is $N
\ge M^2$ (this was in fact not the case in the simple example we
presented).

The order-finding algorithm for arbitrary $m$ is summarized as
follows. Let the first register be of size $n=2m$, and the second register
of size $m$. Let furthermore $y=0$, without loss of generality.

\begin{enumerate}
\item Initialize $$\ket{0}\ket{0}$$
\item Create a superposition using $H^{\otimes n}$ 
$$ \mapsto \frac{1}{\sqrt{2^n}} \sum_{j=0}^{2^n-1} \ket{j}\ket{0} $$
\item Apply $f(j)$ (where $f(j)=\pi^j(0)$)
$$ \mapsto \frac{1}{\sqrt{2^n}} \sum_{j=0}^{2^n-1} \ket{j}\ket{f(j)} $$
\item Apply the QFT$_N$ to the first register (Eq.~\ref{eq:QFT_1})
$$ \mapsto \frac{1}{2^n} \sum_{j=0}^{2^n-1} \sum_{k=0}^{2^n-1} 
e^{2\pi i j k/N} \ket{k} \ket{f(j)} $$
$$ = \frac{1}{2^n} \sum_{k=0}^{2^n-1} \sum_{j=0}^{2^n-1}  
e^{2\pi i j k/N} \ket{k} \ket{f(j)} $$
\item Measure the first register
\end{enumerate}

The probability for obtaining $\ket{k}$ upon measurement is the square
of the amplitude of $\ket{k}$ in the output state of the QFT,
\be
\left( \;\sum_{j=0}^{2^n-1} e^{2\pi i j k/N} \; \right)^2 \,.
\ee
Due to interference of the terms in this summation, the probability is
high for values of $k$ which are an integer multiple of $N/r$ and very
small or zero for other values of $k$. In fact, if $r$ divides $N$
exactly, the probability for obtaining a multiple of $N/r$ upon
measuring the output state is equal to 1 (this was the case in the
examples for the FFT of
Eqs.~\ref{eq:qft_example_1}-\ref{eq:qft_example_8}).

In section~\ref{expt:order}, we will present the first
implementation of the order-finding algorithm.  We will now describe
a specific instance of the order-finding algorithm, namely Shor's
famous algorithm for prime factorization.

\subsubsection{Factoring}

The most famous application of the order-finding algorithm, and
historically the first one to be discovered, is to decompose large
integer numbers into their prime factors.  Quantum factoring consists
of finding the order of the permutation
\be
\pi(y) = a y \; \mbox{mod}\, L \,,
\ee
for $y=1$. $L$ is the integer we want to factor and $a$ can be any integer $<L$
that is coprime with $L$ (i.e. $a$ and $L$ should have no
common factors other than $1$). In words, the permutation consists of
multiplying $y$ by $a$, and taking the remainder of the division of $a
y$ by $L$.

Given the order $r$ of the permutation $a y \; \mbox{mod}\, L$ for
$y=1$, at least one prime factor of $L$ is given by 
\be
\mbox{gcd}(a^{r/2}-1,L) \quad \mbox{or} \quad \mbox{gcd}(a^{r/2}+1,L)
\label{eq:gcd}
\ee
with high probability. Computing the greatest common denominator of
two integers can be done efficiently on a classical computer, using
Euclid's algorithm. These are results from number theory
\cite{Hardy60a,Koblitz94a}.

An important difference between our description of the order-finding
problem and the factoring problem is that for order-finding we assumed
that an oracle was available which we can ask queries of the type
$\pi^x(y)$. As was pointed out in section~\ref{qct:grover}, we
must in practice implement oracle calls ourselves, and we must
therefore consider whether this can be done efficiently. For
factoring, calling the oracle means implementing the permutation
$\pi^x(y) = a^x y\; \mbox{mod}\, L$ with $y=1$, which is equivalent to
evaluation of the function
\be
a^x \; \mbox{mod} \, L \,,
\ee
known as the {\em modular exponentiation} function. 

Since $a^x\; \mbox{mod}\, L$ is by definition a number
between 0 and $L-1$, the second register must be of size $m=\lceil
\log_2 L \rceil$. The first register must be at least twice
as large, so $n=2m$. Since $x = x_{n-1} 2^{n-1} + \ldots + x_1 2^1 +
x_0 2^0$, we have
\begin{eqnarray}
&a^x y \; \mbox{mod}\, L = a^{2^{n-1} x_{n-1}} \ldots a^{2 x_1}
\; a^{x_0} y \; \mbox{mod}\, L & \nonumber \\
&= [ a^{2^{n-1} x_{n-1}} \ldots
[a^{2 x_2}\; [a^{x_0} y\; \mbox{mod}\, L] \; \mbox{mod}\, L] \ldots
\mbox{mod} \, L] &
\label{eq:modexp_factor}
\end{eqnarray}
In words, this means that we first multiply $y$ by $a$ modulo $L$, if
and only if $x_0=1$; then we multiply the result by $a^2$ modulo $L$,
if and only if $x_1=1$ and so forth, until we finally multiply by
$a^{2^{n-1}}$ modulo $L$ if and only if $x_{n-1}=1$.  

Thus, modular exponentiation is reduced to $n =2m \approx 2 \log_2 L$
multiplications modulo $L$, each controlled by just a single qubit
$x_i$. The numbers $a^2, \ldots , a^{2^{n-1}} \mbox{mod}\; L$ by which
we need to multiply can be found efficiently on a classical computer
by {\em repeated squaring}, and multiplication of $m$-bit numbers take
${\cal O}(m^2)$ elementary operations.

The modular exponentiation step can thus be done efficiently, in
${\cal O}((\log_2 L)^3)$ one- and two-qubit gates, and we showed
earlier that the other key step in Shor's factoring algorithm, the
quantum Fourier transform, can also be realized efficiently.  The
factoring problem, widely believed to be intractable on classical
computers,~\footnote{We note that classically, there are many
approaches known to factoring integers besides finding the order of $a
y \; \mbox{mod} \, N$, but all of them are inefficient. It is
possible, although unlikely, that one day an efficient classical
factoring algorithm will be found. Or perhaps a proof will be
constructed that such an algorithm is not possible.} is thus tractable
on quantum computers.

In section~\ref{expt:shor}, we present the first experimental
realization of Shor's algorithm. It is an implementation of the
simplest possible instance for which the algorithm can be
non-trivially demonstrated, namely factorization of the number
fifteen~\footnote{The algorithm fails for even $L$ and for $L$ which
are powers of prime numbers (e.g. the number nine), and factoring is
obviously not applicable to prime $L$. This includes all numbers
smaller than 15. The next simplest instance of factoring is 21.}.

%%%%%%%%%%%%%%%%%%%%%%%%%%%%%%%%%%%%%%%%%%%%%%%%%%%%%%%%%%%%%%%%%%%%%%%%%

\subsection{Quantum simulations}

The possibility of using quantum computers to solve problems in
quantum physics was conjectured by Feynman~\cite{Feynman82a} long
before quantum algorithms for solving mathematical problems such as
factoring were discovered. Seth Lloyd proved this conjecture in
1996~\cite{Lloyd96a}.

The general procedure is to map the possible states of the simulated
system onto the states of a set of qubits, and then to apply a
sequence of quantum gates which produce qubit dynamics analogous to
the dynamics of the simulated system. The final state of the qubits is
then mapped back onto the state of the simulated system.

Explicit protocols have been worked out for several realistic physical
problems. These include

\begin{itemize}
\item Estimation of the eigenvalues and eigenvectors of a 
Hamiltonian (this algorithm invokes the QFT), which can be used to
find the energy levels of an atom for example~\cite{Abrams99a}.
\item Simulation of the dynamics of many-body Fermi systems, using 
either first or second quantized descriptions~\cite{Abrams97a}.
\item Simulation of quantum chaos and localization~\cite{Georgeot01a}.
\end{itemize}

\subsection{Other quantum algorithms and perspectives}

The three main classes of quantum algorithms are (1) those based on
the quantum Fourier transform, (2) search algorithms, and (3) quantum
simulations. All of these were invented in the mid-nineties. Since
then, the range of applications of these algorithms has been extended
and refined. However, virtually no fundamentally new algorithms for
quantum computers have been discovered, despite intense effort all
over the world.

It would be disappointing if no other applications were found than
those currently known. Nevertheless, existing algorithms are already
of significant practical interest. In particular, quantum simulations
may address a broad range of physics problems which are otherwise
intractable. For example, as transistors and other devices continue to
shrink, simulation of their operation at the quantum level may be
crucially important but impossible to do on classical computers.

The impact of quantum computing on the fundamentals of computer
science may be just as profound and long-lasting, as it appears that
the strong Church-Turing thesis must be revised. This thesis states
that any two universal Turing machines (a general computational
device~\cite{Turing36a}) are polynomially equivalent; in other words,
the Church-Turing thesis says that problems which can be efficiently
computed on one Turing machine can always be efficiently computed on
another Turing machine. However, quantum Turing machines, proposed by
David Deutsch~\cite{Deutsch85a}, appear to be capable of efficiently
solving problems which are intractable on classical Turing machines.

%%%%%%%%%%%%%%%%%%%%%%%%%%%%%%%%%%%%%%%%%%%%%%%%%%%%%%%%%%%%%%%%%%%%%%%%%
%%%%%%%%%%%%%%%%%%%%%%%%%%%%%%%%%%%%%%%%%%%%%%%%%%%%%%%%%%%%%%%%%%%%%%%%%

\section{Quantum error correction}
\label{qct:qec}

Quantum computers, like any machine, may have faulty or unreliable
components. In order to nevertheless perform reliable computations,
errors in the state of the qubits must be corrected.

Quantum error correction is similar to its classical analogue in many
respects.  Input states are encoded in a larger system which is more
robust against noise or other error processes than unencoded states,
in the sense that the original information can be retrieved with
greater likelyhood if the input states are encoded than if they are
not. 

For example, a simple classical code encodes 0 as 000 and encodes 1 as
111. If we send an uncoded bit of information through a noisy channel
which flips a bit with probability $p$, then the probability of error
per use of the channel is also $p$. However, if we send the same bit
of information in encoded form (three physical bits), and take a
majority vote between the three bits after the noise process, we can
correctly guess the bit of information unless two of the physical bits
are flipped, which happens with probability $3(1-p)p^2$, or all three
physical bits are flipped, which occurs with probability $p^3$. The
resulting probability of error per bit of information is thus only
$3p^2 -2p^3$, which is smaller than $p$ if $0<p<1/2$. Classically,
encoding can thus decrease the error probability at the expense of
using extra bits and encoding/decoding operations.

We would like to use similar schemes to encode quantum information, in
order to protect quantum computers and quantum communication channels
against errors caused by decoherence. However, quantum error correction
is much more tricky than classical error correction, because

\begin{enumerate}
\item it is not possible to build up redundancy by simply making copies
    of the quantum state, due to the no-cloning
    theorem~\cite{Dieks82a,Wootters82a}.
\item measurement of a quantum system destroys its state, so it is not
    possible to check the state of a qubit,
\item quantum errors can be arbitrary rotations in the Bloch sphere,
    while in classical computers only bit flips ($0 \leftrightarrow
    1$) can occur. 
\end{enumerate}

The remarkable and surprising achievement of Shor~\cite{Shor95a} and
Steane~\cite{Steane96a} was to show that it is nevertheless possible
to reverse the truly random errors due to decoherence. The underlying
principles are

\begin{enumerate}
\label{eq:qec_rules}
\item to use entanglement in order to realize a quantum analogue of
    redundancy,
\item to measure errors without measuring the quantum state itself,
\item to digitize the errors (force an arbitrary error to collapse into
    either no error or a full bit flip, a phase flip or a bit-and-phase
    flip).
\end{enumerate}

There are many excellent references on the theory of quantum error
correction~\cite{Ekert96b,Knill96b}.  We here describe in detail a
two-qubit error detection code, which nicely illustrates how the three
guiding principles of quantum error correction are put into
practice. Next, we give a brief overview of larger codes, capable of
detecting and correcting more errors, and introduce the notion of
fault-tolerancy.

%%%%%%%%%%%%%%%%%%%%%%%%%%%%%%%%%%%%%%%%%%%%%%%%%%%%%%%%%%%%%%%%%%%%%%

\subsection{The two-qubit phase error detection code}
\label{qct:2bitcode}
The two-qubit phase error detection code~\cite{Chuang95c} encodes one qubit
information in the joint state of two qubits. After decoding, it is
possible to tell whether or not a phase error occurred on one of the
qubits. If an error is detected, the state is rejected; otherwise it
is kept.

\begin{figure}[h]
\bcen
\vspace*{1ex}
\includegraphics*[width=10cm]{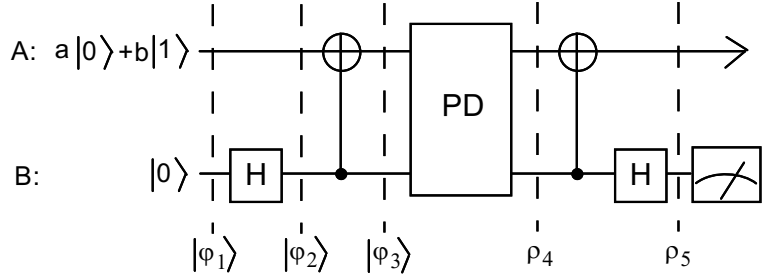} 
\vspace*{-2ex}
\ecen
\caption{Encoding and decoding quantum circuit for the two-qubit code.
In between encoding and decoding, phase damping may disturb the qubit
states.}
\label{fig:2bitcode_circuit}
\end{figure}

The two-qubit phase error detection code encodes logical 0 and 1 as
\begin{eqnarray}
	|0_L \rangle & = &\frac{1}{\sqrt{2}} (|00 \rangle +|11 \rangle )
\\
	|1_L \rangle & = &\frac{1}{\sqrt{2}} (|01 \rangle +|10 \rangle )
\,, 
\label{eq:code}
\end{eqnarray}
where the subscript $L$ denotes logical states.  An arbitrary qubit
state $a \ket{0} + b \ket{1}$ is encoded as
\begin{eqnarray}
	|\psi_3 \rangle 
	& = & a |0_L \rangle + b |1_L \rangle \\
	& = & \frac{1}{\sqrt{2}} \left[\rule{0pt}{2.4ex} 
	a (|00 \rangle +|11 \rangle ) 
      + b (|01 \rangle +|10 \rangle ) \right]
\label{eq:psi}
\,,
\end{eqnarray}
via the quantum circuit of Fig.~\ref{fig:2bitcode_circuit}.  We thus
build up {\em redundancy} by encoding the logical qubit in the state
of two qubits using entanglement (principle 1). 

Now let us first consider an error process which causes a complete
phase flip of one or more qubits, where we define a phase flip of
qubit $i$ as $\sigma_z^i \, \rho \, \sigma_z^i$.  For a code to detect
errors, it suffices that all errors to be detected map the codeword
space ${\cal C}$ (the space spanned by $\ket{0_L}$ and $\ket{1_L}$)
onto its orthogonal complement.  In this way, {\em detection of
errors} can be done unambiguously by a projection onto ${\cal C}$ {\em
without distinguishing individual codewords}, hence without disturbing
the encoded information (principle 2). This is precisely what the code
of Eq.~\ref{eq:code} achieves.

The four possible outcomes of the error process are
\begin{eqnarray}
	|\psi_{II} \rangle = & I \otimes I~ |\psi \rangle  = & 
	a~\frac{ |00 \rangle + |11 \rangle }{\sqrt{2}} 
	+ b~\frac{ |01 \rangle +|10 \rangle }{\sqrt{2}}
\label{eq:iipsi} \\ 
	|\psi_{ZI} \rangle = & \sigma_z \otimes I~ |\psi \rangle = & 
	a~\frac{|00 \rangle -|11 \rangle }{\sqrt{2}} 
	+ b~\frac{|01 \rangle -|10 \rangle }{\sqrt{2}}
\label{eq:zipsi} \\
	|\psi_{IZ} \rangle = & I \otimes \sigma_z |\psi \rangle = & 
	a~\frac{|00 \rangle -|11 \rangle }{\sqrt{2}} 
	+ b~\frac{ -|01 \rangle +|10 \rangle }{\sqrt{2}} 
\label{eq:izpsi} \\
	|\psi_{ZZ} \rangle = & \sigma_z \otimes \sigma_z |\psi \rangle = & 
	a~\frac{|00 \rangle +|11 \rangle }{\sqrt{2}} 
	+ b~\frac{-|01 \rangle -|10 \rangle }{\sqrt{2}}
\label{eq:zzpsi}
\,,
\end{eqnarray}
with the erroneous states $|\psi_{ZI} \rangle $ and $|\psi_{IZ}
\rangle $ orthogonal to the correct state $|\psi_{II} \rangle $. After
decoding, which is the inverse of the encoding operation (see
Fig.~\ref{fig:2bitcode_circuit}):
\begin{eqnarray}
	|\psi_{II} \rangle & \stackrel{dec}{\Rightarrow} 
				& (a |0 \rangle + b |1 \rangle ) |0 \rangle 
\label{eq:phi} 
\\
	|\psi_{ZI} \rangle &  \Rightarrow  
				&(a |0 \rangle - b |1 \rangle ) |1 \rangle 
\label{eq:ziphi} 
\\
	|\psi_{IZ} \rangle &  \Rightarrow 
				&(a |0 \rangle + b |1 \rangle ) |1 \rangle 
\label{eq:izphi} 
\\
	|\psi_{ZZ} \rangle &  \Rightarrow 
				&(a |0 \rangle - b |1 \rangle ) |0 \rangle 
\label{eq:zzphi}
\,.
\end{eqnarray}

We note that the ancilla (the auxiliary qubit) becomes $|1 \rangle $
upon decoding if and only if a {\em single} phase error has occurred.
Furthermore, the two qubits are in a product state after decoding, so
it is possible to read out the syndrome (which indicates whether or
not a phase error occurred) by a projective measurement on the ancilla
without measuring the encoded state, which is held in the first
qubit. If the final state of the ancilla is $\ket{0}$, we trust the
state of the first qubit is properly preserved, and accept it. If the
ancilla is in $\ket{1}$, we distrust the state of the first qubit and
reject it.

Clearly, the error $\sigma_z\otimes\sigma_z$ cannot be detected, but
this occurs only with probability $p^2$, where we let $p$ be the
probability that the phase of a qubit is flipped in the error
process. Furthermore, the code can detect phase errors but cannot
reveal which qubit has the error, so it cannot correct
errors.  Moreover, $|\psi_{IZ} \rangle $ decodes to a correct state in
the first qubit which is rejected.  This affects the absolute fidelity
(the overall probability of successful recovery) but not the
conditional fidelity (the probability of successful recovery if the
state is accepted). All of these properties are intrinsic limitations
of using an error detection code as opposed to an error correction
code. With only two physical qubits per qubit of information, an error
detection code is the best we can do.

So far we have considered only complete phase flip errors, but in
reality phase shifts over arbitrary angles may occur. Fortunately, we
can {\em discretize} such {\em errors} (principle 3) as
follows. Suppose the error is an arbitrary phase shift on the first
qubit: $|0 \rangle \rightarrow |0 \rangle $, $|1 \rangle \rightarrow
e^{i \theta}|1 \rangle $.  Then, the encoded state becomes
\begin{eqnarray}
	& & \frac{1}{\sqrt{2}} \left[\rule{0pt}{2.4ex} 
	  a (|00 \rangle + e^{i \theta} |11 \rangle ) 
	+ b (e^{i \theta} |10 \rangle + |01 \rangle ) \right]
\\
	& = & \frac{1}{\sqrt{2}}\frac{1+e^{i \theta}}{2} 
	\left[\rule{0pt}{2.4ex} a (|00 \rangle +|11 \rangle ) 
	+ b (|10 \rangle + |01 \rangle ) \right]  
\nonumber
\\
 	& + &  \frac{1}{\sqrt{2}}\frac{1-e^{i \theta}}{2} 
	\left[\rule{0pt}{2.4ex} a (|00 \rangle -|11 \rangle ) 
	+ b (-|10 \rangle + |01 \rangle ) \right]
\,.
\end{eqnarray}
The decoded state is now a superposition of the states given by  
Eqs.(\ref{eq:phi}) and (\ref{eq:ziphi}):
\begin{equation}
	\frac{1}{\sqrt{2}}\frac{1+e^{i \theta}}{2}
	(a |0 \rangle + b |1 \rangle ) |0 \rangle + 
	\frac{1}{\sqrt{2}}\frac{1-e^{i \theta}}{2} 
	(a |0 \rangle - b |1 \rangle ) |1 \rangle 
\nonumber
\,. 
\end{equation}
Measurement of the second qubit projects it to either $|0 \rangle$ or
$|1 \rangle$. {\em Because of entanglement}, the first qubit is
projected accordingly to having no phase error, or a complete phase
flip! An arbitrary phase shift of the second qubit is discretized in
the same manner.

In section~\ref{expt:2bitcode}, we will present an NMR demonstration
of the operation of the two-qubit phase error detection code.

\subsection{Error correction codes and fault-tolerancy}
\label{qct:fault-tolerancy}

Codes with more than two physical qubits per logical qubit are capable
not only of detecting errors but also of correcting
errors. Historically, the first two quantum codes capable of
correcting arbitrary single-qubit errors were a nine-qubit
code~\cite{Steane96a} and a seven-qubit code~\cite{Shor95a}. A
five-qubit code followed later~\cite{Laflamme96a}.

Using counting arguments based on the (quantum) Hamming bound, we can
see that five is the minimum number of qubits that can be used to
correct arbitrary errors.  A code which encodes one logical qubit in
$k$ physical qubits has $k-1$ ancillae which can represent up to
$2^{k-1}$ orthogonal error syndromes. There are three types of errors
which can occur on each of the physical qubits: a phase flip, a bit
flip and a combined phase and bit flip (those can be expressed
mathematically as $\sigma_x$, $\sigma_z$ and $\sigma_y$ errors
respectively). It is also possible that no error occurs on any of the
$k$ qubits. Thus, assuming that the errors acting on different qubits
are uncorrelated, we need $3k+1$ orthogonal error syndromes, to be
represented by the $2^k-1$ ancillae. This way, we can simply measure
the ancilla qubits upon decoding, infer precisely what error
occurred and on which qubit it acted, and then correct the error. The
requirement for the length $k$ of the codewords is therefore that
\be
2^{k-1} \ge 3k+1 \,.
\label{eq:syndrome_number}
\ee
If only one type of error is expected to occur, say phase errors,
Eq.~\ref{eq:syndrome_number} can be relaxed to $2^{k-1} \ge k+1$, and
a three-qubit code is sufficient. Also, analogous to the case of
classical codes, more efficient codes can be constructed by encoding
several logical qubits per codeword.\\

Implicit in our discussion of quantum error correction so far is the
assumption that the encoding and decoding operations are perfect, and
that the qubits are simply sent through a noisy channel. In the
context of quantum computation, such a channel would correspond to a
quantum memory device. However, in realistic quantum computers,
information must also be protected in the course of a computation, and
furthermore encoding and decoding operations are themselves subject to
errors. Would it still be helpful to apply quantum error correction
under these circumstances?

The surprising answer is yes, provided the decoherence rate is below a
certain level, the {\em accuracy threshold}, expressed as the
probability of error per elementary logic operation on one or two
qubits~\cite{Aharonov97a,Kitaev97b,Knill98c}. 

This notion is developed in the theory of {\em fault-tolerant} quantum
computation~\cite{Gottesman98a}.  Using concatenated quantum codes and
quantum circuits which minimize error propagation between qubits, the
net error rate can be made {\em arbitrarily small}, provided the
``raw'' error rate is below the accuracy threshold.  Thus, a reliable
quantum computer can be constructed from unreliable components.

%%%%%%%%%%%%%%%%%%%%%%%%%%%%%%%%%%%%%%%%%%%%%%%%%%%%%%%%%%%%%%%%%%%

\section{Summary}

The combination of fundamental concepts in quantum physics, computer
science and information theory have led to the rich field of quantum
computation. Three main theoretical results of this field are that
\begin{enumerate}
\item the complexity of quantum systems grows exponentially with 
the number of elementary quantum systems involved,
\item certain problems which appear intractable on any classical
computer are tractable on a quantum computer,
\item reliable quantum computers can be constructed from unreliable 
components, provided the error rate is below the accuracy threshold.
\end{enumerate}

These results highlight the theoretical potential of quantum
computation for fundamentally new systems and devices, capable of
reliably solving problems beyond the reach of classical machines. From
the presentation in this chapter, requirements for the implementation
of quantum computers naturally emerge. This is the subject of the next
chapter.

%% file: impl.tex
\chapter{Implementation of quantum computers}
\label{ch:impl}

In the first half of this chapter, we ask ourselves what the
fundamental requirements are for building a quantum computer.  In the
second half, we briefly review the state of the art in various
proposed embodiments of quantum computers.

\section{Requirements}

Our understanding of the minimal requirements for quantum computation
has grown considerably over the years. They are often formulated as
the five criteria of David DiVincenzo~\cite{DiVincenzo00a}. These
are

\begin{enumerate}
\item a scalable physical system with well characterized qubits,
\item a universal set of quantum gates,
\item the ability to initialize the state of the qubits to a simple 
fiducial state, such as $\ket{00\ldots0}$,
\item a qubit-specific measurement capability,
\item long relevant decoherence times, much longer than the gate 
operation time.
\end{enumerate}

We shall now explore each of these criteria in detail, in order to
develop a good understanding of their significance as well as their
stringency.

%%%%%%%%%%%%%%%%%%%%%%%%%%%%%%%%%%%%%%%%%%%%%%%%%%%%%%%%%%%%%%%%%%%%%

\subsection{Qubits}
\label{impl:qubits}

The heart of a quantum computer is a set of physical systems which
represent the state of the quantum bits. Since a qubit is by
definition a {\em two-level} quantum system, spin-1/2 particles and
polarized photons are natural realizations of qubits.  In practice, a
qubit may also be represented by two levels of a large manifold of
levels, for example two energy levels of an atom. Proposed qubit
embodiments range from trapped atoms and ions to nuclear and electron
spins, electric charge, magnetic flux and photons (see
section~\ref{impl:state_art}).

\subsubsection{Size of the qubit register}

The complexity of just a 40-qubit quantum computer far exceeds the
complexity of most classical computers. That is, a classical computer
would need vastly more time to simulate the dynamics of a $40$ qubit
computer than it would take the quantum computer itself to run, even
if the clock speed of the classical computer is much greater than the
clock speed of the quantum computer, because the quantum computer can
explore $2^{40}$ computational paths in parallel.

However, in order to factor a 400 digit number (a task well beyond the
capability of classical computers for the foreseeable future), a few
thousand logical qubits are required.\footnote{The second register in
the order-finding algorithm must be large enough to represent the
integer to be factored, which would be $\log_2 10^{400} = 400 \log_2
10 \approx 1329$ bits, and the first register must be at least twice
as large as the second register. In practice, additional scratchpad
qubits may be desired.} With error correction, $10$ to $100$ times
more qubits would be needed.

Fortunately, some other interesting applications require much more
moderate numbers of qubits. For example, a quantum computer with $50$
to $100$ qubits (not counting a possible overhead for error
correction) could compute the electronic orbitals of a small atom such
as boron with greater accuracy than any classical simulation performed
to date~\cite{Abrams99a}.

\subsubsection{Alternatives to qubits}

In principle, a three-level (or higher) quantum system could also
serve as the basic unit of quantum information. However, this offers
no computational advantage and all current proposals for the physical
implementation of quantum computers are based on qubits.

Could we use one $2^n$-level system instead of $n$ two-level systems?
Both have a $2^n$ dimensional Hilbert space and are mathematically
perfectly equivalent. However, in a $2^n$-level system, the levels
must either extend over an exponentially large range of energies or
must lie exponentially close together.  Clearly, this is not a
scalable approach to quantum computing.

Finally, could we use continuous variables, such as the position or
momentum of a particle, to represent quantum information?  Properties
such as entanglement of continuous variables have been explored both
theoretically and experimentally~\cite{Braunstein98b}. However, the
precision of any physical device is limited by noise, and it thus
appears always necessary for a realistic computer to discretize the
continuous variables in order to perform computations \cite{Lloyd99a}.

%%%%%%%%%%%%%%%%%%%%%%%%%%%%%%%%%%%%%%%%%%%%%%%%%%%%%%%%%%%%%%%%%%%%%%

\subsection{Quantum gates}
\label{impl:gates}

The time evolution of a closed quantum system is determined by its
Hamiltonian (section~\ref{qct:dyn&rev}). In order to realize quantum
logic gates, we must therefore be able to control the Hamiltonian over
time such that the resulting time evolutions correspond to the
computational steps of an algorithm~\cite{Benioff80a}. A tremendously
helpful theoretical result is that there exist small sets of unitary
transformations which are universal for quantum computation
(section~\ref{qct:universality}). We shall here single out one
particular universal set of gates, the combination of arbitrary
single-qubit rotations with the {\sc cnot} gate, which has also been
the set of choice in the literature. The starting point for our
discussion of quantum gates is that we require\\

{\em the ability to manipulate the Hamiltonian accurately, quickly and
selectively such that we can perform a universal set of one- and
two-qubit gates on individual qubits or pairs of qubits.}\\

\subsubsection{Coupling network}
\label{page:indirect_couplings}

We must be able to perform two-qubit gates between any two qubits in a
quantum computer. However, a computer with pairwise couplings between
all the qubits (Fig.~\ref{fig:coupling_networks} a) is nearly
impossible to build, as interactions between physical particles tend
to rapidly decrease with the separation between the particles. For
example, nuclear spins in a molecule are well coupled if they are
united in a chemical bond, but the couplings are weaker or sometimes
zero for two- and three-bond couplings.

A common architecture in proposed quantum computer realizations uses
{\em exclusively} nearest-neighbour couplings between qubits placed in a
linear or two-dimensional array (Fig.~\ref{fig:coupling_networks} b).
This is the case for many solid-state proposals, such as inductively
coupled SQUID loops, quantum dots connected by tunneling barriers and
nuclear spins of donor atoms, coupled via the overlap between the
respective electron clouds.

Fortunately, we can effect two-qubit gates between any pair of qubits
even if they aren't all directly coupled to each other, as long as
there exists a path of couplings that indirectly connects any two 
qubits. For example, in order to perform a {\sc cnot}$_{13}$ gate with
the coupling network of Fig.~\ref{fig:coupling_networks} (b), we can
first swap the state of qubits $1$ and $2$ (this can be done via the
sequence {\sc cnot}$_{12}$ {\sc cnot}$_{21}$ {\sc cnot}$_{12}$), then
perform a {\sc cnot}$_{23}$, and finally swap qubits $1$ and $2$
again.  The net effect is a {\sc cnot}$_{13}$.

\bfig
\bcen
\vspace*{1ex}
\includegraphics*[width=12cm]{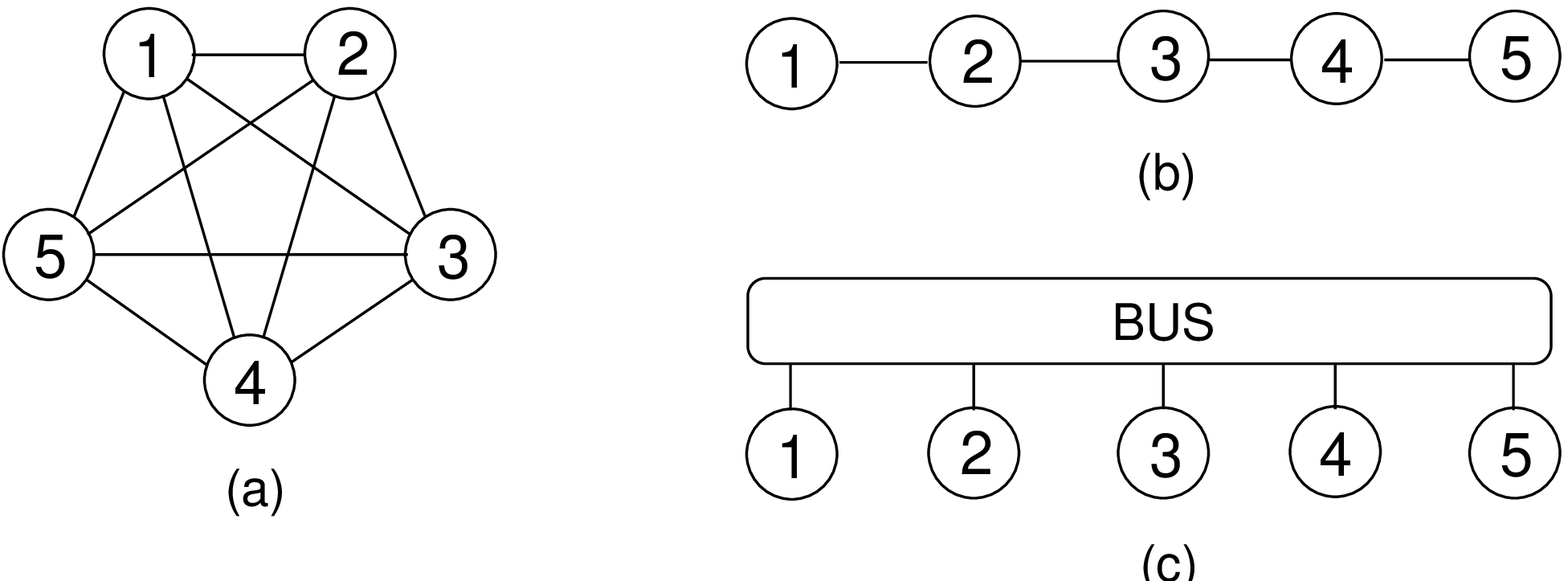} 
\vspace*{-2ex}
\ecen
\caption{Three extreme coupling networks between five qubits. (a) A full 
coupling network. (b) A nearest-neighbour coupling network. (c)
Coupling via a bus.}
\label{fig:coupling_networks}
\efig

Even if none of the qubits are directly coupled to each other, it is
still possible to perform universal computation provided an external
(but still quantum mechanical) degree of freedom can be selectively
coupled to any one qubit in a controlled manner
(Fig.~\ref{fig:coupling_networks} c). The external degree of freedom
serves as a {\em bus} and can facilitate two-qubit gates between any
pair of qubits in much the same way intermediary qubits can facilitate
two-qubit gates between any two qubits in
Fig.~\ref{fig:coupling_networks}.

Examples of such bus degrees of freedom are the collective motional
state of an array of trapped ions, a photon inside an optical cavity
which holds several trapped atoms or a photon which couples to several
quantum dots embedded in a semiconductor structure. Typically, the
spatial range over which collective degrees of freedom can extend
while maintaining the ability to significantly couple to the qubits,
is also limited.  Therefore, such architectures may have to be
supplemented by additional means for interactions between different
groups of qubits.

\subsubsection{Time-dependent control over the Hamiltonian}

It is not always necessary that we can turn on and off all the terms
in the Hamiltonian at will, as long as we can reverse undesired time
evolutions at a later stage. This is commonly done for example in NMR,
by so-called refocusing techniques (similar to spin-echoes), where the
$\omega\sigma_{z}$ term in the Hamiltonion cannot be switched off but
it is possible to perform a $R_x(180)$ or {\sc not} gate. By applying
the {\sc not} gate halfway a time interval $t$ and again at the end,
the $\omega\sigma_z$ evolution which took place in the first time
interval is unwound in the second interval, so there is no net
evolution (up to an overall and thus irrelevant phase shift). We can
verify this mathematically using
Eqs.~\ref{eq:pauli_rel1},~\ref{eq:pauli_rel2},~\ref{eq:R_x}
and~\ref{eq:R_z}:

$$
\hspace{-5cm} R_x(180) e^{-i \omega \sigma_z t/2} R_x(180) e^{-i \omega \sigma_z t/2} =
$$
\vspace{-2em}
\be
(-i \sigma_x)
\left[\cos(\omega t/4) \sigma_I - i \sin(\omega t/4) \sigma_z \right] 
(-i \sigma_x)   
\left[\cos(\omega t/4) \sigma_I - i \sin(\omega t/4) \sigma_z \right] 
= - \sigma_I
\ee
for any $t$. The same refocusing technique can be used to time-reverse
evolution under the Hamiltonian $\sigma_z^i \sigma_z^j$, as also
\be
\sigma_x^i \; \exp(-i t \sigma_z^i \sigma_z^j /2) \;
\sigma_x^i \; \exp(-i t \sigma_z^i \sigma_z^j /2) \;
= \; - \sigma_I^i \sigma_I^j \,,
\ee
for any $t$. This is pictorially illustrated for the case of NMR in
Fig.~\ref{fig:refocusing2}. Of course, it isn't always so easy to
reverse undesired evolutions, as we will see for example in
section~\ref{nmrqc:1bitgates}, but nevertheless the use of unitary
operations to time-reverse specific undesired evolutions can
tremendously reduce the fabrication requirements for quantum
computers.

\subsubsection{Selective addressability}

We have started from the assumption that we must be able to {\em
selectively} perform one- and two-qubit logic gates on arbitrary
qubits. However, we can circumvent the need for selective control of
single- and two-particle Hamiltonians by storing each logical qubit
in the state of three physical qubits.

Seth Lloyd showed that universal quantum computation is possible using
a {\em cellular automata}~\footnote{Computer scientists actually use
this term to describe an even more stringent computational model,
where {\em the same operation} is applied over and over on all the
bits in the computer. Surprisingly, even such a stringent model can be
used to perform meaningful (classical) computations.}  architecture of
the form

$$
D-ABC-ABC-...-ABC
$$

\noindent where the different physical qubits $A$ cannot be 
distinguished from each other nor selectively addressed, and likewise
for the $B$'s and $C$'s~\cite{Lloyd93b}. Each unit cell $ABC$ is used
to represent one qubit. In the default situation, the qubits of
information are stored in the physical qubits $B$, and the $A$'s and
$C$'s are set to $0$, except in one unit cell $i$, where $A$ is set to
$1$.

A single-qubit operation $U$ on qubit $i$ and qubit $i$ only is
obtained via a controlled-$U$ operation of $A$ on $B$ when $A_i$ is
set to 1 while the $A_j$ ($j \neq i$) are set to 0. By applying {\sc
cnot} gates on the whole array, we can move up or down the index of
the unique $A$ which is marked as $1$.  A controlled-$U$ operation of
qubit $i$ onto qubit $i-1$ can be realized using {\sc cnot}s and a
doubly-controlled $U$ gate for which $A_i$ (set to 1) acts as one of
the control qubits.  The special site $D$ is needed to load the
initial state.  The same site $D$ can be used to read out the answer
to the computation one bit at a time.

This model is reasonably well approximated by nuclear spins in long
polymers. It may also ease the fabrication requirements for many
solid-state proposals, including electron spins in quantum dots,
nuclear spins in donor atoms, or nuclear spins in a crystal.

%\bfig
%\caption{Cellular automata D-ABC-ABC-...-ABC}
%\label{fig:automata}
%efig

\subsubsection{Universality without single-qubit terms in the 
Hamiltonian}

Single-qubit operations are not strictly necessary for universal
quantum computation; in fact, it has been shown that ``almost any''
two-qubit gate is universal~\cite{Deutsch95a,Lloyd95a}. Unfortunately,
virtually none of these universal two-qubit gates can be generated by
naturally occurring two-particle Hamiltonians. The underlying reason
is that most interaction Hamiltonians provided by nature exhibit too
much symmetry in the two qubits.  The exchange Hamiltonian, 
\be
{\cal H} = \vec{\sigma}^1 \cdot \vec{\sigma}^2 = \sigma_x^1 \sigma_x^2
+ \sigma_y^1 \sigma_y^2 + \sigma_z^1 \sigma_z^2 \,,
\ee
which describes the interaction between electron spins in quantum dots
coupled through a tunneling barrier (and related proposals), is an
example of such a symmetrical two-qubit Hamiltonian.

However, by encoding each logical qubit in the state of three
physical qubits, universal quantum computation is nevertheless
possible~\cite{DiVincenzo00b}. The physical qubits must be placed in a
linear array and it must be possible to selectively switch on and off
the interactions between neighbouring physical qubits (by tweaking the
tunneling barrier in the example). State initialization to an encoded
basis state is accomplished by cooling the system down to its ground
state in the presence of select couplings.

Encoding each logical qubit in the state of two physical qubits is
also possible but only if the two physical qubits have a different
effective magnetic moment~\cite{Levy01a}. The two encoded basis states
are $\ket{01}$ and $\ket{10}$; they are clearly susceptible to the
exchange interaction.  An additional advantage is that fewer
operations are needed.

\subsubsection{Universality without two-qubit terms in the 
Hamiltonian}
\label{impl:no_twobit_gates}

Quantum computation appears to inherently require non-linear
interactions, i.e. two-qubit terms in the Hamiltonian, in order to
realize two-qubit gates. However, universal quantum computation is
possible with only linear (i.e. single-qubit) quantum gate elements,
supplemented by measurements and classical feedback, which provide the
needed non-linearity~\cite{Knill01a} and allow one to create entangled
qubits via a probabilistic scenario. Using the entanglement thus
created, two-qubit gates are accomplished by a clever combination of
teleportation and single-qubit gates~\cite{Gottesman99a}.

This proposal has been developed in the context of optical
realizations of quantum computers, where it is difficult to obtain
appreciable non-linear interactions between quantum bits (photons)
without too much absorption. Potentially, the scheme may be translated
into other physical systems.

\subsubsection{Accuracy and speed}

An obvious requirement is that the quantum gates be executed with high
fidelity, i.e. that the resulting unitary transformation be as close
as possible to the desired unitary evolution. If that is not possible,
we must have sufficient information over the actual evolution and
sufficient control over the Hamiltonian such that we can undo
erroneous evolutions at a later point in time. 

Erroneous unitary evolutions that cannot be unwound, have the same
detrimental effect on the computation as random errors due to
decoherence and thus effectively increase the decoherence rate. While
they can still be corrected using quantum error correction, this is
associated with a large overhead. Furthermore, quantum error
correction is only effective if the total error rate (due to
decoherence and erroneous unitary evolutions combined) is below the
accuracy threshold, introduced in section~\ref{qct:fault-tolerancy}.

In order to achieve the accuracy threshold, the duration of a typical
logic operation $\tau_{op}$ must be short compared to the coherence
time $\tau_c$. Obviously, the speed and accuracy of logic gates do
usually not go hand in hand, so it is key to make an optimal
trade-off.

Finally, whereas the clock speed is irrelevant from a complexity
theory point of view --- only the scaling of the number of operations
with the problem size is relevant ---, a quantum computer with a clock
speed of 20 Hz (a realistic number for solution NMR for example) may
appear to be of little practical use.  However, let us say that
factoring a 400 digit number would take two hundred thousand quantum
operations, which is a reasonable estimate, and further that quantum
error correction would increase this number by a factor of one
hundred. The resulting twenty million operations would take a
``lousy'' 20 Hz quantum processor one million seconds, which is not
even 12 days, still quite fast given that the same problem would
classically take even the fastest supercomputer longer than the age of
the universe to solve.

%%%%%%%%%%%%%%%%%%%%%%%%%%%%%%%%%%%%%%%%%%%%%%%%%%%%%%%%%%%%%%%%%%%%%%%%%

\subsection{Initialization}
\label{impl:init}

A fundamental condition for computation which is often taken for
granted is the ability to reliably prepare a known input state. If the
input state of a computation is random, the output is of little
use~\footnote{Classically, the output for a random input value may
sometimes be of interest, for example if the task is to determine
whether a certain output value occurs at all. However, since quantum
computations are reversible, the output is always a permutation of the
input, and therefore the output for a random input holds no
information at all.}. Thus, we demand \\

{\em the ability to reliably prepare a {\em pure} input state}.\\

We recall from section~\ref{qct:qubits},
page~\pageref{qct:mixed_vs_pure} that the term ``pure'' implies that
the state is known. If all we know about a qubit is that it is in one
of several states $\ket{\psi_i}$, with certain probabilities, then the
qubit is said to be in a {\em statistical mixture} of states, as
opposed to a pure state.

Unlike for the case of classical computers where it is usually easy to
reset or initialize bits, state initialization can be very difficult
in quantum computers, depending on the physical realization of the
qubits. In qubit implementations where the $\ket{0}$ and $\ket{1}$
states have distinct energies, we can prepare the qubits in their
ground state by letting them equilibrate at a low enough temperature
$T$ such that the energy difference $\Delta E$ between the ground and
excited states satisfies the condition
\be
\Delta E \gg k_B T \,,
\ee
where $k_B$ is Boltzmann's constant. Otherwise, the thermally equilibrated qubits  are in a statistical mixture of
states, described by the density matrix $\rho_{\mathrm{eq}} =
\exp(-{\cal H}/k_B T) / {\cal Z}$, where ${\cal Z}$ is a normalization
factor. At room temperature, $k_B T\approx 26$ meV, which is much
larger than realistic values of $\Delta E$ for many quantum systems.
Therefore, many proposed realizations of quantum computers require
cryogenic temperatures, often in the $10$ or $100$ mK range.

We note that the ability to perform a hard measurement (introduced on
page~\pageref{text:hard_meas} and further discussed in
section~\ref{impl:readout}), automatically leads to the ability to
prepare a pure input state: it suffices to measure the qubit's state,
and change it if needed.

Unfortunately, in many proposed realizations of quantum computers, it
is impossible or very difficult to set up the qubits in a pure initial
state. However, might it not be possible to access the computational
power of quantum systems as long as the state of the qubits is not
completely random ($\rho \neq I/2^n$) ?

\subsubsection{Effective pure states}
\label{page:eff_pure}

Remarkably, it is indeed possible to perform arbitrary quantum
computations on a mixed state, provided the mixed state is {\em
effective pure}, or {\em pseudo-pure} and the observables are
traceless~\cite{Gershenfeld97a,Cory97b} (see
section~\ref{impl:readout}).  Effective pure states are mixed states
described by a density matrix of the form
\be
\rho_{\mathrm{eff}} \; = \; 
\frac{1-\alpha}{2^n}\; I 
\;+\; \alpha\; \ket{\psi}\bra{\psi} \,.
\label{eq:eff_pure}
\ee
For traceless observables, the identity component $I$ does not produce
any signal at all. Furthermore, $I$ doesn't evolve under unitary
transformations, as $U I U^\dag = I$ for all unitary operations $U$. A
system in an effective pure state $\rho_{\mathrm{eff}}$ thus has the
same dynamical behavior and produces the same signal (up to a
proportionality factor $\alpha$) as a system in the corresponding pure
state $\ket{\psi}$.

It is crucial that $\alpha$ not decrease exponentially with the
number of qubits $n$, as $\alpha$ is directly proportional to the
signal strength. If it did, the precision of the measurement would
have to increase exponentially with $n$ or the signal would have to be
averaged over exponentially many experiments. Such an exponential cost
would obviously offset the exponential benefit of quantum computers.

Several methods are known for the preparation of effective pure states
starting from a state in thermal equilibrium at high temperatures, all
of which have been developed in the context of liquid state NMR
quantum computation.  Unfortunately, all these methods have in common
that $\alpha \propto n/2^n$. Then, is such an exponential overhead
inevitably associated with the use of high temperature qubits?

\subsubsection{Cooling down a subset of hot qubits}

The truly surprising answer is no: there exists a {\em scalable}
algorithm for obtaining $k$ pure qubits starting from $n$ partly mixed
qubits. This algorithm, invented by Schulman and
Vazirani~\cite{Schulman99a}, has only a linear overhead in the number
of qubits ($k \propto n$) and a quasi-linear, ${\cal O}(k \log k)$,
overhead in the number of operations. A related algorithm was devised
earlier by Cleve and DiVincenzo in the context of Schumacher
compression; this algorithm requires ${\cal O}(k^3)$ operations and
${\cal O}(\sqrt{k})$ zero entropy auxiliary qubits (which are
recovered at the end of the procedure). The Schulman-Vazirani
algorithm, in contrast, is capable of bootstrapping in the sense that
it can create bits with (near) zero entropy starting from only high
entropy bits.

The idea behind the Schulman-Vazirani scheme is to redistribute the
entropy over the qubits, such that the entropy of a subset of the
qubits approaches zero while the entropy of the remaining qubits
increases (the total entropy is preserved).  In order to calculate the
maximum possible $k$ as a function of $n$, let us define the {\em
polarization} $\epsilon$ of a qubit as the difference in probabilities
between the ground and excited state, tracing out any other
qubits. Mathematically, the polarization of qubit $i$ is defined as
\be
\epsilon = \, \mathrm{Tr} (\rho \sigma_z^i) \,.
\label{eq:polarization}
\ee
A qubit is thus in $|0\rangle$ with probability $\frac{1+\epsilon}{2}$
and in $|1{\rangle}$ with probability $\frac{1-\epsilon}{2}$.  The
theoretical maximum $k_{max}$ of zero temperature $(\epsilon = 1)$
bits that can be extracted from $n$ bits with initial polarization
$\epsilon_0$ is given by entropy conservation,
\be
n H\left(\mbox{$\frac{1+\epsilon_0}{2}$}\right) = k H(1) + (n-k) H(1/2) \,,
\label{eq:entropy_cons}
\ee
where the entropy $H$ is given by
\be 
H(p)=-p\log_{2}p - (1-p) \log_2(1-p) \,,  
\ee
so $H(1)=H(0)=0$ and $H(1/2)=1$. From Eq.~\ref{eq:entropy_cons}, we
find
\be
k_{\mathrm{max}} =
\left[1-H\left(\frac{1+\epsilon_0}{2}\right)\right]n \,,
\ee
which for small $\epsilon_0$ is well approximated by
\be
k_{\mathrm{max}} \approx \epsilon_0^2 \,n \,.
\label{eq:k_max}
\ee
Schulman and Vazirani showed that their scheme is not only efficient
(the overhead is only polynomial in $n$) but is also optimal, in that
it achieves the entropic bound in the limit of large $n$.

We will present an NMR implementation of the elementary building block
of the Schulman-Vazirani scheme in section~\ref{expt:cooling}, and
describe the operation of this building block when we discuss state
initialization in NMR quantum computation
(section~\ref{nmrqc:cooling}).

%%%%%%%%%%%%%%%%%%%%%%%%%%%%%%%%%%%%%%%%%%%%%%%%%%%%%%%%%%%%%%%%%%%%%%%%%

\subsection{Read-out}
\label{impl:readout}

A computation can only be useful if we can access the final result. In
a quantum computer, the final result is represented by the final state
of one or more qubits. We recall (section~\ref{sec:parallel},
page~\pageref{qct:measurement}) that it is not possible to obtain full
information about unknown qubit states, but also
(section~\ref{qct:alg}) that projective measurements in the $\{
\ket{0}, \ket{1} \}$ basis are sufficient if we use quantum
algorithms. Furthermore, a different measurement basis is fine too,
since we can always change basis via a unitary transformation just
before the measurement. In summary, we need\\

{\em the ability to perform accurate projective measurements of the
qubit states.}\\

\subsubsection{Strong and weak measurements}

Read-out of the state of a quantum system requires some form of
coupling of the quantum system to a classical measuring device, such
that at the end of the measurement process, the meter indicates the
state of the quantum system, projected onto the measurement basis. For
example, measurement of the state of a qubit represented by two energy
levels ($\ket{0}$ and $\ket{1}$) in an atom can be done by pumping the
$\ket{1}$ state and looking for fluorescence. If the qubit was in
$\ket{1}$, the atom will fluoresce, a stream of electrons will flow in
a nearby photomultiplier tube, and a signal will appear on the display
of an electrometer. If the atom was in $\ket{0}$, the electrometer
will show no signal. If the qubit state was $a \ket{0} + b \ket{1}$,
the measurement process will collapse the state into $\ket{0}$ or
$\ket{1}$ and the observer will either see a signal or see no signal,
with probabilities $|a|^2$ and $|b|^2$.

%\bfig
%\caption{Meter if (a) qubit is in computational eigenstate, or (b) in 
%a superposition state.}
%\label{fig:meter}
%\efig

Clearly, if the coupling with the measuring apparatus is so strong
that the qubit states instantaneously collapse --- a scenario known as
a {\em hard} or {\em strong} measurement --- we must be able to switch
off the measuring device during the computation. However, it is also
possible to never switch off the measurement provided the quantum
system is only weakly coupled to the meter. In this scenario, called a
{\em weak} measurement, information only very slowly leaks out of the
quantum system. On the one hand, the qubits therefore decohere only
very gradually, as opposed to instantaneously~\footnote{Weak and
strong measurements are understood as taking place on a timescale much
slower or faster than the duration of the computation.}, but on the
other hand, a weak measurement implies that we cannot learn much about
the state of the qubit.

Weak measurements therefore require the use of signal averaging,
either over a large ensemble of identical computers or over
time-sequential experiments performed on a single computer. Signal
averaging may also be needed to boost the reliability of the
measurement if the detector efficiency and/or accuracy is limited.

\subsubsection{Averaged measurements}

Averaging the result of quantum computations poses a specific
difficulty. For example, we recall that in the order-finding algorithm
(section~\ref{qct:shor}), the measurement will with high probability
return an integer multiple of $N/r$, but we don't know {\em which}
multiple. From any multiple $lN/r$, we can determine $r$ with high
probability of success via a classical computation called the
continued fraction expansion. A time averaged measurement, however,
gives $\approx \sum_{l=0}^{r} l N/r = \langle l \rangle N/r$, and an
ensemble averaged measurement gives $\langle l_n \rangle \ldots
\langle l_1 \rangle N/r$ (where the $\langle l_i \rangle$ are the 
bitwise averages of $l$). Either way, it is not possible to compute
$r$ from the averaged measurement outcome except in a few special
cases.

This problem can be circumvented by calculating the continued fraction
expansion on the quantum computer without first measuring the output
state of the order-finding quantum algorithm~\cite{Gershenfeld97a}.
Since the output of the continued fraction expansion is deterministic
(it is $r$), a time or ensemble averaged measurement of the qubits
after the continued fraction expansion will always give $r$ as well.
For all the known quantum algorithms with probabilistic outputs, we
can use similar {\em derandomization} procedures which permit the use
of averaged measurements.

\subsubsection{Other remarks}

$\bullet $ Do we have to be able to read out each individual physical
qubits (the implicit assumption so far)? The answer is no. It is
sufficient to be able to measure the state of just one physical qubit:
at the end of the computation, we can swap one after the other logical
qubit into that physical qubit and measure it. Note that it would not
be good to measure one qubit, then repeat the computation and measure
the next qubit, and so forth (unless the output is derandomized
\cite{Gershenfeld97a}). Again taking order-finding as an example, we
see that at the end of each experiment, the one bit may come from to a
different multiple of $N/r$ every time.\\

\noindent $\bullet$
Do we need to be able to measure qubits during the computation?
Again, the answer is no.  If no subsequent operations depend on the
measurement outcome, such measurements can be left out altogether. If
subsequent operations on the remaining qubits do depend on the
measurement outcome, we can use controlled quantum gates instead.\\

%\bfig
%\caption{A classically controlled $U$ gate controlled by the outcome 
%of a measurement can be replaced by quantum controlled $U$ gates.}
%\label{fig:class_controlled_U}
%\efig

\noindent $\bullet$
Does it matter if a measurement destroys the physical qubit that was
measured? \label{page:destructive_meas} Here also, the answer is
no. In fact, after we measure a qubit, we can do almost anything we
want to it, including throw it away, because this will not affect the
state of the remaining qubits.  However, interactions between the
measured qubits and the remaining qubits must be avoided or
neutralized, because such interactions would alter the state of the
remaining qubits.

%%%%%%%%%%%%%%%%%%%%%%%%%%%%%%%%%%%%%%%%%%%%%%%%%%%%%%%%%%%%%%%%%%%%%%%%%

\subsection{Coherence}
\label{impl:coherence}

The final requirement is that the qubits have a sufficiently long
coherence time $\tau_c$ such that quantum mechanical superposition
states can be preserved throughout the execution of a quantum
algorithm (section~\ref{sec:corr_q_errors}). Or alternatively, and
more realistically given the large number of qubits and operations
involved in realistic computations, \\

{\em the decoherence rate must lie below the threshold error rate for
fault-tolerant quantum computation~\footnote{The estimated values for
the accurracy threshold depend on the error model and on the
architecture of the quantum computer (e.g. whether or not there are
only nearest-neighbour interactions between the qubits and whether or
not parallel operations are allowed). Most estimates lie in the range
of $10^{-6}$ to $10^{-4}$.}:}
\be
\tau_{op}/\tau_c < 10^{-4} \,,
\ee
where $\tau_{op}$ is the typical duration of a quantum
gate. Obviously, this is only a crude measure, as $\tau_{op}$ covers a
wide range of logic gates and $\tau_c$ lumps together a broad range of
errors due to a variety of decoherence processes and systematic
errors which somehow cannot be reversed.

Mathematically, the effect of decoherence can be conveniently
described in the {\em operator sum representation}
\be
\rho \; \mapsto \; \sum_k E_k \rho E_k^\dagger \,.
\label{eq:opsumrep}
\ee
where the $E_k$ are operators acting on the Hilbert space of the
system. The interpretation of this expression is that $\rho$ is
transformed to
\be
\frac{E_k \rho E_k^\dagger}{\mbox{Tr}(E_k \rho E_k^\dagger)}
\ee 
with probability 
\be
p(k) = \mbox{Tr}(E_k \rho E_k^\dagger) \,.
\ee
For trace-preserving processes, the operators $E_k$ must satisfy the
completeness relation
\be
\sum_k E_k^\dagger E_k = I \,.
\ee
The operator sum representation encompasses all physical processes
which can act on a quantum system, including non-unitary processes. We
point out that any given physical process has many possible operator
sum representations, and can thus also be interpreted in different
ways. We also note that for unitary processes, there is only one term
in Eq.~\ref{eq:opsumrep}, $E_0=U$.\\

The general rule for obtaining long coherence times is that the qubits
be highly isolated from the environment, because interactions with the
environment and information leakage to the environment are the cause
of decoherence~\cite{Zurek82a,Zurek91a}. We will now review several
ways in which decoherence can manifest itself.

\subsubsection{Energy dissipation}

Energy dissipation is a decoherence process caused by the exchange of
energy between a quantum system and the environment (the bath).
Physical examples of this process are spontaneous emission in atoms or
semiconductor structures, nuclear or electron spins which return to
the thermal equilibrium with the environment, heating of the motional
state of trapped ions, and so forth.

Energy dissipation to a bath at {\em zero temperature} is described by
\be
\rho \; \mapsto \; E_0 \rho E_0^\dagger + E_1 \rho E_1^\dagger \,,
\label{eq:amp_damp}
\ee
where
\be
E_0 = \left[\matrix{1 & 0 \cr 0 & \sqrt{1-\gamma} }\right] \,,
\hspace*{1.5cm}
E_1 = \left[\matrix{0 & \sqrt{\gamma} \cr 0 & 0 }\right] \,.
\ee
The $E_0$ operation preserves a qubit in the ground state $\ket{0}$
but attenuates the excited state $\ket{1}$; the $E_1$ operation
changes the $\ket{1}$ state into the $\ket{0}$ state with probability
$\gamma$. The overall result of this process, known as {\em amplitude
damping}, is that a qubit in the excited state decays into the ground
state with probability $\gamma$, thereby loosing a quantum of energy
to the environment. As a result, an arbitrary one-qubit density matrix
is transformed as
\be
\left[\matrix{a	& b^* \cr b & c}\right] \; \mapsto \;
\left[\matrix{1-(1-\gamma)(1-a) & b^* \sqrt{1-\gamma} \cr
	      b \sqrt{1-\gamma} & c(1-\gamma)}\right] \,.
\ee

If the environment is at {\em finite temperature}, the process of
Eq.~\ref{eq:amp_damp} must be generalized to
\be
\rho \; \mapsto \; \sum_{k=0}^3 E_k \rho E_k^\dagger \,,
\ee
where
\begin{eqnarray}
\nonumber
\hspace*{-1cm}
E_0 = \sqrt{p} \left[\matrix{1 & 0 \cr 0 & \sqrt{1-\gamma} }\right] 
&\,,&
\hspace*{1.5cm}
E_1 = \sqrt{p} \left[\matrix{0 & \sqrt{\gamma} \cr 0 & 0 }\right] \,,\\
%\ee
%\be
E_2 = \sqrt{1-p} \left[\matrix{\sqrt{1-\gamma} & 0 \cr 0 & 1}\right] 
&\,,&
\hspace*{1.5cm}
E_3 = \sqrt{1-p} \left[\matrix{0 & 0 \cr \sqrt{\gamma} & 0 }\right] \,.
\label{eq:opsumrep_gen_ad}
\end{eqnarray}
Thus, a qubit in the excited state decays to the ground state with
probability $\gamma p$, and a qubit in the ground state is lifted to
the excited state with probability $\gamma (1-p)$. The parameter $p$
depends on the temperature of the environment and the energy
difference between $\ket{0}$ and $\ket{1}$.  The stationary state of
this process, called {\em generalized amplitude damping}, is the mixed
state
\be
\rho_\infty = \left[\matrix{p & 0 \cr 0 & 1-p}\right] \,.
\ee
We can geometrically visualize the effect of generalized amplitude
damping via the transformation of an arbitrary vector on the surface
of the Bloch sphere
\be
(r_x,r_y,r_z) \mapsto \left(r_x \sqrt{1-\gamma}, \, r_y \sqrt{1-\gamma}, 
\, r_z (1-\gamma) + \gamma (2p-1)\right) \,.
\ee
In many physical systems, $\gamma$ is a time-varying function of the
form $\gamma = 1-e^{-t/T_1}$, where $T_1$ is a characteristic time
constant, which was first introduced in NMR~\cite{Bloch46a}. It
corresponds to the {\em lifetime} of excited states.  In real physical
systems, non-unitary exchange of energy can also take place between
different qubits in the system.  This random process also represents a
form of decoherence, just like energy exchange between qubits and the
bath.

Finally, we point out that in most proposed quantum bit
implementations, energy dissipation on the one hand adversely affects
quantum computations, but on the other hand also represents a natural
mechanism for state initialization. After waiting for a sufficiently
long time (several times the $T_1$), the qubits approach thermal
equilibrium with the environment. The thermal equilibrium state thus
constitutes a reproducible and fiducial initial state. If the
environment is close enough to zero Kelvin such that $\Delta E \gg k_B
T$, the thermal equilibrium state is very close to being pure. For
qubits in equilibrium with a high temperature bath, additional state
preparation operations must be performed in order to distill pure
qubits (section~\ref{impl:init}).

\subsubsection{Phase randomization}

Phase randomization results in the loss of coherence (the phase
relationship) between different basis states, and can caused by
interactions with the environment. For example, local fluctuations in
the magnetic field randomize the phase of nuclear or electron spins;
and scattering with defects randomly disturbs the phase of free
electrons in solids. In many systems, phase randomization is the
dominant decoherence process and the coherence time $\tau_c$ is
therefore often loosely taken to be the characteristic phase
randomization time.

A phase shift over some random angle $\theta$ 
changes an arbitrary pure one-qubit state $|\psi \rangle = a |0
\rangle + b |1\rangle $ into $R_z(\theta) |\psi\rangle = a e^{-i
\theta /2} |0 \rangle + b e^{i \theta /2} |1 \rangle $ (see Eq.~\ref{eq:R_z}). The density matrix changes accordingly:
\be
	\rho = |\psi \rangle \langle \psi| = \left[ \begin{array}{cc}
	{|a|^2}&{a b^*}\\{a^* b}&{|b|^2} \end{array} \right]
\mapsto 
	R_z(\theta) \rho R_z(\theta)^{\dagger} 
	= \left[ \begin{array}{cc}
	{|a|^2}			&	{a b^*  e^{-i \theta}}
\\	{a^* b e^{i \theta}}	&	{|b|^2} 
	\end{array} \right]  \,.
\ee
If we model phase randomization as a stochastic process with $\theta$
drawn from a normal distribution with variance $2 \lambda$, the
density matrix resulting from averaging over $\theta$ is
\be
	 \langle \rho' \rangle _{\theta} 
	= \int \frac{1}{\sqrt{4 \pi \lambda}}  
	e^{-\theta^2 / 4 \lambda} R_z(\theta) \rho R_z(\theta)^{\dagger} d\theta 
	= \left[ \begin{array}{cc}
	{|a|^2}				& {a b^*  e^{-\lambda}}
\\	{a^* b e^{-\lambda}}		& {|b|^2} 
	\end{array} \right]. 
\label{eq:phasedamp}
\ee
A mixed initial density matrix is transformed similarly: 
\be
	\left[ \begin{array}{cc}
	{a}	& {b^{*}}
\\	{b}	& {c} 	\end{array} \right] ~~~ \mapsto ~~~ 
	\left[ \begin{array}{cc}
	{a}			& {e^{-\lambda} b^{*}}
\\	{e^{-\lambda} b}	& {c} 	\end{array} \right] 
\label{eq:phasedampmix}
\,,
\ee
since it is a weighted average of the constituent pure states.

The off-diagonal elements of the density matrix thus decay
exponentially over time, and phase randomization is therefore also
called {\em phase damping}. Since the diagonal elements, which
represent the populations of the basis states, remain unaffected,
phase randomization signifies the {\em loss of coherence without net
change of energy}.

The operator sum representation of the phase damping process is given
by the operators
\be
E_0 = \sqrt{\gamma} \left[\matrix{1 & 0 \cr 0 & 1}\right] \,,
\hspace*{1.5cm}
E_1 = \sqrt{1-\gamma} \left[\matrix{1 & 0 \cr 0 & -1}\right] \,,
\label{eq:opsumrep_pd}
\ee
so phase randomization is equivalent to a phase flip which occurs with
probability $1-\gamma$, where $\gamma = (1 + e^{-\lambda})/2$.

We can geometrically visualize the effect of phase randomization via
the transformation of an arbitrary vector on the surface of the Bloch
sphere. Using Eq.~\ref{eq:phasedampmix}, we find that
\be
(r_x,r_y,r_z) \mapsto 
\left(r_x e^{-\lambda}, \, r_y e^{-\lambda},\, r_z \right) \,.
\ee

In many physical systems, $\lambda$ increases linearly over time,
$\lambda = t/T_2$. Like $T_1$, the characteristic time constant $T_2$
originated in NMR \cite{Bloch46a}.  Intrinsic phase randomization
$T_2$ must be distinguished from systematic dephasing $T_2^{sys}$
where the information about the erroneous evolution is known, as
opposed to lost in the environment. Systematic dephasing can in
principle be reversed without quantum error correction, e.g. spin-echo
techniques can reverse dephasing of spins in an inhomogeneous magnetic
field.  The decay rate of the off-diagonal elements due to the
combined effects of systematic and random loss of phase coherence is
often described via the time constant $T_2^*$, given by
\be
\frac{1}{T_2^*} = \frac{1}{T_2} + \frac{1}{T_2^{sys}} \,.
\ee

A clean measurement of the phase damping time constant is complicated
by the fact that amplitude damping also results in loss of coherence
and the decay of off-diagonal entries in the density matrix. The
measured ``$T_2$'' therefore usually includes contributions from
amplitude damping (those contributions are negligible only if $T_1 \gg
T_2$), in addition to phase damping. On the other hand, phase damping
does not affect the diagonal entries of the density matrix; those
entries relax solely due to amplitude damping. Therefore, a clean
measurement of $T_1$ can be done by measuring the decay rate of the
diagonal entries of the density matrix.

For multiple qubits, the effect of phase randomization is often more
pronounced the more qubits are entangled with each other, and the
stronger the degree of entanglement. For example, random phase kicks
acting on qubit $i$ will change the maximally entangled state of $n$
qubits $\ket{\psi_n} = (\ket{00\ldots0}+\ket{11\ldots1})/\sqrt{2})$
into $(e^{-\theta/2}\ket{00\ldots0} +
e^{\theta/2}\ket{11\ldots1})/\sqrt{2}$. If random phase kicks occur on
all qubits, $\ket{\psi_n}$ becomes $(e^{-n\theta/2}\ket{00\ldots0} +
e^{n\theta/2}\ket{11\ldots1})/\sqrt{2}$. Therefore the decoherence
rate of an $n$-qubit maximally entangled state is $n$ times higher
than for a 1-qubit system.~\footnote{This is obviously a crucially
important consideration for quantum error correction, which relies
precisely on entanglement to combat decoherence.}

\subsubsection{Qubit disappearance}

In some realizations of qubits, the physical qubit itself is very
fragile (not just the qubit state). For example, an atom may easily
escape from an optical trap, a photon may leak out of an optical
cavity, or may be absorbed when it is sent through a non-linear
medium, and so forth. The qubit itself may thus disappear or be
annihilated altogether. The resulting error is called an erasure
error.

Quantum error correction can deal with erasure errors if it is
possible to detect the abscence or presence of a qubit and to replace
a missing qubit with a qubit in some standard state. This state is
generally different than the state of the original qubit, but it can
be corrected using standard quantum error correction. In fact, the
overhead can be smaller than usual because the error occurred in a
known location.

We note that it is not a problem if qubits are destroyed in the
measurement process, provided a fresh supply of qubits is readily
available. Traditional single photon detectors are an example of such
destructive meters. Non-demolition photon detection is in
principle also possible and has been demonstrated in the lab, but
remains very hard.

\subsubsection{Leakage outside the qubit manifold}

A related problem arises if a qubit is embodied by two levels which
are part of a larger manifold of levels. If the quantum system
transitions outside the qubit manifold, the computation will obviously
go awry.

This can occur in trapped ions and atoms, electronic states in quantum
dots, magnetic flux states in SQUIDS, and so forth. In contrast, the
Hilbert space of spin-1/2 particles and polarized photons is naturally
confined to two dimensions, so here it is impossible for the qubit to
leak into extraneous levels.

The extra degrees of freedom can also represent an advantage. For
example, it is common to temporarily take the state of an atom or ion
out of the qubit subspace, in order to facilitate quantum logic
operations. This may be needed because of selection rules, or it may
simply be technologically easier to perform the desired logic
operations in this way.  Extra levels also lie at the basis of 
selective measurement schemes, such as fluorescence measurements
in trapped ions.

%%%%%%%%%%%%%%%%%%%%%%%%%%%%%%%%%%%%%%%%%%%%%%%%%%%%%%%%%%%%%%%%%%%%%

\section{State of the art}
\label{impl:state_art}

This section gives only a quick survey of some of the possibilities
and challenges that characterize various proposed quantum computer
implementations. A recent collection of articles \cite{Braunstein00a}
reviews the operation, feasability and state of the art of these
schemes in much more detail. Many important original papers are
collected in \cite{Macchiavello00a}.

\subsection{Trapped ions}
\label{impl:ions}

\subsubsection{Concept}

Cirac and Zoller~\cite{Cirac95a} showed in a seminal paper that a set
of cold ions interacting with laser light and moving in a linear trap
provides a realistic physical system in which to implement a quantum
computer.  Two internal states of the ion serve as the $\ket{0}$ and
$\ket{1}$ levels; they may be electronic levels, hyperfine levels or
Zeeman levels, depending on the ion (e.g. $^{9}$Be$^+$, $^{25}$Mg$^+$,
$^{40}$Ca$^+$ or $^{138}$Ba$^+$), and all of these internal states can
have coherence times of several seconds.

A register of $n$ quantum bits is obtained by loading $n$ ions into an
RF trap. Usually the trapping potential is such that the ions are held
in a linear array (Fig.~\ref{fig:iontrap}), spaced apart by several
micrometers due to the ion-ion Coulomb repulsion. The ions are cooled
down to their motional ground state, usually in two stages, first by
Doppler cooling and then by sideband cooling. For weak traps, cooling
via electromagnetically induced transparency techniques can be used
instead of sideband cooling.

\bfig
\vspace{1ex}
\bcen
\includegraphics*[width=7cm]{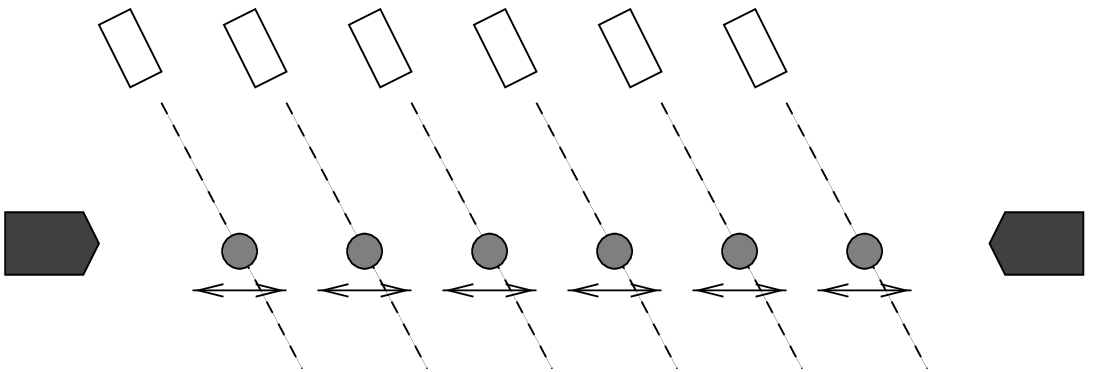}
\ecen
\vspace*{-2ex}
\caption{Schematic diagram (after ~\cite{Cirac95a}) of an ion trap 
containing six ions (the electrodes needed to keep the ions on one
line are not shown). Each ion can be individually addressed with laser
pulses, and the collective vibrational motion of the ions serves as a
bus qubit.}
\label{fig:iontrap}
\efig

Single qubit operations are accomplished using laser pulses focused
tightly on the desired ion. Either the pulses are resonant with the
$\ket{0} \leftrightarrow \ket{1}$ transition, or alternatively Raman
transitions are used, depending on the selection rules, energy
differences and available lasers.

Two qubit operations are accomplished by using the collective
quantized motion of the ions as a bus. First a laser is turned on,
acting on the $k$th ion, and detuned by an amount equal to the energy
of one center-of-mass phonon. The interaction Hamiltonian is then a
Jaynes-Cummings type Hamiltonian of the form
\be
{\cal H}_{k,q} \propto \Omega
\left[ \ket{1_q}_k\bra{0} a e^{-i \phi} + 
       \ket{0}_k\bra{1_q} a^\dagger e^{i \phi} \right] \,,
\ee
where $a^\dagger$ and $a$ are the creation and annihilation operator
of the center-of-mass phonons, $\Omega$ is the Rabi frequency and
$\phi$ is the laser phase. The subscript $q=0,1$ refers to the
transition excited by the laser light; $q=0$ excites the $\ket{0}
\leftrightarrow \ket{1}$ transition whereas $q=1$ excites the $\ket{0}
\leftrightarrow \ket{1'}$ transition, where $\ket{1'}$ is an 
auxiliary energy level (Fig.~\ref{fig:ionlevels}).  A three-step
process then results in a conditional phase shift between two ions $k$
and $l$~\cite{Cirac95a}. First apply a $\pi$ pulse on ion $k$, with
$q=0$ and $\phi = 0$ ($U_1$). Then apply a $2\pi$ pulse on ion $l$,
with $q=1$ and $\phi = 0$ ($U_2$). Finally apply the first pulse again
($U_1$). The resulting operation $U_1 U_2 U_1$ can be converted into a
{\sc cnot} gate via additional single qubit rotations.

\bfig
\vspace*{1ex}
\bcen
\includegraphics*[width=8cm]{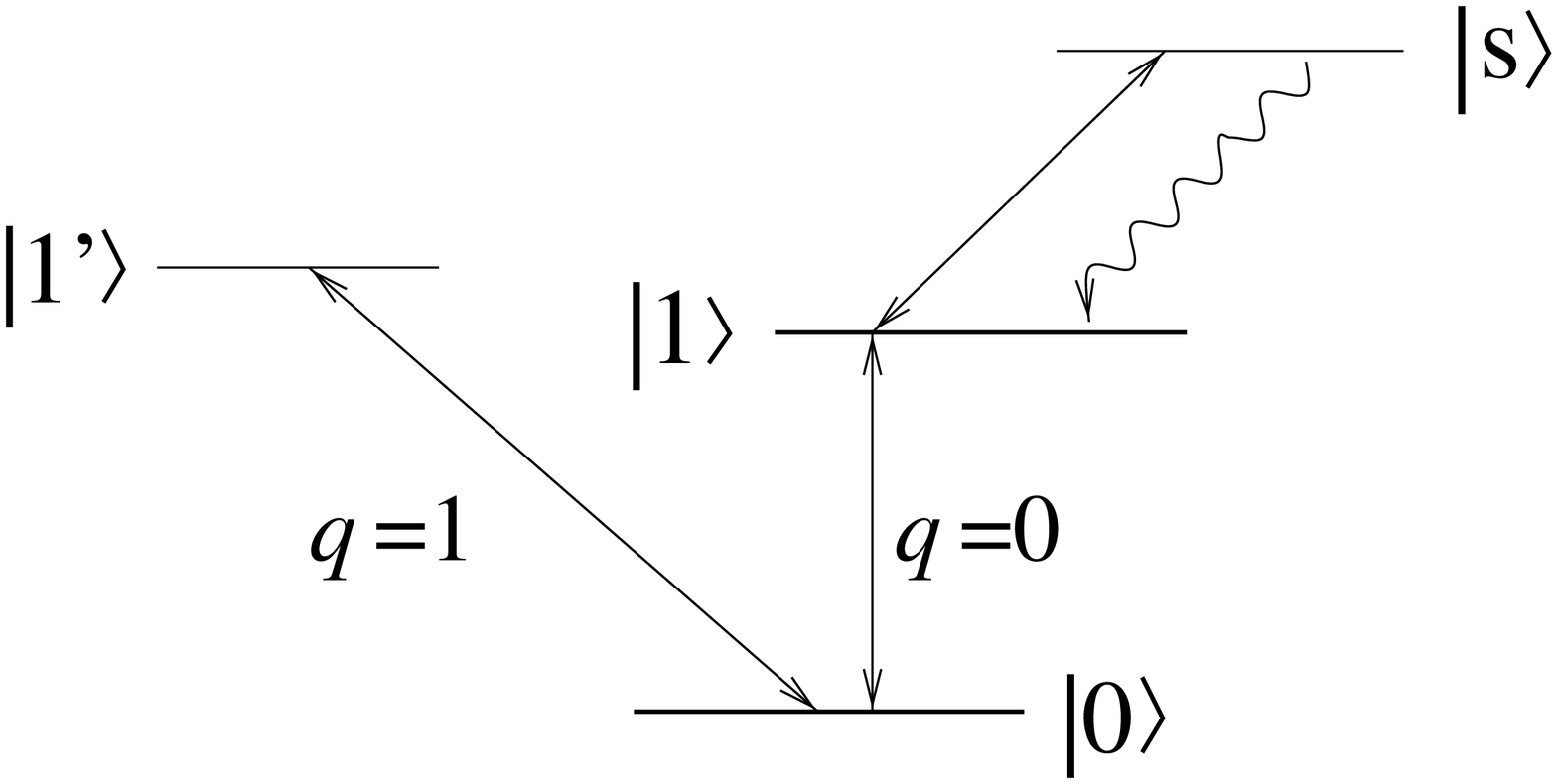}
\ecen
\vspace*{-2ex}
\caption{Model energy level diagram of the relevant internal states 
in a trapped ion. The qubit is embodied by $\ket{0}$ and
$\ket{1}$. Level $\ket{1'}$ assists in two qubit operations and level
$\ket{s}$ is used during read-out.}
\label{fig:ionlevels}
\efig

Measurement is done by exciting the transition between $\ket{1}$ and
an auxiliary ``shelving'' level $\ket{s}$, which exhibits strong
spontaneous decay. Thus, if fluorescence is observed during excitation
of the $\ket{1} \leftrightarrow \ket{s}$ transition, the ion was in
$\ket{1}$; if no fluorescence is observed, we conclude that it was in
$\ket{0}$.\footnote{The assumption is that the ion remains in the
$\{\ket{0},\ket{1}\}$ manifold except temporarily during the two-qubit
gates.}

Inspired by the original Cirac-Zoller scheme, many variations of ion
trap quantum computation have been proposed. Most notably, M{\o}lmer
and S{\o}renson~\cite{Molmer99a} proposed a technique for entangling
$n$ ions using a wide laser beam which covers all ions (instead of a
set of tightly focused beams). Furthermore, this scheme relaxes the
requirement for cooling of the motion.

\subsubsection{Experiments}

Ion traps have been used as frequency standards for a number of
years. Several groups across the world have trapped a single ion and
cooled it down to its motional ground state
\cite{Diedrich89a,Roos00a}. The state of single ions can be well
controlled using laser pulses, and measurement can be done with near
$100 \%$ efficiency \cite{Nagourney86a,Bergquist86a}. David Wineland's
group at NIST in Boulder, CO, and Raineir Blatt's group in Innsbruck,
Austria, have cooled more than one trapped ion to the ground state of
the collective motion~\cite{King98a,Rohde01a}. Only the NIST group has
reported the realization of two-qubit gates with trapped ions. In an
impressive series of experiments, they first demonstrated a
controlled phase shift between an ion and the motion~\cite{Monroe95a},
then entangled two ions~\cite{Turchette98a}, and later used the
M{\o}lmer-S{\o}renson scheme to entangle four ions ~\cite{Sackett00a}.

One of the major challenges in the experiments is heating of the
motional state, and this issue is only partly understood and resolved
\cite{Wineland98a}. Also, additional know-how must be built up such
that universal quantum logic gates can be implemented, which would
allow the realization of simple quantum algorithms. Finally, even
though several proposals for ``scalable'' arrayed approaches to ion
traps exist (e.g. ~\cite{Cirac00a}), it remains unclear how many ions
could be held in a single trap, or how ions in different traps could
be made to communicate in a practical and coherent way.

%%%%%%%%%%%%%%%%%%%%%%%%%%%%%%%%%%%%%%%%%%%%%%%%%%%%%%%%%%%%%%%%%%%%%

\subsection{Neutral atoms}

\subsubsection{Concept}

We will distinguish two very different strategies for trapping atoms:
cavity quantum electrodynamics (QED) and optical lattices. Both have
been proposed as the basis for quantum computers.

In {\em cavity QED} \cite{Grangier00a}, an atom interacts with a
single-photon mode in an optical or microwave cavity
(Fig.~\ref{fig:cavity_QED}). The atom-cavity interaction Hamiltonian
is a Jaynes-Cummins Hamiltonian
\be
{\cal H} \propto g 
\left(a^\dagger \ket{g}\bra{e} + a \ket{e}\bra{g} \right) \,,
\ee
where $a^\dagger$ and $a$ are the raising and lowering operator of the
cavity, $\ket{g}$ and $\ket{e}$ are the ground and excited states of
the atom and $g$ is the vacuum Rabi frequency. The coherent
interaction of the atom and the cavity competes with spontaneous decay
of the atom at a rate $\gamma$ and with cavity decay at a rate
$\kappa$. In order to obtain ``strong'' coupling between the atom and
the cavity, we need $g \gg (\gamma, \kappa)$, and the dwell time of
the atom in the cavity must be long compared to the inverse of these
rates. A high value of $\gamma$ requires a small cavity (only about 10
$\mu$m long) so the electric field of the single photon is very
intense; $\kappa$ is determined by the reflectivity of the mirrors and
$\gamma$ depends on the atom and also on the cavity.

Such cavities, in particular microwave cavities, could in principle be
used for quantum computation, with two internal levels of the atom
($\ket{g}$ and $\ket{e}$) as a qubit. A register of several qubits
could be realized by trapping several atoms in a cavity. Tightly
focused laser beams coming in from the sides of the cavity could
address individual atoms for one-qubit gates, and the cavity mode
could serve as a bus qubit for two-qubit gates, as it interacts with
all the atoms. However, such schemes appear very difficult to realize.

\bfig
\vspace*{1ex}
\bcen
\includegraphics*[width=9cm]{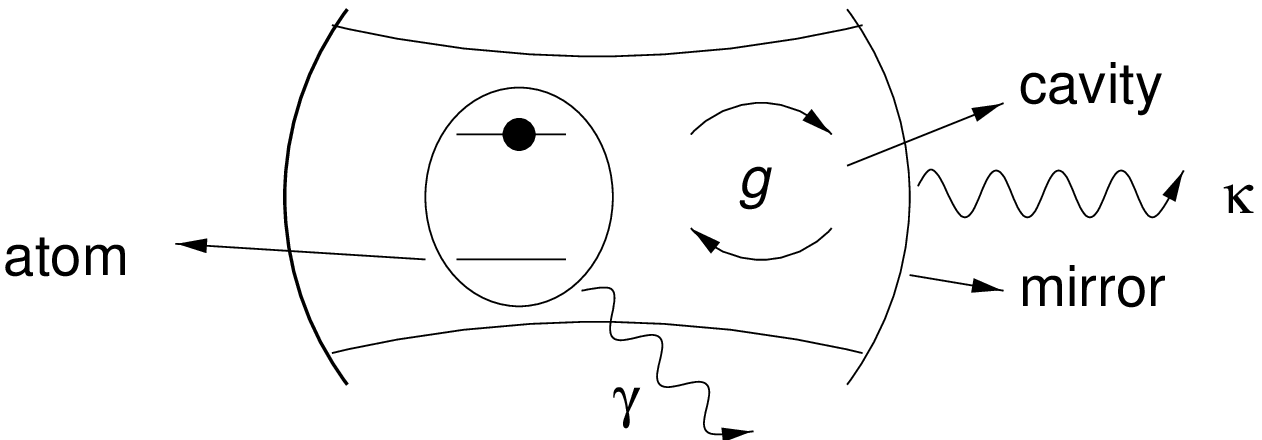}
\ecen
\vspace*{-2ex}
\caption{A single two-level atom is trapped by the cavity mode of a 
single photon. The cavity consists of two curved mirrors. (after
Mabuchi in \cite{Macchiavello00a})}
\label{fig:cavity_QED}
\efig

A more suitable application of cavity-QED, especially for optical
cavities, may lie in the area of quantum communication. A trapped atom
could mediate an interaction between ``flying'' (photon)
qubits. Furthermore, appropriate excitation of an atom trapped in a
cavity could emit single photons in a controlled way, and thus serve
as a source for quantum cryptography. In microwave cavities, an atom
travelling through the cavity could transfer its quantum information
to the cavity, which in turn could transfer the information to a
second atom travelling through the cavity at a later time.  All these
schemes illustrate that atom-cavity interactions provide a rich system
for the exchange of quantum information between ``flying'' qubits and
``standing'' qubits.

Atoms trapped in far off-resonance {\em optical lattices} have also
been proposed as qubits \cite{Deutsch00a,Brennen99a}. An optical
lattice is created by three sets of optical standing waves at right
angles, created by laser beams. A large number of neutral atoms can be
loaded into the optical lattice for example from a magneto-optical
atom trap or from a Bose-Einstein condensate; the atoms are trapped by
the lattice at regular spacings in the ``wells'' of the standing
waves. On-resonance laser beams can produce single-qubit rotations. By
varying the polarization of the trapping lasers, two sets of atoms in
adjacent wells can be made to pairwise occupy the same well so
electric dipole-dipole interactions between the two atoms in each pair
are induced. Two-qubit gates are thus be performed on many pairs of
atoms in parallel, which can be advantageous in some cases but also
constitutes a limitation.  Good ways to measure the internal state of
atoms trapped in an optical lattice are currently being investigated.

\subsubsection{Experiments}

Strong atom-cavity coupling has been achieved in both optical
cavities, most notably with Cs atoms in Kimble's group at Caltech,
\cite{Mabuchi96c} and with Rydberg atoms in microwave cavities, in 
Haroche's lab at the ENS in Paris \cite{Brune96a}. An atom trapped in
an optical cavity has been used to cause a conditional phase shift
between two photons flying through the cavity and interacting with the
atom \cite{Turchette95a}. This experiment represented the first
explicit realization of a two-qubit gate. An optical cavity has been
used as a single-photon source \cite{Law97a} and quantum memory
operation of a single photon mode has been accomplished in a microwave
cavity \cite{Maitre97a}. Also via a microwave cavity, three Rydberg
atoms have been entangled with each other \cite{Rauschenbeutel00a}.

Compared to trapped ions, coherent control and readout are more
complicated in trapped atoms, since the trapping potentials are much
weaker for neutral atoms than for charged ions.  For this and other
reasons, it appears that the potential of cavity QED for quantum
computing is limited, but optical cavities may find good use in
quantum communication. Both optical and microwave cavities also
provide a beautiful testbed for the study of decoherence and quantum
(non-demolition) measurements, as in \cite{Nogues99a}.

The state of the art in optical lattices is still extremely limited.
On the order of $10^6$ neutral Cs atoms have been trapped in a
two-dimensional lattice \cite{Hamann98a}, but it is not currently
possible to reliably create a completely full lattice, much less to
selectively control or read out the state of atoms trapped in an
optical lattice. 

%%%%%%%%%%%%%%%%%%%%%%%%%%%%%%%%%%%%%%%%%%%%%%%%%%%%%%%%%%%%%%%%%%%%%

\subsection{Quantum dots}
\label{impl:qdots}

Loss and DiVincenzo \cite{Loss98a} worked out a proposal for quantum
computing based on ``artificial atoms'', created via semiconductor
structures. In such quantum dots, the qubit is given by the spin of
the excess electron on a single-electron quantum dot, placed in a
static magnetic field. The confinement of the electron in the dot must
be strong enough such that excited excitonic and electronic states
have much higher energies than the spin Zeeman energy, and thus have
negligable occupancy. The lay out of a possible device is shown
schematically in Fig.~\ref{fig:quantum_dots}.

\bfig
\vspace*{1ex}
\bcen
\includegraphics*[width=8cm]{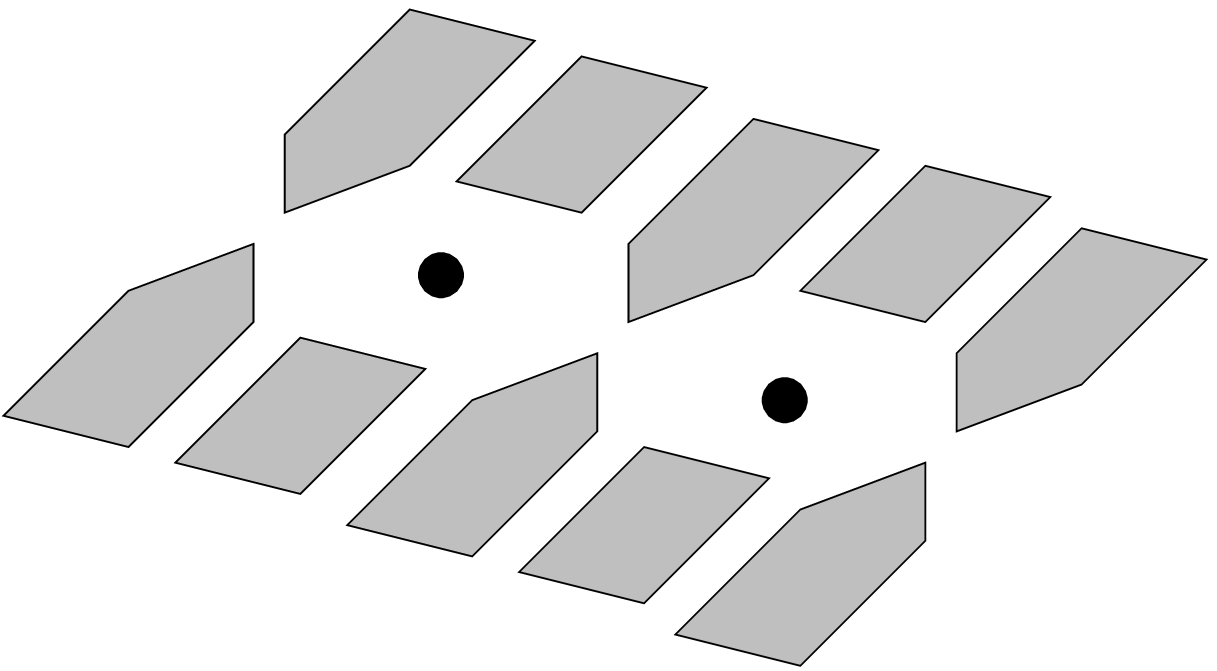}
\ecen
\vspace*{-2ex}
\caption{Conceptual schematic (after \protect\cite{DiVincenzo99a}) of 
one variant of a quantum dot quantum computer.  Lateral side gates on
top of a two-dimensional electron gas (created for example via a
AlGaAs/GaAs/AlGaAs quantum well) confine the motion of an electron to
a very small area (the quantum dot). The tunneling barrier between
neighbouring quantum dots can be controlled via the voltages on the
gates.}
\label{fig:quantum_dots}
\efig

One-qubit rotations could be realized in several ways. Via advanced
scanning probe techniques, it may be possible to locally and
selectively manipulate the electron spin in just one quantum
dot. Alternatively, the Larmor frequency of the electron could be
shifted selectively by changing the bias on the side gates of a
specific dot, or by opening a tunneling barrier to an auxiliary dot
which is ferro-magnetic. A narrowband microwave transverse magnetic
field can then selectively rotate the spin of the electron on a
specific dot.

Two-qubit gates rely on the exchange interaction between the electrons
in adjacent dots, which arises from tunneling through the barrier
between the dots:
\be
{\cal H} = J \, \vec I^1 \cdot \vec I^2 \,.
\ee
By gating the tunneling barrier via the side gates, this interaction
can be turned on and off in a controlled way.

Qubit measurement could be done using a spin-valve tunneling barrier
between the quantum dot and an auxiliary dot. Such spin-dependent
barriers let spin-up electrons pass but block spin-down electrons (or
vice-versa). The presence of an electron in the auxiliary dot can be
detected with a single-electron transistor. Depending on whether or
not we detect an electron on the measuring dot after opening the
spin-dependent barrier, we can conclude that the electron spin was up
or down before measurement.

Highly polarized spins can be obtained by going to very low
temperatures (say 100 mK), by injection from a nearby ferromagnetic or
paramagnetic material or by irradiation with circularly polarized
laser light.  The spin states could also be initialized if a good
measuring device is available.

This concept of a quantum computer has been further studied and worked
out \cite{DiVincenzo99a}; it has also inspired several detailed
related proposals for quantum computing, for example based on
ferro-electrically coupled Si/Ge quantum dots \cite{Levy01a}.

A very different approach \cite{Imamoglu99a} to using electron spins
in quantum dots consists of creating quantum dots in a high finesse
microdisk cavity, so a single optical mode in the cavity can act as a
``bus'' qubit. Near-field laser techniques would enable
qubit-selective one-qubit rotations as well as selective coupling of
the electron spin of a specific quantum dot to the optical cavity
mode.

Finally, other degrees of freedom than spin could serve as quantum bit
levels in quantum dots, such as the spatial coordinate (e.g. an e$^-$
on dot 1 represents $\ket{0}$ and an e$^-$ on dot 2 represents
$\ket{1}$), or excitonic or electronic energy levels. However, the
expectation is that for these degrees of freedom, it may not be
possible to obtain coherence times sufficiently long for meaningful
quantum computation.

\subsubsection{Experiments}

Tarucha's group at the University of Tokyo has fabricated quantum dots
with a small and controlled number (0-20) of free electrons using
vertical structures made of multiple III-V heterostructure quantum
wells
\cite{Tarucha96a}. Coupled quantum dots have also been created using
lateral side gates on top of a GaAs/AlGaAs two-dimensional electron
gas, and molecule-like behavior of such coupled dots has been
exhibited \cite{Livermore96a}. The advantage of this type of quantum
dots over the vertically stacked dots is that it is easier to gate the
dot potentials and the inter-dot coupling. However, smaller structures
($< 0.01 \mu$m$^2$) than those currently available with sidegates must
be constructed in order to obtain sufficiently strong
confinement. Also, charge fluctuations in the electrodes may cause
substantial decoherence, so the vertical dots appear to be
intrinsically more suitable for quantum computing.

The Awschalom group at UC Santa Barbara measured coherence times
(specifically $T_2^*$) of electron spins in GaAs/AlGaAs quantum wells
which approach $1 \mu$s \cite{Kikkawa98a}. Furthermore, they observed
the preservation of spin coherence as electrons were dragged across a
GaAs/ZnSe interface \cite{Malajovich00a}. Coherence time measurements
of the spin of a single excess electron in a quantum dot still need to
be done, and are much needed in order to assess the viability of the
quantum dot electron spin approach to quantum computation.

Spin injection from a paramagnetic semiconductor into GaAs has been
observed to produce nearly $90 \%$ spin polarized electrons in a
magnetic field of 3 T \cite{Fiederling99a}. Spin polarized holes have
been injected from a ferromagnetic semiconductor into a quantum well,
without a magnetic field \cite{Ohno99a}. Finally, circularly polarized
laser light has been used to create spin polarization in the coherence
time measurements \cite{Kikkawa98a}.

Spin-filters (or spin-valves) have been first demonstrated for a
variety of materials in the 70's \cite{Tedrow73a}. Modern
spin-dependent tunneling barriers, with polarizations of up to $85
\%$, are made of metal-EuS-metal junctions \cite{Hao90a}.

Self-assembled InAs quantum dots have been embedded in microdisk
structures (2 $\mu$m in diameter and $0.1 \mu$m thick) with a cavity
quality factor $Q \approx 12000$ \cite{Gerard99a}, but the short
photon lifetime in state-of-the-art cavities forms an important
technological limitation. Furthermore, the need to address each
quantum dot selectively via the tip of an optical fiber and near-field
techniques constrains the density of the quantum dots in the microdisk
to a separation of about 1000 \AA. This in turn limits the number of
quantum dots that can be coupled to a single cavity-mode to a few
dozen.

It is clear that the realization of any of the quantum dot based
proposals requires significant advances in semiconductor
nanofabrication, magnetic semiconductor synthesis and high frequency
measurement techniques. Still, the inherent scalability of the gated
quantum dot proposals combined with the robustness of the spin degree
of freedom make them good long-term candidates for practical quantum
computers.

%%%%%%%%%%%%%%%%%%%%%%%%%%%%%%%%%%%%%%%%%%%%%%%%%%%%%%%%%%%%%%%%%%%%%

\subsection{Superconducting qubits}

\subsubsection{Concept}

The superconducting qubit proposals differ from all the other
proposals discussed here in that the qubit is represented by two
quantized states which are collective states of a ``macroscopic''
number of particles: flux states resulting from the motion of millions
to billions of Cooper pairs through a SQUID or charge states produced
by millions of Cooper pairs in a ``box''.

Josephson junctions play a central role in both approaches
\cite{Averin00a}: if the charging energy $E_C = e^2/2C_J$ (with $C_J$
the capacitance of the Josephson junction) is much larger than the
Josephson coupling energy $E_J$, the device is charge-dominated. In
contrast, if $E_C \ll E_J$, the device is dominated by the
phase across the junction, which is the conjugate variable of the
charge (or equivalently the number) of the excess electrons on one
side of the junction.

{\em Flux qubits} \cite{Mooij99a} are given by two energy levels of
the quantized flux through a superconducting ring with one or more
Josephson junctions in the phase regime (Fig.~\ref{fig:squid}). In a
micrometer sized loop, each flux state arises from the collective
motion of up to $10^9$ Cooper pairs producing a $\mu$A current through
the loop.

\bfig
\vspace*{1ex}
\bcen
\hspace*{1cm} \includegraphics*[width=9cm]{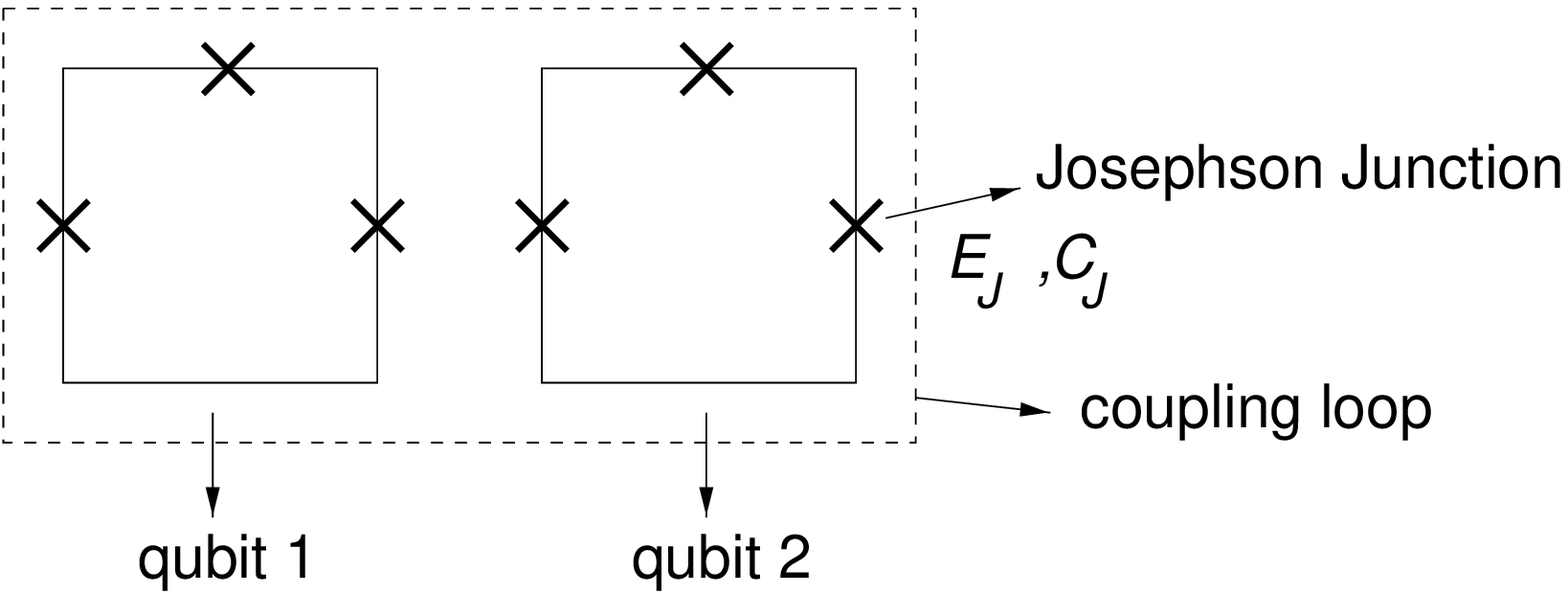}
\ecen
\vspace*{-2ex}
\caption{Schematic of two superconducting flux qubits, each 
embodied by a small superconducting loop interrupted by Josephson
junctions (small barriers made of a resistive material). The devices
shown contain three Josephson junctions, as in
\protect\cite{Vanderwal00a}.}
\label{fig:squid}
\efig

Several schemes have been proposed for logic gates. One-qubit
rotations involve the application of local magnetic fields which
change the environment of a specific qubit. In addition, by changing
the parameters of the Josephson junctions, one-qubit operations can be
performed in different basis.  The interaction mechanism for two-qubit
gates is inductive coupling between neighbouring loops. This coupling
can be enhanced via a separate superconducting coupling loop with
encloses the two qubits (Fig.~\ref{fig:squid}), or alternatively the
two loops can be part of the same superconducting circuit.

Measurement of the flux state of a qubit can be done using a DC
measuring SQUID (not shown in the figure), either enclosing the qubit
or placed next to the qubit. Unfortunately, it is not possible with
current technology to switch the measuring SQUID off by opening the
measuring loop, so the coupling between the measuring SQUID and the
qubit must be very weak in order to prevent excessive decoherence. Of
course, the flip side of very weak coupling is that extensive signal
averaging is required in the measurement.

{\em Charge qubits} \cite{Makhlin99a} rely on the quantized number of
excess electrons on a small superconducting island when it is coupled
to the ground by a number-state dominated Josephson junction
(Fig.~\ref{fig:cooperbox}). As for flux qubits, one-qubit operations
can be realized via local magnetic fields. Two qubit operations can be
done by embedding multiple charge qubits in parallel in a larger
circuit, so the different qubits are coupled. The number of excess
Cooper pairs on the box can be measured via a probe Josephson junction
through which Cooper pairs can tunnel out of the box, depending on the
charge state of the box and on the probe voltage $V_p$
(Fig.~\ref{fig:cooperbox}).

\bfig
\vspace*{1ex}
\bcen
\includegraphics*[width=6cm]{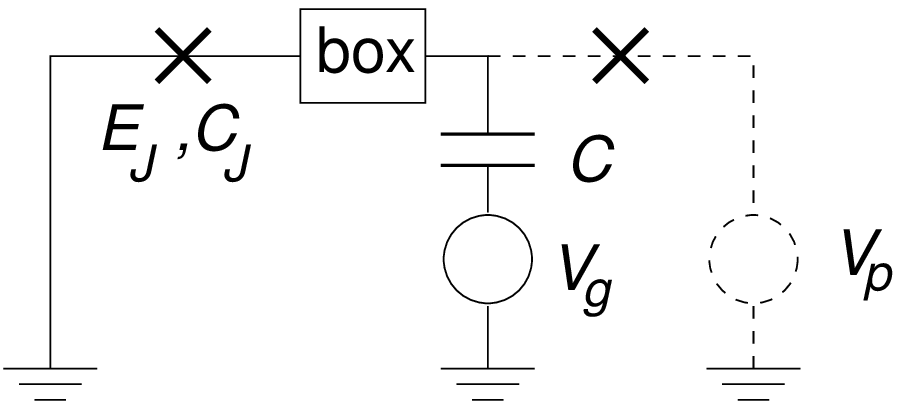}
\ecen
\vspace*{-2ex}
\caption{Schematic of a superconducting charge qubit, realized by a 
small superconducting island or ``box'', coupled to the ground via a
Josephson junction. The electrostatic potential of the island is
controlled by the gate voltage $V_g$. The dashed part of the circuit
serves for readout; it is not part of the actual qubit. In practice,
improved variations of this design are used, which use more Josephson
junctions in order to be able to have more control over the qubit
parameters.}
\label{fig:cooperbox}
\efig

Both the charge qubit and the flux qubit devices must be operated at
low temperature (say 20 mK, which is well below the critical
temperature of Al or other superconducting materials of choice), such
that the ground state is occupied with probability near 1.

\subsubsection{Experiments}

Nakamura and coworkers at NEC in Tsukuba have observed evidence for
coherent superpositions of two charge states in a single-Cooper-pair
box \cite{Nakamura97a}. The energy levels were shifted via a
gate-induced charge $e$ (Fig.~\ref{fig:cooperbox}) such that the
energies of the two lowest levels (and excess charge of $0$ and $2e$)
would become equal (while the next lowest level is at a much higher
energy); however, the Josephson energy splits these two levels and the
resulting eigenstates are coherent superpositions of two collective
states of a macroscopic number of Cooper pairs, which differ in charge
by $2e$. Energy level splitting was experimentally observed, which
suggests that the box was in a coherent superposition of two charge
states \cite{Nakamura97a}.  Later, the same group demonstrated
coherent control of the quantum states in the single-Cooper-pair box,
via pulsed experiments and time domain measurements of multiple Rabi
oscillations
\cite{Nakamura99a}.

The groups of Mooij at Delft and Lukens at SUNY Stony Brook both
observed similar evidence for macroscopic superpositions of flux
states in small SQUID loops \cite{Friedman00a,Vanderwal00a}. Here the
classical energy of two flux states was made equal via an externally
applied static magnetic field. When quantum tunneling between these
two states of equal energy is possible, the loop's eigenstates become
the symmetric and antisymmetric superposition of the original flux
states. The creation of such superposition states was deduced from the
observation of energy level splitting. No time domain measurements
demonstrating coherent control have been performed to date in these
systems.

Even though the state of the art is currently more advanced in charge
qubits than in flux qubits, the expectation is that flux qubits have
longer coherence times than charge qubits, so flux states may be
better qubits than charge states. Measurements of coherence times in
either system would represent significant progress in evaluating
superconducting qubit proposals.

%%%%%%%%%%%%%%%%%%%%%%%%%%%%%%%%%%%%%%%%%%%%%%%%%%%%%%%%%%%%%%%%%%%%%%

\subsection{Solid-state NMR}
\label{impl:solid_nmr}

\subsubsection{Concept}

In a very different solid-state approach, Yamaguchi and Yamamoto
\cite{Yamaguchi99a} propose the use of nuclear spins in a crystal 
lattice as quantum bits (Fig.~\ref{fig:solid_nmr}).  In this proposal,
which was further developed by the same group \cite{Ladd00a}, a
quantum computer consists of a one-dimensional array of spin-1/2
nuclei spaced by a few \AA$\,$ along the $\hat{z}$ axis; the
presence of a magnetic field with a strong gradient along $\hat{z}$
(on the order of 1 T/$\mu$m) separates the Larmor frequency of the
qubits so they can be individually distinguished and addressed.

\bfig
\vspace*{1ex}
\bcen
\includegraphics*[width=7cm]{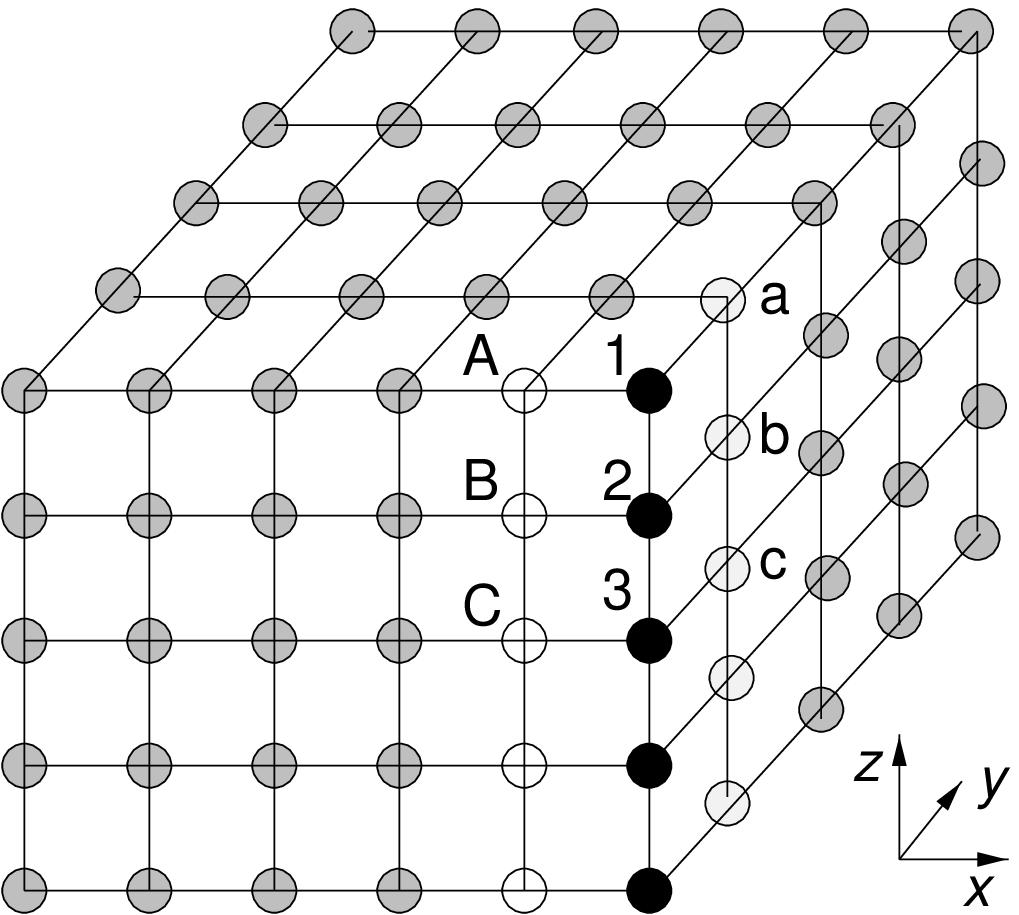}
\ecen
\vspace*{-2ex}
\caption{Model of a crystal lattice quantum computer. The crystal 
shown here has a simple cubic lattice; in practice other lattices have
been proposed, but the idea is the same. The nuclei $1, 2, 3, \ldots$
form one quantum computer, the nuclei $a, b, c, \ldots$ form an
independent computer, and so forth. Nuclei $1$, $a$ and $A$ represent
the analogous qubit in the respective computers.}
\label{fig:solid_nmr}
\efig

In the transverse direction, the magnetic field must be homogeneous
so the crystal contains many identical copies of this one-dimensional
chain of nuclei. All the copies operate as independent quantum
computers provided the interactions between them are switched off. An
important advantage of such a scheme is that the measurement of a
qubit can take place over a large ensemble of nuclear spins instead of
just a single nuclear spin as is required in the Kane proposal
(section~\ref{impl:kane}).

Two qubit gates rely on the magnetic dipole-dipole coupling between
spins in the same computer. With both the magnetic field and the
strong field gradient along the direction of the chain of spins
$\hat{z}$, the coupling Hamiltonian between two spins $i$ and $j$
within the same chain is of the form
\be
{\cal H}_{i,j} \propto \frac{I_z^i I_z^j}{(|j-i|a)^3} \,,
\label{eq:solid_nmr_H1}
\ee
where $a$ is the distance between neighbouring nuclei. This
Hamiltonian is easy to work with and can be selectively suppressed
simply by applying a suitable periodic train of narrowband $180^\circ$
pulses \cite{Leung00a}.

The coupling between spins in different copies has a different
form. For two spins with the same Larmor frequency but located in
different chains $m$ and $n$ (i.e. analogous qubits in different
computers), the coupling Hamiltonian is of the form
\be
{\cal H}_{mn} \propto \frac{1-3\cos^2\theta_{mn}}{(a \lambda_{mn})^3}
\left( 3 I_z^m I_z^n - \vec{I}^m \cdot \vec{I}^n \right) \,,
\ee
where $\lambda_{mn}$ is the distance between the two nuclei in units
of $a$ and $\theta_{mn}$ is the angle between the vector which
connects them and the direction of the applied field $\hat{z}$.  This
coupling can be largely switched off as well, using well-known
solid-state NMR broadband decoupling pulse sequences, such as the
WAHUHA sequence \cite{Mehring83a}.

The WAHUHA pulse sequence also affects the qubit-qubit coupling of
Eq.~\ref{eq:solid_nmr_H1} but this coupling can still be sufficiently
controlled for two-qubit gates using additional narrowband
pulses. However, the coupling between qubits with different resonance
frequencies ($i \neq j$) and in different chains ($m \neq n$) is
partially reintroduced during such two-qubit gates, and causes
decoherence. Fortunately, this effect can be kept quite small by
choosing a very one-dimensional crystal, for example fluorapatite,
Ca$_5$F(PO$)_4$)$_3$, where the $^{19}$F nuclei serve as qubits
\cite{Ladd00a}.

Unlike for liquid NMR, state initialization can in principle be
done by cooling down the sample to the milli-Kelvin regime. In
practice, such low temperature may be difficult to maintain given the
many RF pulses involved in broadband decoupling. Optical pumping or
polarization transfer from electron spins may then be needed.

Readout is much helped by the ensemble nature of the experiment; even
for a relatively small crystal, there are on the order of $10^7$
members in the ensemble. Given the presence of a strong magnetic field
gradient, magnetic resonance force microscopy using microcantilevers
\cite{Rugar92a} has been proposed as a natural way to measure the spin
states.

The main source of decoherence in this system is residual dipolar
couplings. Furthermore, magnetic impurities and cantilever drift (as
the sample would be mounted on the cantilever) also contribute to 
decoherence.

An independent but less detailed proposal for solid-state NMR
quantum computing was presented by Cory {\em et al.}
\cite{Cory00a}. The main difference with the crystal lattice proposal 
is that in the Cory scheme the quantum computer would be an ensemble
of specially designed molecules, held and aligned in a solid state
lattice.

\subsubsection{Experiments}

Solid-state NMR has a tradition of over 50 years, and coherence times
and decoherence mechanisms have been studied since the early days
\cite{Bloembergen49a}. The $T_1$ of nuclear spins in reasonably pure 
crystals is limited by thermal fluctuations of paramagnetic impurities
but can easily be several hours. While $T_2$ is typically only on the
order of milliseconds due to dipolar broadening, it can be lengthened
by several orders of magnitude using well-established broadband
decoupling techniques which have proven their effectiveness.

Growth of high purity crystals of fluorapatite is relatively well
understood. Furthermore, force microscopy with sufficient sensitivity
to detect $10^7$ nuclear spins has been demonstrated \cite{Stipe01a}.

No actual quantum logic gates have been implemented yet, but the
crystal lattice quantum computer has the potential for scaling up to
several hundred qubits, and experiments are underway. The main
challenge lies in the integration and alignment of the different
components, and the design of a micromagnet which produces a strong
magnetic field which is uniform along two axes but has a steep
gradient along the third axis.

The approach based on molecules in solid solutions aims at
intermediate sized quantum computers containing several tens of
qubits. Here also, experiments are underway.

%%%%%%%%%%%%%%%%%%%%%%%%%%%%%%%%%%%%%%%%%%%%%%%%%%%%%%%%%%%%%%%%%%%%%%%%%

\subsection{Dopants in semiconductors}
\label{impl:kane}

Bruce Kane, of the University of Maryland, proposed an approach which
integrates solid-state NMR on donor atoms with semiconductor
electronics \cite{Kane98a}; this scheme has some conceptual
similarities with the quantum dot scheme of
section~\ref{impl:qdots}. The quantum bits are embodied by the
spin-1/2 nucleus of $^{31}$P dopants arranged in a regular array below
the surface of a silicon substrate (Fig.~\ref{fig:kane}), placed in a
static magnetic field perpendicular to the surface.

\bfig
\vspace*{1ex}
\bcen
\includegraphics*[width=8cm]{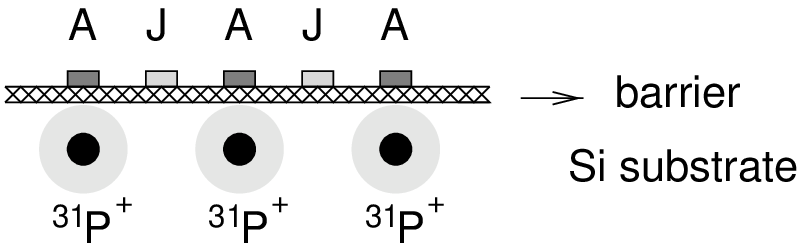}
\ecen
\vspace*{-2ex}
\caption{Model of a Kane-type quantum computer (cross-section), after 
\protect\cite{Kane98a}.}
\label{fig:kane}
\efig

Phosphorus in silicon is an electron donor at room temperature, but at
low temperatures (100 mK), the electron is weakly bound to the
phosporus ion.  A voltage applied to the $A$ gate located about 200
\AA$\,$ above a specific $^{31}$P ion, shifts the electron away from that
nucleus and thereby reduces the hyperfine interaction. As a result,
the energy difference between the spin-up and spin-down state of the
nuclear spin changes and a radio-frequency pulse can then selectively
rotate the state of the one $^{31}$P nuclear spin for which the $A$
gate is biased.

Two-qubit gates between adjacent phosphorus nuclei rely on an indirect
coupling mechanism mediated by their respective electrons. The
coupling Hamiltonian of two donor-electron spin systems is
\be
{\cal H} = A_1 \vec I^{\;1n} \cdot \vec I^{\;1e} 
         + A_2 \vec I^{\;2n} \cdot \vec I^{\;2e} 
         + J   \vec I^{\;1e} \cdot \vec I^{\;2e} 
\ee
where $A_1$ and $A_2$ are the hyperfine interaction energies and $J$
is the exchange energy. In order to obtain reasonable values of $J$
($> 10$ GHz), the donor separation must be no more than 100-200
\AA. A negative voltage applied to the $J$ gate in between two
$^{31}$P nuclei repels the electrons and dimishes their
overlap. Because $J$ is proportional to the electron wavefunction
overlap, the e$^-$-e$^-$ interaction can thus be turned off at will
via the $J$ gates.

Measurement of the $^{31}$P nuclear spin states is done in an
innovative two-step process. First, the $A$ gate of the nuclear
spin we want to measure, say spin $i$, is biased, while the $A$ gate
of a neighbouring nuclear spin, $j$, is not ($A_i > A_j$). Under these
conditions, the $J$ gate between the two dopant atoms $i$ and $j$ is
ramped up such that $J > \mu_BB/2$ and therefore the singlet state
$(\ket{\!\!\uparrow\downarrow} -
\ket{\!\!\downarrow\uparrow})/\sqrt{2}$ becomes the lowest energy state of
the two electron spins (during the computation, $J < \mu_BB/2$ so the
ground state of the two electron spins is
$\ket{\!\downarrow\downarrow}$).  The electron spins will then
adiabatically evolve into the singlet state if spin $i$ is in
$\ket{0}$, whereas they remain in the metastable state
$\ket{\!\!\downarrow\downarrow}$ if spin $i$ is in $\ket{1}$; the
state of spin $j$ is inconsequential. In the second step, the electron
spin state is measured electronically. Both electrons can be bound to
the same donor by biasing the $A$ gates appropriately provided the
electron spins are in $(\ket{\!\!\uparrow\downarrow} -
\ket{\!\!\downarrow\uparrow})/\sqrt{2}$; if the electron spins are in 
$\ket{\!\!\downarrow\downarrow}$, the electrons cannot be bound to the
same donor. Thus, by alternating the voltage applied to the two
respective $A$ gates, charge motion can be induced between the two
donors if and only the electrons are in the singlet state, which in
turns depends on the state of nuclear spin $i$. The charge motion can
be detected via a single-electron transistor.

The electron spins are initialized to the ground state by working at 2 T
and 100 mK. The nuclear spin can be initialized via the measurement
process: measure the nuclear spin and flip it if necessary.

Voltage fluctuations in the control gates, especially the $A$ gates,
are expected to contribute to decoherence, as the Larmor precession
frequency of the nuclear spins is affected by the voltage on the $A$
electrodes. More signficantly, the presence of a single-electron
transistor also induces relaxation, at an estimated rate of 1 kHz.

The Kane proposal inspired a related proposal by Yablonovitch and
coworkers \cite{Vrijen99a}. There are two main differerences with the
Kane scheme: (1) the donor electron spin represents the qubit, instead
of the dopant nuclear spin; (2) there are no $J$ gates; both one-qubit
and two-qubit gates could be realized using $A$ gates, which is made
possible by using silicon-germanium heterostructures, bandgap
engineering and $g$-factor engineering.

\subsubsection{Experiments}

The fabrication technology required for a Kane-type computer is still
beyond the state of the art: it is not currently possible to deposit
individual phosphorus atoms over large areas and with atomic precision
below a silicon surface, nor can we pattern many electrodes of only 50
\AA$\,$ wide and spaced by 200 \AA. Alignment of the gates with
respect to the buried dopant atoms represents an additional
challenge. Finally, further advances in materials technology would be
needed to obtain highly pure $^{28}$Si (the natural abundance of
$^{29}$Si is about $5\%$), and nearly defect-free oxide barriers.

Nevertheless, Kane's proposal is very appealing for its scalability
and elegance, and provides a strong motivation for developing the
necessary technology. In a first step towards the fabrication of a
Kane-type quantum computer, atomic hydrogen has been adsorbed on a
silicon surface, hydrogen desorbed with an STM tip over an area of 1
nm across, and single phosphine molecules have been adsorbed onto the
silicon substrate through the 1 nm holes in the hydrogen layers
\cite{Obrien01a}.

Measurements of $T_1$ have been done long ago \cite{Feher59a}.  The
electron spin $T_1$ at low $^{31}$P concentrations in pure $^{28}$Si
and at 1.5 K has been measured to be thousands of seconds; the
phosphorus spin $T_1$ was over 10 hours. At 100 mK, even longer
$T_1$'s are expected. Finally, Rabi oscillations have been observed
between two low-lying hydrogen-atom like states of an electron weakly
bound to a donor impurity in GaAs \cite{Cole01a}.

The Yablonovitch variant puts less demands on lithography, as only one
gate must be fabricated per qubit, and in addition the spacing between
the qubits can be much larger than in the original proposal (up to 200
nm). On the other hand, there are additional demands on epitaxial
growth techniques due to the need for bandgap and $g$-factor
engineering.

%%%%%%%%%%%%%%%%%%%%%%%%%%%%%%%%%%%%%%%%%%%%%%%%%%%%%%%%%%%%%%%%%%%%%%%%%

\subsection{Other proposals}

Several other proposals exist for the implementation of quantum
computers, in addition to those we have discussed in the preceding
sections. We will just mention two of them.

Platzman and Dykman suggested the use of a quasi-two dimensional set
of electrons floating in vacuum above liquid helium
\cite{Platzman99a}. Individual electrons are laterally confined by
$\mu$m sized electrodes below the helium, and strongly interact
with neighbouring electrons. The qubit states are given by the lowest
hydrogenic levels, at 10 mK. Locally applied electric fields would
produce one-qubit gates and read-out would be done by selectively
releasing excited electrons from the surface, and absorbing them in an
electrometer placed above the surface.

There also exist proposals for all-optical quantum computers
\cite{Milburn89a}. The main obstacle to such devices, the high losses
associated with sufficiently non-linear optical elements, was recently
circumvented by a proposed scheme for efficient quantum computation
with just linear optics \cite{Knill01a} (see page
\pageref{impl:no_twobit_gates}).

%%%%%%%%%%%%%%%%%%%%%%%%%%%%%%%%%%%%%%%%%%%%%%%%%%%%%%%%%%%%%%%%%%%%%%%%%

\section{Summary}

Today's technology enables physicists to separately meet any one of
the five requirements for building a quantum computer exceedingly
well. For example, it is possible already to
\begin{enumerate} 
\item integrate tens of thousands of quantum dots on a single chip 
using semiconductor technology,
\item realize extremely precise two-qubit gates using the 
natural coupling between neighbouring spins,
\item reliably initialize an atom to its internal ground state by 
atomic cooling techniques,
\item perform near-ideal measurements of the internal state of trapped
ions using fluorescence techniques,
\item obtain lifetimes of several days for nuclear spins in solids.
\end{enumerate}
Furthermore, we have seen that many of the traditional requirements
for the physical requirements of quantum computers can be relaxed or
circumvented. 

Nevertheless, satisfying all five criteria within one device still
remains an extraordinary challenge which hasn't been met in any of
these systems. In fact, it has not been possible so far to realize
even the simplest quantum algorithms with any of the proposed
implementations we have discussed.

The crucial difficulty can be summarized as follows: on the one hand,
long coherence times require that the qubits be highly isolated from
the environment; on the other hand, we must have external access to
the qubits in order to initialize, control and read out their
state. The system which best reconciles these opposing requirements
will eventually come out as the ``winning'' quantum computer
realization.

In the next chapter, we describe in detail an experimental system
which we haven't yet discussed, but which, as we shall see, truly
stands out in its accessibility: nuclear spins in molecules disolved
in liquid solution, and manipulated by magnetic resonance techniques.

%% file: nmrqc.tex
\chapter{Liquid-state NMR quantum computing}
\label{ch:nmrqc}

In this chapter, we will examine whether nuclear spins in molecules in
liquid solution satisfy the five requirements for the implementation
of quantum computers. One section will be devoted to each requirement,
except that single-qubit gates and two-qubit gates are treated in
separate sections. Based on this discussion, we will close with
guidelines for molecule design and NMR pulse sequence design.

\section{Qubits}
\label{nmrqc:qubits}

The qubits in NMR quantum computing are given by the spins of suitable
atomic nuclei, placed in a static magnetic field $\vec{B}_0$.

\begin{quote}
{\em We shall here be exclusively interested in spin-1/2 nuclei, such
as $^1$H, $^{13}$C, $^{15}$N, $^{19}$F and $^{31}$P, as they have two
discrete eigenstates.}
\end{quote}

\noindent 
Spin-0 nuclei, for example $^{12}$C and $^{16}$O, are not magnetic and
therefore not detectable with NMR.  Nuclei with spin quantum number
greater than 1/2, such as $^2$H, $^{14}$N, $^{35}$Cl, $^{37}$Cl,
$^{79}$Br and $^{81}$Br, don't make for good qubits either; mapping
the larger number of states (e.g. the spin quantum number of a
spin-3/2 particle can be $-3/2, -1/2, 1/2$ or $3/2$) onto qubit
states, and performing quantum logic gates on them, introduces
additional complications. More significantly, nuclear spins with spin
$> 1/2$ tend to have very short coherence times.

\subsection{Single-spin Hamiltonian}

The Hamiltonian of a spin-1/2 particle in a magnetic field of strength
$B_0$ along the $\hat{z}$ axis
is~\cite{Freeman97a,Ernst87a}~\footnote{Some authors use a different
convention, leaving out the minus sign in Eq.~\ref{eq:1spin_ham}.}
\be
{\cal H}_0 = -\hbar \gamma B_0 \, I_z 
= - \hbar \, \omega_0 \; I_z =
\left[\matrix{-\hbar \omega_0 /2 & 0 \cr 0 & \hbar \omega_0 /2}\right]
\,, \label{eq:1spin_ham}
\ee
where $\gamma$ is the gyromagnetic ratio of the nucleus and $\omega_0
/ 2\pi$ is the Larmor frequency of the spin (we will sometimes leave
the factor of $2\pi$ implicit and call $\omega_0$ the Larmor
frequency). $I_z$ is the angular momentum operator in the $\hat{z}$
direction, which relates to the well-known Pauli matrix as $2 I_z =
\sigma_z$; similarly, we will later use $2 I_x = \sigma_x$ and $2 I_y
= \sigma_y$.

The interpretation of Eq.~\ref{eq:1spin_ham} is that the energy of the
$\ket{0}$ or $\ket{\!\!\uparrow}$ state (given by $\bra{0}{\cal
H}\ket{0}$, the upper left element of ${\cal H}$) is lower than the
energy of $\ket{1}$ or $\ket{\!\!\downarrow}$ ($\bra{1}{\cal H}\ket{1}$)
by an amount $\hbar \omega_0$, as illustrated in the energy diagram of
Fig.~\ref{fig:energy_1spin}.
The energy splitting is known as the {\em Zeeman splitting}.

\bfig
\bcen
\vspace*{1ex}
\includegraphics*[width=2cm]{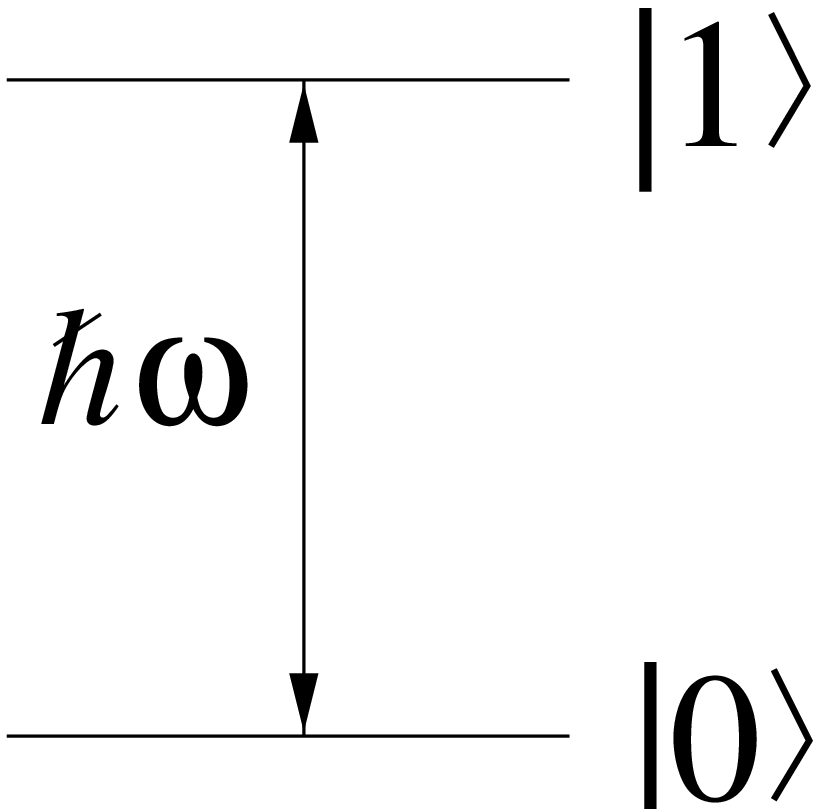} 
\vspace*{-2ex}
\ecen
\caption{Energy diagram for a single spin-1/2.}
\label{fig:energy_1spin}
\efig

The time evolution $e^{-i {\cal H} t/\hbar}$ of the spin state under
the Hamiltonian of Eq.~\ref{eq:1spin_ham} corresponds to a precession
motion in the Bloch sphere (Fig.~\ref{fig:bloch_sphere}) about the
axis of the static magnetic field, similar to the precession of a
spinning top about the axis of gravitation, as shown in
Fig.~\ref{fig:precession}. The $B_0$ field is typically on the order
of 10 Tesla, resulting in precession frequencies $\omega_0$ of a few
hundred MHz, which is in the radio-frequency range .

\bfig
\bcen
\vspace*{1ex}
\includegraphics*[width=2cm]{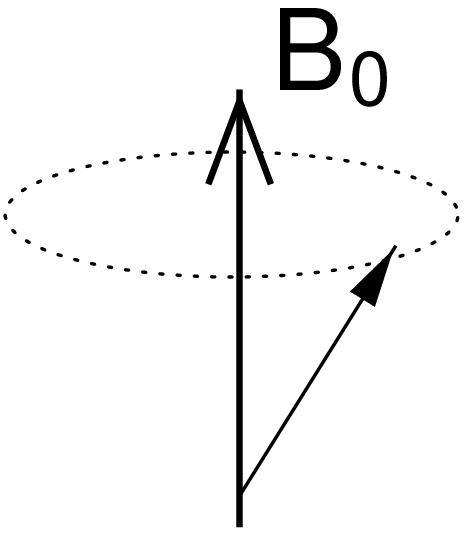} 
\vspace*{-2ex}
\ecen
\caption{Precession of a spin-1/2 about the axis of a static 
magnetic field.}
\label{fig:precession}
\efig

\subsubsection{Distinguishing the nuclear spins in a molecule}

A molecule with $n$ distinguishable spin-1/2 nuclei constitutes an
$n$-qubit quantum computer.  Spins of different nuclear species ({\em
heteronuclear} spins) can be easily distinguished spectrally, as they
generally have very distinct values of $\gamma$ and thus also very
different Larmor frequencies (Table~\ref{tab:larmor_freq}).
Furthermore, spins of the same nuclear species ({\em homonuclear}
spins) which are part of the same molecule can also have distinct
frequencies, due to chemical shifts $\sigma^i$: the electron clouds
slightly shield the nuclei from the externally applied magnetic field
so a different electronic environment leads to a different degree of
shielding and hence different Larmor frequencies
(Fig.~\ref{fig:chem_shifts}).

\begin{table}[h]
\vspace*{1ex}
\begin{center}
\begin{tabular}{c|ccccccc}
nucleus    & $^1$H & $^2$H & $^{13}$C & $^{15}$N & $^{19}$F & $^{31}$P\\\hline
$\omega_0$ & 500   & 77    & 126      & -51      & 470      & 202 
\end{tabular}
\end{center}
\vspace*{-1ex}
\caption{Larmor frequencies [Mhz] of some relevant nuclei, at
11.74 Tesla.}
\label{tab:larmor_freq}
\end{table}

The nuclear spin Hamiltonian for a molecule with $n$ nuclei with
different chemical shifts is thus
\be
{\cal H}_0 = - \sum_{i=1}^n \hbar \, (1 - \sigma_i) \gamma B_0 \; I_z^i
= - \sum_{i=1}^n \hbar \, \omega_0^i \;I_z^i \,.
\label{eq:ham_0_n}
\ee
The range of typical chemical shifts $\sigma_i$ varies from nucleus to
nucleus: it is $\approx 10$ parts per million (ppm) for $^1$H,
$\approx 200$ ppm for $^{19}$F and $\approx 200$ ppm for $^{13}$C. For
a $B_0$ field of about 10 Tesla ($\omega_0$'s of several hundred MHz),
this corresponds to a few kHz to tens of kHz.  Pronounced
asymmetries in the molecular structure and strong differences in the
electronegativity of the atoms in the molecule promote strong chemical
shifts.

\bfig
\bcen
\vspace*{1ex}
\includegraphics*[width=11cm]{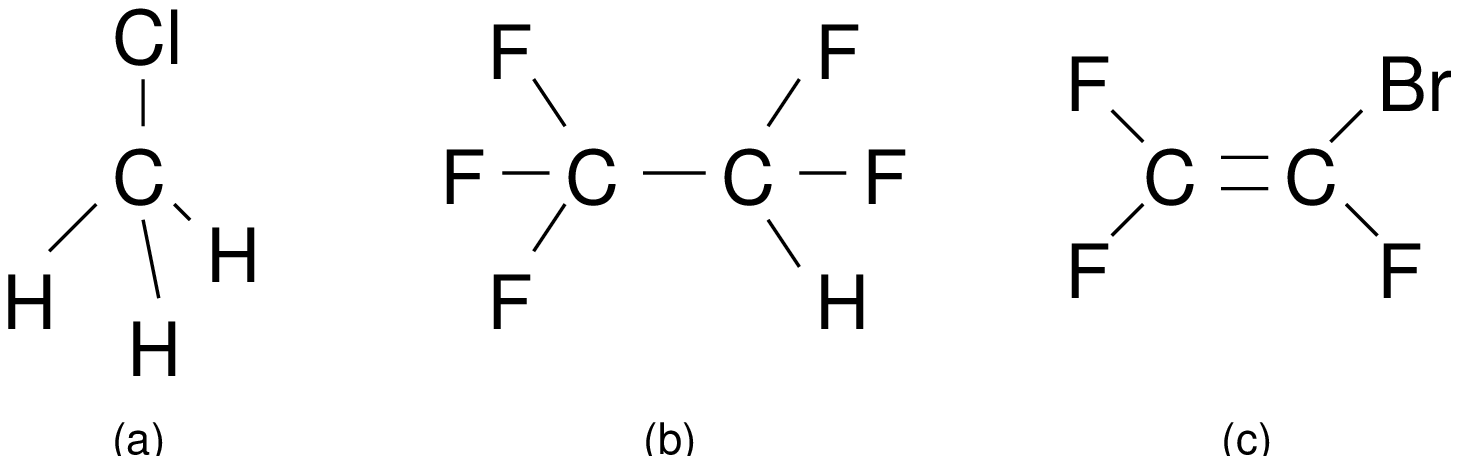} 
\vspace*{-2ex}
\ecen
\caption{(a) The three H atoms in this tetrahedral molecule are in 
equivalent locations with respect to the C and Cl atoms; their Larmor
frequencies are thus identical. (b) The two F nuclei on the right have
a different chemical shift from the three F nuclei on the left,
because the H atom makes both sides inequivalent. However, the three F
nuclei on the left hand side of the molecule are chemically equivalent
to each other, because both ends of the molecule rapidly rotate with
respect to each other around the single C-C bond. (c) The double C=C
bond is rigid, so the left and right side cannot rotate with respect
to each other. All three F nuclei have different chemical shifts.}
\label{fig:chem_shifts}
\efig

We shall first describe the nature
of the interactions between nuclear spins, and then discuss the
operation of two-qubit gates in NMR.

\subsection{Spin-spin interaction Hamiltonian}
\label{nmrqc:spin-spin_interaction}

For nuclear spins in molecules, nature provides two distinct
interaction mechanisms \cite{Abragam61a,Slichter96a}.  The first is a
{\em magnetic dipole-dipole} interaction, similar to the interaction
between two bar magnets in each other's vicinity. It takes place purely
{\em through space} --- no medium is required for this interaction
and it is inversely proportional in strength to the distance between the
two nuclei and depends on the relative position of the nuclei with
respect to the magnetic field. Both intramolecular dipolar couplings
(between spins in the same molecule) and intermolecular dipolar
couplings (between spins in different molecules) are present. However,
when the molecules are disolved in an isotropic liquid, all dipolar
couplings are averaged away due to rapid tumbling.

The second mechanism is known as the {\em $J$-coupling} or {\em scalar
coupling}. This interaction is mediated by the electrons shared in the
chemical bonds between atoms in a molecule. The {\em through-bond}
coupling strength $J$ depends on the element and isotope of the
respective nuclei and decreases with the number of chemical bonds
separating the nuclei. The Hamiltonian is
\be
{\cal H}_J = \hbar \sum_{i<j} 2 \pi J_{ij} I^i \cdot I^j
= \hbar \sum_{i<j} 2 \pi J_{ij} 
(I_x^i I_x^j + I_y^i I_y^j + I_z^i I_z^j) \,,
\label{eq:ham_bond}
\ee
where $J_{ij}$ is the coupling between spins $i$ and $j$.  If the
spectra are first-order, i.e. $|\omega_i - \omega_j| \gg 2\pi |J|$,
Eq.~\ref{eq:ham_bond} simplifies to~\cite{Freeman97a,Ernst87a}
\be
{\cal H}_J = \hbar \sum_{i<j}^n 2 \pi J_{ij} I_z^i I_z^j \,,
\label{eq:ham_J}
\ee
which was the case for all the molecules we selected for our
experiments. The complete Hamiltonian of a closed system of $n$
nuclear spins in isotropic solution and with first order spectra is
then (from Eqs.~\ref{eq:ham_0_n} and~\ref{eq:ham_J})
\be
{\cal H} = - \sum_{i=1} \hbar \, \omega_0^i \;I_z^i  
+ \hbar \sum_{i<j} 2 \pi J_{ij} I_z^i I_z^j \,,
\label{eq:ham_iso}\ee

The interpretation of the scalar coupling term is that a spin
``feels'' a static magnetic field along $ \pm \hat{z}$ produced by
neighbouring spins, in addition to the externally applied $\vec{B}_0$
field. This additional field shifts the energy levels as in
Fig.~\ref{fig:energy_2spins} and the Larmor frequency of spin $i$
shifts by $-J_{ij}/2$ if spin $j$ is in $\ket{0}$ and by $+J_{ij}/2$
if spin $j$ is in $\ket{1}$.

\bfig
\bcen
\vspace*{1ex}
\includegraphics*[width=6cm]{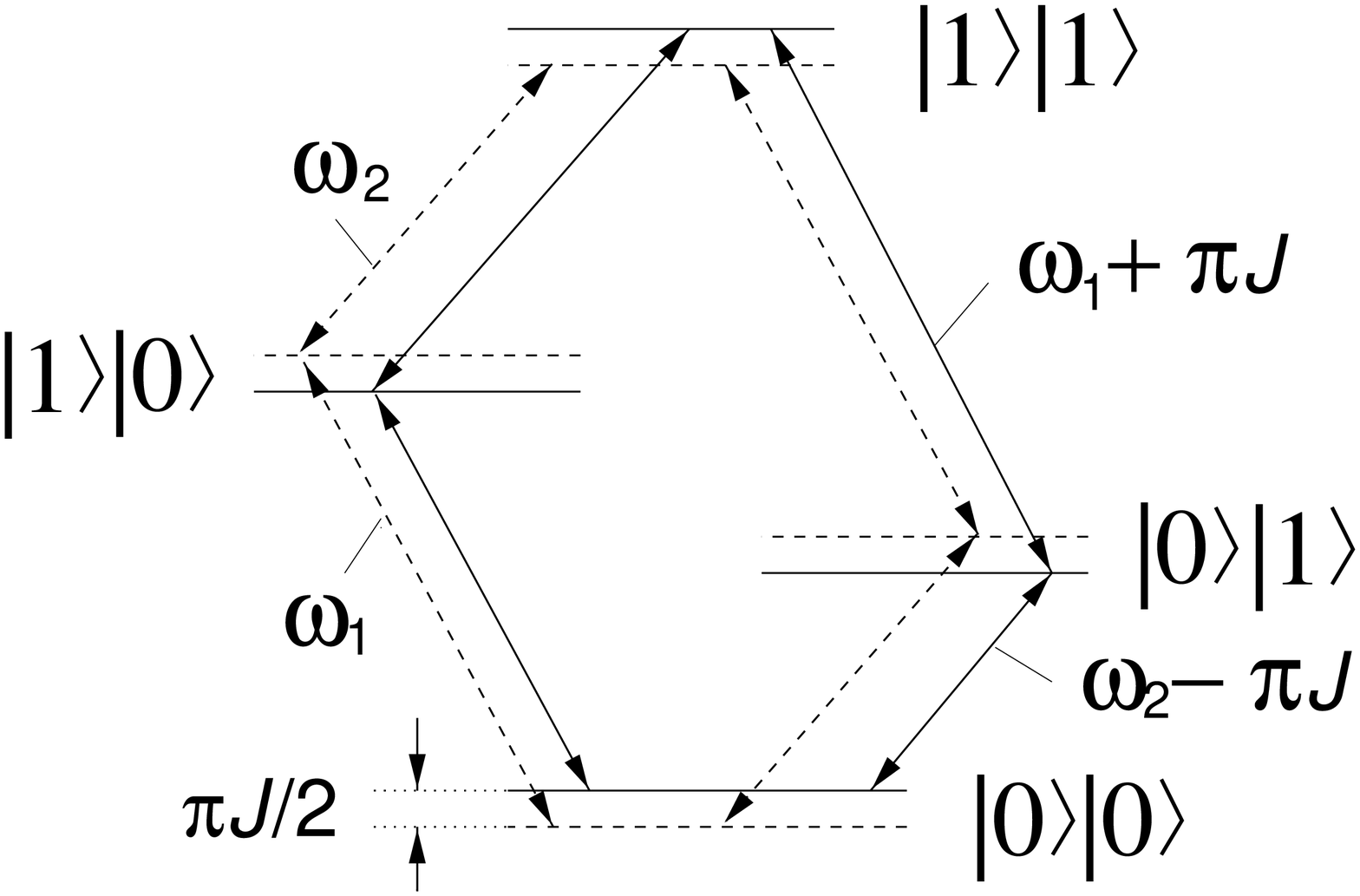} 
\vspace*{-2ex}
\ecen
\caption{Energy level diagram for two $J$-coupled spins in 
isotropic solution (in units of $\hbar$).}
\label{fig:energy_2spins}
\efig

In a system of two coupled spins, the spectrum
(section~\ref{nmrqc:meas}) of each spin therefore actually consists of
two lines, each of which can be associated with the state of the other
spin, $\ket{0}$ or $\ket{1}$.  For three pairwise coupled spins, the
spectrum of each spin contains four lines. For every spin we add, the
number of lines per multiplet doubles, provided all the couplings are
resolved and different lines do not lie on top of each other. This is
illustrated for a five spin system in Fig.~\ref{fig:5spin_multiplet}.

The magnitude of all the pairwise couplings can be found by looking
for common splittings in the multiplets of different spins. Typical
values for $J$ are up to a few hundred Hertz for one-bond couplings
and down to only a few Hertz for three- or four-bond couplings. The
signs of the $J$ couplings can be determined via two-dimensional
correlation experiments with spin-selective pulses
(soft-COSY)~\cite{Bruschweiler87a} or related selective decoupling
experiments; they cannot be derived just from a single spectrum.

Finally, we note that $J$ couplings with spins $> 1/2$ average to
zero, because such nuclei have a quadrupole moment which interacts
with electric field fluctuations and causes the nucleus to rapidly
oscillate between the spin-up and spin-down states. We also point out
that the coupling between magnetically equivalent nuclei is not
observable for symmetry reasons.

%%%%%%%%%%%%%%%%%%%%%%%%%%%%%%%%%%%%%%%%%%%%%%%%%%%%%%%%%%%%%%%%%%%%%%%
%%%%%%%%%%%%%%%%%%%%%%%%%%%%%%%%%%%%%%%%%%%%%%%%%%%%%%%%%%%%%%%%%%%%%%%

\section{Single-qubit operations}
\label{nmrqc:1bitgates}

\subsection{Rotations about an axis in the $\hat{x}\hat{y}$ plane (RF pulses)}
\label{nmrqc:rf_pulses}

We can manipulate the state of a spin-1/2 particle by applying an
electromagnetic field of strength $B_1$ which rotates in the
transverse plane at $\omega_{r\!f}$, at or near the spin precession
frequency $\omega_0$.  The Hamiltonian of the RF field
is~\cite{Freeman97a,Ernst87a}
\be
{\cal H}_{r\!f} = - \hbar \omega_1 
\left[ \cos (\omega_{r\!f} t + \phi) I_x +
\sin (\omega_{r\!f} t + \phi) I_y \right] \,,
\label{eq:ham_rf_lab}
\ee
where $\omega_1 = \gamma B_1$.

In practice, we apply a transverse RF magnetic field which {\em
oscillates} at $\omega_{r\!f}$ along a fixed axis in the lab frame,
rather than {\em rotates}. The oscillating field can be decomposed
into two counter-rotating fields, one of which rotates at
$\omega_{r\!f}$ in the same direction as the spin. We call this
component the $B_1$ field. The other component goes in the opposite
direction and has a negligible effect on the spin
dynamics.\footnote{The presence of this second component produces a
tiny shift in the Larmor frequency, called the Bloch-Siegert
shift\label{page:counterrot_rf}~\cite{Bloch40a}.}

\subsubsection{Nutation under an RF field}

The motion of a nuclear spin subject to both a static and a rotating
magnetic field is rather complex when described in the usual
laboratory coordinate system (the {\em lab frame}). It is much
simplified, however, by describing the motion in a coordinate system
rotating about $\hat{z}$ at or near the spin precession frequency
$\omega_0$ (the {\em rotating frame}).

Suppose we apply the $B_1$ field exactly on resonance with $\omega_0$.
In a frame rotating at $\omega_{r\!f}=\omega_0$, the $B_1$ field then
appears to lie along a fixed axis in the transverse plane, so the RF
field Hamiltonian in the rotating frame becomes
\be
{\cal H}_{r\!f}^{rot} = - \hbar \omega_1 
\left[ \cos (\phi) I_x + \sin (\phi) I_y \right]
\label{eq:ham_rf_rot}
\ee
An observer in this rotating frame will thus see the spin simply
precess about the axis of the $B_1$ field (Fig.~\ref{fig:nutation} a);
this motion is called the {\em nutation}.  The rotation axis is
controlled by the phase of the RF field $\phi$.  An observer in the
lab frame sees the spin spiral down over the surface of the Bloch
sphere, the combined result of precession and nutation
(Fig.~\ref{fig:nutation} b).  In typical NMR experiments, the static
field is much stronger than the RF field, so the precession about
$\hat{z}$ is much faster than the nutation (hundreds of MHz versus
tens of kHz).

\bfig
\bcen
\vspace*{1ex}
\includegraphics*[width=10cm]{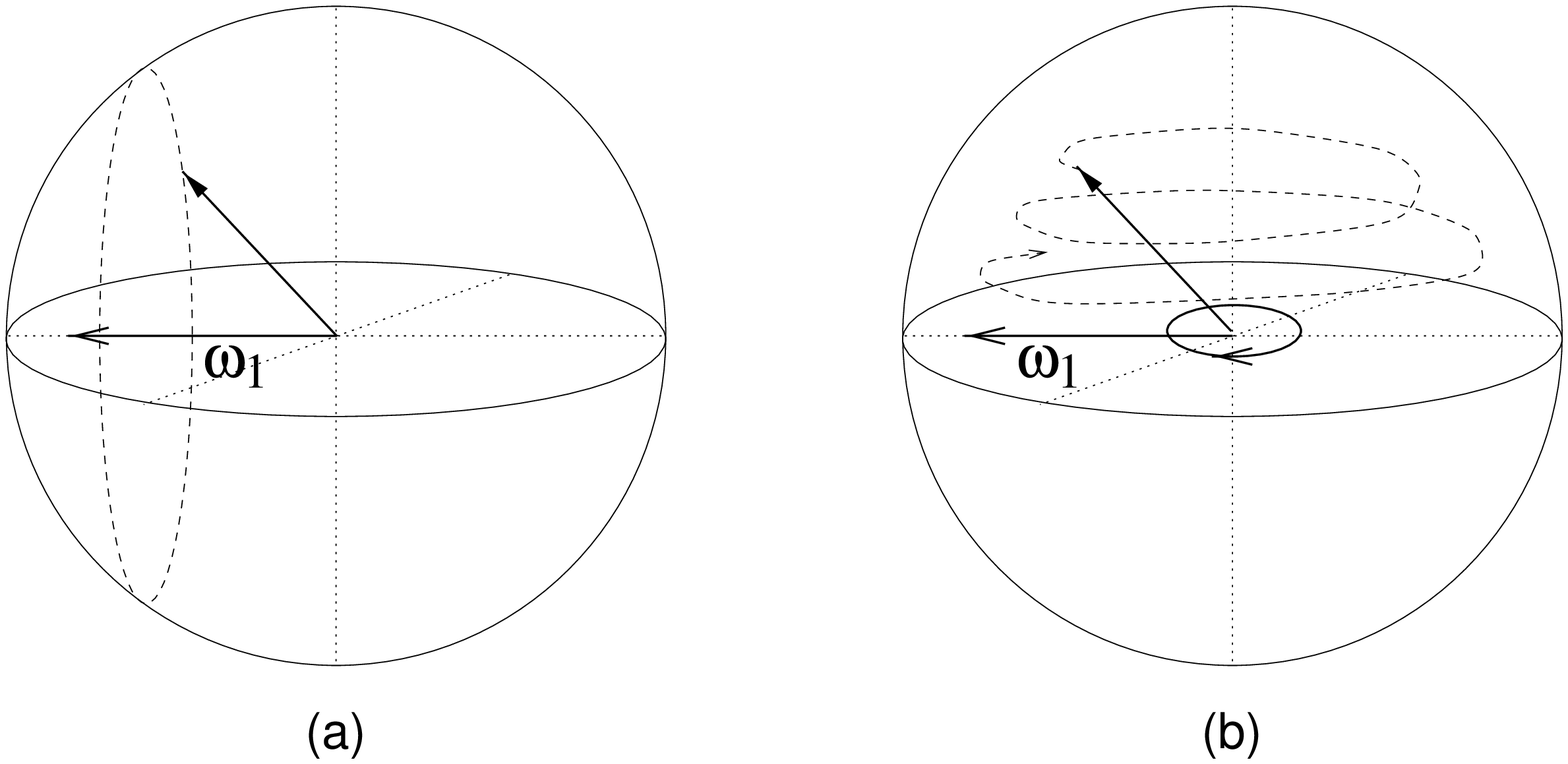} 
\vspace*{-2ex}
\ecen
\caption{Nutation of a spin subject to a transverse RF field (a) 
observed in the rotating frame and (b) observed in the lab frame.}
\label{fig:nutation}
\efig

If the RF field is {\em off-resonance} with respect to the spin
frequency by $\Delta \omega = \omega_0 - \omega_{r\!f}$, the RF
Hamiltonian in the frame rotating at $\omega_{r\!f} (\neq \omega_0)$
becomes
\be
{\cal H}_{r\!f}^{rot} = 
- \hbar \Delta \omega I_z -
\hbar \omega_1 \left[ \cos (\phi) I_x + \sin (\phi) I_y \right]
\label{eq:ham_rf_rot_offres}
\ee
In words, the spin now precesses with frequency
\be
\omega_1' = \sqrt{\Delta\omega^2 + \omega_1^2}
\label{eq:freq_offres_rf}
\ee
about an axis tilted away from the $\hat{z}$ axis by an angle
\be 
\alpha = \mbox{arctan} (\omega_1 / \Delta \omega) \,,
\label{eq:axis_offres_rf}
\ee
as illustrated in Fig.~\ref{fig:offres_rf}. An off-resonant pulse thus
results in a rotation about a different axis and over a different
angle than the same pulse applied on resonance. Off-resonance pulses
can thus be used to effect a rotation about an axis outside the
$\hat{x}\hat{y}$ plane.

\bfig
\bcen
\vspace*{1ex}
\includegraphics*[width=4.5cm]{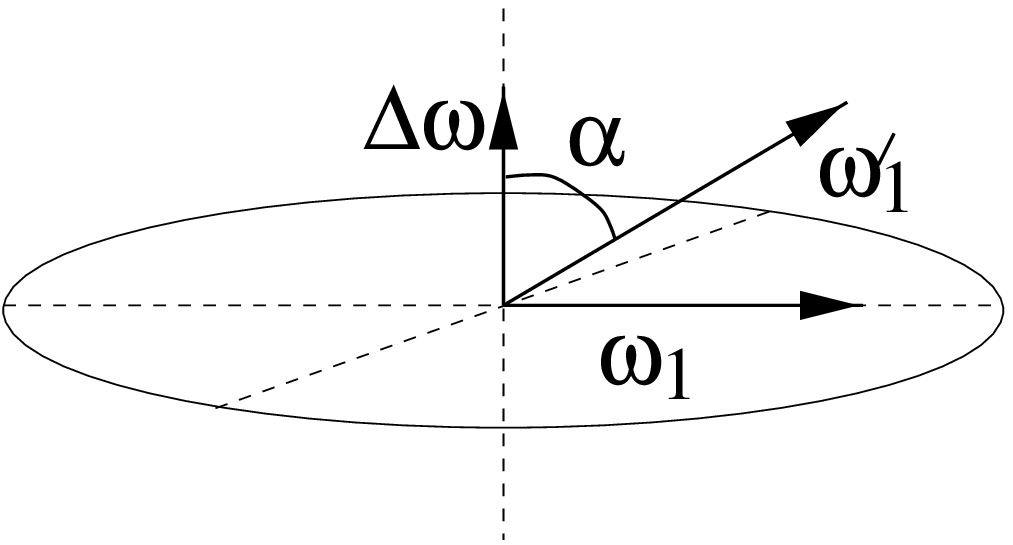} 
\vspace*{-2ex}
\ecen
\caption{Axis of rotation (in the rotating frame) during an 
off-resonant radio-frequency pulse.}
\label{fig:offres_rf}
\efig

\subsubsection{RF pulses}

A resonant RF field gated on for a duration $pw$, nutates a spin in the
rotating frame over an angle
\be
\theta = \gamma \, B_1 \; pw \,.
\ee
The parameter $pw$ is called the {\em pulse width} or {\em pulse
length}. 

Thus, a properly timed and calibrated RF pulse with the right phase
can perform a rotation about $\hat{x}$ of 90$^\circ$, which we will
denote $R_x(90)$ (see Eq.~\ref{eq:R_x}) or for short $X$. A similar
pulse but twice as long realizes a $R_x(180)$ rotation, written for
short as $X^2$. By changing the phase of the RF by 90$^\circ$, we can
similarly implement $Y$ and $Y^2$ pulses.  Changing the phase another
$90^\circ$ gives a negative rotation about $\hat{x}$, denoted
$R_x(-90)$ or $\bar{X}$, and so forth.

We point out that only the relative phase between pulses applied to
the same spin matters.  As soon as we send one pulse on any given
spin, the phase of that pulse sets the phase reference of the
corresponding rotating frame for the remainder of the pulse sequence.

\subsubsection{Quantum picture}

The description of single-spin rotations has been purely classical so
far, and in fact it does not need quantum mechanics at all. For
example, a bar magnet with angular momentum responds in exactly the
same way to magnetic fields as does a nuclear spin.  However, the
quantum nature of spins, and qubits in general, unmistakably emerges
as soon as two or more spins are involved (section~\ref{qct:qubits}).

Underlying the classical Bloch sphere picture is the evolution of a
two-level quantum mechanical system. An RF field induces transitions
between the ground and excited state of the qubit
(Fig.~\ref{fig:energy_1spin}). After applying an RF pulse, a spin
initially in the ground state will upon measurement be found in the
excited state with probability $\sin^2 (\omega_1 p\;\!\!w/2)$. The
projection on the $\hat{z}$ axis of the Bloch sphere oscillates with
$pw$ as $\cos(\omega_1 \; p\!w)$. These oscillations are known as
{\em Rabi oscillations}, and $\omega_1/2\pi$ is the {\em Rabi
frequency}, with typical values of a few hundred Hz to a few hundred
kHz.

%%%%%%%%%%%%%%%%%%%%%%%%%%%%%%%%%%%%%%%%%%%%%%%%%%%%%%%%%%%%%%%%%%%%%

\subsection{Rotations about the $\hat{z}$ axis}
\label{nmrqc:z_rotations}

We recall from Eq.~\ref{eq:universal_1bitgate} that the ability to
implement arbitrary rotations about $\hat{x}$ and $\hat{y}$ is
sufficient for performing arbitrary single-qubit rotations. For
example, two ways to implement a $Z$ rotation using {\em
composite} $X$ and $Y$ pulses are
\be
Z = X Y \bar{X} = Y \bar{X} \bar{Y} \,,
\label{eq:composite_z}
\ee
where {\em time goes from right to left}, as always for concatenated
unitary operations (see section~\ref{sec:remarks_unitaries}). We have
used this technique in our first few experiments
(sections~\ref{expt:dj}-\ref{expt:labeling}). However, there are two
alternative and more convenient ways to implement $Z$ rotations.

The first approach takes place at the pulse sequence design level. The
goal is to move down all the $Z$ rotations to the very end or the
beginning of the pulse sequence. For example, using
Eq.~\ref{eq:composite_z}, we can move a $Z$ rotation past a $X$ or $Y$
rotation,
\be
Z \bar{Y} = X Y \bar{X} \bar{Y} = X Z \,.
\ee
Since $Z$ rotations commute with the Hamiltonian of nuclear spins in
liquid solution, they can be moved across time evolution intervals as
well. Once all $Z$ rotations are gathered at the end of the pulse
sequence, we only need to execute the net remaining $Z$ rotation for
each spin.  $Z$ rotations moved to the start of the sequence have no
effect as the initial state is diagonal (see
section~\ref{nmrqc:init}), so they don't need to be implemented
altogether. This approach was used in the experiment of
section~\ref{expt:grover3}.

The second approach has the same effect, but the experimental
procedure is different. It makes use of an artificial software
rotating frame, on top of the hardware rotating frame provided by a
reference oscillator. A $Z$ rotation is implemented simply by shifting
the software reference frame by 90$^\circ$. Subsequent $X$ and $Y$
pulses are then executed with respect to the new reference frame
(e.g. $X$ in the new frame corresonds to $Y$ in the old frame, and so
forth), and the receiver phase is also set with respect to the new
software frame.  We have used this procedure in our latest experiments
(sections~\ref{expt:cooling}-~\ref{expt:shor}), as it is by far the
easiest to use once the software for the artificial reference frame is
written. Since the $Z$ rotations are now done entirely in the software
and do not require any physical pulses anymore, they are in a sense
``for free'' and perfectly executed. It is in this case advantageous
to convert as many $X$ and $Y$ rotations as possible into $Z$
rotations, using identities similar to Eq.~\ref{eq:composite_z}, for
example
\be
X Y = X Y \bar{X} X = Z X \,.
\ee
We will come back to pulse sequence simplification in
section~\ref{nmrqc:seq_design}.

%%%%%%%%%%%%%%%%%%%%%%%%%%%%%%%%%%%%%%%%%%%%%%%%%%%%%%%%%%%%%%%%%%%%%

\subsection{Selective excitation using pulse shaping}
\label{nmrqc:pulse_shaping}

We can selectively address one spin without exciting any other spins
in the molecule by sending a sufficiently long RF pulse at the
resonance frequency of the desired spin. The frequency selectivity of
RF pulses can be much improved by using so-called {\em soft pulses} or
{\em shaped pulses}, which are designed to excite or invert spins over
a limited frequency region, while minimizing $\hat{x}$ and $\hat{y}$
rotations for spins outside this
region~\cite{Freeman98a,Freeman97a}. Soft pulses start off at low
amplitude $B_1$ (and thus also $\omega_1$), gradually build up to a
maximum amplitude, and taper off again towards the end. Pulse shaping
is usually done by dividing the pulse in a few tens to many hundreds
of discrete time slices, and by changing the amplitude and/or
phase~\footnote{In common pulse shapes, the phase is usually just
$0^\circ$ or $180^\circ$. During phase ramping, the phase is
incremented linearly throughout the pulse profile, as discussed on
page~\pageref{page:phase_ramping}.} slice by slice to create a
tailored amplitude and phase profile.

Fourier theory can give us a rough idea of the frequency response of a
spin to an RF pulse. For example, it tells us that the power of a
pulse of length $pw$ will be confined to a frequency window of roughly
$1/pw$. However, the Fourier transform is a linear transformation
whereas the spin response to an RF field is not linear (it is
sinusoidal); it must thus be calculated with different methods.

For a rectangular (constant amplitude) pulse, the spin response as a
function of $\Delta \omega = \omega_{r\!f} - \omega_0$ is easy to
calculate analytically from Eqs.~\ref{eq:freq_offres_rf}
and~\ref{eq:axis_offres_rf}, or numerically by computing the unitary
operator $e^{-i {\cal H} t/\hbar}$ generated by the Hamiltonian of
Eq.~\ref{eq:ham_rf_rot_offres} (see section~\ref{qct:dyn&rev}). The
response to a shaped pulse is most easily computed by concatenating
the unitary operators of each time slice of the shaped pulse, as the
Hamiltonian is time-independent within each time slice.
Fig.~\ref{fig:profile_pulses} shows the time profile and the
excitation profile for four standard pulse shapes.

\subsubsection{Pulse shape design}

The properties relevant for choosing a pulse shape are:
\begin{itemize}
\item selectivity: product of excitation bandwidth and pulse length 
(lower is more selective),
\item transition range: the width of the transition region between 
the selected and unselected frequency region,
\item power: the peak power required for a given pulse length (low is 
less demanding),
\item self-refocusing behavior (see 
section~\ref{nmrqc:pulse_artefacts}): degree to which the $J$
couplings between the selected spin and other spins are refocused (the
signature for self-refocusing behavior is a flat top in the excitation
profile),
\item robustness: whether the spin response is very sensitive to 
experimental imperfections such as RF field inhomogeneities and
calibration errors,
\item universality: whether the pulse performs the correct rotation 
for arbitrary input states or only for specific input states.
\end{itemize}
Figure~\ref{fig:profile_pulses} strikingly illustrates the difference
in performance between different pulse shapes.
Table~\ref{tab:shaped_pulses} summarizes these properties for a
selection of important pulse shapes. All the pulses in the table are
universal pulses; quantum computations must work for any input state
so we cannot compromise on universality.\footnote{Strictly speaking,
one could use non-universal pulses in the early stages of certain
algorithms, or during the state preparation sequences, but we have
never done this.}  Obviously, no single pulse shape optimizes for all
properties simultaneously, so pulse shape design consists of finding
the optimal trade-off for the desired application. For our
experiments, we have selected molecules with large chemical
shifts, so sharp transition regions are not so important. Furthermore,
the probe and spectrometer can deal with relatively high powers. The
crucial parameters are the effect of coupling during the pulses, the
selectivity (short, selective pulses minimize relaxation) and to some
extent the robustness.

\bfig
\vspace*{-0.7cm}
\begin{center}
%\begin{tabular}{cc}
\rotatebox{90}{\hspace*{8ex} Rectangular} 
\includegraphics*[width=6cm]{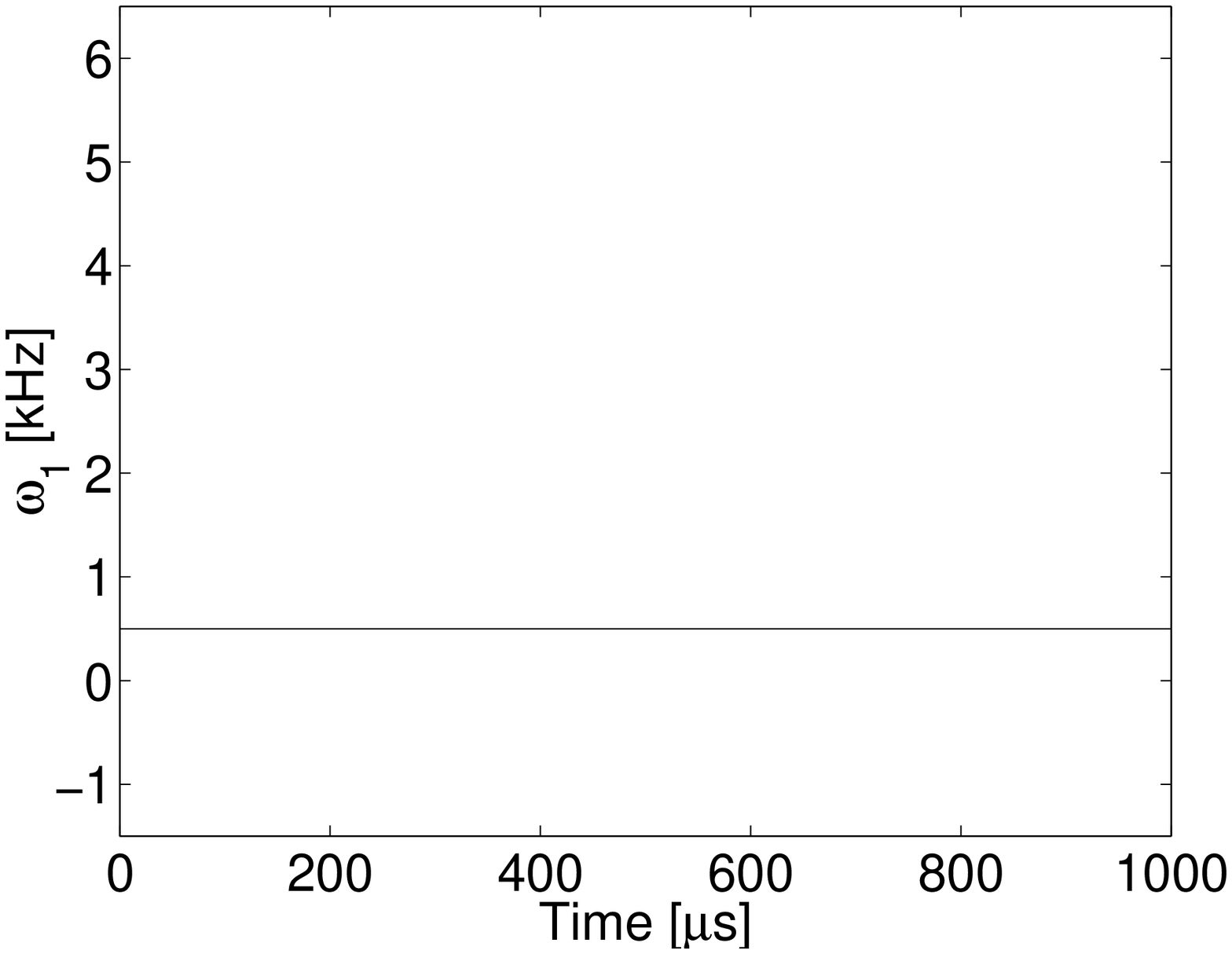} %&
\includegraphics*[width=6cm]{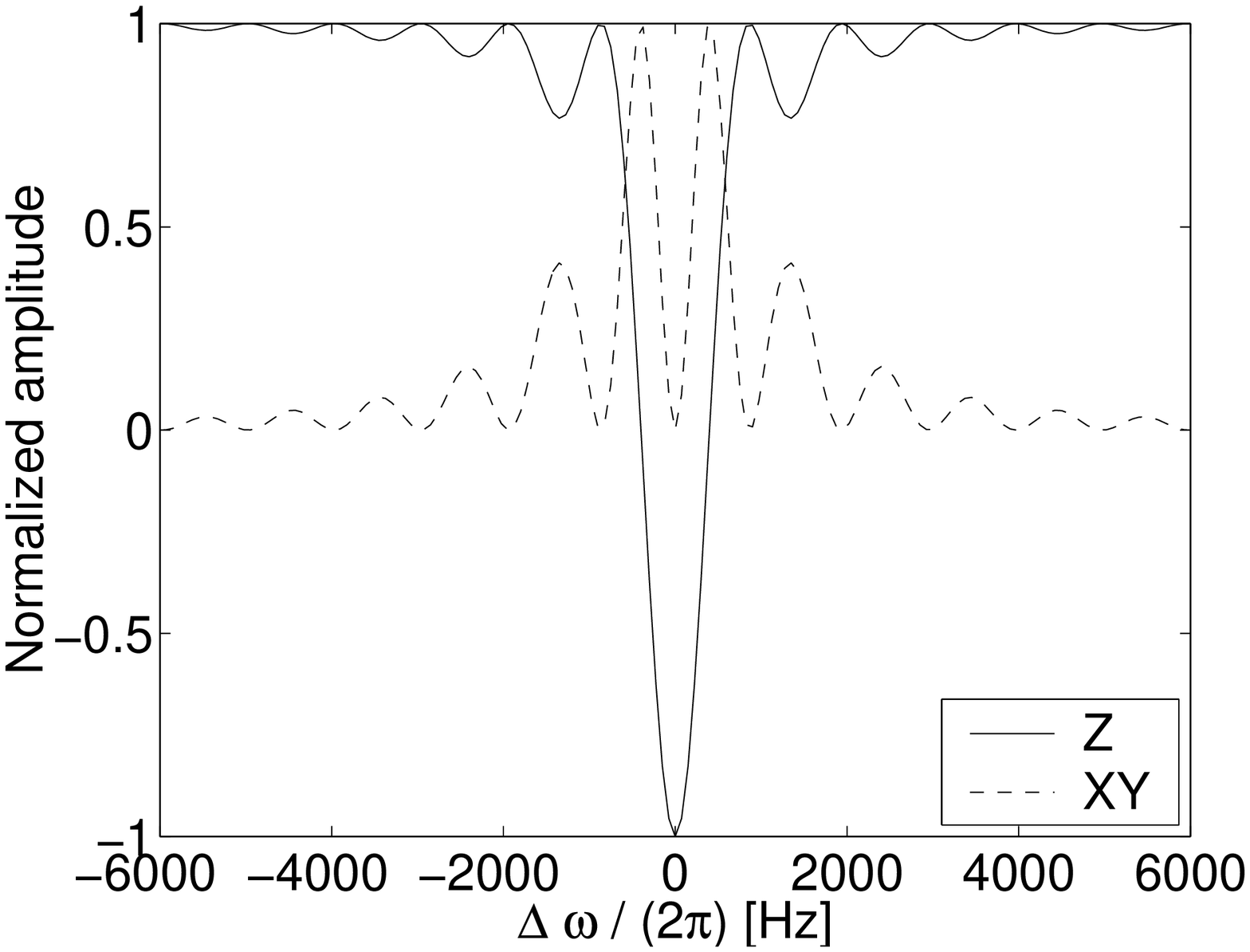} \\
\rotatebox{90}{\hspace*{8ex} Gaussian} 
\includegraphics*[width=6cm]{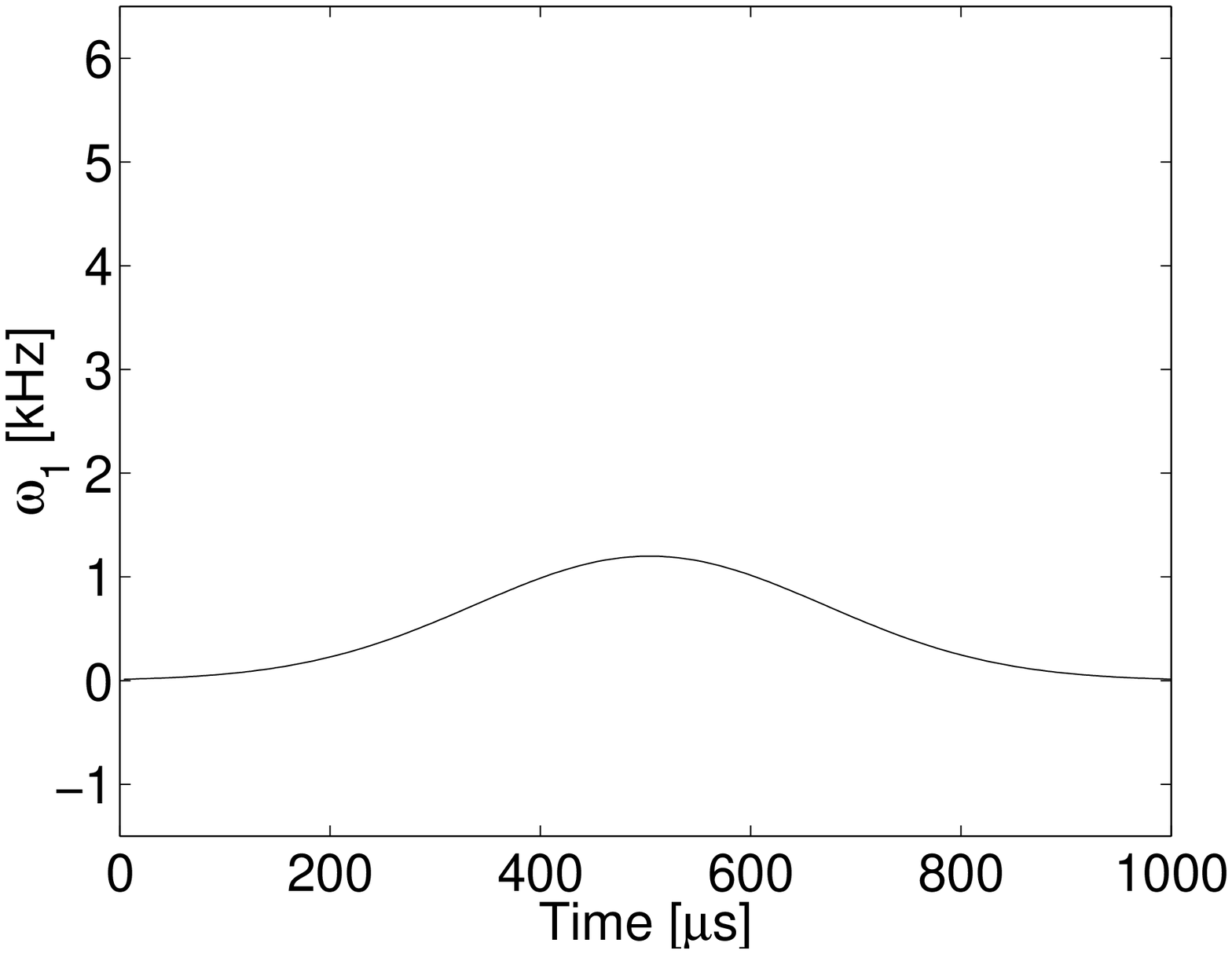} %&
\includegraphics*[width=6cm]{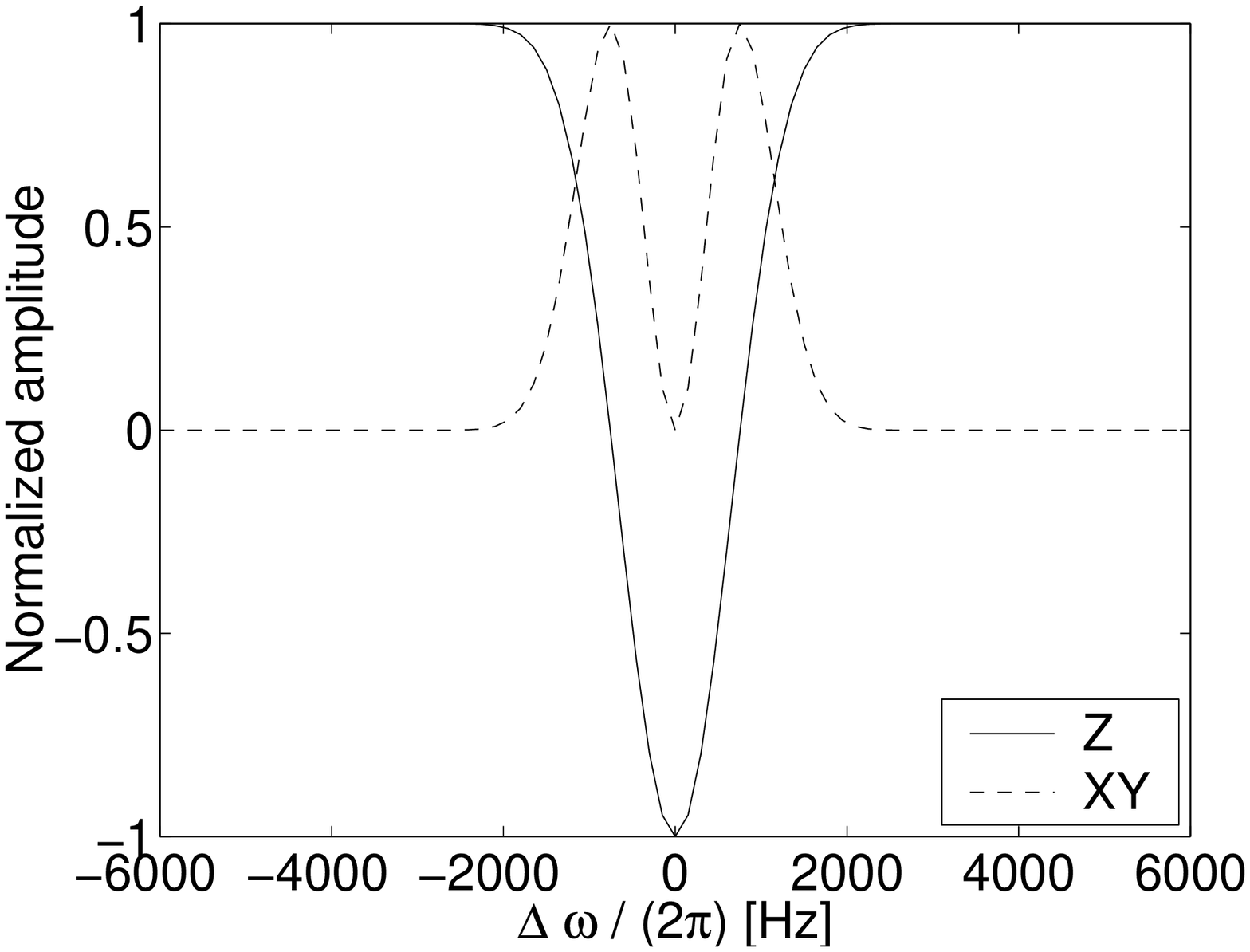} \\
\rotatebox{90}{\hspace*{8ex} Hermite 180} 
\includegraphics*[width=6cm]{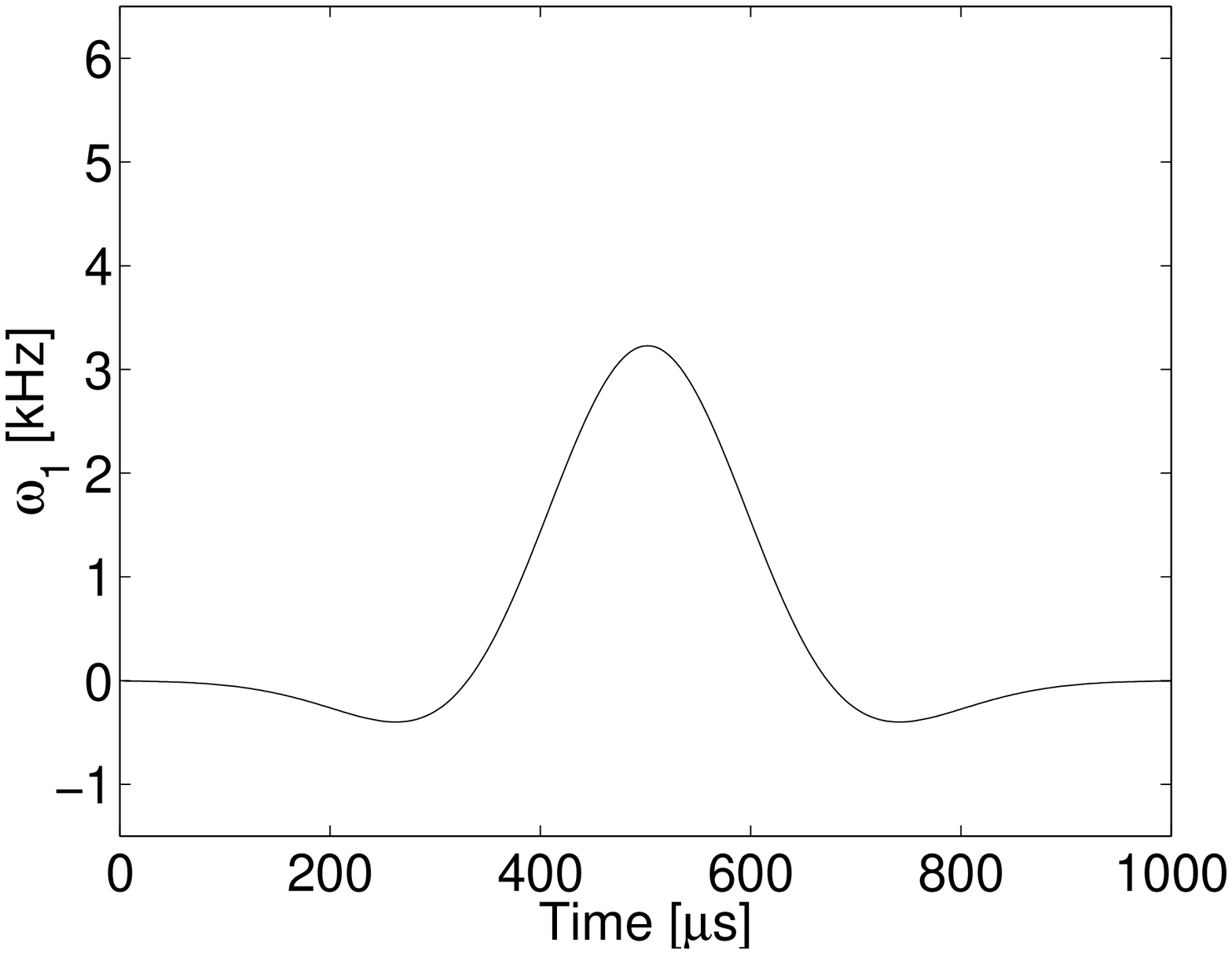} %&
\includegraphics*[width=6cm]{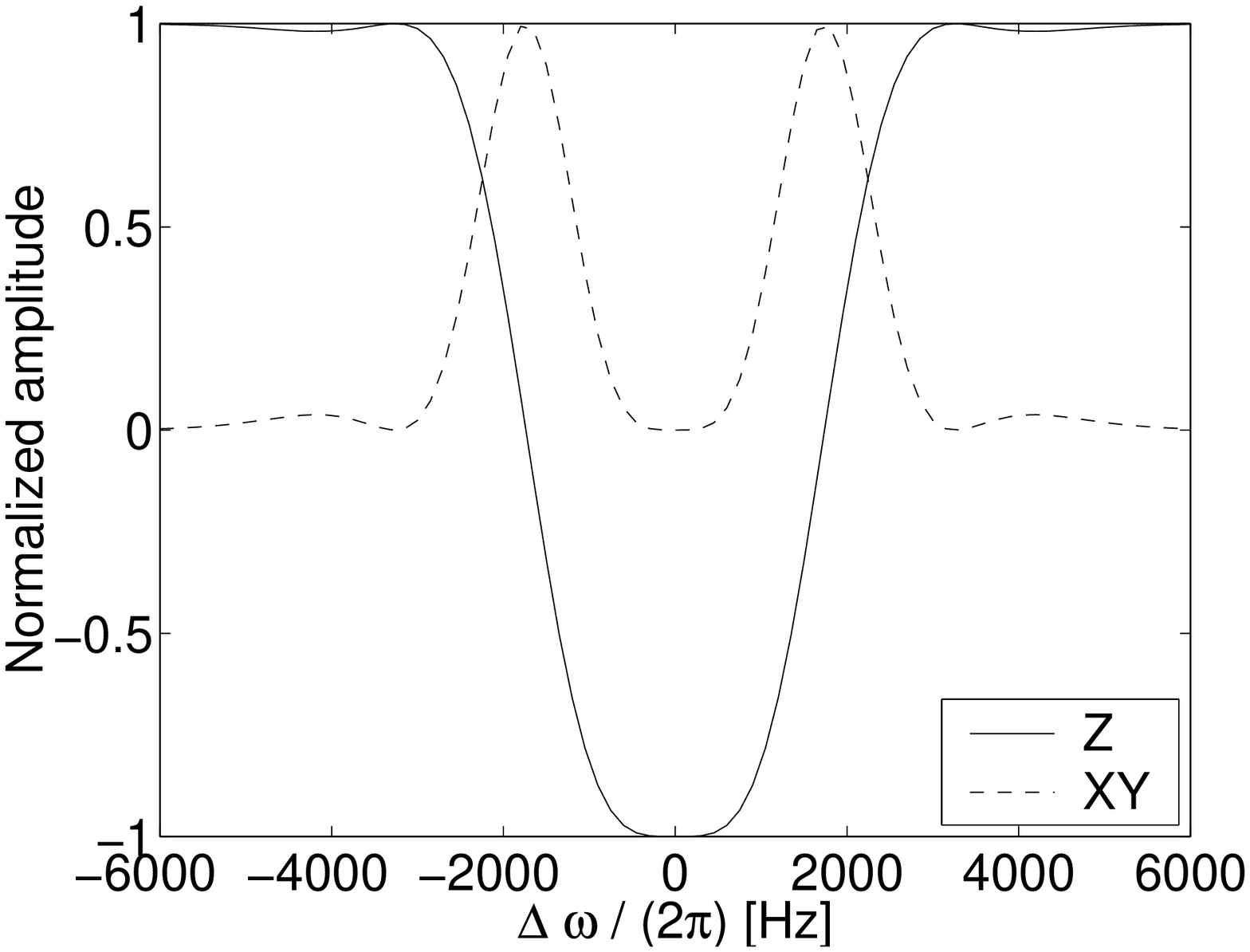} \\
\rotatebox{90}{\hspace*{8ex} {\sc reburp}} 
\includegraphics*[width=6cm]{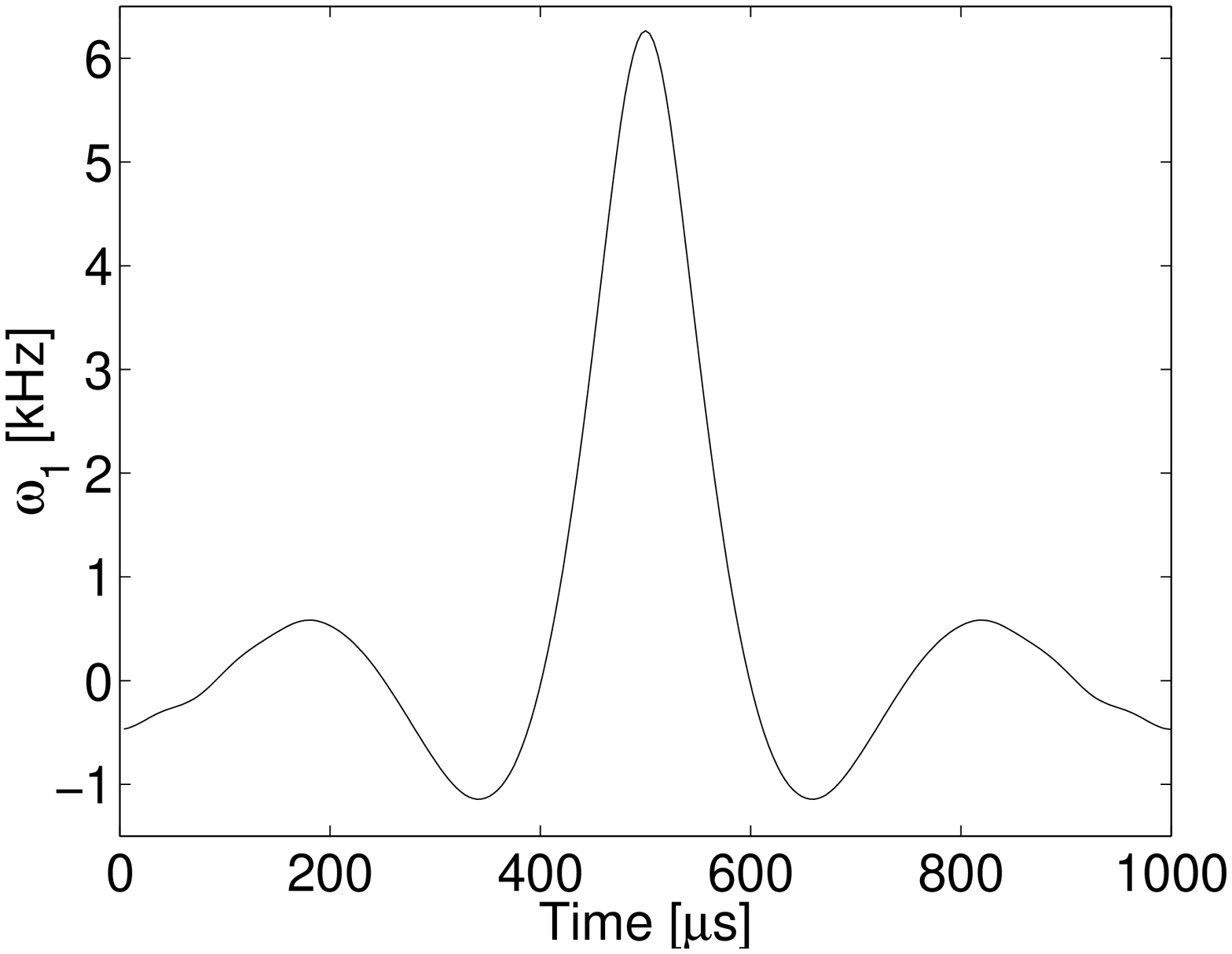} %&
\includegraphics*[width=6cm]{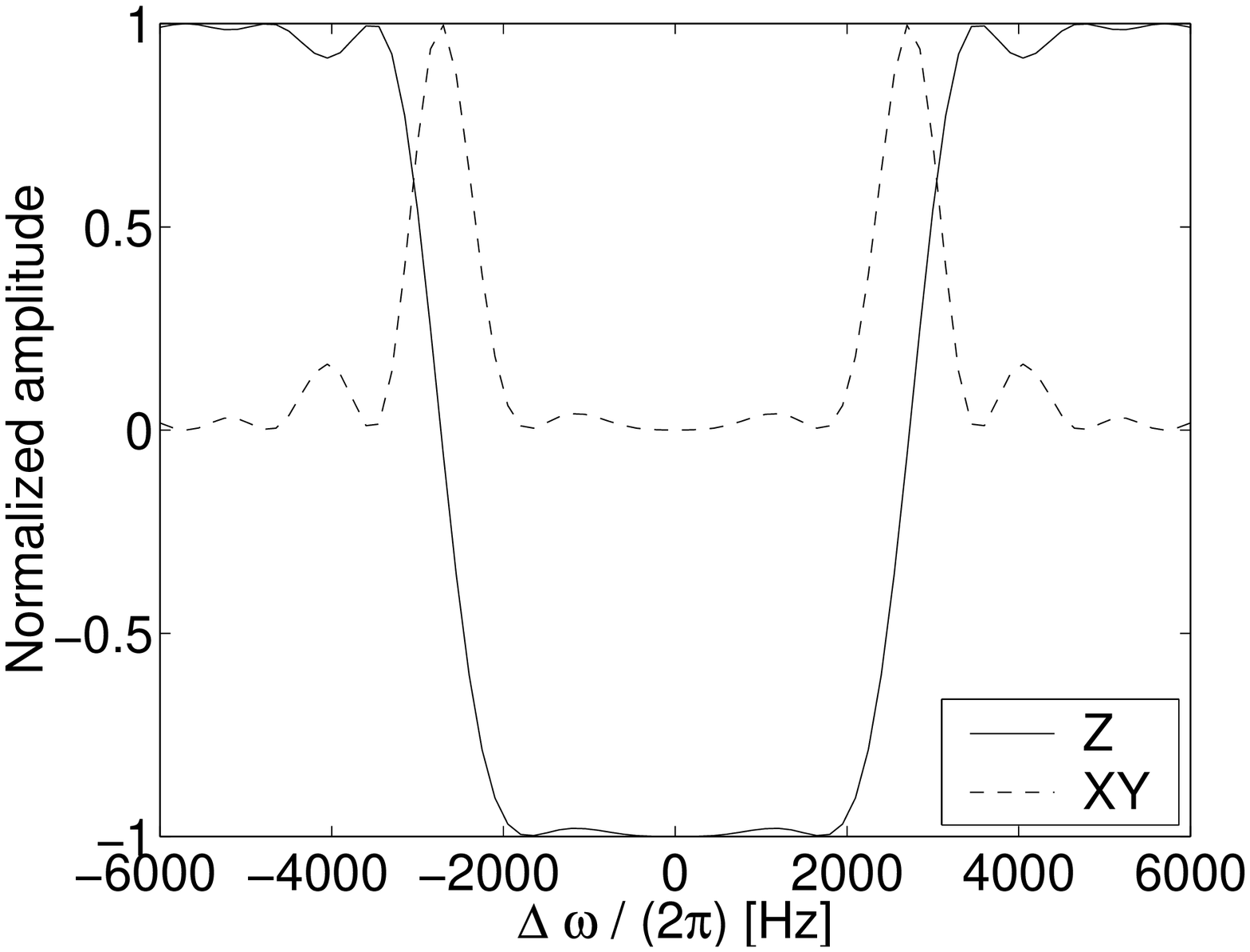} 
%\end{tabular}
\end{center}
\vspace*{-0.7cm}
\caption{(Left) Time profile and (Right) frequency excitation 
profiles (displaying the $z$ and $xy$ component of the Bloch vector
after a pulse when the Bloch vector is along $+\hat{z}$ before the
pulse) for four relevant pulse shapes.}
\label{fig:profile_pulses}
\efig
\afterpage{\clearpage}

\subsubsection{Phase ramping}
\label{page:phase_ramping}

Excitation at a frequency which differs from the RF carrier frequency
$\omega_{r\!f}$ by $\Delta\omega$, is made possible by linearly
incrementing the phase of the pulse during the application of the
pulse, at a rate $\delta \phi / \delta t = \Delta\omega$.  The result
of this procedure, known as {\em phase-ramping}~\cite{Patt91a}, is
that the frequency of the output signal of the phase shifter is
$\delta \phi / \delta t$ higher than $\omega_{r\!f}$, the frequency at
the input of the phase shifter. This is expressed by replacing
Eq.~\ref{eq:ham_rf_lab} by
\begin{eqnarray}
{\cal H}_{r\!f} = \hbar \omega_1 
    \left\{ \cos \left[\omega_{r\!f} t + 
	\left(\phi_0 + \frac{\delta\phi}{\delta t} t\right) \right] I_x +
	   \sin \left[\omega_{r\!f} t + 
	\left(\phi_0 + \frac{\delta\phi}{\delta t} t\right) \right] I_y 
    \right\}    \nonumber \\ 
= \hbar \omega_1 
    \left\{ \cos \left[\left(\omega_{r\!f} +
	   	\frac{\delta\phi}{\delta t}\right) t + \phi_0 \right] I_x +
	    \sin \left[\left(\omega_{r\!f} + 
		\frac{\delta\phi}{\delta t}\right) t + \phi_0 \right] I_y 
    \right\} \,.
\label{eq:ham_phase_ramp_rf}
\end{eqnarray}
In practice, the continuous phase ramp is approximated by discrete
steps $\Delta \phi$ from one time slice of duration $\Delta t$ to the
next, such that $\Delta \phi/\Delta t = \delta \phi/\delta t$. If
$\Delta t$ is short enough such that $\Delta \phi$ is only a few
degrees to about $10$ degrees, this is a good approximation of a
continuous phase ramp.

Excitation at multiple frequencies simultaneously via a single pulse
can be accomplished by an extension of phase ramping.  Within each
time slice, the amplitude and phase of each pulse describe a
vector. In order to merge several pulses into a single pulse, it
suffices to take the vector sum of all the original pulses within each
slice and use this sum to describe the corresponding time slice of the
combined pulse.

\begin{table}[h]
\begin{center}
\begin{tabular}{l|c|c|c|c|c}
	    & selec- &transition&       &  self-   & robust- \\
	    & tivity &range     & power &refocusing& ness\\\hline
Rectangular & poor   & very wide&minimal& no       & good    \\
Gauss 90~\cite{Bauer84a}    &excellent&wide     & low    & fair     & good    \\
Gauss 180~\cite{Bauer84a}   &excellent&wide     & low    & fair     & good    \\
Hrm 90~\cite{Warren84a}      &moderate&moderate  &average& good     & fair    \\
Hrm 180~\cite{Warren84a}     & good   &moderate  &average&very good & fair    \\
{\sc uburp} 90~\cite{Geen91a}    & poor   & narrow   & high  &excellent & poor    \\
{\sc reburp} 180~\cite{Geen91a}  & poor   & narrow   & high  &excellent & poor    \\
{\sc av} 90~\cite{Abramovich93a}       & fair   &moderate  &average& good   & fair    \\
\end{tabular}
\end{center}
\caption{Properties of relevant pulse shapes}
\label{tab:shaped_pulses}
\end{table}

%%%%%%%%%%%%%%%%%%%%%%%%%%%%%%%%%%%%%%%%%%%%%%%%%%%%%%%%%%%%%%%%%%%
\subsection{Single pulses - artefacts and solutions}
\label{nmrqc:pulse_artefacts}

\subsubsection{Bloch-Siegert shifts}

The presence of RF irradiation during pulses causes a shift $\Delta
\omega_{BS}$ in the precession frequency of spins at frequencies well
outside the excitation frequency window~\cite{Emsley90a}. This effect
has become known as a transient (generalized) Bloch-Siegert
shift~\footnote{The original paper by Bloch and
Siegert~\cite{Bloch40a} refers to the frequency shift produced by the
counter-rotating RF field (see page~\pageref{page:counterrot_rf}), but
the term Bloch-Siegert shift has been used in a generalized sense in
the NMR community.}; at a deeper level, the acquired phase can be
understood as an instance of Berry's phase~\cite{Berry84a}. The
magnitude of $\Delta
\omega_{BS}$ is approximately $\omega_1^2/2(\omega_{r\!f} - \omega_0)$ 
(for $\omega_1 \ll \omega_{r\!f} - \omega_0$), where $\omega_0/2\pi$ is the
Larmor frequency in the absence of an RF field. The frequency shifts
can easily reach several hundred Hz and the direction of the shift is
always away from the frequency of the RF field.

Each spin thus accumulates a spurious phase shift during RF pulses
applied to spins at nearby frequencies. Since $\omega_1$ varies over
time for shaped pulses, the Bloch-Siegert shift generally varies
throughout the pulse, but the cumulative phase shifts can be easily
computed in advance for each possible spin-pulse combination, if all
the frequency separations, pulse shapes and pulse lengths are
known. The unintended phase shifts $R_z(\theta)$ can then be
compensated for during the execution of a pulse sequence by inserting
appropriate $R_z(-\theta)$. This is easy to do, especially if software
reference frames are used (section~\ref{nmrqc:z_rotations}).

\subsubsection{Coupled evolution during pulses}

Spins within the same molecule interact with each other via the $J$
coupling (Eq.~\ref{eq:ham_J}). This interaction forms the basis for
two-qubit gates (section~\ref{nmrqc:2bitgates}), but the spin-spin
interactions cannot be turned off and are thus also active during the
RF pulses, which are intended to be just single-qubit
transformations. For short pulses at high power, $J$ is very small
compared to $\omega_1$ so the coupled evolution which takes place
during the pulses is negligible. However, for soft pulses, $\omega_1$
is often of the same order of magnitude as $J$, and in this case the
coupling terms strongly affect the intended nutation, in a way similar
to off-resonance effects (Fig.~\ref{fig:offres_rf}): coupling to
another spin shifts the spin frequency to $\omega_0/2\pi \pm J/2$, so a
pulse sent at $\omega_0/2\pi$ hits the spin off-resonance by $\pm J/2$.

Fortunately, specialized pulse shapes exist which minimize the effect
of coupling during the pulses (see Table~\ref{tab:shaped_pulses}).
Such {\em self-refocusing} pulses~\cite{Geen91a} take a spin over a
complicated trajectory in the Bloch sphere, in such a way that the net
effect of couplings between the selected and non-selected spins is
reduced. It is as if those couplings are only in part or even not at
all active during the pulse (couplings between pairs of non-selected
spins will still be fully active).  Although the self-refocusing
behavior of certain shaped pulses can be intuitively explained to some
degree, many actual pulse shapes have been the result of numerical
optimizations.  Table~\ref{tab:shaped_pulses} summarizes how effective
several common pulse shapes are at refocusing the $J$ couplings. A
general observation is that it is relatively easy to make $180^\circ$
pulses self-refocusing, but much harder for $90^\circ$ pulses.

Complementary to the use of self-refocusing pulses, undesired coupled
evolution that still takes place during a pulse can be (in part) {\em
unwound at a different time} in the pulse sequence via a ``negative''
time evolution (section~\ref{nmrqc:J_2bitgates}).  A complication in
unwinding the coupled evolution is that ${\cal H}_{r\!f}$ and ${\cal
H}_J$ {\em do not commute}; this implies that a real pulse cannot be
perfectly decomposed into an idealized pulse (no coupling present)
followed and/or preceded by a time interval of coupled evolution.

Nevertheless, we found that the coupled evolution is reversed quite
well by a negative time interval {\em both before and after the
pulse},
\be
e^{+i {\cal H}_J\, p\;\!\!w\, \tau /\hbar} e^{-i ({\cal H}_{r\!f} + {\cal
H}_J)\, p\;\!\!w \,/ \hbar} e^{+i {\cal H}_J\, p\;\!\!w\,\tau / \hbar}
\approx e^{-i {\cal H}_{r\!f}\, p\;\!\!w \,/ \hbar} \,,
\label{eq:unwind_J_symm}
\ee
where $\tau$ is chosen in each equation such that the approximations
are as good as possible according to some matrix distance (we have
used the 2-norm distance measure).  A negative time interval only
before {\em or} after the pulse, 
\be
e^{+i {\cal H}_J\, p\;\!\!w\,\tau/\hbar}
e^{-i ({\cal H}_{r\!f} + {\cal H}_J)\, p\;\!\!w\,/\hbar}
\approx e^{-i {\cal H}_{r\!f} \,p\;\!\!w\,/\hbar} \approx
e^{-i ({\cal H}_{r\!f} + {\cal H}_J) \,p\;\!\!w\,/\hbar}
e^{+i {\cal H}_J) \, p\;\!\!w\, \tau/\hbar} \,,
\ee
is much less effective. This is can be seen from
Table~\ref{tab:unwind_J}, which we will now discuss.

For Gaussian $90^\circ$ pulses, the $J_{1i}$ evolution, which doesn't
commute with $X_1$, is unwound much better by two symmetrically placed
negative time evolution intervals of duration $p\!w \;\tau$ than by a
single $\tau$ (we will from now on leave $pw$ implicit; $\tau$ will be
in units of $pw$ throughout) before or after the pulse. Furthermore,
while evolution under $J_{ij}$ ($i,j\neq1$) commutes with a pulse on
spin $1$ and can thus be perfectly reversed, the optimal values of
$\tau$ to unwind $J_{1i}$ and $J_{jk}$ evolution lie much closer
together if the $\tau$'s are placed symmetrically.

Hermite shaped $180^\circ$ pulses are self-refocusing, so $J_{1i}$ is
not active during the pulse and need not be reversed. In the
asymmetric scheme with only a single $\tau$ either before or after the
pulse, $\tau$ must thus be $0$ such that no $J_{1i}$ evolution is
introduced.  However, in order to unwind $J_{ij}$, we need $\tau$ to
be $1$. We can thus not take care of both $J_{1i}$ and $J_{ij}$ in the
asymmetric scheme.  In contrast, a symmetric pair of intervals $\tau$
separated by a $180^\circ$ pulse on spin $1$ gives net zero $J_{1i}$
evolution for any value of $\tau$. We can thus set $\tau$ to the
optimal value for unwinding $J_{jk}$ and obtain an excellent net
single-qubit $180^\circ$ rotation.

\begin{table}[h]
\begin{center}
\begin{tabular}{|c||c|c|}
\hline
\bfseries{gauss90} &    $X_1 \tau$		&$\tau X_1 \tau$   \cr \hline\hline
$J_{1i}$	   & fair, $\tau\approx 0.6$ 	& excellent, $\tau\approx0.57$  \cr
$J_{ij}$	   & perfect, $\tau=1$		& perfect, $\tau=0.5$ 		\cr
$J_{1i}, J_{ij}$   & fair, $(0.6 <) \tau < 1$	& very good, $0.5 < \tau < 0.57$\cr
\hline
\end{tabular}
\end{center}
\begin{center}
\begin{tabular}{|c||c|c|}
\hline
\bfseries{hrm180} &    $X_1 \tau$		&$\tau X_1 \tau$   \cr \hline\hline
$J_{1i}$	  & excellent, $\tau = 0$ 	& excellent, $\forall \tau$  \cr
$J_{ij}$	  & perfect, $\tau=1$		& perfect, $\tau=0.5$ 		\cr
$J_{1i}, J_{ij}$  & poor, $0 < \tau < 1$	& excellent, $\tau=0.5$\cr
\hline
\end{tabular}
\end{center}
\caption{Comparison of the degree to which $J$-coupled evolution during 
a single pulse on spin 1 is unwound via asymmetric versus symmetric
negative time intervals $\tau$ (expressed as a fraction of the
duration of $pw$), for various coupling scenarios. The optimal $\tau$
is indicated in each case.}

\label{tab:unwind_J}
\end{table}

In practice, we have used the symmetrized decomposition of
Eq.~\ref{eq:unwind_J_symm} in order to obtain a first order
improvement in the unitary evolution.  Further fine-tuning can be done
if different negative evolution times are allowed to unwind different
coupling terms. Higher-order corrections are in principle possible
because the undesired evolution is known and can be fully
characterized, although such a scheme would be substantially more
complicated to implement. Also, with the degree of control in current
implementations, it is unclear if such higher order corrections would
be effective.  Finally, if the input state is diagonal, as is the case
in many input state preparation pulse sequences, couplings that do not
involve the selected spin (e.g. $J_{23}$ during $X_1$ pulses) have no
effect, and need not be unwound. In this case, $\tau$ should be
optimized to refocus just the $J_{1i}$ couplings.

It is clear that properly controlling non-commuting terms in the
Hamiltonian will be a recurring challenge for virtually any proposed
quantum computer implementation. We believe that the development of a {\em general} (as
opposed to an ad-hoc) and {\em practical} method for removing the
effect of select terms in the Hamiltonian constitutes an important and
interesting open problem.

%%%%%%%%%%%%%%%%%%%%%%%%%%%%%%%%%%%%%%%%%%%%%%%%%%%%%%%%%%%%%%%%%%%
\subsection{Simultaneous pulses - artefacts and solutions}
\label{nmrqc:simpulse_artefacts}

Simultaneous pulses, as opposed to consecutive pulses, are desirable
for quantum computation because they help keep the pulse sequence
within the coherence time. However, the effects of Bloch-Siegert
shifts, discussed in the previous section for single pulses, are
aggravated during simultaneous pulses. The effect of $J$ couplings
during simultaneous pulses also deserves a separate discussion.

\subsubsection{Bloch-Siegert shifts}

The Bloch-Siegert shifts introduced in the previous section result in
additional problems during {\em simultaneous pulses} applied to two or
more spins {\em at nearby frequencies}.  If we apply spin-selective
pulses simultaneously to two spins $1$ and $2$ with resonance
frequencies $\omega_0^1$ and $\omega_0^2$ (say $\omega_0^1 <
\omega_0^2$), the pulse at $\omega_0^1$ temporarily shifts the
frequency of spin $2$ to $\omega_0^2 + \Delta \omega_{BS}$. As a
result, the pulse on spin $2$, which is still applied at $\omega_0^2$,
will be off-resonance by an amount $-\Delta \omega_{BS}$. Analogously,
the pulse at $\omega_0^1$ is now off the resonance of spin 1 by
$\Delta \omega_{BS}$. The resulting rotations of the spins deviate
significantly from the intended rotations.

\begin{figure}[htbp]
\begin{center}
\includegraphics*[width=8cm]{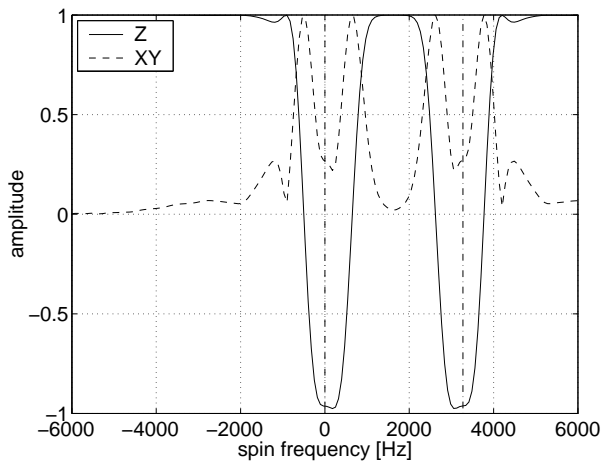} 
\end{center}
\vspace*{-2.0ex}
\caption{Simulation of the amplitude of the $z$ and $xy$ component of 
the magnetization of a spin as a function of its frequency. The spin
starts out along $+\hat{z}$ and is subject to two simultaneous
Hermitian shaped pulses with carrier frequencies at 0 Hz and 3273 Hz
(vertical dashed lines), with a calibrated pulse length of 2650$\mu$s
(ideally 180$^{\circ}$).}
\label{fig:simwithoutbs}
\end{figure}

Fig.~\ref{fig:simwithoutbs} shows the simulated inversion profiles for
a spin subject to two simultaneous Hermite 180$^{\circ}$
pulses separated by 3273 Hz. The centers of the
inversion profiles have shifted away from the intended frequencies and
the inversion is incomplete, which can be seen most clearly from the
substantial residual $xy$-magnetization ($> 30\%$) over the whole
region intended to be inverted. Note that since the frequencies of the
applied pulses are off the spin resonance frequencies, perfect
rotations cannot be achieved no matter what tip angle is chosen. In
practice, simultaneous soft pulses at nearby frequencies have been
avoided in NMR~\cite{Linden99a} or the poor quality of the spin
rotations was accepted.

We have developed an effective, intuitive, albeit simple
procedure~\cite{Steffen00a}\footnote{The idea for this technique is
due to Matthias Steffen, inspired by discussions between Matthias and
myself. We worked out and refined this method together.} to address
this problem, as an alternative to the existing brute force
optimizations~\cite{Pauly91a}: the rotations of the spins can be
significantly improved simply by shifting the carrier frequencies (in
practice most easily done via the phase-ramping techniques described
in section~\ref{nmrqc:pulse_shaping}) such that they track the shifts
of the corresponding spin frequencies. This way, the pulses are always
applied on-resonance with the corresponding spins. The calculation of
the frequency shift throughout a shaped pulse is straightforward and
needs to be done only once, at the start of a series of experiments.

\begin{figure}[htbp]
\begin{center}
\includegraphics*[width=8cm]{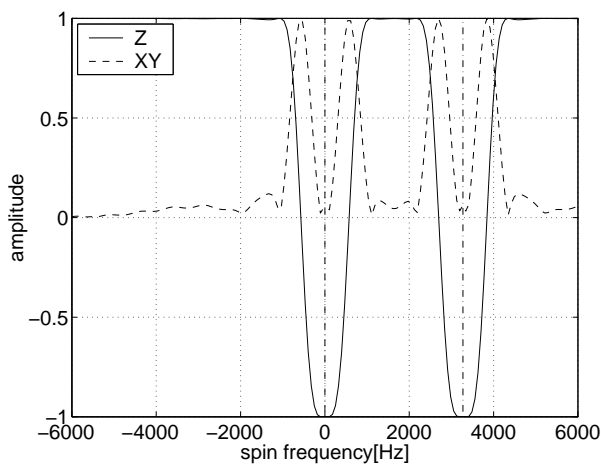} 
\end{center}
\vspace*{-2.0ex}
\caption{Similar to Fig.~\ref{fig:simwithoutbs} but {\em with} the 
frequency shift correction.}
\label{fig:simwithbs}
\end{figure}

Fig.~\ref{fig:simwithbs} shows the simulated inversion profiles for
the same conditions as in Fig.~\ref{fig:simwithoutbs}, but this time
using the frequency shift corrected scheme. The inversion profiles are
much improved and there is very little left-over $xy$ magnetization.

We simulated the inversion profiles for a variety of pulse widths and
frequency separations, for hermite shaped, gaussian shaped and {\sc
reburp} pulses, and verified that the same technique can be used to
correct the frequency offsets caused by three or more simultaneous
soft pulses at nearby frequencies.  The improvement is particularly
pronounced when the frequency window of the shaped pulse is two to
eight times the frequency separation between the pulses, with
improvements in the accuracy of the unitary operator up to a factor of
fifteen.  In all cases the improvements are impressive, illustrating
the robustness and versatility of this method~\cite{Steffen00a}.

We have experimentally confirmed the improvements predicted by the
simulations, and used this technique in the experiments of
sections~\ref{expt:order} and~\ref{expt:shor}.

%%%%%%%%%%%%%%%%%%%%%%%%%%%%%%%%%%%%%%%%%%%%%%%%%%%%%%%%%%%%%%%%%%%%%%

\subsubsection{Coupled evolution during pulses}

The evolution of two coupled spins which are pulsed simultaneouly
leads to multiple-quantum coherences, even if the pulses are
self-refocusing, because the interaction between the two pulses
disturbs the self-refocusing behavior~\cite{Kupce95a}.  Can we
compensate for this coupled evolution using negative time evolution
before and after? Or is it better to send the pulses back to back?
Table~\ref{tab:unwind_J_sim} reviews these questions for the case of
gaussian $90^\circ$ and Hermite $180^\circ$ pulses.

The back to back Gaussian $90^\circ$ pulses very poorly unwind any
$J$'s involving one or both selected spins.  Using simultaneous pulses
and symmetrically placed $\tau$'s, $J_{12}$ is still not very well
reversed, but all the $J_{1i}$ and $J_{2i}$ ($i\neq1,2$) are unwound
very well. $J_{ij}$ ($i,j\neq1,2$) always commutes with pulses on
spins 1 and 2, and can thus be perfectly unwound, with an optimal
value of $\tau$ close to the optimal value for reversing the evolution
under other $J$'s.  In the end, the achieved unitary evolution is
quite good when using simultaneous pulses with negative evolution
before and after, as long as $J_{12}$ is not too strong.

Because Hermite shaped $180^\circ$ pulses are self-refocusing,
$J_{12}$ is not active during the back to back pulses and need not be
reversed; however, all the other couplings are very poorly unwound
when using back to back pulses. During two simultaneous pulses,
$J_{12}$ is almost fully active, but can be unwound quite well; the
$J_{1i}$ and $J_{2i}$ can be reversed very well too. As always,
$J_{ij}$ commutes with pulses on spins 1 and 2, and can thus be
perfectly unwound.  If the negative evolution is arranged
symmetrically (and only then), the optimal values of $\tau$ for
unwinding the evolution under the respective couplings are all
approximately the same, resulting in a very good unitary
transformation for simultaneous pulses with symmetric unwinding.

The general conclusion is that it is much better to send the two
pulses simultaneously than back to back. Furthermore, as was the case
for single pulses, it is by far better to have negative evolution
both before and after the pulses.

\begin{table}[h]
\begin{center}
\begin{tabular}{|c||c|c|c|}
\hline
\bfseries{gauss90}	& $\tau X_1 X_2 \tau$	  & $\tau X_{1,2} \tau$		 & $X_{1,2} \tau$ \cr \hline\hline
$J_{12}$		& poor, $\tau\approx0.57$ & fair, $\tau\approx 0.5$      & poor, $\tau\approx 0.9$ \cr
$J_{1i}, J_{2i}$	& poor, $\tau\approx1.06$ & excellent, $\tau\approx0.57$ & fair, $\tau=\approx 0.57$ \cr
$J_{12}, J_{1i}, J_{2i}$& poor, $0.57<\tau<1.15$  & good, $(0.5 <) \tau < 0.57$  & poor, $0.45 < \tau <0.65$ \cr
$J_{ij}$		& perfect, $\tau=1$	  & perfect, $\tau=0.5$		 & perfect, $\tau=1$ \cr
\hline
\end{tabular}
\end{center}
\begin{center}
\begin{tabular}{|c||c|c|c|}
\hline
\bfseries{hrm180}	& $\tau X_1 X_2 \tau$	& $\tau X_{1,2} \tau$ 	   & $X_{1,2} \tau$ \cr \hline\hline
$J_{12}$		& excellent, $\tau=0$ 	& good, $\tau\approx 0.45$ & good, $\tau\approx 0.9$ \cr
$J_{1i}, J_{2i}$	& poor, $\tau > 0$	& excellent, $\forall\tau$ & excellent, $\tau=0$ \cr
$J_{12}, J_{1j}, J_{2i}$
			& poor, $\tau=0$	&very good, $\tau\approx0.45$ & poor, $0 < \tau < 0.9$ \cr
$J_{ij}$		& perfect, $\tau=1$	& perfect, $\tau=0.5$ 	   & perfect, $\tau=1$ \cr
\hline
\end{tabular}
\end{center}
\caption{Comparison of the degree to which $J$-coupled evolution
during two pulses is unwound for three scenario's: (1) two pulses back
to back preceeded and followed by negative evolution, (2) two
simultaneous pulses preceeded and followed by negative evolution, and
for comparison (3) two simultaneous pulses only followed by negative
evolution.  The optimal $\tau$ is indicated in each case.}
\label{tab:unwind_J_sim}
\end{table}

Even with symmetrically placed negative time intervals and using
simultaneous pulses, the coupled evolution which takes place during
the pulses is often (depending on the value of the respective
$J_{ij}$) not unwound to the same degree as in the case of single
pulses, and $\tau X_1 2\tau X_2 \tau$ may in some cases give better
results than simultaneous pulses. However, implementing the negative
time evolution in between the two pulses makes the pulse sequence much
longer.

In the experiments, we have also used three simultaneous
pulses. Similar arguments for the compensation of $J$-coupled
evolution as those we made for two simultaneous pulses, hold for three
or more simultaneous pulses.

%%%%%%%%%%%%%%%%%%%%%%%%%%%%%%%%%%%%%%%%%%%%%%%%%%%%%%%%%%%%%%%%%%%%%%%
%%%%%%%%%%%%%%%%%%%%%%%%%%%%%%%%%%%%%%%%%%%%%%%%%%%%%%%%%%%%%%%%%%%%%%%

\section{Two-qubit operations}
\label{nmrqc:2bitgates}

\subsection{The controlled-{\sc not} in a two-spin system}
\label{nmrqc:J_2bitgates}

The basis for two-qubit gates in NMR is the pairwise interaction
between spins in the same molecule, described in
section~\ref{nmrqc:spin-spin_interaction}.  The most natural two-qubit
gate between nuclear spins in a molecule is therefore an evolution
under the coupling Hamiltonian of Eq.~\ref{eq:ham_J} for a duration
$t$,
\be
U_J(t) = \exp[-i 2\pi J I_z^1 I_z^2 t] =
\left[\matrix{	e^{-i \pi J t/2} & 0 & 0 & 0 \cr
		0 & e^{+i \pi J t/2} & 0 & 0 \cr
		0 & 0 & e^{+i \pi J t/2} & 0 \cr
		0 & 0 & 0 & e^{-i \pi J t/2} }\right] \,.
\ee
Because of the central importance of the controlled-{\sc not} gate in
the theory of quantum computation (section~\ref{qct:gates}), we shall
now discuss the implementation of the {\sc cnot} gate using the $J$
coupling.

A first possible implementation of the {\sc cnot} gate consists of
applying a line-selective 180$^\circ$ pulse at $\omega_0^2 +
J_{12}/2$. This pulse inverts spin 2 (the target qubit) if and only if
spin 1 (the control) is $\ket{1}$~\cite{Cory97b}. In general, if a
spin is coupled to more than one other spin, half the lines in the
multiplet must be selectively inverted. This is usually very
impractical and it may in fact be impossible when some of the lines in
the multiplet fall on top of each other.

\begin{figure}
\bcen
\vspace*{1ex}
\includegraphics*[width=12cm]{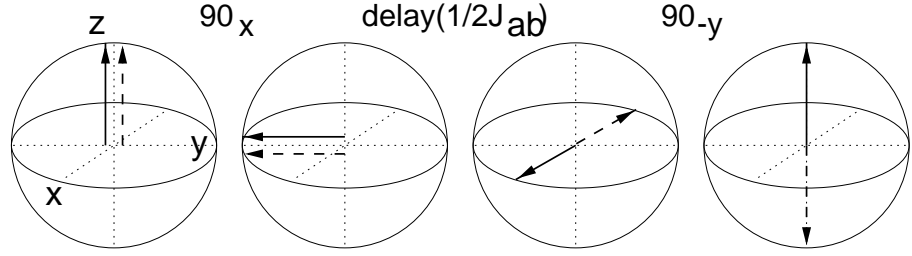} 
\vspace*{-2ex}
\ecen
\vspace*{-1.0ex}
\caption{Bloch-sphere representation of the operation of the 
{\sc CNOT}$_{12}$ gate between two nuclear spins $1$ and $2$ in a
molecule. Spin $2$ is shown in a reference frame rotating about
$\hat{z}$ at $\omega_0^2/2\pi$, in case spin $1$ is
$\ket{0}$ (solid line) and \ket{1} (dashed line).}
\label{fig:inept}
\end{figure}

An alternative and more widely used implementation of the {\sc cnot}
gate is illustrated in Fig.~\ref{fig:inept}~\cite{Gershenfeld97a}.
First, a spin-selective pulse on spin $2$ about $\hat{x}$ (an rf pulse
centered at $\omega_0^2/2\pi$ and of a spectral bandwidth such that it
covers the frequency range $\omega_0^2/2\pi \pm J_{12}/2$ but not
$\omega_0^1/2\pi$), rotates spin 2 from $+\hat{z}$ to $-\hat{y}$.
Then the spin system is allowed to freely evolve for a duration of
$1/2J_{12}$ seconds.  Because the precession frequency of spin 2 is
shifted by $\pm J_{12}/2$ depending on whether spin 1 is in $\ket{1}$
or $\ket{0}$, after $1/2J$ seconds spin 2 will have rotated to either
$+\hat{x}$ or to $-\hat{x}$ (in the reference frame rotating at
$\omega_0^2/2\pi$), depending on the state of spin 1.  Finally, a
90$^\circ$ pulse on spin 2 about the $-\hat{y}$ axis (still in the
rotating frame) rotates spin 2 back to $+\hat{z}$ if spin 1 is
$\ket{0}$, or to $-\hat{z}$ if spin 1 is in $\ket{1}$.  The net result
is that spin 2 is flipped if and only spin 1 is in $\ket{1}$, which
corresponds exactly to the classical truth table for the {\sc cnot}
presented in Fig.~\ref{fig:cnot_table}.

However, a sequence as in Fig.~\ref{fig:inept} actually implements the
unitary transformation
\be
X_2 \, U_J(1/2J) \, Y_2 =
\tilde{U}_{\mbox{\sc cnot}_{12}}= 
	  \left[\matrix{1 & 0 & 0 & 0 \cr
			0 & i & 0 & 0 \cr 
			0 & 0 & 0 &-i \cr 0
			& 0 & 1 & 0 }\right] \,,
\label{eq:U_cnot_tilde}
\ee
which is similar to but different from $U_{\mbox{\sc cnot}}$ of
Eq.~\ref{eq:U_cnot}. An additional phase shift on both spins is needed
in order to obtain $U_{\mbox{\sc cnot}}$ exactly:
\be
Z_1 \bar{Z_2} X_2 \, U_J(1/2J) \, Y_2 =
U_{\mbox{\sc cnot}_{12}} = 
	  \left[\matrix{1 & 0 & 0 & 0 \cr
			0 & 1 & 0 & 0 \cr
			0 & 0 & 0 & 1 \cr
			0 & 0 & 1 & 0 }\right] \,.
\label{eq:U_cnot_2}
\ee

%%%%%%%%%%%%%%%%%%%%%%%%%%%%%%%%%%%%%%%%%%%%%%%%%%%%%%%%%%%%%%%%%%%%

\subsection{Refocusing select $J$ couplings}

The coupling terms in the Hamiltonian of nuclear spins in a molecule
are given by nature and cannot be turned off. Therefore, in order
to implement a {\sc cnot}$_{ij}$ in a molecule with $n$ coupled spins,
we need a means to effectively deactivate all couplings except
$J_{ij}$. This is done in NMR by {\em refocusing} the undesired
coupled evolutions via a sequence of $180^\circ$ pulses, in a similar
way as is done in spin-echo experiments.

Fig.~\ref{fig:refocusing2} pictorially shows how refocusing pulses
can neutralize the $J$ coupling between two spins. In (a), the
evolution of spin 1 which takes place in the first time interval is
reversed in the second time interval, due to the $180^\circ$ pulse on
spin $2$. In (b), spin 1 continues to evolve in the same direction the
whole time, but still comes back to its initial position thanks to the
$180^\circ$ pulse on spin 1. The second $180^\circ$ pulse is needed to
ensure that both spins return to their initial state regardless of
the initial state. We note that if refocusing pulses
are sent on both spins simultaneously, the coupling is active again.

\bfig
\bcen
\vspace*{1ex}
\includegraphics*[width=12cm]{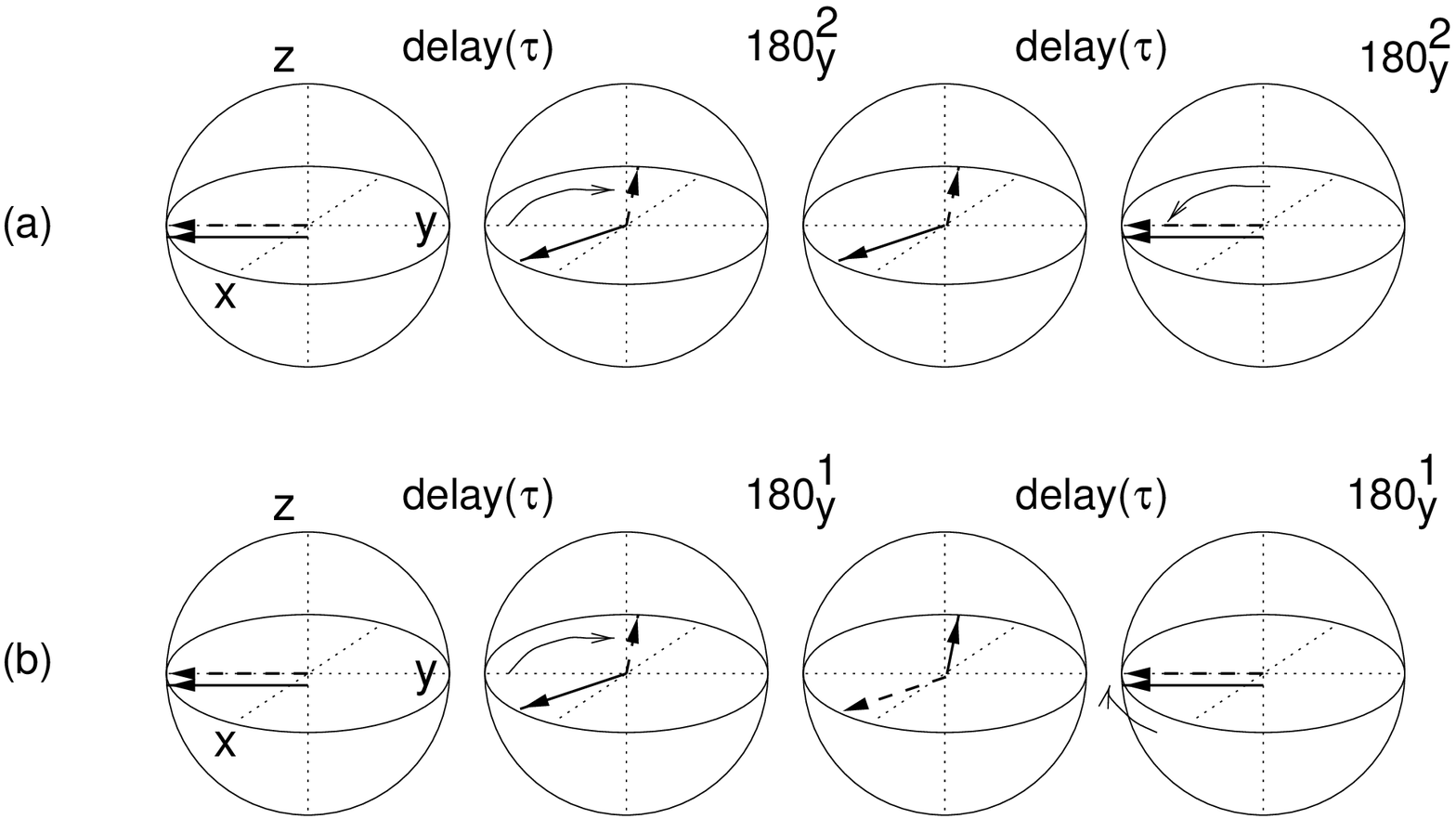} 
\vspace*{-2ex}
\ecen
\vspace*{-1.0ex}
\caption{Bloch-sphere representation of the operation of a simple 
scheme to refocus the coupling between two coupled spins. The diagram
shows the evolution of spin $1$ (in the rotating frame) initially
along $-\hat{y}$, when spin 1 is in $\ket{0}$ (solid) or in $\ket{1}$
(dashed).}
\label{fig:refocusing2}
\efig

Mathematically, we can see how refocusing of $J$ couplings works via
\be
X_1^2 \, U_J(\tau) \, X_1^2 
= U_J(-\tau) =
X_2^2 \, U_J(\tau) \, X_2^2 \,,
\ee
which leads to
\be
X_1^2\, U_J(\tau) \, 
X_1^2\, U_J(\tau)  = I =
X_2^2\, U_J(\tau)\, 
X_2^2\, U_J(\tau) 
\ee
for all values of $\tau$ (all $X_i^2$ may also be $Y_i^2$).

Fig.~\ref{fig:refocusing3} shows a refocusing scheme which preserves
the effect of the $J_{12}$ coupling in a four spin system, while
effectively inactivating all the other couplings. The underlying idea
is that a coupling between spins $i$ and $j$ acts ``forward'' during
intervals where both spins have the same sign in the diagram, and acts
``in reverse'' whenever the spins have opposite signs. Whenever a
coupling acts forward and in reverse for the same duration over the
course of a refocusing scheme, it has no net effect. If the forward
and reverse evolutions are not balanced in duration, a net coupled
evolution takes place corresponding to the excess forward or reverse
evolution.

\bfig
\bcen
\vspace*{1ex}
\includegraphics*[width=8cm]{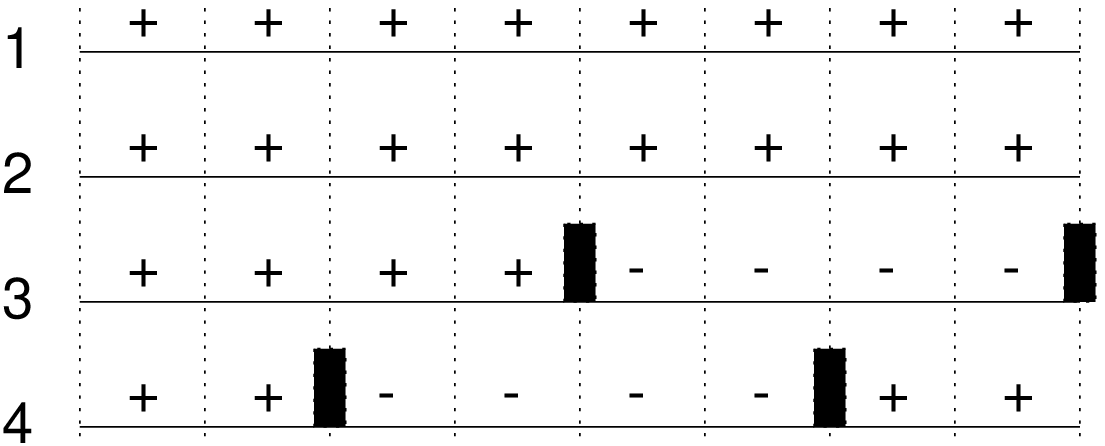} 
\vspace*{-2ex}
\ecen
\vspace*{-1.0ex}
\caption{Refocusing scheme for a four spin system. $J_{12}$ is active 
the whole time but the effect of the other $J_{ij}$ is neutralized.
The interval is divided into slices of equal duration, and the ``+''
and ``-'' signs indicate whether a spin is still in its original
position, or upside down. At the interface of certain time slices,
$180^\circ$ pulses (assumed to be instantaneous, and shown as black
retangles) are sent on one or more spins; the pulsed spins transition
from $+$ to $-$ or back.}
\label{fig:refocusing3}
\efig

Systematic methods for designing refocusing schemes for multi-spin
systems have been developed specifically for the purpose of quantum
computing. The most compact scheme is based on Hadamard
matrices~\cite{Leung00a,Jones99a}, but this is also the experimentally
most demanding, as it requires that many spins be pulsed
simultaneously. On the other extreme are schemes without any
simultaneous pulses~\cite{Linden99c}, but which take significantly
longer.

Finally, we point out that refocusing schemes can be considerably
simplified if we know that certain spins are along the $\hat{z}$ axis,
because the $J$ coupling does not affect spins along the $\hat{z}$
axis. This is a common situation in the early stages of a quantum
computation. Fig.~\ref{fig:refocusing4} gives such a simplified
refocusing scheme for five coupled spins.

\bfig
\bcen
\vspace*{1ex}
\includegraphics*[width=8cm]{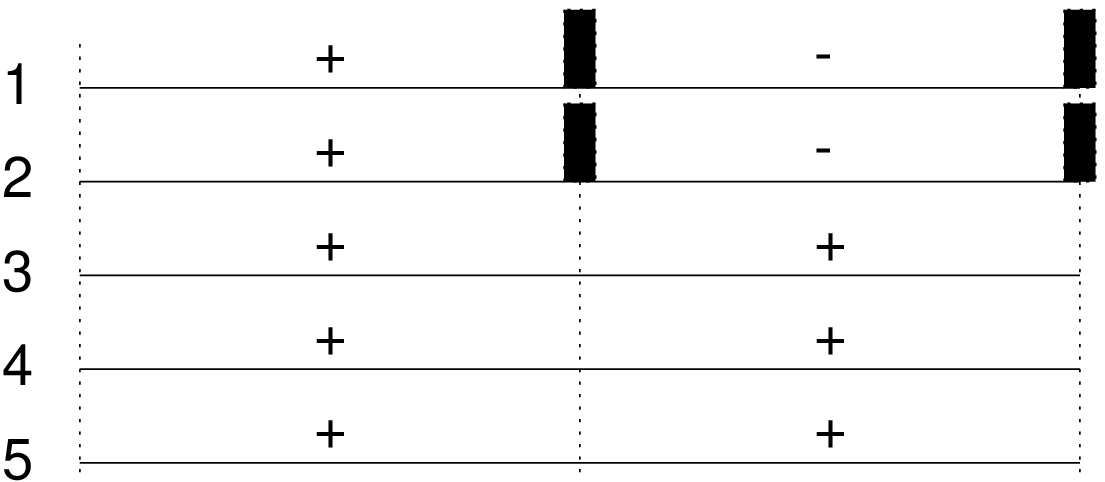} 
\vspace*{-2ex}
\ecen
\vspace*{-1.0ex}
\caption{Simplified refocusing scheme for five spins, which can be 
used if we know in advance that spins $3$, $4$ and $5$ are along $\pm
\hat{z}$. $J_{12}$ is active, but $J_{13},$ $J_{14},$ $J_{15},$ 
$J_{23},$ $J_{24},$ $J_{25}$ are inactive. The remaining couplings are
active but have no effect given the initial state.}
\label{fig:refocusing4}
\efig

%%%%%%%%%%%%%%%%%%%%%%%%%%%%%%%%%%%%%%%%%%%%%%%%%%%%%%%%%%%%%%%%%%

\subsubsection{Negative time evolution}

It is sometimes necessary to implement a unitary transformation which
corresponds to a free evolution under certain couplings for a negative
time. From the preceding paragraphs, we see that this can be achieved
using a refocusing sequence where the desired couplings evolve in
reverse for a longer time than they act forward.

Starting from an existing refocusing scheme, negative evolutions can
thus be obtained simply by {\em reducing the appropriate delay times}.
Of course, the effective negative time evolution obtained in this way
cannot be longer than the original delay times. If longer negative
delay times are needed, we can {\em increase the remaining delay
times} in order to increase the imbalance between forward and reverse
evolutions for the desired couplings.

If there is no refocusing sequence already in place which can be
changed, realizing negative time evolution under certain couplings
requires an additional refocusing sequence. This is obviously to be
avoided if possible.

%%%%%%%%%%%%%%%%%%%%%%%%%%%%%%%%%%%%%%%%%%%%%%%%%%%%%%%%%%%%%%%%%%

\subsubsection{Coupling network considerata}

For systems with more than two spins, two complications arise when
performing two-qubit gates. They are illustrated by the two extreme
coupling networks in Fig.~\ref{fig:coupling_networks} (a)
and (b). First, if two spins are not directly coupled to each other, a
{\sc cnot} between these two spins must be done using intermediary
spins (see section~\ref{impl:gates},
p.~\pageref{page:indirect_couplings})~\cite{Collins00a}.  Second, if every spin is
coupled to every other spin (this is possible only in relatively small
molecules), the pulse sequence of Fig.~\ref{fig:inept} must be
supplemented by a complex refocusing scheme which selects only the
desired coupling.

Clearly, both scenarios are associated with a considerable
overhead. It is important to note that this overhead is only
polynomial (at most quadratic) in the number of qubits. A {\sc cnot}
between any two qubits along a chain of $n$ spins with just
nearest-neighbour couplings takes at most $2(n-2)$ swap operations,
and a Hadamard based refocusing scheme between $n$ fully coupled spins
takes at most $n$ time segments and no more than $n$ $180^\circ$
pulses per segment.

From a computer science point of view, the overhead during two-qubit
gates is thus almost irrelevant, as it does not affect the efficiency
(polynomial versus exponential cost) of an algorithm.  From an
experimentalist's point of view, the overhead is of course significant
given the limited state-of-the-art in experimental quantum
computing. Any possibility for minimizing the number of refocusing or
swap operations, for example by mapping the coupling network onto the
particular algorithm at hand, should therefore be exploited.
Furthermore, this overhead does potentially negate the benefits of
quadratic speed-ups obtained in algorithms such as Grover's.

%%%%%%%%%%%%%%%%%%%%%%%%%%%%%%%%%%%%%%%%%%%%%%%%%%%%%%%%%%%%%%%%%%%%%%%
%%%%%%%%%%%%%%%%%%%%%%%%%%%%%%%%%%%%%%%%%%%%%%%%%%%%%%%%%%%%%%%%%%%%%%%
\section{Qubit initialization}
\label{nmrqc:init}

\subsection{The initial state of nuclear spins}

\subsubsection{Nuclear spins in thermal equilibrium}

The experimentally most accessible state is the state where the spin
is in {\em thermal equilibrium with the environment}, described by
\be
\rho_{eq} = \frac{\exp-{\cal H}_0/k_B T}{\cal Z}
= \frac{1}{\cal Z}
\left[\matrix{ e^{-\hbar\omega_0/2k_BT} & 0 \cr
	         0 & e^{+\hbar\omega_0/2k_BT}}\right]
\label{eq:boltzman_1spin}
\ee
\noindent so the spin statistics are given by the Boltzman 
distribution,
\be
\mbox{Pr}[\ket{0}] = \frac{e^{-\hbar \omega_0 / 2k_B T}}{\cal Z}
= \frac{e^{-\hbar \omega_0 / 2k_B T}} {e^{-\hbar \omega_0 / 2k_B T} +
e^{+\hbar \omega_0 / 2k_B T}}
\ee
\be
\mbox{Pr}[\ket{1}] = \frac{e^{+\hbar \omega_0 / 2k_B T}}{\cal Z}
= 1 - \mbox{Pr}[\ket{0}] \,.
\label{eq:Pr1_nmr}
\ee
For typical magnetic field strengths (about $10$ Tesla), $\hbar
\omega_0 / k_B T \approx 10^{-5} \ll 1$, so we can very well 
approximate the exponentials in
Eqs.~\ref{eq:boltzman_1spin}-\ref{eq:Pr1_nmr} via the first order
Taylor expansion,
\be
\rho_{eq} \approx \frac{1}{2}
\left[\matrix{ 1+\hbar\omega_0/2k_BT & 0 \cr
	       0 & 1-\hbar\omega_0/2k_BT}\right] \,,
\ee
and the spin polarization (Eq.~\ref{eq:polarization}) in thermal
equilibrium, $\epsilon_0$, is
\be
\epsilon_0 = \hbar\omega_0/2kT \ll 1 \,.
\ee
Since $\hbar
\omega_0 / k_B T \approx 10^{-5} \ll 1$, we have that 
$\mbox{Pr}[\ket{0}] \approx \mbox{Pr}[\ket{1}]$.

Similarly, the state of $n$ spins in thermal equilibrium is described
by
\be
\rho_{eq} \approx \frac{1}{2^n}
\left[\matrix{
1+\sum_k^n \frac{\hbar\omega_0^k}{2k_BT}  \cr 
& 1 - \frac{\hbar\omega_0^n}{2k_BT}
    + \sum_{k}^{n-1} \frac{\hbar\omega_0^k}{2k_BT} \cr 
& & \ddots \cr
& & & 1-\sum_k^n \frac{\hbar\omega_0^k}{2k_BT} }\right] \,,
\label{eq:rho_eq_n}
\ee
where we have neglected the effect of the coupling energies, a
perfectly valid approximation at typical magnetic fields (10 Tesla),
as $\hbar \omega_0$ is about $10^6$ times larger than $\hbar 2\pi
J_{ij}$.

The $2^n$ possible states of $n$ spins then occur with almost equal
probabilities; we cannot know in which state a thermally
equilibrated $n$-spin system really is~\footnote{It may appear that we
do know the state very well, namely $\rho_{eq}$ of
Eq.~\ref{eq:rho_eq_n}, but it is important to note that this density
matrix represents a statistical mixture of states. The mixedness
expresses precisely our uncertainty about whether each spin is up or
down.}.  The situation we desire is very diffirent: a single and known
state (say the $\ket{00\ldots0}$ state) should be occupied with
probability 1.

\subsubsection{Hyperpolarization}

Various physical cooling methods could be used to boost the
polarization of the spins.  Cooling the liquid NMR sample down to the
milli-Kelvin regime would result in very high polarizations. However,
the sample would be frozen, so the molecules wouldn't be able to
tumble around as is the case in liquids, and as a result, dipolar
couplings would be reintroduced. Intramolecular dipolar couplings
complicate the spin dynamics and intermolecular couplings result in
broad spectral lines.  Several
proposals~\cite{Yamaguchi99a,Ladd00a,Cory00a} exist to address these
complications (see also section~\ref{impl:solid_nmr}) and it is
conceivable that quantum computers will be realized using solid-state
NMR in the future.

The use of optical pumping~\cite{Fitzgerald98a} for polarization
enhancement has already been demonstrated in a two-qubit molecule
($^{13}$CHCl$_3$) in liquid solution, which was then used for a
quantum computation~\cite{Verhulst01a}. The pumping procedure consists
of several steps. First, the spin of an unpaired electron in vaporized
rubidium is hyperpolarized by shining circularly polarized laser light
on the D1 electronic transition of Rb. Then the Rb vapor is mixed with
xenon gas, and polarization is transferred from the Rb electron spin
to the $^{129}$Xe nuclear spins as Van der Waals molecules are formed
or two-body collisions take place. Finally, the hyperpolarized
$^{129}$Xe is mixed with the quantum computer molecule and
polarization is transferred (via SPINOE cross-relaxation) to the spins
which serve as quantum bits. Qubit polarization enhancements by a
factor of $10$ to $100$ have already been achieved, but the resulting
polarization of $10^{-4}$ to $10^{-3}$ is still several orders of
magnitude away from full polarization.

Two proton spins have recently been polarized to an estimated $10\%$
polarization using {\em para} hydrogen, and subsequently used in a
quantum computation~\cite{Hubler00a}. In thermal equilibrium at 20.4 K
(the boiling point of H$_2$), H$_2$ contains more than $99\%$ {\em
para} hydrogen. These are dihydrogen molecules with the two $^1$H
spins in the singlet state. Activated charcoal or some other catalyst
is needed to accelerate the {\em ortho/para} conversion. By reacting
$n/2$ {\em para} hydrogen molecules with an appropriate precursor
molecule, an $n$ qubit quantum computer molecule with highly polarized
spins can be formed.  With this method, polarizations of about $50\%$
should be within reach. However, while finding suitable quantum
computer molecules is difficult in itself
(section~\ref{nmrqc:molecule}), the additional requirement that the
molecule must be easily formed from a precursor and H$_2$ presents a
substantial limitation.

Other hyperpolarization techniques used in NMR are dynamic nuclear
polarization (DNP) \cite{Jeffries63a} and chemically induced dynamic
nuclear polarization (CIDNP). In DNP, polarization is transferred from
the electron spin in a free radical to the nuclear spins. Since the
magnetic moment of an electron spin is about $1800$ times stronger
than that of nuclear spins, its equilibrium polarization is
accordingly higher.  These techniques have not yet been demonstrated
in combination with a quantum computation.\\

Clearly, the state of the art in any of the hyperpolarization
techniques is still far from producing fully polarized spins useful
for quantum computing.  Even though these techniques may be
significantly further developed in the future, we must look for other
state initialization procedures if we want to study quantum
computation with room temperature spins today.

%%%%%%%%%%%%%%%%%%%%%%%%%%%%%%%%%%%%%%%%%%%%%%%%%%%%%%%%%%%%%%%%%

\subsection{Effective pure states}
\label{nmrqc:eff_pure}

The use of room temperature nuclear spins for quantum computing has
been made possible by the invention of {\em effective pure} states, or
{\em pseudo-pure} states, briefly introduced on
page~\pageref{page:eff_pure}. This surprising concept makes an
ensemble of nuclear spins at room temperature look as if it were at
zero temperature, up to a decrease in signal strength.

The starting point is a well-known fact, namely that the NMR signal is
proportional to population {\em differences}, irrespective of the
populations themselves. Thus, a density matrix proportional to the
identity matrix does not produce a signal --- for every molecule in
which a spin points one way, there is another molecule where the
corresponding spin points the opposite way, so their signals cancel
out. Mathematically, we say that the observables in NMR are traceless
(see section~\ref{nmrqc:meas}). Furthermore, $U I U^\dagger = I$, that
is the identity matrix does not transform under unitary
transformations. We thus need to concern us only with the {\em
deviation density matrix} $\rho_{\Delta}$, the component of the
density matrix which deviates from the identity background:
\be
\rho_\Delta = \rho - I/2^n \,,
\ee
where we assume that $\rho$ is normalized, that is $\mbox{Tr}(\rho)=1$.

Gershenfeld and Chuang~\cite{Gershenfeld97a}, and independently Cory,
Havel and Fahmy~\cite{Cory97a,Cory97b}, then observed that a density
matrix of the form of Eq.~\ref{eq:eff_pure},
\be
\rho_{\mathrm{eff}} = 
\frac{1-\alpha}{2^n} I + \alpha \ket{\psi}\bra{\psi}
\,.
\label{eq:eff_pure_2}
\ee
gives the signal and has the dynamical behavior of just the second
term, $\ket{\psi}\bra{\psi}$, which represents a pure state.  We
therefore call $\rho_{\mathrm{eff}}$ an effective pure state, or
pseudo-pure state.

Written out in matrix form for $\ket{\psi}=\ket{00\ldots0}$,
Eq.~\ref{eq:eff_pure_2} becomes
\be
\rho_{\mathrm{eff}} \; = \; 
\frac{1-\alpha}{2^n} 
\left[\matrix{1 \cr  & 1 \cr & & \ddots \cr & & & 1}\right]
\;+\; \alpha\; 
\left[\matrix{1 \cr  & 0 \cr & & \ddots \cr & & & 0}\right]
\,.
\ee
We see that the signature of an effective pure basis state of $n$
spins is that the density matrix $\rho_{\mathrm{eff}}$ is diagonal and
that {\em all the diagonal entries (populations) are equal, except
one}, in this case the first entry (the ground state population).

How do we obtain $\rho_{\mathrm{eff}}$ (Eq.~\ref{eq:eff_pure_2}) from
$\rho_{eq}$ (Eq.~\ref{eq:rho_eq_n}) ?  The procedure for preparing
effective pure states must incorporate a non-unitary step one way or
another as the eigenvalues of $\rho_{eq}$ and $\rho_{\mathrm{eff}}$
are not the same. Three methods are known to do this: logical
labeling\cite{Gershenfeld97a}, spatial averaging\cite{Cory97b} and
temporal averaging\cite{Knill98b}.  Logical labeling consists of the
selection of a subspace of the Hilbert space, in which all subsequent
computations take place. In temporal averaging, the output spectra of
separate, consecutive experiments are added together (each with a
different state preparation sequence). Spatial averaging is similar to
temporal averaging, but averaging takes place over space instead of
over time. 

These three methods will be explained in detail in the next three
sections.  To date, temporal and spatial averaging have been the most
widely used techniques for preparing effective pure states. Several
hybrid schemes~\cite{Knill98b,Knill00a} have also been developed which
trade off complexity of the preparation steps for the number of
experiments.

Unfortunately, as we shall see, all these state preparation schemes
have in common that creating effective pure states incurs an
exponential cost either in the signal strength or in the number of
consecutive experiments involved.  The reason for this cost is that
effective state preparation techniques simply select out the signal
from the ground state population present in thermal equilibrium and
the fraction of the molecules in the ground state is proportional to
$n/2^n$.  Such an exponential overhead obviously defeats the purpose
of quantum computation, but is not problematic for experiments with
small numbers of qubits.

%%%%%%%%%%%%%%%%%%%%%%%%%%%%%%%%%%%%%%%%%%%%%%%%%%%%%%%%%%%%%%%%%%%%%%

\subsection{Logical labeling}
\label{nmrqc:loglab}

Logical labeling~\cite{Gershenfeld97a,Vandersypen99a} consists of
applying a pulse sequence which rearranges the thermal equilibrium
populations such that a subset of the spins is in an effective pure
state, conditioned upon the state of the remaining spins.  Then the
computation is carried out within this embedded subsystem. This
concept of embedding was previously used to observe Berry's phase in
NMR spectroscopy~\cite{Suter86a}.

For example, the thermal equilibrium deviation density matrix for a
homonuclear three-spin system is approximately
\begin{eqnarray}
&&\hspace{.5ex}{ 000\;\, 001\;\, 010\;\, 011\;\, 
100\;\, 101\;\, 110\;\, 111} 
\vspace*{-2ex}\nonumber \\
\rho_{eq} = &
\frac{1}{2^3}\; \frac{\hbar \omega_0}{2k_BT}&
\left[\matrix{3 \cr
	       & 1 \cr
	       &  & 1 \cr
	       &  &  &-1 \cr  
	       &  &  &  & 1 \cr
	       &  &  &  &  &-1 \cr
	       &  &  &  &  & &-1 \cr
	       &  &  &  &  & & &-3}\right] \,,
\label{eq:thermal_3}
\end{eqnarray}
where the labels above the density matrix help identify the
populations of the respective states. We note that the populations
within the {\em subspace} spanned by the states $\ket{000},$
$\ket{011},$ $\ket{101}$ and $\ket{110}$ naturally have the signature
of an effective pure state.

In order to simplify subsequent logical operations and to separate the
signals of the effective pure subspace and its complement, the
populations can be rearranged by a sequence of 1 and 2-qubit unitary
operations to obtain
\begin{eqnarray}
&&\hspace{.5ex}{ 000\;\, 001\;\, 010\;\, 011\;\, 
100\;\, 101\;\, 110\;\, 111} 
\vspace*{-2ex}\nonumber \\
\rho_{\mathrm{eff}} = &
\frac{1}{2^3}\; \frac{\hbar \omega_0}{2k_BT} &
\left[\matrix{3 \cr
	       & 1 \cr
	       &  & 1 \cr
	       &  &  & 1 \cr  
	       &  &  &  &-1 \cr
	       &  &  &  &  &-1 \cr
	       &  &  &  &  & &-1 \cr
	       &  &  &  &  & & &-3}\right] \,.
\label{eq:eff_pure_3}
\end{eqnarray}
Now the subspace $\{ \ket{000},$ $\ket{001},$ $\ket{010},$ $\ket{011}
\}$ is in an effective pure state. This subspace corresponds to just 
spins $2$ and $3$ conditioned on or {\em labeled} by the state of the
spin $1$ being $\ket{0}$ (we will call this the $\ket{0}_1$
subspace). The logical labeling procedure, combined with removal of
coupling to spin $1$ for the remainder of the pulse sequence, thus
allows 2-qubit quantum computations on an effective pure state of
spins $2$ and $3$.

The subspace dimension is limited by the number of equally populated
states in equilibrium, which is $C_n^{n/2} = n!/[(n/2)!]^2$ (for even
$n$) in a homonuclear system, giving $k = {\rm
log}_2(1+C_n^{n/2})$. Thus for large $n$, $k/n$ tends to 1 ($n$=40 for
$k$=37). For heteronuclear spin systems, the analysis is more complex
and $k/n$ is generally smaller, but the number $k$ of cold qubits that
logical labeling can extract from $n$ hot spins still scales
favorably. The number of operations required to rearrange the
populations also scales polynomially with $n$.

What is the signal strength obtained via logical labeling? We recall
that the NMR signal strength is proportional to population
differences. For homonuclear systems with even $n$, the $C_n^{n/2}$
equal entries in $\rho_{eq}$ (and thus $\rho_{\mathrm{eff}}$) are all
zero (for large odd $n$, they are very close to zero). The largest
entry in $\rho_{eq}$ is $n \hbar\omega_0 / 2^n 2k_BT$. The maximum
signal strength $S$ obtainable from a logically labeled state thus
scales as $n/2^n$. Since only one experiment is involved, the noise
$N$ is independent of $n$. The dependence of the signal-to-noise ratio
on $n$ is thus
\be
\frac{S}{N} \propto \frac{n}{2^n}
\ee

In section~\ref{expt:labeling}, we will present an implementation of
logical labeling on a three-spin system. Despite its elegance, the
logical labeling procedure has not been used much in practice. The
main reason is that one or more spins must be sacrificed as labeling
spins, and extra spins are still very ``expensive'' due to the
difficulty of finding large molecules with suitable properties for
quantum computing.

%%%%%%%%%%%%%%%%%%%%%%%%%%%%%%%%%%%%%%%%%%%%%%%%%%%%%%%%%%%%%%%%%%%%%%

\subsection{Temporal averaging}
\label{nmrqc:templab}

Temporal labeling consists of adding up the spectra of multiple
experiments, where each experiment starts off with a different state
preparation pulse sequence which permutes the populations. The
preparation sequences are designed such that the sum of the resulting
input states has the effective pure state signature. By the linearity
of quantum mechanics, the sum of the output states of the respective
experiments corresponds to the output which would be obtained if the
input state were the sum of the respective input states. This will
become clear as we discuss three variations of temporal averaging.

\subsubsection{Cyclic permutations}

The original temporal averaging scheme takes a sum over $2^n-1$
experiments for an $n$ spin molecule. Each of the state preparation
sequences implements a different cyclic permutation of all populations
except the ground state population.

For example, suppose the thermal equilibrium density matrix of two
spins is
\be
\rho_1 = \rho_{eq} =
\left[\matrix{ a \cr
	       & b \cr
	       &  & c \cr
	       &  &  & d }\right] \,,
\label{eq:templab_eq}
\ee
where we use $a, b, c$ and $d$ in order to emphasize that this method
works for arbitrary initial population distributions.  If we cyclicly
permute the last three diagonal entries via a unitary transformation
$U_p=U_{\mathrm{\sc cnot}_{12}} U_{\mathrm{\sc cnot}_{21}}$, we obtain
\be
\rho_2 = U_p \rho_{eq} U_p^\dagger =
\left[\matrix{ a \cr
	       & d \cr
	       &  & b \cr
	       &  &  & c }\right] \,,
\label{eq:templab_perm1}
\ee
and if we permute $\rho_{eq}$ with $U_p^2=U_{\mathrm{\sc cnot}_{21}}
U_{\mathrm{\sc cnot}_{12}}$, we get
\be
\rho_3 = U_p^2 \rho_{eq} {U_p^\dagger}^2
\left[\matrix{ a \cr
	       & c \cr
	       &  & d \cr
	       &  &  & b }\right] \,.
\label{eq:templab_perm2}
\ee
We see that with $\rho_{\mathrm{eff}} = \rho_1 + \rho_2 + \rho_3$ and
$e=b+c+d$,
\be
\rho_{\mathrm{eff}} =
\left[\matrix{ 3a \cr
	       &  e \cr
	       &  & e \cr
	       &  &  & e }\right] =
e \left[\matrix{ 1 \cr
	         & 1 \cr
	         &  & 1 \cr
	         &  &  & 1 }\right] +
(3a-e) \left[\matrix{1 \cr
	             &  0 \cr
	             &  & 0 \cr
	             &  &  & 0 }\right] \,.
\label{eq:templab_sum}
\ee

How does the signal-to-noise ratio (SNR) obtained from the resulting
summation scale with $n$ ? The ground state populations from all
$2^n-1$ experiments simply add up, and the ground state population of
any one experiment goes as $n/2^n$; the noise $N$ increases as the
square root of the number of experiments. Thus, with the number of
experiments
\be
l=2^n-1
\ee
\noindent the signal-to-noise ratio goes as
\be
\frac{S}{N} \propto \frac{n}{2^n}\; \frac{2^n-1}{\sqrt{2^n-1}} 
= \frac{n}{2^n} \,\sqrt{2^n-1} = \frac{n}{2^n} \, \sqrt{l} \,.
\ee
This is the same $S/N$ we would obtain if we signal averaged over
$l$ identical logical labeling experiments.

Since the implementation of cyclic permutations becomes rapidly very
complex for $n>2$, we have used this temporal averaging scheme only
for experiments on two qubits
(sections~\ref{expt:dj}-\ref{expt:2bitcode}). We have developed the
following more practical approach for larger $n$.

\subsubsection{Linearly independent permutations}

The purpose of temporal averaging is just to average out differences
between $2^n-1$ populations. This can be done in many ways besides
doing cyclic permutations. In fact, for any set of $2^n-1$ linearly
independent population distributions $\mathrm{diag}(\rho_i)$, we can
solve for a set of weights $v_i$ such that
\be
\rho_{\mathrm{eff}} = \sum_{i=1}^l v_i \rho_i \,.
\ee

The main advantage is that each of the state preparation pulse
sequences can be kept much simpler than the sequences needed for
cyclic permutations.  Furthermore, while this approach may still
require up to $2^n-1$ experiments to get exactly
$\rho_{\mathrm{eff}}$, it is flexible enough that
$\rho_{\mathrm{eff}}$ can be well approximated using far fewer
experiments.

The main disadvantage is that $S/N$ is suboptimal. For $l$ experiments
with $v_i$,
\be
\frac{S}{N} \propto \frac{n}{2^n} \; 
\frac{\sum_{i=1}^l v_i}{\sqrt{\sum_{i=1}^l v_i^2}} 
\quad \le \quad
\frac{n}{2^n} \; \frac{l}{\sqrt{l}} = \frac{n}{2^n}\, \sqrt{l} \,,
\ee
with equality only if all the weights, $v_i$, are equal to $1$.
Especially if some of the $v_i$ are negative, the $S/N$ can be quite
poor. Nevertheless, we successfully used this method to prepare an
effective pure state of $n=3$ spins (section~\ref{expt:grover3}), and
then used this state as the input state for Grover's algorithm.

\subsubsection{Product operator approach}

Temporal averaging can be simplified significantly further by taking
advantage of the {\em structure} in the thermal equilibrium and
effective pure state density matrices. This structure is most easily
understood not in terms of the density matrices themselves but instead
of their Pauli matrix expansion. In this description, the thermal
equilibrium deviation density matrix for five homonuclear spins is
\be
\rho_{eq} = ZIIII + IZIII + IIZII + IIIZI + IIIIZ
\label{eq:rho_eq_prodop}
\ee
where we use $IIIIZ$ instead of the more cumbersome notation $\sigma_I
\otimes \sigma_I \otimes \sigma_I \otimes \sigma_I \otimes \sigma_z$.
For $n$ spins, $\rho_{eq}$ thus consists of $n$ product operator
terms.  The five-spin effective pure ground state is~\footnote{We chose to use $Z=\sigma_z$ instead of $I_z=\sigma_z/2$ in 
order not to have different powers of two in front of the respective
terms ($ZZ = 4 I_zI_z$, $ZZZ = 8 I_zI_zI_z$ and so forth).}
\begin{eqnarray}
&\rho_{\mathrm{eff}} = ZIIII + \ldots + IIIIZ\; +\; ZZIII + \ldots +
IIIZZ\; +& \nonumber \\ &ZZZII + \ldots + IIZZZ\; +\; ZZZZI + \ldots +
IZZZZ\; +\; ZZZZZ&\,,
\label{eq:rho_eff_prodop}
\end{eqnarray}
a total of $31 = 2^n-1$ terms.  Using short sequences of {\sc cnot}
operations, the $n=5$ terms obtained in equilibrium can be transformed
into different sets of five terms, according to the following simple
transformation rules, which follow from the definition of the
controlled-{\sc not}:
\begin{eqnarray}
II & \stackrel{\mathrm{\sc CNOT}_{12}}{\longrightarrow} & II  \,,\\
IZ & \stackrel{\mathrm{\sc CNOT}_{12}}{\longrightarrow} & ZZ  \,,\\
ZI & \stackrel{\mathrm{\sc CNOT}_{12}}{\longrightarrow} & ZI  \,,\\
ZZ & \stackrel{\mathrm{\sc CNOT}_{12}}{\longrightarrow} & IZ  \,.
\end{eqnarray}
For homonuclear $n$ spin systems, the summation of as few as $\lceil
(2^n-1)/n \rceil$ different experiments thus suffices to create all
$2^n-1$ terms.

This scheme achieves a savings in the number of separate experiments
$l$ by a factor of $n$, compared to cyclic permutations.  Furthermore,
the $S/N$ is optimal because all the terms are added up with equal and
positive weights:
\be
l = \frac{2^n-1}{n} \,,
\ee
\be
\frac{S}{N} = \frac{n}{2^n}\; \frac{2^n-1/n}{\sqrt{2^n-1/n}} 
= \frac{n}{2^n} \, \sqrt{l} \,.
\ee

In practice, it may be advantageous to use slightly more experiments
in order to keep the preparation sequences as short as possible.  In
the five-qubit experiment presented in section~\ref{expt:order}, we
used nine experiments, giving a total of $9\times5=45$ product
operator terms in the summation.  The fourteen extra terms were
canceled out pairwise, using {\sc not} ($X_i^2$) operations to flip
the sign of selected terms, using
\begin{eqnarray}
I & \stackrel{\mathrm{\sc NOT}}{\longrightarrow}& I  \,,\\
Z & \stackrel{\mathrm{\sc NOT}}{\longrightarrow}& -Z  \,.
\end{eqnarray}
Of course, terms which are canceled out do no contribute to the
signal, but they do still contribute to the noise, so this diminishes
$S/N$.

For heteronuclear systems, the situation is slightly more complex, as
the terms in Eq.~\ref{eq:rho_eq_prodop} must be weighted by the
respective $\omega_i$. For example, for a fully heteronuclear
five-spin system,
\be
\rho_{eq} = \omega_1 ZIIII + \omega_2 IZIII + \omega_3 IIZII 
+ \omega_4 IIIZI + \omega_5 IIIIZ \,.
\label{eq:rho_eq_prodop_weight}
\ee
The density matrix obtained via temporal averaging contains the same
weights, and may thus not be effective pure.  Nevertheless, for partly
heteronuclear, partly homonuclear molecules, significant reductions in
the number of experiments can be achieved while preserving a good
$S/N$, as we demonstrated in a seven-spin experiment
(section~\ref{expt:shor}).

%%%%%%%%%%%%%%%%%%%%%%%%%%%%%%%%%%%%%%%%%%%%%%%%%%%%%%%%%%%%%%%%%%%%%%

\subsection{Spatial averaging}
\label{nmrqc:spatial}

Spatial averaging~\cite{Cory97b} uses a pulse sequence containing {\em
magnetic field gradients} to equalize all the populations except the
ground state population. The magnetic field gradient causes spins in
different regions of the sample to precess at different frequencies,
so their phases are apparently randomized. In fact, the dephasing is
not really random and can be undone by applying a reverse field
gradient, as long as molecules haven't randomly diffused too far
through the sample volume to a region of different magnetic field
strength. Either way, the effect on the density matrix is that all the
off-diagonal entries (except zero quantum coherences) are erased.

Spatial averaging pulse sequences are most easily understood in terms
of product operators too. A possible procedure for two homonuclear spins,
in a similar notation as in
Eqs.~\ref{eq:rho_eq_prodop}-\ref{eq:rho_eq_prodop_weight}, goes as
follows~\cite{Cory97b}:
\begin{eqnarray}
&					& ZI + IZ 	\nonumber\\
& \stackrel{R_x^2(60)}{\begin{CD} @>>>\end{CD}}    
	& ZI + \frac{1}{2} IZ - \frac{\sqrt{3}}{2} IY	\\
& \stackrel{\mathrm{grad}_z}{\begin{CD} @>>>\end{CD}}    
	& ZI + \frac{1}{2} IZ 				\\
& \stackrel{R_x^1(45)}{\begin{CD} @>>>\end{CD}}    
	& \frac{\sqrt{2}}{2} ZI + \frac{1}{2} IZ - \frac{\sqrt{2}}{2} YI \\
& \stackrel{d(1/2J_{12})}{\begin{CD} @>>>\end{CD}}    
	& \frac{\sqrt{2}}{2} ZI + \frac{1}{2} IZ + \frac{\sqrt{2}}{2} XZ \\
& \stackrel{R_y^1(-45)}{\begin{CD} @>>>\end{CD}}    
	& \frac{1}{2} ZI - \frac{1}{2} XI + \frac{1}{2} IZ 
			 + \frac{1}{2} XZ + \frac{1}{2} ZZ \\
& \stackrel{\mathrm{grad}_z}{\begin{CD} @>>>\end{CD}}
	& \frac{1}{2} ZI +\frac{1}{2} IZ + \frac{1}{2} ZZ
\end{eqnarray}

The last term ($ZZ$) contains no net polarization for either spin, and
the total final polarization is thus a factor of two lower than the
initial polarization; half of the initial polarization has been erased
by the gradient fields. For every additional spin involved in the
spatial averaging procedure, the signal decreases by another
factor of two, similar to the case of logical labeling. Thus,
\be
\frac{S}{N} \propto \frac{n}{2^n} \,.
\ee

Only one experiment is involved, but the preparation sequence quickly
becomes unwieldy for large spin systems, although methods for
designing the spatial averaging sequence for arbitrary $n$
exist~\cite{Sharf00b,Sakaguchi00a}.  Also, since the signal strength
decreases rapidly, signal averaging the same experiment many times may
be required anyways, so the number of experiments needed may in the
end be comparable to the case of temporal averaging. This technique
has been successfully used by several groups for state preparation on
two or three spins, but we have never used it.

%%%%%%%%%%%%%%%%%%%%%%%%%%%%%%%%%%%%%%%%%%%%%%%%%%%%%%%%%%%%%%%%%%%%%%

\subsection{Efficient cooling}
\label{nmrqc:cooling}

We recall from section~\ref{impl:init} that surprisingly, the
exponential cost characteristic of effective pure states is not
inherent to the use of ``high temperature'' qubits ($\hbar \omega \ll
k_BT$). Schulman and Vazirani, invented an algorithm to cool a subset
of the spins in a molecule down to the ground state without any
exponential overhead~\cite{Schulman99a,Chang01a}.

\begin{figure}[h]
\bcen
\vspace*{1ex}
\includegraphics*[width=5.5cm]{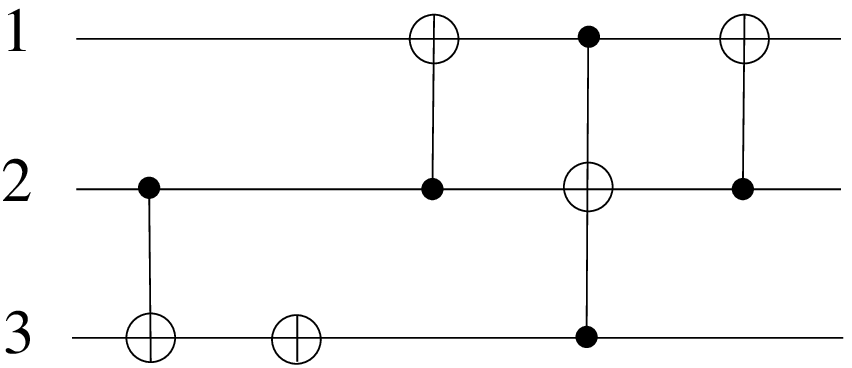} 
\vspace*{-2ex}
\ecen
\caption{A quantum circuit that implements the Schulman-Vazirani 
boosting procedure. The controlled-swap (Fredkin) gate has been
replaced by two {\sc cnot}'s and a {\sc Toffoli} gate, as in
Fig.~\protect\ref{fig:fredkin_circuit}.}
\label{fig:PTcircuit1}
\end{figure}

The following ``boosting procedure'' serves as the building block for
this algorithm (Fig.~\ref{fig:PTcircuit1}). Given
three qubits $1, 2,$ and $3$ with identical $\epsilon=\epsilon_0$, the
initial state $\ket{x_1} \ket{x_2} \ket{x_3}$ is one of the eight
possible states $\ket{0}\ket{0}\ket{0}, \ket{0}\ket{0}\ket{1}, \ldots,
\ket{1}\ket{1}\ket{1}$, with respective probabilities
$(\frac{1+\epsilon_0}{2})^3$, $(\frac{1+\epsilon_0}{2})^2
(\frac{1-\epsilon_0}{2})$, $\ldots,$ $(\frac{1-\epsilon_0}{2})^3$.\\

First perform a {\sc cnot} operation on $3$ conditioned on the state
of $2$.  The new state of the three qubits is $\ket{x_1'}
\ket{x_2'} \ket{x_3'} = \ket{x_1} \ket{x_2} \ket{x_2 \oplus x_3}$,
where $\oplus$ denotes addition modulo $2$.  Note that {\em
conditioned} on $\ket{x_3'}=\ket{0}$, the polarization of $2$ is now
$\frac{2\epsilon_0}{1+{\epsilon_0}^2}$ ($2$ is almost twice as
``cold'' as before); {\em conditioned} on $\ket{x_3'}=\ket{1}$, the
polarization of $2$ is $0$ ($2$ is at infinite temperature). However,
{\em overall}, the polarization of $2$ is still the same as before,
$\epsilon_0$. The polarization of $1$ is of course also still
$\epsilon_0$.  We then perform a {\sc not} operation on $3$ followed
by a {\sc Fredkin} gate with $3$ as the control qubit. The result is
that $1$ and $2$ are swapped if and only if $\ket{x_3'}=\ket{0}$ (and
thus if and only if $2$ has been cooled): $\ket{x_1''} \ket{x_2''}
\ket{x_3''} = \ket{x_2'} \ket{x_1'} \ket{x_3'} $ if
$\ket{x_3'}=\ket{0}$, and $\ket{x_1''} \ket{x_2''} \ket{x_3''} =
\ket{x_1'} \ket{x_2'} \ket{x_3'}$ otherwise. On average, $1$ will thus
be colder than before. The resulting polarization of $1$ is $\epsilon
= \frac{3\epsilon_0}{2}+\mathcal{O}\mathnormal{({\epsilon_0}^3)}$,
where the higher order terms are negligible, so the polarization of
spin $1$ is enhanced by a factor of $3/2$.

In order to achieve increasingly higher polarizations, this boosting
procedure must be applied repeatedly, whereby a fraction of the cold
spins obtained from one round is made progressively colder in the
next.  Spins of little or no polarization are discarded in each
round. Analyzing the polarization transfer using energy and
temperature considerations, Schulman and Vazirani showed
\cite{Schulman99a} that the progression of rounds can be arranged so
that $k$ bits with nearly optimal enhancement can be extracted,
approaching the entropy bound of Eq.~\ref{eq:entropy_cons}.
Furthermore, the number of elementary operations (pulses and delay
times in NMR) required to accomplish the entire process is only
${\mathcal{O}} (n \, \mbox{log}
\, n)$.  In summary, \\

{\em the highly random initial state of room temperature nuclear spins
represents no fundamental obstacle to scalable quantum computation.}\\

However, the prefactor in the overhead in the number of spins is $n/k
\approx 1/\epsilon_0^2$ (Eq.~\ref{eq:k_max}, for small $\epsilon_0$),
which is unreasonably high (about $10^{9}$) for thermally equilibrated
nuclear spins at room temperature with current magnetic field
strengths. It means we would need a molecule with at least $k 10^9$
spins in order to obtain a $k$-qubit computer. This is clearly
impractical.

Until hyperpolarization techniques become much more advanced, the
significance of the Schulman-Vazirani scheme for NMR quantum computing
is thus purely at a fundamental level: NMR quantum computing is in
principle scalable.  However, for systems
with much higher initial polarizations, Schulman-Vazirani cooling can
be very useful if it is difficult to otherwise obtain completely pure
qubits.

%%%%%%%%%%%%%%%%%%%%%%%%%%%%%%%%%%%%%%%%%%%%%%%%%%%%%%%%%%%%%%%%%%%%%%
%%%%%%%%%%%%%%%%%%%%%%%%%%%%%%%%%%%%%%%%%%%%%%%%%%%%%%%%%%%%%%%%%%%%%%

\section{Read-out}
\label{nmrqc:meas}

\subsection{NMR spectra}

\subsubsection{Measurement procedure}

The magnetic signal of a {\em single} nuclear spin is to weak to be
directly detected\footnote{Under certain circumstances, the spin
states can be inferred via optical techniques. This is the case for
example in ion traps.}. Therefore, NMR experiments are done using a
large {\em ensemble} of identical molecules, typically on the order of
$10^{18}$, disolved in a liquid solvent. The same\footnote{This
requires extremely homogeneous magnetic fields (both DC and RF). In
practice, the operations applied to different molecules are only
approximately the same (see section~\ref{expt:apparatus}).} operations
are applied to all the molecules in the ensemble, so the final state
of the spins is the same in all molecules.

The measurement is done with an RF coil mounted next to the sample
(section~\ref{expt:apparatus}), which records the oscillating magnetic
signal produced by the transverse component of the magnetic moment of
the precessing spins (the longitudinal component does not precess and
is not picked up by the coil); this time-domain signal is
Fourier-transformed in order to obtain a {\em spectrum}.

Different spins (qubits) in a molecule are spectrally distinguishable
via their Larmor frequencies $\omega_i$ (section~\ref{nmrqc:qubits}),
and the amplitude and phase of the different spectral lines give
information about the respective spin states.  Mathematically, the
time-domain signal of spin $i$ can be expressed as
\be
V(t) = 2 V_0 \mbox{Tr} 
\left[e^{-i{\cal H}t/\hbar} \rho(0) e^{i{\cal H}t/\hbar} (-i I_x^i - I_y^i) \right]
\,, \label{eq:FID} 
\ee
where $\rho(0)$ is the density matrix at the start of the measurement
and $V_0$ is the maximum signal strength (discussed on
page~\pageref{nmrqc:snr}). The phases of the observable $(-i I_x^i -
I_y^i)$ are chosen\footnote{Other authors have adopted different
conventions.} such that a positive absorptive line corresponds to a
spin along $-\hat{y}$, a negative absorptive line to a spin along
$+\hat{y}$, and positive and negative dispersive lines to a spin along
$+\hat{x}$ and $-\hat{x}$ respectively. Eq.~\ref{eq:FID} represents
the signal in the lab frame, but by mixing the signal with a reference
oscillator at $\omega_0^i$, we obtain instead the {\em expectation
value} of $-i I_x^i - I_y^i$ in the rotating frame, which is the
relevant reference frame for quantum computing.  If $\rho$ is mixed,
as is the case in room temperature experiments, the expectation value
represents an averaged read out over the statistical mixture of
states. What we observe is the excess of spins in the most populated
state(s).

Since a spin along the $\pm \hat{z}$ axis of the Bloch-sphere, which
corresponds to the $\{\ket{0},\ket{1}\}$ basis, does not produce an
NMR signal, we have to {\em change basis} via a $R_x(90)$ {\em
read-out pulse} in order to perform a measurement in the
$\{\ket{0},\ket{1}\}$ basis.  With the above phase conventions, a spin
in $\ket{0}$ before the read-out pulse will give a positive absorptive
line after the read-out pulse, and a spin in $\ket{1}$ will give a
negative line. Inspection of the spectrum acquired after a read-out
pulse thus immediately reveals the projection of the spin state onto
the $\{\ket{0},\ket{1}\}$ basis just before the read-out pulse.

\subsubsection{Measurement process}

What is really happening to the spins during the measurement process?
What difference does it make whether or not an observer looks at the
signal, or even whether the signal is recorded? And why can we
accurately measure both the (non-commuting) $\hat{x}$ and $\hat{y}$
components of the state of a quantum mechanical object ?

The measurement of the spin states in NMR is a weak measurement (see
section~\ref{impl:readout}): the measuring apparatus, the RF coil, is
present all the time, but it is only very weakly coupled to the
nuclear spins and contributes very little to decoherence. Of course,
the spins still decohere through interactions with other spins and
with the ``bath'', and in addition the spins dephase due to
$\vec{B}_0$ inhomogeneities. The oscillation of the magnetic signal
therefore decays over time (usually exponentially). The decaying time
domain signal picked up by the the RF coil is called the {\em free
induction decay} (FID).

If the envelope of the FID decays as $e^{-t/T}$ (usually $T=T_2^*$,
defined in section~\ref{nmrqc:decoherence}), the Fourier transform of
the FID is a Lorentzian line, 
\be
\propto \frac{1}{1+(\omega-\omega_0)^2} - \frac{i\omega}{1+(\omega-\omega_0)^2} \,,
\ee
which has a linewidth at half height of
\be
\Delta f = \frac{\Delta \omega}{2\pi} = \frac{1}{2\pi T} \,.
\label{eq:linewidth}
\ee

Since the measurement is weak, only very little information can be
obtained about the state of individual spins. However, thanks to the
large number of identical molecules in an NMR sample, we can acquire
much more information about the (average) spin state than we could
even in principle ever obtain about the state of an individual
spin. For example, the built-in averaging nature of ensemble
measurements allows us to directly measure the expectation value of
two non-commuting observables.

Ensemble averaged measurements can thus in some respects provide more
information than projective measurements on single quantum systems. At
the same time, averaging erases certain information which quantum
algorithms rely on. Fortunately, all current quantum algorithms can be
modified to circumvent this difficulty, as explained in
section~\ref{impl:readout}.

\subsubsection{Multiplet fine structure}

Extra information about the spin states is contained in the multiplet
fine structure of the spectra. The spectrum of each spin in a $n$-spin
molecule may be split in up to $2^{n-1}$ lines, due to $J$-coupling
terms in the Hamiltonian which modulate the time domain signal of
Eq.~\ref{eq:FID} by $J$ Hertz. After we have determined the magnitude
and sign of the $J$ couplings, we can then associate each line in the
multiplet with the state of the other spins, as shown in
Fig.~\ref{fig:5spin_multiplet} for a five-spin molecule.

\bfig
\bcen
\vspace*{1ex}
\includegraphics*[width=6cm]{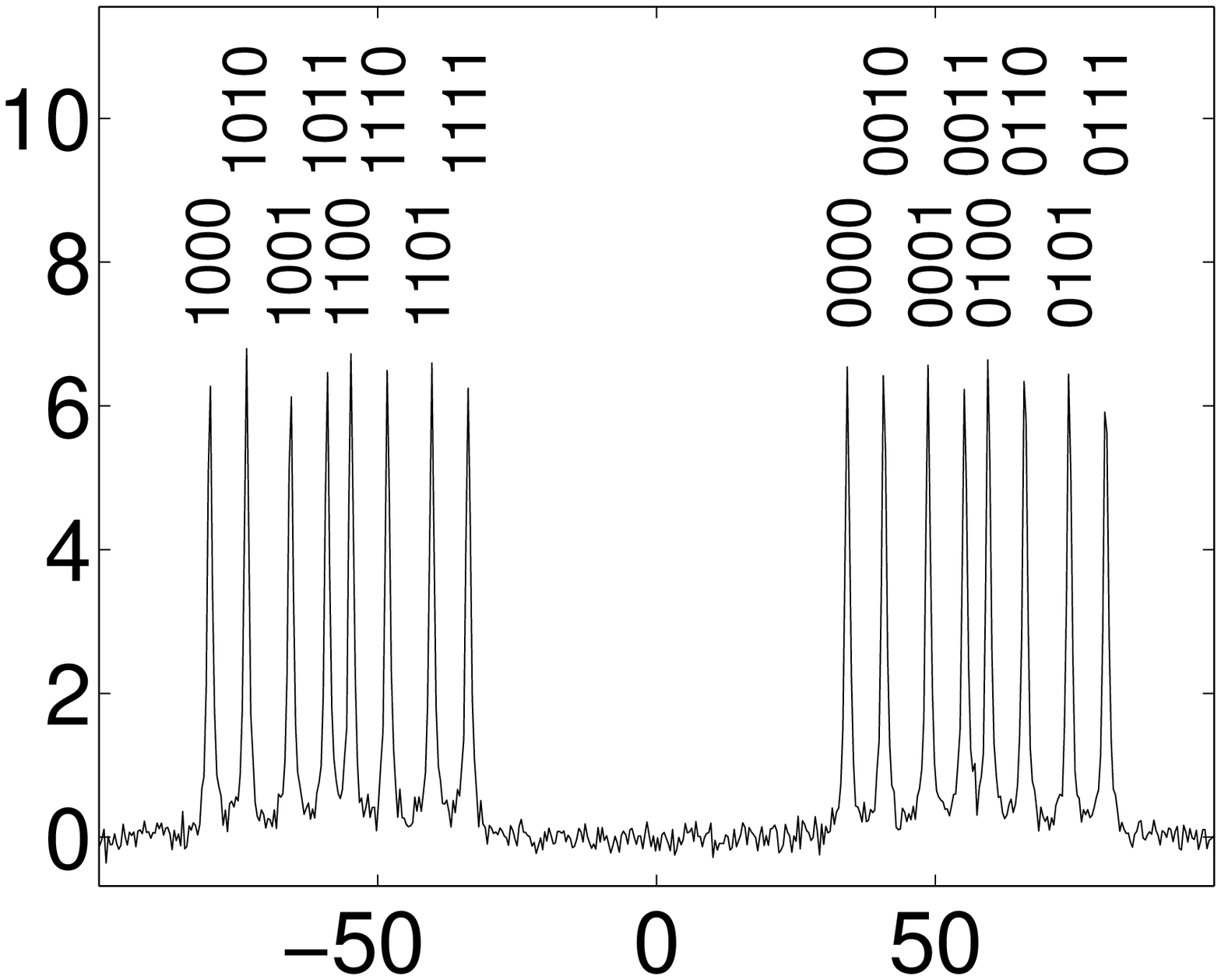} 
\vspace*{-2ex}
\ecen
\caption{The thermal equilibrium spectrum (amplitude of the real part) 
of spin $1$ in a molecule of five coupled spins (more details on this
molecule are given in section~\ref{expt:order}).  Frequencies are
given in units of Hz, with respect to $\omega_0^1$. The state of the
remaining spins is as indicated, based on $J_{12}<0$ and $J_{13},
J_{14}, J_{15} > 0$; furthermore, $|J_{12}| > |J_{13}| > |J_{15}| >
|J_{14}|$.}
\label{fig:5spin_multiplet}
\efig

The presence or abscence of specific lines in a multiplet of one spin
can thus reveal information about the other spins. For example, for an
effective pure ground state (section~\ref{nmrqc:eff_pure}), the only
line we expect to see in the spectrum of
Fig.~\ref{fig:5spin_multiplet} is the line labeled $0000$. Similarly,
in logical labeling (section~\ref{nmrqc:loglab}), the multiplet
structure can be used to identify the logically labeled subspace.  We
have used the extra information in the fine structure in many of the
experiments presented in chapter~\ref{ch:expt}.

\subsubsection{Signal-to-noise ratio}
\label{nmrqc:snr}

The maximum NMR signal, $V_0$, measured via a pick-up coil after
applying a read-out pulse to a spin in thermal equilibrium, is
proportional to
\begin{enumerate}
\item the number of molecules, which is linear in the volume, $V$, and 
the concentration, $n_c$,
\item the number of equivalent sites, $n_e$, in the molecule for the 
spin (e.g. this number is three for $^1$H in CH$_3$Cl),
\item the equilibrium polarization, $\epsilon_0$, which is proportional 
to $\gamma$, $B_0$ and $1/T_s$, where $T_s$ is the absolute
temperature of the sample,
\item $\omega_0$ (because the measurement is inductive), which is 
proportional to $\gamma$ and $B_0$,
\item the quality factor, $Q$, of the coil,
\item the filling factor, $\eta$, (the fraction of the coil volume 
occupied 
by the sample),
\item a factor, $K$, which depends on the coil geometry and reflects the 
coupling of the spins to the coil.
\end{enumerate}

The {\em noise} in NMR measurements is normally dominated by the
thermal noise of the coil. The rms noise amplitude is proportional to
the square root of
\begin{enumerate}
\item the absolute temperature of the coil, $T_c$,
\item the shunt resistance of the tuned circuit, $R=QL\omega_0$, ($L$ 
is the inductance),
\item the width of the narrow band audio-filter, $\Delta f$.
\end{enumerate}

How much the lines in an NMR spectrum rise above the noise level
depends not only on the actual signal strength and noise level, but
also on the degree to which the signal is spread out in
frequency. Thus, the signal-to-noise ratio of an NMR spectrum is also
proportional to $1/m$, where $m$ is the multiplicity of the multiplet,
and to $T_2^*$, as $\int_0^\infty \exp(-t/T_2^*) = T_2^*$ (a long
$T_2^*$ gives narrow and thus tall lines).

In summary, the signal-to-noise ratio can be expressed as
\be
\frac{S}{N} \propto 
\frac{n_c V n_e \gamma^2 B_0^2 Q \eta K T_2^*}
{m T_s (T_c Q L \gamma B_0 \Delta f)^{1/2}}
= \frac{n_c V n_e \gamma^{3/2} B_0^{3/2} Q^{1/2} \eta K T_2^*}
{m T_s (T_c L \Delta f)^{1/2}}
\,.
\label{eq:SNR}
\ee

In practice, many of these parameters are interdependent. For a more
detailed discussion of the signal-to-noise ratio, see
Ref~\cite{Hoult76a}.

%%%%%%%%%%%%%%%%%%%%%%%%%%%%%%%%%%%%%%%%%%%%%%%%%%%%%%%%%%%%%%%%%%%%%%%%%%%

\subsection{Quantum state tomography}

The spectra of a few select spins suffice to obtain the answer to a
computation. Nevertheless, the full density matrix conveys a lot of
extra information, which can be used to expose the presence of errors
not visible in the single output spectra and furthermore is a useful
tool for debugging pulse sequences.

The procedure for reconstructing the density matrix is called quantum
state tomography \cite{Chuang98b,Chuang98a,Chuang98c}. In order to
explain the idea behind this procedure, we take another look at the
signal of Eq.~\ref{eq:FID}. The operator $-i I_x^i - I_y^i$ selects
specific entries in the density matrix, called {\em single quantum
coherence} (SQC) elements.  The SQC elements ``connect'' basis states
which differ by only one quantum of energy (for example $\ket{00}
\leftrightarrow \ket{01}$ but not $\ket{00} \leftrightarrow \ket{11}$). 
The SQC elements of a two-spin density matrix are
\be
\left[\matrix{	. & . & \times & . \cr
		. & . & . & \times \cr
		\times & . & . & . \cr 
		. & \times & . & . \cr
		}\right]
\quad \mbox{and} \quad
\left[\matrix{	. & \times & . & . \cr
		\times & . & . & . \cr 
		. & . & . & \times \cr 
		. & . & \times & . \cr }\right]
\label{eq:SQC}
\ee
for spins $1$ and $2$ respectively. We recall that the density matrix
is Hermitian, so the entries above and below the diagonal (going from
upper left to lower right) are each other's complex conjugate. In
Eq.~\ref{eq:SQC}, there are thus only two independent SQC elements for
each spin. Those complex numbers are directly proportional to the area
underneath the two spectral lines in each of the doublets of the
two-spin spectrum.  For an $n$ spin system, the $2^{n-1}$ lines within
each multiplet can be identified with $2^{n-1}$ SQC elements per spin.

Quantum state tomography then consists of repeating the computation
many times, each time looking at the final state of the spins ``from a
different angle'', by applying different sets of read-out pulses which
rotate different elements of the density matrix into observable
positions. In an idealized experiment (no imperfections), the density
matrix uniquely follows from the area underneath the individual lines
within all the multiplets. In real experiments, all the spectral
information may not be compatible with each other, but we can obtain a
good estimate of the actual density matrix via a least-squares
fit.

Since quantum state tomography involves on the order of $4^n$
experiments (the number of degrees of freedom in the density matrix),
it is practical only for experiments involving a few spins; we have
reconstructed density matrices only for experiments with two or three
spins (sections~\ref{expt:dj}-~\ref{expt:grover3}).

%%%%%%%%%%%%%%%%%%%%%%%%%%%%%%%%%%%%%%%%%%%%%%%%%%%%%%%%%%%%%%%%%%%%%%
%%%%%%%%%%%%%%%%%%%%%%%%%%%%%%%%%%%%%%%%%%%%%%%%%%%%%%%%%%%%%%%%%%%%%%

\section{Decoherence}
\label{nmrqc:decoherence}

The decoherence process of uncoupled nuclear spins is well described
by a combination of two phenomena: longitudinal and transverse
relaxation\footnote{For coupled spins (dipole coupled or $J$
coupled), the decoherence process also includes cross-relaxation and
the nuclear Overhauser effect (NOE), but we shall not discuss those
here.}~\cite{Abragam61a,Slichter96a}. These two processes are closely
related to generalized amplitude damping and phase damping
respectively, which have been described mathematically in
section~\ref{impl:coherence}.  We will first present the main
decoherence mechanisms and then describe standard methods to measure
the characteristic relaxation time constants.

%%%%%%%%%%%%%%%%%%%%%%%%%%%%%%%%%%%%%%%%%%%%%%%%%%%%%%%%%%%%%%%%%%%%

\subsection{Principal mechanisms}
\label{nmrqc:dec_mech}

Relaxation of nuclear spins is caused by fluctuations in the magnetic
field experienced by the spins. Whether the magnetic field
fluctuations contribute to energy exchange with the bath or phase
randomization depends on the {\em time scale} of the
fluctuations. Roughly speaking, we have that

\begin{itemize}
\item fluctuations at $\omega_0$ lead to efficient energy
exchange with the spins (the bath and the spins act as RF transmitters
and receivers tuned to the same frequency),
\item fluctuations at zero frequency, i.e. slow
fluctuations, give rise to phase randomization.
\end{itemize}

Depending on the mechanism, however, fluctuations at $\omega_0$ and at
$2\omega_0$, may also contribute to phase randomization. Similarly,
fluctuations at $2\omega_0$ may also contribute to energy
exchange. Finally, if two or more coupled spins are present, energy
exchange is also promoted by fluctuations at the sum and difference
frequencies of the spins ($\omega_0^i \pm \omega_0^j$)
\cite{Abragam61a}.

The following mechanisms at a microscopic scale contribute to
relaxation of nuclear spins in liquid solution:

\begin{enumerate}
\item Intermolecular dipole-dipole interactions with nuclear spins. \\ 
This interaction is modulated by molecular translation and rotation
and contributes to phase randomization ($T_2$). It can be dominant in
relatively large molecules, when all the high-$\gamma$ nuclei are well
separated in the molecule.
\item Intramolecular dipole-dipole interactions with nuclear spins. \\
This interaction fluctuates due to molecular tumbling; its
contribution to $T_1$ scales as $T_1^{-1} \propto \gamma_i \gamma_j /
r_{ij}$, where $\gamma_i$ and $\gamma_j$ are the gyromagnetic ratios
for spins $i$ and $j$, and $r_{ij}$ is the distance between the two
nuclei.
\item Intra- and intermolecular dipole-dipole interactions with 
electron spins.\\ If unpaired electrons are present, such as in
paramagnetic ions and free radicals, this decoherence mechanism will
usually dominate because electrons have a much large magnetic
moment than the nuclei.
\item Chemical shift anisotropy.\\ If the chemical shift is
anisotropic, it rapidly fluctuates due to molecular tumbling. This
effect increases with magnetic field strength as $T_1^{-1} \propto
B_0^2$ (chemical shifts are linear in the field strength).
\item Spin-rotation interaction.\\  Molecular rotations create 
magnetic fields which are modulated due to collisions. This mechanism
is important especially in small, symmetric molecules.
\item Scalar coupling. \\ Rapid fluctuations in the $J$ coupling
contribute to relaxation.
\item Quadrupolar coupling. \\ Nuclei with a spin quantum number larger
than 1/2 don't have a spherically symmetric nuclear charge. As a
result, such nuclei interact with electric field
gradients. Fluctuations in the electric field gradient due to
molecular tumbling cause quadrupolar nuclei to relax very fast.
\item Coupling to quadrupolar nuclei. \\ The rapidly 
fluctuating spin state of quadrupolar nuclei contributes to relaxation
of other spins.
\item Chemical exchange. \\ Fast chemical exchange of part of a
molecule causes the chemical shifts of nuclei in the remaining part of
the molecule to rapidly jump back and forth between two or more values.
\end{enumerate}

In addition, fluctuations due to noisy RF amplifiers or other external
sources which emit electro-magnetic fields at $\omega_0$ shorten
$T_1$. Similarly, magnetic field inhomogeneities (in $B_0$ or $B_1$)
shorten the apparent $T_2$. However, magnetic field inhomogeneities
can in principle be easily unwound via refocusing pulses, provided
diffusion rates are slow compared to the time scale of the
operations. The literature therefore distinguishes between $T_2$, the
intrinsic transverse relaxation time constant, and $T_2^*$, which
incorporates both intrinsic relaxation and inhomogeneous broadening.

\subsubsection{Minimizing relaxation}

To some degree, relaxation is influenced by parameters under the
control of the experimenter. For quantum computation, it is crucial to
maximally take advantage of the possibilities to reduce relaxation.

A first set of guidelines hinges on the idea of {\em motional
narrowing}. Rapid molecular tumbling shortens the correlation times of
many fluctuations and therefore tends to lengthen $T_2$ (which is
usually much shorter than $T_1$). The tumbling rate depends on the
following parameters:

\begin{enumerate}
\item Molecule size: small molecules tumble more easily.
\item Viscosity of the solvent: lower viscosity obviously promotes 
rapid tumbling (supercritical solvents are ideal from this point of
view, as they combine the high density and solubility of liquids with
the low viscosity of gases).
\item Temperature: higher temperatures provide more thermal energy for 
tumbling and also tend to reduce solvent viscosity.
\end{enumerate}

\noindent Additional guidelines for sample preparation and molecule selection are:

\begin{enumerate}
\item Remove oxygen and other paramagnetic impurities from the solution.
\item Avoid quadrupolar nuclei in the molecule.
\item Reduce the solute concentration in order to reduce intermolecular 
relaxation, and choose solvents preferrably with nonmagnetic nuclei or
low-$\gamma$ nuclei, or else with different nuclear species than those
in the solute molecule, because like nuclei relax each other more
efficiently than unlike nuclei.
\item Remove reagents with which the molecule may exchange chemically.
\end{enumerate}

\noindent Finally, non-intrinsic relaxation can be minimized by 

\begin{enumerate}
\item making the $B_0$ field as homogeneous as possible.
\item Spinning the sample about the $\hat{z}$ axis in order to average 
out the remaining transverse field inhomogeneities.
\item Filter out particles in the solvent, as they create magnetic 
field 
inhomogeneities (assuming their magnitic susceptibility is different
than that of the solvent).
\item Using RF coils with good homogeneity.
\item Blanking the amplifiers in between RF pulses.
\item Reduce radiation damping by reducing the sample concentration or 
lowering the $Q$ of the probe (during a pulse, the spins are tipped
into the transverse plane, so they induce a voltage in the coil which
in turn tips the spins back).
\end{enumerate}

%%%%%%%%%%%%%%%%%%%%%%%%%%%%%%%%%%%%%%%%%%%%%%%%%%%%%%%%%%%%%%%%%%%%

\subsection{Characterization}

We have used the following (standard) procedures for measuring
the $T_1$, $T_2$ and $T_2^*$~\cite{Freeman97a}.

{\em Inversion recovery} constitutes a clean measurement of
$T_1$. First, a $180^\circ$ pulse inverts the spin from $+\hat{z}$ to
$-\hat{z}$; then the spin is allowed to relax back to its equilibrium
state $+\hat{z}$ for a variable duration $t$; finally, a $90^\circ$
read out pulse tips the spin into the $\hat{x}\hat{y}$ plane and the
signal is recorded. The pulse sequence is thus
\be
X^2 - t - X - \mbox{acquisition} \,.
\ee
With properly set receiver phase settings, the peak height of the measured spectrum varies with
$t$ as
\be
S = S_0 \left[1 - \alpha e^{-t/T_1}\right] \,,
\label{eq:T1}
\ee
where $\alpha$ is a fitting parameter which compensates for incomplete
inversion due to RF field inhomogeneities (ideally $\alpha=2$).
Typical values for $T_1$ are a few seconds to a few tens of seconds.\\

$T_2^*$ is the time constant of the free induction decay, so $T=T_2^*$
in Eq.~\ref{eq:linewidth} (assuming the line is Lorentzian, which is
not necessarily the case in an inhomogeneous magnetic field), and
\be
\Delta f = \frac{\Delta \omega}{2\pi} = \frac{1}{2\pi T_2^*} \,.
\ee
We can thus easily derive $T_2^*$ from the {\em linewidth} at half
height.  Measurement of $T_2$ requires that dephasing due to magnetic
field inhomogeneities be refocused. The standard measurement for $T_2$
is the {\em Carr-Purcell-Meiboom-Gill} (CPMG) pulse sequence. First a
$Y$ pulse takes the spin to $+\hat{x}$. Then, the subsequence
\be
\frac{\tau}{4} \; X^2 \; \frac{\tau}{2} \; X^2 \; \frac{\tau}{4}
\ee
is repeated $k$ times, where $k$ is arrayed, and the signal is
recorded for each value of $k$. Typical values of $\tau$ are 1-10 ms,
short enough such that minimal diffusion takes place during the delay
times, and long enough such that the duty cycle (the ratio of the
duration of the pulses over the delay times) is not too high.  The
measured signal will decay exponentially with the total decay time
$t=k\tau$,
\be
S = S_0 \; e^{-t/T_2} \,.
\label{eq:T2}
\ee
In theory for small molecules $T_2 \approx T_1$, although in practice
$T_2$ values were a few tenths of a second to a few seconds in the
molecules we have used, substantially shorter than $T_1$.

We note that the measured values of $T_2$ (and $T_2^*$) do not
correspond exactly to the phase damping time constants defined in
section~\ref{impl:coherence}. The CPMG measurement gives the decay
rate of the single quantum coherence elements of the density matrix,
to which both amplitude damping and phase damping contribute. However,
in practice, often $T_1 \gg T_2$ in which case the CPMG measurement
does give the phase damping time constant, to good approximation.

Finally, the $T_2$ measurement is affected by coupled evolution, in
particular when $\tau$ is on the order of $1/2J$. In a multi-spin
system, it becomes difficult to choose $\tau$ so it is different
enough from $1/2J$ for the various $J$-coupling strengths. Different
choices of $\tau$ give considerably different measured $T_2$'s, so
their meaning is diminished \cite{Vold78a}.

%%%%%%%%%%%%%%%%%%%%%%%%%%%%%%%%%%%%%%%%%%%%%%%%%%%%%%%%%%%%%%%%%%%%
%%%%%%%%%%%%%%%%%%%%%%%%%%%%%%%%%%%%%%%%%%%%%%%%%%%%%%%%%%%%%%%%%%%%

\section{Molecule design}
\label{nmrqc:molecule}

The choice of a suitable molecule is crucial for the success of NMR
quantum computing experiments. The fundamental properties which make a
molecule suitable for quantum computation follow from the preceding
sections.

First, the number of spin-1/2 nuclei in the molecule must be equal or
larger than the required number of qubits. Reasonable choices for
qubits include $^1$H, $^{13}$C, $^{15}$N, $^{19}$F and $^{31}$P, as
they all have a spin-1/2 nucleus, and are found relatively easily in
small organic molecules (however, isotopic labeling is needed to
obtain $^{13}$C and $^{15}$N in high concentration).

Second, in order to be able to complete a large number of two-qubit
operations within the coherence time, we desire
\be
|J_{ij}| \gg \frac{1}{T_2}, \frac{1}{T_1} \,.
\label{eq:J_gg_1/T2}
\ee

Third, in order to have sufficiently slow coupled evolution during
spin-selective shaped pulses, we need $|\omega_1| \gg |J_{ij}|$ and
since spin-selectivity requires $|\omega_0^i - \omega_0^j| >
|\omega_1|$, we desire
\be
|\omega_0^i - \omega_0^j| \gg |J_{ij}| \,.
\label{eq:Domega_gg_J}
\ee
This condition at the same time ensures that the spectra are first
order. We note that Eqs.~\ref{eq:J_gg_1/T2} and~\ref{eq:Domega_gg_J}
automatically guarantuee that $|\omega_0^i - \omega_0^j| \gg 1/T_2,
1/T_1$, such that many one-qubit operations can be done within the
coherence time as well.

Eq.~\ref{eq:Domega_gg_J} is exceedingly well satisfied in
heteronuclear molecules.  In homonuclear molecules, strong chemical
shifts are promoted by strong asymmetries in the molecule.  The normal
range of chemical shifts is about 200 ppm for $^{19}$F (about 100 kHz
at 10 Tesla), 200 ppm for $^{13}$C nuclei (about 25 kHz),
10 ppm for $^1$H (about 5 kHz) and $> 300$ ppm for
$^{15}$N (about 7.5 kHz). In homonuclear molecules, $^{19}$F and
$^{13}$C are thus preferred. $^{19}$F and $^{13}$C also tend to have
strong $J$ couplings, needed to satisfy Eq.~\ref{eq:J_gg_1/T2} while
$^1$H often has smaller $J$ couplings. 

On the one hand, low-$\gamma$ nuclei such as $^{15}$N and $^{13}$C
tend to have longer coherence times than high-$\gamma$ nuclei such as
$^1$H and $^{19}$F.  On the other hand, high-$\gamma$ nuclei such as
$^1$H and $^{19}$F have the advantage that they give the strongest
signals (recall Eq.~\ref{eq:SNR}).  $^{13}$C, $^{15}$N have a low
$\gamma$, and the $\gamma$ of $^{31}$P is intermediate (see
Table~\ref{tab:larmor_freq}).

Section~\ref{nmrqc:dec_mech} discusses several other elements related
to molecule design which affect the coherence time.

We have already seen (section~\ref{nmrqc:2bitgates}) that
Eq.~\ref{eq:J_gg_1/T2} is not binding. This condition can be extended
to say that a sufficient network of $J$'s larger than $1/T_2$ must be
available , such that a two-qubit gate between any pair of spins
(implemented directly or indirectly) takes a short time compared to
the coherence time.

Similarly, Eq.~\ref{eq:Domega_gg_J} assumes that we need to
individually address all the spins, but this isn't always necessary
either (section~\ref{impl:gates}). A polymer with a unit cell $ABC$
which repeats itself $n$ times and terminates on a $D$ (where $A, B,
C$ and $D$ have distinct chemical shifts) can possibly serve as an $n$
qubit computer (Fig.~\ref{fig:coupling_networks}).  The caveat is that
it isn't known how to set up a proper initial state when using nuclear
spins at room temperature in such an architecture.

Finally, there are several more mundane but even more important
practical requirements for quantum computer molecules: they must be
stable (i.e. not decompose) for a reasonably long time, disolve in an
NMR solvent (chloroform, acetone, ether, DMSO, benzene, toluene, among
others), be available or possible to synthesize, be affordable ($99\%$
$^{13}$C or $^{15}$N enriched compounds can be very expensive) and
safe. Indeed, many molecules which one could draw on the board for
their beautiful presumed NMR properties turn out to be unstable, very
hard to synthesize, or toxic.

%%%%%%%%%%%%%%%%%%%%%%%%%%%%%%%%%%%%%%%%%%%%%%%%%%%%%%%%%%%%%%%%%%%%%%
%%%%%%%%%%%%%%%%%%%%%%%%%%%%%%%%%%%%%%%%%%%%%%%%%%%%%%%%%%%%%%%%%%%%%%

\section{Pulse sequence design}
\label{nmrqc:seq_design}

A computation with nuclear spins consists of a carefully designed
sequence of RF pulses separated by delay times, corresponding to
computational steps. Those elementary instructions, pulses and delay
times, can be viewed as the {\em machine language} of an NMR quantum
computer.

The goal of pulse sequence design is to translate a high-level
description of a quantum algorithm into unitary transformations acting
on one or several qubits, then to decompose each unitary operation
into one- and two-qubit gates, and finally into pulses and delay
times. This process is analogous to {\em compiling} code on
traditional computers.

We know from previous sections on quantum gates (\ref{qct:gates}) and
their implementation in NMR
(\ref{nmrqc:1bitgates}-\ref{nmrqc:2bitgates}) that many pulse
sequences result in exactly the same unitary transformation. Good
pulse sequence design therefore attempts to find the {\em shortest and
most robust} pulse sequence that implements the desired
transformations.

A key point in pulse sequence design is that the process must itself
be {\em efficient}. For example, suppose an algorithm acts on five
qubits with initial state $\ket{00000}$ and that the final state is
$(\ket{01000}+\ket{01100}/\sqrt{2}$. The overall result of the
sequence of unitary transformations is thus that qubit $2$ is flipped
and that qubit $3$ is placed in an equal superposition of $\ket{0}$
and $\ket{1}$. This net transformation can obviously be obtained
immediately by the sequence $X^2_2 Y_3$. However, the effort needed to
compute this net transformation generally increases exponentially with
the problem size, so such extreme simplifications are not practical.

\subsection{Simplification at three levels}

At the most abstract level of pulse sequence simplification, careful
study of a quantum algorithm can give insight in how to reduce the
resources needed. For example, we recall that a key step in both the
Deutsch-Jozsa algorithm and the Grover algorithm can be described as
the transformation $\ket{x}\ket{y} \rightarrow \ket{x}\ket{x \oplus
y}$ (see Eqs.~\ref{eq:dj_U_f} and~\ref{eq:grover_U_f}), where
$\ket{y}$ is set to $(\ket{0}-\ket{1})/\sqrt{2}$, so that the
transformation in effect is $\ket{x}(\ket{0}-\ket{1})/\sqrt{2}
\rightarrow (-1)^{f(x)} \ket{x}(\ket{0}-\ket{1})/\sqrt{2}$. We might 
thus as well leave the last qubit out as it is never changed.

At the next level, that of quantum circuits, we can use the
simplification rules such as those illustrated in
Fig.~\ref{fig:simplify_circuits}. In this process, we can fully take
advantage of commutation rules to move building blocks around, as
illustrated in Fig.~\ref{fig:commutations} (see also
page~\pageref{page:commutation}). Commutation rules can also tell us
which gates can in principle be executed simultaneously.  Furthermore,
we can use the fact that $U$ acting on a diagonal density matrix
doesn't need to have the right phases (e.g. compare
Eq.~\ref{eq:U_cnot_tilde} and Eq.~\ref{eq:U_cnot_2}). Finally, we can
take advantage of the fact the most building blocks have many
equivalent implementations, as shown in
Fig.~\ref{fig:equivalent_circuits}.

\bfig
\bcen
\vspace*{1ex}
\includegraphics*[width=13cm]{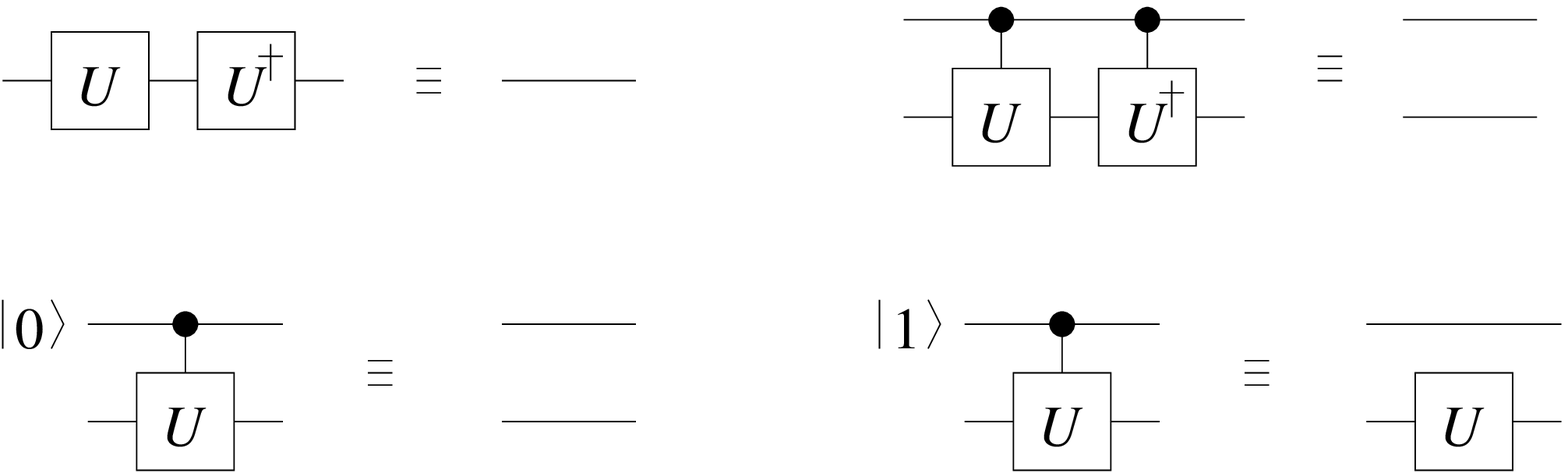} 
\vspace*{-2ex}
\ecen
\caption{Simplification rules for quantum circuits}
\label{fig:simplify_circuits}
\efig

\bfig
\bcen
\vspace*{1ex}
\includegraphics*[width=11cm]{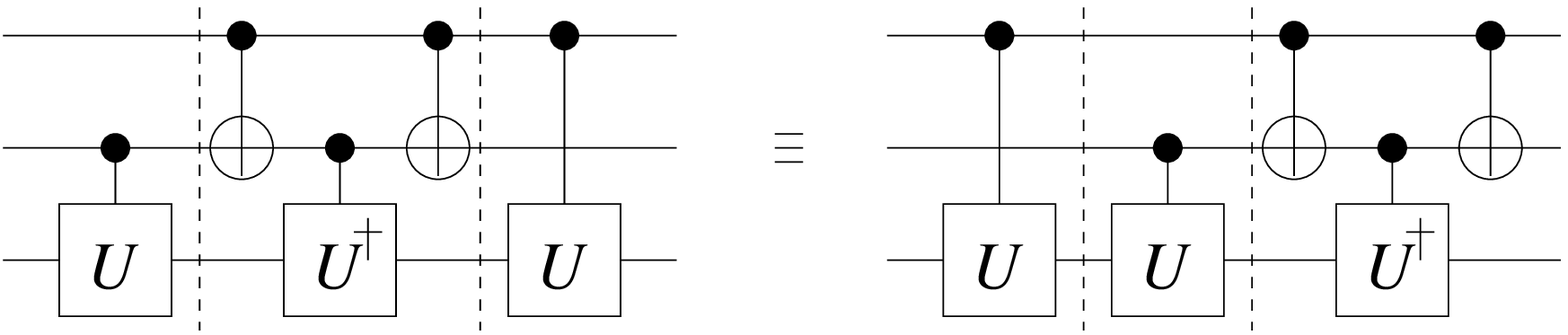} 
\vspace*{-2ex}
\ecen
\caption{Commutation of unitary operators can help simplify quantum 
circuits by moving building blocks around such that cancellations of
operations as in Fig.~\protect\ref{fig:simplify_circuits} become
possible. For example, the three components (separated by dashed
lines) in these two equivalent realizations of the {\sc Toffoli} gate
commute with each other and can thus be executed in any order.}
\label{fig:commutations}
\efig

\bfig
\bcen
\vspace*{1ex}
\includegraphics*[width=11cm]{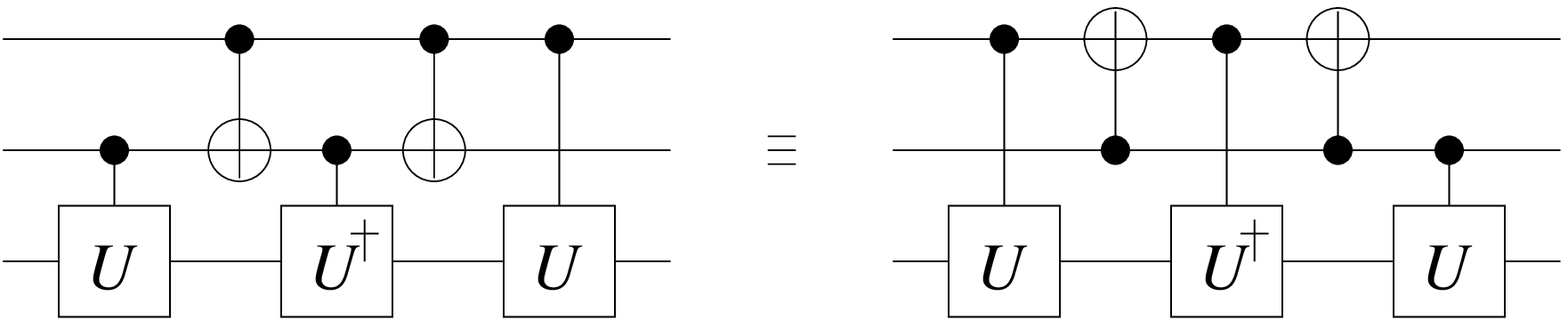} 
\vspace*{-2ex}
\ecen
\caption{Choosing one of several equivalent implementations can help 
simplify quantum circuits, again by enabling cancellation of
operations as in Fig.~\protect\ref{fig:simplify_circuits}. The {\sc
Toffoli} gate has two control qubits, whose role is symmetric and can
thus be swapped.}
\label{fig:equivalent_circuits}
\efig

At the lowest level, that of pulses and delay times, further
simplification is possible by taking out adjacent pulses which cancel
out, such as $X$ and $\bar{X}$. In fact, we can go one step further
and choose those pulses sequence for each building block which will
give the most cancellation of pulses. For this purpose, it is
convenient to have a library of equivalent implementations for the
most commonly used quantum gates. For example, two equivalent
decompositions of a {\sc cnot}$_{12}$ gate (with $J_{12} > 0$) are
\be
Z_1 \, \bar Z_2 \, X_2 \; 1/2J \;Y_2 \,,
\label{eq:cnot_decomposition}
\ee
where time goes {\em from right to left}, and
\be
\bar Z_1 \, \bar Z_2 \, \bar X_2 \; 1/2J \; \bar Y_2 \,,
\ee
and two equivalent implementations of the {\sc hadmard} gate on qubit 2 are
\be
X_2^2 \, Y_2
\ee
and 
\be
\bar Y_2 \, X_2 \,.
\label{eq:had_decomposition}
\ee
Then, if we want to perform a {\sc hadamard} operation on qubit 2
followed by a {\sc cnot}$_{12}$ gate, it is best to choose the
decompositions of Eqs.~\ref{eq:cnot_decomposition}
and~\ref{eq:had_decomposition}, such that the resulting pulse
sequence,
\be
Z_1 \, \bar Z_2 \, X_2 \; 1/2J \; Y_2 \quad \bar Y_2 \, X_2
\ee
simplifies to
\be
Z_1 \, \bar Z_2 \, X_2 \; 1/2J \; X_2 \,.
\ee

Furthermore, refocusing sequences can be kept as simple as
possible by examing which couplings really need to be refocused. Early
on in a pulse sequence, several qubits may still be along $\pm
\hat{z}$ in which case their mutual coupling has no effect and thus
need not be refocused. Similarly, if a subset of the qubits can be
traced out at some point in the sequence, the mutual interaction
between these qubits does not matter anymore, so only their coupling
with the remaining qubits must be refocused.

There is of course some interplay between the three levels of pulse
sequence simplification. For example, the value of individual $J$
couplings doesn't come in explicitly until the lowest level, but it is
possible (and important) to work around small or zero couplings
already at the level of quantum circuits.

Finally, we note that pulse sequence design the way we have described
it assumes that the quantum computer molecule is fully known and
characterized in advance. In contrast, conventional NMR pulse
sequences must work for any molecule, because the spectral properties
of the molecule are usually not known in advance. Exact knowledge of
the Larmor frequencies and $J$-coupling constants allows one not only
to greatly simplify the pulse sequences, but also to achieve much more
accurate unitary transformations than would otherwise be possible.

\subsection{Design for robustness}

The exact choice of pulse sequence greatly affects the robustness
against erroneous unitary evolutions, in particular those due to
coupled evolution during pulses and the inhomogeneity of the RF field
used to pulse the spins. In addition to keeping pulse sequences short,
robustness is thus an important consideration in the process of
designing pulse sequences.

Undesired coupled evolution can be minimized by choosing suitable
pulse shapes (section~\ref{nmrqc:pulse_artefacts}) but also through
pulse sequence design, at the lowest level. Simultaneous pulses,
especially $90^\circ$ pulses, on spins with a large mutual $J$
coupling should be avoided (section~\ref{nmrqc:simpulse_artefacts})
and coupled evolution during pulses can be unwound by adjusting the
adjacent refocusing sequences
(section~\ref{nmrqc:simpulse_artefacts}).

Erroneous evolution because of RF field inhomogeneity can be very
substantial (section~\ref{expt:apparatus}), but can in principle be
unwound: a $X^2$ pulse causes some spread in the spin states and we
expect a subsequent $\bar{X^2}$ pulse to unwind this spread quite
well, definitely much better than another $X^2$ pulse.  For longer
trains of $180^\circ$ pulses, it isn't always so easy to predict which
choice of phase for the pulses is most robust to RF field
inhomogeneities. For example, contrary to our intuition, $X^2 X^2
\bar{X^2} \bar{X^2}$ performs much better than $X^2 \bar{X^2} X^2
\bar{X^2}$ and similar extensions exist for longer trains of 
$180^\circ$ pulses~\cite{Levitt82a}.
%(Fig.~\ref{fig:cancel_pi_pulses})

%\bfig
%\caption{Spectra after four $180^\circ$ pulses. The phases of the 
%pulses are indicated for each case. (UNCOUPLED SPIN and COUPLED SPIN)}
%\label{fig:cancel_pi_pulses}
%\efig

Quantum computing pulse sequences are hardly ever so transparant,
unfortunately. Actual refocusing sequences are complicated by the fact
that spin-selective $180^\circ$ pulses on different spins are
interspersed with each other. Furthermore, $90^\circ$ pulses disturb
possible cancellation between preceding and subsequent $180^\circ$
pulses.

A general framework for undoing systematic errors such as those due to
RF field inhomogeneities is highly desirable. This is clearly an
ambitious undertaking, but it is encouraging to know that very strong
cancellation of such errors {\em has} been observed, even in complex
quantum computing sequences (see sections~\ref{expt:labeling}
and~\ref{expt:grover3}).

%%%%%%%%%%%%%%%%%%%%%%%%%%%%%%%%%%%%%%%%%%%%%%%%%%%%%%%%%%%%%%%%%%%%%%%

\section{Summary}

The main message of this chapter is that nuclear spins in molecules in
liquid solution largely satisfy the five requirements for the
implementation of quantum computers:

\begin{enumerate}
\item $\surd$ spin-1/2 nuclei in a molecule are well-defined qubits,
\item $\surd$ the dynamics of coupled nuclear spins can be 
controlled via RF pulses and delay times, even though certain terms in
the Hamiltonian cannot be switched off,
\item ($\surd$ ) room temperature nuclear spins can be made to look 
like they are at zero temperature, although currently only at an
exponential cost,
\item $\surd$ the state of each qubit can be read out 
spectroscopically, provided a large ensemble of molecules is used,
\item $\surd$ nuclear spins have long coherence times (easily a few 
seconds).
\end{enumerate}

\noindent In the next chapter, we will present a series of 
experiments in which we explore how these methods and concepts
translate into the reality of actual quantum computations.

%% file: expt.tex
    \chapter{Experimental realization of NMR quantum computers}
\label{ch:expt}
\chaptermark{Experimental realization}

After a description of the experimental apparatus\footnote{The
experiments of sections~\ref{expt:dj} and~\ref{expt:2bitcode} took
place in the Chemistry Department at Stanford University. The
remaining experiments took place at the IBM Almaden Research Center.
Both NMR spectrometers are largely identical, but where they differ,
the description is for the instrument at IBM.}, we will give a brief
overview of the experimental NMR quantum computing work performed to
date (section~\ref{expt:overview}). We then present in detail a series
of eight experiments in which we explored quantum computing in
practice. These are an early quantum computation (\ref{expt:dj}), an
early quantum error detection experiment (\ref{expt:2bitcode}), an
explicit demonstration of cold dynamics using room temperature spins
(\ref{expt:labeling}), a quantum computation performed in a liquid
crystal solvent (\ref{expt:lc}), a study of systematic errors with
three spins (\ref{expt:grover3}), an implementation of efficient
cooling of one out of three spins (\ref{expt:cooling}), a realization
of the order-finding algorithm with five qubits (\ref{expt:order}) and
prime factorization of the number fifteen using seven spins and Shor's
algorithm (\ref{expt:shor}).

%%%%%%%%%%%%%%%%%%%%%%%%%%%%%%%%%%%%%%%%%%%%%%%%%%%%%%%%%%%%%%%%%%%%%%
%%%%%%%%%%%%%%%%%%%%%%%%%%%%%%%%%%%%%%%%%%%%%%%%%%%%%%%%%%%%%%%%%%%%%%

\section{Experimental apparatus}
\label{expt:apparatus}

Figure~\ref{fig:apparatus} schematically shows the main components of
an NMR spectrometer. A sample containing a large number of identical
molecules disolved in liquid solution is placed in a strong magnetic
field. Radio-frequency pulses are applied to the sample via a
radio-frequency coil and the same coil is used to detect the magnetic
signal of the spins during read out.  The whole experiment is
controlled by a workstation. We now describe each component in
more detail.

\bfig
\bcen
\vspace*{1ex}
\includegraphics*[width=14cm]{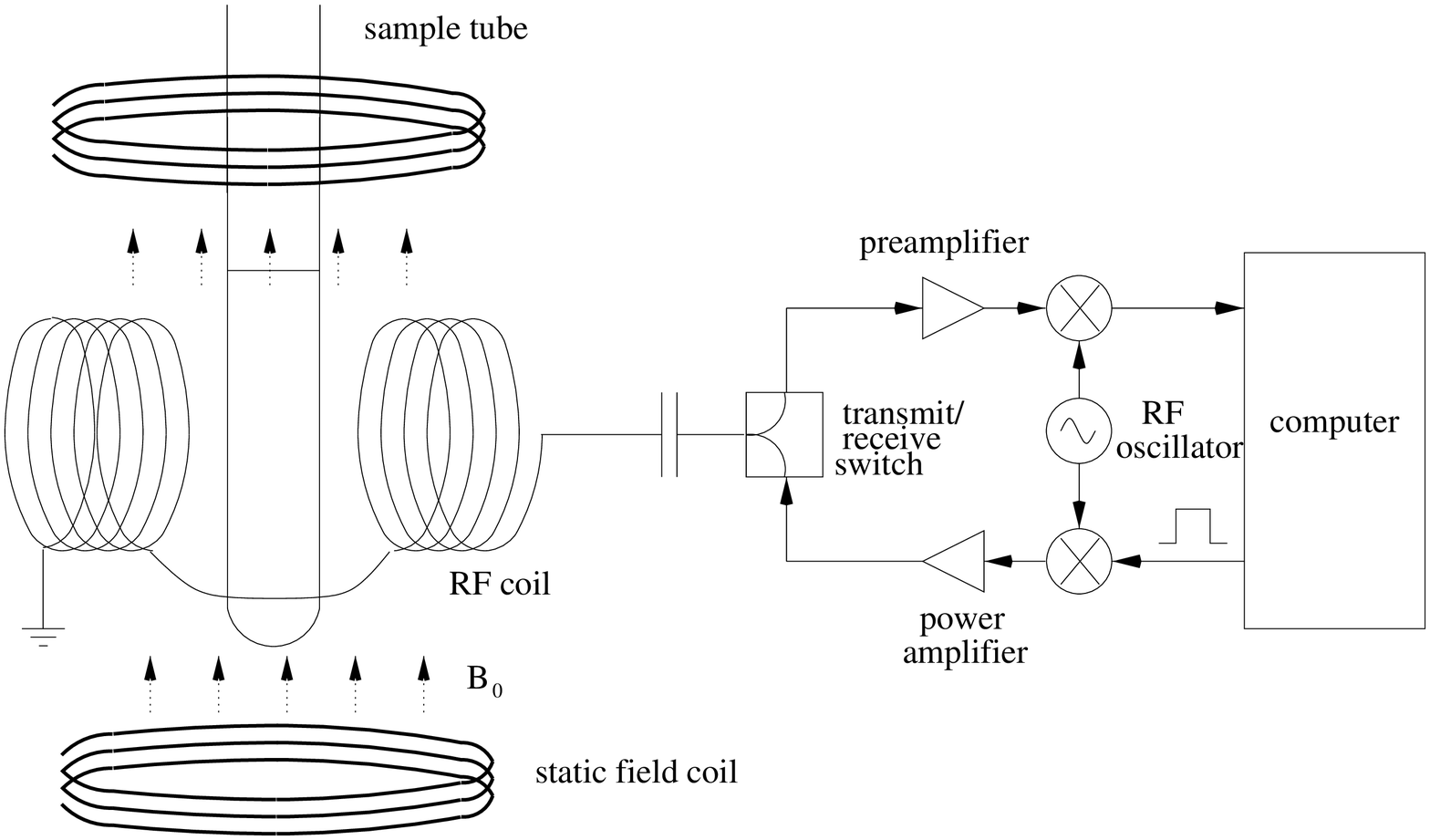} 
\vspace*{-2ex}
\ecen
\caption{Schematic overview of an NMR apparatus.}
\label{fig:apparatus}
\efig

\subsection{Sample}
\label{nmrqc:sample}

The heart of an NMR quantum computer is a molecule containing several
atoms with spin-1/2 nuclei. In practice, the signal from a
single molecule is too weak to be detected with current techniques, so
on the order of $10^{18}$ molecules are used in order to boost the
signal. Each molecule in the ensemble acts as an individual quantum
computer, and all $10^{18}$ quantum computers go through the same
operations.  The fact that there are many molecules in the sample does
not increase the power of the computer; it just increases the signal
strength. The power of the computer depends only the number of spins
per molecule (see section~\ref{nmrqc:molecule} for a discussion of
molecule design).

The molecules are disolved in a liquid solvent at room temperature and
atmospheric pressure.  Solvent selection is based on the solubility of
the quantum computer molecule in the solvent and on the coherence time
of the qubits obtained in the solvent, which depends on residual
couplings between spins in the solvent and in the solute (see
section~\ref{nmrqc:decoherence}). The solute concentration is a
trade-off between signal strength and coherence times.

The liquid solution is held in a thin-walled glass NMR sample tube
(5mm outer diameter, 4.2 mm inner diameter), filled to about 5 cm from
the bottom of the tube (Fig.~\ref{fig:sample}). The walls of the glass
vial must be very straight and of uniform thickness, in order to
minimize magnetic susceptibility variations. We have used high quality
sample tubes purchased from New Era Enterprises and from Wilmad.

\bfig
\bcen
\vspace*{1ex}
\includegraphics*[width=7cm,angle=-90]{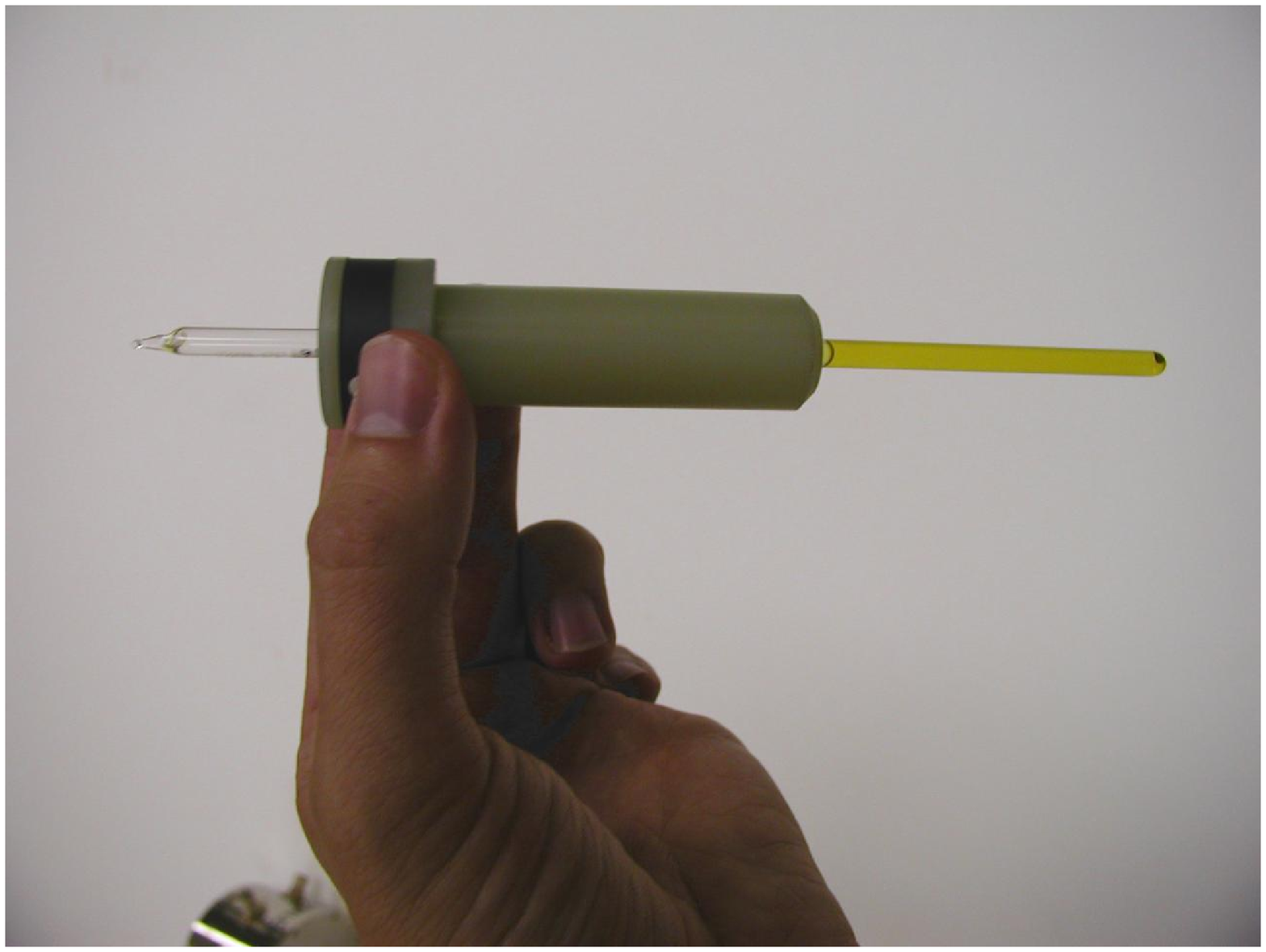} 
\vspace*{-2ex}
\ecen
\caption{A typical NMR sample. The sample tube is held by a sample 
holder when it is inserted in the superconducting magnet.}
\label{fig:sample}
\efig

Sample preparation includes careful removal of oxygen (O$_2$ is
paramagnetic and causes rapid relaxation), water (needed if the
quantum computer molecules react with H$_2$O) and particulates (they
degrade the magnetic field homogeneity). Afterwards, the open end of
the glass sample tube is flame sealed so that water, oxygen and other
impurities cannot leak in.

NMR solvents are usually deuterated. The deuterium NMR signal is used
as part of a feedback loop which keeps the magnetic field strength
constant over the course of a series of experiments
(section~\ref{nmrqc:magnet}). We have purchased deuterated solvents
from Cambridge Isotopes Laboratories and Aldrich.

%%%%%%%%%%%%%%%%%%%%%%%%%%%%%%%%%%%%%%%%%%%%%%%%%%%%%%%%%%%%%%%%%%%%

\subsection{Magnet}
\label{nmrqc:magnet}

The sample tube is placed in the room temperature bore of a
superconducting magnet built by Oxford Instruments
(Fig.~\ref{fig:magnet}). The magnet consists of a superconducting
solenoid immersed in a bath of liquid Helium (at 4.2 Kelvin).  The
Helium vessel is surrounded by a vacuum seal, a liquid Nitrogen vessel
and another vacuum seal. The whole magnet is mounted on air-cushioned
vibration isolation legs.

A persistent current of about 100 A through the windings of the
solenoid produces a magnetic field in the bore of 11.7 Tesla,
reasonably strong for an NMR magnet and about 200,000 times the
strength of the earth's magnetic field. The resulting Larmor
frequencies are in the range of 50 to 500 MHz (see
Table~\ref{tab:larmor_freq}).  About three meters away from the center
of the magnet, the stray magnetic field is still about five Gauss (10
times the earth's magnetic field). Clearly, it is important to keep
all magnetic objects away from the magnet, as they may otherwise be
pulled in and damage the magnet.

\bfig
\bcen
\vspace*{1ex}
\includegraphics*[width=8cm,angle=-90]{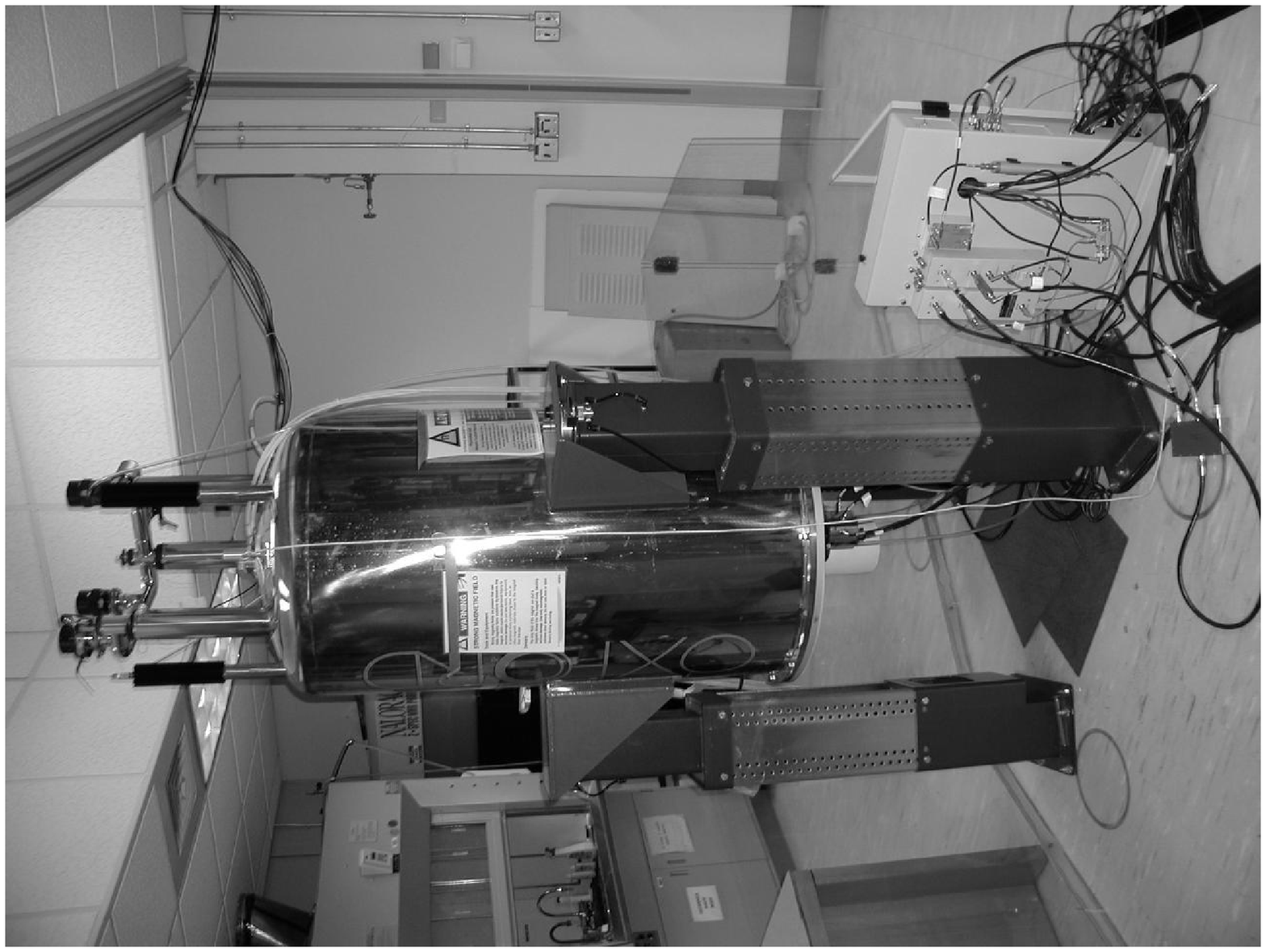} 
\vspace*{-2ex}
\ecen
\caption{Oxford Instruments 500 MHz wide-bore {NMR} magnet. Fill ports 
for liquid nitrogen and helium stick out from the top. The cabinet
near one of the magnet legs contains transmit/receive switches,
preamplifiers and mixers. The probe is inserted in the bore of the
magnet from below and the sample is inserted from the top. It sits in 
the probe in the center of the solenoid.}
\label{fig:magnet}
\efig

Strong fields are advantageous because the separation between the
spectral lines of nuclei of the same isotope (the chemical shift)
increases linearly with the field strength. Large frequency
separations make it easier to address each qubit
individually. However, spin coherence times may decrease as the field
goes up (relaxation due to chemical shift anisotropies increases as
the field increases), so it isn't clear that an even stronger field
would be better for quantum computing.

The bore diameter of our magnet is 89 mm, compared to 52 mm for
standard magnets intended for liquid state NMR. The extra space
permits the use of custom-built probes with better performance than
the commercially available liquids probes, although we did use
narrow-bore probes in all the experiments presented in this work.\\

Of crucial importance is the {\em homogeneity} of the magnetic field,
as it directly affects the spectral linewidths and thereby both the
signal-to-noise ratio and the overlap of lines within a multiplet. An
inhomogeneous field also causes dephasing in the course of a pulse
sequence, although this can in large part be refocused.

Two sets of {\em shimming coils} mounted around the bore produce
magnetic fields which even out any inhomogeneities in the field of the
main solenoid. One set consists of about ten superconducting coils,
which are energized upon installation and never readjusted.  The
current of the second set of about $25$ room temperature coils can be
adjusted by the user. Each shimming coil creates a magnetic field with
a specific spatial variation in strength: the field strength of the
$Z_1$ coil varies linearly along the $\hat{z}$ axis, the $Z_2$ coil
varies quadratically along $\hat{z}$ and so forth. A typical set of
shimming coils contains $Z$ coils up to fifth or sixth order,
transverse ($X$ and $Y$) coils up to fourth order, and combined coils
(e.g. $X_2 Z$) up to fourth order as well.

The optimal shim settings are sensitive to the RF coil geometry, the
solvent susceptibility, the sample height, the glass tube dimensions
and susceptibility, the temperature and the presence of magnetic
objects in the vicinity of the magnet. The homogeneity of the coil can
be assessed via the lock signal strength (see below), the shape and
decay rate of the FID and the lineshape and linewidth. With a lot of
effort, variations in the strength of the static magnetic field can be
made better than 1 part in $2
\times 10^9$ over the active region of the sample (4.2 mm in diameter
by 1.5 to 2 cm in height), giving linewidths of only 0.2 Hz at 500
MHz~\footnote{At this point, the intrinsic $T_2$ of most samples
dominates the linewidth. In fact, with most samples it is not possible
to obtain such narrow lines.}, a truly extraordinary homogeneity.\\

A second important consideration is that the field strength of a
superconducting magnet slowly {\em drifts} over time, as the current
through the windings does dissipate power, albeit only a tiny
amount. For a good NMR magnet, the drift is below one Hertz per
hour. To put this in perspective, at a drift rate of 1 Hz/hr the
field decreases by only 8.76 kHz per year, which is 8.76 kHz / 500 MHz
$< 0.002 \%$ per year.  An NMR magnet can thus easily be used for
several decades without any substantial loss in field strength.

Even though the field drift is very slow, it is still appreciable in
experiments where precise control over the spin dynamics is required,
as is the case of quantum computation. The spin Larmor frequencies
slowly drift away from the RF source frequencies so pulses will be
off-resonance. Furthermore, the rotating reference frame provided by
the RF sources gets progressively out of phase with the actual
rotating frame of the spins.

The drift of the magnetic field is therefore compensated for via a
room temperature $Z_0$ coil which superimposes a magnetic field on top
of the field produced by the main solenoid. The current through the
compensating coil is regulated via a feed-back loop aimed at {\em
locking} the frequency of the deuterium signal of the solvent (and
thus also the field strength) to a prescribed value; the deuterium
nuclei in the solvent are pulsed every few seconds, and the deuterium
signal is monitored (the deuterium frequency is 77 MHz, far away from
other frequencies of interest). At the start of a series of
experiments, the user must set up the lock power and the gain and
phase of the lock feed-back signal.  From then on, the lock mechanism
operates automatically in the background.

Other possible solutions for $B_0$ drift include making the RF source
frequencies track the drifting Larmor frequencies, or the use of
additional $180^\circ$ pulses to refocus chemical shift evolution.
The latter involves additional pulses and is not desirable. 

%%%%%%%%%%%%%%%%%%%%%%%%%%%%%%%%%%%%%%%%%%%%%%%%%%%%%%%%%%%%%%%%%%%%%

\subsection{Probe}

The probe is in a cylindrical aluminum housing (Fig.~\ref{fig:probe})
which contains the RF coils, a tuning and matching electrical circuit,
a temperature control system, a sample spinning mechanism and
sometimes gradient coils.

\subsubsection{RF coils and tune/match circuits}

Saddle-shaped Helmholtz RF coils mounted near the top of the probe
closely surround the glass sample tube over a height of about 1.5
cm. The region of the sample which is well coupled to the coils,
called the {\em active region}, is a little larger than the region
surrounded by the RF coil, usually about 2cm. The coils typically have
only one to three windings, and are made of low-resistivity metals
such as copper or Pd-plated copper foil or Al filled copper wire in
order to compensate for susceptibility differences.

\bfig
\bcen
\vspace*{1ex}
\includegraphics*[width=9cm]{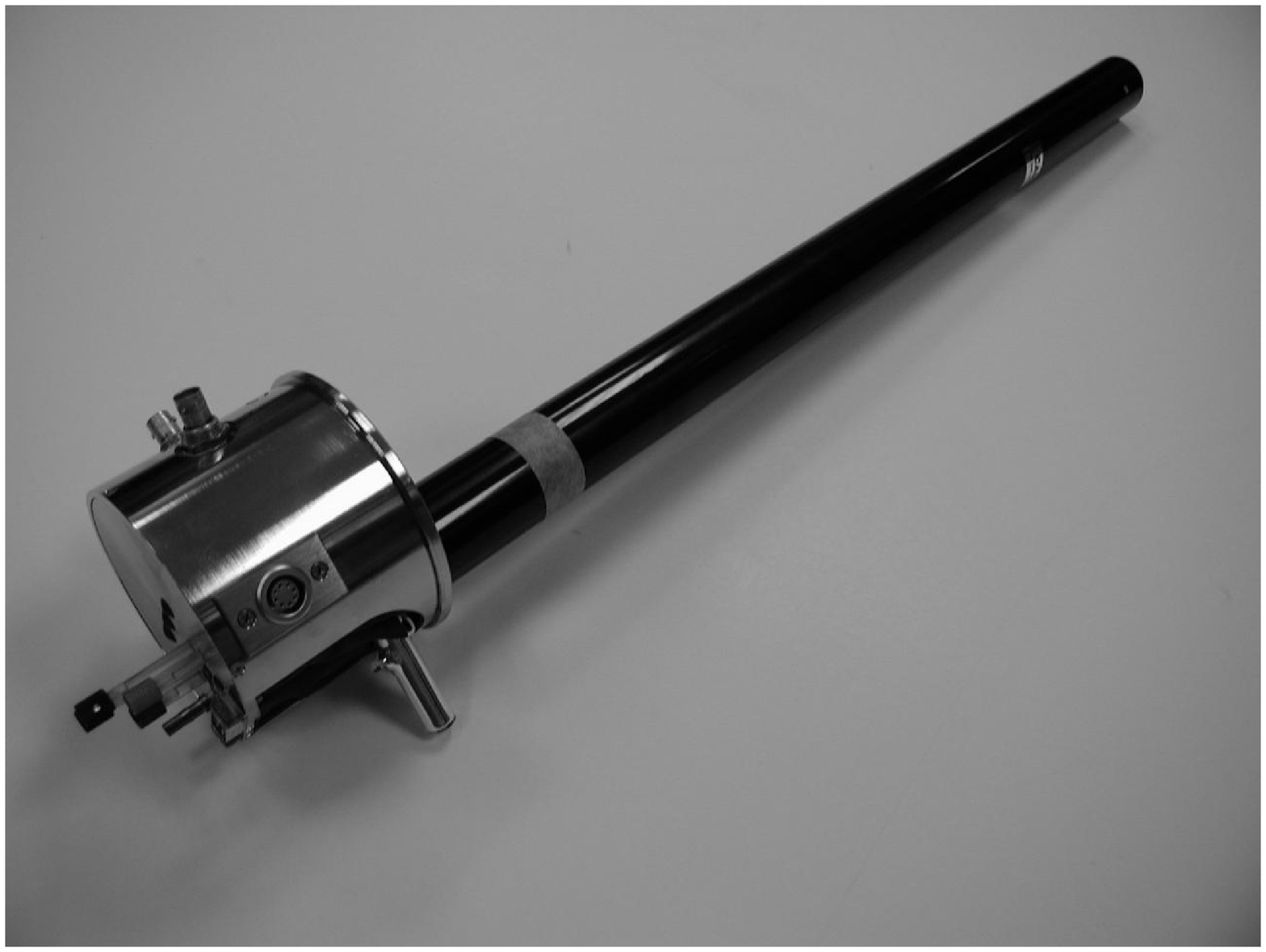} 
\vspace*{-2ex}
\ecen
\caption{Nalorac HFX Probe. The RF coils sit near the top of the probe. 
BNC connectors, a cooling air inlet, a connector for the gradient
coils and knobs to adjust to tune/match capacitors are visible at the
bottom of the probe.}
\label{fig:probe}
\efig

The coils are incorporated in a resonant circuit tuned to the Larmor
frequency of one or several nuclei, in order to obtain a high quality
factor $Q$ (values of 100 to 300 are typical), and thus a high
signal-to-noise ratio (Eq.~\ref{eq:SNR}).  The exact resonance
frequency of the circuit can be adjusted via a mechanically variable
capacitor. Using a second variable capacitor, the impedance of the
circuit is matched to $50 \Omega$.
%Fig.~\ref{fig:tune_match} shows the tuning and matching circuit. 
The tune and match capacitors are usually mounted close by the coil,
and are adjustable via long mechanical rods which stick out from the
bottom of the probe. Probe tuning and matching is done by minimizing
the reflected power for the desired frequencies.

%\bfig
%\caption{Electronic circuit to tune the frequency and match the 
%impedance of the resonant circuit which incorporates the RF coils.
%Actual tuning circuits are more complex and have multiple resonance
%frequencies.}
%\label{fig:tune_match}
%\efig

Because of the difficulty of building high $Q$ resonant circuits with
multiple resonances over a wide frequency range, many commercial
probes contain two pairs of Helmholtz coils, mounted at right angles
with little overlap such that there is little cross-talk between the
two sets of coils. One coil then serves the {\em high-band} nuclei
$^1$H and $^{19}$F (500 and 470 MHz at 11.7 T) and the other coil
serves the {\em low-band} nuclei (the highest of which is $^{31}$P, at
202 MHz). The lock ($^2$H, at 77 MHz) is usually on the high-band coil
such that the lock signal interferes as little as possible with the
other low-band signals. Our probe is a ``normal'' probe, with the
high-band coil on the inside; ``inverse'' probes have the high-band
coil on the outside.

The sensitivity of a probe depends not only on the $Q$ but also on the
filling factor $\eta$ and the geometrical coupling $K$ between the
coils and the spins (Eq.~\ref{eq:SNR}). Because of the reciprocity
between transmitting and receiving RF signals, a convenient measure
for the sensitivity is the minimum $90^\circ$ pulse length for a given
power and coil volume. The absolute minimum achievable $90^\circ$
pulse length (typically $6-15 \mu$s) depends also on how much power
the probe can take before the coil windings or the capacitors arc.

The RF field homogeneity of saddle shaped Helmholtz coils is quite
poor:
%, as illustrated in Fig.~\ref{fig:rf_homo}. 
the envelope of the Rabi oscillation decays by about $5\%$ per
$90^\circ$ rotation. In other words, the error of a single one-qubit
rotation just due to RF coil inhomogeneity is on the order of
$5\%$. Fortunately, the effects of this error can, at least in
principle, be largely undone by clever pulse sequence design
(sections~\ref{nmrqc:seq_design} and~\ref{expt:grover3}).

%\bfig
%\caption{The Rabi oscillation is severely damped because of RF field 
%inhomogeneity.}
%\label{fig:rf_homo}
%\efig

More homogeneous RF coils could be easily designed, for example, by
using a solenoidal geometry. However, this would sacrifice $B_0$
homogeneity. The $B_0$ homogeneity is several orders of magnitude
better than the $B_1$ homogeneity, and this is needed because in
typical pulse sequences the number of revolutions about $\vec{B}_0$ is
also many orders of magnitude larger than the number of Rabi
oscillations about $\vec{B}_1$.

Another possibility to improve the RF field homogeneity would be to
limit the sample volume to the homogeneous region of the RF
coils. However, the abrupt change in magnetic susceptibility at the
interface of the liquid sample and the glass or gas would then distort
the $B_0$ field in the active region. We have experimented with
specially designed plugs with a susceptibility matched to that of the
solvent, but such plugs give only modest improvements and are hard to
use in combination with flame sealed sample tubes.

\subsubsection{Other functions of the probe}

The sample temperature is regulated and under user control via a
temperature controlled nitrogen flow inside the probe, directed over
the sample tube. Seperate nitrogen flows suspend the sample holder on
a thin layer of nitrogen gas and make the sample spin about the
$\hat{z}$ axis. The spinning rate is regulated and under user control
over the range of 0 to 50 Hertz.

In addition to the RF coils, some NMR probes contain also either one
($Z$) or three ($X,Y,Z$) gradient coils. These coils produce a static
magnetic field in the $\hat{z}$ direction, but the strength of this
field varies linearly along the $\hat{x},\hat{y}$ or $\hat{z}$ axis.
\\

For the first two experiments (\ref{expt:dj}-\ref{expt:2bitcode}), we
used a Varian made tripple resonance HCN probe
(i.e. simultaneously tuned to $^1$H, $^{13}$C and $^{15}$N). For all
the other experiments, we have used a tripple resonance HFX probe made
by Nalorac ($X$ means that the low band coil is tunable over a wide
range). Both probes are equipped with gradient coils but we have not
used them in any of the experiments of chapter~\ref{ch:expt}. 
%The specifications for the Nalorac probe are given in
%Appendix~\ref{app:specs}.

%%%%%%%%%%%%%%%%%%%%%%%%%%%%%%%%%%%%%%%%%%%%%%%%%%%%%%%%%%%%%%%%%%%%%

\subsection{Transmitter}

The function of the transmitter is to send RF pulses to the probe. We
used a custom-modified Varian $^{\sc UNITY}\!$ {\sc INOVA}
spectrometer, equipped with four transmitter channels.
%Fig.~\ref{fig:xmit_schematic} gives a schematic overview of the 
%transmitter chain, and 
Fig.~\ref{fig:xmit_picture} shows a photograph of the spectrometer
electronics cabinet.

%\bfig
%\caption{Schematic overview of the transmitter side of the 
%spectrometer.}
%\label{fig:xmit_schematic}
%\efig

A {\em master oscillator} crystal (a temperature controlled crystal
oscillator) provides four {\em frequency sources} {(PTS 620
RKN2X-62/X-116)} with a {10 MHz} reference signal. From the {10 MHz}
input signal, each PTS source creates a continuous wave signal (up to
{1 V$_{rms}$}) in the range of {1-620 MHz} via direct synthesis. The
resolution of the sources is {0.01 Hz} (we didn't set the last digit,
though), the phase noise is {$- 63$ dBc}) and the stability is as good
as that of the master oscillator. The frequency sources are set {20
MHz} higher than the frequency desired for the RF pulses.

\bfig
\bcen
\vspace*{1ex}
\includegraphics*[width=10cm,angle=-90]{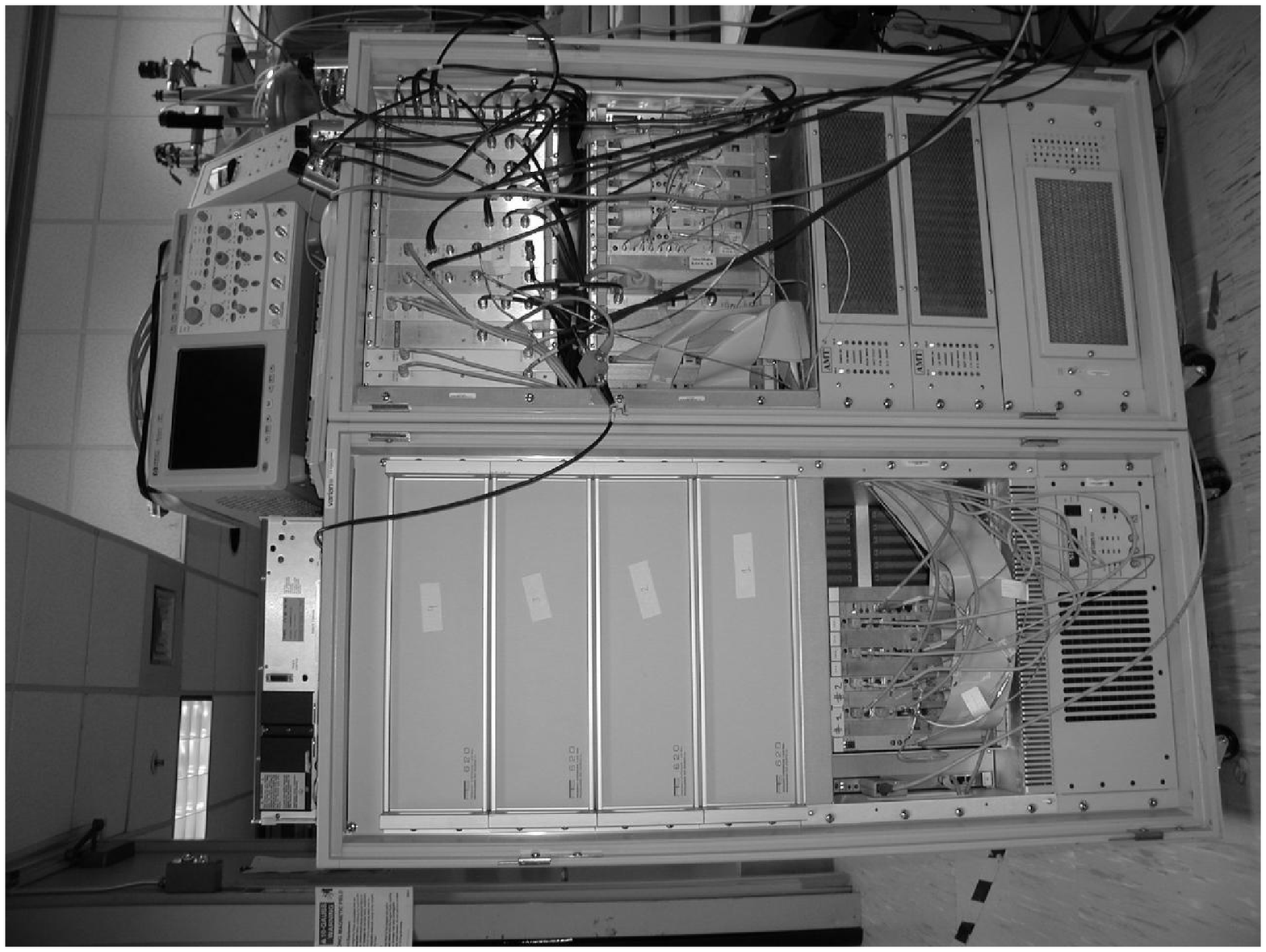} 
\vspace*{-2ex}
\ecen
\caption{Spectrometer electronics cabinet. The magnet is visible behind 
the cabinet.}
\label{fig:xmit_picture}
\efig

The four resulting CW signals are input to a set of four {\em
transmitter boards}. These boards gate the signals in order to create
pulses of the intended duration. The minumum pulse length is {100 ns}
and the resolution is {50 ns}. The phase of the pulses can be set in
steps of $0.5^\circ$. This is implemented in two stages: a $90^\circ$
step coarse phase shifter is complemented by a fine phase shifter
which achieves a resolution of $0.5^\circ$ by mixing two quadratures
with adjustable amplitudes. Each transmitter board contains a linear
attenuator, which controls the amplitude of the outgoing signal from
zero amplitude to full amplitude in 4095 steps. The transmitter boards
also contain a {\em single-sideband mixer} which mixes the gated
signals with a {20 MHz} signal so the outgoing signals have the
frequency desired for the RF pulses sent to the probe.

The amplitude and phase control of the transmitter boards can be used
to create shaped pulses. A set of four fast memory boards, called {\em
waveform generator boards} is used to load all the information needed
for several consecutive shaped pulses quickly enough onto the
respective transmitter boards.

From the transmitter boards, the signals are routed to four coarse
{\em attenuators}, which can attenuate the signals over a range of {79
dB} in steps of {1 dB}. The coarse attenuators thus have a far greater
dynamic range than the linear attenuators in the transmitter boards,
but lack the fine control needed to create shaped pulses. During a
pulse, the coarse attenuator is kept at a fixed setting, but the
setting can be changed from one pulse to the next.

Next, a set of {\em linear amplifiers} turns the signal-level pulses
into high power RF pulses. Two amplifier units (AMT model 3900-15)
each contain a low-band amplifier ({6-200 MHz}, {300 W} maximum pulse
power, {60 dB} gain) and a high band amplifier ({200-500 MHz}, 100~W
maximum pulse power, {50 dB} gain). In CW mode, the maximum power is
{30 W} and {15 W} respectively. These amplifiers are especially
designed for NMR experiments, with rise and fall times of 200~ns and
fast blanking circuits ({$< 2 \mu$s} on/off, TTL signal) which are
crucial to avoid the amplifiers putting out excessive noise in
between pulses. The blanked output noise is {$< 20$~dB} over the
thermal noise.

In the standard configuration, the spectrometer automatically routes
the signals from transmitter boards $1$ and $2$ (via the coarse
attenuator) to the high- or low band amplifier within the first dual
amplifier unit, depending on whether the signals are high or low
band. Similarly, the signals from transmitter boards $3$ and $4$ go to
the correct side of the second dual amplifier unit. In some
experiments, we have used different configurations, using external
combiners, in order to extend the routing capabilities of the
spectrometer as needed. We have also inserted stepper attenuators with
a range of 1~dB and a resolution of 0.1 dB in order to even out
differences in output power between the four transmitter boards.

The output of the two high- and low-band power amplifiers are combined
as needed with high power combiners, and then routed to the high and
low band coil of the probe via active PIN diode {\em transmit/receive
switches}. In transmit mode (during the pulse sequence), these
switches connect the probe to the power amplifier output with less
than 0.5~dB loss and isolate the power amplifiers from the receiver
preamplifiers (see section~\ref{expt:receiver}). Extra protection for
the preamplifiers is provided by quarter wave length cables and shunt
diodes.

In order to attenuate broadband noise put out by the amplifiers,
narrow band, high-pass or low-pass {\em filters} are inserted between
the power amplifier and the transmit/receive switch as needed.

The pulse amplitude and duration are calibrated via a series of
experiments in which the amplitude and/or duration are systematically
varied. The amplitude of the resulting output spectra varies
sinusoidally as a function of the pulse amplitude and duration. The
settings which give the first zero crossing of the output signal are
the optimal $180^\circ$ pulse settings. For a $90^\circ$ pulse, either
the amplitude or the duration must be halved.
 
For the seven-qubit experiment of section~\ref{expt:shor}, we
installed an additional frequency source {(PTS 620 MHO 2YX-62}),
gating circuit and power amplifier (ENI model {500 LA}, {1 V} max
input, {27 dB} gain) in order to be able to send CW power at the $^1$H
frequency during the pulse sequence (but not during the read-out)
without sacrificing any of the four transmitter channels which were
used to pulse the $^{19}$F and $^{13}$C spins. A narrow-band $^1$H
filter at the output of the ENI amplifier ensured a low noise level
going into the probe at the $^{19}$F and $^{13}$C frequencies.

%%%%%%%%%%%%%%%%%%%%%%%%%%%%%%%%%%%%%%%%%%%%%%%%%%%%%%%%%%%%%%%%%%%%%

\subsection{Receiver}
\label{expt:receiver}

The function of the receiver is to record the voltage induced in the
coil by the oscillating magnetic signals from the spins.  We have
tested and used a prototype four-channel receiver system designed by
Varian NMR. Conventional spectrometers have only one receiver channel.

%\bfig
%\caption{Schematic overview of the receiver side of the spectrometer.}
%\label{fig:rcv_schematic}
%\efig

With the transmit/receive switch in receive mode, the NMR signal is
routed from the high-band and low-band RF coils to a high- or low-band
{\em preamplifier}, with a typical loss of 0.1-0.2 dB. The overall
noise figure of the preamplifiers is 1.7 dB for the high band preamp
and 1.2-1.6 dB for the low band preamp, low enough such that the total
noise level is dominated by the coil rather than by the preamp. The
preamp gain is about 35 dB.

The amplified RF signals are then {\em mixed} with the output of the
PTS sources to an intermediate frequency (IF) around 20 MHz. The four
IF signals are routed to the receiver boards in the electronics
cabinet, where the signals are mixed with a 20 MHz signal down to
audio-frequencies, separated into two quadratures, and sent through
audio {\em filters} (the filter bandwidth can be adjusted from 1000 Hz
to 256 kHz). Both quadratures are then amplified to the desired level
and {\em digitized} (the maximum number of points is 524288 and the
maximum sampling rate is 1 MHz).  Finally, the digitized signals are
uploaded to a workstation.

We note that the digitized signal is phase referenced against the same
frequency sources (PTS RF sources and the 20 MHz source) as are used
in the transmitter chain. For each channel, the phase of the receiver
is thus coherent with the phase of the transmitter, and the phase of
the output spectra will be exactly the same every time the same pulse
sequence is executed.

%%%%%%%%%%%%%%%%%%%%%%%%%%%%%%%%%%%%%%%%%%%%%%%%%%%%%%%%%%%%%%%%%%%%%%

\subsection{Workstation}

The spectrometer is operated via Vnmr (Varian software) running on a
Sun Ultra 10 workstation. Once the hardware is configured properly
for a certain type of experiment and for a certain number and kind of
spins, the user can set up new experiments entirely by computer.

Pulse sequences are written in C, with additional commands such as
``send a pulse on channel 2'', provided by Varian. Such a command must
be accompanied by parameters and extra commands to specify, for
example, that the pulse must last $426~\mu$s, be phase shifted by
$-29^\circ$ with respect to the $\hat{x}$ axis of the oscillator
reference frame (the lab frame), be 14~dB below full power, have a
gaussian shaped profile, and also that it causes a $12^\circ$ phase
shift on spin $4$ and a $-25^\circ$ phase shift on spin $1$, has a
pre-pulse delay of $10~\mu$s, a post-pulse delay of $20~\mu$s, and so
forth.

For each experiment, we wrote a {\em framework} with convenient macros
which can be called in actual pulse sequence programs. For example,
all of the information of the preceding paragraph can then be replaced
by a simple statement of the type ``send a $90^\circ$ pulse about
$\hat{y}$ on spin 2'' (where $\hat{y}$ is now understood to be in the
rotating frame of the spin). Based on the preceding pulses in the
sequence, on correction factors computed in advance, and on
calibration values, the computer will then automatically find and set
the right values for all the parameters.

Each pulse sequence and framework must be compiled, and the compiled
code is submitted to the spectrometer. A FIFO buffer absorbs timing
differences between how long the hardware takes to execute specific
instructions and how long the workstation takes to process and submit
the instructions. The FIFO buffer can hold only a few simple shaped
pulses, so pulse shaping instructions are loaded onto dedicated
waveform generator boards instead.

The Varian software can also be used to Fourier transform the FID, and
to display the output spectra. The spectra can then be further
processed, for example by applying line-broadening, zeroth and first
order phase corrections, and baseline corrections.

In addition, we wrote extensive MATLAB routines which interface with
the standard Varian software. These routines make it easier to set up
a large number of different experiments in an automated way. The data
is automatically stored in the desired directory on the hard disk and
processed by another set of MATLAB routines.  For example, in temporal
labeling experiments (section~\ref{nmrqc:templab}), these routines add
up the data from multiple experiments with precomputed phase settings,
and in quantum state tomography experiments
(section~\ref{nmrqc:meas}), the density matrix is derived from a large
set of output spectra.  The versatility and generality of MATLAB thus
easily allows us to process the data in a specialized way.\\

Clearly, an NMR quantum computer, or any quantum computer, is not a
stand-alone unit. Its operation must be controlled by a (powerful)
classical computer. Similar to pulse sequence design, it is key that
the classical resources needed to control the quantum computer not
increase exponentially with the problem size, or with the size of the
quantum computer. This condition is indeed met in the NMR
experiments.\\

Later in this chapter, we shall present eight experiments in which we
explore the use of the apparatus described here. First we give an
overview of NMR quantum computing experiments by our and other groups.
All the experiments mentioned in this overview were done using an
apparatus similar to ours.

%%%%%%%%%%%%%%%%%%%%%%%%%%%%%%%%%%%%%%%%%%%%%%%%%%%%%%%%%%%%%%%%%%%%

\section{Overview of NMR quantum computing experiments}
\label{expt:overview}

Nuclear magnetic resonance spectroscopy, invented in 1946, developed
from a method to study magnetism into a powerful and versatile tool
for the study of molecular structure and reaction dynamics. The first
25 years of NMR were dominated by CW slow passage experiments. In the
1970's, pulsed Fourier transform spectroscopy was developed, which led
to an unprecedented expansion of the field and its applications.  Many
of the pulsed NMR protocols have a structure which we now recognize is
similar to quantum computing pulse sequences; for example, the INEPT
pulse sequence for polarization transfer is in essence the same as the
sequence for a {\sc cnot} gate.

Only in the last four years have researchers begun to implement NMR
pulse sequences with the explicit purpose of studying quantum
computation. Several groups besides our own have pursued liquid NMR
quantum computing very actively and they continue to implement a variety
of quantum information processing tasks. We will now give a very brief
overview of this work.

\subsubsection{Quantum algorithms}

The first quantum algorithms ever implemented experimentally were
Grover's algorithm for two qubits~\cite{Chuang98a,Jones98b} and the
Deutsch-Jozsa algorithm for two qubits~\cite{Chuang98c,Jones98a}
(section~\ref{expt:dj}). These experiments were performed
in the Fall of 1997 and Winter of 1998, by Ike Chuang's group at UC
Berkeley and Stanford University using the $^{13}$C and $^1$H spins of
$^{13}$C-labeled chloroform and in Jonathan Jones's lab at Oxford
University using two $^1$H spins of cytosine.

Later, the quantum counting algorithm (an extension of Grover's search
algorithm) was also implemented on the two spins of
cytosine~\cite{Jones99b}.  The two-qubit Grover algorithm was
implemented again on a subspace of two spins out of the three $^{19}$F
spins of bromotrifluoroethylene, in the first demonstration of logical
labeling~\cite{Vandersypen99a} (section~\ref{expt:labeling}).
Finally, the three-qubit Grover algorithm was realized using the
$^1H$-$^{13}$C-$^{19}$F spin system of dibromofluoromethane, with up to 28
Grover iterations, involving a record 280 two-qubit gates
\cite{Vandersypen00a} (section~\ref{expt:grover3}).

The Deutsch-Jozsa algorithm for three qubits was implemented in Ray
Freeman's lab at Cambridge University using the three $^1$H nuclei in
2,3-dibromoproponic acid, and exploring the use of transition
selective pulses~\cite{Linden98a}. The same molecule was used again
later in a similar experiment~\cite{Dorai00a}. A more advanced version
of the algorithm was demonstrated using the three $^{13}$C nuclei in
fully labeled alanine, and swap gates to realize two-qubit gates
between the two weakly coupled $^{13}$C spins~\cite{Collins00a}. The
same molecule was used without swap gates for another three-qubit
Deutsch-Jozsa experiment~\cite{Kim00a}.  A partial (particularly
simple) implementation of the five-qubit Deutsch-Jozsa algorithm was
carried out using one $^1$H, $^{15}$N and $^{19}$F nucleus and two
$^{13}$C nuclei in a molecule derived from
glycine~\cite{Marx00a}. This experiment, done by Steffen Glaser's
group in Franfurt, was the first demonstration of coherent control
over five qubits.

The implementation of quantum algorithms was taken to a new level of
complexity by the first implementation of a Shor-type quantum
algorithm for order-finding on a five-fluorine spin system, carried
out at IBM/Stanford~\cite{Vandersypen00b}
(section~\ref{expt:order}). This algorithm combined exponentiated
permutations with the three-qubit quantum Fourier transform; the
latter had been implemented earlier in itself using $^{13}$C labeled
alanine~\cite{Weinstein01a}.  The five-qubit experiment was followed
by a seven-qubit demonstration, also at IBM/Stanford, of the simplest
instance of Shor's quantum factoring algorithm, the prime
factorization of the number 15 \cite{Vandersypen01a}
(section~\ref{expt:shor}).

\subsubsection{Quantum error correction}

The first demonstration of quantum error correction was done using
alanine and trichloro-ethylene, in David Cory's group at MIT/Harvard
and by Raymand Laflamme and Emmanual Knill at Los
Alamos~\cite{Cory98a}. They implemented the three-qubit phase error
correction code and studied its operation for one particular input
state in the presence of gradient fields to introduce artifical
errors, and also when subject to just intrinsic decoherence.  A more
complete version of this experiment was carried out later by the same
groups, using gradient fields~\cite{Sharf00a}. Meanwhile, a complete
experiment for the two-bit phase error detection code had been
implemented by our group, for intrinsic decoherence~\cite{Leung99a}
(section~\ref{expt:2bitcode}). Recently, the Los Alamos group
demonstrated the five-bit phase and amplitude error correction code
for full bit and phase flip errors that were artificially
introduced~\cite{Knill01b}.

\subsubsection{Quantum simulations}

Relatively little but very interesting work has been done on quantum
simulations. David Cory's group first simulated the
dynamics of truncated quantum harmonic and anharmonic oscillators,
using the two proton spins of
2,3-dibromothiophene~\cite{Somaroo99a}. Later, the same group
simulated a non-physical three-body interaction using the three carbon
spins in fully labeled alanine~\cite{Tseng99a}.

\subsubsection{Other quantum protocols}

The group at Los Alamos prepared an effective pure GHZ state (a GHZ
state is a maximally entangled state of three particles)
\cite{Laflamme98a}, and later performed a similar experiment on seven
spins \cite{Knill00a}. Even though the claim that entangled states or
cat states had been prepared has been refuted on the basis that the
spin states at room temperature are too mixed to be entangled
\cite{Braunstein99a} (rather than entangled states, a three spin and
seven spin coherence has been observed), these experiments remain the
first, albeit relatively simple, experiments with three respectively
seven qubits.  GHZ correlations on mixed states have been studied
further in an experiment by the MIT group \cite{Nelson00a}.

Also at Los Alamos, a teleportation protocol has been carried out
using two $^{13}$C nuclei and one $^1$H nucleus in $^{13}$C labeled
trichloroethylene \cite{Nielsen98b}. Superdense coding was also
demonstrated, with $^{13}$C labeled chloroform \cite{Fang00a}. Just
like the work on pseudo-entangled states, the significance of these
experiments is limited by the mixedness of the states and furthermore
by the fact that the nuclei are separated from each other by only a
few Angstroms.

\subsubsection{Polarization enhancement}

It is clear that the tiny polarizations obtained for nuclear spins in
thermal equilibrium at room temperature severely limit the usefulness
of NMR quantum computers. Several experiments have been done to study
the feasability of boosting the nuclear spin polarization. In an
algorithmic approach, the building block of the Schulman-Vazirani
cooling scheme has been demonstrated at IBM/Stanford using the three
fluorine spins of bromotrifluoroethylene~\cite{Chang01a}
(section~\ref{expt:cooling}). However, this scheme is impractical as
long as the starting polarization remains very low. Also at
IBM/Stanford, the initial polarization of the $^1$H and $^{13}$C spins
in labeled chloroform has been increased by a factor of about 10 using
optical pumping techniques \cite{Verhulst01a}. Using a different
approach, namely the transfer of {\em para} hydrogen into a suitable
molecule, a molecule with two spins with $10\%$ polarization (compared
to $10^{-5}$) has been created \cite{Hubler00a}. In both the optical
pumping and the {\em para} hydrogen experiment, a quantum algorithm
was executed on the hyperpolarized qubits.

\subsubsection{Solvent work}

The use of liquid crystal solvents for NMR quantum computing was first
demonstrated at IBM/Stanford \cite{Yannoni99a}
(section~\ref{expt:lc}), and further studied at
IBM/ Berkeley \cite{Marjanska00a}. Liquid crystal solvents have also
been used as a solvent for a molecule containing $^{133}$Cs
atoms, which have a spin-7/2 nucleus \cite{Khitrin01a}.

\subsubsection{Perspective}

Figure~\ref{fig:comparework} puts some of the work that has been done
in perspective. This chart is not exhaustive but is certainly
representative. The most striking feature is that except for very
simple protocols requiring very few gates, all the experiments have
been based on nuclear spins in liquid solution. Of the other
implementations, trapped ions have made the most progress,
demonstrating entanglement of four ions (NIST) \cite{Sackett00a}.
Using cavity quantum electrodynamics, a two-qubit phase gate acting on
photons has been realized (Caltech) \cite{Turchette95a}. Finally, Rabi
oscillations have been observed in a superconducting charge qubit
(NEC) \cite{Nakamura99a}.

\bfig
\begin{center}
\vspace{1ex}
\raisebox{2.5cm}{\includegraphics*[width=6cm,angle=90]{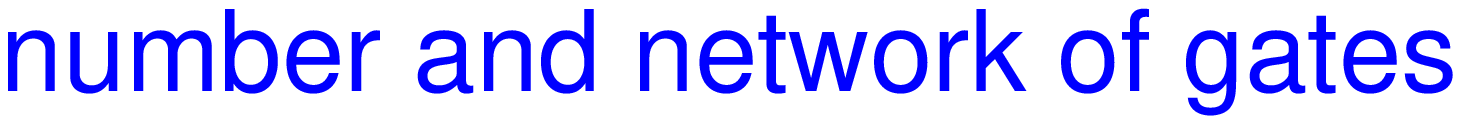}} 
\includegraphics*[width=11.2cm]{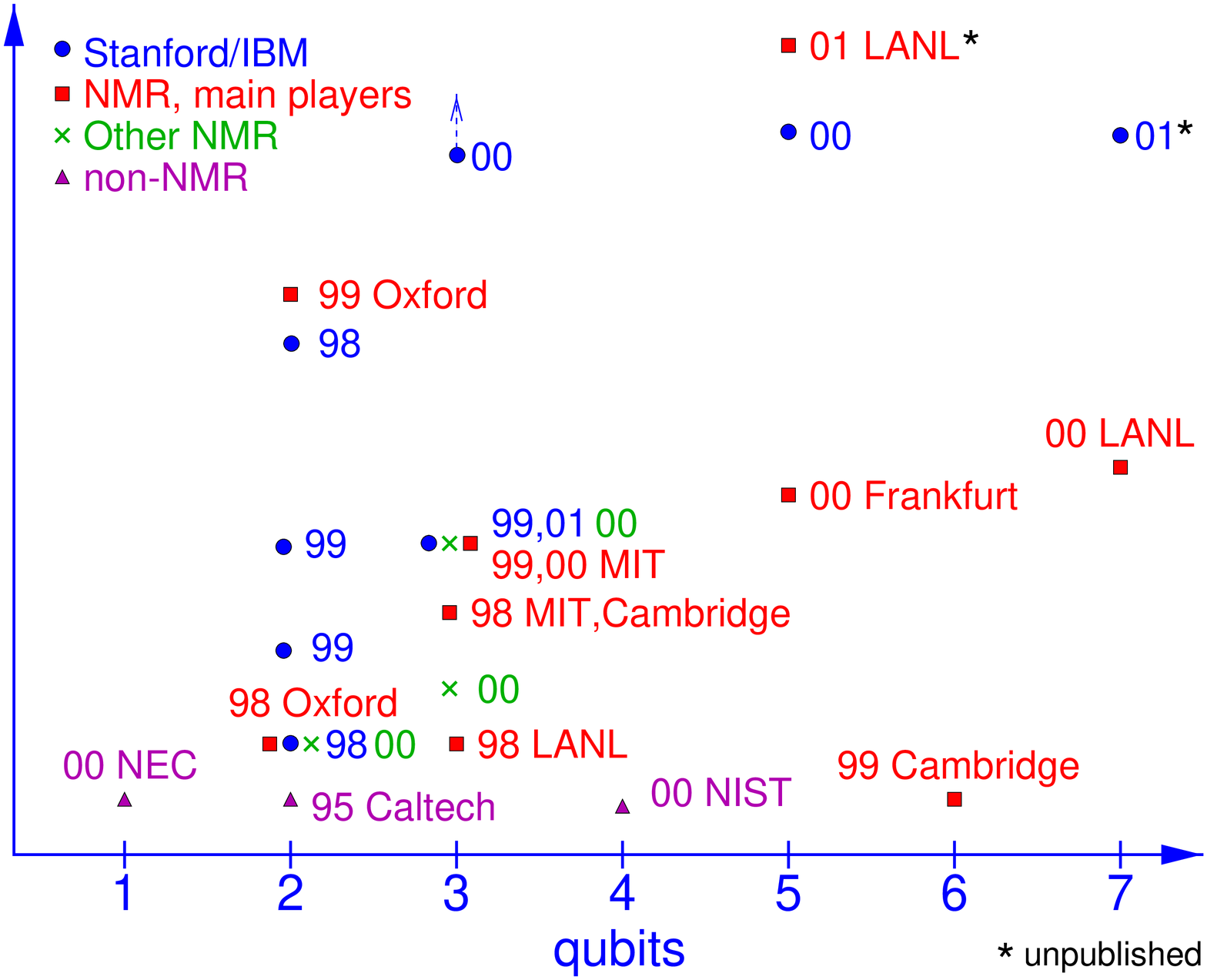} 
\end{center}
\vspace*{-2ex}
\caption{Overview of quantum computing experiments. The difficulty of 
an experiment depends mainly on two parameters: (1) the number of
qubits involved and (2) the complexity of the protocol executed with
those qubits, in terms of the number of gates and the demands on the
coupling network (e.g. experiments using only nearest neighbour
couplings are easier than experiments which need a complete or nearly
complete coupling network). Numbers next to the data are the year
published.}
\label{fig:comparework}
\efig

The general trend in the NMR work is towards more qubits and more
complex quantum algorithms and other quantum information processing
tasks. Nevertheless, the bulk of the experiments so far have been done
on only two or three spins.

%%%%%%%%%%%%%%%%%%%%%%%%%%%%%%%%%%%%%%%%%%%%%%%%%%%%%%%%%%%%%%%%%%%%%%

\section{A first quantum algorithm (2 spins)}

\label{expt:dj}

\subsection{Problem description}

In this first experiment~\cite{Chuang98c}\footnote{ The theory for
this experiment was worked out by Ike Chuang and Seth Lloyd. Ike also
wrote the framework for the pulse sequences and an interface with
MATLAB. The actual experiments were carried out by myself. The data
analysis and discussion of errors was the joint work of myself, Xinlan
Zhou, Debbie Leung and Ike Chuang.},
we implemented the simplest possible version of the Deutsch-Jozsa
algorithm (see section~\ref{qct:dj}), which determines whether an
unknown function $f$ with one input bit and one output bit is constant
or balanced.  There are four possible such functions, two of which are
constant, $f_1(x)=0, f_2(x)=1$ and two of which have an equal number
of 0 and 1 outputs: $f_3(x)=x, f_4(x)=$ {\sc not}$x$.

\begin{figure}[htbp]
\vspace*{1ex}
\begin{center}
\includegraphics*[width=10cm]{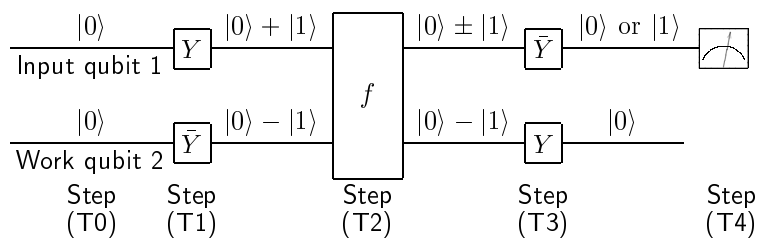}
\vspace*{-2ex}
\end{center}
\caption{Quantum circuit for performing the simplest instance of the 
Deutsch-Jozsa algorithm.}
\label{fig:dj_outline}
\end{figure}

To determine whether such a function is constant or balanced is
analogous to determining whether a coin is fair, with heads on one
side and tails on the other, or fake, with heads or tails on both
sides.  Classically, one must look at the coin twice, first one side
then the other, to determine if it is fair or fake.  The Deutsch-Jozsa
algorithm exploits quantum coherence to determine if a quantum `coin'
is fair or fake while looking at it only once.  This simplest instance
of the algorithm requires one `input' spin and one `work' spin, and is
schematically represented by the quantum circuit shown in
Fig.~\ref{fig:dj_outline}.

\subsection{Experimental procedure}

Experimentally, this quantum algorithm was implemented using the
nuclear spins of the $^1$H and $^{13}$C atoms in a chloroform molecule
(CHCl$_3$) as the input and
work qubits.  The sample contained a 200 mM, 0.5 ml solution of $99
\%$ $^{13}$C enriched chloroform (purchased from Cambridge Isotope
Laboratories, Inc. [CLM-262]) dissolved in deuterated acetone, at room
temperature and standard pressure.

The five theoretical steps (T0)--(T1) shown in
Fig.~\ref{fig:dj_outline} were experimentally implemented as follows:

\noindent {\bf (E0)} 
We prepared an effective pure ground state using temporal averaging
with cyclic permutations (section~\ref{nmrqc:templab}). For two spins,
this involves the summation of three experiments in which the
populations of the $\ket{01}$, $\ket{10}$, and $\ket{11}$ states are
cyclically permuted before performing the computation, as in
Eqs.~\ref{eq:templab_eq}-\ref{eq:templab_sum}.

Note that while this method requires $f(x)$ to be evaluated 3 times,
it is actually not necessary.  Although step (T0) stipulates a pure input
state $\ket{00}$, the algorithm works equally well if the input
qubit is initially $\ket{1}$; furthermore, when the work qubit is
initially $\ket{1}$, it fails, and cannot distinguish constant from
balanced functions, but this does not interfere with other computers
which have worked (Fig.~\ref{fig:dj_spectra}). Thus, a thermal state
is a good input for this algorithm, and only one experiment needs to
be performed.  We will present data from both thermal and effective
pure input states.

\noindent {\bf (E1)}
The Hadamard operations in the general description of the
Deutsch-Jozsa algorithm (section~\ref{qct:dj}) can be implemented by
$Y_1$ and $\bar{Y}_2$ rotations since we know the initial state of
each qubit is $\ket{0}$.

\noindent {\bf (E2)} The function $y \rightarrow y \oplus f(x)$ is 
implemented using RF pulses and the spin-spin interaction.  Recall
that spin $1$ represents the input qubit $x$, and spin $2$ the work
qubit $y$ where $f$ stores its output.  $f_1$ is then implemented as
$\tau/2 \; X_2^2 \;\tau/2 \; X_2^2$, to be read from left to right,
where $\tau/2$ represents a time interval of $1/4J \approx 1.163$ ms
($J=215$ Hz in chloroform), during which coupled spin evolution
occurs.  This is a well known refocusing pulse sequence which performs
the identity operation (section~\ref{nmrqc:J_2bitgates}). $f_2$ is
$\tau/2 \; X_2^2 \;
\tau/2$, similar to $f_1$ but without the final $180^\circ$ pulse, 
so that spin $2$ is inverted.  $f_3$ is $Y_2 \; \tau \; \bar Y_2 X_2
\; \bar Y_1 \bar X_1 Y_1$, which implements a {\sc cnot} 
operation, in which $2$ is inverted if and only if $1$ is in the
$\ket{1}$ state.  Finally, $f_4$ is implemented as $Y_2 \; \tau \;
\bar Y_2 \bar X_2 \; \bar Y_1 \bar X_1 Y_1$, which is similar to $f_3$ but leaves spin $2$ inverted.

\noindent {\bf (E3)} The inverse of {(E1)} is done by applying the 
RF pulses $\bar{Y}_1 Y_2$ to take both spins back to $\pm \hat
z$. Spin $1$, which was $\ket{0}$ at the input, is thus transformed
into $\ket{0}$ or $\ket{1}$ for constant or balanced functions
respectively.

\noindent {\bf (E4)} The result is read out by applying a read-out 
pulse $X_1$ to bring spin $1$ back into the $\hat x-\hat y$ plane.

\subsection{Experimental results}

\begin{figure}[htbp]
\bcen
\includegraphics*[width=14cm]{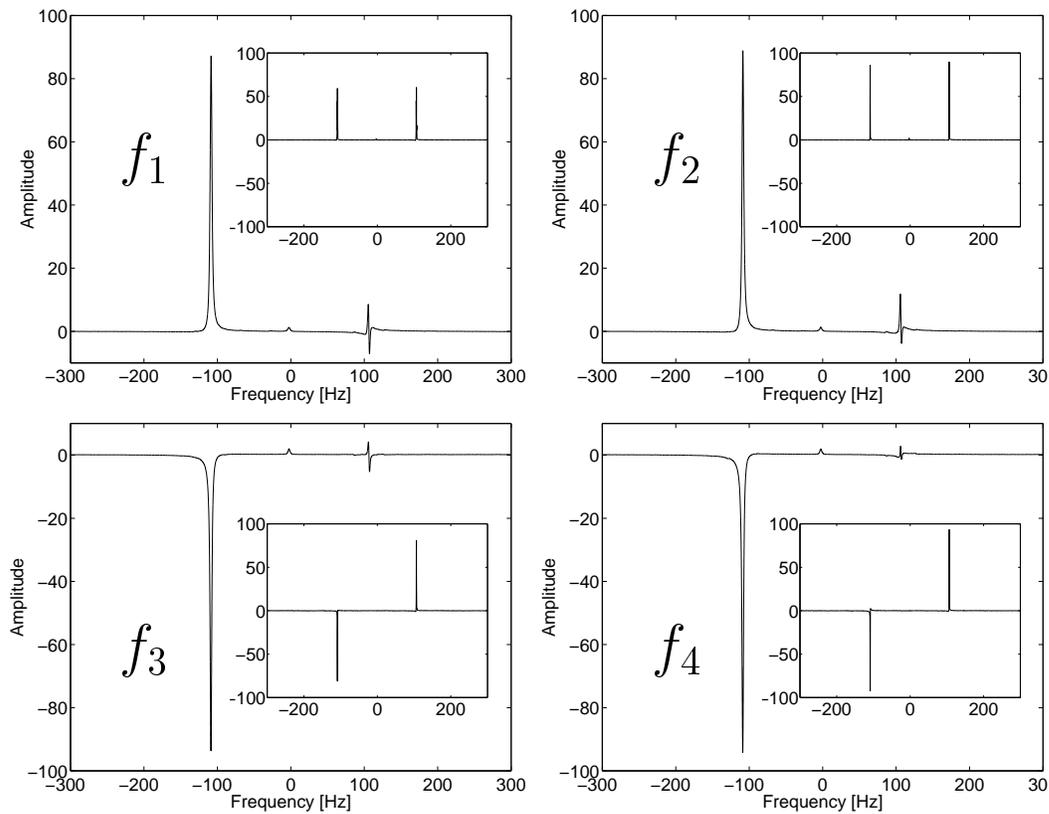}
\ecen
\caption{Proton spectrum after completion of the Deutsch-Jozsa 
algorithm and a single read-out pulse $X_1$, with an effective pure
input state $\ket{00}$ and with a thermal input state [Inset].  The
low (high) frequency lines correspond to the transitions
$\ket{00}\leftrightarrow\ket{10}$
($\ket{01}\leftrightarrow\ket{11}$). The frequency is relative to
$\omega_0^1/2\pi$ (the Larmor frequency of spin 1), and the amplitude has
arbitrary units. The phase is set such that a spectral line is real
and positive (negative) when spin 1 is $\ket{0}$ ($\ket{1}$) right
before the read-out pulse.}
\label{fig:dj_spectra}
\end{figure}

The prediction is that the spectral line of spin 1 will be up for
constant $f$ and down for balanced $f$.  The experimentally measured
spectra obtained with an effective pure input state immediately reveal
whether $f(x)$ is constant or balanced
(Fig.~\ref{fig:dj_spectra}). For the thermal input state [inset], the
left line is from molecules with the proper ($\ket{0}$) input state
for spin 2 and gives the answer to Deutsch's problem. The right line
is from molecules which started off with spin 2 in $\ket{1}$, and with
that input state the algorithm fails in distinguishing constant from
balanced $f$, as predicted.

We also characterized the entire deviation density matrix $\rho_\Delta
\equiv \rho - \mbox{Tr} (\rho)\, I/4$ describing the final 2-qubit state 
(Fig.~\ref{fig:dj_denmat}). The deviation density matrix was
obtained from the integrals of the proton and carbon spectral lines,
acquired for a series of nine experiments with different read-out
pulses for each spin (quantum state tomography, see
section~\protect\ref{nmrqc:meas}). 

\bfig
\bcen
\includegraphics*[width=12cm]{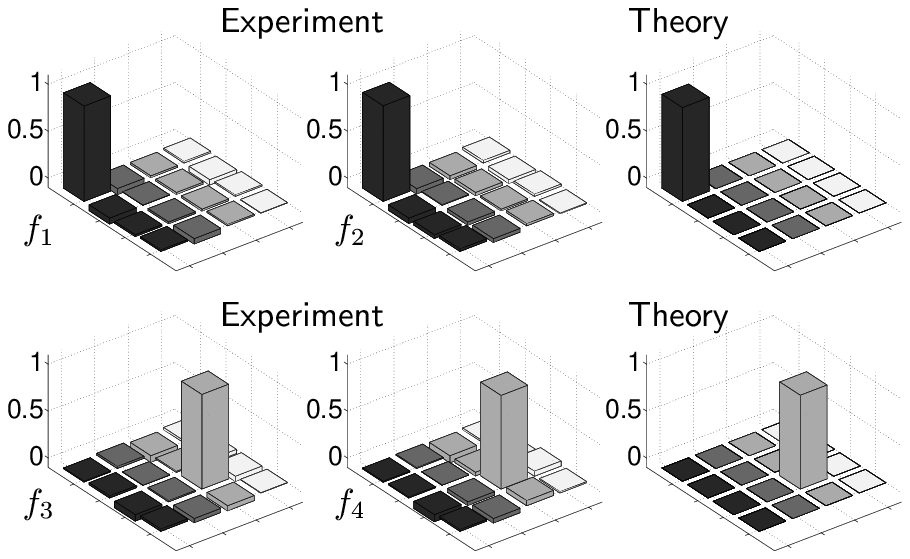}
\ecen
\caption{Experimentally measured and theoretically expected deviation 
density matrices after completion of the Deutsch-Jozsa algorithm.  The
diagonal elements represent the normalized populations of the states
$\ket{00}, \ket{01}, \ket{10}$ and $\ket{11}$ (from left to
right). The off-diagonal elements represent coherences between
different states. The magnitudes are shown with the sign of the real
component; all imaginary components were small.}
\label{fig:dj_denmat}
\end{figure}

\subsection{Discussion}

The experimental results unambiguously demonstrate the complete proper
functioning of the quantum algorithm and provide data for the
following error analysis.  In the experiments, the normalized
pure-state population measured from the deviation density matrix
(ideally equal to $1$), varied from $0.998$ to $1.019$. The other
deviation density matrix elements (ideally $0$), were smaller than
$0.075$ in magnitude.  The relative error on the experimental
pure-state output density matrix $\rho_{exp}$, defined as
\be
 \parallel\rho_{exp} - \rho_{theory}\parallel_2 /
 \parallel \rho_{theory}\parallel_2  \;,
\ee
where $\parallel \cdot \parallel_2$ is the 2-norm matrix
distance\footnote{The 2-norm gives the absolute value of the largest
eigenvalue. It is a pessimistic measure, compared to the traditional
$1- \mbox{Tr}(\sqrt{\rho_1} \rho_2
\sqrt{\rho_1})$ (the latter is defined only for non-negative matrices,
though).}, varied between $8$ and $12\%$.

Quantum computation requires that a coherent superposition be
preserved for the duration of the computation.  The relaxation time
constants for proton and carbon were $T_1\approx 19$ and $25$~s, and
$T_2\approx 7$ and $0.3$~s respectively; these were much longer than
required for our experiment, which finished in about $7$~ms, so
relaxation introduced negligible errors.

The single most important source of errors in the experiments was the
RF field inhomogeneity and pulse length calibration imperfections.  A
direct measure of this inhomogeneity is the $\approx 200$ $\mu$s time
constant of the exponentially decaying envelope observed from applying
a single pulse, as a function of pulse width.  Including the
population permutation sequence, about 7 pulses are applied to each
nucleus, with a cumulative duration of $\approx 70-100 \mu s$.

The second most important contribution to errors is the low carbon
signal-to-noise ratio: the carbon $\mbox{signal peak
height}/{\mbox{RMS noise}}$ was about 35, versus $\approx 4300$ for
the proton. The carbon signal was much weaker because the carbon
gyromagnetic ratio is 4 times smaller, and the carbon receiver coil is
mounted farther away from the sample.  Smaller contributions to errors
came from incomplete thermalization between subsequent experiments,
carrier frequency offsets, and numerical errors in the data analysis.

In summary, the quantum computation succeeded, and the quantum
computer solved a problem in fewer steps than is possible
classically. Furthermore, for this small-scale quantum computer,
imperfections were dominated by technology, rather than by fundamental
issues.

%%%%%%%%%%%%%%%%%%%%%%%%%%%%%%%%%%%%%%%%%%%%%%%%%%%%%%%%%%%%%%%%%%%%%%
%%%%%%%%%%%%%%%%%%%%%%%%%%%%%%%%%%%%%%%%%%%%%%%%%%%%%%%%%%%%%%%%%%%%%%

\section{Quantum error detection (2 spins)}
\label{expt:2bitcode}

\subsection{Problem description}

The goal of this experiment~\cite{Leung99a}\footnote{This experiment
was proposed by Debbie Leung and Ike Chuang. They also worked out the
theory. Samples were selected and prepared by Mark Sherwood and Nino
Yannoni. I performed the actual experiment. The data analysis and
numerical simulations were mostly the work of Debbie, Ike and Xinlan
Zhou.} was to study the effectiveness of quantum error correction in a
real experimental system, focusing on effects arising from
imperfections of the logic gates. We did this by testing the two-qubit
error detection code of section~\ref{qct:2bitcode} on a two-spin
molecule.

We designed the experiment such that potential artificial origins of
(favorable) bias were eliminated in the following ways.  First, we
compared the preservation of arbitrary states stored with and without
coding (the latter is unprotected but not affected by coding
operations). Second, by ensuring that all qubits used in the code
decohere at nearly the same rate, we eliminated apparent improvements
brought about by having an ancilla with a lifetime much longer than
the original unencoded qubit. Third, our experiment utilized only
naturally occurring error processes.

\subsection{Experimental procedure}

The molecule selected for this experiment was $^{13}$C-labeled sodium
formate (CHOO$^-$Na$^+$) at 15$^\circ$C. The sample was a 0.6 ml, 1.26
molar solution (8:1 molar ratio with anhydrous calcium chloride) in
deuterated water.  The proton and carbon were used as input (qubit 1)
and ancilla (qubit 2) respectively, and have relaxation time constants
of $T_1^H = 9$~s, $T_1^C = 13.5$~s, $T_2^H = 0.65$~s and $T_2^C =
0.75$~s. The fact that $T_2 \ll T_1$ ensures that the effect of
amplitude damping was small compared to that of phase
damping. Furthermore, $T_2^H \approx T_2^C$, as desired.\footnote{We
performed a second set of experiments with $^{13}$CHCl$_3$, for which
$T_2^H \gg T_2^C$. Those were also published in~\cite{Leung99a} but we
shall not present them here.}

For the preparation of arbitrary input states, we took advantage of
the axisymmetry of phase damping, by which it is sufficient to
prepare a set of states in one half of a vertical cross section
through the center of the Bloch sphere. The input state was therefore
prepared by a $Y_1(\theta)$ pulse, where $\theta$ was arrayed from
$0^\circ$ to $180^\circ$ in steps of $18^\circ$.

The state of the ancilla must be effective pure in order for the code
to work. The desired state ($\ket{0}$) was obtained by temporal
averaging, via a summation of two experiments: one experiment started
off with $\rho_{eq}$ and in the other experiment the populations of
$\ket{01}$ and $\ket{11}$ were interchanged at the start, via a {\sc
cnot}$_{21}$ ($Y_1 \tau X_1$, from left to right, with $\tau = 1/2J$).

In the coding experiment, we performed the encoding ($Y_2 \bar X_1
\bar Y_1 \tau Y_1$) and decoding ($Y_1 \tau$ $\bar Y_1 X_1 \bar Y_2$)
operations before and after phase damping, whereas in the control
experiment, these operations were omitted (see
Fig.~\ref{fig:2bit_overview}). The output state of the stored qubit was
read out on spin 1 via a $X_1$ pulse.

\bfig
\bcen
\vspace*{1ex}
\includegraphics*[width=12cm]{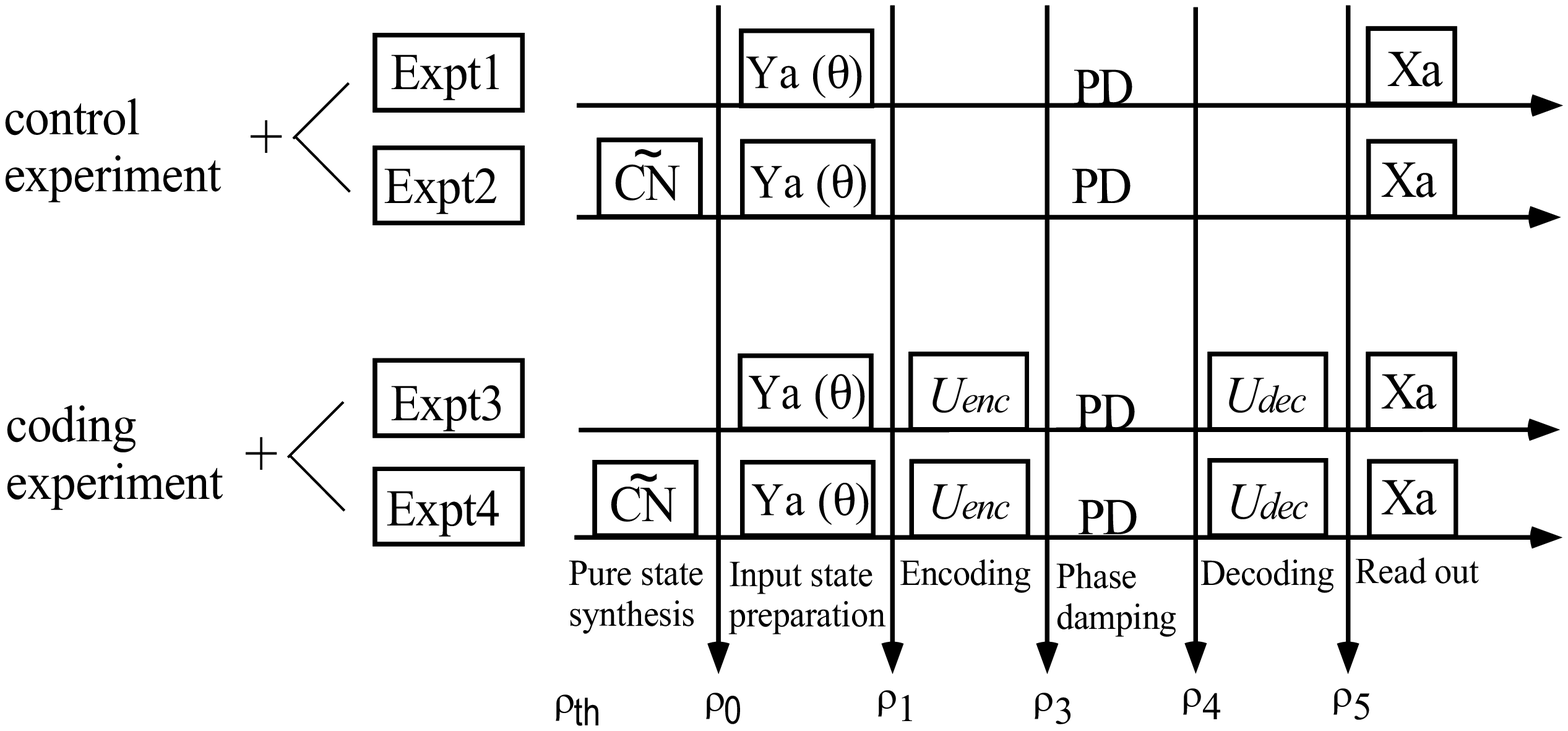} 
\vspace*{-2ex}
\ecen
\caption{Schematic diagram for the two-bit code experiment.}
\label{fig:2bit_overview}
\efig

If the ancilla spin is in the effective pure state $\ket{0}$, only the
low-frequency line in the doublet of spin 1 is present. However, if a
phase error occurs on the encoded state, the ancilla will be $\ket{1}$
after decoding. Thus, after decoding, the low-frequency line of the
doublet of spin 1 corresponds to ``accepted'' states and the
high-frequency line corresponds to ``rejected'' states.

\subsection{Experimental results}

The predicted accepted output states with and without coding are shown
in Fig.~\ref{fig:2bit_simulation}. These plots are the result of
numerical simulations which include the phase damping model of
Eq.~\ref{eq:opsumrep_pd} in section~\ref{impl:coherence}.  The main
feature is that the Bloch sphere becomes ellipsoidal without encoding
but remains largely spherical (i.e. the state is better preserved)
with encoding. We note that the amplitude of the accepted
states is shrunk with respect to the original state; this is because
the two-bit code can only detect, not correct errors.

\bfig
\bcen\vspace*{1ex}
\includegraphics*[width=12cm]{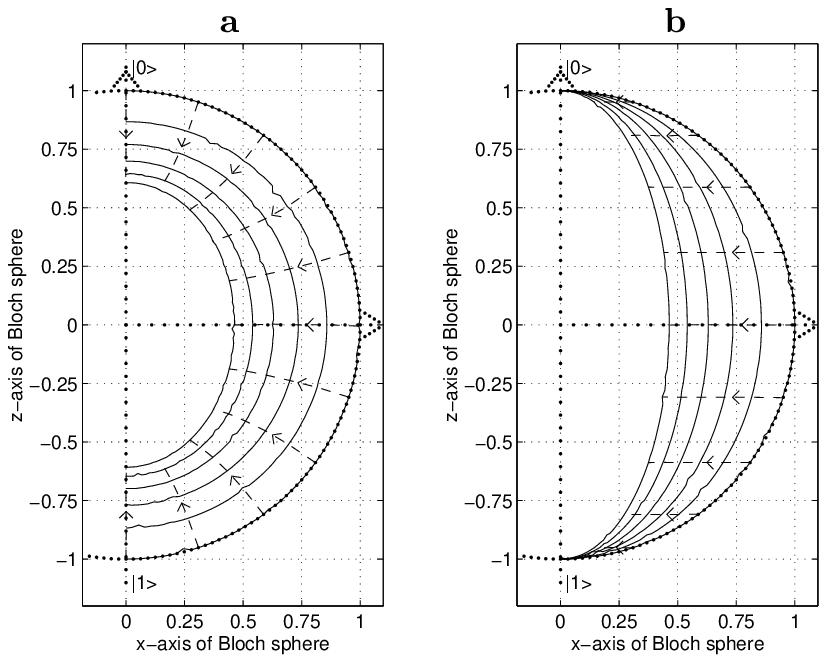} 
\vspace*{-2ex}
\ecen
\caption{Predicted Bloch spheres (a) with and (b) without encoding,
for a set of equally spaced storage times ($k \times 61.5$~ms
for $k=0,1\ldots,$5), corresponding to a probability of phase error
(without encoding) of $p = 0,$ $0.071,$ $0.133,$ $0.185,$ $0.230$ and
$0.269$.}
\label{fig:2bit_simulation}
\efig

Figure~\ref{fig:2bit_expt} shows the experimentally measured accepted
output states, again with and without coding. The agreement of the
main features between Figs.~\ref{fig:2bit_simulation}
and~\ref{fig:2bit_expt} is striking.

\bfig
\bcen
\vspace*{1ex}
\includegraphics*[width=12cm]{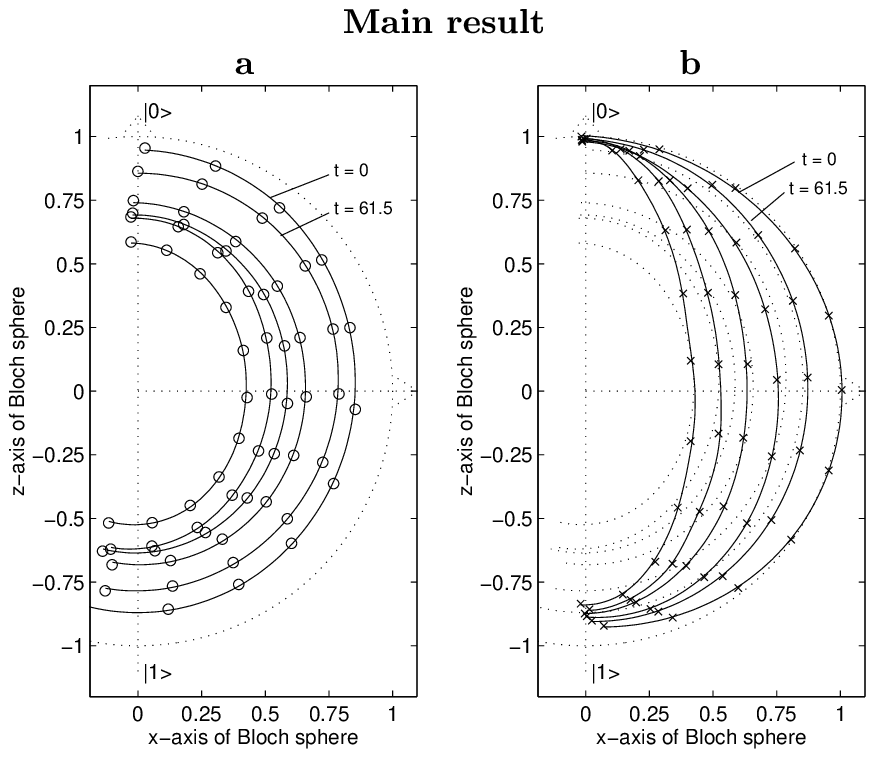} 
\vspace*{-2ex}
\ecen
\caption{Experimentally measured Bloch spheres (a) with and (b) 
without encoding, for the same storage times as in
Fig.~\protect\ref{fig:2bit_simulation}. The circles are experimental
data points and the solid lines are least square fitted ellipses.}
\label{fig:2bit_expt}
\efig

\subsection{Discussion}

We quantified how well a quantum state is preserved experimentally via
the ellipticity of the Bloch sphere, which we define as
\be
\sqrt{\frac{I(\theta = 0)}{I(\theta=\frac{\pi}{2})}}
\ee
where the intensity $I$ as a function of $\theta$ is ideally of the
form
\be
I_{ideal}(\theta)  =  A + B \sin^2\theta  \,.
\ee
In order to include signal strength attenuation with increasing
$\theta$ and constant offsets in the angular positions, we actually
fitted (non-linear least-squares fit) the experimental data points to
the expression
\be
I_{exp}(\theta) =  (A + B \sin^2(\theta + D)) (1- C (\theta + D)) 
\ee
instead.

\bfig
\bcen
\vspace*{1ex}
\includegraphics*[width=12cm]{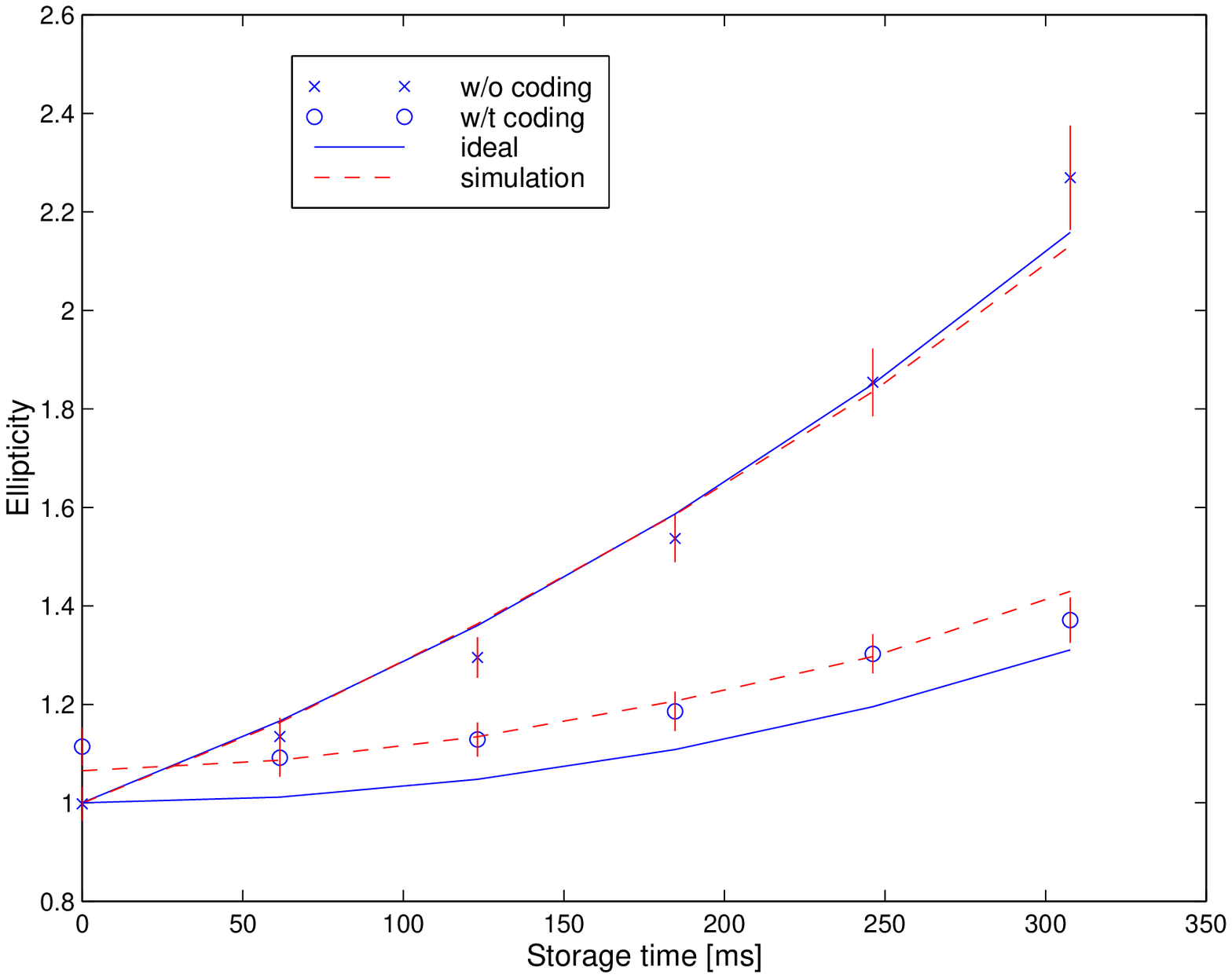} 
\vspace*{-2ex}
\ecen
\caption{Ellipticity as a function of storage time. The experimental 
datapoints are given for the case with and without coding, along with
ideal predictions as well as simulations which take the effect of RF
inhomogeneities into account.}
\label{fig:2bit_ellipt}
\efig

The data of Fig.~\ref{fig:2bit_ellipt} demonstrate that coding removes
the first order term in the growth of the Bloch sphere ellipticity as
a function of storage time, which corresponds to the first order
improvement in the conditional fidelity expected from the theory
(section~\ref{qct:2bitcode}). However, imperfections in the logic
gates caused the ellipticity to increase by $10 \%$ for the case of
zero storage time (i.e. the zeroth order term). This number represents
the cost of ``noisy'' gates.

While the data of Fig.~\ref{fig:2bit_expt} exhibit a clear correction
effect, there are notable deviations from the ideal case of
Fig.~\ref{fig:2bit_simulation}. First, the ellipses with coding are
smaller than their counterparts without coding (this reduces the
absolute fidelity but not the conditional fidelity).  This is most
obvious when the storage time is zero, in which case the coding and
the control experiments should produce equal outputs.  Second, the
signal strength is attenuated with increasing $\theta$ relative to
ideal ellipses.  Third, although the data points are well fitted by
ellipses, their angular positions are not exactly as expected
(``$\theta$-offsets'').  Finally, the spacings between the ellipses
deviate from expectation.

A major cause of experimental errors was RF field inhomogeneity, which
causes gate imperfections. This was determined by a series of
simulations of RF inhomogeneity effects, which reproduced the reduced
amplitude in the coding experiments.  The asymmetry in the
experimental Bloch spheres is well explained by amplitude damping.
We do not have a convincing explanation for the other two
discrepancies.

In summary, we have demonstrated experimentally that using a two bit
phase damping detection code, the coherence time of qubits in bulk NMR
systems can be conditionally lengthened.  These experimental results
also provide quantitative measures of the major imperfections in the
system.  The principle source of errors, RF field inhomogeneity, was
studied and a numerical simulation was developed to model our data.
Despite the imperfections, a net amount of error reduction was
observed, when comparing cases with and without coding, including
gate errors in both cases.

%%%%%%%%%%%%%%%%%%%%%%%%%%%%%%%%%%%%%%%%%%%%%%%%%%%%%%%%%%%%%%%%%%%%%%

\section{Logical labeling (3 spins)}
\label{expt:labeling}

\subsection{Problem description}

The goal of this experiment~\cite{Vandersypen99a}\footnote{This
experiment was proposed by myself. I also worked out the theory, wrote
the framework for dealing with homonuclear spins, invented the
``uncoupling frame'', wrote the pulse sequences, carried out the
experiment and did the data analysis, under the guidance of Ike
Chuang. Nino Yannoni and Mark Sherwood came up with the molecule,
prepared the sample and gave advice on NMR techniques.}  was
threefold: (1) to take a step in complexity from two spins to three
spins, (2) to explore the use of homonuclear spin systems and (3) to
test the concept of logical labeling for the first time.  These three
goals come together naturally as at least three spins are required for
a meaningful demonstration of logical labeling, and all three spins
must be homonuclear (section~\ref{nmrqc:loglab}).\\

The operations needed for logical labeling follow from comparison of
Eq.~\ref{eq:thermal_3} with Eq.~\ref{eq:eff_pure_3}: the populations of the
states $\ket{001} \leftrightarrow \ket{101}$ and $\ket{010} \leftrightarrow
\ket{110}$ must be interchanged, while the remaining four populations must
be unaffected. This requires a {\sc cnot}$_{21}$ and a {\sc cnot}$_{31}$.

As a test of logical labeling and subsequent control over the
dynamical behavior of the logically labeled spins, we chose to
implement Grover's algorithm on the effective pure two-qubit subspace
within the three-spin molecule. With the two-qubit version of this
algorithm, one can find the unique but unknown $x_0$ among $N=4$
possible values of $x$ which satisfies $f(x_0)=1$ in just one query,
compared to on average 2.25 queries classically.\\

A fourth, additional, goal was to study the preservation of the
effective pure state after many operations. Grover's algorithm lends
itself perfectly to such studies in the form of many repeated Grover
iterations (section~\ref{qct:grover}).

\subsection{Experimental procedure}

We selected bromotrifluoroethylene (Fig.~\ref{fig:chem_shifts} c)
dissolved in deuterated acetone (10 mol$\%$) as the central molecule
in our experiments, because the spin-1/2 $^{19}$F nuclei have large
$J$-couplings and chemical shifts, as well as long coherence times,
which make it suitable for quantum computation. The $^{12}$C nuclei
are non-magnetic and the interaction of the spin-3/2 Br nucleus with
the fluorine spins is averaged out due to fast Br relaxation
(chapter~\ref{ch:nmrqc}).

The $^{19}$F Larmour frequencies are $\approx$ 470 MHz (at 11.7 T) and
the spectrum is first order, consisting of three well-separated
quadruplets, with $\omega_1 - \omega_2 / 2\pi \approx$ 13.2 kHz and
$\omega_3 - \omega_1 / 2\pi \approx$ 9.5 kHz. The coupling constants
are measured to be $J_{12} = -122.1$ Hz, $J_{13} = 75.0$ Hz and
$J_{23}= 53.8$ Hz (see also~\cite{Elleman62a}).  In order to
simultaneously address the four lines in one quadruplet without
affecting the other two quadruplets, the envelope of the RF pulses was
Gaussian shaped and the RF power was adjusted to obtain pulses of
$\approx 300 \mu$s.

The two {\sc cnot} gates of the logical labeling step commute and can
thus be executed simultaneously. This was done via the pulse sequence
of Fig.~\ref{fig:loglab_seq}. This sequence implements the {\sc cnot}
gates only up to single-spin $Z$ rotations, which suffices for a
diagonal initial state as used here. Furthermore, the $I_z^2$, $I_z^3$
and $I_z^2 I_z^3$ terms in the Hamiltonian have no effect, since spins
$2$ and $3$ remain along $\pm \hat{z}$. $I_z^1$ can be ignored because
the pulses were applied in a reference frame in resonance with each
spin.

\bfig
\vspace*{1ex}
\bcen
\includegraphics*[width=8cm]{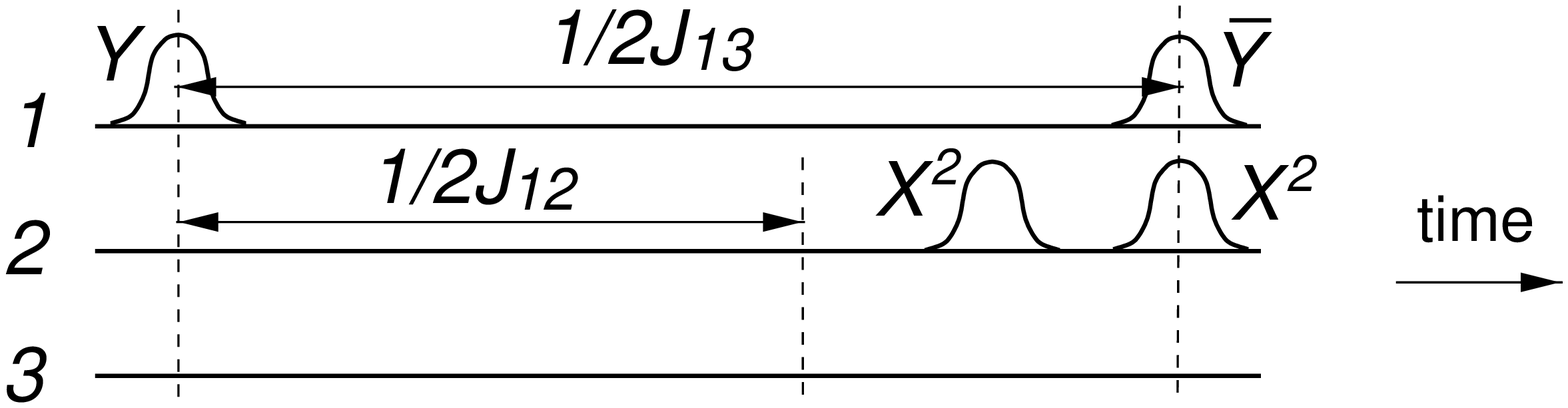}
\ecen
\vspace*{-2ex}
\caption{The logical labeling pulse sequence. The {\sc cnot}$_{21}$
and {\sc cnot}$_{31}$ are merged.
% by $Y_A - \tau_1 - X_B^2 - \tau_2 - X_B^2 \bar{Y}_A$, where 
%$Y$ and $\bar{Y}$ denote a $90 ^{\circ}$ rotation about the $\hat{y}$
%and $-\hat{y}$ axis (right hand rule). $X^2$ are 180$^{\circ}$
%rotations.
%and $\tau_1$ and $\tau_2$ are delay times of $1/4J_{AC} - 1/4J_{AB}$ 
%and $1/4J_{AC} + 1/4J_{AB}$.
}
\label{fig:loglab_seq}
\end{figure}

\begin{figure}[t]
\vspace*{1ex}
\begin{center}
\includegraphics*[width=7cm]{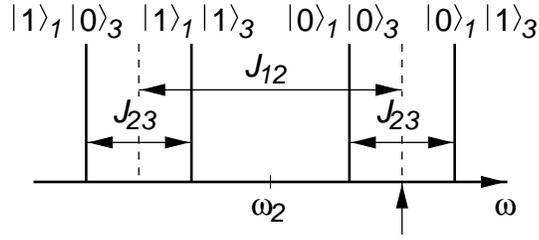}
\end{center}
\vspace*{-2ex}
\caption{Spectrum of the transitions of spin $2$, with the states of 
the other two spins as indicated. With respect to a reference frame
rotating at $\omega_2 / 2\pi$ $- J_{12}/2$ (indicated by an arrow),
spins $2$ which see a spin $1$ in $\ket{0}$, evolve under $J_{23}$
only and are thus {\em uncoupled} from $1$. $J_{12}$ does affect $2$'s
evolution in the $\ket{1}_1$ subspace, but the signal of this subspace
does not interfere with that of the $\ket{0}_1$ subspace. Similarly,
spin $3$'s rotating frame must be moved to $\omega_3/2\pi -
J_{13}/2$. A separate channel was used for spins $2$ and $3$.}
\label{fig:loglab_uncoupling}
\end{figure}

After logical labeling, the state must remain effective pure
throughout the subsequent computation. This requires that while a
computation is carried out using spins $2$ and $3$, spin $1$ must ``do
nothing'', which is non-trivial in a system of coupled
spins~\cite{Linden99c}: the effect of $J_{12}$ and $J_{13}$ must be
removed. This could be done by using two refocusing $X^2_1$ pulses
during every logical operation between $2$ and $3$, but we have
devised a different method, which exploits the fact that it suffices
to remove the effect of $J_{12}$ and $J_{13}$ {\em within the
$\ket{0}_1$ subspace}. This uncoupling frame method requires no pulses
at all and is described in
Fig.~\ref{fig:loglab_uncoupling}.\footnote{We note that the uncoupling
frame is somewhat related to selective
decoupling~\protect\cite{Maher61a}, a technique to determine the
relative sign of $J$-couplings by moving the reference frame of one
nucleus (not several, as here) to the center of a submanifold.}

Mathematically, the transformation of the state of spins 2 and 3 to a
reference frame offset by $\Delta \omega_2 = -\pi J_{12}$ and $\Delta
\omega_3 = - \pi J_{13}$ with respect to $\omega_2$ and $\omega_3$ 
respectively, gives
\be
{\cal H}' = 
2\pi \left[ \; (I_I^1+I_z^1) \; J_{23} I_z^2 I_z^3 
   + \; (I_I^1 - I_z^1) \;
	  (J_{23} I_z^2 I_z^3 - J_{12} I_z^2 - J_{13} I_z^3) \right]
	  \,,
\label{eq:frame_hamiltonian}
\ee
where $I_I = \sigma_I/2$ (one half times the identity matrix). The
first (second) term in the expression of $\cal{H}'$ acts exclusively
on the $\ket{0}_1$ ($\ket{1}_1$) subspace. Any state within the
$\ket{0}_1$ subspace evolves only under $J_{23} I_z^2 I_z^3$ and will
remain within the $\ket{0}_1$ subspace. Within the $\ket{0}_1$
subspace and using the uncoupling reference frame, spins $2$ and $3$
are thus uncoupled from $1$. 

For the implementation of the Grover algorithm on the logically
labeled spins in the uncoupling frame, we used a pulse sequence
similar to the one described in~\cite{Chuang98a}. First, $Y_2 Y_3$
rotates both spins from $\ket{00}$ into
$(\ket{00}+\ket{01}+\ket{10}+\ket{11})/2$, an equal superposition of
the four possible inputs. The amplitude of the $\ket{x_0}$ term is
then amplified in two steps (see section~\ref{qct:grover}). First, one
of four functions $f_{x_0}(x)$ is evaluated, flipping the sign of the
$\ket{x_0}$ term. This is done by one of four conditional phase flips
$Y_2 Y_4 \; \Phi_2 \Theta_3 \;
\bar{Y}_2 \bar{Y}_3 \; 1/2J_{23}$, where $\Phi = X$ for $f_{00}$ and
$f_{10}$ and $\Phi = \bar{X}$ for $f_{01}$ and $f_{11}$. $\Theta = X$
for $f_{00}$ and $f_{01}$ and $\Theta = \bar{X}$ for $f_{10}$ and
$f_{11}$. Second, inversion about the average is implemented by a
Hadamard gate on both spins, followed by the conditional phase flip
corresponding to $f_{00}$, and another Hadamard gate. The pulse
sequence for this inversion step can be reduced to $X_2 X_3 \; Y_2 Y_3
\; 1 / 2 J_{23} \; \bar{Y}_2 \bar{Y}_3$.

The entire sequence for Grover's algorithm takes approximately 20 ms
and the labeling step takes about 7 ms. The coherence time for the
three $^{19}$F spins, expressed as the measured transverse relaxation
time constant T$_2$ $\approx$ 4-8 s, is sufficiently long for
coherence to be maintained throughout the labeling and computation
operations.

\subsection{Experimental results}

Fig.~\ref{fig:loglab_popul} shows the measured populations of the
eight basis states, before and after the sequence of
Fig.~\ref{fig:loglab_seq}. The results agree with the theoretical
predictions of Eqs.~\ref{eq:thermal_3} and~\ref{eq:eff_pure_3}.

\bfig
\vspace*{1ex}
\bcen
\includegraphics*[width=13cm]{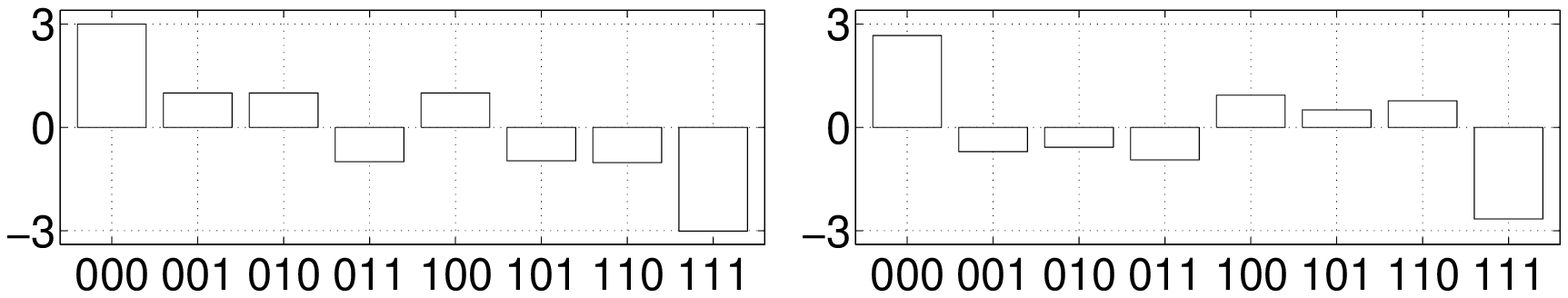}
\ecen
\vspace*{-2ex}
\caption{Experimentally determined populations (in arbitrary units, 
and relative to the average) of the states $\ket{000}, \ldots
,\ket{111}$ (Left) in thermal equilibrium and (Right) after logical
labeling. The populations were determined by partial state
tomography~\protect\cite{Chuang98b}.}
\label{fig:loglab_popul}
\end{figure}

We experimentally confirmed that spins 2 and 3 are uncoupled from spin
1 by reconstructing the deviation density matrix of the three-spin
system after creating the state $(\ket{00} + \ket{11})/\sqrt{2}$ in
the $\ket{0}_1$ subspace. For this three spin system, quantum state
tomography involved a series of 27 consecutive experiments with
different sets of read-out pulses.  Fig.~\ref{fig:loglab_denmat} shows
the $\ket{0}_1$ subsystem in the predicted effective pure state and
uncoupled from the $\ket{1}_1$ subspace. The relative error in the
state is $\parallel\rho_{\rm exp} - \rho_{\rm
th}\parallel_2$/$\parallel\rho_{\rm th}\parallel_2$ = 19 $\%$.

\bfig
\vspace*{1ex}
\begin{center}
\includegraphics*[width=9cm]{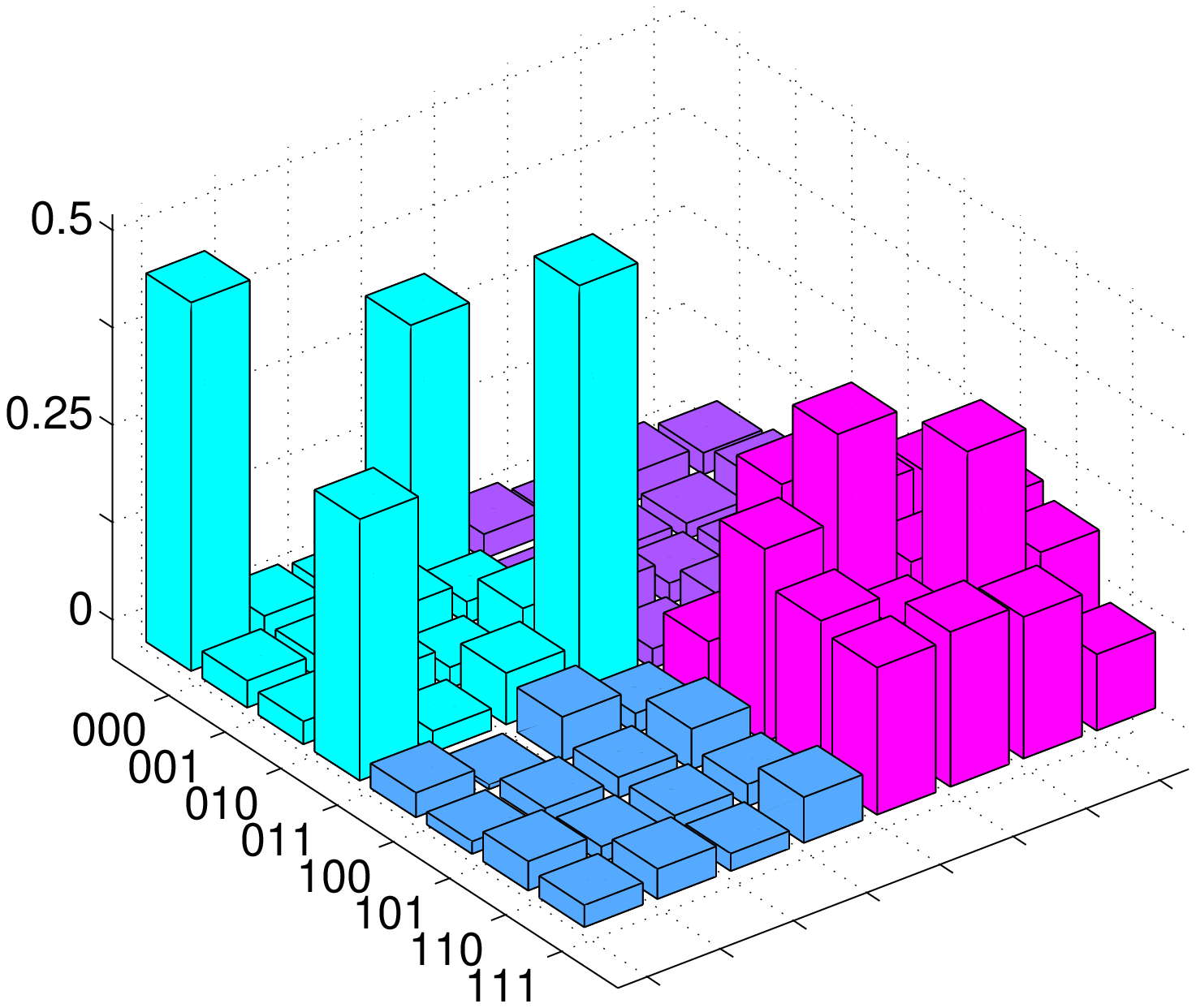}
\end{center}
\vspace*{-2ex}
\caption{Normalized experimental deviation density matrix (with the 
diagonal shifted to obtain unit trace for the effective pure state),
shown in absolute value. The entries in the second quadrant are very
small, which means that the $\ket{0}_1$ and $\ket{1}_1$ subspaces are
uncoupled. The four density matrix elements which stick out (in the
first quadrant) are, in the logically labeled subspace, the
$\ket{00}\bra{00}$ and $\ket{11}\bra{11}$ entries (which represent
populations) and the $\ket{00}\bra{11}$ and$\ket{11}\bra{00}$ entries
(which represent double quantum coherences).}
\label{fig:loglab_denmat}
\end{figure}

The theoretical prediction for the Grover algorithm is that the output
state of qubits $2$ and $3$ is the (effective pure) state
$\ket{x_0}$. This can be determined by a measurement of spins 2 and 3
after a read-out pulse on each spin.  The experimental spectra
(Fig.~\ref{fig:loglab_spectra}, Left) as well as the deviation density
matrices of the logically labeled subspace before, during and after
the computation, confirm that the state remains an effective pure
state throughout the computation and that the final state is
$\ket{x_0}$.

An interesting question is how many logical operations can be executed
while preserving the effective pure character of the spins, and
further, how quickly errors accumulate during longer pulse
sequences. We study this by iterating the conditional flip and
inversion steps in Grover's algorithm, which ideally gives rise to a
periodic pattern: for $N=4$, the amplitude of the $x_0$ term is
expected to be 1 after 1 iteration, and again after $4, 7, \ldots$
iterations~\cite{Grover97a}. Fig.~\ref{fig:loglab_spectra}
demonstrates the expected periodic behavior in the output state in
experiments with up to 37 iterations, which requires 448 pulses and
takes about 700 ms.

\bfig
\vspace*{1ex}
\begin{center}
\includegraphics*[width=4cm]{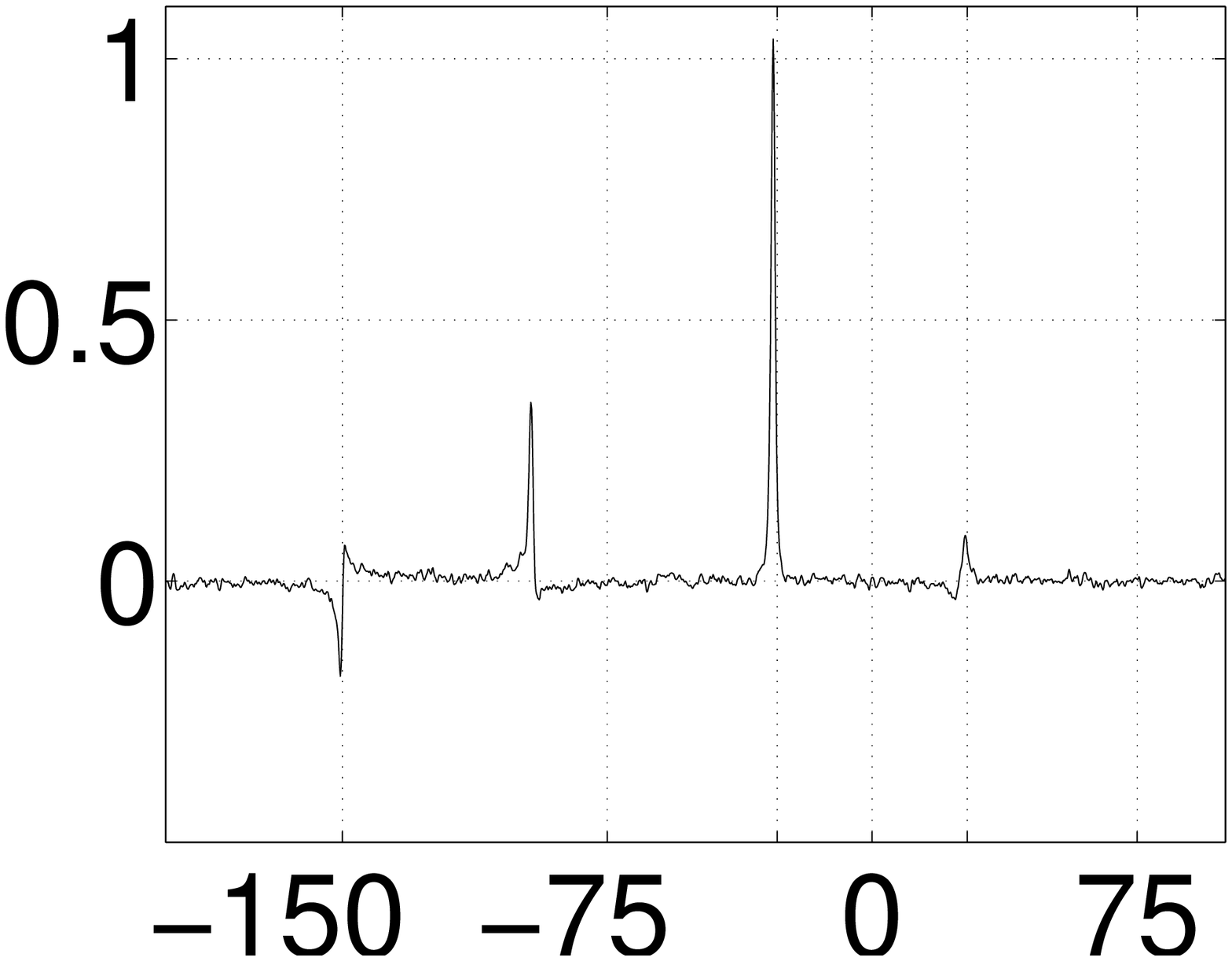}
\includegraphics*[width=4cm]{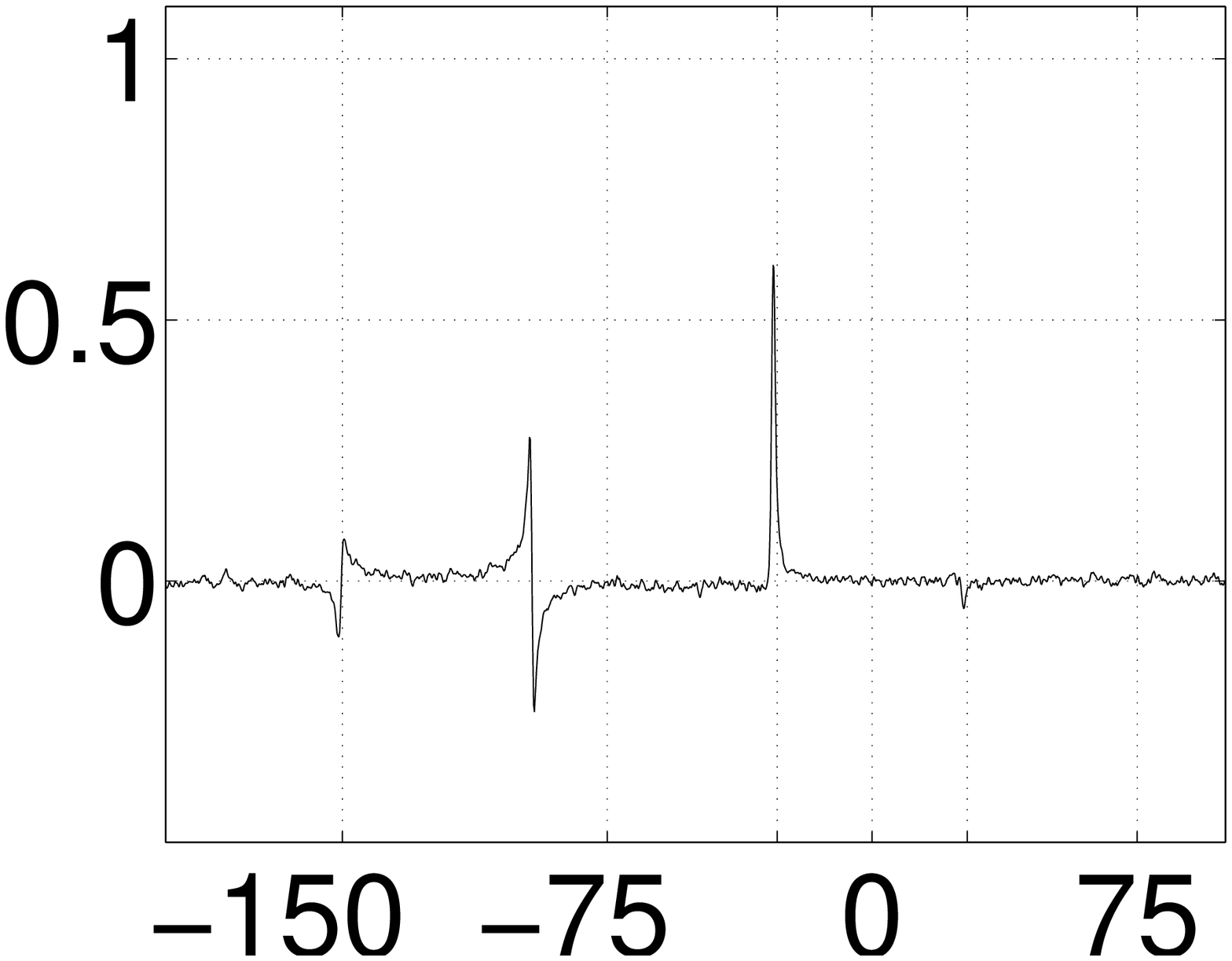}
\includegraphics*[width=4cm]{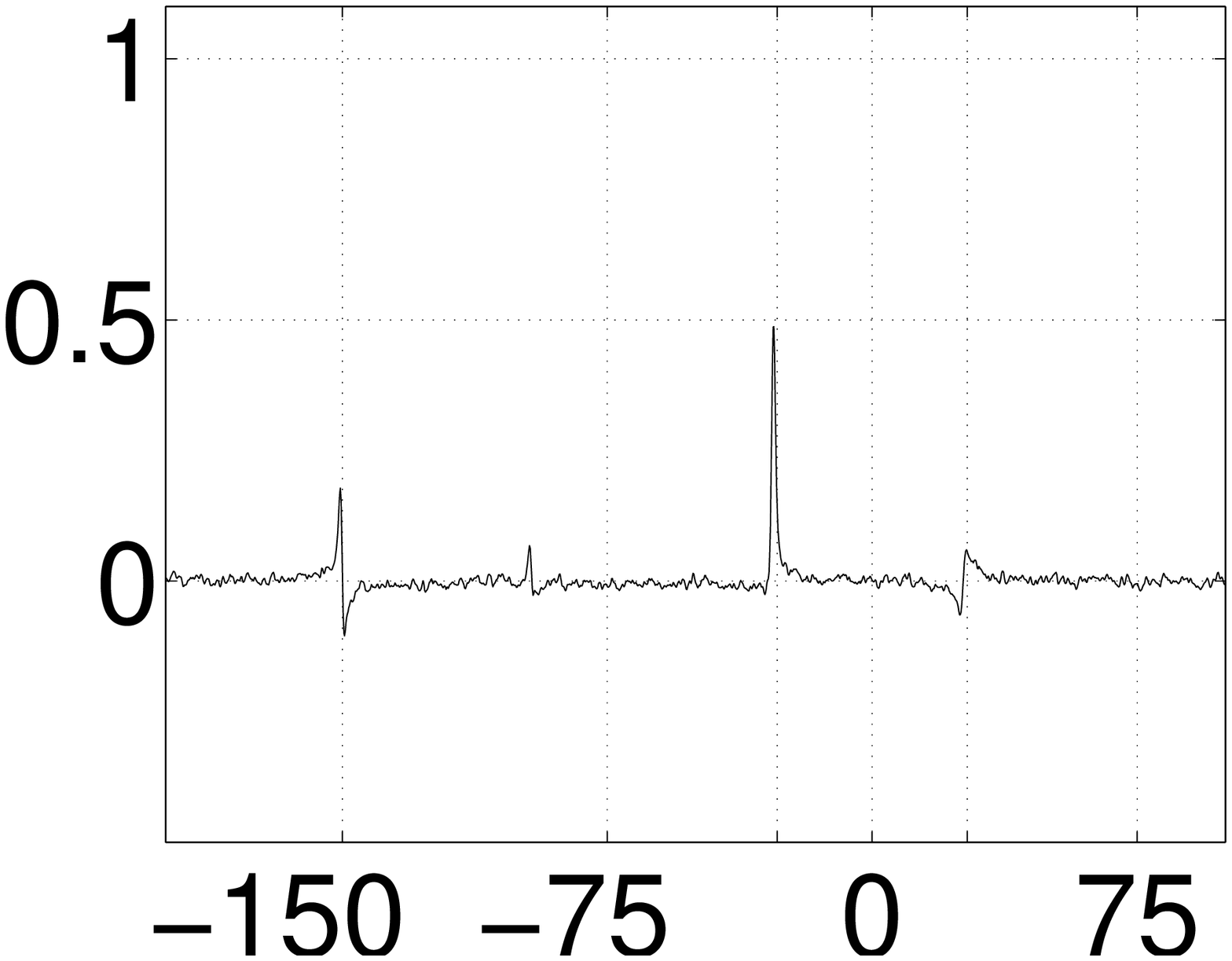}\\
\includegraphics*[width=4cm]{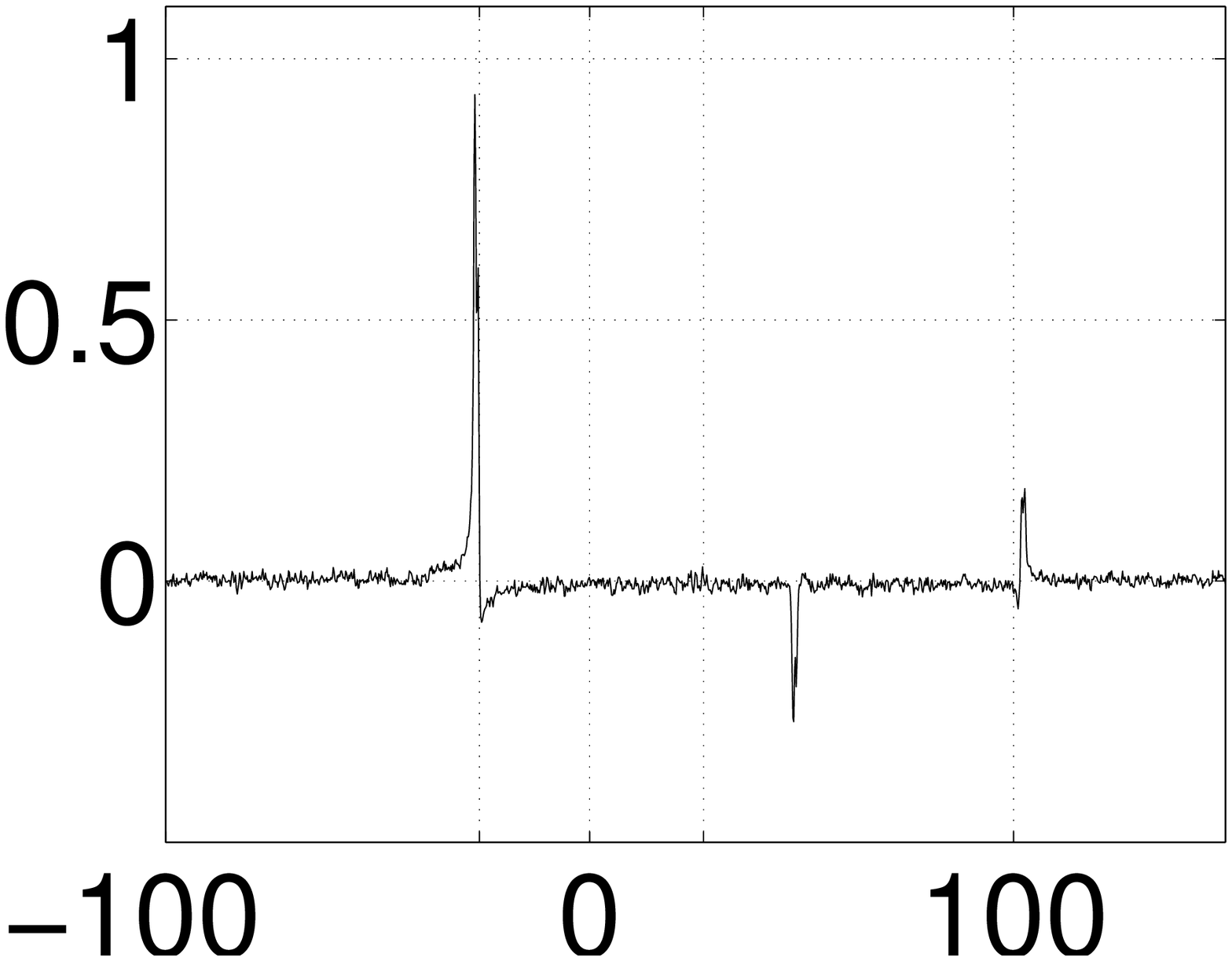}
\includegraphics*[width=4cm]{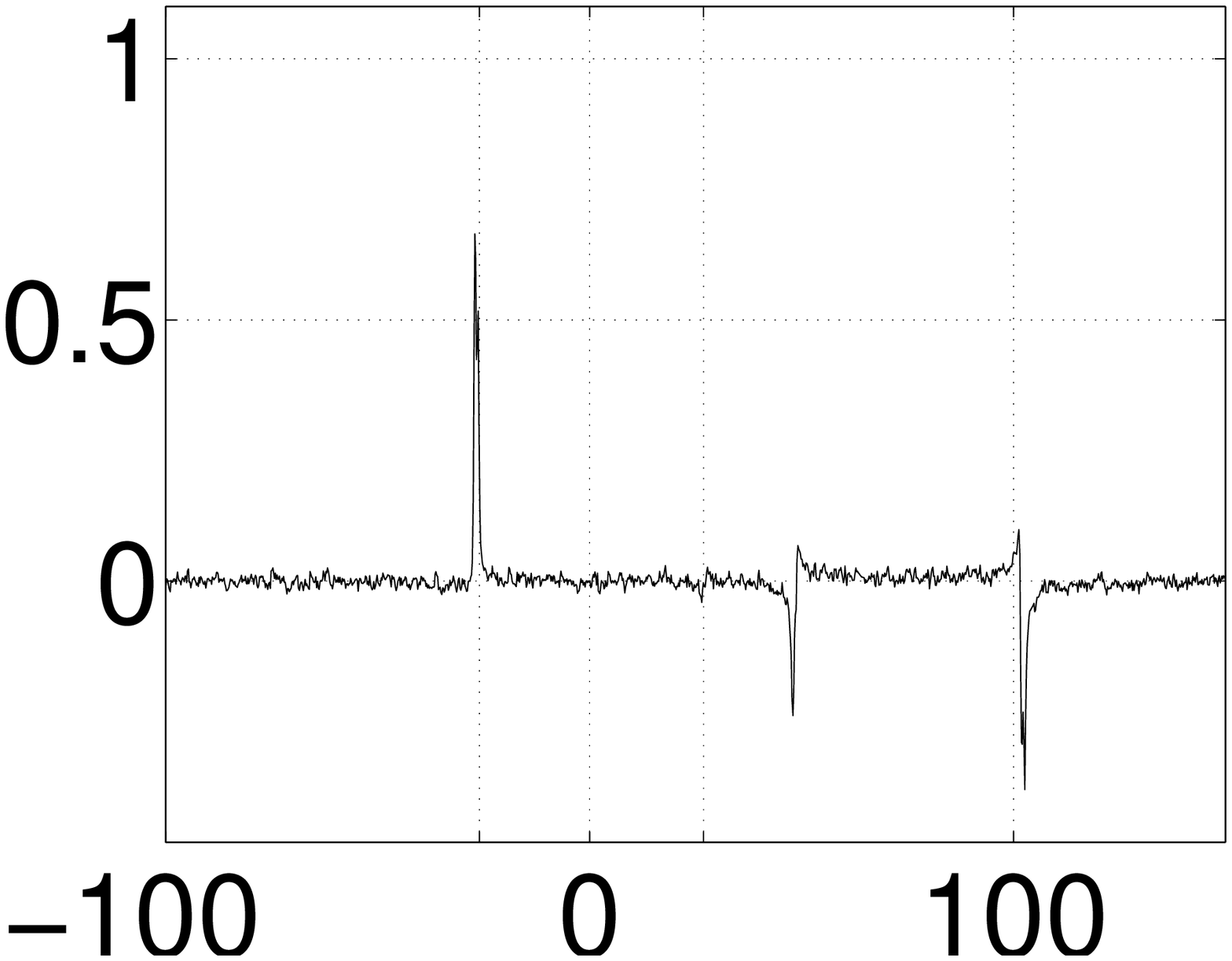}
\includegraphics*[width=4cm]{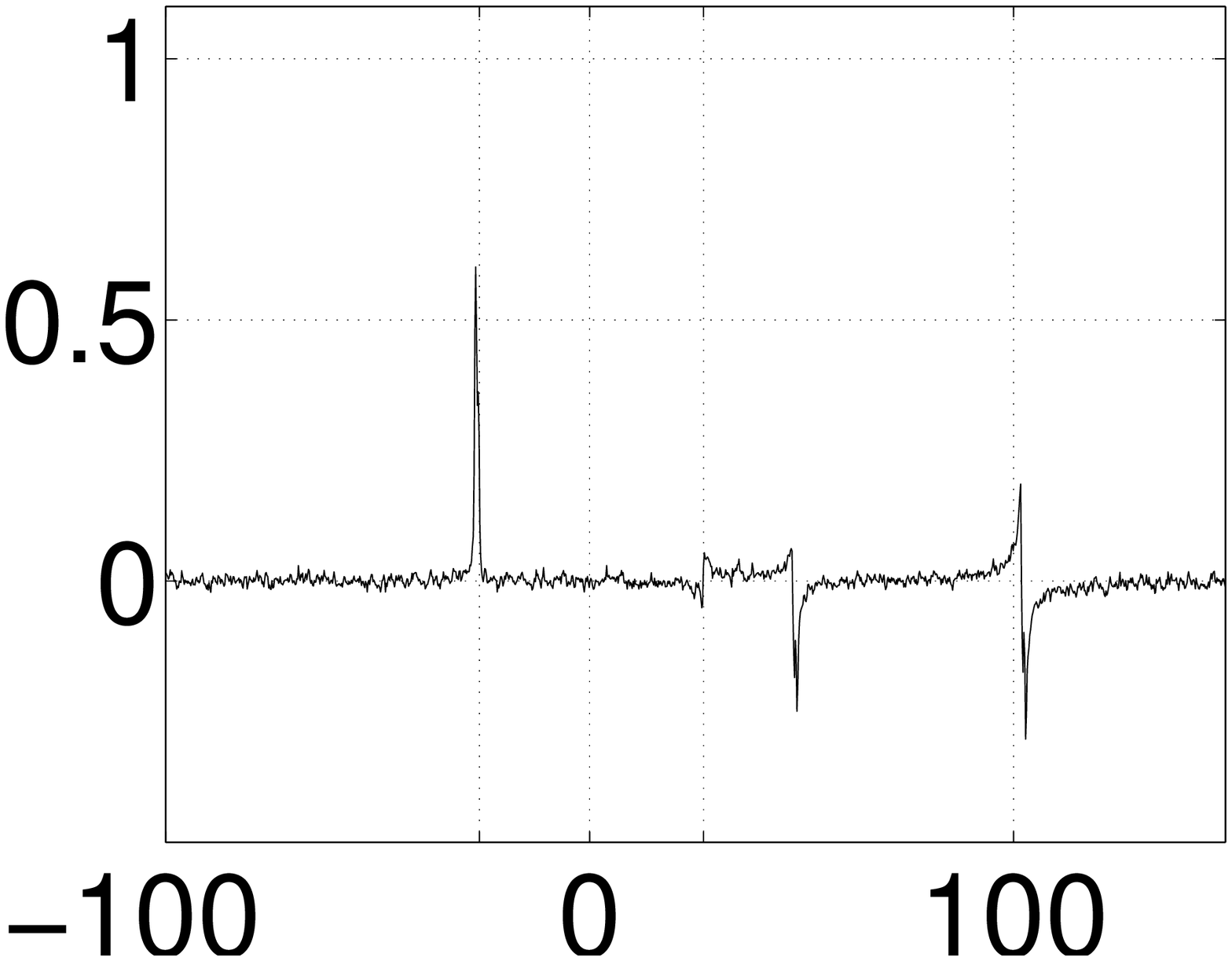}
\end{center}
\vspace*{.5ex}
\caption{Real part of experimental spectra (frequencies in the 
uncoupling reference frame) for spin $2$ (Top) and $3$ (Bottom), after
executing Grover's algorithm 1 (Left), 19 (Center) and 37 (Right)
times, with $\ket{x_0} = \ket{00}$. The $\ket{0}_1$ subspace
corresponds to the spectral lines at $\pm J_{23}/2 = \pm 26.9$
Hz. Ideally, the line at -26.9 Hz is positive and absorptive, with
unit amplitude, while the line at +26.9 Hz is zero. Even after 37
iterations, $x_0$ can be unambiguously determined.}
\label{fig:loglab_spectra}
\end{figure}

\subsection{Discussion}

The experiment met all four goals we set.  It demonstrated that a
$k$-qubit room temperature system can behave as if it were very cold,
up to an exponential decrease in signal strength, when it is properly
embedded in an $n$-spin system.  Furthermore, we demonstrated coherent
control over a homonuclear three-spin system.

The effective pure states were preserved for an unexpectedly long time
and a surprisingly large number of pulses.  We attribute this in part
to the use of the uncoupling frame, which provides an elegant and
simple alternative to refocusing schemes involving $180^\circ$
pulses. Multiple couplings with ancillae spins can be neutralized
simply by moving the carrier frequencies of the computation spins by
the appropriate $\sum \pm J/2$. In contrast, refocusing pulses would
have to be applied during every single evolution interval and their
complexity rapidly increases as more $J$-couplings are to be
refocused~\cite{Linden99c}. This technique may find application in
future experiments using logical labeling, as well as in quantum error
detection experiments~\cite{Leung99a}, where the computation must only
proceed within the subspace labeled error-free by the
ancillae. However, refocusing pulses are still required whenever a
coupling must be removed over an entire system rather than in a
subspace only.

In addition to decoherence, errors mainly arise from imperfections in
the pulses and coupled evolution during the 300 $\mu$s pulses
(reduction of $\tau_1$ and $\tau_2$ partially compensated for this
effect). Spectra of the quality of Fig.~\ref{fig:loglab_spectra} were
obtained by choosing a particular implementation of the composite
$\hat{z}$ rotation in the phase flip step, such that the errors it
introduces partially cancel with the errors of the inversion step. For
example, while $Y X \bar{Y}$, $\bar{Y} \bar{X}$ Y, $X \bar{Y} \bar{X}$
and $\bar{X} Y X$ are all mathematically equivalent, the errors may in
practice add up or cancel out with the errors from previous or
subsequent pulse sequence segments. Clearly, a general optimization
procedure will be very helpful for designing effective pulse sequences
in future experiments involving more qubits.

%%%%%%%%%%%%%%%%%%%%%%%%%%%%%%%%%%%%%%%%%%%%%%%%%%%%%%%%%%%%%%%%%%%%%%

\section{Liquid crystal solutions (2 spins)}
\label{expt:lc}

\subsection{Problem description}

The goal of this experiment~\cite{Yannoni99a}\footnote{Nino Yannoni
proposed to use liquid crystal solvents instead of liquid solvents. He
also selected a suitable liquid crystal solvent and prepared the
sample. Mark Kubinec (at UC Berkeley) and I (at IBM) did
experiments. The published data were taken at IBM. Mark Sherwood and
Dolores Miller simulated the spectra and verified
first-orderness. This work was done under the guidance of Ike Chuang.}
was to explore the use of liquid crystal solvents for quantum
computing. In principle, liquid crystals offer several advantages over
liquids as solvents for molecules used for NMR quantum computing.  A
liquid crystal solvent partly orients the solute molecules
(Fig.~\ref{fig:liquid_crystal}) and as a result, the dipolar coupling
between nuclear spins in molecules is not averaged out anymore as it
is in isotropic solution.  Liquid crystal solvents therefore permit a
significant increase in clock frequency, while short spin-lattice
relaxation times permit fast succession of experiments.  Even more
importantly, the clock frequency may increase by more than the
relaxation rates, so more operations may be completed within the
coherence time.

\bfig
\bcen
\includegraphics*[width=4cm]{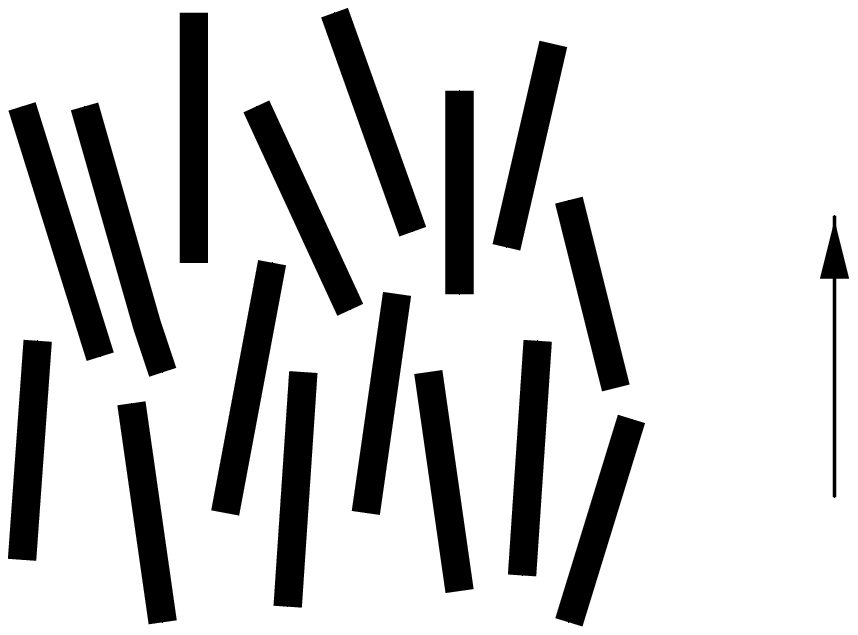}
\ecen
\caption{Liquid crystals are a phase of matter whose order is 
intermediate between that of a liquid and that of a crystal.  The
molecules are typically rod-shaped organic moieties about 25 Angstroms
in length and their ordering is a function of temperature. The liquid
crystal shown here is in the nematic phase. The degree of
orientational order of the constituent molecules decreases with
decreasing temperature.}
\label{fig:liquid_crystal}
\efig

The coupling Hamiltonian for $n$ spins in a molecule dissolved in a
liquid crystal solvent is
\be
{\cal H}^{\rm lc} / \hbar
  =  - \sum_{i}^n {\omega'}_0^i \; I_z^i 
    + \sum_{i<j}^n J_{ij} (I_x^i I_x^j + I_y^i I_y^j + I_z^i I_z^j)
    + \sum_{i<j}^n D_{ij} \left[ { 2 I_z^i I_z^j
          - \frac{1}{2} (I_x^i I_x^j + I_y^i I_y^j) }\right].
\label{eq:ham_lc}
\ee
A dipolar term, which was absent from the coupling Hamiltonian in
liquid solution (Eq.~\ref{eq:ham_iso}) now appears because the
molecules are partially oriented. Also the resonance frequency
${\omega'}_0^i$ includes the effects of molecular orientation and
chemical shift anisotropy~\cite{Emsley75a}. The dipolar coupling
strength $D$, which also depends on the orientation, is typically
$100$ Hz to $10$ kHz.  

Unfortunately, the pulse sequences used for NMR quantum computing in
isotropic liquids can not be applied if the Hamiltonian takes the form
of Eq.~\ref{eq:ham_lc}. However, for an $n$-spin system with {\em first
order} spectra, Eq.~\ref{eq:ham_lc} becomes~\cite{Saupe67a}
\be
     H^{\rm lc}  =
     - \sum_{i=1}^n \hbar \, {\omega'}_0^i \;I_z^i  
+ \hbar \sum_{i<j}^n 2 \pi (J_{ij} + 2D_{ij}) I_z^i I_z^j \,.
\label{eq:ham_lc_iso}
\ee
This Hamiltonian has the same form as the Hamiltonian of
Eq.~\ref{eq:ham_iso}, so the pulse sequences that have been used
successfully for NMR quantum computing in isotropic solution can now
be applied directly to liquid crystal solutions, permitting
computations with $f_{clock} = 2|(J + 2D)|$ Hz, a frequency that can
be much higher than $2|J|$.

\subsection{Experimental approach and results}

We chose $^{13}$C-labeled chloroform (CHCl$_3$) as the quantum computer
molecule. This is the molecule we had used successfully in the first
quantum computing experiments. The liquid crystal we selected was
ZLI-167 (EMI Industries, Hawthorne, NY).

Table~\ref{tab:relax_lc} shows the $^{13}$C - $^{1}$H coupling
strength, the spin-lattice relaxation time ($T_1$) and the spin-spin
relaxation time ($T_2$) for $^{13}$C and $^{1}$H in
chloroform ($^{13}$CHCl$_3$) in liquid crystal solution and for
comparison also in isotropic solution, both at ambient
temperature.

\begin{table}[htbp]
\begin{center}
%\vspace*{1cm}
\begin{tabular}{c|cccccc}
solvent & $J$ & $J+2D$ & $T_1$ ($^{13}$C)   &  $T_1$ ($^{1}$H)
     & $T_2$ ($^{13}$C) &  $T_2$ ($^{1}$H)
\\\hline
acetone-$d_6$ & 215 & \rule{2ex}{0.2pt} & 25 & 19 & 0.3 & 7 \\
ZLI-1167 & \rule{2ex}{0.2pt} & 1706 & 2 & 1.4 & 0.2 & 0.7 \\
\end{tabular}
\end{center}
\vspace*{-1ex}
\caption{$^{13}$C-$^{1}$H spin couplings [Hz] and relaxation times [s] for 
$^{13}$C$^{1}$HCl$_3$ in isotropic (deuterated acetone) and liquid
crystal (ZLI-1167) solution.}
\label{tab:relax_lc}
\end{table}

In order to show that quantum computations can be done successfully
using liquid-crystal solution NMR, we have implemented the Grover
search algorithm for two qubits using $^{13}$CHCl$_3$ dissolved
in ZLI-1167. The carbon and proton spins were first prepared in an
effective pure state created by temporal labeling (via cyclic
permutations, see section~\ref{nmrqc:templab}) followed by the Grover
protocol used in the logical labeling experiment of
section~\ref{expt:labeling}.

\begin{figure}[h]
\bcen
\vspace*{1ex}
\includegraphics*[width=5cm]{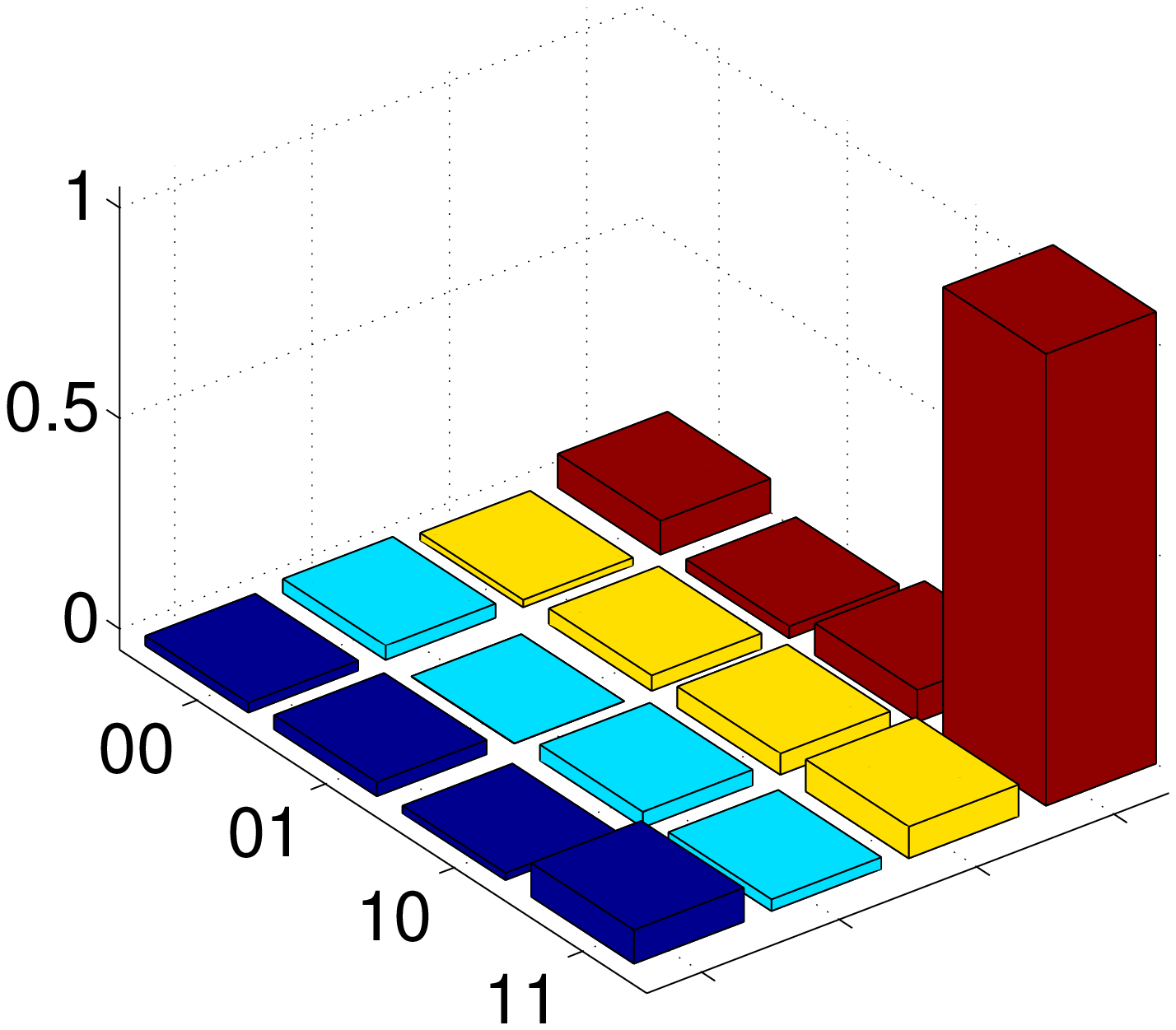} 
\vspace*{-2ex}
\ecen
\caption{Experimentally measured deviation density matrix elements for 
the $x_0=11$ case.}
\label{fig:lcgrover_denmat}
\efig

Complications do arise with the use of liquid crystal solvents: (1)
the NMR lines of small molecules dissolved in liquid crystal solvents
are susceptible to a broadening mechanism not found in isotropic
solution, most likely due to variations in the degree of orientation
caused by thermal gradients in the sample.  Nonetheless, resonance
line widths $<2$ Hz ($^{13}$C) and $<3$ Hz ($^{1}$H) were obtained for
$^{13}$C$^{1}$HCl$_3$ in ZLI-1167; (2) the large dipolar couplings may
cause unwanted evolution of the spins during the RF pulses, especially
in homonuclear spin systems.

We have not used liquid crystal solvents in any of the later
experiments. However, due to their many advantages and only moderate
disadvantages, we believe it is worth revisiting this possibility in
the context of larger spin systems.

%%%%%%%%%%%%%%%%%%%%%%%%%%%%%%%%%%%%%%%%%%%%%%%%%%%%%%%%%%%%%%%%%%%%%%

\section{Cancellation and prevention of systematic errors (3 spins)}
\label{expt:grover3}

\subsection{Problem description}

In this experiment~\cite{Vandersypen00a}\footnote{I worked out the
theory and many of the pulse sequence simplification ideas, along with
Matthias Steffen and Ike Chuang. The simplified temporal averaging
scheme is due to Ike. The experiment and data analysis was done by
myself and Matthias. The molecule was synthesized by Greg Breyta, and
proposed by Nino Yannoni and Mark Sherwood, who also gave advice on
NMR techniques.}  we studied the effect of systematic errors in the
one-qubit gates. Aside from the well-understood scaling limitations
due to the use of a high-temperature (almost random) system instead of
a low-temperature (low entropy) polarized spin system
(section~\ref{nmrqc:init}), such errors represent an important
limitation in using nuclear spins in molecules to implement larger
quantum algorithms.

Single-qubit gates are implemented by applying RF pulses of precise
duration and phase, but which in practice greatly vary in strength
over the sample volume, causing the gate fidelity~\cite{Schumacher96b}
to be less than $95$\% (section~\ref{expt:apparatus}).  Producing a
homogeneous RF field is difficult because of the sample geometry and
the necessity of keeping the $B_1$ field transverse to the $B_0$
field.  If such systematic errors simply accumulated, these
observations would imply that for a success rate of only 1$\%$, fewer
than 90 gates (0.95$^{90} \approx$ 0.01) applied to any one spin could
ever be cascaded in these systems.

One technique which has been proposed for controlling errors in
quantum gates is quantum error correction (section~\ref{qct:qec}), but
this is associated with a large overhead. The principle attribute of
quantum error correction techniques is their ability to correct
completely random errors which originate from fundamentally
irreversible {\em decoherence} phenomena; in principle, systematic
errors, which are inherently {\em reversible} --- at least on an
appropriate time scale --- should be much easier to control, given
knowledge about their origin.

We tested the extent to which systematic errors can (1) be avoided by
simplifying the pulse sequences and (2) be made to cancel out in
practice. The concrete experiment for this test consisted of repeated
executions of the two main steps in Grover's algorithm (the oracle
call and the inversion about the average described in
section~\ref{qct:grover}) for three spins.

The three-qubit Grover algorithm can find a ``marked'' element $x_0$
among $N=8$ possible values of $x$ in only two oracle queries
(evaluations of $f(x)$), whereas a classical search needs 4.375
attempts on average to find $x_0$. The oracle call requires one {\sc
Toffoli} gate plus several one-qubit gates, and so does the inversion
about the average. The amplitude of $\ket{x_0}$ is predicted to reach
a first maximum after two iterations, and oscillates as the number of
iterations increases.

\subsection{Experimental approach}

The first method we developed to reduce the number of one- and
two-spin gates needed to implement arbitrary unitary operations was
the general pulse simplification methodology of
section~\ref{nmrqc:seq_design}. As a result, the pulse sequence for
each Toffoli gate used in the algorithm was reduced from 70.5
$90^\circ$ pulses and 8 evolutions of $1/2J$ (if only {\sc cnot}'s and
1-qubit gates would have been used to implement the controlled-$V$'s,
as in the standard methods of~\cite{Barenco95a}) to 19 pulses, 2
evolutions of $1/2J$ and 3 of $1/4J$.

The second method to reduce the complexity concerns the initialization
of the qubits to an effective ground state. We generalized the
temporal averaging procedure from a scheme based on cyclic
permutations to a scheme based on arbitrary linearly independent
experiments (section~\ref{nmrqc:init}).  Whereas cyclic permutations
would require seven experiments with very complex state preparation
sequences, all the data shown here were obtained using just three
experiments with much simpler preparation sequences. The expected
variance of the $2^n-1$ populations obtained with this state
preparation procedure amounts to only 7$\%$ of their average value.

The actual pulse sequences used in the experiment are collected in
Appendix~\ref{app:grover3}.

\subsection{Experimental Results}

We selected $^{13}$C-labeled CHFBr$_2$~\footnote{Synthesized by
heating a mixture of $^{13}$CHBr$_3$ (2.25 g, CIL) and HgF$_2$ (2.8 g,
Aldrich) in increments (5$^{\circ}$C for 15 min) from 70$^{\circ}$C to
85$^{\circ}$C in a Kugelrohr apparatus and condensing the product into
a cooled bulb.  This material was re-distilled bulb-to-bulb at
65$^{\circ}$C to give 750 mg (99 $\%$ purity) of $^{13}$CHFBr$_2$,
which was dissolved in d6-acetone.} for our experiments. The $^1$H,
$^{19}$F and $^{13}$C spins served as the quantum bits. The coupling
constants in this heteronuclear spin system are $J_{HC}$ = 224 Hz,
$J_{HF}$ = 50 Hz and $J_{FC}$ = $-311$ Hz. As always, the scalar
interaction with the Br nuclei is averaged out and only contributes to
decoherence.

In order to read out the final state of the three spins, we used the
extra information given by the multiple lines in the spectrum of each
spin.  Given that all three spins are in an effective pure energy
eigenstate and that they are all mutually coupled, each spectrum
contains only a single line, the frequency of which, combined with the
knowledge of the $J_{ij}$, reveals the state of the remaining spins.
The inset of Fig.~\ref{fig:grover_denspect} (a) gives the
experimentally measured $^{13}$C output spectrum after two Grover
iterations. Fig.~\ref{fig:grover_denspect} (a) also gives the complete
output {\em deviation} density matrix.

\begin{figure}[h]
\vspace*{1ex}
\bcen
\begin{overpic}[width=7cm]{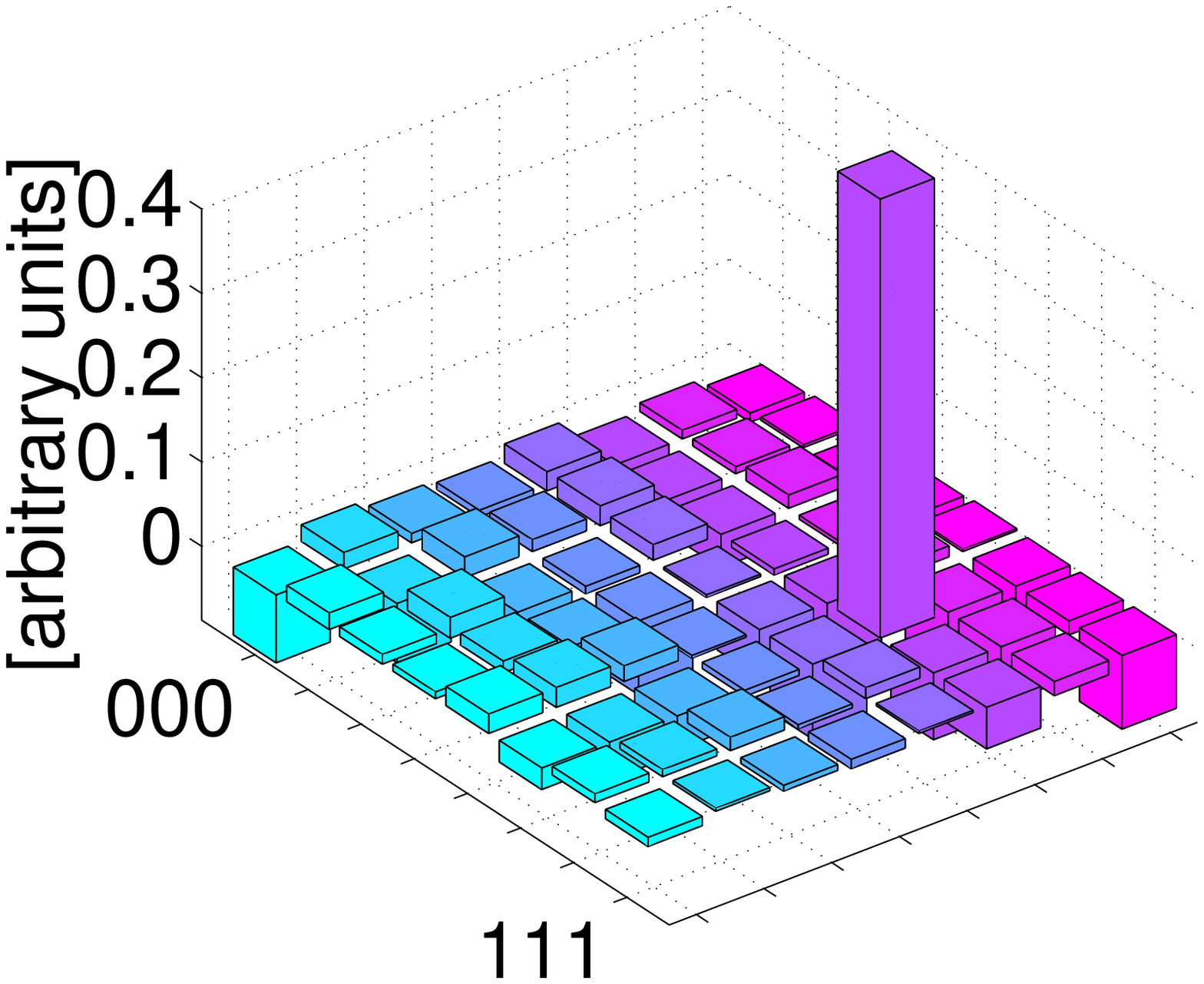}
\put(18,40){\includegraphics*[width=3.1cm]{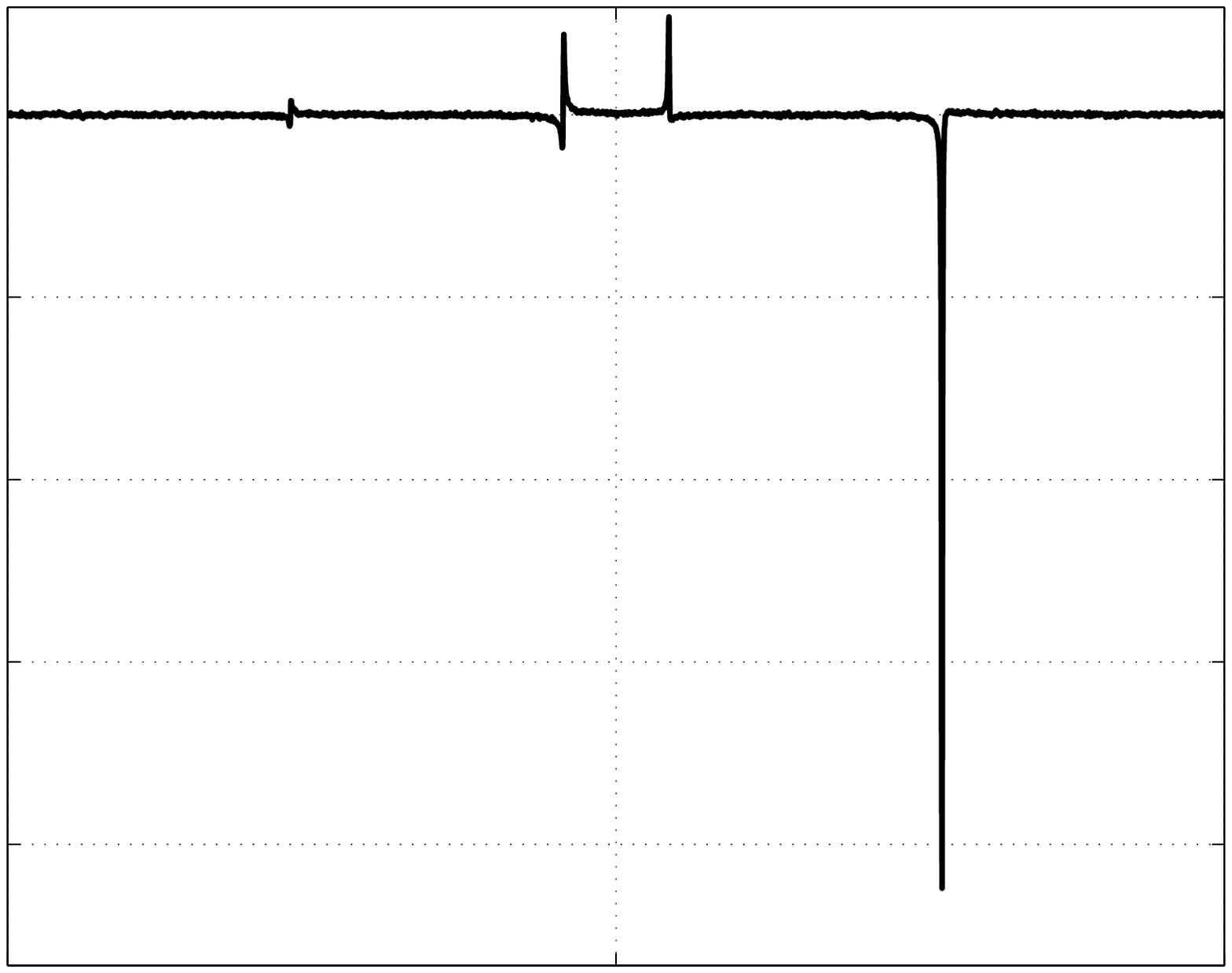}}
\end{overpic}
\hspace*{.5cm}
\begin{overpic}[width=7cm]{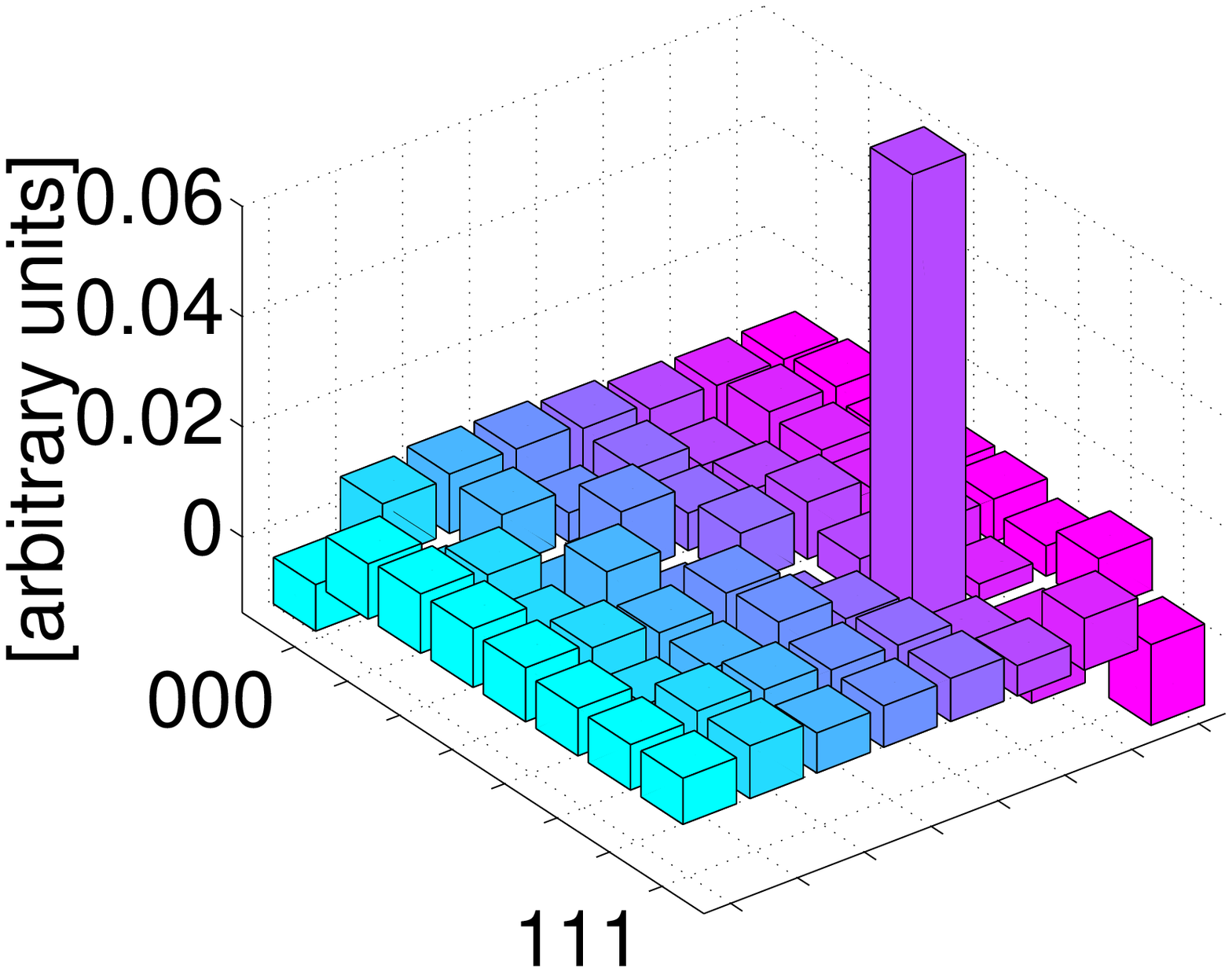}
\put(20,40){\includegraphics*[width=3.1cm]{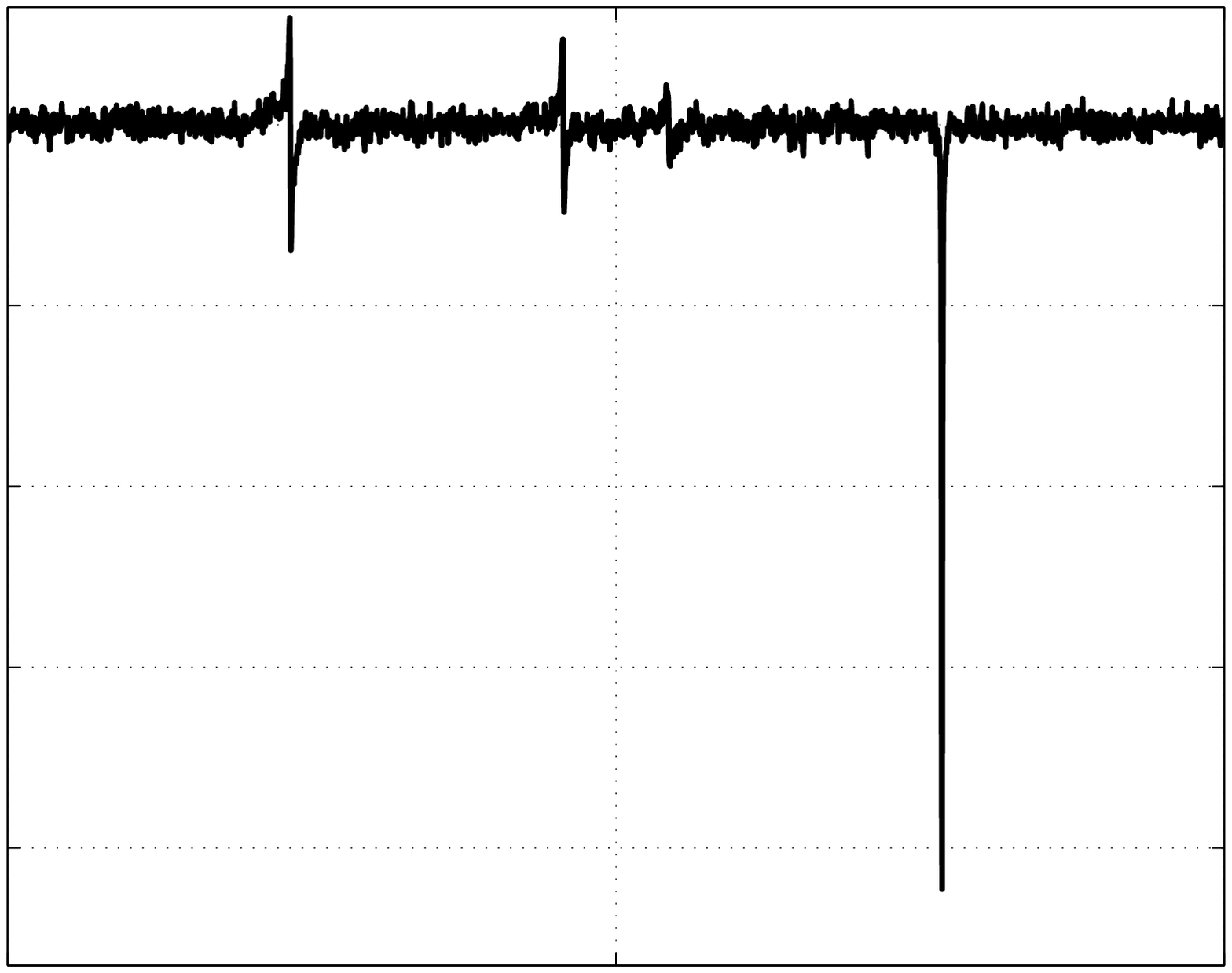}}
\end{overpic}
\ecen
\vspace*{-2ex}
\caption{(Top) Experimental deviation density matrices $\rho_{exp}$ for 
$\ket{x_0}=\ket{1} \ket{0} \ket{1}$, shown in magnitude with the sign
of the real part (all imaginary components were small), after (a) 2
and (b) 28 Grover iterations. (Bottom) The corresponding $^{13}$C
spectra ($^{13}$C was the least significant qubit). The receiver phase
and read-out pulse are set such that the spectrum be absorptive and
positive for a spin in $\ket{0}$.}
\label{fig:grover_denspect}
\end{figure}

The agreement between experimental results and theoretical predictions
is good, considering that about 100 pulses were used and that the
systematic error rate exceeds $5\%$ per RF pulse (the measured signal
loss due to RF field inhomogeneity after applying $X_i$). This
suggests that the systematic errors cancel each other out to some
degree. We examined this in more detail in a series of experiments
with increasingly longer pulse sequences executing up to 28 Grover
iterations (Fig.~\ref{fig:grover_iter}).

\begin{figure}[h]
\vspace*{1ex}
\bcen
\includegraphics*[width=6.5cm]{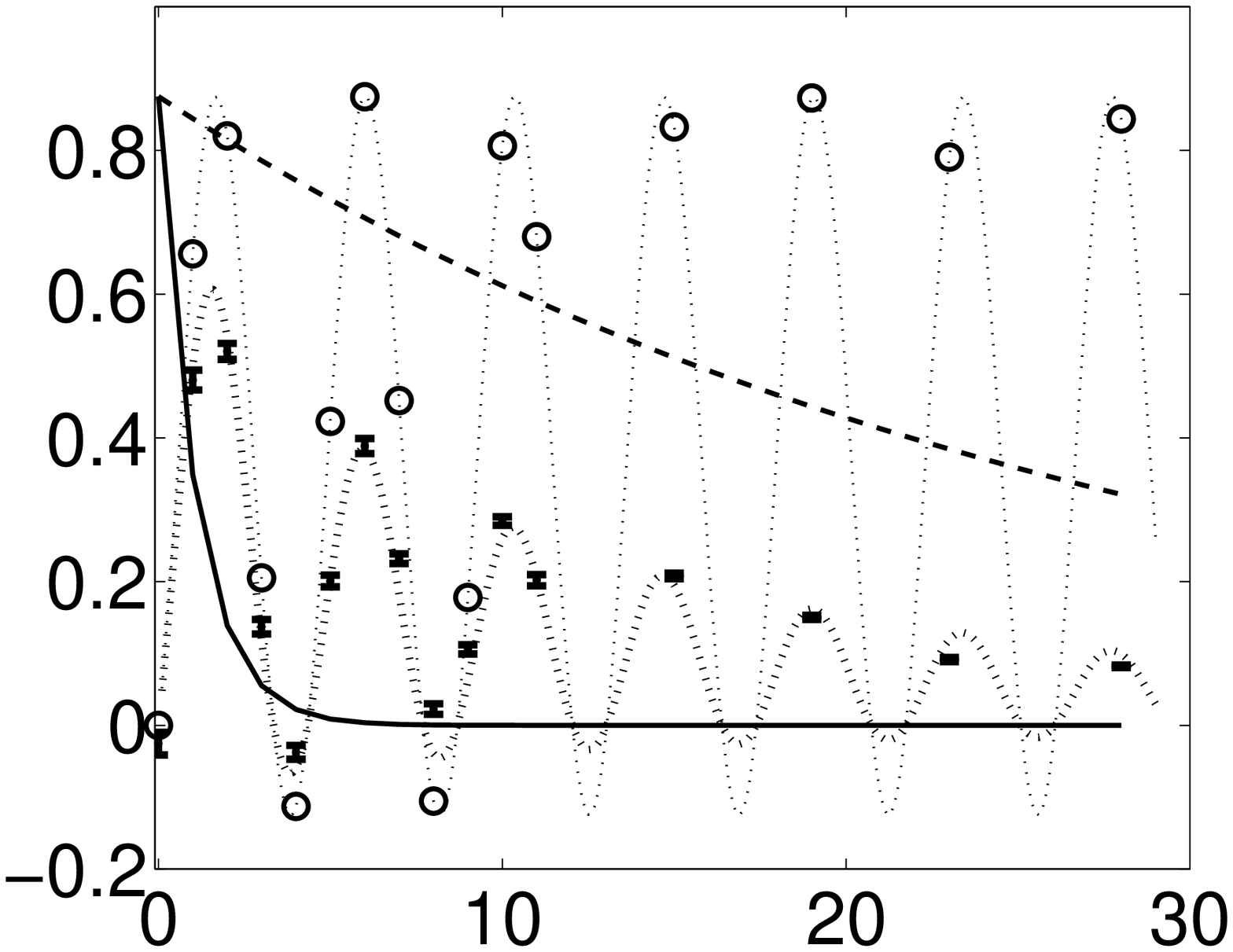}
\hspace*{.7cm}
\includegraphics*[width=6.5cm]{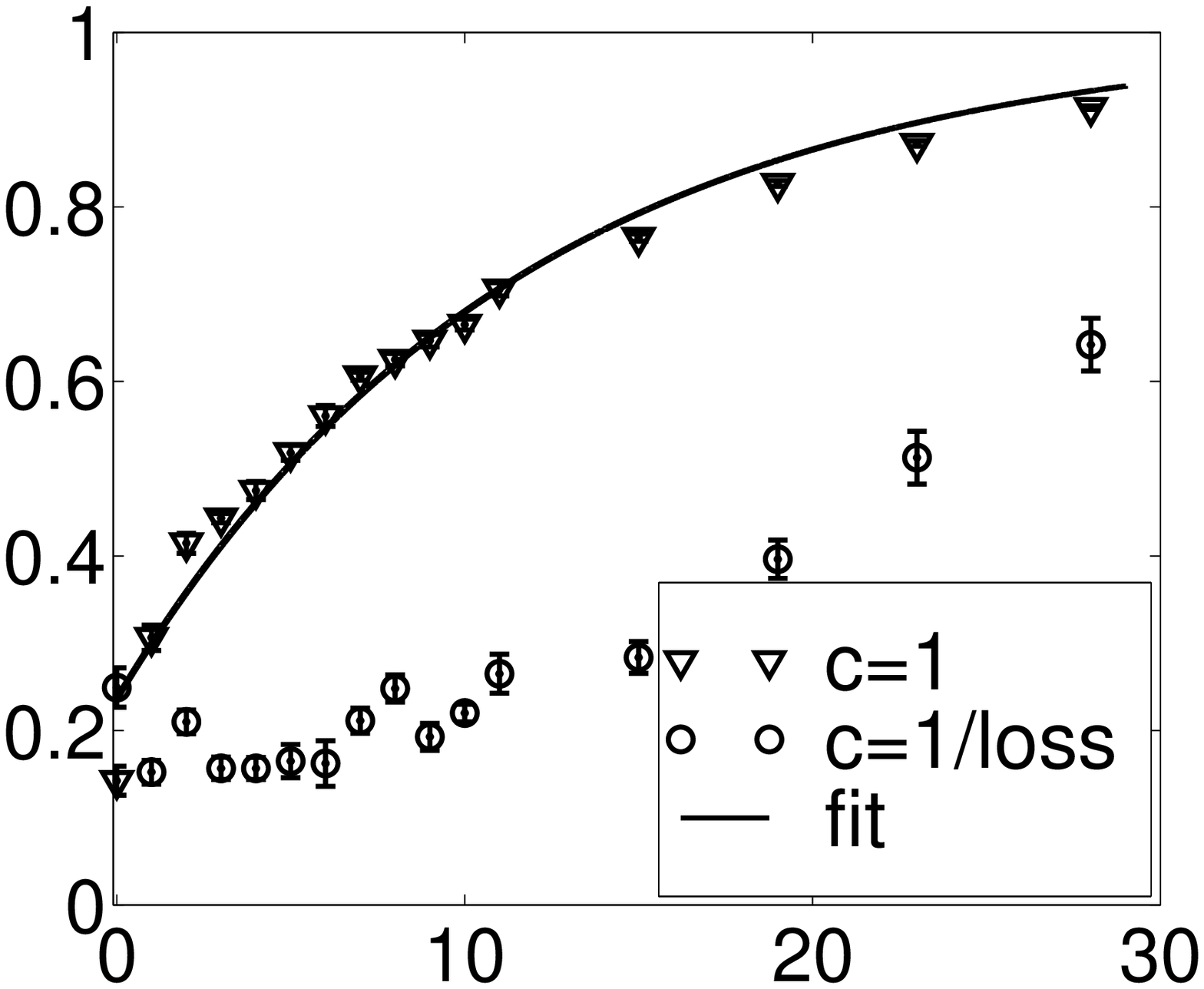}
\ecen
\vspace*{-2ex}
\caption{(a) Experimental (error bars) and ideal (circles) 
amplitude of $d_{x_0}$, with fits (dotted) to guide the eye. Dashed
line: the signal decay for $^{13}$C due to intrinsic phase
randomization or decoherence (for $^{13}$C, $T_2 \approx 0.65$
s). Solid line: the signal strength retained after applying a
continuous RF pulse of the same cumulative duration per Grover
iteration as the pulses in the Grover sequence (averaged over the
three spins; measured up to 4 iterations and then extrapolated). (b)
The relative error $\epsilon_r$.}
\label{fig:grover_iter}
\end{figure}

\subsection{Discussion}

Fig.~\ref{fig:grover_iter} (a) shows that the diagonal entry $d_{x_0}$
of $\rho_{exp}$ oscillates as predicted but the oscillation is damped
as a result of errors, with a time constant $T_d$ of 12.8 iterations.
However, $T_d$ would have been smaller than 1.5 if the errors due to
just the RF field inhomogeneity were cumulative
(Fig.~\ref{fig:grover_iter} (a), solid line).  Remarkably, after a
considerable initial loss, $d_{x_0}$ decays at a rate close to the
$^{13}$C $T_2$ decay rate (dashed line), which can be regarded as a
lower bound on the overall error rate.  

A more complete measure to quantify the error and benchmark results is
the relative error $\epsilon_r = \parallel c \rho_{exp} - \rho_{th}
\parallel_2 / \parallel \rho_{th} \parallel_2$, where $\rho_{exp}$ and 
$\rho_{th}$ are the experimental and theoretical (traceless) deviation
density matrices. Comparison of $\epsilon_r$ with $c=1$ and $c$ equal
to the inverse of the signal loss (Fig.~\ref{fig:grover_iter} (b))
reveals that signal loss dominates over other types of
error. Furthermore, the small values of $\epsilon_r$ with $c>1$
suggest that $\ket{x_0}$ can be unambiguously identified, even after
almost 1350 pulses. This is confirmed by the density matrix measured
after 28 iterations, which has a surprisingly good signature
(Fig.~\ref{fig:grover_denspect} (b)). Given the error of $>5\%$ per
single $90^\circ$ rotation, all these observations demonstrate that
substantial cancellation of errors took place in our experiments.

The error cancellation achieved was partly due to a judicious choice
of the phases of the refocusing pulses, but a detailed mathematical
description in terms of error operators is needed to fully exploit
this effect in arbitrary pulse sequences. This difficult undertaking
is made worthwhile by our observations. This conclusion is
strengthened by a similar observation in the experiment of
section~\ref{expt:labeling}. Also, we believe that error cancellation
behavior is {\em not} just a property of the Grover iterations,
because we found experimentally that the choice of implementation of
the building blocks dramatically affects the cancellation
effectiveness.

In summary, more than 280 two-qubit quantum gates involving 1350 RF
pulses were successfully cascaded, which far exceeds not only the
number of gates used in all previous NMR quantum computing experiments
but also the limitation of 90 pulses, imposed by cumulative systematic
errors.  Whereas the cancellation of systematic errors makes it
possible to perform such a surprisingly large number of operations,
the methods for simplifying pulse sequences reduce the number of
operations needed to implement a given quantum circuit. This
combination permitted the observation of 28 full cycles of the Grover
algorithm with 3 spins, and suggests that many other interesting
quantum computing experiments may be within reach.

%%%%%%%%%%%%%%%%%%%%%%%%%%%%%%%%%%%%%%%%%%%%%%%%%%%%%%%%%%%%%%%%%%%%%%

\section{Efficient cooling (3 spins)}
\label{expt:cooling}

\subsection{Problem description}

The goal of this experiment~\cite{Chang01a}\footnote{I devised this
experiment. Darrick Chang, a summer student, worked out the pulse
sequences and took the data under my guidance, and with the help of
Matthias Steffen.} was to cool down the spin temperature of one out of
three spins using the Schulman-Vazirani scheme for efficient cooling
\cite{Schulman99a}. A secondary goal was to draw the attention of the
quantum computing community to the Schulman-Vazirani algorithm.

The ideas underlying Schulman-Vazirani cooling have been introduced in
section~\ref{impl:init}. The quantum circuit which summarizes the
steps in the Schulman-Vazirani boosting procedure, was described in
section~\ref{nmrqc:cooling}, Fig.~\ref{fig:PTcircuit2}. We include it
here again for convenience.

\begin{figure}[h]
\bcen
\vspace*{1ex}
\includegraphics*[width=5.5cm]{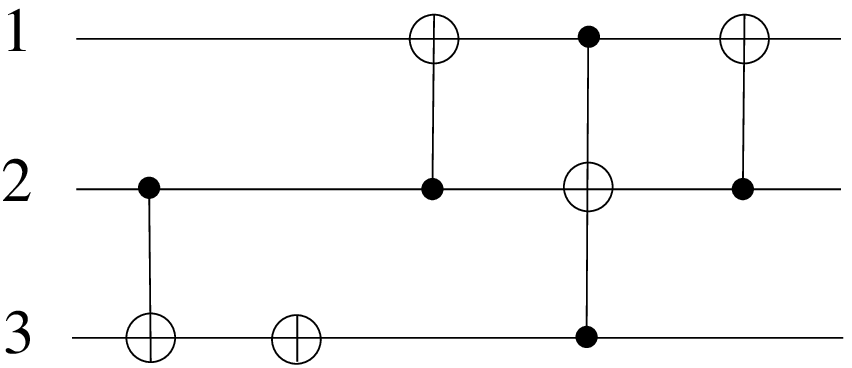} 
\vspace*{-2ex}
\ecen
\caption{A quantum circuit that implements the Schulman-Vazirani 
boosting procedure for cooling one out of three qubits.}
\label{fig:PTcircuit2}
\end{figure}

While it is common in NMR to enhance the polarization of a
low-$\gamma$ nucleus by polarization transfer from a high-$\gamma$
nucleus~\cite{Ernst87a,Freeman97a}, in this experiment the goal was to
enhance the polarization of one out of three {\em equally} polarized
spins. Any polarization gain is thus exclusively the result of
bootstrapping using the Schulman-Vazirani cooling scheme.  In this
case, the theoretical maximum achievable polarization enhancement is
by a factor of 3/2; this is equivalent to lowering the spin
temperature of that spin by the same factor.

\subsection{Experimental procedure}

A 2 mol $\%$ solution of $\mathrm{C_{2}F_{3}Br}$ in deuterated acetone
provided three spins with equal polarizations. We chose this
particular molecule for its remarkable spectral properties (we had
successfully used it in the logical labeling experiment of
section~\ref{expt:labeling}): strong chemical shifts (0, 28.2, and
48.1 ppm, arbitrarily referenced) and large scalar couplings
($J_{ab}=-122.1$ Hz, $J_{ac}=75.0$ Hz, and $J_{bc}=53.8$ Hz) combined
with long relaxation times ($T_2$'s $\approx 4$-$8$ s). The
experiments were conducted at $30.0^{\circ}$C.

The quantum circuit of Fig.~\ref{fig:PTcircuit2} results in the
unitary operation (with qubit $1$ the most significant qubit)
\be
U=
\left( \begin{array}{p{4mm}p{4mm}p{4mm}p{4mm}p{4mm}p{4mm}p{4mm}p{4mm}}
0 & 1 & 0 & 0 & 0 & 0 & 0 & 0 \\
1 & 0 & 0 & 0 & 0 & 0 & 0 & 0 \\
0 & 0 & 1 & 0 & 0 & 0 & 0 & 0 \\
0 & 0 & 0 & 0 & 1 & 0 & 0 & 0 \\
0 & 0 & 0 & 0 & 0 & 1 & 0 & 0 \\
0 & 0 & 0 & 1 & 0 & 0 & 0 & 0 \\
0 & 0 & 0 & 0 & 0 & 0 & 1 & 0 \\
0 & 0 & 0 & 0 & 0 & 0 & 0 & 1
\end{array} \right)
\,,
\label{eq:U_SV}
\ee
which transforms the thermal density matrix as 
\be
I_z^1+I_z^b+I_z^3 \,\, {\rightarrow} \,\,
\frac{3}{2} \, I_z^1 + \frac{1}{2} \, I_z^2 - I_z^1I_z^3 - I_z^2I_z^3
\,.
\label{eq:thermal_SV}
\ee
The propagator $U$ thus redistributes the populations in such a way
that the highest populations are moved to states where qubit 1 is in
$|0{\rangle}$.  This can be clearly seen by expressing the
resulting density matrix in explicit matrix form:
\be
{\rho}_{f}=\frac{1}{2}
\left( \begin{array}{p{4mm}p{4mm}p{4mm}p{4mm}p{4mm}p{4mm}p{4mm}p{4mm}}
1 & 0 & 0 & 0 & 0 & 0 & 0 & 0 \\
0 & 3 & 0 & 0 & 0 & 0 & 0 & 0 \\
0 & 0 & 1 & 0 & 0 & 0 & 0 & 0 \\
0 & 0 & 0 & 1 & 0 & 0 & 0 & 0 \\
0 & 0 & 0 & 0 & -1 & 0 & 0 & 0 \\
0 & 0 & 0 & 0 & 0 & -1 & 0 & 0 \\
0 & 0 & 0 & 0 & 0 & 0 & -1 & 0 \\
0 & 0 & 0 & 0 & 0 & 0 & 0 & -3
\end{array} \right)
\,.
\label{eq:rho_f_SV}
\ee
Because the density matrix remains in a diagonal state after
application of each quantum gate in Fig.~\ref{fig:PTcircuit2}, the
boosting procedure can actually be implemented using a simplified
quantum circuit: replacing each gate with a gate whose unitary matrix
is correct up to phases preserves the transformation given by
Eq.\ref{eq:U_SV}.  Consequently, the Toffoli gate, for which the
fastest known implementation takes on the order of $7/4J$ seconds
(taking all $J_{ij}$ to be $\approx J$), can be substituted with a
Toffoli gate correct up to phases --- consisting of a $90^{\circ}$
$\hat{y}$ rotation of qubit $2$ when qubit $3$ is in $\ket{1}$,
followed by a $180^{\circ}$ $\hat{z}$ rotation of $2$ when $1$ is in
$\ket{1}$ and a $-90^{\circ}$ $\hat{y}$ rotation of $2$ when $3$ is in
$\ket{1}$ --- which takes only $1/J$ seconds.  The actual pulse
sequence used in the experiment is given in
Fig.~\ref{fig:cooling_seq}.  This sequence was designed by standard
pulse sequence simplification techniques supplemented by Bloch-sphere
intuition.  The resulting unitary operator is
\be
\tilde{U}=
\left( \begin{array}{p{4mm}p{4mm}p{4mm}p{4mm}p{4mm}p{4mm}p{4mm}p{4mm}}
0 & 1 & 0 & 0 & 0 & 0 & 0 & 0 \\
1 & 0 & 0 & 0 & 0 & 0 & 0 & 0 \\
0 & 0 & -1 & 0 & 0 & 0 & 0 & 0 \\
0 & 0 & 0 & 0 & -1 & 0 & 0 & 0 \\
0 & 0 & 0 & 0 & 0 & 1 & 0 & 0 \\
0 & 0 & 0 & -1 & 0 & 0 & 0 & 0 \\
0 & 0 & 0 & 0 & 0 & 0 & 1 & 0 \\
0 & 0 & 0 & 0 & 0 & 0 & 0 & 1
\end{array} \right)
\,.
\ee

\begin{figure}[h]
\bcen
\vspace*{1ex}
\includegraphics*[width=16cm]{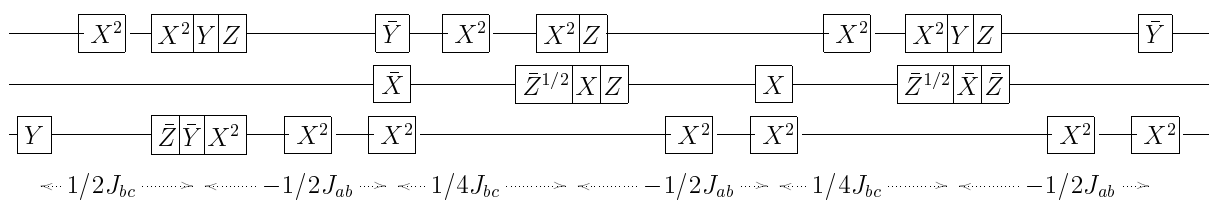} 
\vspace*{-2ex}
\ecen
\caption{Pulse sequence used to implement the boosting procedure. 
This pulse sequence is designed for molecules with
$J_{ab}<0$ and $J_{ac}, J_{bc}>0$.}
\label{fig:cooling_seq}
\end{figure}

All pulses were spin-selective, and varied in duration from $1$ to $3$
ms.  Hermite 180 and av90 shaped
pulses were employed for $180^\circ$ and $90^\circ$ rotations
respectively, in order to minimize the effect of the $J$ couplings
between the selected and non-selected spins during the
pulses. Couplings between the unselected spins are irrelevant whenever
those spins are along $\pm \hat{z}$, as is the case here.

Bloch-Siegert shifts (sections~\ref{nmrqc:pulse_artefacts}
and~\ref{nmrqc:simpulse_artefacts}) were accounted for in the pulse
sequence out of necessity: they resulted in an extra phase acquired by
the non-selected spins in their respective on-resonance reference
frames, in some cases by more than $90^\circ$ per pulse, while even
phase shifts on the order of $5^\circ$ are unacceptable.
Bloch-Siegert corrections and other $\hat{z}$ rotations were
implicitly performed by changing the phase of subsequent RF pulses.
The duration of the pulse sequence of Fig.~\ref{fig:cooling_seq} is
about $70$ ms.

\subsection{Experimental results}

The theoretical predictions for the spectrum of each spin after the
boosting procedure can be derived most easily from
Eq.~\ref{eq:rho_f_SV}, taking into account the sign and magnitude of
the $J$-couplings. After a readout pulse on spin $1$, the four
spectral lines in the spectrum of $1$ should ideally have normalized
amplitudes $1:2:1:2$, compared to $1:1:1:1$ for the thermal
equilibrium spectrum (for spins $2$ and $3$, the boosting procedure
ideally results in normalized amplitudes of $0:1:0:1$ and $-1:0:0:1$,
respectively).  So the prediction is that the boosting procedure
increases the signal of spin $1$ averaged over the four spectral lines
by a factor of 3/2, equal to the bound for polarization enhancement
established in section~\ref{nmrqc:cooling}.  The experimentally
measured spectra before and after the boosting procedure are shown in
Fig.~\ref{fig:cooling_spect}. The operation of the boosting procedure
was further validated by measuring the full three-spin deviation
density matrix, shown in Fig.~\ref{fig:cooling_denmat}.

\bfig
\vspace*{1ex}
\begin{center}
\includegraphics*[width=4.2cm]{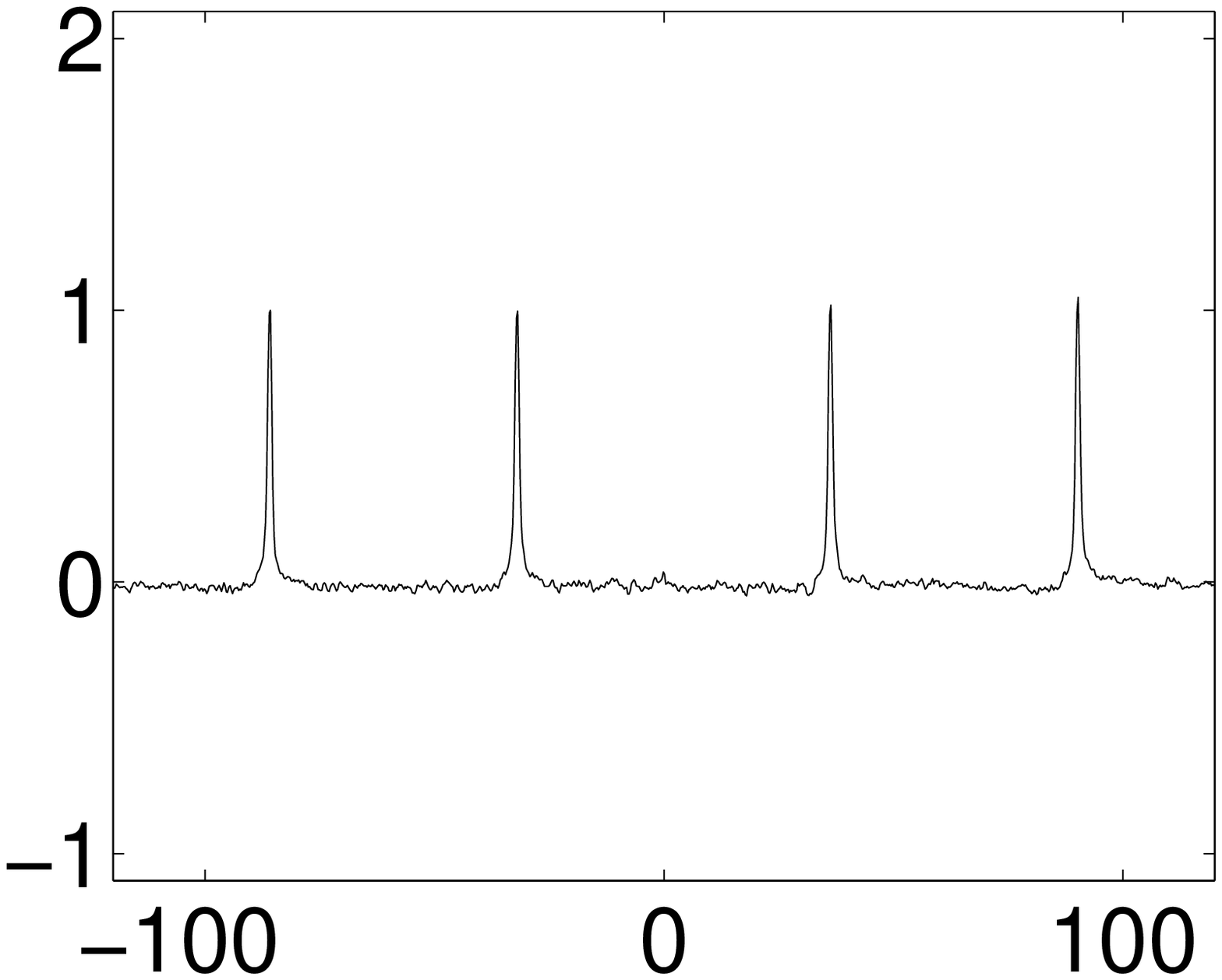} \hspace*{-1ex}
\includegraphics*[width=4.2cm]{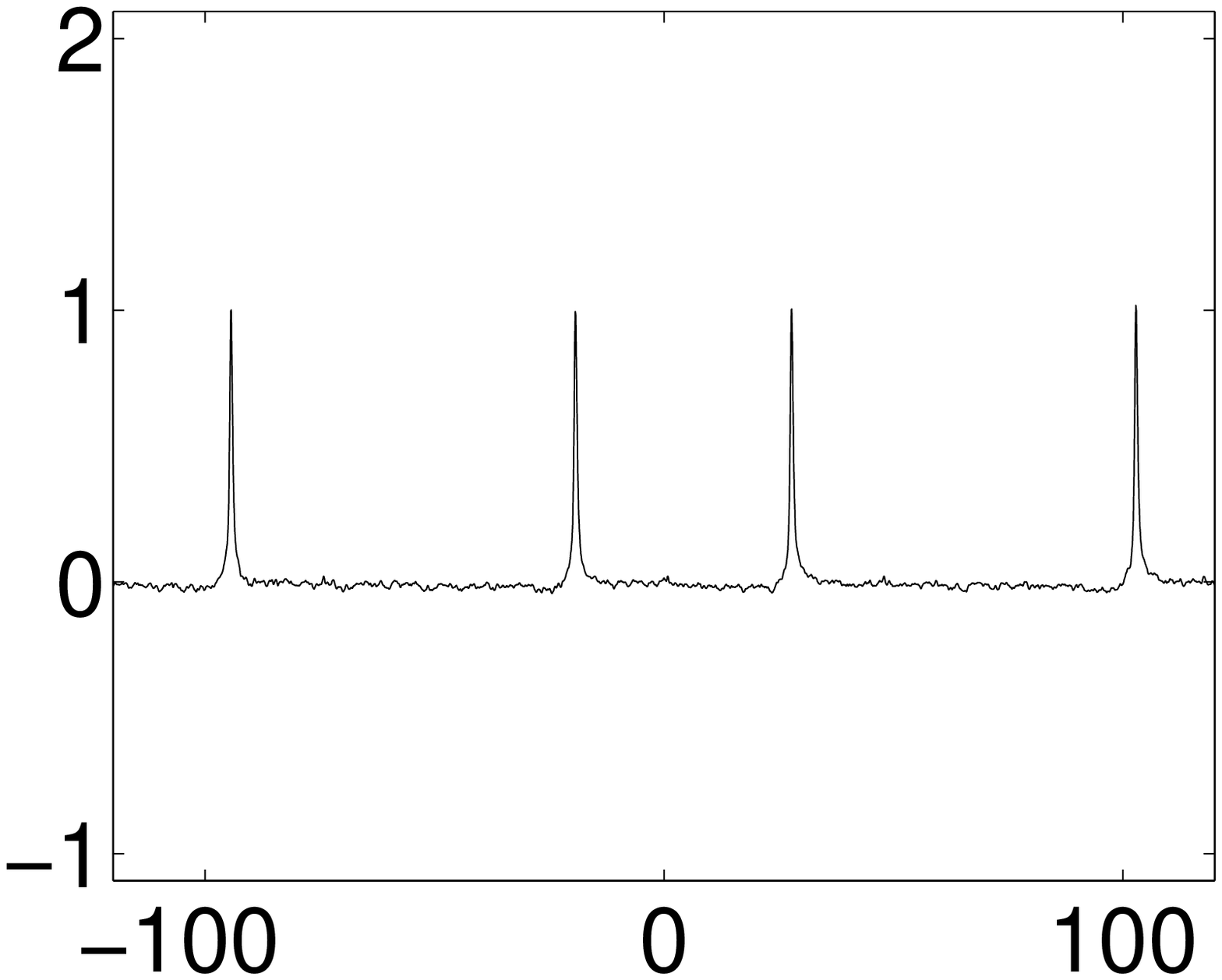} \hspace*{1ex}
\includegraphics*[width=4.2cm]{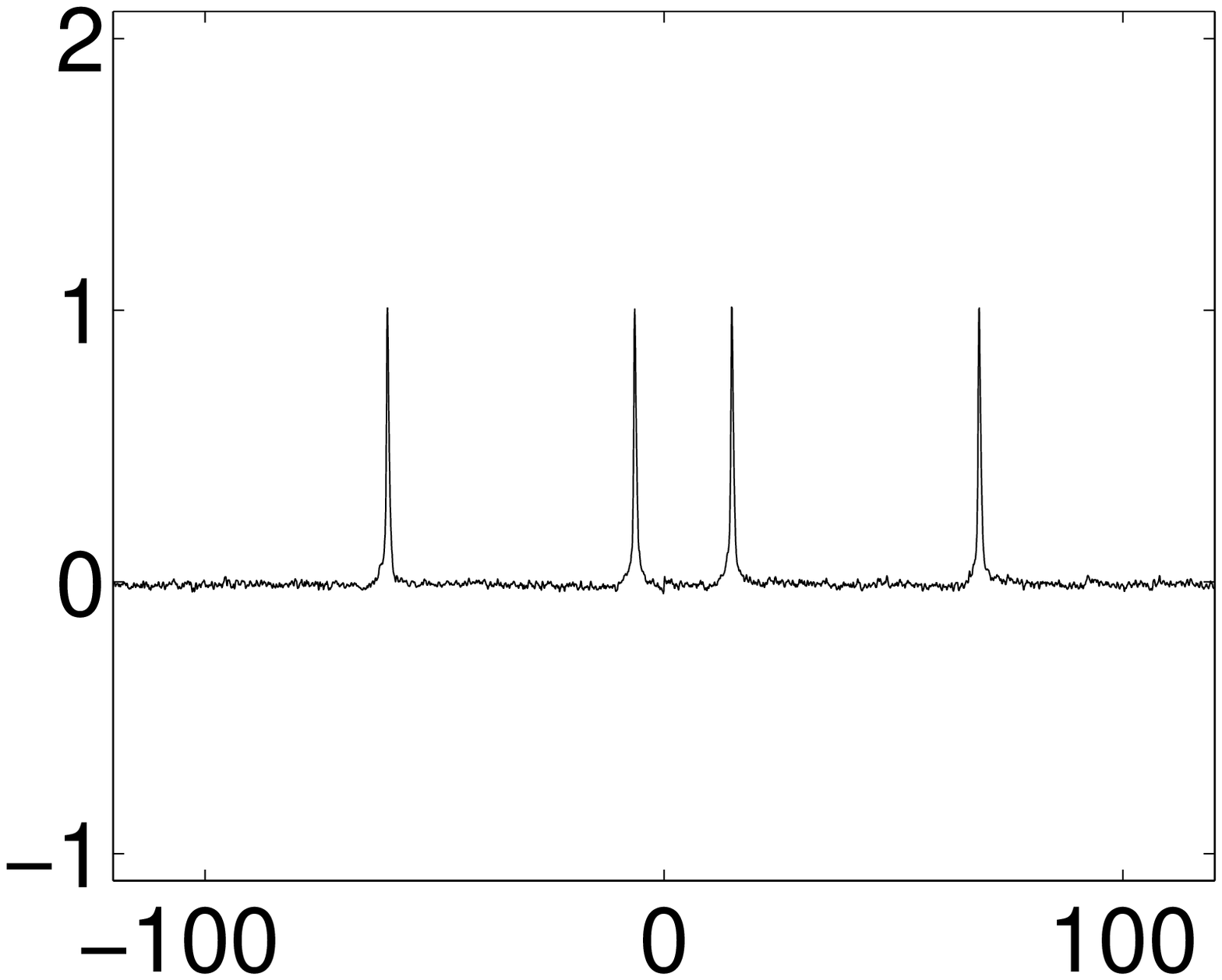}
\includegraphics*[width=4.2cm]{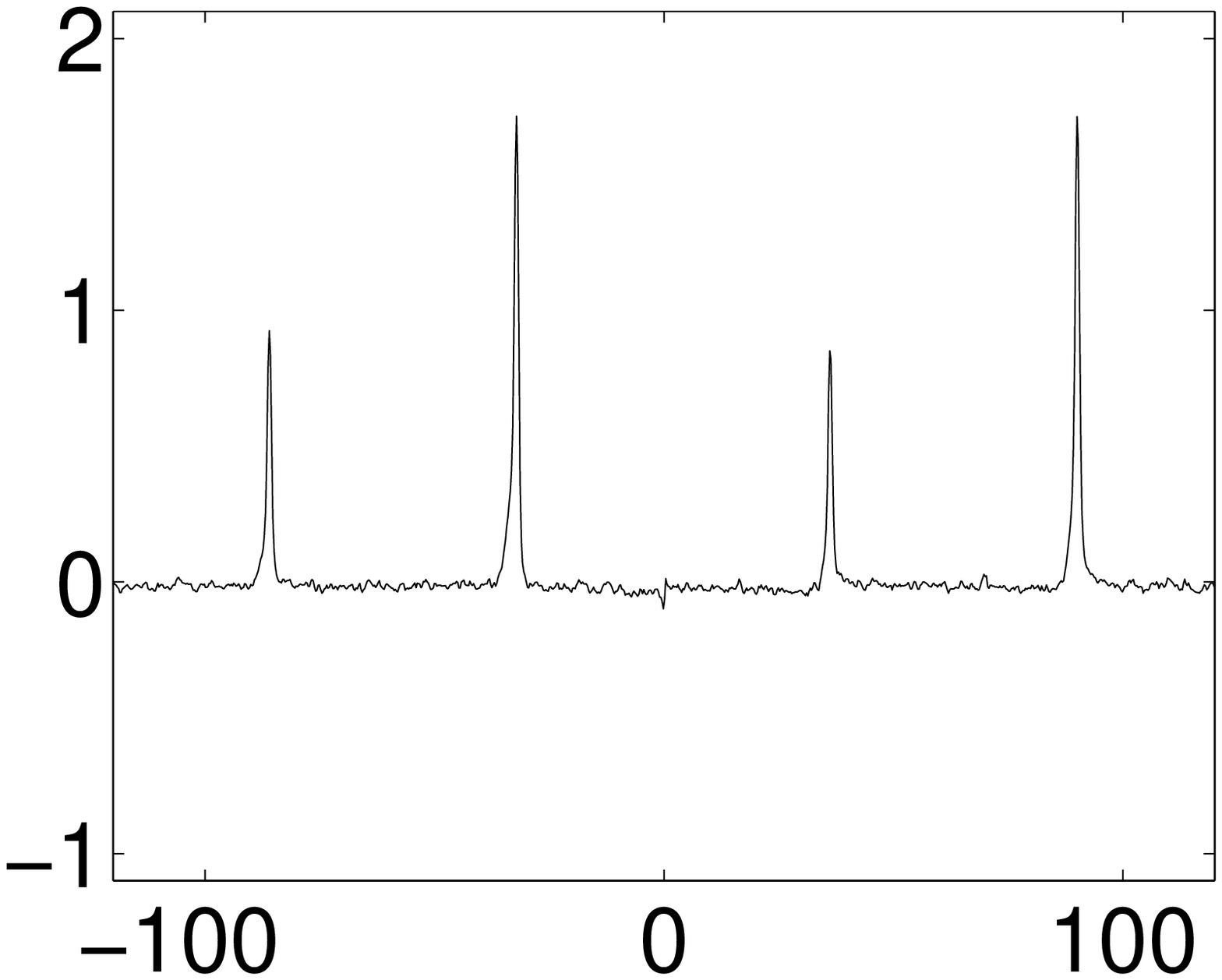} \hspace*{-1ex}
\includegraphics*[width=4.2cm]{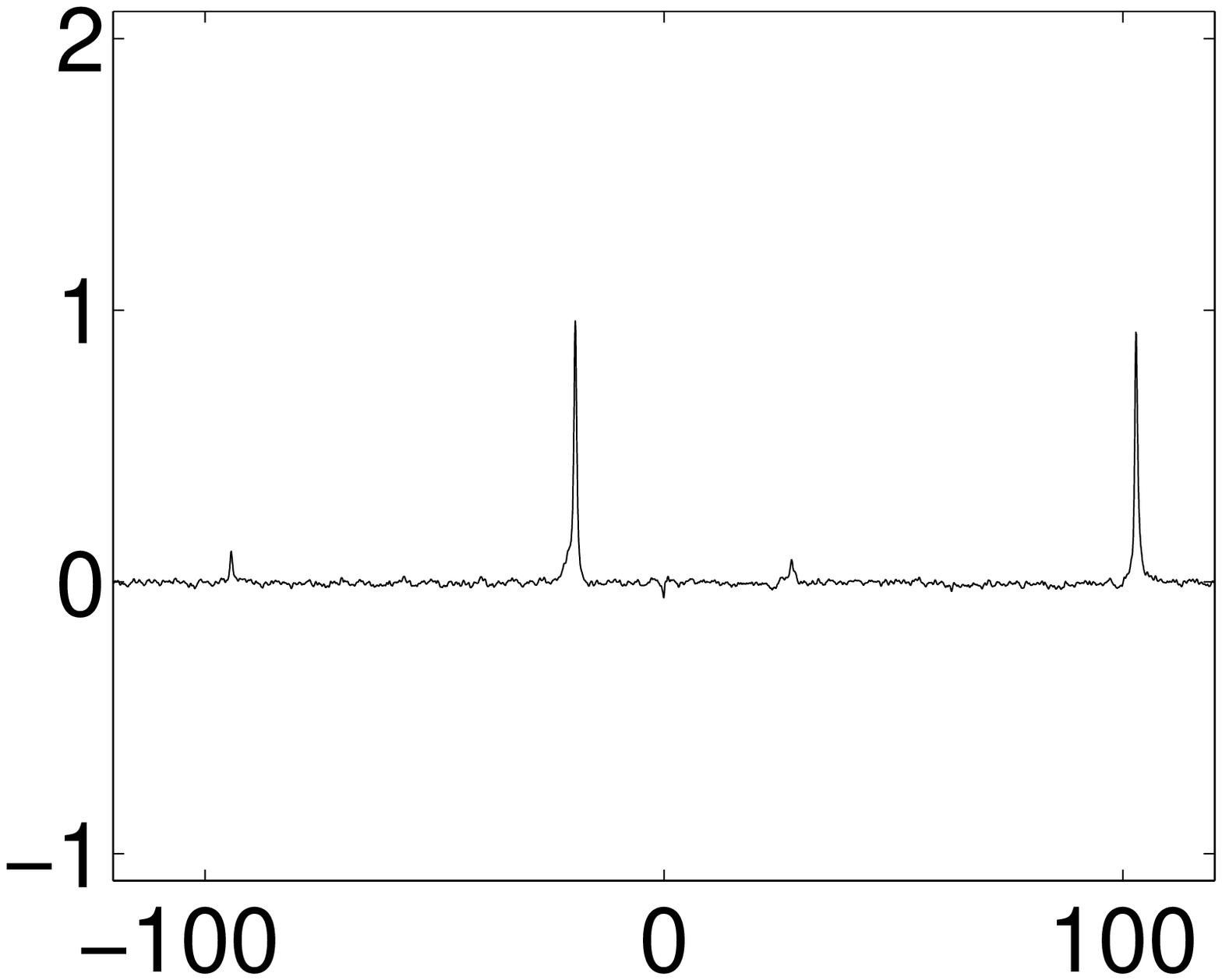} \hspace*{1ex}
\includegraphics*[width=4.2cm]{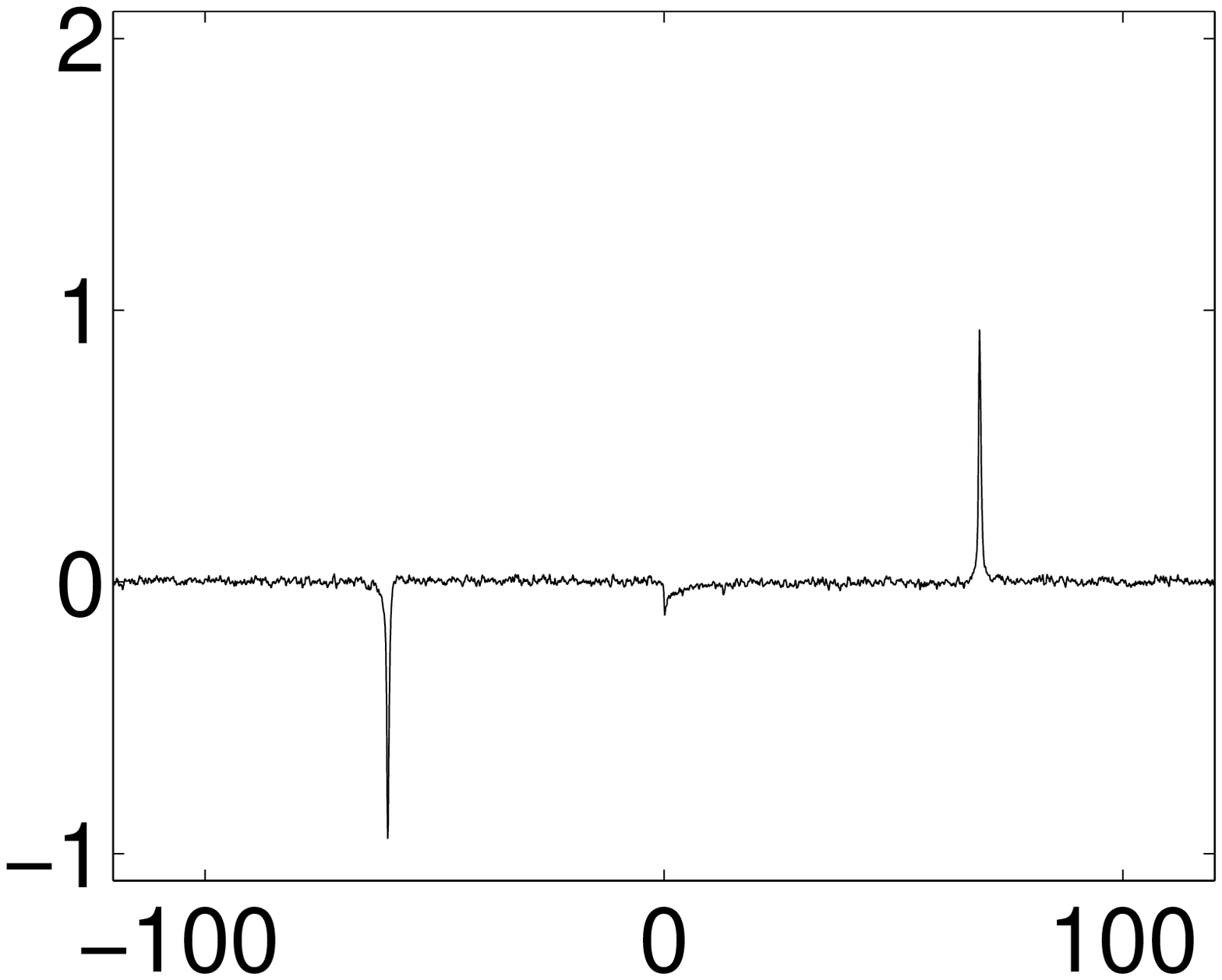}
\end{center}
\vspace*{-2ex}
\caption{Experimentally measured spectra of spin $1$ (Left), spin $2$
(Center) and spin $3$ (Right), after a readout pulse on the
corresponding spin, for the spin system in thermal equilibrium (Top)
and after applying the boosting procedure (Bottom). The real part of
the spectra is shown, and the spectra were rescaled in order to obtain
unit amplitude for the thermal equilibrium spectra. Frequencies are in
Hz with respect to the Larmour frequency of the respective spins.}
\label{fig:cooling_spect}
\efig

\bfig
\vspace*{1ex}
\bcen
\includegraphics*[width=5.5cm]{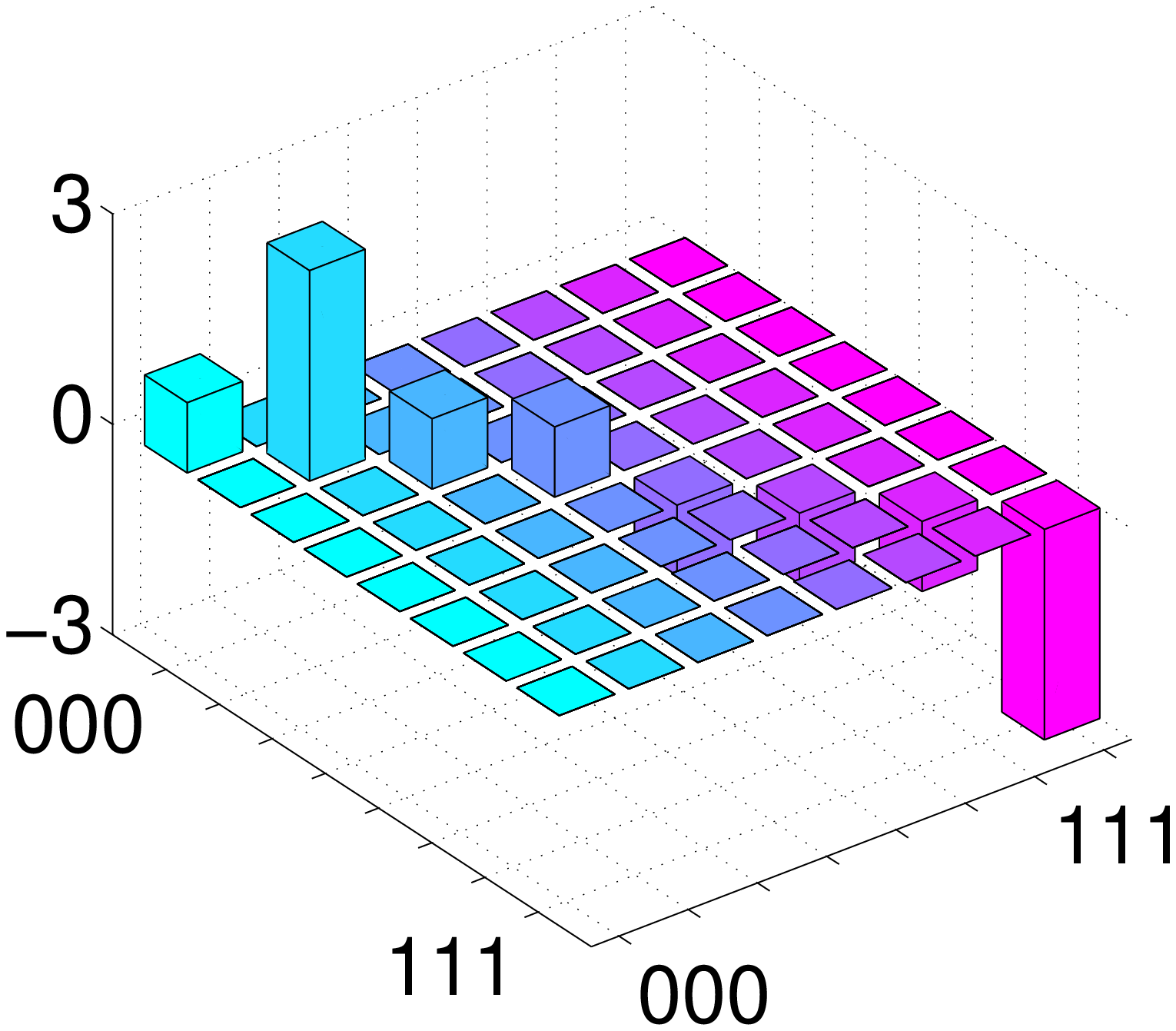}
\includegraphics*[width=5.5cm]{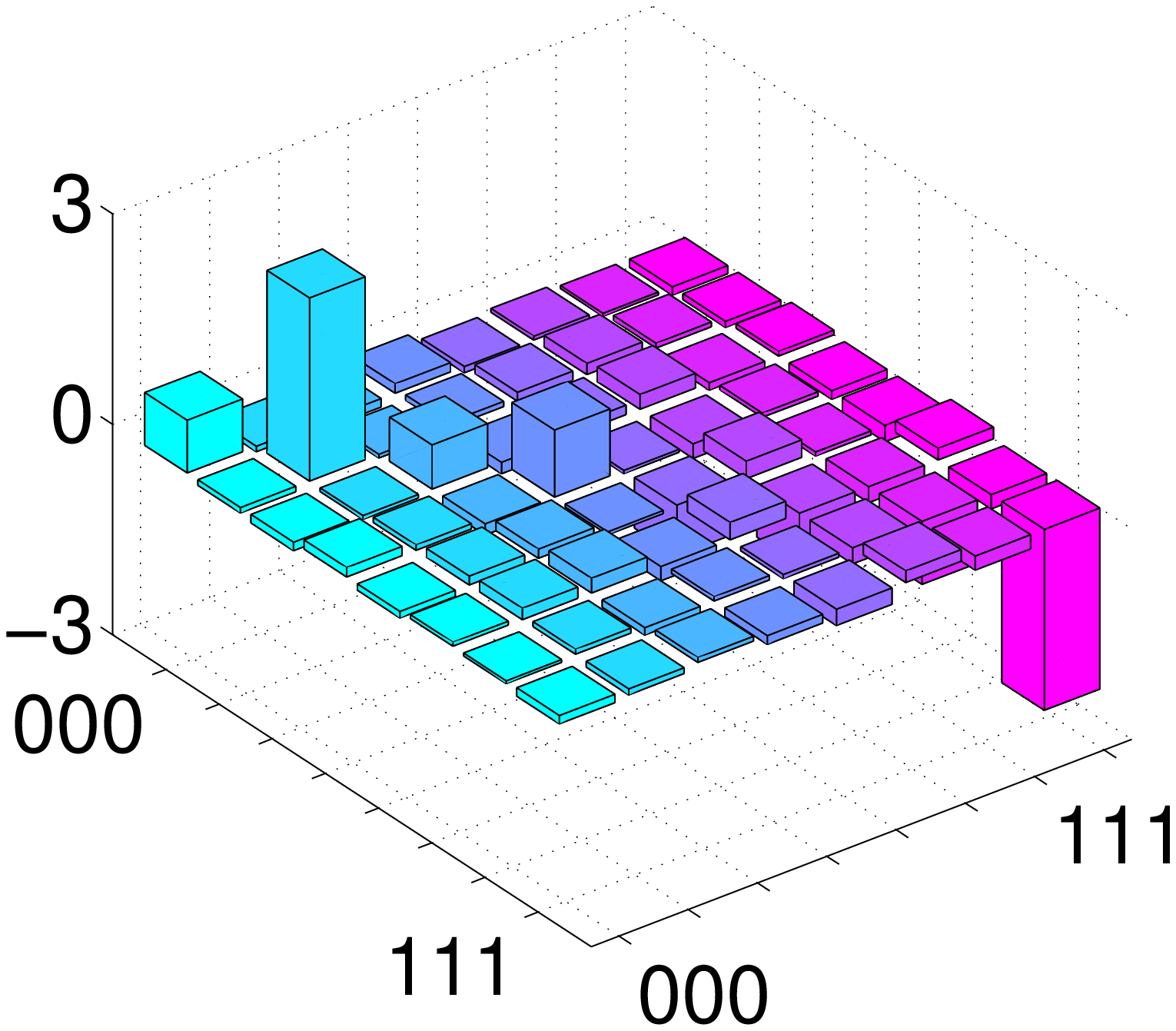}
\ecen
\vspace*{1ex}
\caption{Pictorial representation of the theoretical (left) and 
experimentally measured (right) density matrices, shown in magnitude
with the sign of the real part (all imaginary components were very
small).}
\label{fig:cooling_denmat}
\efig

\subsection{Discussion}

Clearly, the signal of spin $1$ has increased on average as a result
of the boosting procedure, and the relative amplitudes of the four
lines are in excellent agreement with the theoretical predictions. The
measured areas under the four peaks combined before and after
polarization transfer have a ratio of $1.255{\pm}0.002$.  The spectra
of spins $2$ and $3$ after the boosting procedure are also in
excellent agreement with the theoretical predictions, up to a small
overall reduction in the signal strength. 

The experimentally measured density matrix demonstrates not only that
the boosting procedure exchanges the populations as intended, but also
that it doesn't significantly excite any coherences.  The
experimentally measured
$\mathrm{Tr}({\rho}_{f}I_z^1)/\mbox{Tr}\;( {\rho}_{i}I_z^1)$ gives a
polarization enhancement factor of $1.235{\pm}0.016$, consistent with
the enhancement obtained just from the peak integrals of spin $1$. The
experimental implementation of the boosting procedure thus
successfully increased the polarization of spin $1$.

Despite the excellent qualitative agreement between the measured and
predicted data, the quantitative polarization enhancement of spin $1$
is lower than ideally achievable.  Given the absence of substantial
coherences (Fig.~\ref{fig:cooling_denmat}), we attribute this
suboptimal enhancement primarily to signal attenuation due to RF field
inhomogeneity and, to a lesser extent, to transverse relaxation. The
minor excitation of coherences is attributed mostly to incomplete
removal of undesired coupled evolution during the RF pulses.

In summary, we have experimentally demonstrated the building block for
the hyperpolarization procedure outlined by Schulman and Vazirani on a
homonuclear three-spin system.  However, the repeated boosting
required in a much larger spin system would be counteracted by
relaxation and other causes of signal decay, such as RF field
inhomogeneity. Also, when starting from thermal equilibrium at room
temperature, the overhead in the number of nuclear spins required for
the complete Schulman-Vazirani scheme is impractically large, despite
its linear scaling (see section~\ref{nmrqc:cooling}). Nevertheless, for higher
initial polarizations, Schulman-Vazirani cooling may be a very
valuable tool.

%%%%%%%%%%%%%%%%%%%%%%%%%%%%%%%%%%%%%%%%%%%%%%%%%%%%%%%%%%%%%%%%%%%%

\section{Order-finding (5 spins)}
\label{expt:order}

\subsection{Problem description}

At the time, the quest for the experimental realization of quantum
computers had resulted in the creation of specific entangled quantum
states with four qubits using trapped ions~\cite{Sackett00a}, and in
the creation of a seven-spin coherence using nuclear
spins~\cite{Knill00a}. Also using nuclear spins, Grover's search
algorithm~\cite{Chuang98a,Jones98b,Vandersypen00a} and the
Deutsch-Jozsa algorithm~\cite{Chuang98c,Jones98a,Marx00a} on two,
three and five qubit systems had been demonstrated.

However, a key step which remained yet to be demonstrated was a
computation with the structure of Shor's factoring algorithm, which
appears to be common to all quantum algorithms that achieve an
exponential speedup~\cite{Cleve98a}. Implementing the two main
components of this structure, exponentiated unitary transformations
and the quantum Fourier transform (section~\ref{qct:shor}), is
challenging because they require not just the creation of {\em static}
entangled states, but also precise {\em dynamic} quantum control over
the evolution of multiple entangled qubits, over the course of tens to
hundreds of quantum gates for the smallest meaningful instances of
this class of algorithms.  The evolution of the states is precisely
where NMR quantum computers appear to have an exponential advantage
over classical computers~\cite{Schack99a}.

The goal of this experiment~\cite{Vandersypen00b}\footnote{I was
planning a five qubit experiment to demonstrate period finding for
some test functions when Richard Cleve proposed to do order-finding,
which greatly enhanced the meaningfulness of the experiment. I worked
out the theory and invented the much improved temporal labeling
scheme. Matthias Steffen and I did the actual experiment
together. Matthias wrote most of the pulse sequence framework. The
five-spin molecule was discovered by Nino Yannoni and synthesized by
Greg Breyta. Dolores Miller and Mark Sherwood did spectral simulations
of the molecule. All this work was done under the guidance of Ike
Chuang.}  was to experimentally implement a quantum algorithm for
finding the order of a permutation (section~\ref{qct:shor}); its
structure is the same as for Shor's factoring algorithm and it scales
exponentially faster than any classical algorithm for the problem.

Specifically, we set out to implement the smallest instance of
order-finding for which quantum computers outperform classical
computers: the case of permutations $\pi$ on four elements ($M =
4$), using a total of five qubits ($m=2$ and $n=3$).

It can be proven that in this case the best classical algorithm needs
two queries of the oracle to determine $r$ with certainty, and that
using only one query of the oracle, the probability of finding $r$
using a classical algorithm can be no more than $1/2$. One optimal
classical strategy is to first ask the oracle for the value of
$\pi^3(y)$: when the result is $y$, $r$ must be $1$ or $3$; otherwise
$r$ must be $2$ or $4$.  In either case, the actual order can be
guessed only with probability $1/2$. 

In contrast, the probability of success is $\sim 0.55$ after only one
oracle-query using the quantum circuit of Fig.~\ref{fig:order_circuit}
on a single quantum computer: depending on the measurement outcome
after running the algorithm, we can make a probabilistic guess $r'$ as
shown in Fig.~\ref{fig:order_guess}. The probability of success
$\mbox{Pr}[r'=r]$ is independent of the distribution of $r$ or $\pi$.

\bfig
\vspace*{1ex}
\bcen
\includegraphics[height=2.4cm]{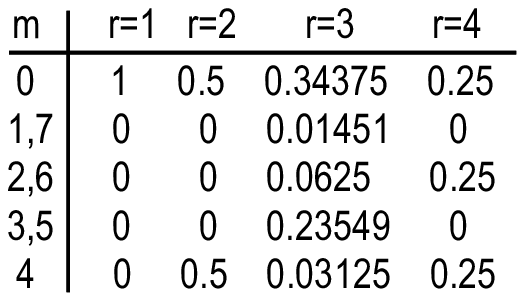} \hspace{1cm}
\includegraphics[height=2.4cm]{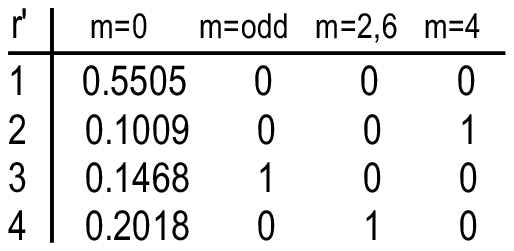}
\end{center}
\vspace*{-2ex}
\caption{(Left) The probabilities that the measurement result $m$ 
is $0, 1, \ldots,$ or $7$, given $r$ (for an ideal single quantum
computer). (Right) The optimal probabilities with which to make a
guess $r'$ for $r$, given $m$.}
\label{fig:order_guess}
\end{figure}

The prediction is that the algorithm will put the spins in the state
$\ket{x_0}$.  The $^{13}$C and $^1$H readout spectra for the four
possible $x_0$ are shown in Fig.~\ref{fig:lcgrover_spect}.  As
predicted for two spins in an effective pure state, the value of $x_0$
is clearly indicated by the amplitude and phase of the two resonance
lines in the $^{13}$C and $^1$H spectra.  Measurement of the deviation
density matrix using quantum state tomography confirms that the output
states are as theoretically predicted, as illustrated in
Fig.~\ref{fig:lcgrover_denmat} for $x_0=11$.

\subsection{Discussion}

The $^{13}$C-$^{1}$H coupling in the liquid crystal (ZLI-1167) is
eight times larger than the scalar coupling in acetone-d$_6$,
corresponding to a computer with a clock that is eight times faster.
The product of the shortest coherence time and the clock frequency
$T_2 f_{clock} = 2T_2J$, which approximates the number of gates that
can be executed while maintaining coherence, is about five times
higher in ZLI-1167 than in deuterated acetone, meaning that more
complex algorithms could be implemented using the liquid crystal
solvent.

Furthermore, the chloroform $^{13}$C and $^1$H spin-lattice relaxation
times are about $12$ times shorter in ZLI-1167 than in acetone-$d_6$.
Since all experiments (including most calibration experiments) require
an equilibration time of $5T_1$, an order of magnitude savings in
time can be significant, and will become more so as the number of
qubits increases.  This advantage will be especially important for
experiments requiring temporal averaging (or also just signal
averaging), or in quantum state tomography experiments.

Another advantage of using liquid crystal solvents is that they permit
a different choice of spin-bearing molecules that may be suitable for
quantum computing.  Dipolar coupling, which is manifest in the NMR
spectra of oriented molecules, requires only proximity between the
spins of interest.  As a result, two spins that are separated by
several bonds and which have no scalar coupling may, if spatially
proximate, have dipolar coupling sufficiently large for quantum
computation.  The ability to control the degree of orientation of the
solute molecule by varying the solvent temperature and solute
concentration~\cite{Emsley75a} provides the experimentalist with means
of tailoring the NMR spectrum to meet the requirements for quantum
computing.  In addition, magic-angle spinning and multiple pulse
methods can be used to preferentially scale the dipolar splitting in
the spectrum of a liquid-crystal-oriented molecule to convert it to
first order~\cite{Ouvrard91a}.

\clearpage
\bcen
\vspace*{1ex}
\includegraphics*[width=4.7cm]{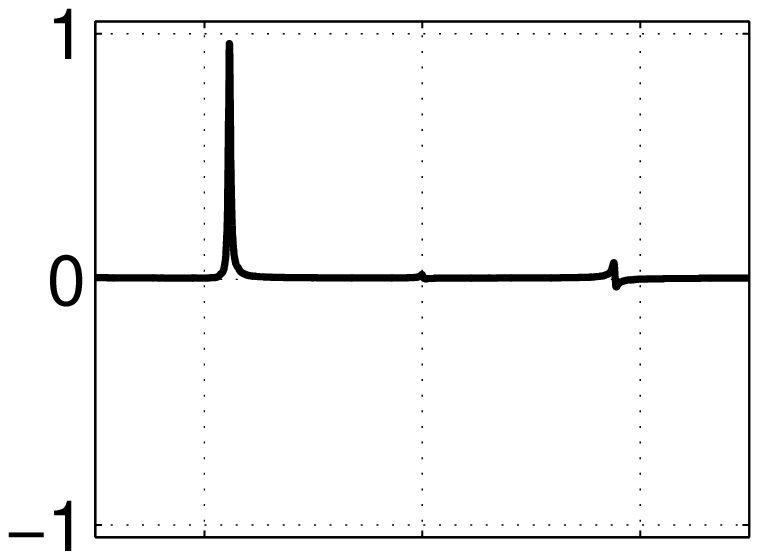}
\includegraphics*[width=4.7cm]{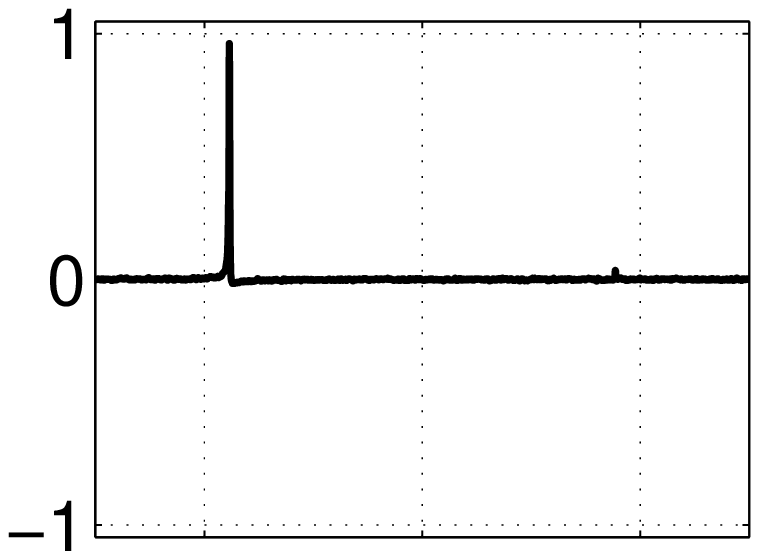}\\
\includegraphics*[width=4.7cm]{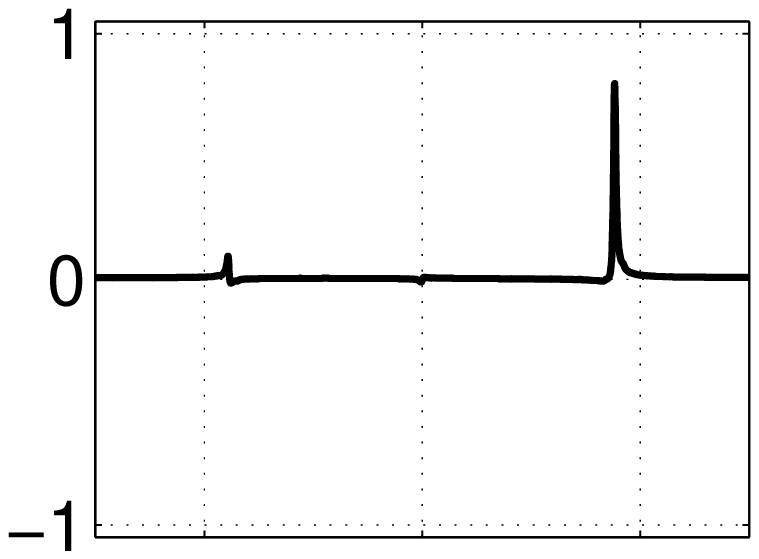}
\includegraphics*[width=4.7cm]{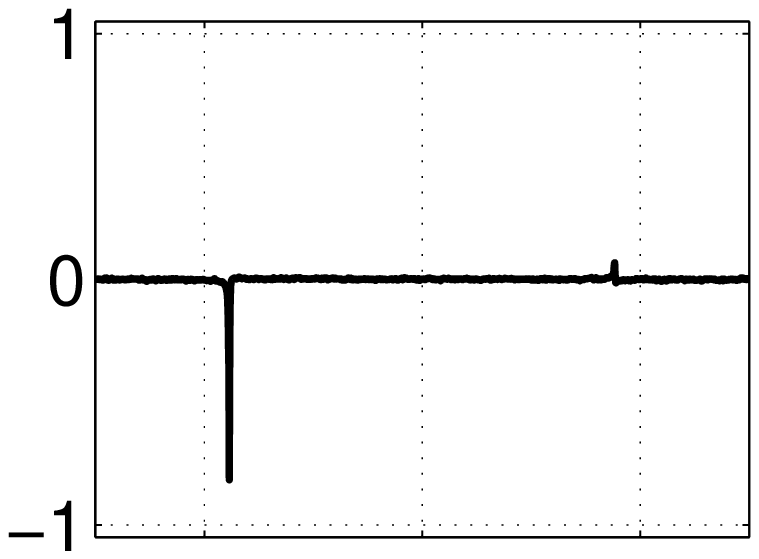}\\
\includegraphics*[width=4.7cm]{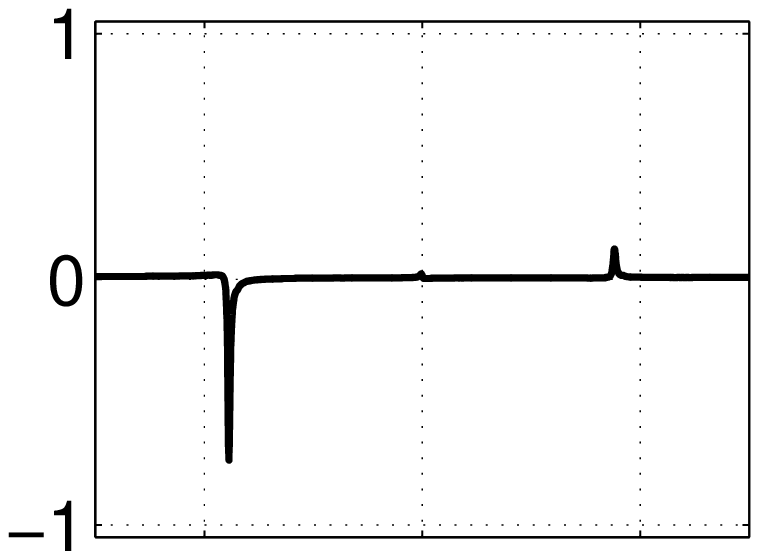}
\includegraphics*[width=4.7cm]{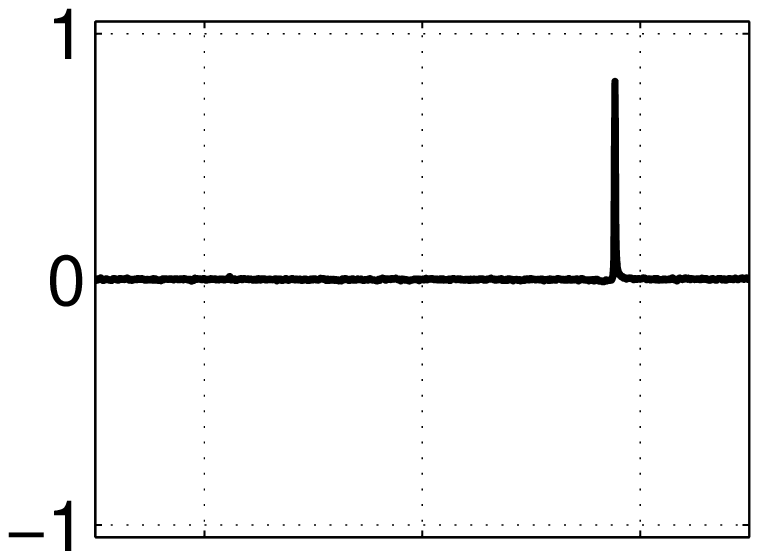}\\
\includegraphics*[width=4.7cm]{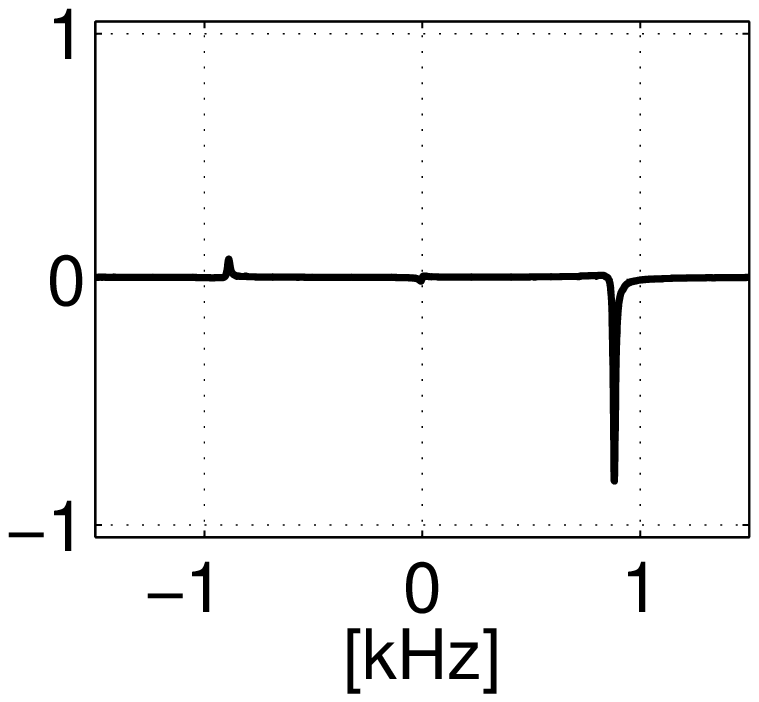}
\includegraphics*[width=4.7cm]{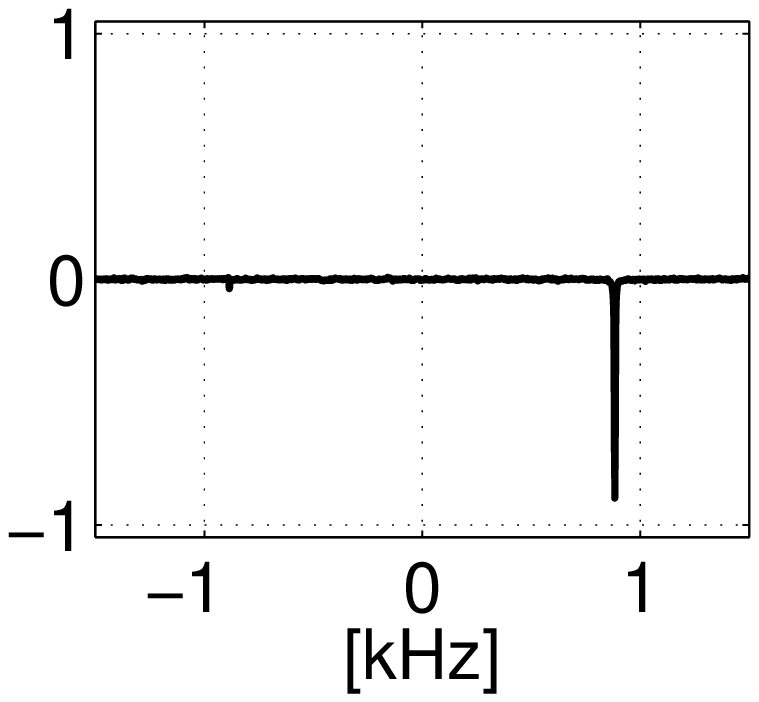}
\vspace*{-2ex}
\ecen
\bfig
\caption{Spectral readout of the results of the 2-qubit Grover 
search using $^{13}$C$^{1}$HCl$_3$ in a liquid crystal solvent showing
absorption and emission peaks which clearly indicate the value of
$x_0$ ($00$, $01$, $10$, and $11$, from top to bottom).  The real part
of the $^1$H (left) and $^{13}$C (right) spectra are shown, with
frequencies relative to $\omega_0^H/2\pi$ and $\omega_0^C/2\pi$).
The vertical scale is arbitrary.}
\label{fig:lcgrover_spect}
\efig

Note that since an ensemble of $\sim 10^{18}$ quantum computers
contribute to the signal in our experiment, the order will follow
from the output data with virtual certainty.

\subsection{Experimental approach}
\label{expt:shor_expt}

\begin{figure}[h]
\vspace*{1ex}
\begin{center}
\includegraphics[width=4cm]{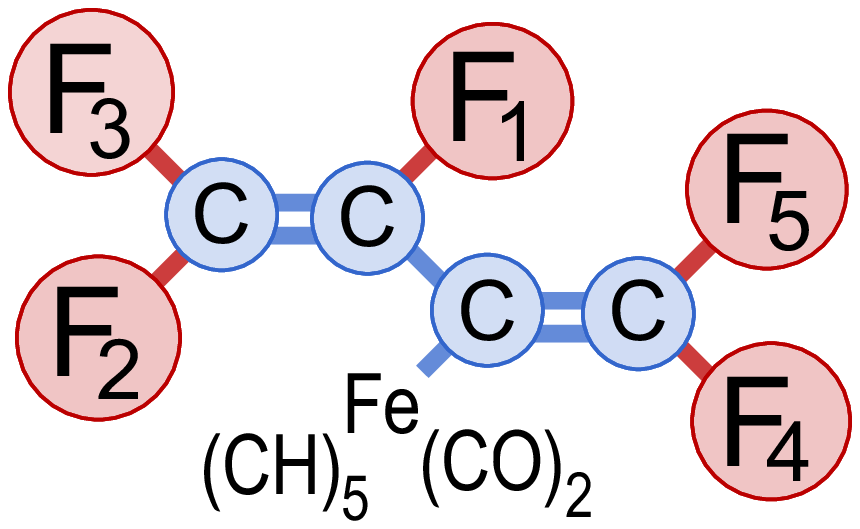} \hspace{1cm}
\includegraphics[width=4cm]{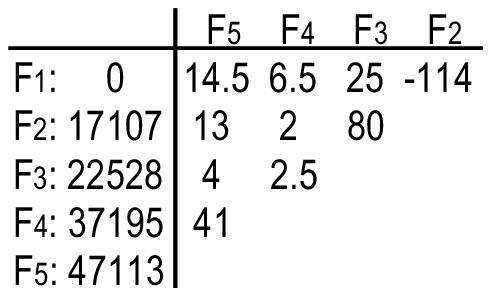}
\end{center}
\vspace*{-2ex}
\caption{Structure of the pentafluorobutadienyl 
cyclopentadienyl dicarbonyl iron complex, with a table of the relative
chemical shifts of the $^{19}$F spins at $11.7$ T [Hz], and the
$J$-couplings [Hz]. A total of $76$ out of the $80$ lines in the $5$
spectra are resolved.}
\label{fig:order_molec}
\end{figure}

We custom synthesized a molecule~\cite{Green68a} containing five
fluorine nuclear spins which served as the qubits. The molecule as
well as the chemical shifts and $J$-coupling constants are shown in
Fig.~\ref{fig:order_molec}.  The linewidths of the NMR transitions
were $\sim$ 1 Hz, so the $T_2^*$ dephasing times of the spins were
$\approx$ 0.3~s (the $T_2$ are longer). The $T_1$ time
constants were measured to be between $3$ and $12$ s.

All five spins were prepared in an effective pure state via the
product operator approach to temporal averaging
(section~\ref{nmrqc:init}). We used $9$ experiments (the theoretical
minimum is $\lceil 2^n-1/n\rceil = 7$), giving a total of $45$ product
operator terms. The $45 - (2^n-1) = 14$ extra terms were canceled out
pairwise, using {\sc not} operations, which flip the sign of selected
terms. The nine state preparation sequences were\\

{\sc cnot}$_{51}$ {\sc cnot}$_{45}$ {\sc cnot}$_{24}$ {\sc not}$_3$, 
$\quad$ $\quad$ 
{\sc cnot}$_{21}$ {\sc cnot}$_{52}$ {\sc cnot}$_{45}$ {\sc cnot}$_{34}$,
\\ \indent 
{\sc cnot}$_{14}$ {\sc cnot}$_{31}$ {\sc cnot}$_{53}$ {\sc not}$_2$, 
$\quad$ $\quad$ 
{\sc cnot}$_{12}$ {\sc cnot}$_{15}$ {\sc cnot}$_{13}$ {\sc cnot}$_{41}$,
\\ \indent 
{\sc cnot}$_{32}$ {\sc cnot}$_{13}$ {\sc cnot}$_{25}$ {\sc not}$_4$, 
$\quad$ $\quad$ 
{\sc cnot}$_{31}$ {\sc cnot}$_{43}$ {\sc cnot}$_{23}$ {\sc not}$_5$,
\\ \indent 
{\sc cnot}$_{53}$ {\sc cnot}$_{25}$ {\sc cnot}$_{12}$ {\sc not}$_4$, 
$\quad$ $\quad$ 
{\sc cnot}$_{54}$ {\sc cnot}$_{51}$ {\sc not}$_2$,
\\ \indent 
{\sc cnot}$_{35}$ {\sc cnot}$_{23}$ {\sc not}$_1$.\\

The actual computation was realized via a sequence of $\sim 50$ to
$\sim 200$ radio-frequency (RF) pulses, separated by time intervals of
free evolution under the Hamiltonian, for a total duration of $\sim
50$ to $\sim 500$ ms, depending on $\pi$.  The pulse sequences for the
order-finding algorithm were designed by translating the quantum
circuits of Fig.~\ref{fig:order_circuit} into one- and two-qubit
operations, employing the simplification methods of
section~\ref{nmrqc:seq_design}.  The transformation $\ket{x}\ket{y}
\mapsto \ket{x}\ket{\pi^x(y)}$ (Eq.~\ref{eq:oracle_orderfinding}) is
realized by one of the following sequences, which form a representative subset of all possible permutations:\\

$\bullet$ \noindent $r=1$: c{\sc z}$_{54}$ {\sc cnot}$_{35}$ c{\sc
z}$_{54}^\dagger$ {\sc cnot}$_{35}$ c{\sc z}$_{34}$ (c{\sc z}$_{ij}$
rotates spin $j$ by $90^\circ$ about $\hat{z}$ if and only if spin $i$
is $\ket{1}$).

$\bullet$ \noindent $r=2$: {\sc cnot}$_{35}$.  

$\bullet$ \noindent $r=3$: {\sc cnot}$_{32}$ {\sc cnot}$_{25}$ {\sc
cnot}$_{32}$ {\sc cnot}$_{21}$ c{\sc z}$_{14}$ {\sc cnot}$_{51}$ c{\sc
z}$_{14}^\dagger$ {\sc cnot}$_{51}$ c{\sc z}$_{54}$ {\sc cnot}$_{21}$
c{\sc z}$_{15}$ {\sc cnot}$_{41}$ c{\sc z}$_{15}^\dagger$ {\sc
cnot}$_{41}$ c{\sc z}$_{45}$. (this sequence does the transformation
$\pi^x(y)$ for $y=2$ only; sequences for $r=3$ that would work for any
$y$ are prohibitively long).

$\bullet$ \noindent $r=4$: {\sc cnot}$_{24}$ c{\sc z}$_{34}$ c{\sc
z}$_{54}$ {\sc cnot}$_{35}$ c{\sc z}$_{54}$.\\

Each transformation was tested independently to confirm its proper
operation. All pulse sequences were implemented on the four-channel
spectrometer described in section~\ref{expt:apparatus}, and using the
methods of section~\ref{nmrqc:1bitgates} to serve two qubits with one
channel.  The frequency of one channel was set at $(\omega_2 +
\omega_3)/2$, and the other three channels were set on the resonance
of spins $1, 4$ and $5$.  The chemical shift evolutions of spins $2$
and $3$ were calculated with the help of a time-counter, which kept
track of the time elapsed from the start of the pulse
sequence. On-resonance excitation of spins $2$ and $3$ was achieved
using phase-ramping techniques. All pulses were spin-selective and
Hermite shaped.  Rotations about the $\hat{z}$-axis were implemented
by adjusting the phases of the subsequent pulses. Unintended phase
shifts of spins $i$ during a pulse on spin $j \neq i$ were calculated
and accounted for by adjusting the phase of subsequent pulses. During
simultaneous pulses, the effect of these phase shifts was largely
removed by shifting the frequency of the pulses via phase-ramping, in
such a way that they track the shifting spin frequencies and thereby
greatly improve the accuracy of the simultaneous rotations of two or
more spins.

Upon completion of the pulse sequence, the states of the three spins
in the first register were measured and the order $r$ was determined
from the read-out. Since an ensemble of quantum computers rather than
a single quantum computer was used, the measurement gives the bitwise
average values of $m_i\, (i=1,2,3)$, instead of a sample of $m=m_1 m_2
m_3$ with probabilities given in
Fig.~\ref{fig:order_guess}~\footnote{It is not clear that
the bitwise average outputs of the QFT suffice to determine $r$ for
permutations on arbitrary $n$. Instead, the continued fraction
expansion can be ran on the quantum computer to compute $r$.}.
Formally, measurement of spin $i$ returns $O_i = 1 - 2 \langle m_i
\rangle = 2\, \mbox{Tr}(\rho I_{zi})$, where $\rho$ is the final density
operator of the system. The $O_i$ are obtained experimentally from
integrating the peak areas in the spectrum of the magnetic signal of
spin $i$ after a $90^\circ$ read-out pulse on spin $i$, phased such
that positive spectral lines correspond to positive $O_i$. The
theoretically predicted values of $O_i$ ($i=1,2,3$) for each value of
$r$ follow directly from the probabilities for $m$ in
Fig.~\ref{fig:order_guess}. For reference, we also include the values
of $O_4$ and $O_5$ (for $y=0$; if $y \neq 0$, $O_4$ and $O_5$ can be
negative): for the case $r=1$ the $O_i$ are $1, 1, 1, 1, 1$; for $r=2$
they are $1, 1, 0, 1, 0$; and for $r=4$ they are $1, 0, 0, 0, 0$.  For
$r=3$, the $O_{i}$ ($i=1,2,3$) are $0, 1/4, 5/16$, and $O_{4}$ and
$O_5$ can be $0$, $\pm 1/4$ or $\pm 1/2$, depending on $y$. The value
of $r$ can thus be unambiguously determined from the spectra of the
three spins in the first register. This was confirmed experimentally
by taking spectra for these three spins, which were in good
agreement with the theoretical expectations.

In fact, the complete spectrum of any one of the first three spins
uniquely characterizes $r$ by virtue of extra information contained in
the splitting of the lines. For the spectrum of spin $1$ the values of
$O_i$ given above indicate that for $r=1$, only the $0000$ line (see
Fig.~\ref{fig:order_labeling}) will be visible since spins $2-5$ are
all in $\ket{0}$. Furthermore, this line should be absorptive and
positive since spin $1$ is also in $\ket{0}$.  Similarly, for $r=2$
the $0000, 0001, 0100$ and $0101$ lines are expected to be positive,
and for $r=4$ all $16$ lines should be positive. Finally, for $r=3$,
the {\it net} area under the lines of spin $1$ should be zero since
$O_1=0$, although most individual lines are expected to be non-zero
and partly dispersive.

\subsection{Experimental results}

\begin{figure}[h]
\bcen
\includegraphics*[height=2in]{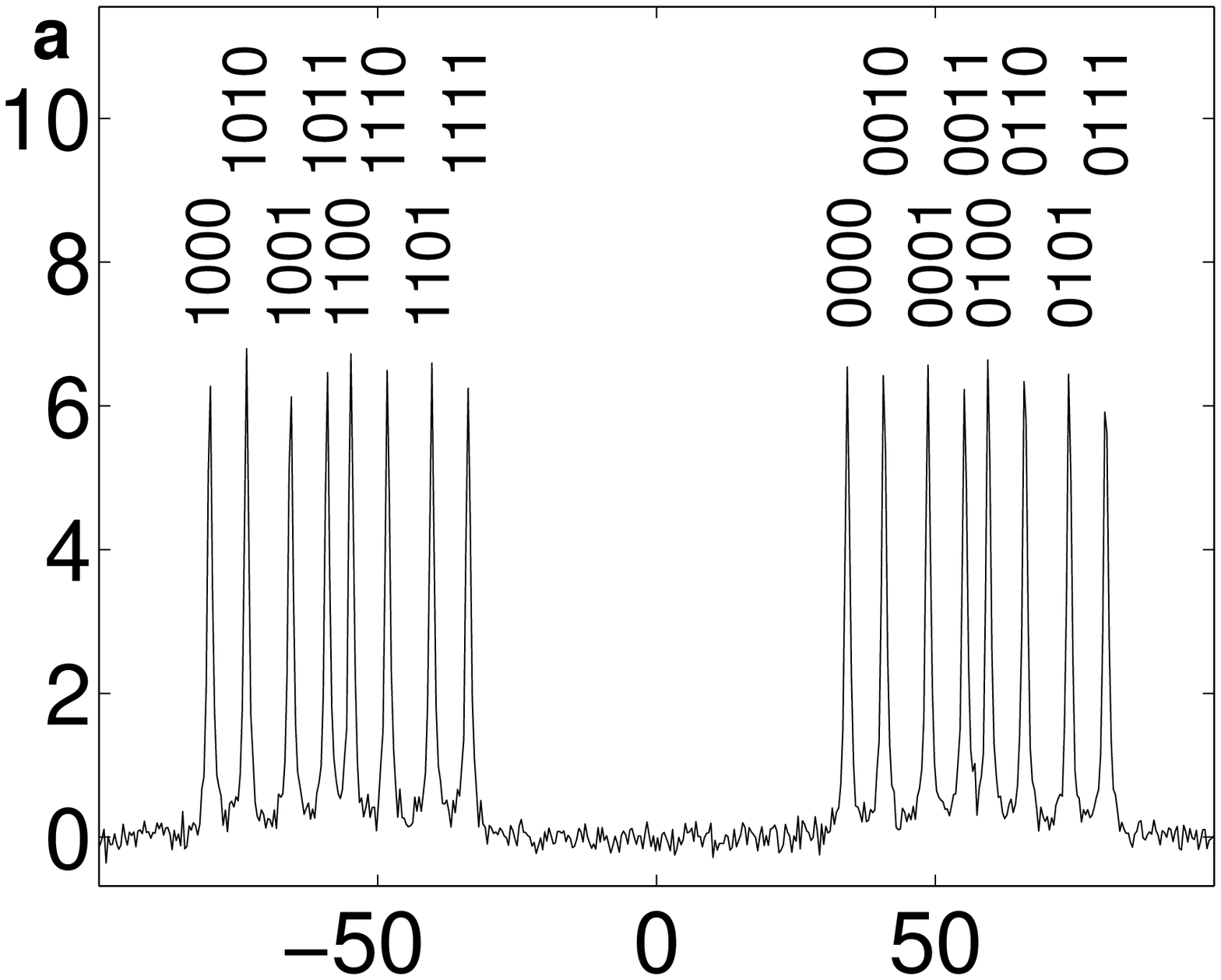} \hspace{.5cm}
\includegraphics*[height=2in]{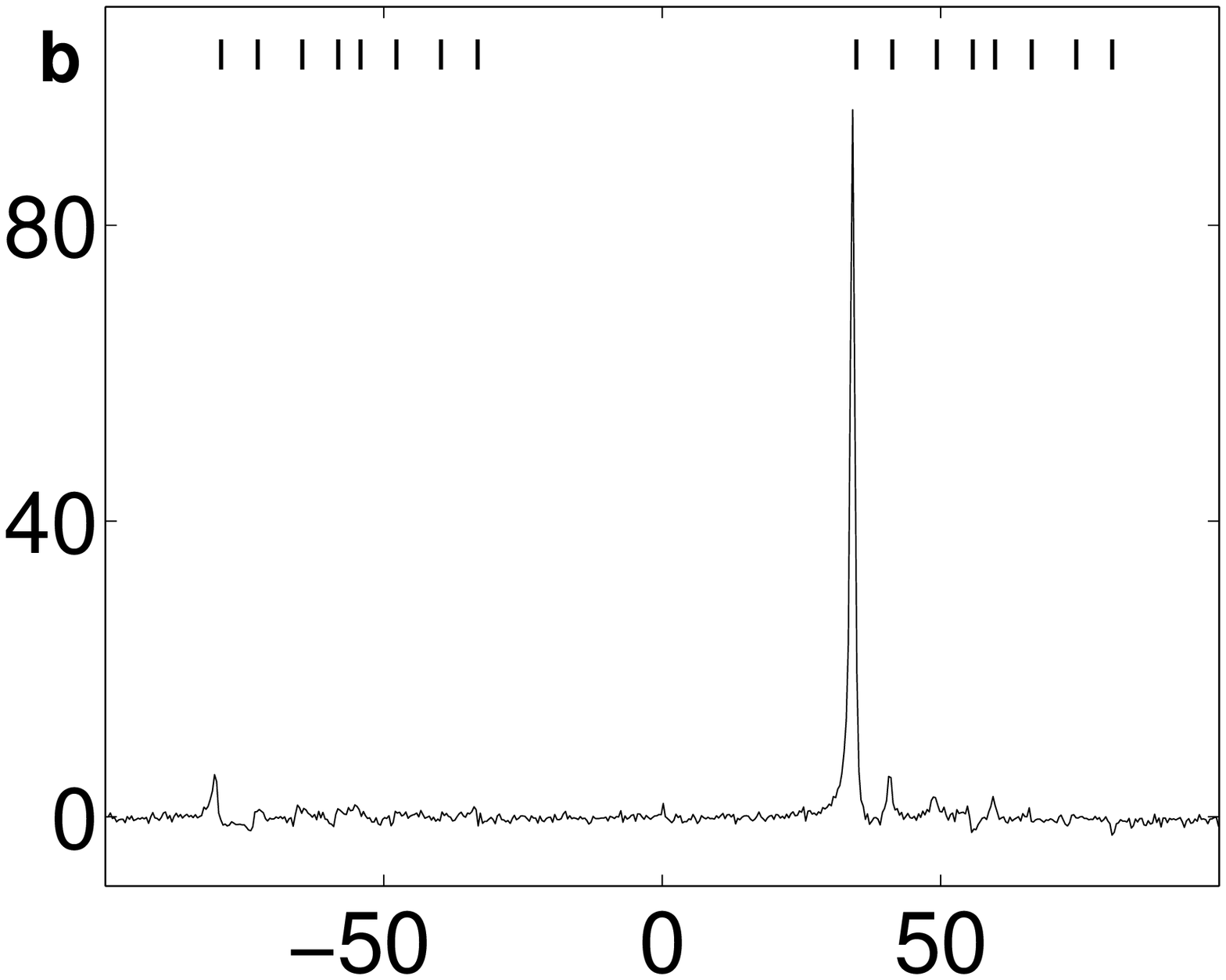}
\ecen
\vspace*{-2ex}
\caption{(a) The spectrum of spin $1$ in thermal
equilibrium.  Taking into account the sign and magnitude of the
$J_{1,j}$, the $16$ lines in the spectrum of spin $1$ can be labeled
as shown. (b) The same spectrum when the spins are in an
effective pure state. Only the line labeled $0000$ is present. All
spectra shown here and in Fig.~\protect\ref{fig:order_spectra} display
the real part of the spectrum in the same arbitrary units, and were
obtained without phase cycling or signal-averaging (except for
Fig.~\protect\ref{fig:order_spectra} c, where 16 identical scans were
averaged). A $0.1$ Hz filter was applied. Frequencies are in units of
Hz with respect to $\omega_1/2\pi$. }
\label{fig:order_labeling}
\end{figure}

The experimentally measured thermal equilibrium spectrum for spin 1 is
shown in Fig.~\ref{fig:order_labeling}a.  After the state
preparation, only the $0000$ line should remain visible in the
spectrum of each spin, reflecting that only molecules with all other
spins in the ground state contribute to the signal. The resulting data
are remarkably clean, as illustrated for spin 1 in
Fig.~\ref{fig:order_labeling}b.

\vspace*{1ex}

\begin{figure}[h]
\begin{center}
\hspace*{-2ex} $ \noindent \begin{array}{ll}
\includegraphics*[height=2in]{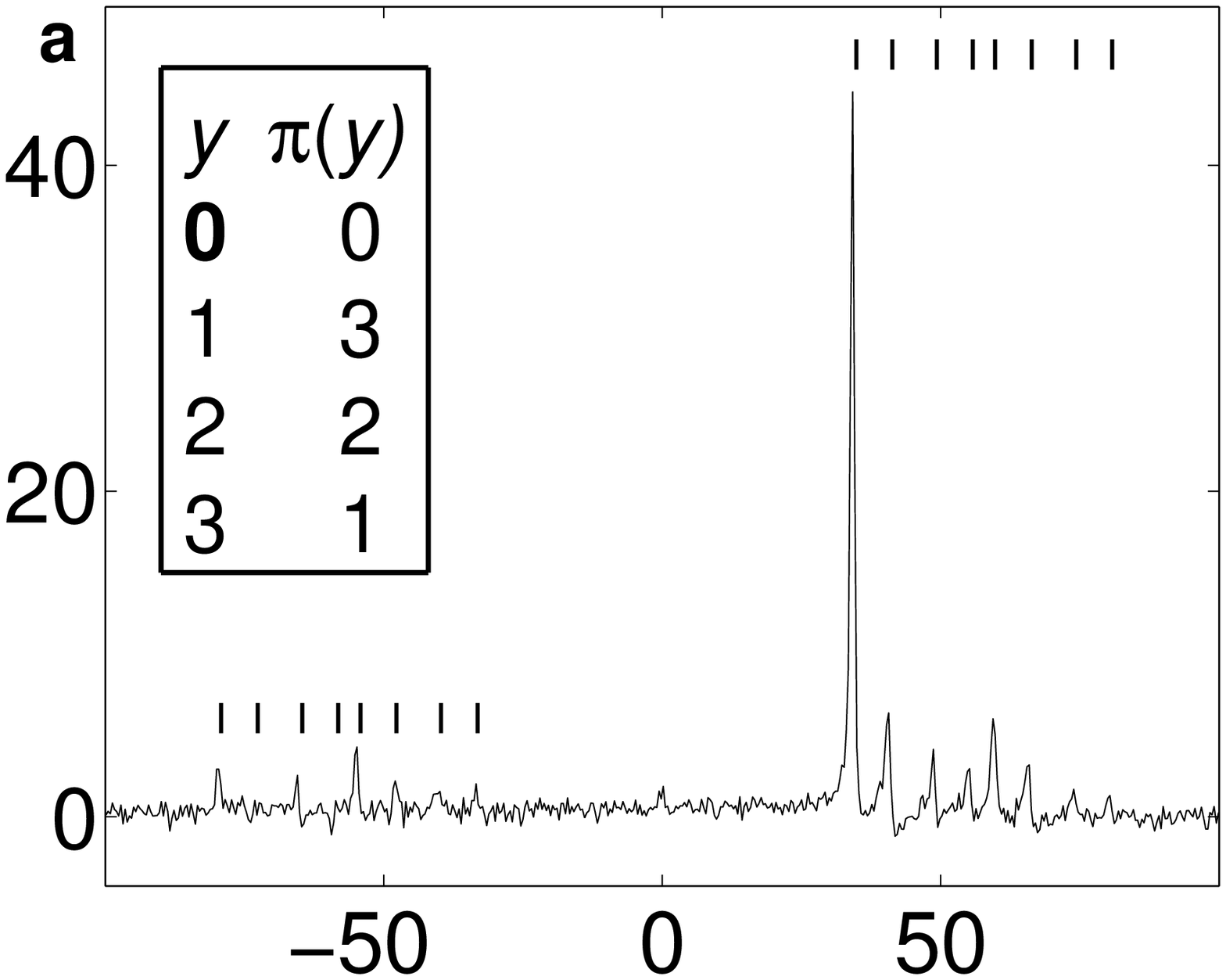} &
\includegraphics*[height=2in]{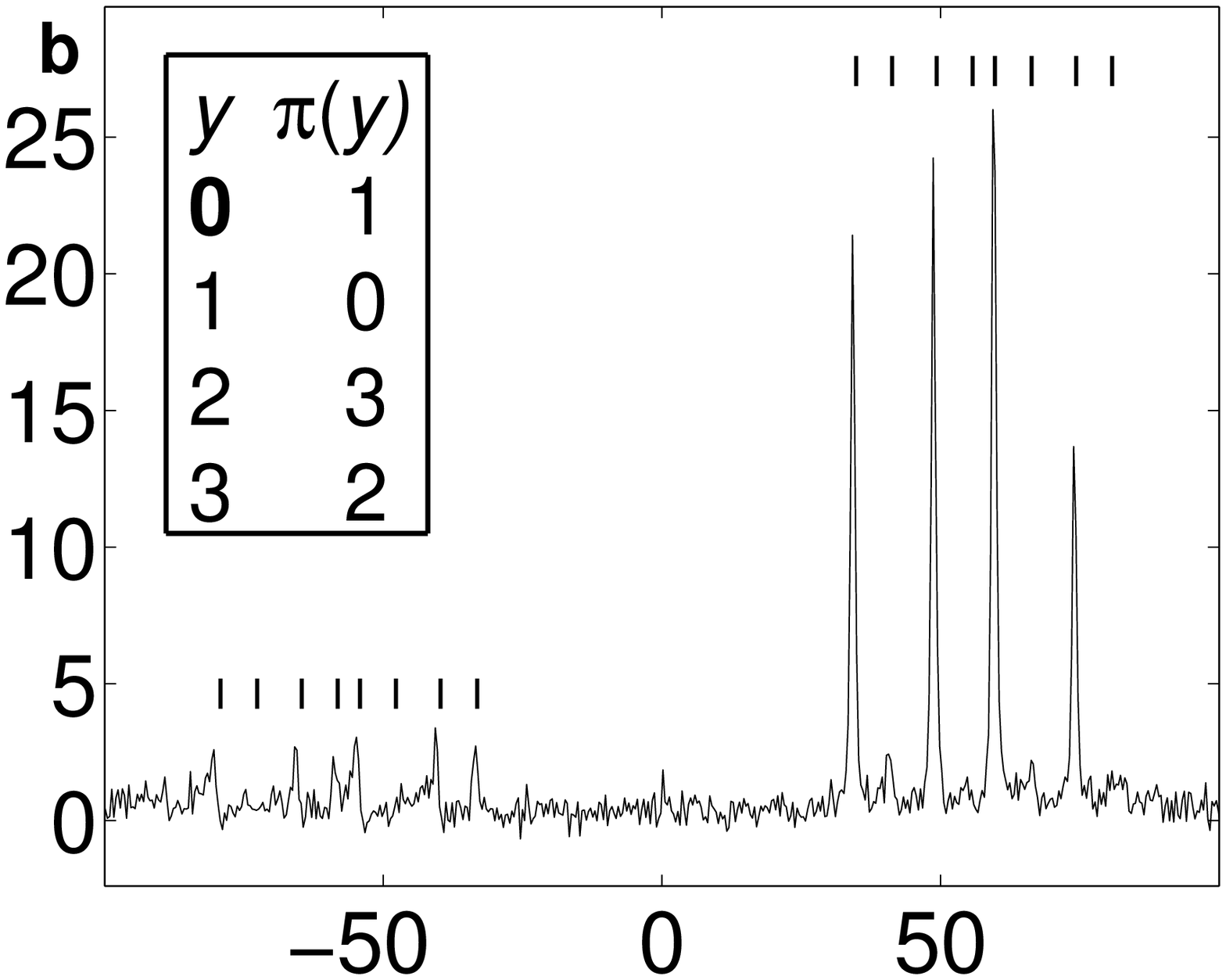} \\
\includegraphics*[height=2in]{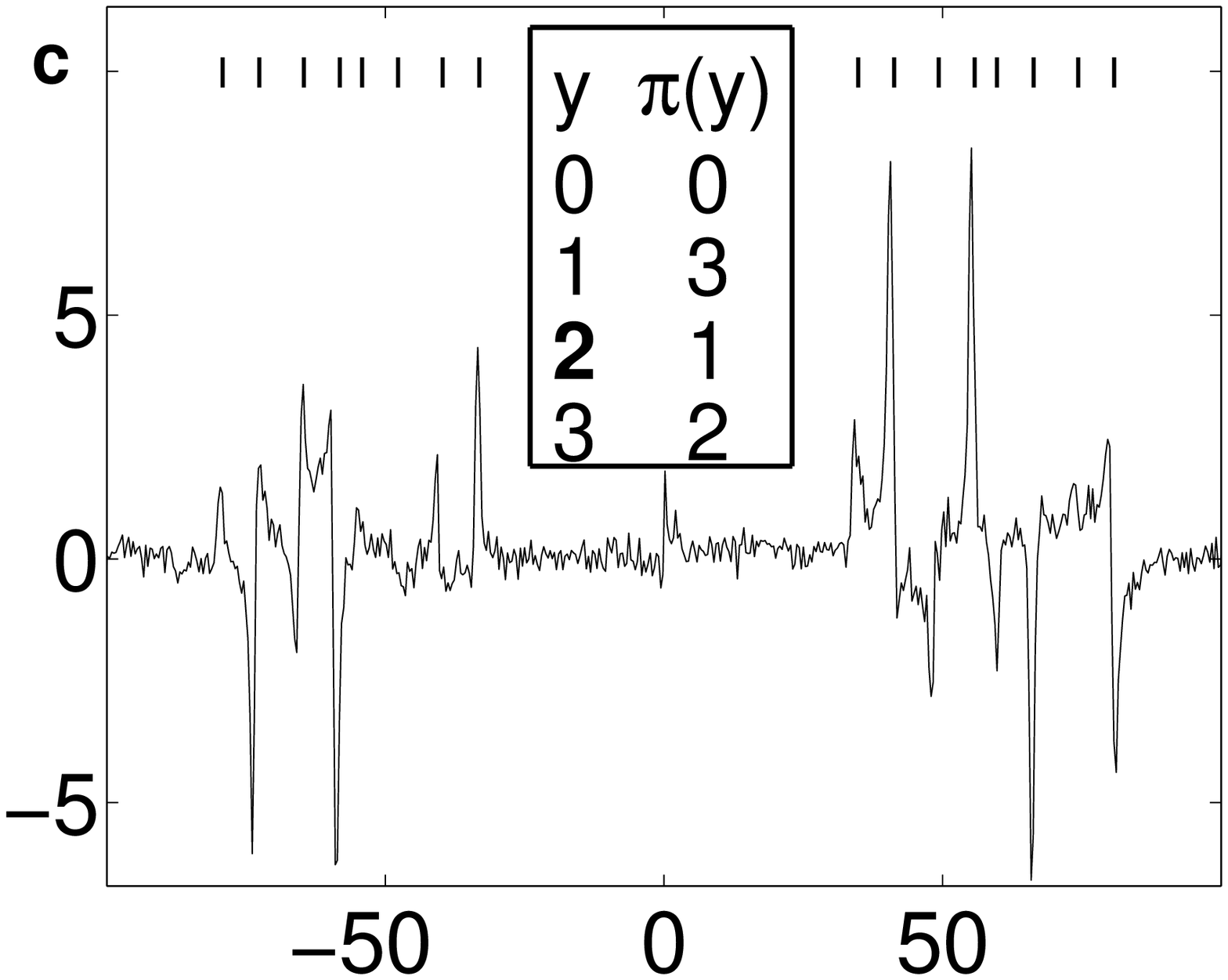} &
\includegraphics*[height=2in]{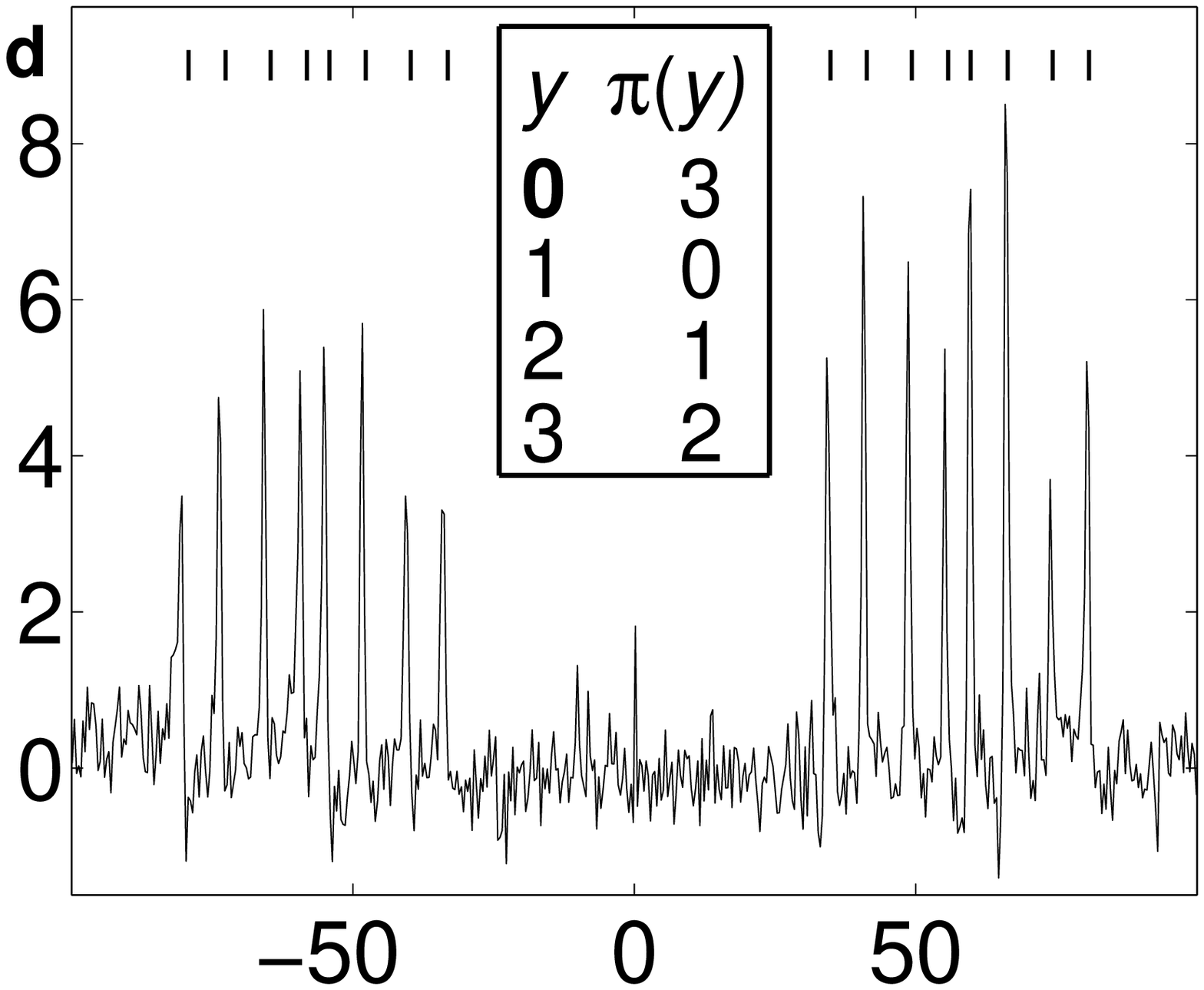} 
\end{array} $
\end{center}
\vspace*{-2ex}
\caption{Spectra of spin $1$ acquired after executing the 
order-finding algorithm. The respective permutations are shown in
inset, with the input element highlighted.  The $16$ marks on top of
each spectrum indicate the position of the $16$ lines in the thermal
equilibrium spectrum.}
\label{fig:order_spectra}
\end{figure}

Figure~\ref{fig:order_spectra} shows the spectrum of spin 1 after
running the order-finding algorithm with an effective pure input
state, for the pulse sequences given in section~\ref{expt:shor_expt}.
In all cases, the spectrum is in good agreement with the predictions,
both in terms of the number of lines, and their position, sign and
amplitude.  Slight deviations from the ideally expected spectra are
attributed mostly to incomplete refocusing of undesired coupled
evolutions and to decoherence.

\vspace*{1ex}

\subsection{Discussion}

The success of the order-finding experiment required the synthesis of
a molecule with unusual NMR properties and the development of several
new methods to meet the increasing demands for control over the spin
dynamics. The major difficulty was to address and control the qubits
sufficiently well to remove undesired couplings while leaving select
couplings active --- much of the more advanced techniques for
single-qubit rotations presented in section~\ref{nmrqc:1bitgates} were
developed for this experiment. Furthermore, the pulse sequence had to
be executed within the coherence time. Clearly, the same challenges
will be faced in moving beyond liquid state NMR, and we anticipate
that solutions such as those reported here will be useful in future
quantum computer implementations, in particular in those involving
spins, such as solid state NMR~\cite{Kane98a}, electron spins in
quantum dots~\cite{Loss98a} and ion traps~\cite{Sackett00a}.

%%%%%%%%%%%%%%%%%%%%%%%%%%%%%%%%%%%%%%%%%%%%%%%%%%%%%%%%%%%%%%%%%%%%
%%%%%%%%%%%%%%%%%%%%%%%%%%%%%%%%%%%%%%%%%%%%%%%%%%%%%%%%%%%%%%%%%%%%

\section{Shor's factoring algorithm (7 spins)}
\label{expt:shor}

\subsection{Problem description}

Prime factorization of a composite number using a quantum computer has
been the ``Holy Grail'' of the early exploration of small scale
quantum computers. The fundamental interest in quantum factoring
combined with the unprecedented complexity of the experiment make the
experimental demonstration of quantum factoring a landmark
achievement.  Our goal has been to accomplish this feat
\cite{Vandersypen01a}.\footnote{Ike Chuang encouraged me to think
about what it would take to do a quantum factoring experiment, and to
lead the work. I worked out the theory and requirements under Ike's
guidance. Xinlan Zhou suggested the method for multiplication by four
modulo fifteen. Nino Yannoni found the molecule based on my input
regarding molecule requirements and Greg Breyta synthesized the
molecule. Mark Sherwood helped with the molecule work and with liquid
NMR techniques. The software framework for the experiment was written
by Matthias Steffen, with my input. He and I did the experiment
together and came up with several new techniques for coherent control
of multiple coupled spins. I wrote the decoherence model for seven
spins.}

\begin{figure}[h]
\bcen
\includegraphics*[width=13cm]{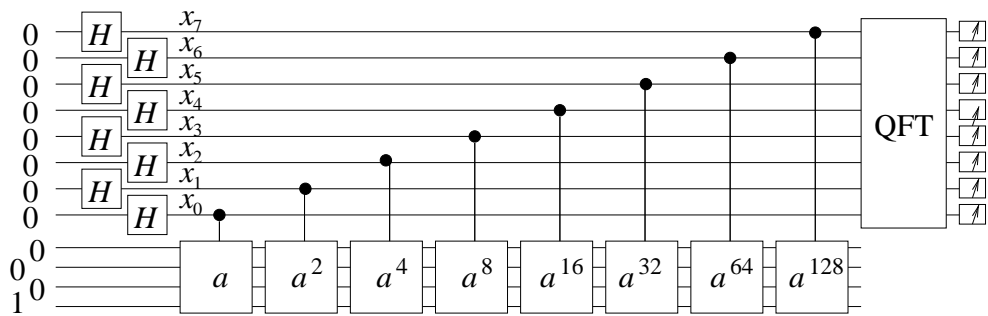}
\ecen
\vspace*{-2ex}
\caption{Outline of the quantum circuit for quantum factorization
of the number fifteen. The first register of $n=8$ qubits is
initialized to $\ket{0}$ and then put into the equal superposition
$\sum_{x=0}^{2^n-1} \ket{x}$ using $n$ Hadamard operations. The second
register is initialized to $\ket{y} = \ket{1}$. Then we multiply the
second register by $a^x \mbox{mod}\, 15$ via $n$ controlled
multiplications by $a^{2^k}$ modulo 15. Next the quantum Fourier
transform is applied to the first register and the qubits of the first
register are measured.}
\label{fig:shor_circuit_big}
\end{figure}

The smallest number $L$ for which Shor's algorithm can be meaningfully
implemented is $L=15$, given that the algorithm fails for $L$ even or
a prime power ($L = p^\alpha$, with $p$ prime). For $L=15$,
the size of the second register must be at least $m = \lceil \log_2 15
\rceil = 4$ and the first register must in principle be of size at 
least $n=2m=8$ (section~\ref{qct:shor}).  The quantum circuit is shown
in Fig.~\ref{fig:shor_circuit_big}, where we used
Eq.~\ref{eq:modexp_factor} to decompose the modular exponentiation
into a sequence of multiplications controlled by one qubit each
\cite{Beckman96a} .

\vspace*{1ex}

Repeated squaring of $a$ on a classical computer efficiently gives the
numbers $a, a^2$ through $a^{2^{n-1}}$. These numbers are
summarized in Table~\ref{tab:repeated_squaring} for all $a < 15$ and
coprime with 15 (in a practical application, it suffices to do this
for just one value of $a$).  The table also gives $a^x$ for values of
$x$ which are not a power of two, so we can see in advance what the
period $r$ of $f(x) = a^x \mbox{mod}\, 15$ is, although this is not
needed for Shor's algorithm.

\begin{table}[h]
$$
\begin{array}{c@{\quad}c@{\quad}|@{\quad}cccccccc@{\quad}|@{\quad}c@{\quad}c@{\quad}c}
 &  & \multicolumn{8}{c|@{\quad}}{x} & & & \mbox{gcd} \vspace*{-1ex}\\       
 &  & 0 & 1 & 2 & 3 & 4 & 5 & 6 & 7 & \raisebox{2ex}{r} & \raisebox{2ex}{$a^{r/2} \pm 1$} & \mbox{with 15}\\ \hline
 &2 & 1 & 2 & 4 & 8 & 1 & 2 &\ldots&& 4 & 2^{4/2} \pm 1 = 3,5   &3,5\\
 &4 & 1 & 4 & 1 & 4 &\ldots&&   &   & 2 & 4^{2/2} \pm 1 = 3,5   &3,5\\
 &7 & 1 & 7 & 4 &13 & 1 & 7 &\ldots&& 4 & 7^{4/2} \pm 1 = 48,50 &3,5\\
a&8 & 1 & 8 & 4 & 2 & 1 & 8 &\ldots&& 4 & 8^{4/2} \pm 1 = 63,65 &3,5\\
 &11& 1 &11 & 1 &11 &\ldots&&   &   & 2 &11^{2/2} \pm 1 = 10,15 &5,3\\
 &13& 1 &13 & 4 & 7 & 1 & 13&\ldots&& 4 &13^{4/2} \pm 1 =168,170&3,5\\
 &14& 1 &14 & 1 &14 &\ldots&&   &   & 2 &14^{2/2} \pm 1 = 13,15 &-,5\\
\end{array}
$$
\caption{The table gives  $f(x) = a^x \mbox{mod} \,15$ for all $a < 15$ 
coprime with 15 and for successive values of $x$. For each value of
$a$, the period $r$ emerges from the sequence of output values
$f(x)$. Calculation of the greatest common denominator of $a^{r/2} \pm
1$ and 15 then gives at least one prime factor of 15.}
\label{tab:repeated_squaring}
\end{table}

From repeated squaring, we see that $a^4 \mbox{mod} \,15 = 1$ for all
valid $a$ (Table~\ref{tab:repeated_squaring}). This means for the
quantum circuit of Fig.~\ref{fig:shor_circuit_big} that the
multiplications by $a^4, a^8, \ldots, a^{128}$ are trivial: if
$\ket{x_k} = \ket{1}$ ($k \ge 2$), we multiply by $a^{2^k} = 1$,
i.e. we do nothing, and if $\ket{x_k} = \ket{0}$ we also do
nothing. Therefore, all the controlled multiplications except the ones
by $a$ and $a^2$ can be left out.  For $a=4,11$ or $14$, we even have
$a^2 \mbox{mod} \,15 = 1$, so in this case we only need to keep the
controlled multiplication by $a$. We thus see that at most two qubits
of the first register act non-trivially during the modular
exponentiation, and we might as well leave out the other qubits
altogether.\footnote{In reality, if in the process of repeated
squaring we find that $a^{2^k} \mbox{mod} \,15 =1$ for some $k$, while
$a^{2^{k-1}} \neq 1$, we know that the period $r$ must be
$r=2^k$. There is then really no need anymore to run the quantum
algorithm. This is the case for $L=15$ for any choice of
$a$. Nevertheless, the non-trivial operation of Shor's algorithm can
still be demonstrated~\cite{Beckman96a}.} Since the essence of Shor's
algorithm lies in the interplay between modular exponentiation and the
QFT, we chose to retain $n=3$ qubits to represent $x$.

\begin{figure}[h]
\bcen
\includegraphics*[width=8cm]{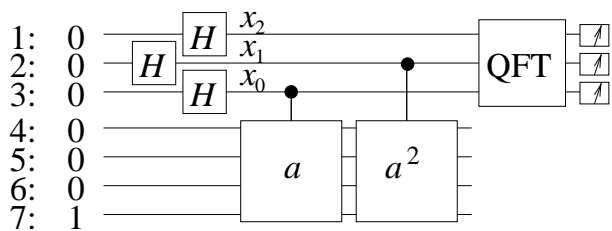}
\ecen
\vspace*{-2ex}
\caption{Simplified quantum circuit for quantum factorization 
of the number fifteen.}
\label{fig:shor_circuit_small}
\end{figure}

In total, we shall thus use seven qubits ($n=3$ and $m=4$), as in
Fig.~\ref{fig:shor_circuit_small}. The possible choices of $a$ break
down into two groups. The first group ($a=4, 11$ and $14$) is ``easy''
as only multiplication by $a$ is needed; we will refer to the second
group ($a=2,7,8$ and $13$) as the ``difficult'' case. We will
implement the algorithm both for the easy and the difficult case,
using $a=11$ and $a=7$ respectively.

\vspace*{1ex}

\begin{figure}[h]
\bcen
\raisebox{2.0ex}{\includegraphics*[width=3.4cm]{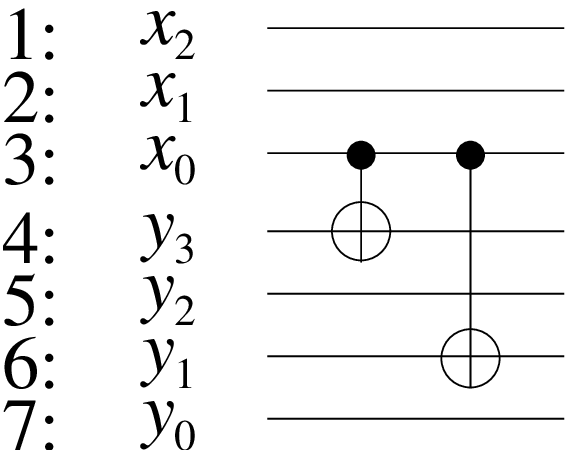}}
\hspace*{1.5cm}
\includegraphics*[width=7cm]{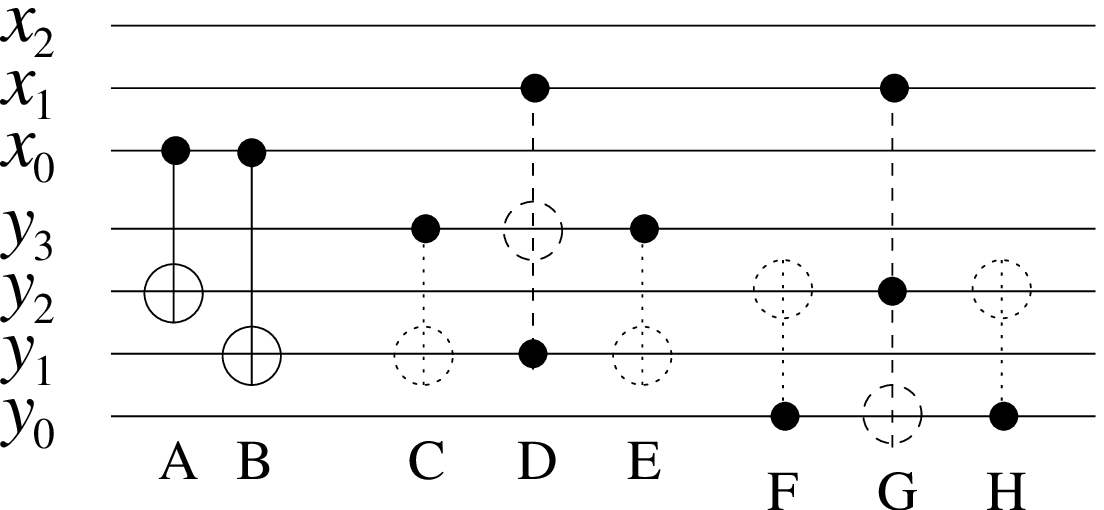}
\ecen
\vspace*{-2ex}
\caption{Quantum circuit for the modular exponentiation. (Left) 
for $a=11$; controlled multiplication of $y=1$ by 11 is replaced by
controlled addition of 10 to $y=1$. (Right) for $a=7$; gates $A$ and
$B$ correspond to addition of 6 to $y=1$ controlled by $x_0$, and
gates $C$ through $H$ multiply the result by 7 controlled by $x_1$. As
we will see in section~\protect\ref{expt:shor_appr}, the gates shown
in dotted lines can be left out and the gates shown in dashed lines
can be replaced by similar but simpler gates.}
\label{fig:shor_circuit_modexp}
\end{figure}

The controlled multiplications are done as follows. Multiplication of
$y=1$ by $a$ is equivalent to adding $a-1$ to $y=1$. The controlled
multiplication by $a$ can thus be implemented via a few {\sc cnot}s of
qubit 3 onto select qubits in the second register, as shown in
Fig.~\ref{fig:shor_circuit_modexp}.  Multiplication of $y$ by 4 shifts
the bits of $y$ over two places:

$$
\begin{array}{cccccc}
       &    & y_3 & y_2 & y_1 & y_0 \\
\times &    &     &     &     &  4  \\ \hline
   y_3 & y_2 & y_1 & y_0 &  0  &  0  
\end{array} 
$$

In order to take the remainder of $y_3 y_2 y_1 y_0 0 0$ divided by 15,
we note that the weight of bit $y_2$ is now 16 and the weight of $y_3$
is now 32; furthermore, we note that $16 \mbox{mod} \,15 = 1$ and that
$32 \mbox{mod} \,15 = 2$. In other words, $y_3 y_2 y_1 y_0 0 0\,
\mbox{mod} \,15 = y_1 y_0 y_3 y_2$. In effect, multiplication of $y$ by
$4$ modulo 15 comes down to swapping bit $y_3$ with $y_1$ and $y_2$
with $y_0$. Controlled multiplication by four can thus be accomplished
via two controlled-{\sc swap} or {\sc Fredkin} gates, which can be
decomposed into {\sc cnot}'s and {\sc toffoli}'s using the
construction of Fig.~\ref{fig:fredkin_circuit}. The resulting quantum
circuit is shown in Fig.~\ref{fig:shor_circuit_modexp}.

The resulting quantum circuits for the modular exponentiation are
shown in Fig.~\ref{fig:shor_circuit_modexp}, both for the easy and the
difficult case. The Hadamard gates don't need to be broken up further
and we have already discussed the quantum Fourier transform on three
qubits in section~\ref{qct:shor} as well as in
section~\ref{expt:shor}.

\vspace*{1ex}

%%%%%%%%%%%%%%%%%%%%%%%%%%%%%%%%%%%%%%%%%%%%%%%%%%%%%%%%%%%%%%%%%%%%

\subsection{Experimental approach}
\label{expt:shor_appr}

\subsubsection{Molecule}

\begin{figure}[h]
\bcen
\includegraphics*[width=12cm]{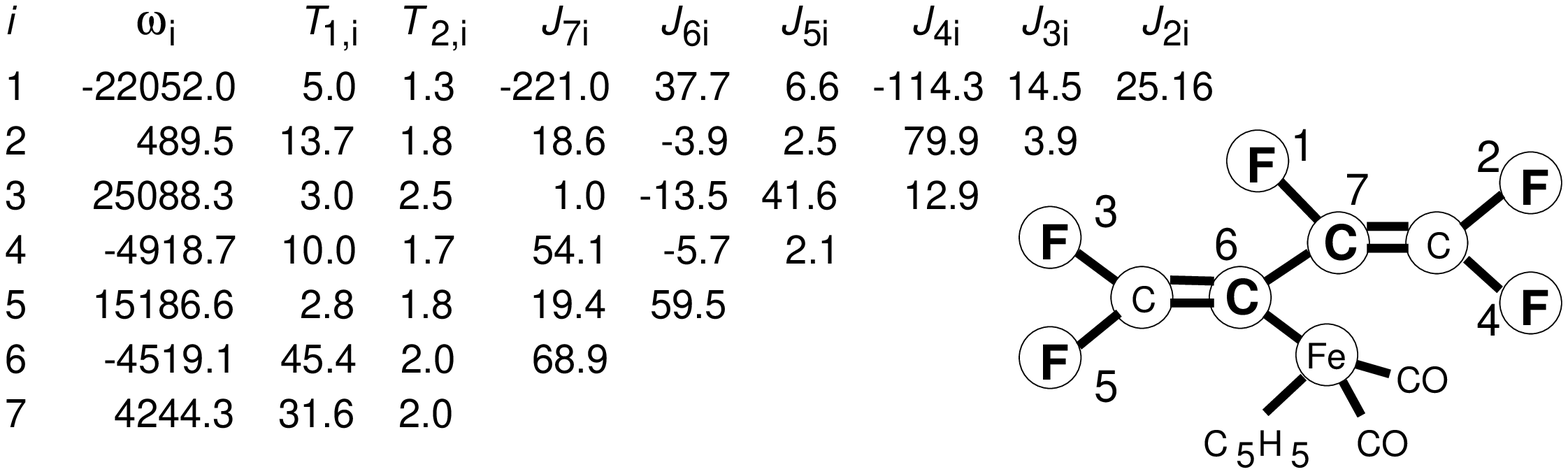}
\ecen
\vspace*{-2ex}
\caption{The seven spin molecule, along with the measured $J$-coupling 
constants [Hz], chemical shifts at 11.7 T [Hz] and relaxation time
constants [s].}
\label{fig:shor_molecule}
\end{figure}

We chose to use the same molecule which worked so well in the
five-qubit experiment (Fig.~\ref{fig:order_molec}), but with the two
inner carbon atoms $99\%$ $^{13}$C enriched, in order to obtain two
extra qubits (from measurements on the $1 \%$ natural abundance
$^{13}$C compound, we had established that those two carbons would be
the best choice for isotopic labeling).  We disolved this molecule in
deuterated ether. The molecular structure, as well as the coupling
constants and chemical shifts, are shown in
Fig.~\ref{fig:shor_molecule}. The assignment of qubits to spins (shown
via the labels next to the spin-1/2 nuclei) is the result of a
trade-off between sensitivity (for which qubits $1,2$ and $3$ should
be $^{19}$F) and the demands on the coupling network. The same
assignment was used in all experiments.  The $^1$H spins in the iron
complex broaden the lines of the nearest $^{13}$C spin, and were
therefore decoupled during the pulse sequence.

The synthesis of this unusual molecule was quite complex, and is
summarized in Fig.~\ref{fig:synthesis}. Ethyl (2-$^{13}$C)bromoacetate
(Cambridge Isotopes) was converted to ethyl 2-fluoroacetate by heating
with AgF followed by hydrolysis to sodium fluoroacetate using NaOH in
MeOH.  The resulting salt was converted to 1,1,1,2-tetrafluoroethane
using MoF$_6$ \cite{Vanderpuy79a} and was subsequently treated with
two equivalents of BuLi followed by I$_2$ to provide
trifluoroiodoethene \cite{Burdon96a}.  Half of the ethene was
converted to the zinc salt which was recombined with the remaining
ethene and coupled using Pd(Ph$_3$P)$_4$ to give
(2,3-$^{13}$C)-hexafluorobutadiene \cite{Dolbier87a}.  The end product
was obtained by reacting this butadiene with the anion obtained from
treating [(p-C5H5)Fe(CO)2]2 with sodium amalgam \cite{Green68a}.

\bcen
\includegraphics*[width=13cm]{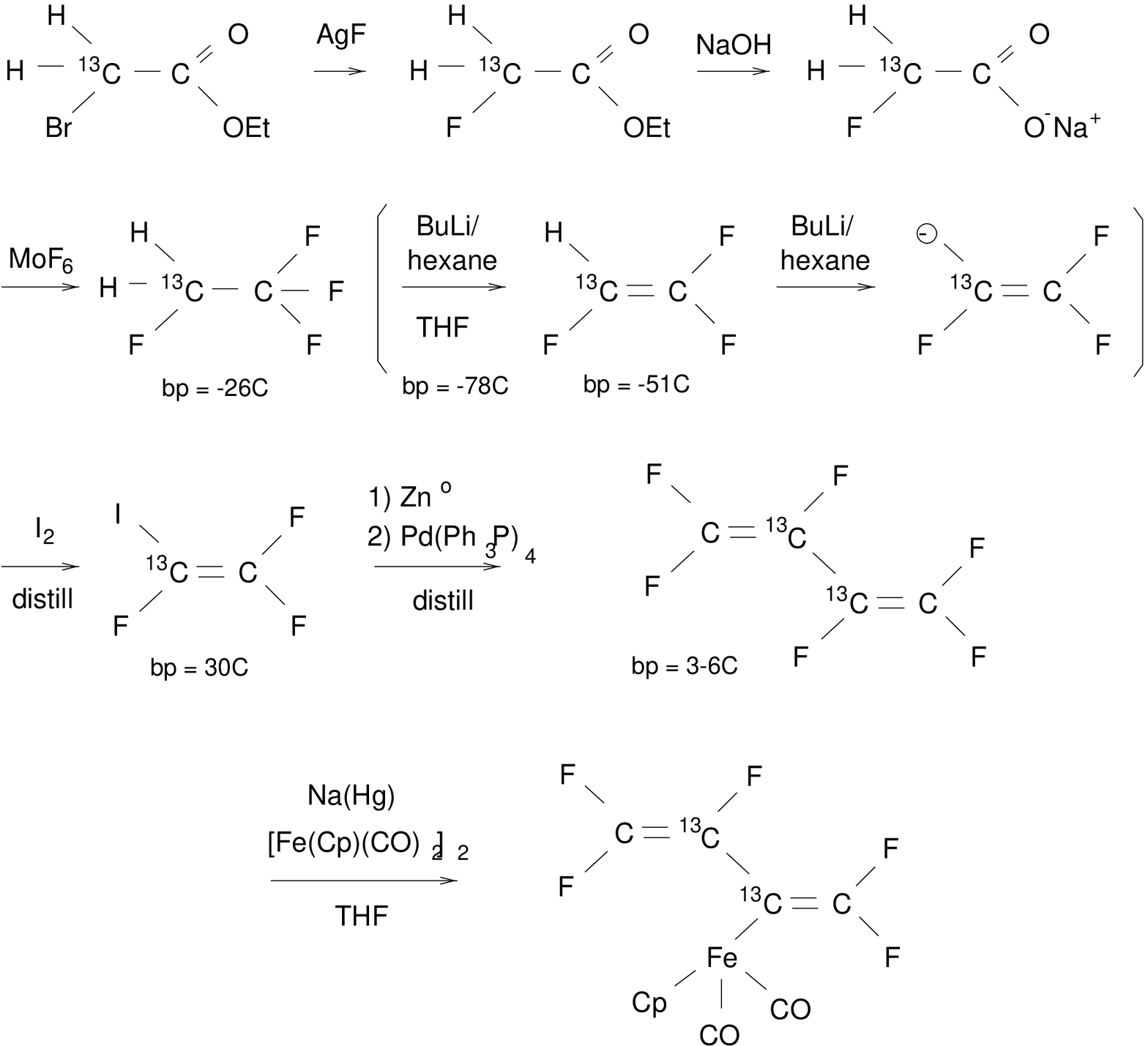}
\ecen
\vspace*{-2ex}
\begin{figure}[h]
\caption{Schematic diagram of the synthesis of the seven-spin molecule.}
\label{fig:synthesis}
\end{figure}

\subsubsection{Input state preparation}

For the creation of an effective pure state of all seven spins, we
used a two-stage extension of the scheme used in the five-spin
experiment. In the first stage, the five fluorine spins are made
effective pure via summation of the nine experiments of
Table~\ref{table:temp_av}, where the first five qubits are fluorine
spins and the last two are carbons.

\begin{table}[h]
$$
\small
\begin{array}{l|ccccccc}
\mbox{Equilibrium}                          & ZIIIIII & IZIIIII & IIZIIII & IIIZIII & IIIIZII \\ \hline
1.\; C_{24} C_{12} C_{31} C_{51}            & ZIIZZII & ZZZZZII & ZIZZZII &-IIZIIII & IIIIZII \\
2.\; C_{35} C_{43} C_{14} N_2               & ZIIIIII & ZZIIIII &-IIZIIII & ZZIZZII & ZZIIZII \\
3.\; C_{43} C_{14} C_{21} C_{31} N_5        & ZIZIZII & ZZZIZII & IIZIIII &-IIIZIII & ZZZIIII \\
4.\; C_{42} C_{14} C_{51} C_{21} N_1 N_5    & ZIZZIII & ZZIZIII & ZZZZIZZ &-IIIZIII & IIIIZII \\
5.\; C_{35} C_{43} C_{24} C_{35}            & ZIIIIII & IZZIIII & IIZIIII & IZZZIII & IZZIZII \\ 
6.\; C_{13} C_{15} C_{21} C_{12}            & ZIZIIII & IZIIIII & IIZIIII & IIZZIII & IIZIZII \\
7.\; C_{42} C_{34} C_{53} C_{31}            & ZIIIZII & IZIZZII & IZZZZII & IIIZIII & IIIZZII \\
8.\; C_{42} C_{34} C_{53} C_{42} C_{34} N_1 &-ZIIIIII & IZIZIII & IIZZZII & IIIZIII & IIIZZII \\
9.\; C_{35} C_{34} C_{51} N_4 N_2 N_3       & ZIIZIII & IZIIZII &-IIZIIII &-IIIZZII &-IIIIZII
\end{array}
$$
\normalsize
\caption{First stage of the temporal averaging procedure to prepare an 
effective pure state of five nuclei of the same species and two other
nuclei of another species. The table shows how the terms in the
thermal equilibrium density matrix are transformed by each temporal
averaging sequence (the $IIIIIZI$ and $IIIIIIZ$ terms are not shown as
they are left unaffected in this first stage).  $C_{ij}$ stands for
{\sc cnot}$_{ij}$ and $N_i$ stands for {\sc not}$_i$.}
\label{table:temp_av}
\end{table}

These state preparation sequences were designed to be as short as
possible by making optimal use of the available coupling network. The
nine experiments are repeated four times, each time with different
additional operations. In this process, all except four of the
$IIIIIZI$ and $IIIIIIZ$ terms are averaged away (these terms have
about one fourth the weight of the other terms, because the carbon
gyromagnetic ratio is four times smaller than the fluorine
gyromagnetic ratio) by applying a $180^\circ$ pulse on each carbon
spin in almost half of the cases. In the first set of nine
experiments, no extra {\sc cnot}s are performed. In the second set,
additional {\sc cnot}s turn the first carbon (spin 6) from $I$ into
$Z$; in the third set, the second carbon (spin 7) is converted from
$I$ into $Z$ and in the fourth set both carbons are converted into
$Z$. The term $IIIIIZZ$ is also created. Summation of the $4
\times 9 = 36$ experiments creates the desired seven-spin effective
pure state.

The pulse sequence for each {\sc cnot}$_{ij}$ in the temporal
averaging sequences was
\begin{eqnarray}
\mbox{for} \;& J_{ij} > 0: \quad &X_j \; 1/4J_{ij} \; X_i^2 X_j^2 \; 1/4J_{ij} \; X_i^2 X_j^2 \bar Y_2   \,,\\
\mbox{for} \;& J_{ij} < 0: \quad &X_j \; 1/4|J_{ij}| \; X_i^2 X_j^2 \; 1/4|J_{ij}| \; X_i^2 X_j^2 \; Y_2   \,,
\end{eqnarray}
which takes advantage of the fact that the input state is diagonal.\\

\subsubsection{Quantum circuit and pulse sequence simplification}

The pulse sequences for the actual Shor algorithm were designed using
the guidelines of section~\ref{nmrqc:seq_design}.  At the level of
quantum circuits, we used the rules of
Fig.~\ref{fig:simplify_circuits} in the following ways:
\begin{itemize}
\item Gate $C$ of Fig.~\ref{fig:shor_circuit_modexp} is a {\sc cnot}
controlled by a qubit which is in $\ket{0}$ and can thus be left out.
\item Gate $F$ of Fig.~\ref{fig:shor_circuit_modexp} is a {\sc cnot}
controlled by a qubit which is in $\ket{1}$ and can thus be replaced
by a {\sc not} gate.
\end{itemize}
Furthermore, the evolution of the second register does not matter
anymore during the QFT on the first register (see
section~\ref{qct:shor}), as the second register is traced out upon
read-out. Therefore,
\begin{itemize}
\item We can leave out gate $H$ of
Fig.~\ref{fig:shor_circuit_modexp}, since it comes at the end of the
modular exponentiation and does not involve any qubits in the first
register.
\item We can also take out gate $E$; it commutes with gates $F$ 
and $G$, so we can move it down to the end of the modular
exponentiation, where it becomes inconsequential.
\end{itemize}

Furthermore, we took advantage of the fact that the target of the two
{\sc Toffoli}'s (gates $D$ and $G$ in
Fig.~\ref{fig:shor_circuit_modexp}) are in a computational basis
state, not in a superposition state. Therefore, the phases of only
half the non-zero entries in the {\sc Toffoli} unitary matrices need
to be the same, and we can use a simplified set of gates to implement
them.  

\vspace*{1ex}
\begin{figure}[h]
\bcen
\includegraphics*[width=8cm]{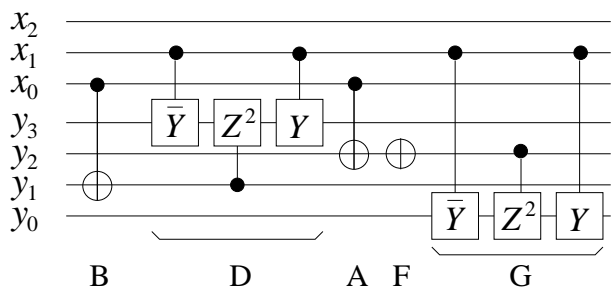}
\ecen
\vspace*{-2ex}
\caption{Simplified quantum circuit for the modular exponentiation 
for the ``difficult'' case ($a=7$).}
\label{fig:shor_circuit_modexp2}
\end{figure}

In addition, we simplified the refocusing schemes in two ways. Early
on in the pulse sequence, certain spins in the second register are
still in $\ket{0}$ or $\ket{1}$, and the couplings between any two
such spins need not be refocused. In order to maximally take advantage
of this simplification, gate $A$ was performed after gate $D$. This is
allowed since $A$ commutes with $B$ and $D$. During the QFT,
the evolution of the second register does not matter anymore, so the
couplings between the spins in the second register need not be
refocused during the QFT (couplings between the first and second
register qubits must still be refocused).  We did refocus
inhomogeneous dephasing for all spins in the transverse plane at all
times.

The quantum circuit resulting from these various
simplifications is shown in Fig.~\ref{fig:shor_circuit_modexp2}.

\subsubsection{Spectrometer}

The four-channel spectrometer was converted into a seven-channel
spectrometer using the same approach as for the five-qubit experiment
of section~\ref{expt:order}: additional rotating reference frames were
created using a software time counter, and phase ramping techniques
enabled excitation of spins away from the signal source frequency. The
frequency source of channel 1 was placed at $\omega_1$, on resonance
with spin 1, source 2 was set at $(\omega_2 + \omega_4)/2$, source 3
at $(\omega_3 + \omega_5)/2$ and source 4 at $(\omega_6 +
\omega_7)/2$. In addition, we installed an extra frequency source, 
power amplifier and power combiner for proton decoupling.

We selected Hermite 180 and Gaussian 90 pulse shapes based on our
previous experience, collected in Table~\ref{tab:shaped_pulses}, and
on the spectral properties of the molecule and the demands of the
algorithm. Generalized Bloch-Siegert effects were precomputed and
compensated for, both for single (section~\ref{nmrqc:pulse_artefacts})
and simultaneous (section~\ref{nmrqc:simpulse_artefacts}) pulses.
Simultaneous pulses on strongly coupled spins were avoided.
Furthermore, coupled evolution during the pulses was unwound using
symmetrically placed negative delay times, following the guidelines of
Table~\ref{tab:unwind_J} for single pulses and
Table~\ref{tab:unwind_J_sim} for simultaneous pulses.

The complete pulse sequence for the difficult case contains about 300
$180^\circ$ pulses and about 30 $90^\circ$ pulses. The duration of the
temporal averaging part is on the order of 200 ms, the modular
exponentiation takes about 400 ms, and the QFT sequence lasts about
120 ms.

\subsubsection{Read-out}

Upon completion of the pulse sequence, the reduced density matrix of
the three qubits in the first register is predicted to be
\be
\rho = \sum_l w_l \ket{l N/r}\bra{l N/r}
= \sum_l w_l \ket{l \frac{8}{r}}\bra{l \frac{8}{r}}
\ee

We shall attempt to deduce $r$ from the output spectra of the first
three qubits, after a readout pulse, which represent bitwise ensemble
averaged values. Once $r$ is known, we can find the prime factors of
$L=15$ as (recall Eq.~\ref{eq:gcd})
\be
\mbox{gcd} (a^{r/2} \pm 1, 15) \,.
\label{eq:factors_15}
\ee

%%%%%%%%%%%%%%%%%%%%%%%%%%%%%%%%%%%%%%%%%%%%%%%%%%%%%%%%%%%%%%%%%%%%%%

\subsection{Experimental results}

\begin{figure}[h]
\bcen
\includegraphics*[width=9cm,height=4.5cm]{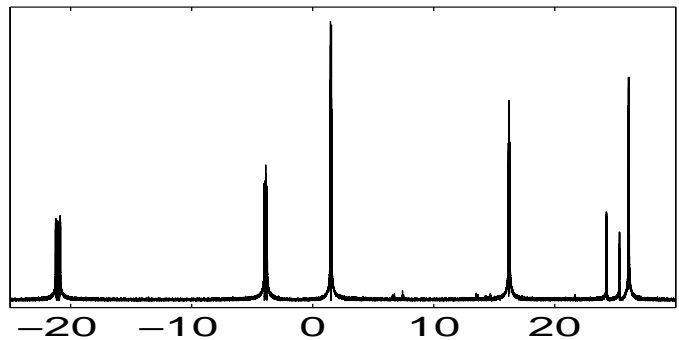}
\ecen
\vspace*{-2ex}
\caption{Fluorine spectrum of the seven-spin molecule of 
Fig.~\protect\ref{fig:shor_molecule}. The five major lines correspond
(from left to right) to qubits 1, 4, 2, 5, 3. In addition, two smaller
lines from impurities are visible around 25 kHz. The spectrum is shown
in absolute value. Frequencies are given in kHz, with respect to an
arbitrary reference frequency near 470 MHz.}
\label{fig:shor_thermal_wide_F}
\end{figure}

\begin{figure}[h]
\hspace*{2.5cm} {\sf Spin 1} 
\hspace*{3.5cm} {\sf Spin 2}
\hspace*{3.5cm} {\sf Spin 3}
\vspace*{-1ex}
\bcen
\includegraphics*[width=4.9cm]{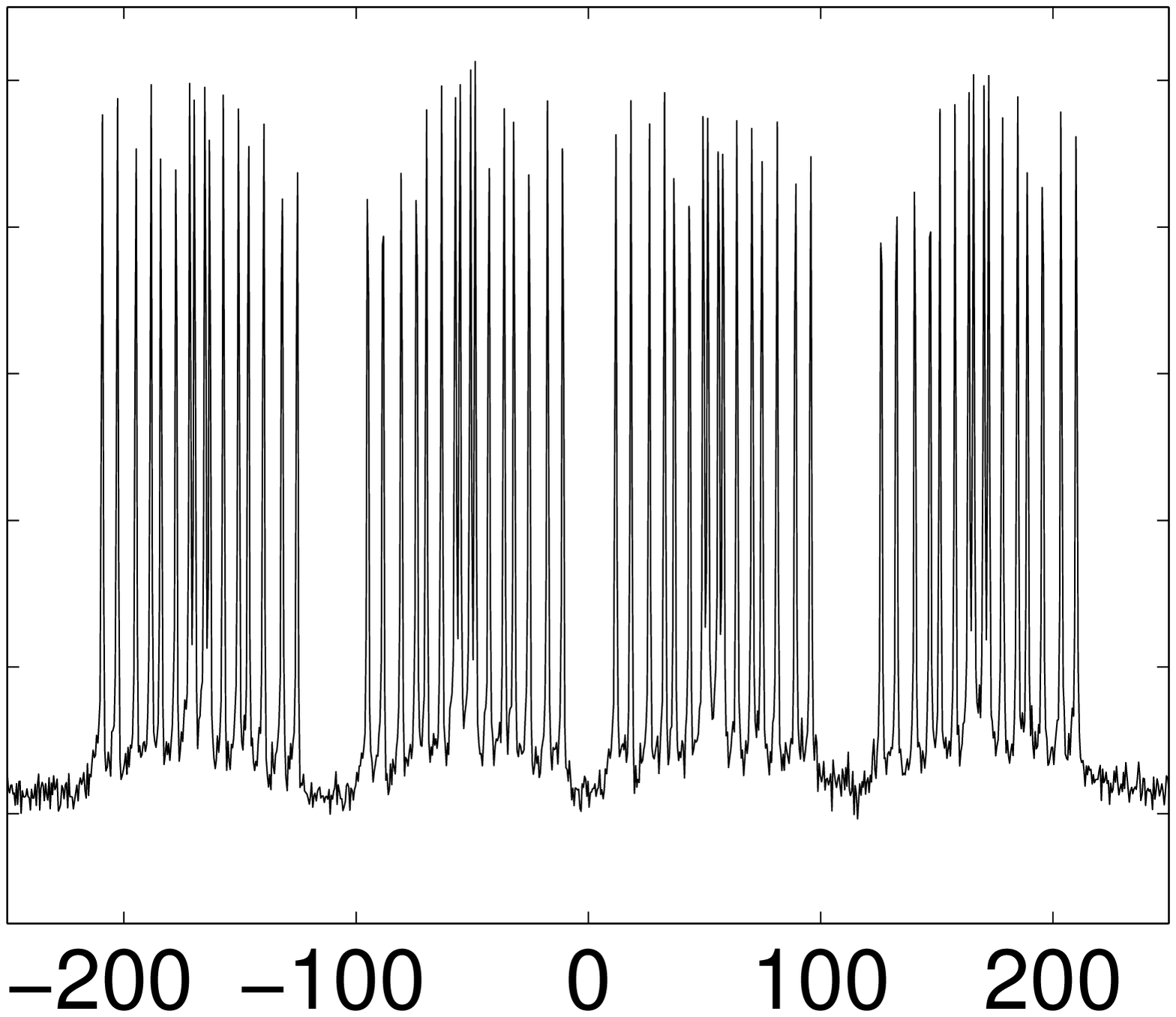}
\includegraphics*[width=4.9cm]{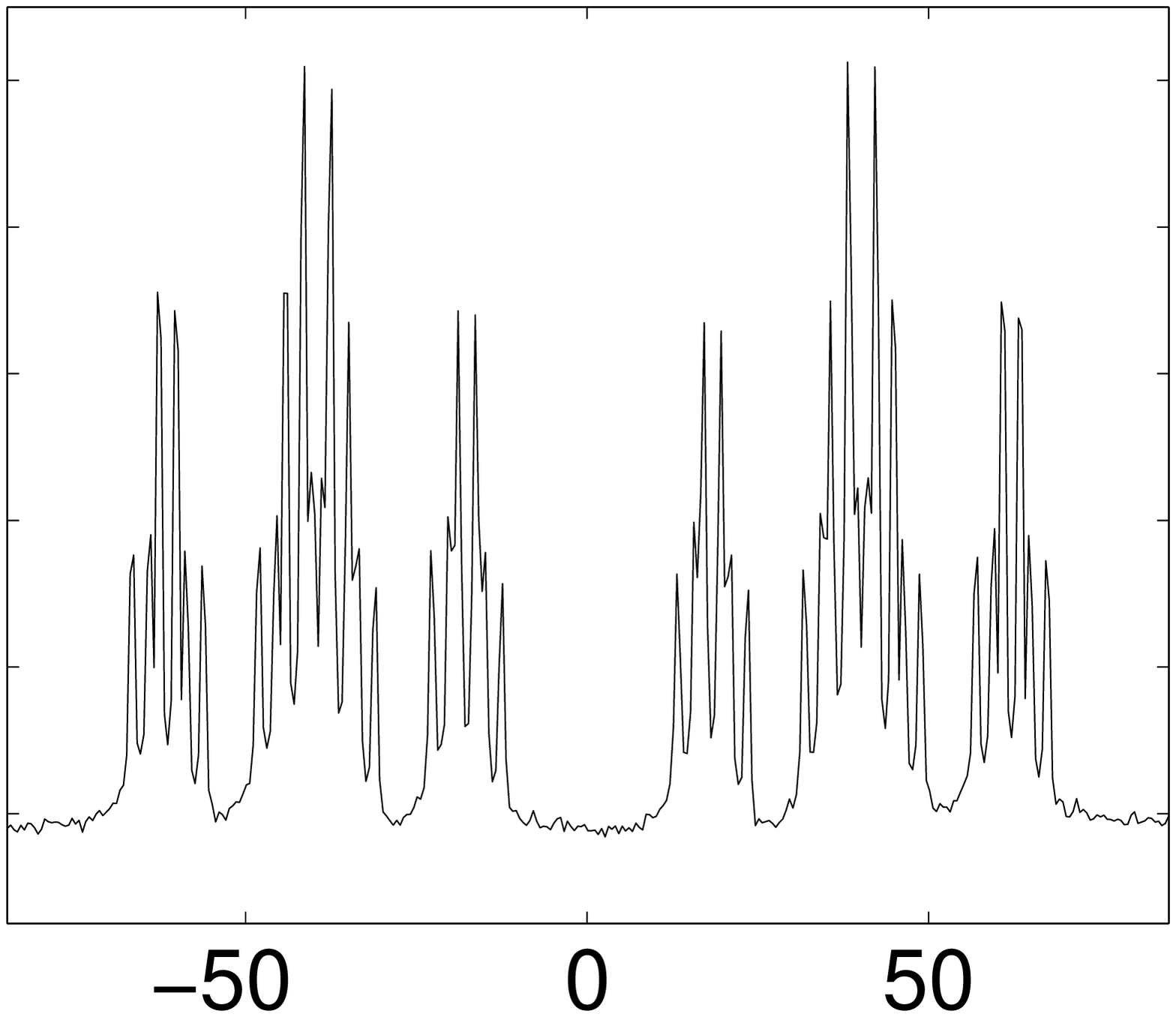}
\includegraphics*[width=4.9cm]{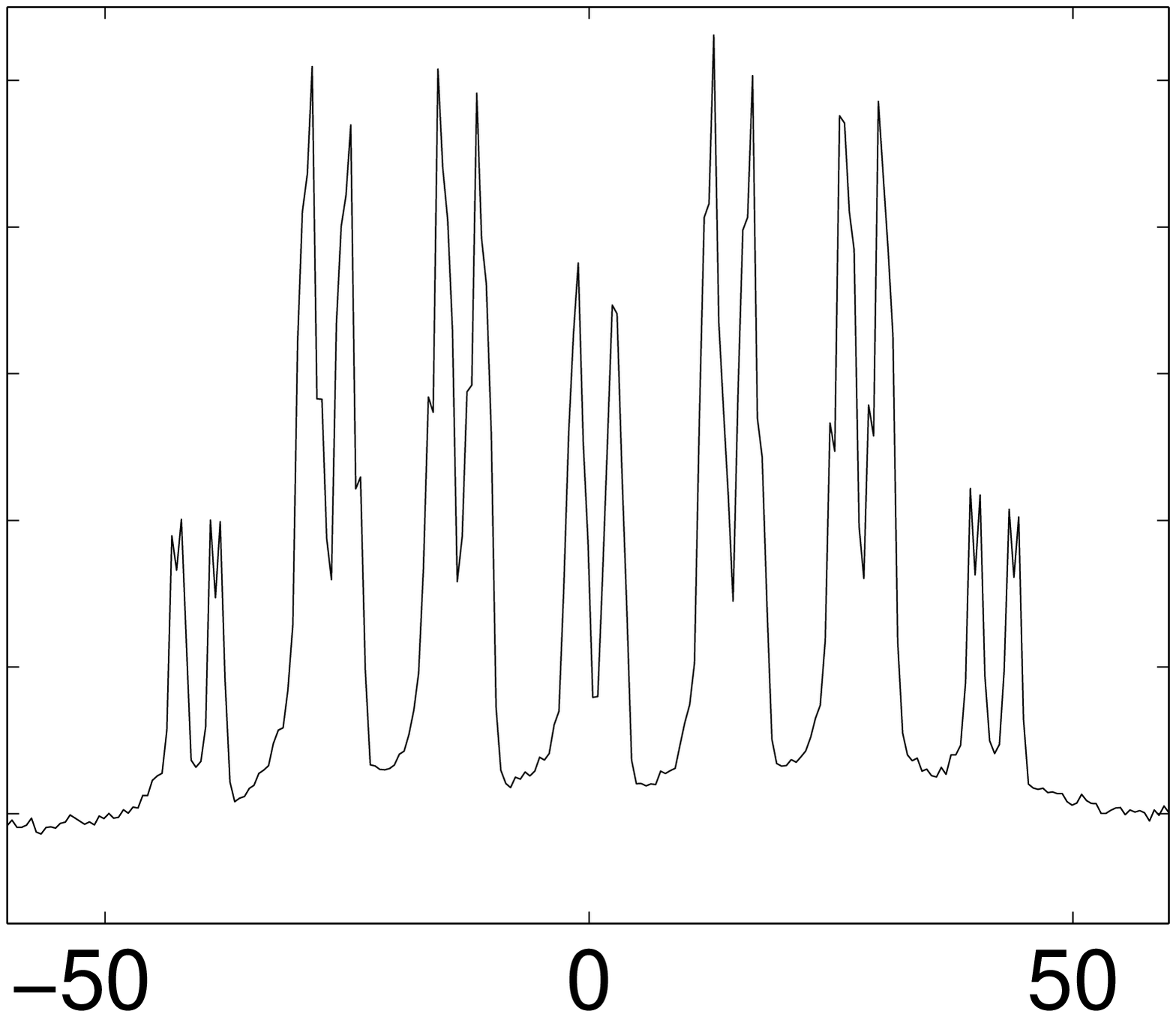}
\ecen
\bcen
{\sf Spin 4} 
\hspace*{3.5cm}
{\sf Spin 5} \vspace*{2ex} \\ 
\includegraphics*[width=4.9cm]{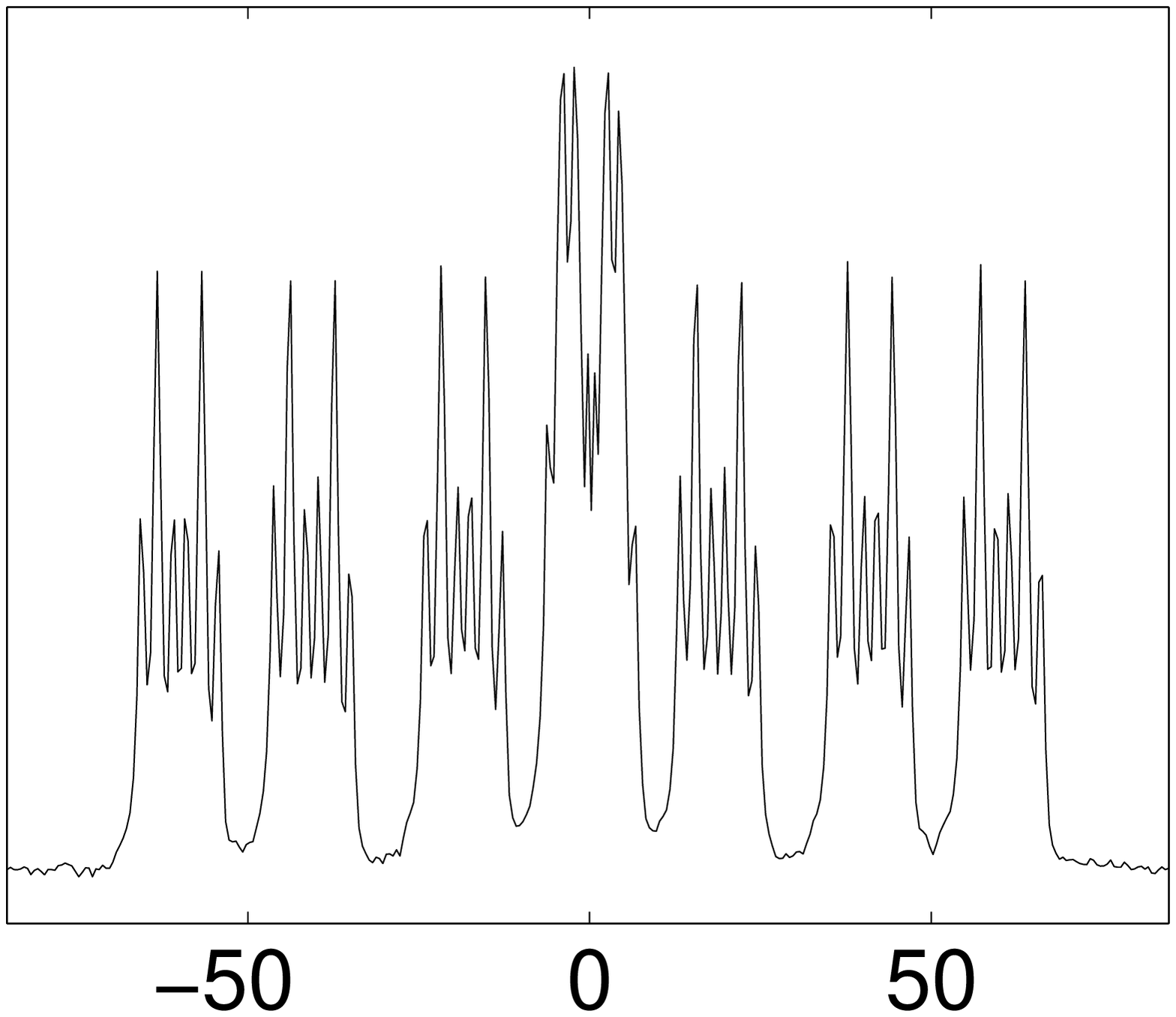}
\includegraphics*[width=4.9cm]{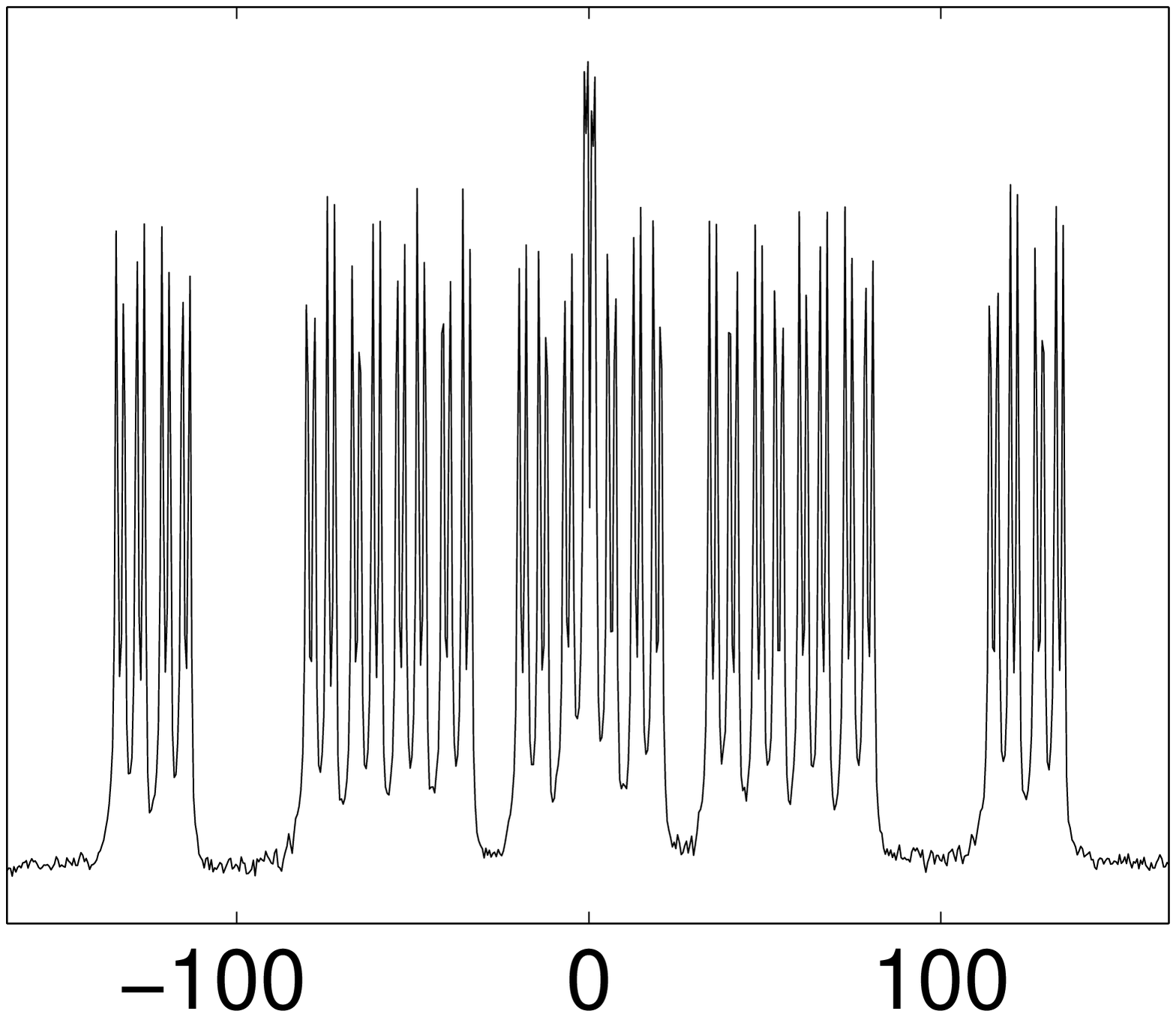}
\ecen
\vspace*{-2ex}
\caption{Experimentally measured spectra for the five fluorine spins  
in thermal equilibrium. The real part is displayed, in arbitrary
units. Frequencies are with respect to $\omega_i/2\pi$, in Hz.}
\label{fig:shor_therm_F}
\end{figure}

\begin{figure}[h]
\bcen
{\sf Spin 6} 
\hspace*{3.5cm}
{\sf Spin 7} \vspace*{2ex} \\
\includegraphics*[width=4.9cm]{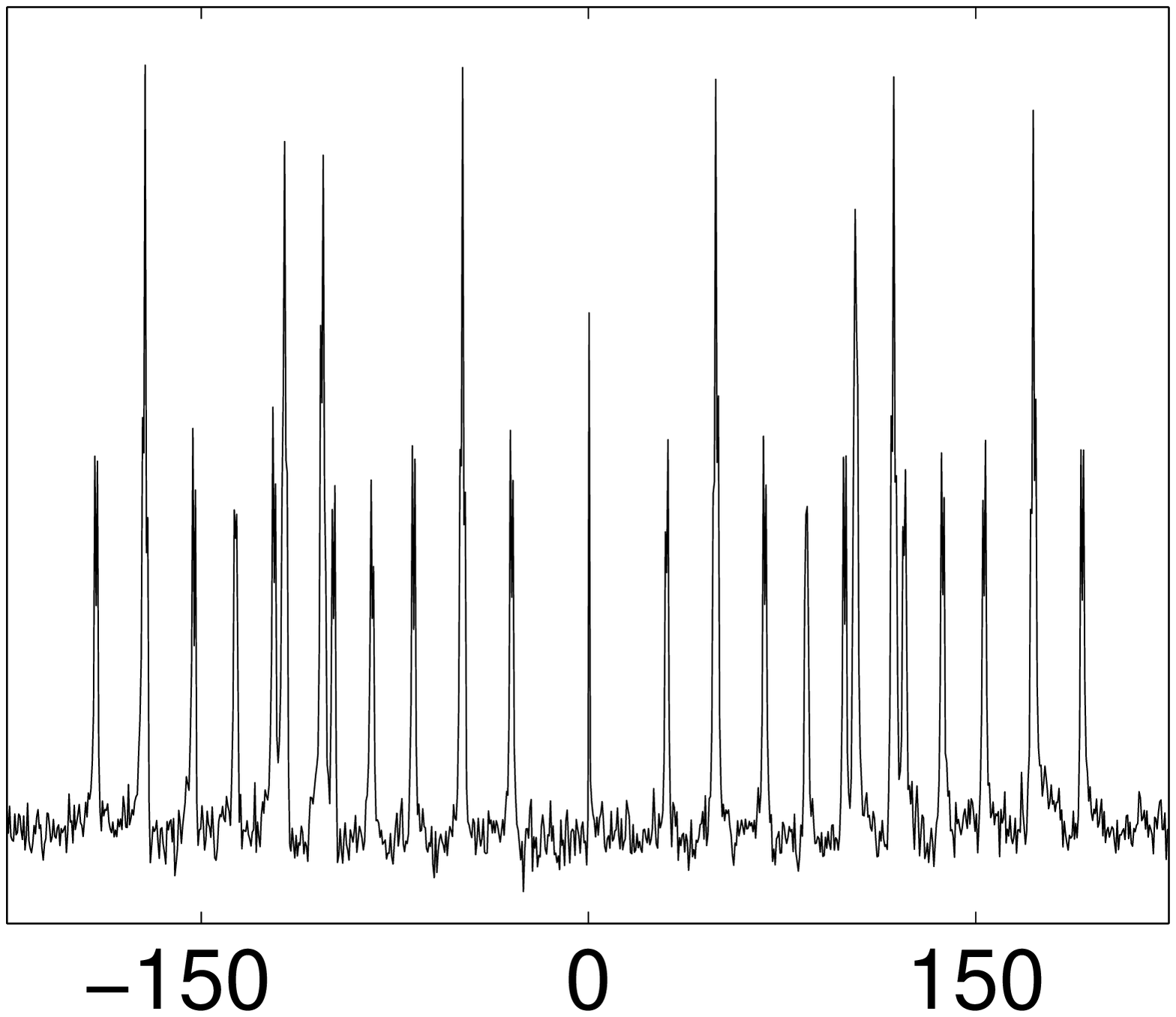}
\includegraphics*[width=4.9cm]{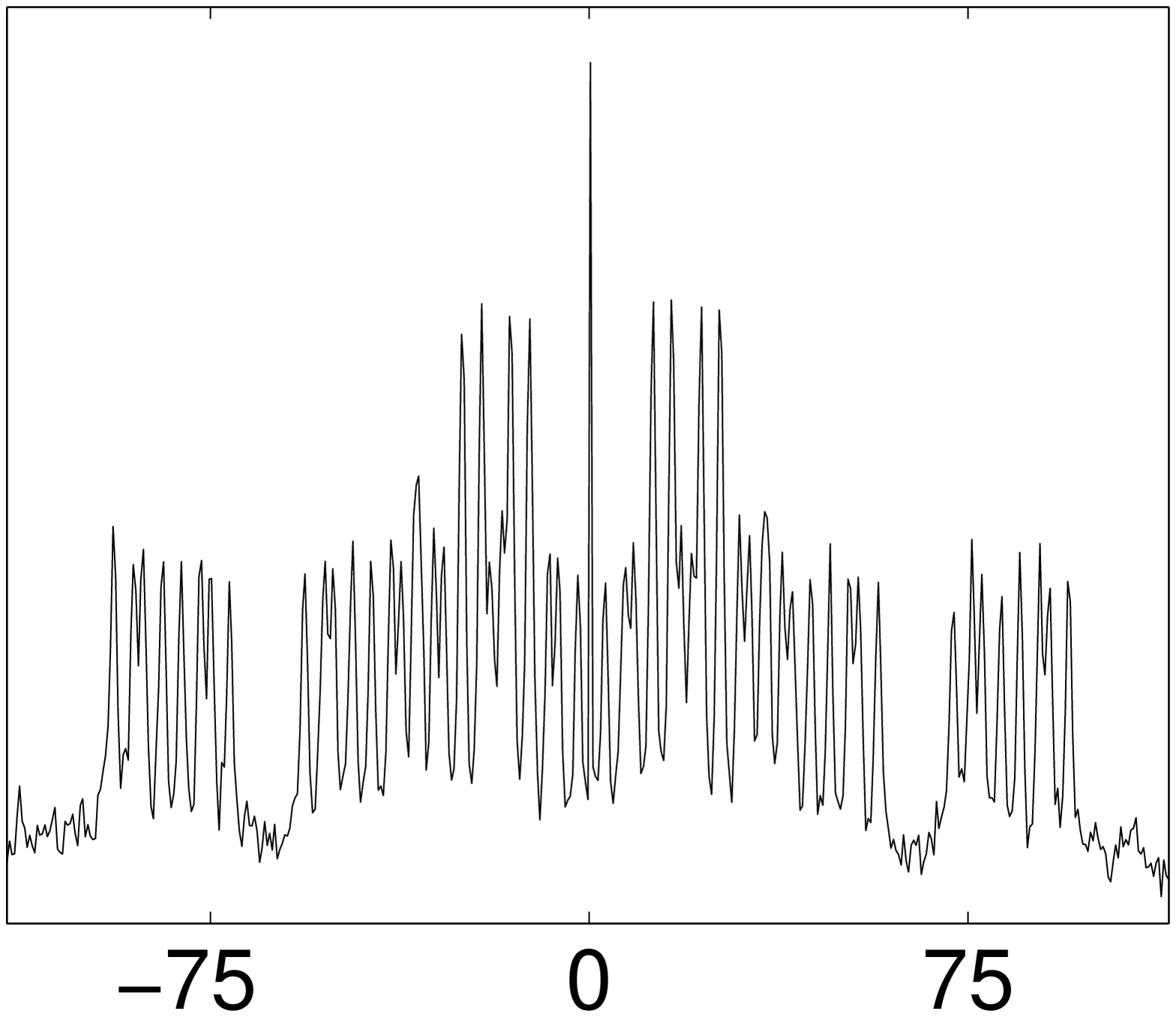}
\ecen
\vspace*{-2ex}
\caption{Experimentally measured spectra for the two carbon spins  
in thermal equilibrium. The real part is displayed, in arbitrary
units. Frequencies are with respect to $\omega_i/2\pi$, in Hz.}
\label{fig:shor_therm_C}
\end{figure}

Fig.~\ref{fig:shor_thermal_wide_F} shows the thermal equilibrium
fluorine spectrum of the sample.  Fig.~\ref{fig:shor_therm_F} zooms in
on five of the lines in this spectrum, corresponding to the five
fluorine spins of the quantum computer molecule of
Fig.~\ref{fig:shor_molecule}.  Fig.~\ref{fig:shor_therm_C} shows the
spectra of the two carbon spins.

\vspace*{1ex}

\newpage

\vspace*{1ex}

\vspace*{1ex}

Each multiplet contains up to $2^6 = 64$ lines, because each spin is
coupled to six other spins. For spin 1, all 64 lines are beautifully
resolved. For the other spins, some of the lines fall on top of each
other, but this does not pose a problem since at the end of Shor's
algorithm we only need to know the overall signature: lines up, lines
down, or partly up / partly down.

\begin{figure}[h]
\hspace*{2.5cm} {\sf Spin 1} 
\hspace*{3.5cm} {\sf Spin 2}
\hspace*{3.5cm} {\sf Spin 3}
\vspace*{-1ex}
\bcen
\includegraphics*[width=4.9cm]{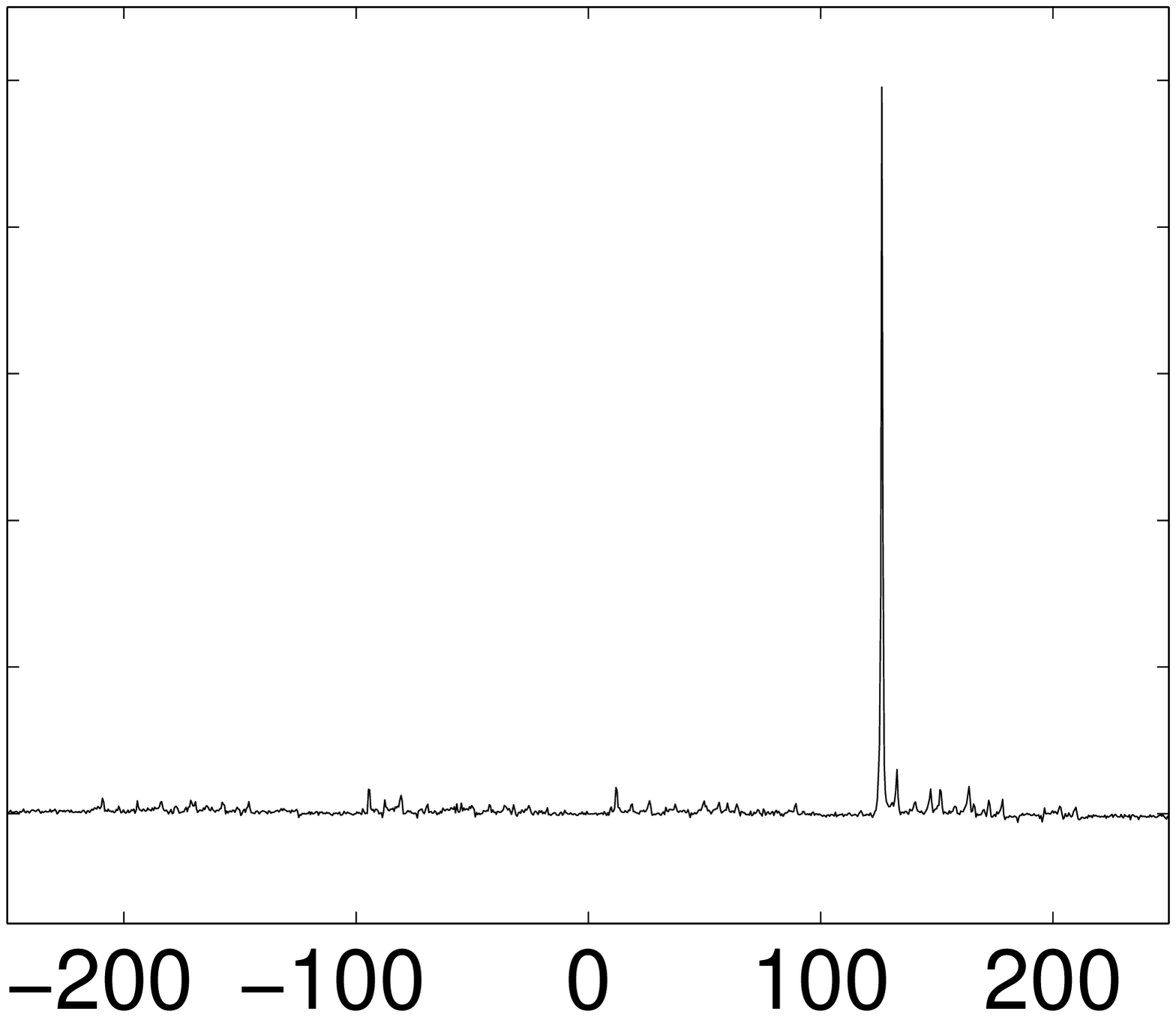}
\includegraphics*[width=4.9cm]{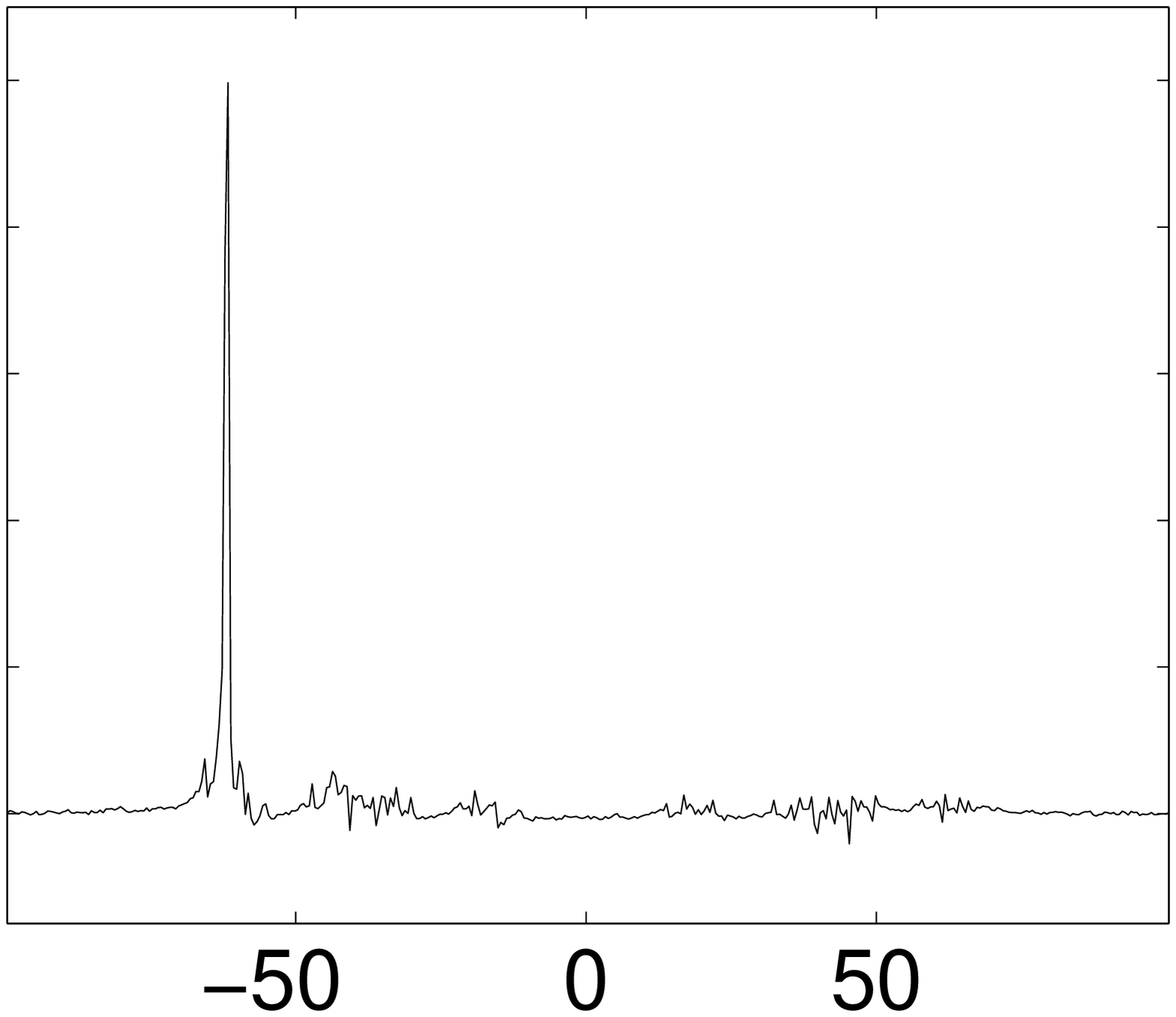}
\includegraphics*[width=4.9cm]{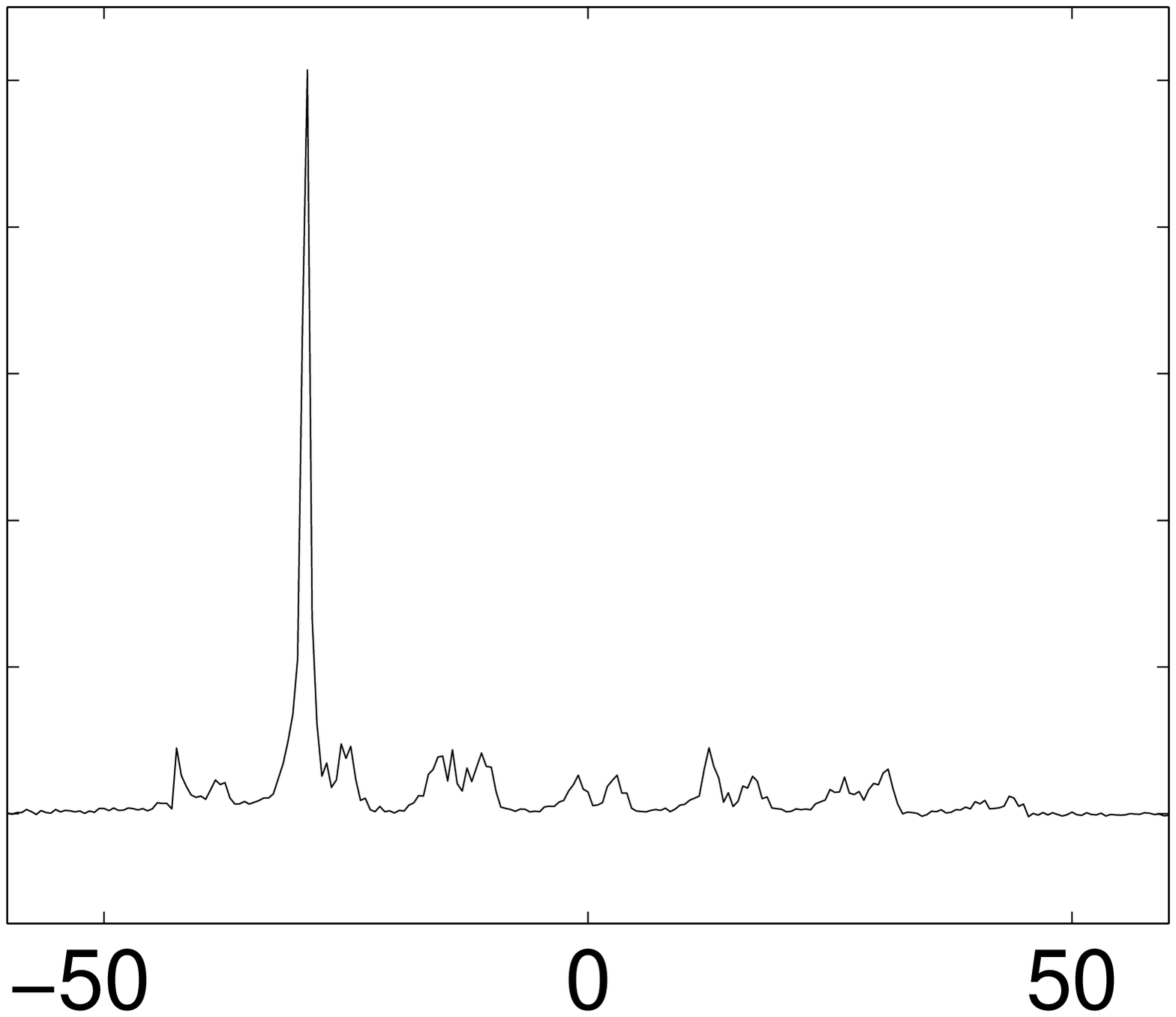}
\ecen
\vspace*{-2ex}
\caption{Experimentally measured spectra, similar to 
Fig.~\protect\ref{fig:shor_therm_F}, after preparing all seven spins in
the effective pure ground state.}
\label{fig:shor_pure}
\end{figure}

The spectra of spins 1, 2 and 3 after preparing a seven-spin
effective pure ground state are shown in Fig.~\ref{fig:shor_pure}.
These spectra are the summation of 36 spectra, each obtained
after a different input state preparation pulse sequence.  As
expected, only one line is retained in each multiplet, which indicates
that we have distilled a suitable initial state for Shor's algorithm.

\vspace*{1ex}

\begin{figure}[h]
\hspace*{2.5cm} {\sf Spin 1} 
\hspace*{3.5cm} {\sf Spin 2}
\hspace*{3.5cm} {\sf Spin 3}
\vspace*{-2ex}
\begin{center}
\rotatebox{90}{\sf \small \hspace*{5ex} Expt. $\quad$ Ideal} 
\includegraphics*[width=4.8cm]{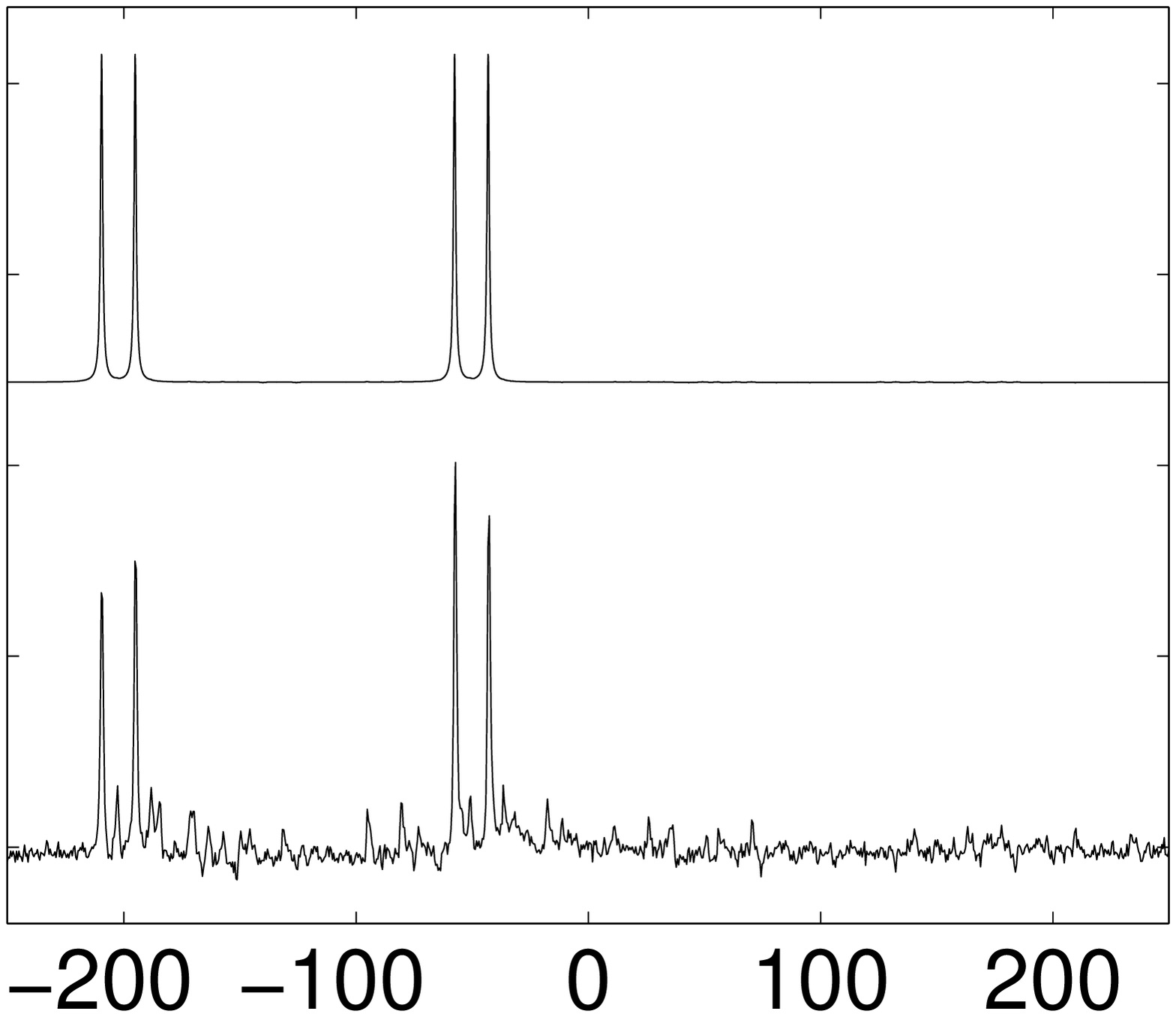}
\includegraphics*[width=4.8cm]{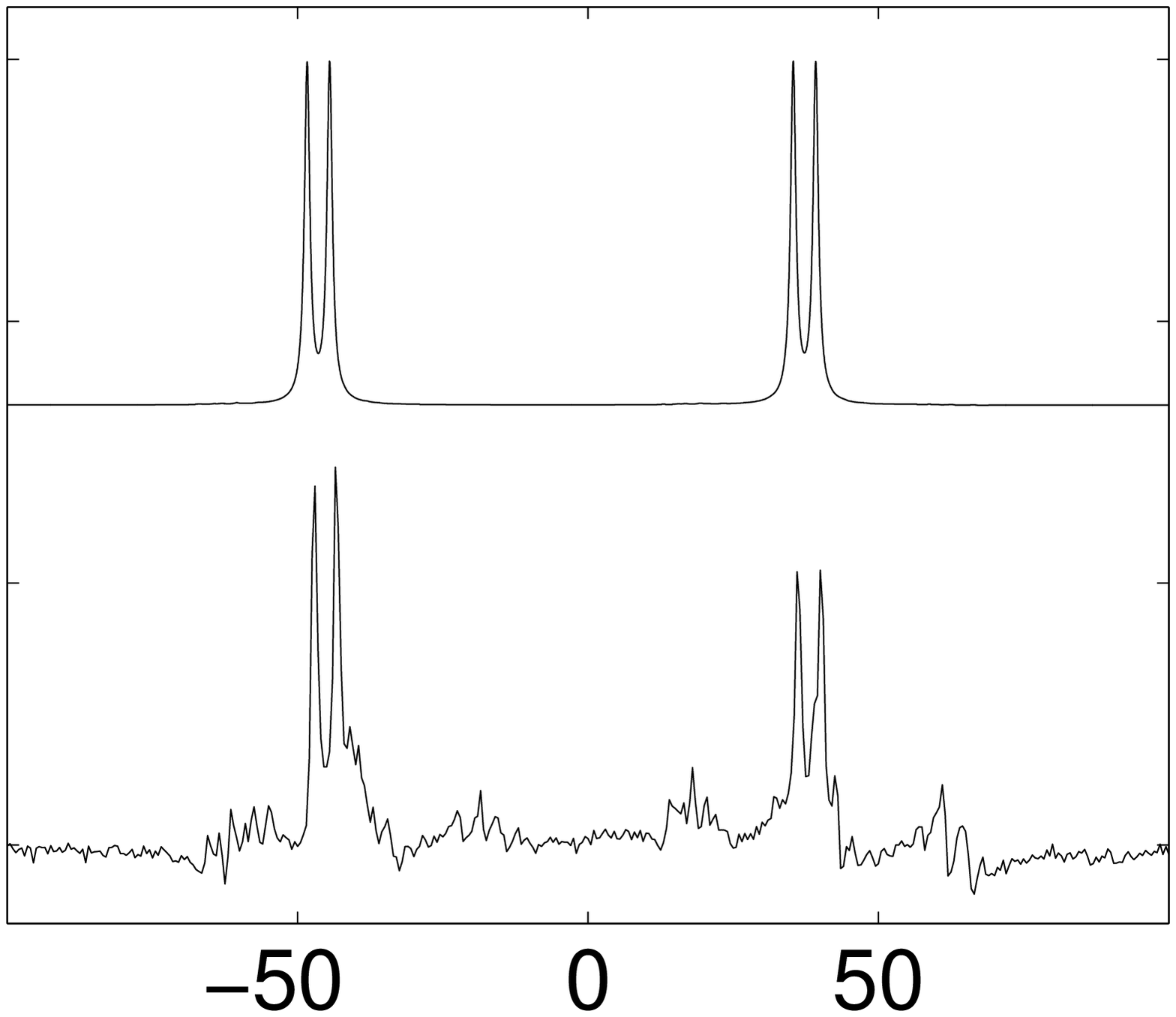}
\includegraphics*[width=4.8cm]{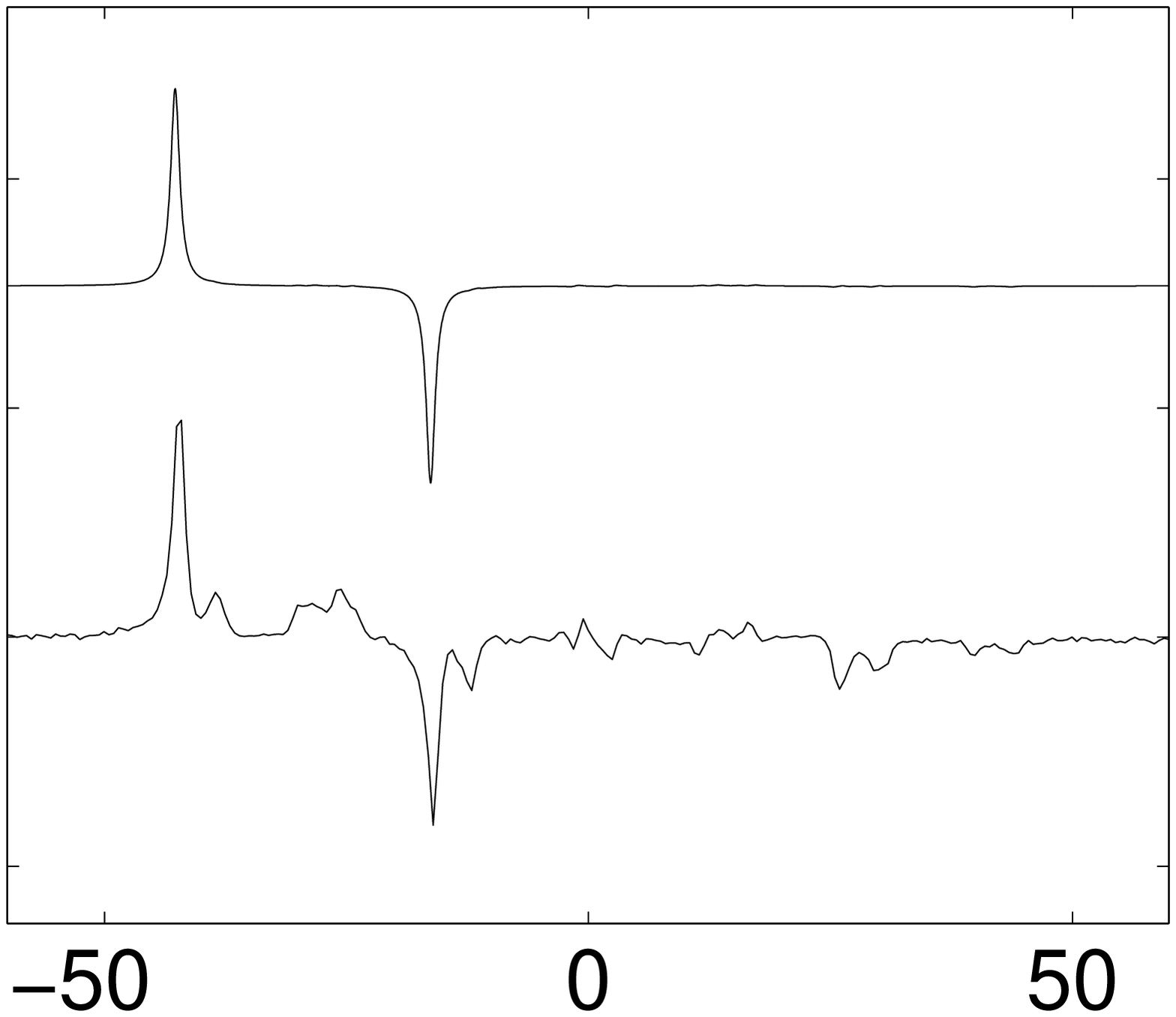}
\ecen
\vspace*{-2ex}
\caption{(Bottom) Experimentally measured and (Top) ideally expected 
spectra of spins 1, 2 and 3 after completion of the ``easy'' case of
Shor's algorithm ($a=11$). Positive and negative lines indicate that
the state of the spin is $\ket{0}$ and $\ket{1}$ respectively.}
\label{fig:shor_easy_i}
\end{figure}

The experimentally measured spectra upon completion of the ``easy''
case of Shor's algorithm are shown in Fig.~\ref{fig:shor_easy_i}.
Clearly, the lines of spins 1 and 2 are up, so qubits 1 and 2 are in
$\ket{0}$; qubit 3 is in a mixture of $\ket{0}$ and $\ket{1}$ as it
has positive and negative lines. With qubit 3 the most significant
qubit after the QFT \cite{Coppersmith94a}, the first register is thus
in a mixture of $\ket{000}$ and $\ket{100}$, or $\ket{0}$ and
$\ket{4}$ in decimal notation. The periodicity in the amplitude of
$\ket{x}$ is thus 4, and therefore $r = 8 / 4 = 2$. If we plug this in
in Eq.~\ref{eq:factors_15}, we obtain $\mbox{gcd} (11^{2/2} \pm 1, 15)
= 3,5$. The prime factors of 15 have thus been unambiguously derived
from the output spectra.

\vspace*{1ex}

\begin{figure}[h]
\hspace*{2.5cm} {\sf Spin 1} 
\hspace*{3.5cm} {\sf Spin 2}
\hspace*{3.5cm} {\sf Spin 3}
\vspace*{-2ex}
\begin{center}
\rotatebox{90}{\sf \small \hspace*{5ex} Expt. $\quad$ Ideal} 
\includegraphics*[width=4.8cm]{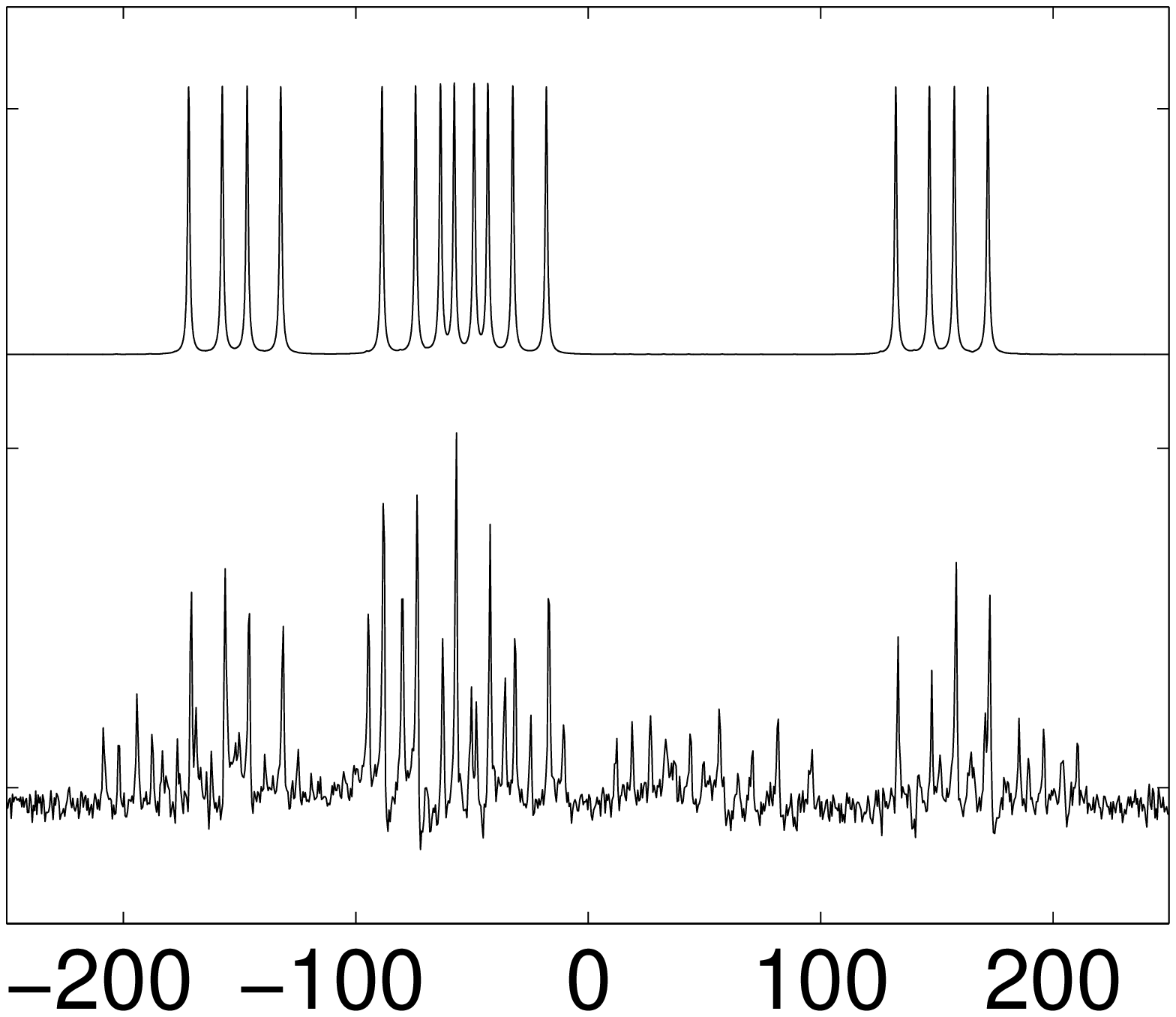}
\includegraphics*[width=4.8cm]{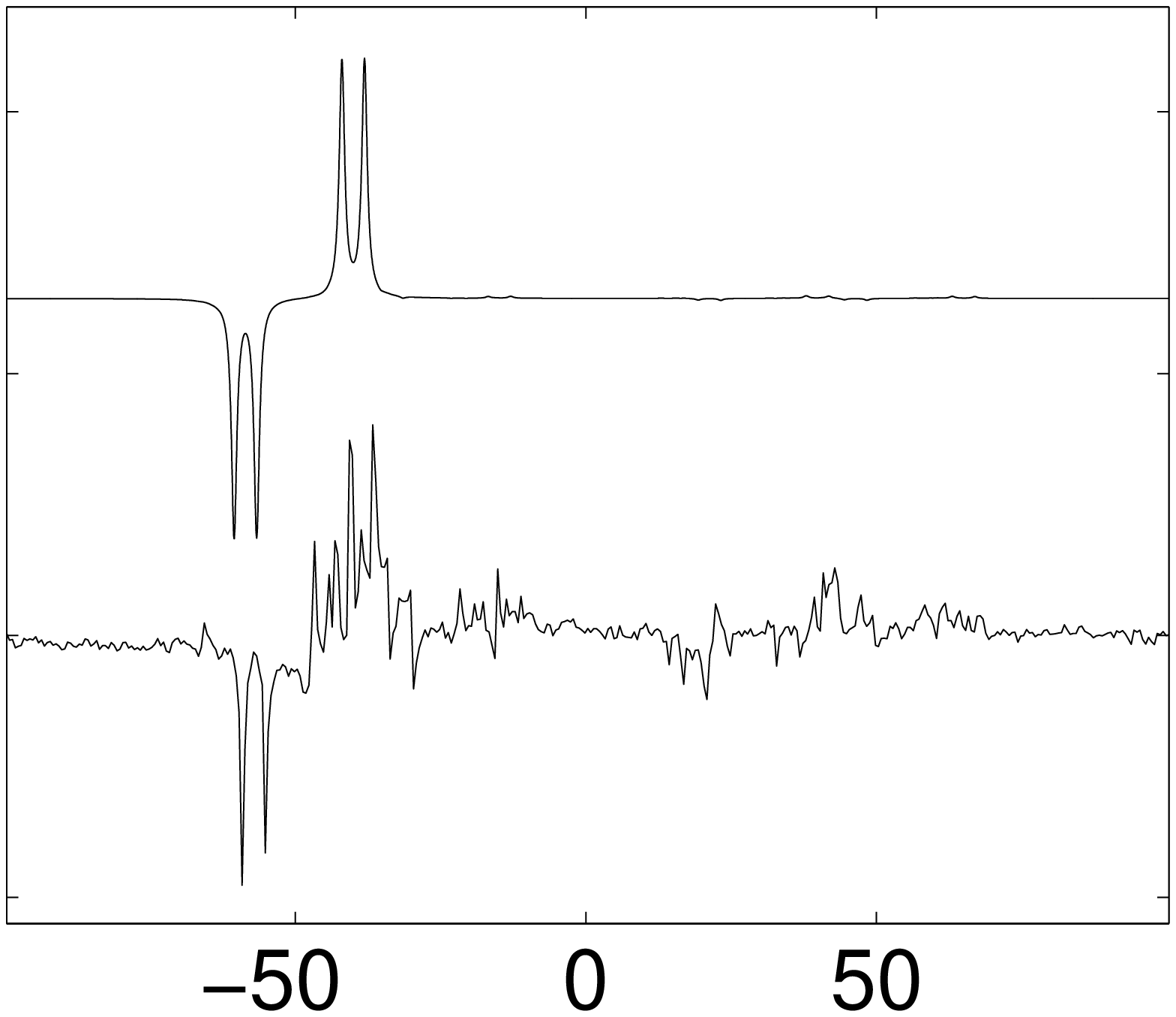}
\includegraphics*[width=4.8cm]{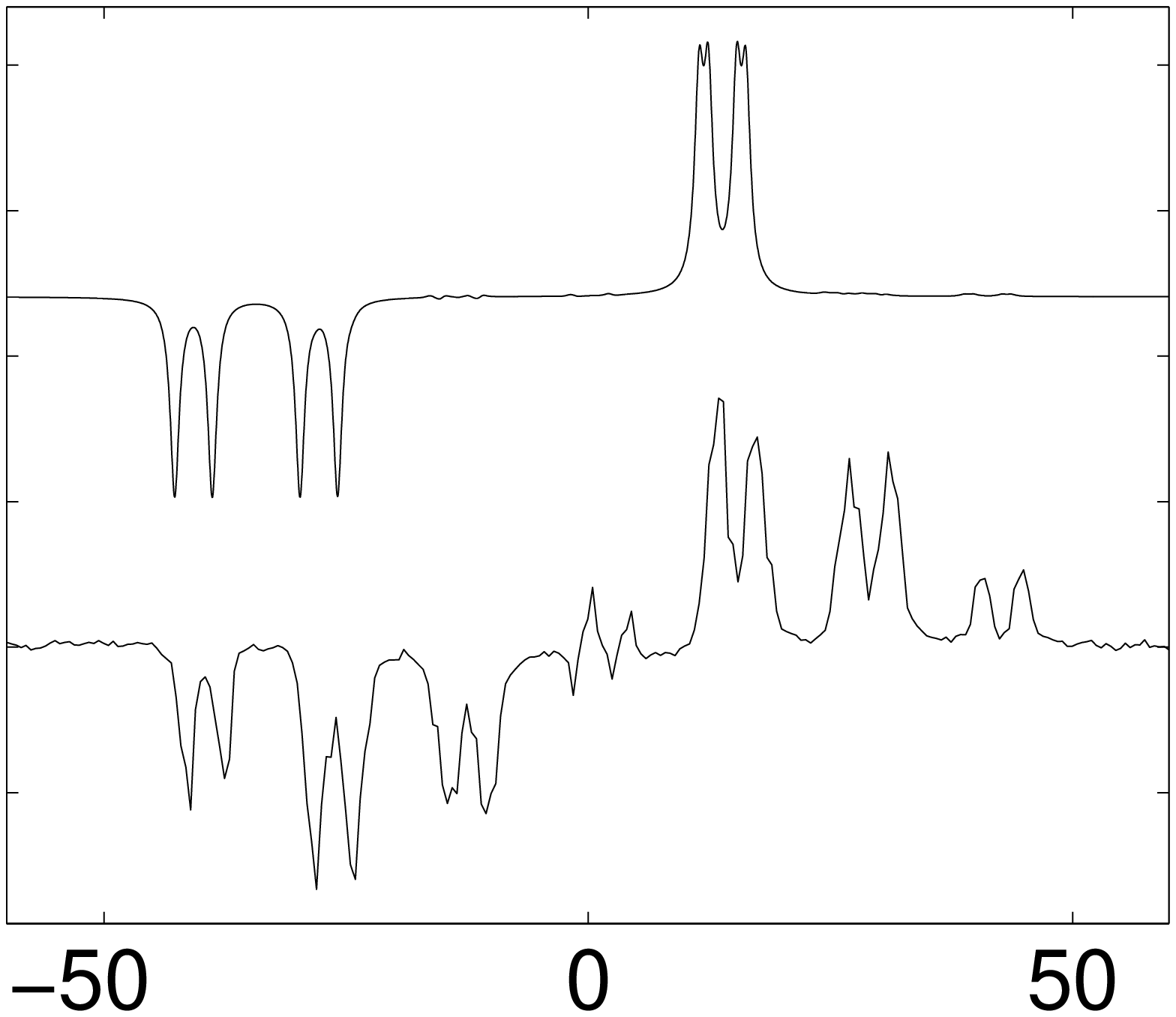}
\ecen
\vspace*{-2ex}
\caption{Similar to Fig.~\protect\ref{fig:shor_easy_i} but for the 
``difficult'' case ($a=7$).}
\label{fig:shor_hard_i}
\end{figure}

Similar spectra for the ``difficult'' case are shown in
Fig.~\ref{fig:shor_hard_i}. From the spectra, we conclude that qubit 1
is in $\ket{0}$, and qubits 2 and 3 are in a mixture of $\ket{0}$ and
$\ket{1}$. The register is thus in a mixture of $\ket{000}$,
$\ket{010}$, $\ket{100}$ and $\ket{110}$, which in decimal is
$\ket{0}$, $\ket{2}$, $\ket{4}$ and $\ket{6}$. The periodicity in the
amplitude of the first register is thus 2, and therefore $r = 8/2 =
4$. Plugging this in in Eq.~\ref{eq:factors_15} gives $\mbox{gcd}
(7^{4/2} \pm 1, 15) = 3,5$. Even after the very long pulse sequence
of the ``difficult'' case, the prime factors of 15 have thus
successfully been found using Shor's algorithm and a quantum computer.

\vspace*{1ex}

There clearly are substantial discrepancies between the measured
spectra and the ideally expected spectra, however, most notably for
the difficult case. The effective pure state spectra also exhibit
small non-idealities. We have worked long and hard trying to improve
the quality of the data, learned a lot in the process and made
substantial progress since the initial experiments. However, it seemed
that something fundamental was preventing us from getting cleaner data
than those of Figs.~\ref{fig:shor_easy_i} and~\ref{fig:shor_hard_i}.

We decided to attempt to model the effect of decoherence, even though
it isn't obvious from the spectra that decoherence would explain the
data. We simply wanted to understand what the effect of decoherence is
throughout the pulse sequences for Shor's algorithm.

\subsection{Decoherence model}

The NMR literature gives very detailed descriptions of decoherence in
coupled spin systems, going back to the ideas of Redfield
\cite{Redfield57a,Redfield65a} and worked out later by many others
\cite{Vold78a,Jeener82a,Ernst87a}. This so-called superoperator
formalism is very general and allows one to simulate decoherence in
nuclear spin systems starting from knowledge of internuclear
distances, chemical shift anisotropy tensors, and so forth, and of
correlations of these mechanism in time and space, which are governed
by molecular tumbling and diffusion rates. While such calculations are
in principle amenable to computer simulations, detailed knowledge about
the correlation functions and the other parameters is not always
available. Furthermore, a full superoperator description of relaxation
is exceedingly complex for the case of seven coupled spins: the
superoperator matrix would be of size $4^7 \times 4^7$, and thus have
268435456 degrees of freedom. This is more than most current computers
can store in memory, and simulation of decoherence in the course of a
sequence of 300 RF pulses appears out of the question with this
approach.

We therefore set out to construct a numerical model for decoherence
that is simple and workable, while still predictive.  The resulting
simplified decoherence model is in essence is an extension of the
Bloch equations \cite{Bloch46a}, an early phenomenological description
of nuclear spin relaxation in terms of just $T_2$ and $T_1$.  Such a
simplified description is justified only if each spin in a molecule
experiences a local magnetic field which randomly fluctuates in time
and there are negligble correlations between the local field
experienced by different nuclei.  It is not clear a priori that this
is the case in our system, but a simple model can at least serve to
give a first idea of the impact of decoherence during the execution of
Shor's algorithm. Comparison of the simulation results with the
experimental data will reveal the usefulness of this simplified model.

In order to make the model easily compatible with our matrix formalism
for unitary operations (section~\ref{qct:gates}), we chose to describe
decoherence in the operator sum representation or Kraus representation
\cite{Kraus83a}. The operator sum representation of both generalized
amplitude damping ($T_1$) and phase damping (related to $T_2$) has
been described in the quantum computing literature \cite{Nielsen00b}
(see also section~\ref{impl:coherence}), although only for single
spins, and for both processes separately.

We have devised and implemented an integrated model of phase damping
and generalized amplitude damping acting on seven coupled spins in
the course of arbitrary pulse sequences. We have been able to keep the
model workable by assuming that each spin decoheres
independently and by making the following observations:

\begin{enumerate}
\item Amplitude damping error operators acting on different spins 
commute.
\item Phase damping error operators acting on different spins commute.
\item Amplitude damping and phase damping commute with each other. 
This follows from the fact that the error operators $E_i$ for
amplitude damping (Eq.~\ref{eq:opsumrep_gen_ad}) commute with the
$E_i$ for phase damping (Eq.~\ref{eq:opsumrep_pd}) when applied to
$\sigma_x, \sigma_y, \sigma_z$ and $\sigma_I$ and thus also when they
act on arbitrary $\rho$.
\item Phase damping commutes with the ideal unitary evolution under 
${\cal H}$, as both processes are described by diagonal matrices.
\end{enumerate}

As a result, there is no need to simulate all these processes
simultaneously; they can be simulated one after the other, in any
order. This prevents an explosion of terms in the operator sum
representation, which would have been the case if cross-products of
all the $E_i$ had been required. However,

\begin{enumerate}
\item Amplitude damping does {\em not} commute with the ideal unitary 
evolution 
under ${\cal H}$.
\item Phase and amplitude damping do {\em not} commute with the ideal 
unitary evolution during RF pulses.
\end{enumerate}

In order to still maintain a workable model, we have nevertheless
treated these processes as if they did commute. This is not
necessarily a good approximation, but it does allow us to get
a first estimate of the effect of decoherence.

Concretely, we modeled a delay time of duration $t$ via $e^{- i {\cal
H} t/\hbar}$ followed by amplitude damping acting on spin 1 for a
duration $t$, then amplitude damping acting on spin 2 and so forth,
followed by phase damping acting for the same duration on each spin
one after the other. Similarly, a shaped pulse of duration $pw$ was
modeled via an ideal shaped pulse, preceded by amplitude damping and
phase damping, acting on each spin separately for a duration $pw$.
Thanks to these approximations, the simulation of the complete Shor
pulse sequence, including 36 temporal averaging sequences, takes
only a few minutes to run on four IBM {\sc power}3-II processors.  We
measured the characteristic amplitude and phase damping time constants
for each spin and plugged those values into the model (excerpts from
the simulation code are given in Appendix~\ref{app:model}).  The model
thus has no free parameters.

\begin{figure}[h]
\hspace*{2.5cm} {\sf Spin 1} 
\hspace*{3.5cm} {\sf Spin 2}
\hspace*{3.5cm} {\sf Spin 3}
\vspace*{-2ex}
\begin{center}
\rotatebox{90}{\sf \small \hspace*{5ex} Expt. $\quad$ Simul.} 
\includegraphics*[width=4.8cm]{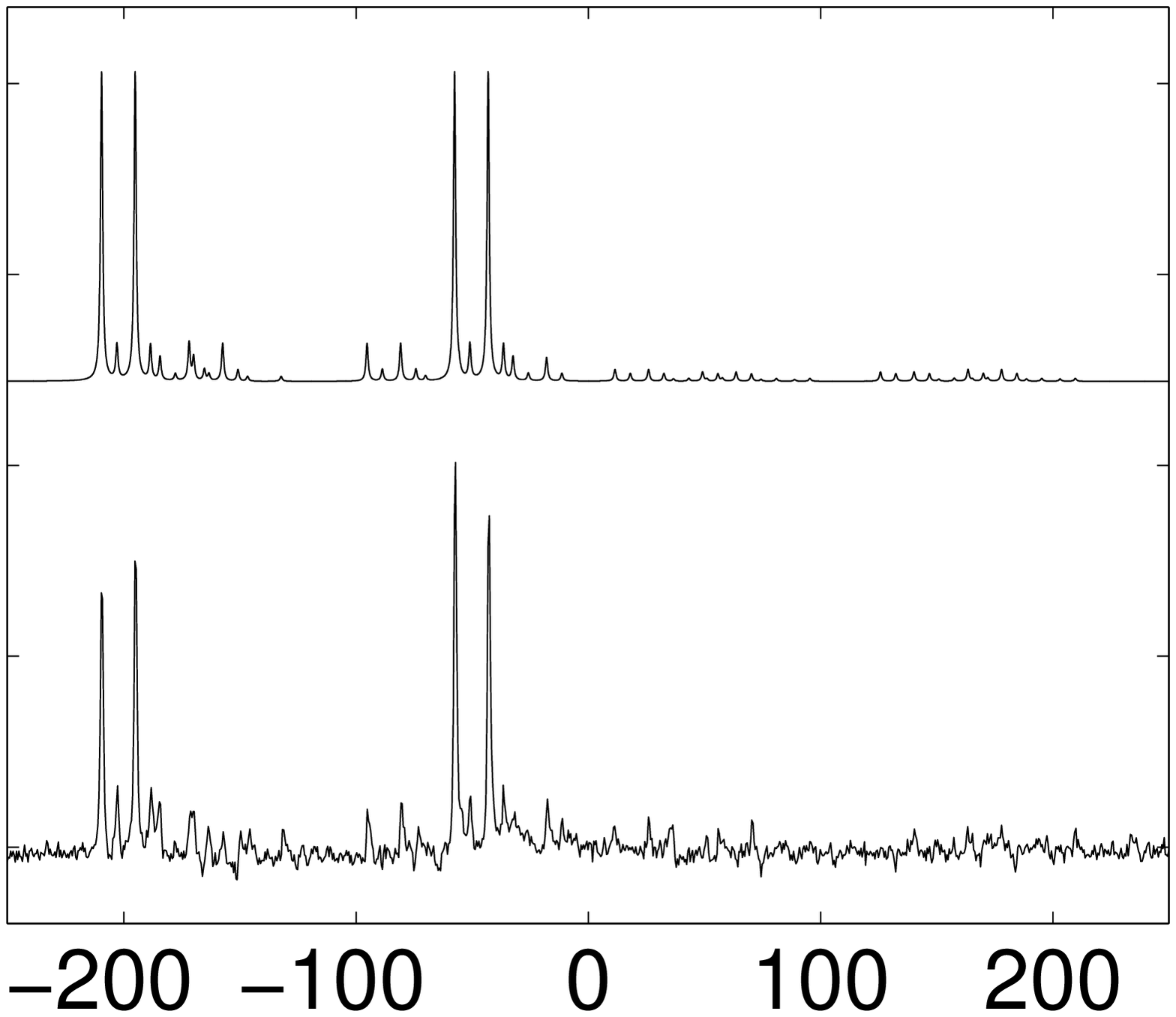}
\includegraphics*[width=4.8cm]{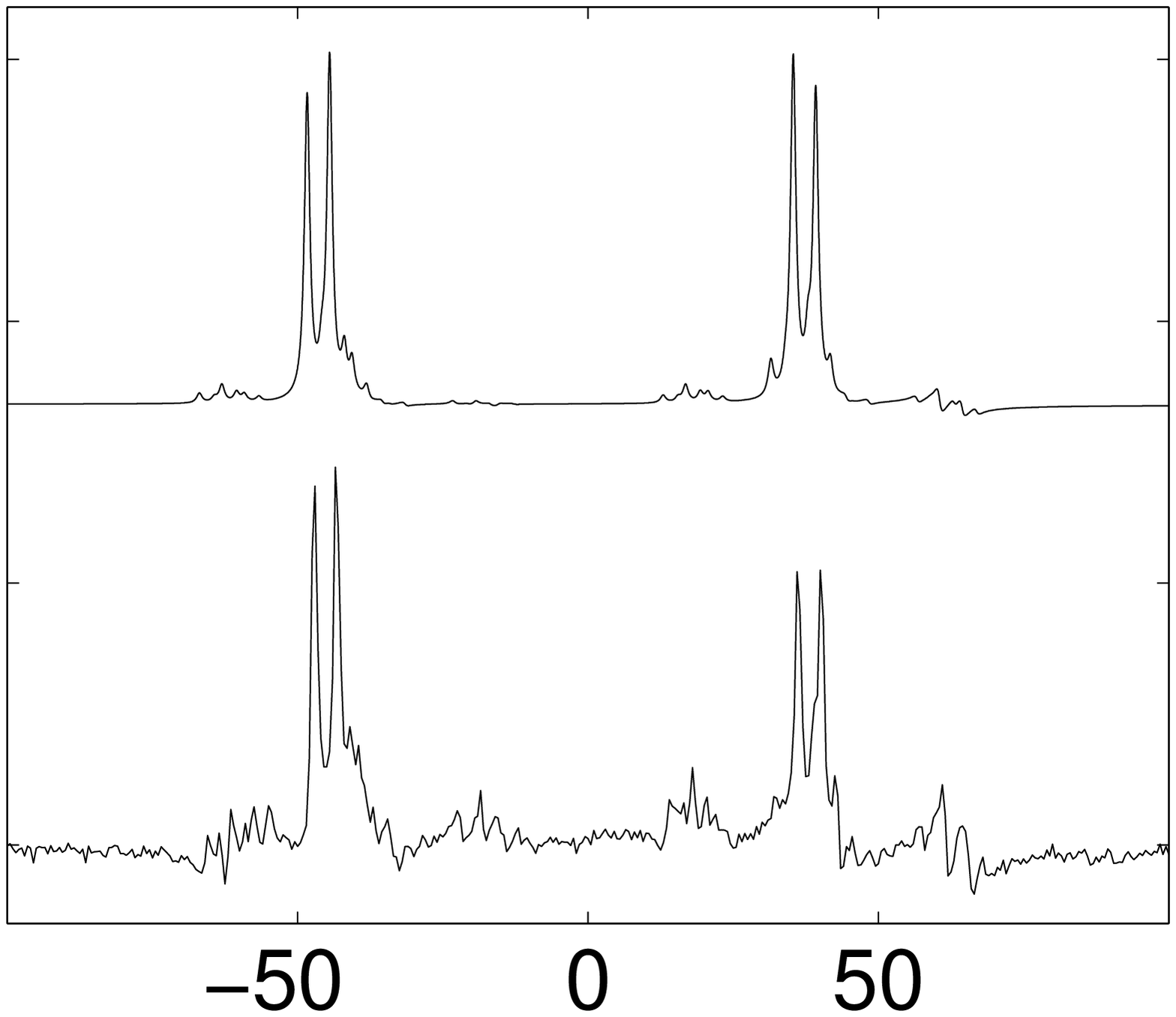}
\includegraphics*[width=4.8cm]{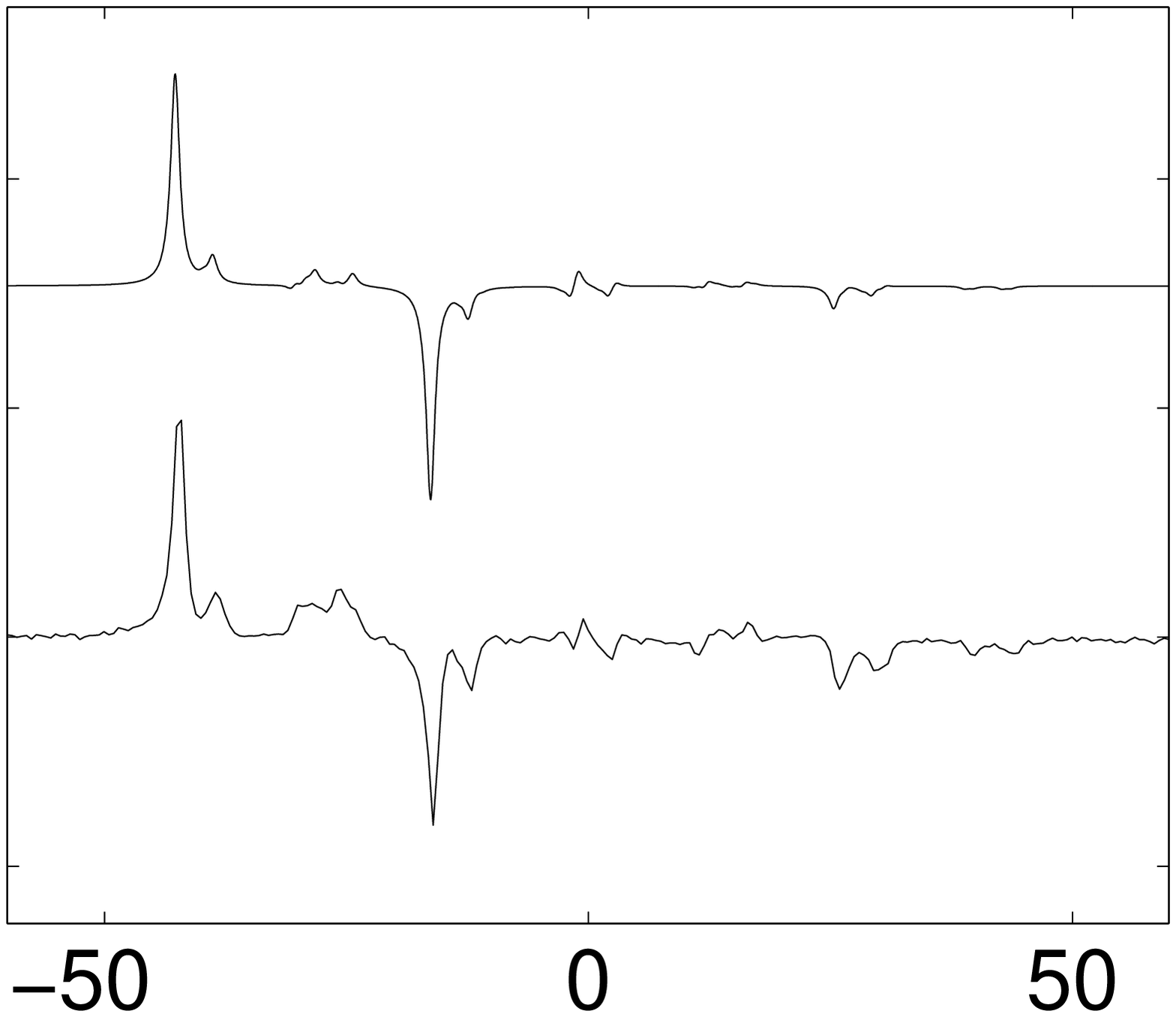}
\ecen
\vspace*{-2ex}
\caption{Comparison between (bottom) the experimentally measured 
spectra, which are the same as in Fig.~\ref{fig:shor_easy_i}, and
(top) simulated spectra based on the decoherence model, for the
``easy'' case ($a=11$).}
\label{fig:shor_easy_d}
\end{figure}

\begin{figure}[h]
\hspace*{2.5cm} {\sf Spin 1} 
\hspace*{3.5cm} {\sf Spin 2}
\hspace*{3.5cm} {\sf Spin 3}
\vspace*{-2ex}
\begin{center}
\rotatebox{90}{\sf \small \hspace*{5ex} Expt. $\quad$ Simul.} 
\includegraphics*[width=4.8cm]{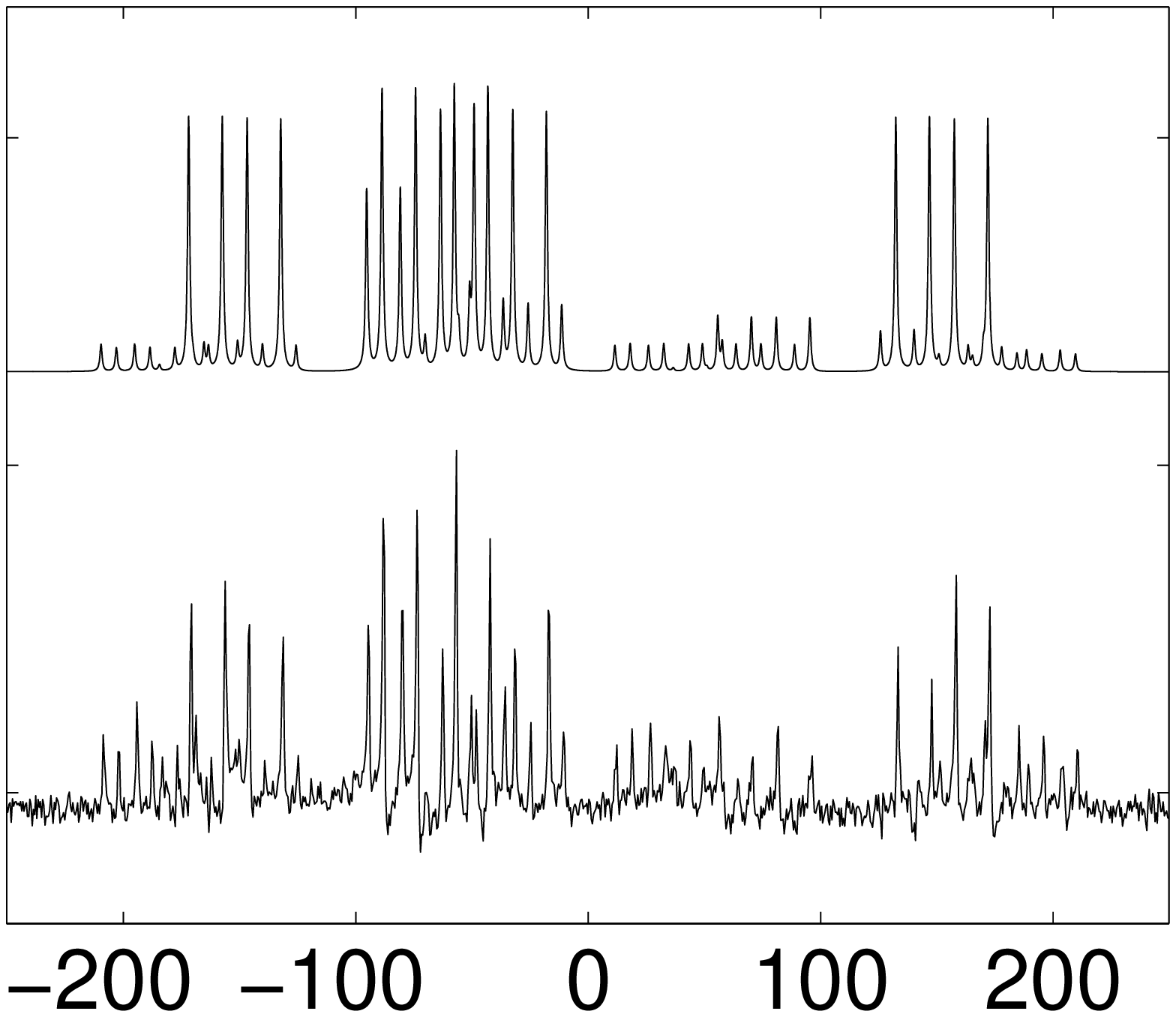}
\includegraphics*[width=4.8cm]{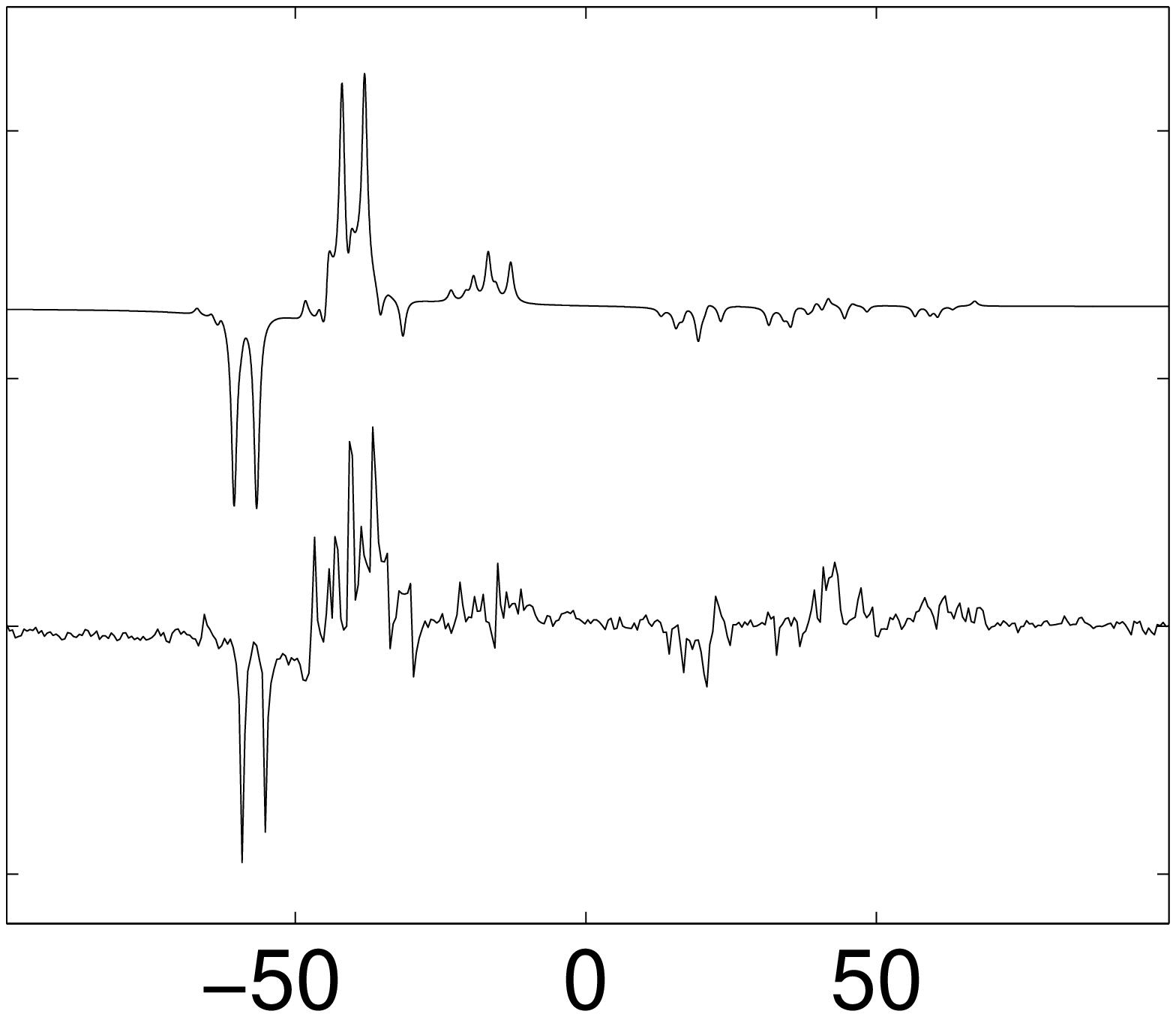}
\includegraphics*[width=4.8cm]{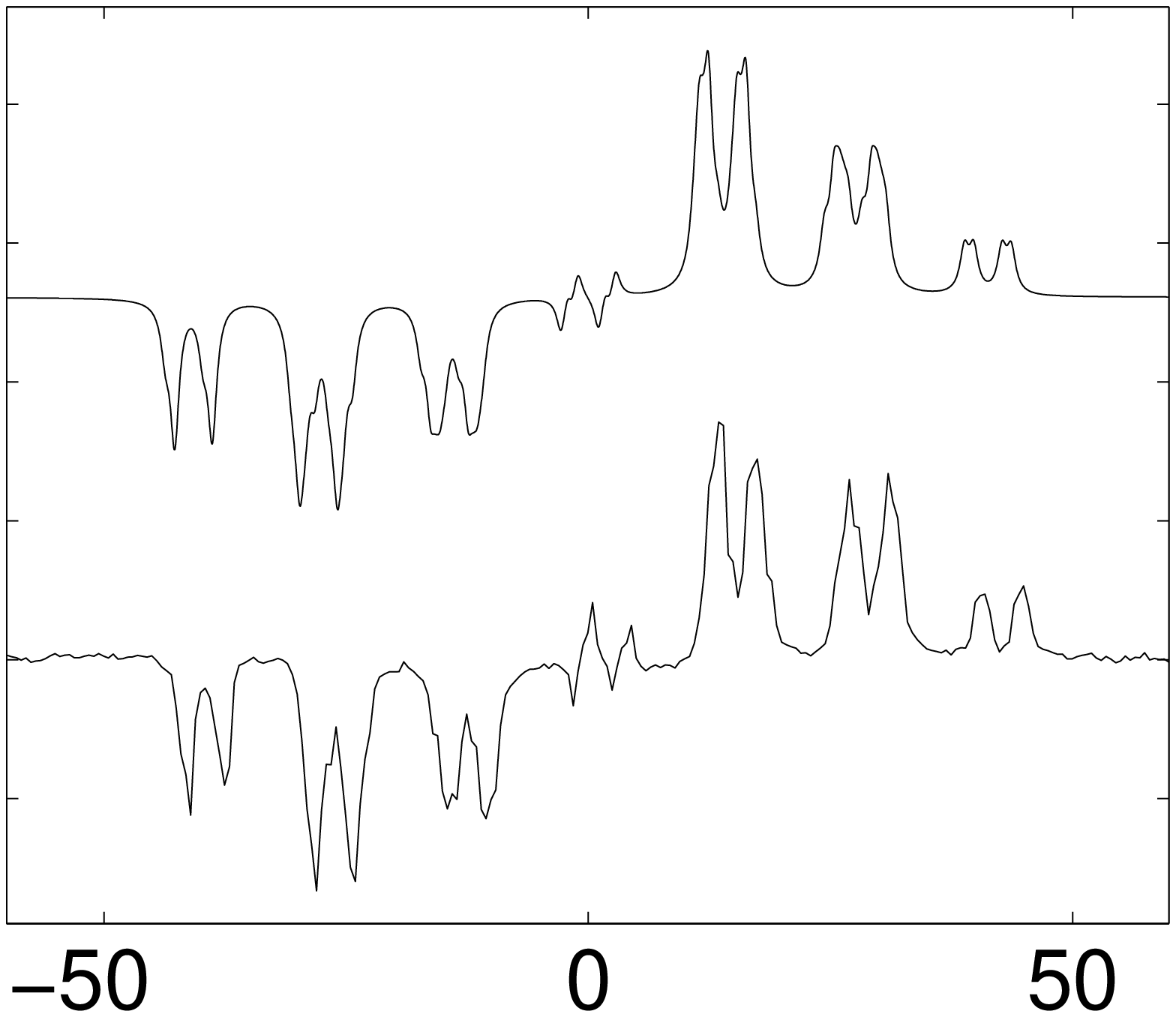}
\ecen
\vspace*{-2ex}
\caption{Similar to Fig.~\protect\ref{fig:shor_easy_d} but for the 
``hard'' case ($a=7$).}
\label{fig:shor_hard_d}
\end{figure}

The output spectra obtained from the simulation are shown in
Figs.~\ref{fig:shor_easy_d} and~\ref{fig:shor_hard_d} for the easy and
difficult case respectively, along with the experimental output
spectra.  In both cases, the model reproduced the main unexpected
observed non-idealities in the data in a remarkably convincing manner.

\vspace*{1ex}

\vspace*{1ex}

The good agreement between the spectra predicted by the relaxation
model and the experimental spectra suggests that the assumptions
underlying the model are valid, at least to a reasonable degree.  We
attribute the remaining discrepancies between the data and the
simulations to the approximations made in the model as well as to
experimental imperfections such as RF inhomogeneity, imperfect
calibrations and incomplete unwinding of coupled evolution during the
pulses.

\subsection{Discussion}

The prime factors of fifteen can be deduced from the output spectra,
both for the easy and the difficult case, demonstrating the proper
operation of the factoring algorithm. This is in itself really
remarkable.

The unexpected non-idealities in the data are well reproduced by a
simple numeric model that incorporates the effect of decoherence. This
is the first NMR quantum computing experiment in which decoherence was
the dominant source of errors; the demands of Shor's algorithm are
clearly pushing the limits of the current molecule, despite its
exceptionally good properties.  Certainly, the predictive but workable
decoherence model for multiple coupled spins is a valuable tool to
assess the feasability of future NMR quantum computing experiments.

Finally, the good agreement between the measured and simulated spectra
suggests that the degree of unitary control in the experiment was very
high, which bodes well for related proposed implementations of quantum
computers~\cite{Kane98a,Loss98a}.

%%%%%%%%%%%%%%%%%%%%%%%%%%%%%%%%%%%%%%%%%%%%%%%%%%%%%%%%%%%%%%%%%%%%

\section{Summary}

The sequence of successful experiments presented in this chapter
clearly demonstrates that nuclear spins in liquid solution represent a
beautiful playground in which to explore quantum computing
experimentally. The main results of these experiments are

\begin{itemize}
\item experimental proof of principle of quantum computing --- the 
solution of mathematical problems in fewer steps than is possible
with any classical machine,
\item experimental proof of principle of quantum error detection ---
coding gives a first order improvement in the conditional fidelity,
\item an explicit demonstration of zero temperature dynamics with 
room temperature nuclear spins,
\item the observation of a surprisingly large degree of cancellation 
of systematic errors --- the concatenation of a record 280 two-qubit
gates on three spins,
\item a demonstration that liquid crystal solvents can be used for 
NMR quantum computing --- such solvents may allow more gates within
the coherence time,
\item the demonstration of coherent control over up to seven coupled 
nuclear spins over the course of tens of two-qubit gates,
\item the demonstration of a workable yet predictive model of 
decoherence for seven coupled nuclear spins.
\end{itemize}

The main experimental challenges we had to address in the course of
these experiments include
\begin{itemize}
\item the identification and synthesis of suitable molecules,
\item the exponential overhead in the creation of effective pure 
states,
\item the limitations of a four-channel spectrometer,
\item "turning off" undesired terms in the Hamiltonian, which is 
especially tricky if they don't commute with the desired terms,
\item developing predictive numerical models of unitary and 
non-unitary spin dynamics.
\end{itemize}

Despite the successful resolution of these challenges in our
experiments, liquid NMR quantum computers are difficult to scale,
in particular because of the exponential overhead in the input state
preparation. It also appears unlikely that the accuracy threshold for
fault-tolerant quantum computation can be reached, except perhaps for
very small molecules.

Nevertheless, full quantum algorithms with 10-15 spins may be
possible, and interesting demonstrations of coherent control over
several dozens of spins may be possible using thermal input states.
Certainly, nuclear spins in liquid solution will continue to provide
an exciting and accessible avenue to study quantum computation
experimentally.

%% file: concl.tex
    \chapter{Conclusions}

Quantum computation has now entered the realm of experimental reality:
simple quantum computers based on nuclear spins in molecules have been
used to solve problems in fewer steps than is possible with any
classical device. We have continuously been pushing the state of the
art in NMR quantum computing by implementing meaningful quantum
algorithms on two, then three, later five and finally seven qubits. In
these experiments, we have not been content with just the simplest
quantum protocols requiring only nearest neighbour couplings, but
instead chose realistic experiments which really put the quantum
computers to test.

In order to make this possible, special molecules were synthesized and
a large number of experimental techniques have been developed for
state initialization, coherent control and read-out. These include
more efficient and effective temporal averaging schemes for state
preparation, methods to reduce cross-talk during spin-selective shaped
pulses, methods to reduce cross-talk between shaped pulses on
different spins, the uncoupling frame for automatic refocusing within
a subspace, pulse sequence simplification procedures at various
levels, software rotating reference frames, refocusing
schemes to remove undesired terms in the Hamiltonian, ways to
interpret ensemble averaged measurements, and readout schemes based on
the multiplet fine structure.

We have gained a good understanding of the main sources of
imperfections in the experiments by developing a set of practical and
predictive simulation tools which allow us to model both unitary and
decoherence processes for molecules with multiple coupled nuclear
spins. In the early heteronuclear experiments, RF field inhomogeneity
was the dominant source of errors. In the later homonuclear
experiments, which require longer pulses, coupled evolution during the
RF pulses became the most significant source of errors. Other sources
of errors included imperfect calibrations, incomplete frequency drift
compensation and limited signal-to-noise. All these imperfections
arise from technological rather than from fundamental issues, and we
have demonstrated that such systematic errors can be cancelled out to
a great extent. Even though intrinsic decoherence played a role in all
the experiments, only in the last experiment did decoherence become
the dominant source of errors.

There exist well-understood scaling limitations for nuclear spin
quantum computers starting off in thermal equilibrium at room
temperature. Most importantly, the probability that all spins in a
quantum computer with $n$ spins start off in the desired ground state
decreases as $n/2^n$. While a subset of the spins can be efficiently
cooled down algorithmically, the overhead for this procedure is
impractically large unless substantial hyperpolarization of the
nuclear spins can be achieved by other means.  An additional problem
is that the coherence time of nuclear spins in molecules tends to
shorten as the number of spins increases.

Nevertheless, I am convinced that solution NMR quantum computing will
continue to help address many open problems and unanswered questions
(and raise many more). Those questions include very fundamental ones,
such as ``where does the power of quantum computing come from?''.  But
as quantum computation has become an experimental reality, many
important open problems are {\em practical} ones. (1) How can we
remove the effect of undesired terms in the Hamiltonian via an
approach that is both general and practical? (2) How do we arrange the
sequence of operations in a systematic way, so as to maximize the
cancellation of systematic errors? (3) How do we create and validate
practical and predictive decoherence models?

The work presented in this thesis work constitutes a first step
towards answering these important questions, but it represents by no
means an endpoint. The significance of the practical open problems
will only increase as other, perhaps more scalable implementations of
quantum computers reach the stage of realizing actual quantum
computations. Similarly, I hope that the significance of the
techniques and concepts we developed for NMR quantum computing will
grow, as they find useful application in other realizations of quantum
computers.

Both the questions and the answers which stemmed from the work on NMR
quantum computers have tremendously increased our understanding of
what it takes to build a quantum computer.  Still, the big question we
asked in the beginning remains open: {\em can} we build a practical
quantum computer?

This question can only be answered by measuring the coherence time and
testing techniques for initialization, control and read-out of the
qubit states in a variety of potential embodiments of quantum
computers. The fundamental requirement for any implementation is that
the accuracy threshold for fault-tolerant quantum computation be
reached\footnote{In order to perform meaningful quantum computations
without using quantum error correction, the probability of error
should be even lower, so the accuracy threshold is a reasonable upper
bound for the error rate.}. Specifically, the probability of error per
elementary logic gate should be less than about $0.01~\%$, in the most
optimistic estimates at present.  It is clear that reaching the
accuracy threshold represents a formidable challenge; it requires not
only coherence times long compared to the gate times, but also
extremely accurate unitary control over the qubits. Nevertheless,
$0.01~\%$ doesn't seem a priori impossible, and in the spirit of
Feynman, we say:

\begin{quote}
{\em We know of no laws of physics that prohibit practical
quantum computers --- we just haven't gotten around to building one
$\ldots$}
\end{quote}

Therefore, while I don't believe that horses can be made to fly, I do
think that practical quantum computers may one day be built. It is my
hope and expectation that this thesis work has contributed to the
realization of this wonderful challenge, which holds such a great
promise.

%% file: app.tex
%\chapter{Appendix}

\chapter{Numerical model}
%\section{Numerical model}
\label{app:model}

The MATLAB model for simulation of the unitary and non-unitary
processes in the course of a pulse sequence contains four primitives:

\begin{enumerate}
\item \verb+d.m+: free evolution under the Hamiltonian
\item \verb+X.m+, \verb+Y.m+ and \verb+Z.m+: ideal single-spin rotations
\item \verb+gad7.m+: generalized amplitude damping (GAD)
\item \verb+pd7.m+: phase damping (PD)
\end{enumerate}

The last two programs, which model non-unitary processes, act directly
on a density matrix. The programs for unitary processes act on density
matrices via the program \verb+rho.m+. The simulation programs
require that the number of qubits \verb+nq+ be declared in advance,
and that the Hamiltonian and the Pauli matrices be set up by calling
\verb+def.m+.

In the next subsections, we shall give excerpts from the MATLAB
code of \verb+def.m+, \verb+rho.m+, \verb+d.m+, \verb+X.m+,
\verb+gad7.m+ and \verb+pd7.m+ and two helper programs.
Finally, we will give an excerpt from a pulse sequence which calls
these primitives.

\section{Set up the Hamiltonian and Pauli matrices}
\begin{singlespace}
\begin{small}
\begin{verbatim}
% File:   def.m
% Date:   1997
% Author: Lieven Vandersypen <lieven@snow.stanford.edu>
% Declares variables used in simulation programs
% Use: declare nq (number of qubits), then type def

global nqubits Si Sx Sy Sz H
nqubits=nq; % number of qubits

% Pauli matrices
Si=eye(2); Sx=[0 1 ; 1 0]; Sy=[0 -i; i 0]; Sz=[1 0 ;0 -1];

switch nqubits
...
case 7

% Pauli matrices
Sziiiiii=mykron(Sz,Si,Si,Si,Si,Si,Si);
...
Siiiiiiz=mykron(Si,Si,Si,Si,Si,Si,Sz);

% J-coupling strengths
% the labeling 1 through 7 is in order of frequency

J12=-114; J23=80;    J34=2.5;   J45=41.6;  J56=1;     J67=69;
J13=25;   J24=2;     J35=3.9;   J46=19.4;  J57=-13.5;
J14=6.6;  J25=13;    J36=18.5;  J47=60;
J15=14.5; J26=54;    J37=-3.8;
J16=-221; J27=-5.7;
J17=38;

% Hamiltonian (in the multiply rotating frame)

H=2*pi*J12*Sziiiiii/2*Siziiiii/2 + 2*pi*J13*Sziiiiii/2*Siiziiii/2 +...
2*pi*J14*Sziiiiii/2*Siiiziii/2 + 2*pi*J15*Sziiiiii/2*Siiiizii/2 + ...
2*pi*J16*Sziiiiii/2*Siiiiizi/2 + 2*pi*J17*Sziiiiii/2*Siiiiiiz/2 + ...
2*pi*J23*Siziiiii/2*Siiziiii/2 + 2*pi*J24*Siziiiii/2*Siiiziii/2 + ...
2*pi*J25*Siziiiii/2*Siiiizii/2 + 2*pi*J26*Siziiiii/2*Siiiiizi/2 + ...
2*pi*J27*Siziiiii/2*Siiiiiiz/2 + 2*pi*J34*Siiziiii/2*Siiiziii/2 + ...
2*pi*J35*Siiziiii/2*Siiiizii/2 + 2*pi*J36*Siiziiii/2*Siiiiizi/2 + ...
2*pi*J37*Siiziiii/2*Siiiiiiz/2 + 2*pi*J45*Siiiziii/2*Siiiizii/2 + ...
2*pi*J46*Siiiziii/2*Siiiiizi/2 + 2*pi*J47*Siiiziii/2*Siiiiiiz/2 + ...
2*pi*J56*Siiiizii/2*Siiiiizi/2 + 2*pi*J57*Siiiizii/2*Siiiiiiz/2 + ...
2*pi*J67*Siiiiizi/2*Siiiiiiz/2;

end
\end{verbatim}
\end{small}
\end{singlespace}

%%%%%%%%%%%%%%%%%%%%%%%%%%%%%%%%%%%%%%%%%%%%%%%%%%%%%%%%%%%%%%%%%%%%%

\section{Action unitary operator on density matrix}
\begin{singlespace}
\begin{small}
\begin{verbatim}
% rho.m
% Lieven Vandersypen <lieven@snow.stanford.edu>
%
% calculates the final density matrix rf for an initial density 
% matrix ri and a unitary operation U acting on ri

function rf = rf(ri,U)
global nqubits
rf=U*ri*U';
\end{verbatim}
\end{small}
\end{singlespace}

%%%%%%%%%%%%%%%%%%%%%%%%%%%%%%%%%%%%%%%%%%%%%%%%%%%%%%%%%%%%%%%%%%%%%

\section{Time evolution under the Hamiltonian}
\begin{singlespace}
\begin{small}
\begin{verbatim}
% File: d.m
% Author: Lieven Vandersypen <lieven@snow.stanford.edu>
% 
% d(t) simulates a free evolution period of t seconds
 
function R=d(t)
global H
R=expm(-i*H*t);
end
\end{verbatim}
\end{small}
\end{singlespace}

%%%%%%%%%%%%%%%%%%%%%%%%%%%%%%%%%%%%%%%%%%%%%%%%%%%%%%%%%%%%%%%%%%%%%
%\clearpage
\section{Single-spin rotations}
Ideal rotations about $\hat{x}$, $\hat{y}$ and $\hat{z}$ are simulated
by the programs \verb+X.m+, \verb+Y.m+ and \verb+Z.m+. We give here
the code for \verb+X.m+ only, as the other programs are analogous.
\begin{singlespace}
\begin{small}
\begin{verbatim}
% File: X.m
% Date: 04-Oct-99
% Author: Lieven Vandersypen <lieven@snow.stanford.edu>
%
% Usage: X(spinname, angle)
% Rotation of spin 'spinname' about X over angle*pi/2 
% (right hand rule)
% example: X(2,3) rotates spin 2 about X over 270 degrees

function X=X(spinname,angle)

global nqubits

if nargin == 1
  if nqubits == 1
    angle=spinname;     spinname=1;
  else
    error('X.m requires two arguments if nqubits > 1 !')
  end
end

operator=expm(-i*angle*pi/2*[0 1;1 0]/2);  % calculate 1-spin operator
X=gop(spinname,operator); % turns 1-spin operator into n-spin operator
\end{verbatim}
\end{small}
\end{singlespace}

%%%%%%%%%%%%%%%%%%%%%%%%%%%%%%%%%%%%%%%%%%%%%%%%%%%%%%%%%%%%%%%%%%%%%
%\clearpage
\section{Generalized amplitude damping}
\begin{singlespace}
\begin{small}
\begin{verbatim}
% file: gad7.m
% April 2001
% Lieven Vandersypen <lieven@snow.stanford.edu>
%
% simulate generalized amplitude damping (GAD), 7 spins
% model assumes no correlation of GAD on different spins
%
% Usage rout=gad(rin,p,t,ratio)
%
%   rin   initial density matrix
%   p     equilibrium polarization
%   t     duration for which GAD acts
%   ratio set this to 1 to simulate GAD, set this to say 1e9 to
%         simulate the same sequence without GAD
%   rout  final density matrix

function rout=gad(rin,p,t,ratio)

T1_1=ratio(1)*5.0;    % the labeling is in order of frequency
T1_2=ratio(1)*10.0;
T1_3=ratio(1)*13.7;
T1_4=ratio(1)*2.8;
T1_5=ratio(1)*3.0;
T1_6=ratio(1)*31.6;
T1_7=ratio(1)*45.4;

p1=p;                     % fluorine polarization
p2=0.5+(p-0.5)*1.25/4.7;  % carbon polarization

% T1 effects on spin 1
g=1-exp(-t/T1_1);
E{1}=sqrt(p1)*[1 0; 0 sqrt(1-g)];   E{2}=sqrt(p1)*[0 sqrt(g);0 0];
E{3}=sqrt(1-p1)*[sqrt(1-g) 0;0 1];  E{4}=sqrt(1-p1)*[0 0 ;sqrt(g) 0];
r1=0;
for k=1:4
r1 = r1 + mykron(E{k},eye(2),eye(2),eye(2),eye(2),eye(2),eye(2))*...
rin*mykron(E{k},eye(2),eye(2),eye(2),eye(2),eye(2),eye(2))':
end

...

% T1 effects on spin 7
g=1-exp(-t/T1_7);
E{1}=sqrt(p2)*[1 0; 0 sqrt(1-g)];   E{2}=sqrt(p2)*[0 sqrt(g);0 0];
E{3}=sqrt(1-p2)*[sqrt(1-g) 0;0 1];  E{4}=sqrt(1-p2)*[0 0 ;sqrt(g) 0];
r7=0;
for k=1:4
r7 = r7 + mykron(eye(2),eye(2),eye(2),eye(2),eye(2),eye(2),E{k})*...
r6*mykron(eye(2),eye(2),eye(2),eye(2),eye(2),eye(2),E{k})';
end

rout=r7;
\end{verbatim}
\end{small}
\end{singlespace}

%%%%%%%%%%%%%%%%%%%%%%%%%%%%%%%%%%%%%%%%%%%%%%%%%%%%%%%%%%%%%%%%%%%%%
\section{Phase damping}
\begin{singlespace}
\begin{small}
\begin{verbatim}
% file: pd7.m
% April 2001
% Lieven Vandersypen <lieven@snow.stanford.edu>
%
% simulates phase damping (PD), 7 spins
% model assumes no correlation of PD on different spins
%
% Usage rout=pd(rin,t,ratio)
%
%   rin   initial density matrix
%   t     duration for which PD acts
%   ratio set this to 1 to simulate PD, set this to say 1e9 to
%         simulate the same sequence without PD
%   rout  final density matrix

function rout=pd(rin,t,ratio)

T2_1=ratio(2)*1.3;    % the labeling is in order of frequency
T2_2=ratio(2)*1.7;
T2_3=ratio(2)*1.8;
T2_4=ratio(2)*1.6;
T2_5=ratio(2)*1.5;
T2_6=ratio(2)*2.0;
T2_7=ratio(2)*2.0;

% T2 effects on spin 1
g=(1+exp(-t/T2_1))/2;
E{1}=sqrt(g)*[1 0; 0 1]; E{2}=sqrt(1-g)*[1 0;0 -1];
r1=0;
for k=1:2
r1 = r1 + mykron(E{k},eye(2),eye(2),eye(2),eye(2),eye(2),eye(2))*...
rin*mykron(E{k},eye(2),eye(2),eye(2),eye(2),eye(2),eye(2))';

end

...

% T2 effects on spin 7
g=(1+exp(-t/T2_7))/2;
E{1}=sqrt(g)*[1 0; 0 1];   E{2}=sqrt(1-g)*[1 0;0 -1];
r7=0;
for k=1:2
r7 = r7 + mykron(eye(2),eye(2),eye(2),eye(2),eye(2),eye(2),E{k})*...
r6*mykron(eye(2),eye(2),eye(2),eye(2),eye(2),eye(2),E{k})';

end

rout=r7;
\end{verbatim}
\end{small}
\end{singlespace}

%%%%%%%%%%%%%%%%%%%%%%%%%%%%%%%%%%%%%%%%%%%%%%%%%%%%%%%%%%%%%%%%%%%%%
%\clearpage
\section{Helper programs}
\begin{singlespace}
\begin{small}
\begin{verbatim}
% File: gop.m   (generalized operator)   
% Date: 08-Oct-99
% Author: Lieven Vandersypen <lieven@snow.stanford.edu>
%
% Usage: gop(s,U)
% In:   single-qubit unitary operator U, qubit name s
% Out:  n-spin unitary operator which acts on qubit s with U and 
%       trivially on the remaining qubits

function gop=gop(s,U)

global nqubits

goplocal=U;

for position=1:(s-1)
   goplocal=kron(eye(2),goplocal);
end
for position=s+1:nqubits
   goplocal=kron(goplocal,eye(2));
end
gop=goplocal;
\end{verbatim}
\end{small}
\end{singlespace}

\begin{singlespace}
\begin{small}
\begin{verbatim}
% File:   mykron.m
% Date:   17-Aug-98
% Author: I. Chuang <ike@isl.stanford.edu>
%
% kronecker product function which accepts multiple arguments.

function out = mykron(ma,mb,varargin)

if (length(varargin) == 0)
  out = kron(ma,mb);
  return;
else
  out = mykron(kron(ma,mb),varargin{:});
  return;
end
\end{verbatim}
\end{small}
\end{singlespace}

%%%%%%%%%%%%%%%%%%%%%%%%%%%%%%%%%%%%%%%%%%%%%%%%%%%%%%%%%%%%%%%%%%%%%

\section{Pulse sequence code in MATLAB}

The following could be an executable in MATLAB which simulates a pulse
sequence, taking into account the effect of decoherence.

\begin{singlespace}
\begin{small}
\begin{verbatim}
nq=7;
def;

ratio=1;  % model GAD and PD (if ratio was set to a larger number,
          % say 1e9, the simulation would use near infinite T1 and T2.

% set up thermal density matrix
p1=(0.5000)+5e-4; p2=p1/4.7*1.25;
rt1=(1-p1)*eye(2) + (2*p1-1)*[1 0 ;0 0];  % fluorine spins
rt2=(1-p2)*eye(2) + (2*p2-1)*[1 0 ;0 0];  % carbon spins
rit=mykron(rt1,rt1,rt1,rt1,rt1,rt2,rt2);  % 7-spin molecule

... % [here would go some code which produces R starting from rit]

% the remainder does a cnot_52, with partial refocusing of J couplings

time=abs(1/8/J25)

R=rho(R,X(2,1)*Z(2,-1));

R=rho(R,d(time)); 
R=gad7(R,p1,time,ratio); R=pd7(R,time,ratio);
R=rho(R,X(1,2)*X(3,2));

R=rho(R,d(time));
R=gad7(R,p1,time,ratio); R=pd7(R,time,ratio);
R=rho(R,X(1,2)*X(2,2)*X(5,2));

R=rho(R,d(time)); 
R=gad7(R,p1,time,ratio); R=pd7(R,time,ratio);
R=rho(R,X(1,2)*X(3,2));

R=rho(R,d(time));
R=gad7(R,p1,time,ratio); R=pd7(R,time,ratio);
R=rho(R,X(1,2)*X(2,2)*X(5,2));

R=rho(R,Y(2,-1)*X(7,1)*Z(7,1)*Z(5,1));
\end{verbatim}
\end{small}
\end{singlespace}

%%%%%%%%%%%%%%%%%%%%%%%%%%%%%%%%%%%%%%%%%%%%%%%%%%%%%%%%%%%%%%%%%%%%%
%%%%%%%%%%%%%%%%%%%%%%%%%%%%%%%%%%%%%%%%%%%%%%%%%%%%%%%%%%%%%%%%%%%%%

\chapter{Pulse sequence three-spin Grover search}
\label{app:grover3}

We here give the final pulse sequences used in the experiment of
Section~\ref{expt:grover3}. They are taken from the C code submitted
to the spectrometer, with additional comments for clarity.

\begin{singlespace}
\begin{small}
\begin{verbatim}
======================================================================
NOTATION
======================================================================

Yb(1) represents a 1*pi/2 = pi/2 pulse on spin b about the Y axis
mXc(0.5) represents a 0.5*pi/2 = pi/4 pulse on spin c about the 
-X axis, etcetera

AB(), AC() and BC() represent simultaneous pulses on two spins. The 
first two arguments are the tip angle in units of pi/2, and the last 
two arguments are the phase of each pulse.
    Example: AC(0.5,2,PHX,PHmY) performs a pi/4 pulse on spin a about 
             the X axis, and a pi pulse on spin c about the -Y axis

ABC() represents a simultaneous pulse on all three spins. The first 
three arguments are the tip angle in units of pi/2, and the last 
three arguments are the phase of each pulse.

=====================================================================
OUTLINE
=====================================================================

delay(d1);                            /* THERMALIZATION        */

tom3htemplab1();                      /* TEMPORAL LABELING */

ABC(1,1,1,PHY,PHY,PHY);               /* Hadamard on each spin */

  /* the parameter ctype determines which will be the marked element x0
     that will be "found" during the execution of the algorithm */

  pfa=2; pfb=2; pfc=2;                  /* default values */

  if (ctype){                           /* if ctype=0, skip search */

  if (((ctype-1)/4)%2) pfa=0;
  if (((ctype-1)/2)%2) pfb=0;
  if (((ctype-1)/1)%2) pfc=0;

loop(v9,v10);                        /* start loop Grover iterations */
/* function evaluation */
  ABC(pfa,pfb,pfc,PHX,PHX,PHX);      /* depends on ctype */
  tom3hphaseflip4();                 /* flip sign 111 term */
  ABC(pfa,pfb,pfc,PHmX,PHmX,PHmX);   /* depends on ctype */
/* inversion about the average */
  ABC(1,1,1,PHY,PHY,PHY);            /* pi/2 Y pulse on each spin */
  tom3hphaseflip4();                 /* flip sign 111 term */
  ABC(1,1,1,PHmY,PHmY,PHmY);         /* pi/2 -Y pulse on each spin */
endloop(v10);

  }    /* close if (ctype) */

tomoPULSE;                            /* TOMOGRAPHY PULSES          */

=====================================================================
tom3hphaseflip4() - this function implements a diagonal unitary 
  operator  with the elements [1 1 1 1 1 1 1 -1] on the diagonal
=====================================================================

Yb(2); delay(1/8/Jbc); Xa(1); Ya(0.5); Xa(1); delay(1/8/Jbc); 
Yb(1); Xb(0.5); delay(1/4/Jab); mYc(2); delay(1/4/Jab); Yb(1); mXb(1);
delay(1/8/Jbc); Ya(2); delay(1/8/Jbc);
mXb(1); Yb(0.5); delay(1/4/Jab); Yc(1); Xc(0.5); Yc(1); delay(1/4/Jab);
Yb(1); delay(1/8/Jac); mYb(2); delay(1/8/Jac);

=====================================================================
tom3htemplab1()  - this function implements one of several sequences,
  each of which transforms the initial (thermal equilibrium) density 
  matrix into one of the terms of the temporal labeling summation
=====================================================================

  switch (ptype){
  case 0: break;                        /* I */
  case 1:{                              /* cnotab */
    Yb(1); delay(1/4/Jab); Xc(2); delay(1/4/Jab); BC(1,2,PHX,PHX);
    break; }
  case 2:{                              /* cnotac */
    Yc(1); delay(1/4/Jac); Xb(2); delay(1/4/Jac); BC(2,1,PHX,PHX);
    break;  }
  case 3:{                              /* cnotbc, Jbc<0 */
    mYc(1); delay(1/4/Jbc); Xa(2); delay(1/4/Jbc); AC(2,1,PHX,PHX);
    break; }
  case 4:{                              /* cnotba */
    Ya(1); delay(1/4/Jab); Xc(2); delay(1/4/Jab); AC(1,2,PHX,PHX);
    break; }
  case 5:{                              /* cnotca */
    Ya(1); delay(1/4/Jac); Xb(2); delay(1/4/Jac); AB(1,2,PHX,PHX);
    break; }
  case 6:{                              /* cnotcb, Jbc<0 */
    mYb(1); delay(1/4/Jbc); Xa(2); delay(1/4/Jbc); AB(2,1,PHX,PHX);
    break; }
  case 7:{                              /* cnotab.cnotca (time -->) */
    Yb(1); delay(1/4/Jab); Xc(2); delay(1/4/Jab); BC(1,2,PHX,PHX);
    Ya(1); delay(1/4/Jac); Xb(2); delay(1/4/Jac); AB(1,2,PHX,PHX);
    break; }
  }
\end{verbatim}
\end{small}
\end{singlespace}